\documentclass[10pt, a4]{article}
\usepackage{amsthm,amsfonts,amssymb,euscript,amsmath,comment,slashed}

\usepackage[affil-it]{authblk}
\usepackage{graphicx}
\usepackage{mathrsfs}
\usepackage{marvosym}
\usepackage{eufrak}
\usepackage[margin=1in,dvips]{geometry}
\usepackage{color}
\usepackage{placeins}
\usepackage{comment}
{
\vskip.5pc}

\newtheorem{proposition}{Proposition}[section]
\newtheorem{theorem}[proposition]{Theorem}
\newtheorem{bigthe}{Theorem}
\newtheorem{lemma}[proposition]{Lemma}
\newtheorem{remark}[proposition]{Remark}
\newtheorem{definition}[proposition]{Definition}
\newtheorem{corollary}[proposition]{Corollary}

\newtheorem{bigconj}{Conjecture}

\newtheorem*{corolnonum}{Corollary}

\usepackage[margin=1in,dvips]{geometry}
\usepackage{graphicx}
\usepackage{enumerate}

\numberwithin{equation}{section}

\usepackage{xcolor} % A package to add color.

\def\ub {\underline{u}}
\def\th {\theta}
\def\Lb {\underline{L}}
\def\Hb {\underline{H}}
\def\chib {\underline{\chi}}
\def\chih {\hat{\chi}}
\def\chibh {\hat{\underline{\chi}}}
\def\omegab {\underline{\omega}}
\def\etab {\underline{\eta}}
\def\betab {\underline{\beta}}
\def\alphab {\underline{\alpha}}

\def\hot{\widehat{\otimes}}

\def\sigmac{\check{\sigma}}

\def\mub{\underline{\mu}}

\def\a {\alpha}
\def\b {\beta}
\def\ab {\alphab}
\def\bb {\betab}
\def\nab {\slashed{\nabla}}
\def\ep {\epsilon}
\def\om {\omega}
\def\omb {\omegab}
\def\Om {\Omega}
\def\rd {\partial}

\def\tpH {\widetilde{\psi}_H}
\def\tp {\widetilde{\psi}}
\def\tpHb {\widetilde{\psi}_{\Hb}}

\def\tg {\widetilde{g}}
\def\tb {\widetilde{b}}
\def\Er {\mathcal E}
\def\de {\delta}
\def\trchb {\slashed{\mbox{tr}}\chib}
\def\trch{\slashed{\mbox{tr}}\chi}
\def\ttrch {\widetilde{\slashed{\mbox{tr}}\chi}}
\def\ttrchb {\widetilde{\slashed{\mbox{tr}}\chib}}
\def\tK {\widetilde{K}}
\newcommand{\ombs}{\underline{\omega} \mkern-13mu /\,}
\newcommand{\tombs}{\widetilde{\underline{\omega}} \mkern-13mu /\,}
\def\f {\frac}
\def\i {\infty}
\def\Ke {\mathcal K}
\def\NS {\mathcal N_{sph}}
\def\NH {\mathcal N_{hyp}}
\def\NI {\mathcal N_{int}}

\def\H {\mathcal H^+}
\def\CH {\mathcal C\mathcal H^+}
\def\alp {\alpha}

\newcommand{\bea}{\begin{eqnarray}}
\newcommand{\eea}{\end{eqnarray}}
\def\beaa{\begin{eqnarray*}}
\def\eeaa{\end{eqnarray*}}

\DeclareMathOperator*{\esssup}{ess\,sup}

\newcommand{\eqs}{=_{\scalebox{.6}{\mbox{S}}}}
\newcommand{\eqrs}{=_{\scalebox{.6}{\mbox{RS}}}}

\renewcommand{\div}{\slashed{\mbox{div }}}
\newcommand{\curl}{\slashed{\mbox{curl }}}

\newcommand{\tr}{\slashed{\mbox{tr }}}

\def\ls{\lesssim}

\newcommand{\Ls}{{\mathcal L} \mkern-10mu /\,}

\title{The interior of dynamical vacuum black holes I: \\ The $C^0$-stability of the Kerr Cauchy horizon}

\author[1]{Mihalis Dafermos\thanks{dafermos@math.princeton.edu}}
\author[2]{Jonathan Luk\thanks{jluk@stanford.edu}}
\affil[1]{\small  Department of Mathematics, Princeton University, Washington~Road,~Princeton~NJ~08544,~United~States~of~America \vskip.1pc \ }
\affil[1]{\small  Department of Pure Mathematics and Mathematical Statistics, University~of~Cambridge,~Wilberforce~Road,~Cambridge CB3 0WB, United Kingdom \vskip.1pc \ }
\affil[2]{\small  Department of Mathematics, Stanford University, 450~Serra~Mall~Building~380,~Stanford~CA~94305-2125,~United~States~of~America \ }

\begin{document}

\maketitle
\vspace{-5ex}
\begin{center}{\it\large Dedicated to our teacher and mentor Demetrios Christodoulou}
\end{center}
\vspace{1ex}

\begin{abstract}
We initiate a series of works where we study
the interior of dynamical rotating vacuum black holes without symmetry.
In the present paper, we take up the problem starting from
appropriate Cauchy data for the Einstein vacuum equations
defined on a hypersurface already within the black hole interior, representing
the expected geometry just inside the event horizon.
We prove that for all such data,
the maximal Cauchy evolution
can be extended across a non-trivial piece of Cauchy horizon
as a Lorentzian manifold with continuous metric. In subsequent
work, we will retrieve our assumptions on data 
assuming only that the black hole event horizon geometry suitably
asymptotes  to a rotating
Kerr solution.
In particular, 
\emph{if the exterior region of the Kerr family is proven to be dynamically stable}---as is widely expected---then it will follow that
 the  $C^0$-inextendibility formulation
of Penrose's celebrated strong cosmic censorship conjecture 
is in fact \emph{false}.
The proof  suggests, however, that the 
$C^0$-metric Cauchy horizons thus arising are generically singular in an essential way, representing so-called
``weak null singularities'',
and thus that   a revised version of strong
cosmic censorship holds.
\end{abstract}

\tableofcontents
\section{Introduction}
The two-parameter family of \emph{Kerr spacetimes} $(\mathcal{M}, g_{a,M})$~\cite{Kerr} constitutes the most celebrated  class of    explicit
solutions to the \emph{Einstein vacuum equations}
\begin{equation}
\label{INTROvace}
{\rm Ric}(g)=0,
\end{equation}
the governing equations of general relativity~\cite{Einstein1915} 
in the absence of matter. The family
includes the
\emph{Schwarzschild solutions}~\cite{schwarzschild1916} 
as the non-rotating one-parameter  $a=0$ subcase.
In general, when the rotation parameter is strictly less than the mass,
i.e.~$|a|<M$, the spacetimes $(\mathcal{M}, g_{a,M})$ contain
a \emph{black  hole} region~\cite{boyer1967maximal, carter1968global}, 
bounded in the past by a null hypersurface 
$\mathcal{H}^+$ known as the \emph{event horizon}.
The Kerr stationary \emph{exterior region}, i.e.~the complement of the black hole, 
is conjectured (see e.g.~\cite{vacuumscatter}) to be dynamically
stable as a solution of $(\ref{INTROvace})$, under small perturbation of initial data,
up to modulation of the parameters. 
The Kerr exterior geometry is moreover expected  to represent
the  endstate of a wide variety of complicated real-world
astrophysical systems which undergo gravitational collapse~\cite{Penroseunsolv}.

In view of the central role of the Kerr exterior for our current world picture, it
 is disturbing to recall that the Kerr black hole region itself, in the strictly rotating case 
$a\ne0$, hides in its \emph{interior} one of the
most  difficult conceptual puzzles of 
general relativity:~the dynamical breakdown of classical ``Laplacian'' determinism for 
all observers
entering the black hole.
This breakdown is manifested by the presence of a so-called \emph{Cauchy horizon}
$\mathcal{CH}^+$ (see~\cite{hawking1967occurrence})
beyond which Cauchy data no longer uniquely determine the solution.
The unsettling nature of this state of affairs has given rise to               a conjecture, \emph{strong cosmic censorship}, first formulated by Penrose~\cite{PenroseSCCrefer},
which postulates that the above puzzling behaviour of Kerr is in fact non-generic.
This  conjecture
would  
imply in particular that the interior region of Kerr must be in some way
unstable (under small perturbation of initial data) in the vicinity
of its Cauchy horizon $\mathcal{CH}^+$. 
The original formulation of the conjecture was modelled on the prototype
of the exceptional non-rotating Schwarzschild case $a=0$, where no Cauchy horizon
is present, but rather, the spacetime can be viewed as terminating at
a ``spacelike singularity'' across which the metric cannot be extended, even
merely continuously~\cite{SbierskiCzero}. 
The singular behaviour of Schwarzschild, 
though fatal for reckless observers entering the black hole, can be thought of as epistemologically preferable
for general relativity as a theory, since this ensures that the future, however bleak, is indeed determined.

\begin{figure}
\centering{
\def\svgwidth{12pc}
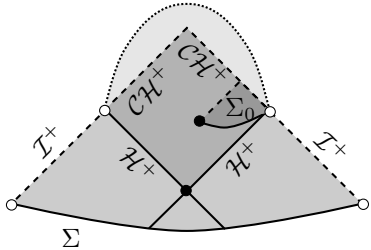}
\caption{The stability of the Cauchy horizon  $\mathcal{CH}^+$ from data on $\Sigma_0$}\label{maintheorfigure}
\end{figure}

We here initiate
a series of works to
address the stability of the Kerr Cauchy horizon for vacuum spacetimes
governed by $(\ref{INTROvace})$, with no symmetry assumed,
and more generally, the question  of the nature of
the boundary of spacetime in dynamical black hole interiors.
The present paper will take up the problem starting from general spacelike initial data
which are meant to represent a hypersurface $\Sigma_0$
already in a black hole interior but close to the event horizon $\mathcal{H}^+$ in a suitable
sense.
For the evolution of such data, 
the $C^0$-stability of a piece of the Kerr Cauchy horizon
$\mathcal{CH}^+$ will be obtained, contrary to the original expectations
motivated by the Schwarzschild picture. See Figure~\ref{maintheorfigure}.
In a subsequent paper~\cite{DafLuk2}, the data considered here  will be
shown to arise from general  characteristic initial data posed on two intersecting
null hypersurfaces, one of which
is meant to represent the event horizon $\mathcal{H}^+$ 
of a dynamical black hole whose
geometry approaches Kerr. In a third paper~\cite{DafLuk3}, 
the case of double null data $\mathcal{H}^+_1\cup\mathcal{H}^+_2$
\emph{both} hypersurfaces $\mathcal{H}^+_i$ ($i=1,2$)
of which are future complete,
globally close to, and asymptote to two nearby Kerr solutions will be considered,
and the global $C^0$-stability of the entire bifurcate  Cauchy horizon of two-ended Kerr
will be obtained. 
In particular, the arising spacetimes will have no spacelike singularities.

Taken together with a future positive resolution of the stability of the exterior of Kerr conjecture, 
{\bf \emph{our results in particular
falsify the $C^0$ formulation of strong cosmic censorship, as well as
the original  expectation that spacetime generically terminates at a spacelike singularity.}}

The  picture  established here of $C^0$-stable Cauchy horizons  was in fact
 first suggested from 
the detailed study of a series of spherically symmetric 
toy models~\cite{Hiscock, PI1, A-Ori, BS} culminating in the Einstein--Maxwell--scalar
field system, for which the analogous result  was 
proven in~\cite{D1, D2, D3}. 
For this latter model,  it has been proven~\cite{LukOh2017one, LukOh2017two} 
moreover 
that, under suitable genericity conditions, the Cauchy horizons that
persist inside black holes are singular
in an essential sense, 
and in particular, that the $C^0$ extensions of the metric cannot be $C^2$, in
fact, the results strongly suggest that they
cannot have locally square integrable Christoffel symbols (and thus cannot be interpreted
even as weak solutions of the Einstein vacuum equations $(\ref{INTROvace})$).
Though in this paper we shall not prove analogues of these \emph{instability}
results in our non-symmetric setting, we note that
our theorem is compatible with the statement
that for generic initial data for $(\ref{INTROvace})$, the $C^0$-stable Cauchy horizons
$\mathcal{CH}^+$ will at the same time
indeed be singular  from the point of view of higher regularity, 
in precisely the sense suggested by
the vacuum ``weak null singularities'' recently constructed
in~\cite{LukWeakNull}.
 (In fact, the expectation that these Cauchy horizons are indeed 
generically singular in an essential way
is one of the fundamental difficulties of our stability proof, and the methods
of~\cite{LukWeakNull} will thus play here a paramount role.)
For earlier heuristic studies of the relevance of the spherically
symmetric toy models for vacuum dynamics without symmetry, 
see~\cite{brady1995nonlinear, Ori1997, FO}.  Thus,
{\bf \emph{our work suggests that an alternative, well-motivated formulation
of strong cosmic censorship, originally due to Christodoulou~\cite{Chr}, could still hold.}}

In the remainder of this introduction, we flesh out the above comments with a more detailed
discussion,
starting in {\bf Section~\ref{SCHintro}}
with a review of the Schwarzschild case 
and its spacelike singularity.
We will then turn in {\bf Section~\ref{KCHintro}}
to the rotating Kerr case and its smooth Cauchy horizon in the black hole interior, 
formulating both the conjectured stability of the exterior ({\bf Conjecture~\ref{stabofkerrconj}}) and 
$C^0$ formulation of strong cosmic censorship ({\bf Conjecture~\ref{C0formSCC}}), which
would imply instability of the interior.
With this background we will present in {\bf Section~\ref{MTtnseisag}} 
 a first version of the main theorem of the present paper on
 the $C^0$ stability of the Cauchy horizon ({\bf Theorem~\ref{PRWTO}}), 
 as well as the upcoming results of~\cite{DafLuk2,DafLuk3} 
 ({\bf Theorems~\ref{DEUTERO}} and~{\bf \ref{TRITO}}).
 In particular, these theorems together with a positive resolution of Conjecture~\ref{stabofkerrconj} imply that
 Conjecture~\ref{C0formSCC} is in fact false.
 In {\bf Section~\ref{INTROprework}},
 we will put these theorems in the context of relevant previous work,
 motivating the revised  formulation of strong cosmic censorship ({\bf Conjecture~\ref{ChrSCC}}).
 (For  a more detailed discussion of previous work, see our survey~\cite{JonathanMesurvey}.)
Finally, in {\bf Section~\ref{firstremarksintro}} we will give a first indication of the proof and a guide
to the rest of the paper.

\subsection{The Schwarzschild $a=0$ case}
\label{SCHintro}
Any discussion of black holes in general relativity must begin with the simplest
example, that of \emph{Schwarzschild}~\cite{schwarzschild1916}. 
This vacuum solution, discovered already in 1915, sits
in the larger Kerr family $(\mathcal{M},g_{a,M})$ (to be reviewed in Section~\ref{KCHintro} below)
as the special one-parameter subfamily $(\mathcal{M},g_{0,M})$ 
corresponding to $a=0$. The early discovery of Schwarzschild
was made possible by its high degree of
symmetry. The metric is both stationary (in fact, static) and spherically symmetric, as is most
easily seen from the following explicit form in a coordinate patch:
\begin{equation}
\label{Schform}
g_{0,M} = -(1-2M/r)dt^2+(1-2M/r)^{-1}dr^2+r^2(d\theta^2+\sin^2\theta \, d\phi^2).
\end{equation}
In fact, according to Birkhoff's theorem~\cite{jebsen1921general}, all spherically symmetric
solutions of $(\ref{INTROvace})$ can be written locally in the form 
$(\ref{Schform})$ for some parameter $M\in \mathbb R$. In what follows,
we shall only consider the case $M>0$, whose global geometry will be seen
in Section~\ref{BHRandEH}
to indeed contain a black hole region.

\begin{figure}
\centering{
\def\svgwidth{12pc}
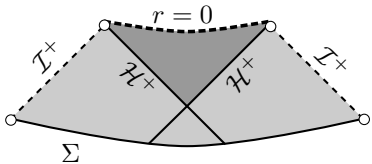}
\caption{Schwarzschild as the maximal future Cauchy development of $\Sigma$}\label{Schwfigure}
\end{figure}

The modern interpretation of ``the Schwarzschild spacetime'' $(\mathcal{M},g_{0,M})$
views it as the maximal Cauchy development (see~\cite{geroch})
of vacuum initial data prescribed on an appropriate two-ended
asymptotically flat hypersurface $\Sigma$.\footnote{This happens to coincide
with the so-called ``maximal analytically extended'' Schwarzschild 
spacetime~\cite{kruskal, synge}, but
it is essential here to think of it in terms of the language
of the global initial value problem of~\cite{geroch} as a maximal \emph{Cauchy
development}. See the discussion in~\cite{JonathanMesurvey}.
The distinction will be fundamental
when we turn to Kerr in Section~\ref{KCHintro}.} 
 The spacetime  is best
illustrated by the  \emph{Penrose diagram}
of Figure~\ref{Schwfigure}, which depicts---as a bounded subset
of the plane---the range of a global double null coordinate system $U$, $\underline{U}$. 
We depict in fact 
only  the maximal \emph{future} Cauchy development of $\Sigma$.\footnote{The specific choice of $\Sigma$, so that the bifurcation sphere (see Section~\ref{BHRandEH} below)
lies in its future
has simply been made for greater generality.}

Astrophysical black holes which arise in gravitational collapse do \emph{not} have two ends!
The potential physical relevance of the extended Schwarzschild metric was first exhibited by its connection with
the Oppenheimer--Snyder model~\cite{oppenheimer1939continued} 
of gravitational collapse of a dust surrounded by vacuum. The latter spacetimes
contain a subset of $(\mathcal{M},g_{0,M})$ which can be thought of as the 
maximal future Cauchy development of the hypersurface-with boundary $\Sigma'$ depicted in Figure~\ref{localSchwarz}. All our statements can be reinterpreted so as to refer only to this
region.
Nonetheless, it is convenient in the meantime for definiteness
to base our discussion on  $(\mathcal{M},g_{0,M})$ as we have defined it.
With reference to Figure~\ref{Schwfigure}, we may now 
 review in Sections~\ref{CI+WCC}--\ref{C0inex} 
 the features of Schwarzschild geometry relevant for us.

\subsubsection{Completeness of $\mathcal{I}^+$ and ``weak cosmic censorship''}
\label{CI+WCC}
The asymptotic boundary $\mathcal{I}^+$ is known as \emph{future null infinity}
and is an idealisation of far away observers in the ``radiation zone''. Note that as $\Sigma$
has two asymptotically flat ends, in our convention $\mathcal{I}^+$ has
two connected components. 
A fundamental property that $\mathcal{I}^+$ enjoys in Schwarzschild 
is that of \emph{completeness}.
(A naive concrete realisation of this is the statement that
the spacetime in a neighbourhood of each component of $\mathcal{I}^+$ can be covered by
coordinates $r, u, \theta, \phi$, related to $(\ref{Schform})$ by 
$u=t+r^*(r)$, with $r^*$ defined
by the equation $\frac{dr^*}{dr}= (1-2M/r)$, and where $u$ takes on all
positive and negative values. See~\cite{Geroch:1978ub} where this issue
was first discussed. For a general definition of what it means
for an asymptotically flat vacuum 
spacetime to ``possess a complete future null infinity $\mathcal{I}^+$'',
see~\cite{Chrmil}.)
The
completeness of $\mathcal{I}^+$
 has then the interpretation that ``far-away observers'' can observe for all retarded 
 time $u$. In view of the properties to be discussed in Sections~\ref{BHRandEH}--\ref{C0inex} below, this completeness is quite
fortuitous!

The conjecture that ``for solutions of $(\ref{INTROvace})$ arising from
\emph{generic} asymptotically flat initial data, null infinity~$\mathcal{I}^+$ is complete'' is
the modern formulation of what is known  as \emph{weak cosmic censorship}.
See~\cite{Chrmil}.\footnote{The original informal statement
of this conjecture, due to Penrose~\cite{Penroseergo}, 
is that ``singularities are cloaked by event horizons''. We will
discuss the notion of event horizon in Section~\ref{BHRandEH}.  The formulation in terms
of
completeness of future null infinity $\mathcal{I}^+$, which is easier
to make precise and already captures the essence of the conjecture, originates
with~\cite{Geroch:1978ub, geroch1979global} and is given a definitive form  in~\cite{Chrmil}. 
See also~\cite{Horowitz1979}.
Note of course that to make sense,
this formulation must be applied to the maximal Cauchy development,
not some non-globally hyperbolic extension thereof, cf.~the example of
negative mass Schwarzschild. See our survey~\cite{JonathanMesurvey}.
The conjecture used to be referred to simply as ``cosmic censorship''.
The term ``weak'' was added to differentiate with \emph{strong cosmic censorship},
discussed in Section~\ref{SCCintro}, although the modern formulations of these
conjectures are
not logically related in the way these adjectives would suggest.}
    The genericity restriction, not present in the original expectation, was motivated from
subsequent work of Christodoulou~\cite{Christodoulou4} on
the coupled Einstein--scalar field equations
\begin{equation}
\label{ESF1}
{Ric}_{\mu\nu}-\frac12 g_{\mu\nu}R =8\pi T_{\mu\nu}\doteq
8\pi (\partial_\mu\psi\partial_\nu \psi-\frac12g_{\mu\nu} \partial^\alpha\psi\partial_\alpha\psi),\qquad \Box_g\psi =0,
\end{equation}
under spherical symmetry. In contrast to the vacuum equations $(\ref{INTROvace})$ which
are constrained by Birkhoff's theorem mentioned above, equations $(\ref{ESF1})$
do admit non-trivial dynamics in spherical symmetry and can be thought
of as one of the simplest toy-models for understanding 
general dynamics of $(\ref{INTROvace})$ 
in a $1+1$ dimensional context, in the astrophysically relevant case of one asymptotically flat end.
For $(\ref{ESF1})$ under spherical symmetry, Christodoulou~\cite{Christodoulou4} proved the analogue of 
weak cosmic censorship, but he also showed 
that the genericity assumption is indeed
\emph{necessary} via explicit examples of ``naked singularities''~\cite{ChrSph2}.
Proving weak cosmic censorship for $(\ref{INTROvace})$ without symmetry assumptions is
a fundamental open problem in classical general relativity with potential
significance for observational astronomy. 
It can be thought of as the ``global existence conjecture'' for the Einstein equations. See~\cite{Chr}
for this terminology. We will contrast this conjecture with \emph{strong cosmic censorship}
in Section~\ref{SCCintro}.

\subsubsection{The black hole region and the event horizon $\mathcal{H}^+$}
\label{BHRandEH}
The
notion of future null infinity $\mathcal{I}^+$ allows us to identify the so-called
 \emph{black hole} region of $(\mathcal{M},g_{0,M})$.
This can be characterized as the complement of the past of $\mathcal{I}^+$, i.e.~$\mathcal{M}\setminus J^-(\mathcal{I}^+)$, and corresponds to 
the darker shaded region of the Penrose diagram of Figure~\ref{Schwfigure}.
The  interior of the black hole  can be represented in the form 
$(\ref{Schform})$ by the range $0<r<2M$, $-\infty<t<\infty$
(together with standard spherical coordinates $\theta$, $\phi$).
The \emph{future event horizon} $\mathcal{H}^+$ (on which $r$ extends smoothly to $2M$
but the coordinates of
$(\ref{Schform})$ break down)
is a bifurcate future complete null hypersurface
separating the black hole region from 
its exterior\footnote{In our convention, with $\Sigma$ as depicted,
the ``exterior'' contains a compact subset of
the so-called white-hole region $\mathcal{M}\setminus J^+(\mathcal{I}^-)$.}. 
Note that the ``stationary'' Killing field $\partial_t$ in the coordinates of
$(\ref{Schform})$ is spacelike
in the interior of the black hole region.

The constant-$(r,t)$ spheres of the interior of the 
black hole region 
are examples of \emph{closed trapped surfaces}~\cite{Penrose.2}, 
i.e.~both future null expansions
are negative. The presence of these trapped surfaces is
related to the geodesic incompleteness property which we will turn to immediately below.

\subsubsection{Trapped surfaces imply geodesic incompleteness}
\label{GeoInc}
Despite being the Cauchy development of complete, asymptotically flat 
initial data, the Schwarzschild
 manifold $(\mathcal{M},g_{0,M})$ is itself \emph{future causally geodesically incomplete}.
More specifically,
timelike geodesics $\gamma$ are future incomplete if and only if they cross $\mathcal{H}^+$
into the black hole region; in particular, observers who stay in the exterior $J^-(\mathcal{I}^+)$ live forever  (cf.~the completeness of $\mathcal{I}^+$ discussed
in Section~\ref{CI+WCC}).

\begin{figure}
\centering{
\def\svgwidth{7pc}
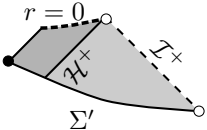}
\caption{The astrophysically relevant maximal future Cauchy development of $\Sigma'$}\label{localSchwarz}
\end{figure}

It was originally speculated that the geodesic incompleteness property
of Schwarzschild was a result of its high degree of symmetry~\cite{RevModPhys.29.432,
lifshitz1963investigations}. This expectation
was spectacularly falsified by
Penrose's celebrated  \emph{incompleteness\footnote{Following~\cite{Chrmil}, we have referred to the result of~\cite{Penrose.2} as an ``incompleteness'' theorem and not a ``singularity'' theorem.
See again the discussion in~\cite{JonathanMesurvey}.
This distinction will be quite clear when the theorem
is applied to Kerr (cf.~Section~\ref{KCH2intro}). 
In the Schwarzschild case,
geodesic incompleteness of $\mathcal{M}$ is indeed related to true ``singular'' behaviour
as will be shown in  Section~\ref{C0inex}.} theorem}~\cite{Penrose.2}.
This theorem states that a globally hyperbolic spacetime with a non-compact
Cauchy hypersurface satisfying ${\rm Ric}(v,v)\ge 0$
for all null vectors $v$ is future causally geodesically incomplete if
it contains a closed trapped surface.  Under additional assumptions of
asymptotic flatness, one can also
infer that necessarily 
$\mathcal{M}\setminus J^-(\mathcal{I}^+)\neq\emptyset$.
Note that by Cauchy stability (see for instance~\cite{Hawking&Ellis}), 
the existence of a closed 
trapped surface in the spacetime is always stable to perturbation of initial data.
Applied to the maximal future Cauchy
development of vacuum initial data sufficiently close to Schwarzschild
data on $\Sigma$, this 
in particular
implies
that the geodesic incompleteness property of Schwarzschild, as well
as the property that $\mathcal{M}\setminus J^-(\mathcal{I}^+)\ne\emptyset$,\footnote{One
cannot, however, infer the completeness of $\mathcal{I}^+$, cf.~Section~\ref{CI+WCC}.
Thus weak cosmic censorship is non-trivial even in a neighbourhood of Schwarzschild.
See also the statement of Conjecture~\ref{stabofkerrconj}.}
is stable to perturbation of
initial data on $\Sigma$.\footnote{So as not to refer to the two-ended case,
note that we could also apply Penrose's theorem to the maximal Cauchy
development of small perturbations of complete, one-ended initial data to
a suitable Einstein--matter system which 
are vacuum when restricted to the subset $\Sigma'$
depicted in Figure~\ref{localSchwarz}. In this way,
the theorem applies to  small perturbations
away from spherical symmetry
of the collapsing dust spacetimes of Oppenheimer--Snyder~\cite{oppenheimer1939continued}.
The theorem can also be non-trivially applied to the  set of solutions 
of Christodoulou's model~\cite{Christodoulou4} 
which are proven to dynamically form a trapped surface, or, the recent 
dynamically forming trapped surfaces in vacuum~\cite{Chr}.}

\subsubsection{$C^0$-inextendibility}
\label{C0inex}
What fate awaits incomplete observers $\gamma$ in the Schwarzschild spacetime?
One easily shows that such 
$\gamma$  
asymptote (in finite proper time)
to what can be considered as a spacelike\footnote{We will discuss in what sense
this boundary can be viewed as spacelike in Section~\ref{space?}.}
 ``singular'' boundary of
spacetime (``$r=0$''), 
where the curvature (in fact the Kretschmann scalar) blows up asymptotically~\cite{Hawking&Ellis}.
This latter property is readily computable from 
the form~$(\ref{Schform})$.
It follows that the manifold $(\mathcal{M},g_{0,M})$ is \emph{inextendible
as a Lorentzian manifold with $C^2$ metric}, i.e.~there
does not exist a connected $4$-dimensional Lorentzian manifold
 $(\widetilde{\mathcal{M}}, \widetilde{g})$ with $C^2$ metric $\widetilde{g}$
 and an isometric embedding $i:\mathcal{M}\to \widetilde{\mathcal{M}}$ such that
 $i(\mathcal{M})\ne\widetilde{\mathcal{M}}$.

Seen from the PDE perspective, it is not clear that
the mere failure of the metric to be $C^2$ is in itself sufficient to justify interpreting
spacetime as having ended at an essential, ``terminal'' singularity.\footnote{For an early discussion
of this issue, see the  comments
in Section~8.4 of~\cite{Hawking&Ellis}.}
Indeed, the Einstein equations can be shown~\cite{L21} 
to be well-posed for general initial data
with curvature only in $L^2$. Moreover, explicit
solutions with $\delta$-function singularities
in curvature  on null hypersurfaces  have been considered in
the physics literature~\cite{Penrose65}, and recently a general well posedness
theorem has been proven allowing data with such singular fronts~\cite{LR}.
It turns out, however, that one can say something much stronger
about the singular behaviour of Schwarzschild.
Physically speaking, incomplete observers $\gamma$ not only measure infinite curvature but
are \emph{torn apart} by infinite tidal deformations as they
approach $r=0$. A related more natural statement is 
 that the spacetime $(\mathcal{M},g)$
is future inextendible as a Lorentzian manifold, in the
sense above, \emph{but now 
with metric only assumed to be continuous}. 
This $C^0$-inextendibility property of Schwarzschild, stated already in~\cite{Hawking&Ellis},
has only recently been proven by Sbierski~\cite{SbierskiCzero}.

The prospect of observers 
being torn apart by infinite tidal deformations may at first seem a rather unpleasant
feature of the Schwarzschild solution. We will see it in a
different light, however, after examining the situation in the rotating case $a\ne 0$,
where it is precisely the \emph{absence} of this feature which turns out to be  problematic!
The Schwarzschild $C^0$-inextendibility property  will then serve as a model for the
formulation of strong cosmic censorship reviewed in Section~\ref{SCCintro}.

\subsection{The rotating $a\ne 0$ Kerr case}
\label{KCHintro}
The larger two-parameter Kerr family of metrics $g_{a,M}$ in which the one-parameter
Schwarzschild subfamily sits was only discovered in 1963~\cite{Kerr}. These metrics retain
a stationary Killing field $\partial_t$ but, for $a\ne 0$,
only one of the generators of spherical symmetry
$\partial_\phi$;
they are thus stationary and axisymmetric.
We
refer the reader to~\eqref{BL} in 
Appendix~\ref{sec.Kerr.geometry} for the form of the Kerr metric
in local coordinates 
 and to~\cite{o'neill}
for  a leisurely discussion of its geometry. We will only
consider the subextremal case $0\le |a|<M$.\footnote{We note that the dynamics near the extremal
case $|a|=M$ are expected to be complicated, see for instance~\cite{Aretakis2}.
We will make some comments regarding extremality at various points further on.}

\begin{figure}
\centering{
\def\svgwidth{12pc}
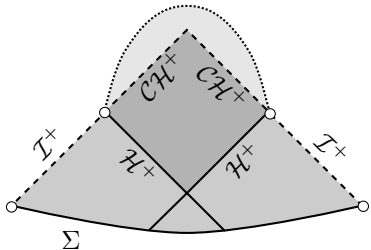}
\caption{Kerr as the maximal future Cauchy development of $\Sigma$, with a non-unique
extension}\label{KerrPD}
\end{figure}

In our convention, ``the Kerr spacetime'' $(\mathcal{M},g_{a,M})$ will be 
the maximal Cauchy development of two-ended asymptotically flat initial data
$\Sigma$. This is a \emph{proper subset} of the maximal analytic extension $\widetilde{\mathcal{M}}$ 
of~\cite{boyer1967maximal, carter1968global}.
Restricting again to the future of $\Sigma$, we may illustrate the Kerr 
spacetime $(\mathcal{M}, g_{a, M})$
in the  rotating case $a\ne 0$  as the two darker shaded regions of 
 the  Penrose diagram of Figure~\ref{KerrPD}.
% \[
%\input{withjonathan1.pstex_t}
%\]
Concretely, as in Schwarzschild, 
we are  simply depicting above the range of a global bounded 
double null coordinate system $U$, $\underline{U}$.
(The existence of such a global coordinate system is now non-trivial,
but was indeed shown by Pretorius--Israel~\cite{Pretorius}. This type of coordinate
system will form the basis of our work in Section~\ref{secsetup}.)
The significance of the lightest shaded region---\emph{not} part
of $(\mathcal{M},g)$ in our convention but part of  $(\widetilde{\mathcal{M}},\widetilde{g})$---will be discussed in 
Section~\ref{KCH2intro} below.  The existence of such a smooth extension when $a\ne0$
 is in complete
contrast with the Schwarzschild case.

Before turning to the Kerr black hole interior and its extendibility properties,
let us quickly remark some fundamental
 similarities with the $a=0$ case.
 The Kerr manifold $(\mathcal{M},g_{a,M})$ in the subextremal range $|a|<M$
shares with Schwarzschild all the  properties described in 
Sections~\ref{CI+WCC}--\ref{GeoInc}, in particular,~it has a complete
null infinity $\mathcal{I}^+$, a nontrivial black hole region
$\mathcal{M}\setminus J^-(\mathcal{I}^+)$, and it is 
 geodesically incomplete.
As with Schwarzschild, timelike geodesics in $(\mathcal{M},g_{a, M})$ are future 
incomplete if and only if they cross $\mathcal{H}^+$. 

\begin{figure}
\centering{
\def\svgwidth{6.5pc}
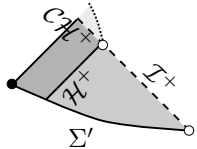}
\caption{The astrophysically relevant  maximal future Cauchy development of $\Sigma'$}\label{KerrphysrelevPD}
\end{figure}

Though basing our discussion on the Cauchy development
$(\mathcal{M},g_{a, M})$ of a two-ended hypersurface $\Sigma$, 
just as in the Schwarzschild case, the potentially 
 astrophysically relevant regime is only the future Cauchy
 development of $\Sigma'$ depicted now in Figure~\ref{KerrphysrelevPD}.
The main result of the present paper, Theorem~\ref{PRWTO}, as well as our upcoming
Theorem~\ref{DEUTERO}, has an interpretation in terms of this region alone,
and it is only the considerations related to Theorem~\ref{TRITO} which are intimately connected
to the unphysical two-ended picture specifically.
We will return to this issue at various points in what follows.

The  Kerr family of spacetimes should not be viewed as representing 
merely a particular class of explicit solutions  generalising
 Schwarzschild. The  family can be uniquely characterized by
certain properties among stationary solutions and,
in view of this, Kerr is conjectured to play a  central role for
 the dynamics of general solutions of $(\ref{INTROvace})$ \emph{without symmetry}.
 Thus, 
 various properties of Kerr---if stable!---would have profound significance for general relativity.
We begin thus with a review of Kerr's role in the dynamics of $(\ref{INTROvace})$.
 \FloatBarrier

\subsubsection{Stationary uniqueness and the stability of the exterior}
\label{edwgiaeusta9eia}
\begin{figure}
\centering{
\def\svgwidth{12pc}
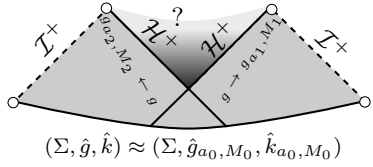}
\caption{The conjectured stability of the Kerr exterior from two-ended data}\label{stabofKerrfig}
\end{figure}

Let us  first remark that the sub-extremal Kerr family
is known to exhaust all
\emph{stationary}, \emph{axisymmetric} vacuum exterior spacetimes bounded by
a connected non-degenerate event
horizon $\mathcal{H}^+$~\cite{carter1971axisymmetric, robinson1975uniqueness}.
Moreover, axisymmetry follows from stationarity in the case where the metric
is close to Kerr, thus showing
that the Kerr family is at the very least isolated in the family of 
stationary, not necessarily axisymmetric,
solutions~\cite{Alexakis3}.
This suggests that it is natural to conjecture the 
asymptotic stability of the Kerr exterior region under evolution by $(\ref{INTROvace})$.
This is a celebrated conjecture in classical general relativity 
(see e.g.~\cite{vacuumscatter}
for background):

\begin{bigconj}(Nonlinear stability of the Kerr exterior)
\label{stabofkerrconj}
Let $(\Sigma, \hat{g}, \hat{k})$  denote an initial data set for $(\ref{INTROvace})$ 
suitably close
to the induced data of a subextremal Kerr solution with parameters $0\le |a_0|< M_0$. 
Then the maximal Cauchy development 
\begin{itemize}
\item[(a)]
will possess a complete null infinity $\mathcal{I}^+$ (cf.~weak cosmic censorship
discussed in Section~\ref{CI+WCC}),
\item[(b)]
will have
a black hole exterior region ${J}^-(\mathcal{I}^+)$ 
bounded to the future by a smooth future affine complete horizon $\mathcal{H}^+$ 
such that the geometry remains close to $g_{a_0, M_0}$ in $J^-(\mathcal{I}^+)$  (See
Figure~\ref{stabofKerrfig})
and 
\item[(c)]
will dynamically approach another Kerr metric with nearby parameters
$0\le |a_i|<M_i$, i=$1$, $2$, in each of the exterior parts $J^-(\mathcal{I}^+_i)$ corresponding
to the two components of null infinity ($\mathcal{I}^+=\mathcal{I}^+_1\cup\mathcal{I}^+_2$),
in particular,  along $\mathcal{H}^+_i$, at an inverse polynomial rate.
For generic initial data, $a_i\ne 0$, for $i=1,2$.
\end{itemize}
\end{bigconj}

Thus, according to the above, for small perturbations of Kerr,
 not only is null infinity $\mathcal{I}^+$ still complete (i.e.~``weak cosmic censorship''
 is true restricted to a neighbourhood of Kerr), 
 but
the Kerr form of the metric is (asymptotically)
stable in the exterior region $J^-(\mathcal{I}^+)$, up to modulation of parameters. 
We emphasise that the conjecture makes no statement about the structure
of the black hole \emph{interior}, i.e.~the non-empty complement
$\mathcal{M}\setminus J^-(\mathcal{I}^+)$!
 One should compare with the celebrated nonlinear stability of
Minkowski space ${\mathbb R}^{3+1}$, proven by Christodoulou--Klainerman~\cite{CK}:
There, 
it is shown  that for spacetimes $(\mathcal{M},g)$ arising from small
perturbations of trivial initial data, then (a) $\mathcal{I}^+$ is complete,  (b)
$\mathcal{M}=J^-(\mathcal{I}^+)$, i.e.~there is \emph{no} black hole region and the
spacetime remains everywhere close to $\mathbb R^{3+1}$ and (c)
the spacetime asymptotically again settles down to $\mathbb R^{3+1}$ at
an at least inverse polynomial rate.
Note that in the Kerr case,
the statement that generically $a_i\ne 0$, $i=1,2$,
simply reflects the 
expectation that the set of initial data for which the final parameter
$a_i$ takes any particular value will be of finite codimension in the set of all data.

\begin{figure}
\centering{
\def\svgwidth{7pc}
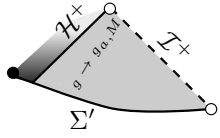}
\caption{The  stability of the Kerr exterior localised to the one-ended case}\label{stabofKerrlocalfig}
\end{figure}

We have stated  Conjecture~\ref{stabofkerrconj} for clarity in terms of two-ended initial data. We remark that,
in view of a trivial domain of dependence argument,
the full content of Conjecture~\ref{stabofkerrconj} would be captured by  statements
(a), (b) and (c) restricted to the Cauchy development of $\Sigma'$, where
$\mathcal{I}^+$ has now only one connected component. The situation
is depicted in Figure~\ref{stabofKerrlocalfig}. It is the assumption that the boundary
of $\Sigma'$ is trapped which would ensure that for sufficiently small perturbations
of Kerr data, the new event horizon $\mathcal{H}^+$ again lies in the domain
of dependence of $\Sigma'$.

\FloatBarrier

It is well known that
the assumption of closeness to Kerr in the stationary
uniqueness theorem of~\cite{Alexakis3} 
can be dropped
if one introduces  the \emph{a priori} assumption that
the  metric is real analytic~\cite{Hawking&Ellis}. Indeed, this
was the original formulation of the ``no-hair'' theorem.
This analyticity assumption is artificial however because  stationarity does not guarantee that
the Killing vector $\partial_t$ is timelike everywhere in the exterior, and it is only
in the latter case where analyticity follows from the ellipticity of the reduced equations.
(Indeed, in Kerr itself with $a\ne0$, there is a region \emph{outside} the black and
white holes where $\partial_t$ becomes spacelike. This is the so-called \emph{ergoregion}.)
If the assumption of analyticity can be replaced by mere smoothness, 
then the resulting uniqueness property  would suggest
the more ambitious conjecture
 that vacuum spacetimes arising from generic asymptotically
 flat Cauchy data (with one end, and not necessarily initially close to Kerr)
 will either disperse or
settle down to finitely many rotating Kerr black holes moving away from each other.
See~\cite{Penroseunsolv}.
According to this then, not only would the Kerr exterior be a stable endstate, but it would be
 \emph{the generic
endstate} of all non-dispersing vacuum solutions.

\subsubsection{The Cauchy horizon $\mathcal{CH}^+$ and the breakdown of determinism}
\label{KCH2intro}
We now turn to the issue of breakdown of determinism in Kerr. As we have remarked already,
 the spacetime $(\mathcal{M},g_{a,M})$  is
  \emph{smoothly extendible} to a larger 
Lorentzian manifold $(\widetilde{\mathcal{M}}, \widetilde{g})$, again
satisfying the Einstein vacuum equations $(\ref{INTROvace})$, 
such that
the boundary $\mathcal{CH}^+$ of $\mathcal{M}$ in $\widetilde{\mathcal{M}}$ 
is a bifurcate null hypersurface. 
Seen from the point of view of the global initial value problem~\cite{geroch}, 
however, these extensions are severely non-unique!

What has happened in the above example is a breakdown of \emph{global hyperbolicity},
the notion first introduced by J.~Leray~\cite{Leray} to characterize (for a general class
of hyperbolic
equations) the region
where the domain of dependence property      allows for proofs
of uniqueness. The 
hypersurface $\Sigma$ is no longer a Cauchy hypersurface
for $\widetilde{\mathcal{M}}$, and this is why data on $\Sigma$ no longer
uniquely determine the solution in $\widetilde{\mathcal{M}}\setminus \mathcal{M}$,
cf.~the lightest
shaded region depicted in  the  Penrose diagram of Figure~\ref{KerrPD}. 
The null boundary $\mathcal{CH}^+$
is thus known as a \emph{Cauchy horizon}. This terminology is originally
due to Hawking~\cite{hawking1967occurrence}.

Kerr in fact exhibits an especially spectacular manifestation of breakdown of
determinism in that predictability fails without any 
local observer directly measuring that the classical
regime has been exited. All incomplete
observers $\gamma$ of the original spacetime cross $\mathcal{CH}^+$ into the extension
$\widetilde{\mathcal{M}}\setminus\mathcal{M}$.\footnote{This is in contrast with the
Cauchy horizons associated with Christodoulou's~\cite{ChrSph2} naked singularity solutions of~$(\ref{ESF1})$. See
the discussion in Section~\ref{SCCintro}.}
In particular,  it is this that justifies our convention, following~\cite{Chrmil}, 
in referring to
 Penrose's  theorem of~\cite{Penrose.2} as an \emph{incompleteness} theorem, and not as a
 \emph{singularity} theorem.\footnote{The situation is confused by the fact that the maximal
 analytic extension $(\widetilde{\mathcal{M}},\tilde{g})$ of~\cite{carter1968global} 
 happens to itself be
 incomplete, on account of what can be viewed as ``timelike singularities''. There
 have been several attempts in the literature to formulate theorems which would imply that
 \emph{any} ``reasonable'' extension of $\mathcal{M}$, \underline{not} necessarily the
  analytic  
 $\widetilde{\mathcal{M}}$, would still have to be incomplete. Applying this to some
 maximal non-globally hyperbolic extension 
 one could infer that there must be  local obstructions
 to extension which could be identified as ``singularities''. Indeed, the paper~\cite{hawking1967occurrence}, where the terminology ``Cauchy horizon'' was introduced,
 proves precisely such a theorem, followed by~\cite{hawking1970singularities}. In view
 of their many assumptions, these 
 theorems are not, however, very satisfactory. In retrospect, it is precisely
 identifying the significance of the
 failure of determinism connected to Cauchy horizons which appears to be the more
 fundamental legacy of~\cite{hawking1967occurrence}, rather than 
 successfully restoring the ``inevitability'' of singularity.}
 
A tempting escape from the epistemological problem of the above breakdown of determinism
associated to  the Kerr Cauchy horizon is 
to appeal to the fact that it is an issue which after
all only concerns observers falling into the black hole. 
Recall from above that inextendible timelike geodesics $\gamma$ of Kerr
are future incomplete if and only if
they enter  $\mathcal{M}\setminus J^-(\mathcal{I}^+)$. The entire future is in particular 
still determined
for observers ``at infinity'', 
in the sense that $\mathcal{I}^+$  is itself complete. 
 Moreover, if ``weak cosmic censorship'' of Section~\ref{CI+WCC} 
 indeed holds,  then this latter completeness property is true for the evolution of generic initial data,
 i.e.~the entire future is safely determined for 
 observers at infinity.  Thus, one could ask: 
 \emph{Why worry about the failure of determinism if it only applies   to observers who enter black holes?}

The above arguments notwithstanding,
there is a consensus that
such an escape is unsatisfactory. 
The question of determinism is meant to be an issue of
principle.
If
determinism is indeed  a fundamental property of the theory, it should 
be independent of our
choice not to recklessly fall into a black hole.
Thus, irrespectively of the validity of ``weak cosmic censorship'', the presence
of the Kerr Cauchy horizon is generally considered to pose 
a serious problem for classical general relativity.

\subsubsection{The blue-shift instability}
\label{BSIintro}
\begin{figure}
\centering{
\def\svgwidth{6pc}
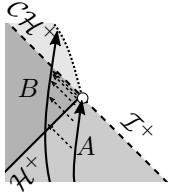}
\caption{The blue-shift instability}\label{blueshiftfigure}
\end{figure}

It was Penrose again who first identified
a key to a promising potential resolution of the conceptual problems generated by
Cauchy horizons.  He   observed~\cite{penrose1968battelle} that
the Kerr Cauchy horizon is associated with an infinite blue-shift effect depicted in
Figure~\ref{blueshiftfigure}.
Here,  $A$ and $B$ denote the projection of two timelike geodesics (i.e.~observers).
The wordline of observer $A$ avoids the black hole and has infinite proper length. 
Observer $B$ enters the black hole and arrives at the Cauchy
horizon in  finite proper time. A signal sent by $A$ at constant frequency will be
infinitely shifted to the blue when received by $B$ as $B$ approaches his
Cauchy horizon crossing time.

The above geometric optics effect is 
indeed manifested as an instability in the behaviour of linear
wave equations like the scalar wave equation
\begin{equation}
\label{linearwavePOLY}
\Box_g\psi=0
\end{equation}
on a fixed Kerr background $(\mathcal{M},g_{a,M})$ for $a\ne0$.
See already Section~\ref{LWERNK} for details.
Equation $(\ref{linearwavePOLY})$ can in turn be viewed as a naive linearisation
of the Einstein vacuum equations $(\ref{INTROvace})$. Thus, this already
suggests that perhaps the Kerr Cauchy horizon $\mathcal{CH}^+$ is in some sense unstable.
Let us emphasise 
that such an instability would still be in principle 
compatible with Conjecture~\ref{stabofkerrconj}, as the latter
stability conjecture only concerns the \emph{exterior} to the black hole.

Of course, linear equations like $(\ref{linearwavePOLY})$ on a fixed globally hyperbolic background can at worst blow up asymptotically (see for instance Chapter~12 of~\cite{RingCauchy}),
i.e.~exactly at the Cauchy horizon $\mathcal{CH}^+$ itself. 
In the full \emph{non-linear} theory, however, governed by $(\ref{INTROvace})$,
one might reasonably expect that this instability 
forces spacetime to generically break down ``before'' a Cauchy horizon has
the chance to form.
This led to 
the further expectation that upon small perturbation, not only is the Cauchy horizon
unstable, but the Penrose    diagram
of the spacetime
would         revert to the Schwarzschild picture (Figure~\ref{Schwfigure}), with  
a strong spacelike singularity across which the metric is 
inextendible as a continuous Lorentzian metric,
just as described in Section~\ref{C0inex}.

\subsubsection{The $C^0$-formulation of strong cosmic censorship}
\label{SCCintro}
The above considerations regarding the blue-shift allowed  Penrose to put forth a  general conjecture---now
known as \emph{strong cosmic censorship}---which tries to salvage Laplacian determinism
for general relativity~\cite{PenroseSCCrefer}. The gist of this conjecture is that, the Kerr
solution notwithstanding, for 
\emph{generic} asymptotically flat initial data,  the type of behaviour exhibited 
in Section~\ref{KCH2intro} does \emph{not} arise.
There have appeared many different formulations in the literature over the years; for
further discussion see the monograph~\cite{earman1995bangs} and 
our survey~\cite{JonathanMesurvey}.
One can distill the following 
mathematical formulation, 
presented in~\cite{Chr}, which captures the spirit of  the original expectation
of the conjecture:
\begin{bigconj}[$C^0$-formulation of strong cosmic censorship]
\label{C0formSCC}
For generic compact or asymptotically flat vacuum initial data, the maximal Cauchy development
is inextendible as a Lorentzian manifold with $C^0$ (continuous) metric.
\end{bigconj}

In the language of PDE's, strong cosmic censorship should be viewed as a statement of global uniqueness.\footnote{Cf.~the discussion
of ``weak cosmic censorship'' as a statement of ``global existence'' in Section~\ref{CI+WCC}. 
We again emphasise that, with these formulations here,
``strong cosmic censorship'' is not
stronger than ``weak cosmic censorship''! See~\cite{Chrmil} for further discussion.}
The above $C^0$ formulation corresponds precisely with the inextendibility statement that indeed
holds
for Schwarzschild as described in Section~\ref{C0inex}, and thus would mean
that the maximal Cauchy development is unique in what would appear to be the largest
possible reasonable class of spacetimes to consider.
A manifestly weaker formulation which has been considered in the literature replaces
$C^0$ with $C^2$~(see~\cite{RingCauchy}). In view of 
our discussion in Section~\ref{C0inex} above, however, the $C^2$-formulation  would fail
to capture uniqueness from the perspective of the modern PDE theory of $(\ref{INTROvace})$, 
for which  well-posedness
has been proven well below the $C^2$ threshold.
We will have to revisit this issue 
in Section~\ref{INTROprework}, after the main results of our paper will force us to rethink
the validity 
of  Conjecture~\ref{C0formSCC}.

Let us remark finally that some assumption of compactness or asymptotic flatness in
Conjecture~\ref{C0formSCC} is clearly necessary in any formulation of strong
cosmic censorship, as otherwise
there exist trivial, stable examples of Cauchy horizons.
For instance, the maximal Cauchy evolution of the Minkowski subsets
$\{t=0\}\cap\{|r|<1\}$, and more interestingly, $\{t^2=r^2+1\}\cap \{t>0\}$, are extendible
beyond smooth Cauchy horizons, and this property can be seen to be stable in a 
suitable sense.  One  should not think, however,
 that Cauchy horizons
are necessarily associated with non-compactness or other causal
pathologies, or necessarily hidden inside black holes as in the case of Kerr. 
Indeed, the examples of naked singularities of Christodoulou~\cite{ChrSph2}, referred
to in Section~\ref{CI+WCC}, in addition to exhibiting spacetimes with an incomplete
$\mathcal{I}^+$, provide examples where spacetime is
 extendible beyond a 
Cauchy horizon $\mathcal{CH}^+$ such that 
 all $p\in \mathcal{CH}^+$ correspond
to TIPs~(see~\cite{geroch1972ideal} for this terminology)
whose past have \emph{compact} closure with initial 
data.\footnote{Before the examples of Christodoulou~\cite{ChrSph2},
one could imagine that 
the Kerr-type Cauchy horizons arising from $i^+$ were the \emph{only} type of smooth Cauchy
horizons which occurred in evolution from smooth asymptotically flat data.
This led to formulations of strong cosmic censorship in the literature 
which did not impose a genericity assumption but focussed only on excluding
completely TIPs  whose
past had compact closure when intersected with $\Sigma$. 
(See for instance the formulation in the textbook~\cite{Wald}.)
Of course,  such formulations simply side-step the problem
posed by the existence of Kerr, for which a separate conjecture would still have
to be formulated. In any case, the examples of~\cite{ChrSph2} would
seem to make this a mute point, though for the vacuum equations $(\ref{INTROvace})$,
it remains an open problem to construct examples analogous to~\cite{ChrSph2}.}
In showing the instability of naked singularities for $(\ref{ESF1})$, Christodoulou has
in fact obtained a version of Conjecture~\ref{C0formSCC} for the system $(\ref{ESF1})$ under
spherical symmetry, together with his proof of
weak cosmic censorship discussed in Section~\ref{CI+WCC}.
This does not have much bearing, however, for the issue of the stability of the Kerr Cauchy
horizon, since for Christodoulou's model,
one can show relatively easily that Kerr-like
Cauchy horizons arising from $i^+$ \emph{cannot} occur under \emph{any} initial conditions. 
 We will discuss a spherically symmetric
model generalising $(\ref{ESF1})$
that does allow for such Kerr-like Cauchy horizons in Section~\ref{modelprobintro}.

\subsubsection{Is the finite boundary generically spacelike?}\label{space?}

Let us comment explicitly on the other expectation mentioned above
connected with
strong cosmic censorship: the statement that, for \emph{generic} initial data, the finite boundary of
the maximal Cauchy development is \emph{spacelike}.

As in our discussion of Conjecture~\ref{C0formSCC}, this  expectation
can be thought of as being
suggested by the Schwarzschild behaviour. One can try to understand this spacelike nature
of the boundary
canonically in the sense of TIPs~\cite{geroch1972ideal}. 
In that language, the Schwarzschild singularity
can indeed be viewed as ``spacelike''
in the sense that the set of TIPs corresponding to $\{r=0\}$
have the property that any two are disjoint after intersecting with the future
of a sufficiently late Cauchy surface. (Note that this property 
yields more  than the statement that
the boundary of the Penrose diagram is spacelike, as the latter suppresses information about
the angular directions.) For general vacuum Cauchy developments,
without symmetry, it is not however straightforward to 
distinguish a priori the TIPs representing the ``finite boundary'' from the TIPs 
representing various points at infinity (e.g.~from $\mathcal{I}^+$). 
Thus we shall not attempt here a general definition
of what it means for the ``finite boundary'' of spacetime to be spacelike.

Of course, if we happen to know that the maximal Cauchy development 
$(\mathcal{M},g)$ is indeed extendible continuously
to $(\widetilde{\mathcal{M}}, \widetilde{g})$, 
then this gives a direct way of identifying the causal structure
of the boundary of 
$\partial\mathcal{M}$ in $\widetilde{\mathcal{M}}$. Thus, if Conjecture~\ref{C0formSCC}
fails, then we can immediately entertain in this sense the causal properties of the part of
the finite boundary corresponding to $\partial\mathcal{M}\subset \widetilde{\mathcal{M}}$.\footnote{Note that if extensions are suitably regular, then general
causality 
theory and well-posedness already tell us that $\partial\mathcal{M}\subset\widetilde{\mathcal{M}}$ 
is almost everywhere a $C^1$ null
hypersurface.}
%
%
%It would be interesting to understand how the ``spacelike'' nature of a singularity may be related to the $C^0$ inextendibility
%of the metric. (Note that  the ``finite'' boundary of the maximal development must be
%in some sense singular at any point at which it is ``spacelike''.) 
%
%While it is difficult in general to define a causal structure for the singular boundary of spacetime,
%when the metric is continuously extendible beyond the singular boundary, then the boundary
%is necessarily null and we can say definitively that part of the boundary fails to be spacelike.
With this,  we shall see in the next
section that the expectation that the finite boundary is generically  spacelike---however
one might have hoped to formulate this precisely---will share the same negative fate
of Conjecture~\ref{C0formSCC}.

\subsection{The main theorems}
\label{MTtnseisag}
In this paper, we will inaugurate a series of  works giving a definitive resolution of   the
$C^0$-metric stability properties of the Kerr Cauchy horizon.
Our results will in particular
imply (see Section~\ref{negativeres} below) that, if the exterior stability of Kerr  (Conjecture~\ref{stabofkerrconj}) is indeed true,
then the formulation of strong cosmic censorship
given in Section~\ref{SCCintro}  (Conjecture~\ref{C0formSCC}) 
is in fact false, and moreover, the entire  finite boundary of spacetime is null
for all spacetimes arising from data sufficiently near two-ended Kerr data.

\subsubsection{The evolution of spacelike data and the non-linear
$C^0$-stability of the Cauchy horizon}\label{sec:MT}
%\FloatBarrier
In the present paper, we take up the problem from initial data for $(\ref{INTROvace})$
posed on a hypersurface $\Sigma_0$ which is modelled on a Pretorius--Israel
$\underline u+u=C$ hypersurface sandwiched between two constant-$r$
hypersurfaces of the Kerr interior with $r$ close to its event horizon value. 
The assumption on our data is that
 they    asymptote to induced Kerr data on $\Sigma_0$, with parameters $0<|a|<M$,
 at an inverse polynomial rate
 in $\underline u$. We think of these (see 
Section~\ref{forthevent} immediately 
below!)~as the ``expected induced data''
from a general dynamical vacuum black hole settling down to Kerr, when viewed
on a suitably chosen spacelike hypersurface ``just inside'' the event horizon.
The hypersurface is in fact foliated by trapped spheres.
The main result of the present paper is then the following

\begin{bigthe}
\label{PRWTO}
Consider general vacuum initial data 
corresponding to the expected induced geometry of a dynamical black hole settling down
to Kerr (with parameters $0<|a|<M$)
on a suitable spacelike hypersurface $\Sigma_0$ in the black hole interior.
Then the maximal future development spacetime $(\mathcal{M},g)$ corresponding to $\Sigma_0$ 
is globally
covered by a double null foliation and has a 
non-trivial Cauchy horizon $\mathcal{CH}^+$
across which the metric is continuously extendible.
\end{bigthe}

\begin{figure}
\centering{
\def\svgwidth{5pc}
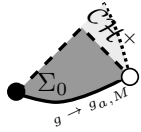}
\caption{Stability of $\mathcal{CH}^+$ from data on $\Sigma_0$}\label{maintheorfigureThe1}
\end{figure}

The domain of the spacetime is depicted in
Figure~\ref{maintheorfigureThe1}.
It turns out that we can in fact retrieve our assumptions on the geometry of $\Sigma_0$
 from an assumption on the
event horizon $\mathcal{H}^+$ and thus directly relate them to the stability of Kerr conjecture 
as
formulated in Conjecture~\ref{stabofkerrconj}, as well as to the expectation that
generic vacuum spacetimes (not necessarily initially close to Kerr) 
must either disperse or eventually
settle down to a number of Kerr black holes. 
This event horizon formulation will be the subject of the forthcoming paper 
discussed below.

\subsubsection{Forthcoming work: event horizon data and the stability of the red-shift region}
\label{forthevent}
\begin{figure}
\centering{
\def\svgwidth{5pc}
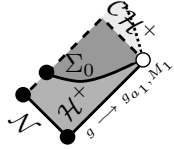}
\caption{Stability of $\mathcal{CH}^+$ from data on $\mathcal{H}^+\cup \mathcal{N}$}\label{maintheorfigureThe2}
\end{figure}

In an upcoming follow-up paper of this series~\cite{DafLuk2}, 
we will obtain the above induced data  of
Theorem~\ref{PRWTO} on a hypersurface $\Sigma_0$ in the interior of a spacetime arising
from a characteristic initial value problem
with data posed on a bifurcate null hypersurface $\mathcal{N}\cup\mathcal{H}^+$. 
See~Figure~\ref{maintheorfigureThe2}.
Here, the initial hypersurface $\mathcal{H}^+$ is meant to represent the event horizon 
of a dynamic vacuum black hole settling down to a rotating Kerr solution.

\begin{bigthe}[to appear \cite{DafLuk2}]
\label{DEUTERO}
Consider vacuum initial data on a bifurcate null hypersurface
$\mathcal{N}\cup \mathcal{H}^+$ such that $\mathcal{H}^+$ 
is future affine complete and the data suitably  approach
the event horizon geometry of 
Kerr (with parameters $0<|a|<M$). Then the maximal future development  $(\mathcal{M},g)$
contains a hypersurface $\Sigma_0$ as in
Theorem~\ref{PRWTO} and thus
again has a
non-trivial Cauchy horizon $\mathcal{CH}^+$
across which the metric is continuously extendible.
\end{bigthe}

It follows from the above that any black hole spacetime suitably settling down to
a rotating Kerr in its exterior will have a non-trivial piece of Cauchy horizon.
Given the conjectured stability of the Kerr exterior (Conjecture~\ref{stabofkerrconj}),
we can already infer a negative result for the $C^0$ formulation of strong cosmic
censorship (Conjecture~\ref{C0formSCC}). We make this explicit in what follows.

\subsubsection{Corollary: The $C^0$-formulation of strong cosmic censorship is false}
\label{C0formfalsecorsec}
\label{negativeres}
\begin{figure}
\centering{
\def\svgwidth{10pc}
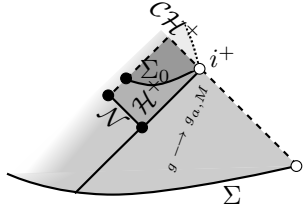}
\caption{The $C^0$ formulation of SCC is false!}\label{theCorfig}
\end{figure}
If the exterior region of Kerr is indeed proven stable up to and including
the event horizon $\mathcal{H}^+$, it will follow
that all spacetimes arising from asymptotically flat 
initial data sufficiently close to 
Kerr data on $\Sigma$ will satisfy the assumptions
of Theorem \ref{DEUTERO}, where $\mathcal{N}$ is simply taken to be an
arbitrary sufficiently short
incoming null hypersurface intersecting $\mathcal{H}^+$.
A corollary of our results is thus
\begin{corolnonum}
If stability of the Kerr exterior (Conjecture~\ref{stabofkerrconj}) 
is true, then the Penrose diagram of Kerr is stable near $i^+$ and the
$C^0$-formulation of strong cosmic censorship (Conjecture~\ref{C0formSCC})
is \underline{false}.
\end{corolnonum}

The Corollary is depicted in Figure~\ref{theCorfig}.
The above statement thus also falsifies the expectation that generically the finite boundary of
spacetime is spacelike (cf.~Section~\ref{space?}).

If the more speculative final state picture of~\cite{Penroseunsolv}, described at the end of
Section~\ref{edwgiaeusta9eia}, indeed holds,  then it will follow from
Theorem~\ref{DEUTERO}
that  generic, non-dispersing solutions of the vacuum equations $(\ref{INTROvace})$
arising from asymptotically
flat initial data have a metric which is $C^0$-extendible across a piece of a non-empty
null Cauchy horizon.
In this sense, null finite boundaries of spacetime are ubiquitous
in gravitational collapse.

\subsubsection{Forthcoming work: The $C^0$-stability of the bifurcation sphere of the Cauchy horizon}
\begin{figure}
\centering{
\def\svgwidth{12pc}
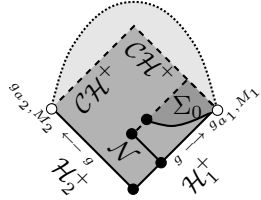}
\caption{Stability of bifurcate $\mathcal{CH}^+$ from data on $\mathcal{H}^+_1\cup\mathcal{H}^+_2$}\label{maintheorfigureThe3}
\end{figure}

For our final forthcoming result, we return  to the conclusion of 
Conjecture~\ref{stabofkerrconj} as applied specifically in the two-ended case.
In that setting, not only does one have a 
single future-affine complete hypersurface asymptoting to 
Kerr, but one has in fact a 
bifurcate future event horizon
$\mathcal{H}^+_1\cup\mathcal{H}^+_2$, both parts of which remain globally close
to a reference Kerr, each moreover asymptoting to two nearby subextremal Kerr's
with parameters $a_i$, $M_i$, $i=1,2$.
Taking such a bifurcate hypersurface as our
starting point, 
we have then the following theorem:

\begin{bigthe}[to appear \cite{DafLuk3}]
\label{TRITO}
Consider vacuum initial data on a bifurcate null hypersurface
$\mathcal{H}^+_1\cup\mathcal{H}^+_2$,
such that both hypersurfaces are future complete, and globally
close to, and asymptote to Kerr metrics with nearby parameters
$0<|a_1|<M_1$, and $0<|a_2|<M_2$, respectively. Then the maximal future development 
$(\mathcal{M},g)$
can be covered by a double null foliation and moreover can be
extended as a $C^0$ metric across a bifurcate  Cauchy horizon $\mathcal{CH}^+$
as depicted in Figure~\ref{maintheorfigureThe3}. All future-inextendible causal geodesics in $\mathcal{M}$ can be extended to
 cross   $\mathcal{CH}^+$.
\end{bigthe}

\begin{figure}
\centering{
\def\svgwidth{12pc}
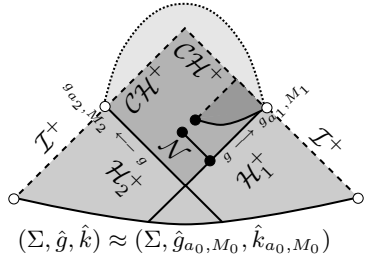}
\caption{The global stability of the Kerr Penrose diagram}\label{globstabfig}
\end{figure}

 Thus, if Conjecture~\ref{stabofkerrconj} is true, then not only is Conjecture~\ref{C0formSCC} 
 false,
 but the entire Penrose diagram of Kerr is stable, up to and including the bifurcation
 sphere of the Cauchy horizon. See Figure~\ref{globstabfig}. It follows in this case
 that  \emph{no part} of the boundary of spacetime  is spacelike,
 a spectacular failure   of the original expectation described in Section~\ref{space?}.

Of course, the above picture where the entire finite boundary of spacetime 
consists of a Cauchy horizon (of TIPs with non-compact intersection with initial data) could be an artifice of the two-ended case.
Thus, one can still hold out hope that in the  standard picture of
gravitational collapse of asymptotically flat, one-ended data, although the
Corollary of Section~\ref{C0formfalsecorsec} 
would apply to yield \emph{part} of the boundary being null,
there would necessarily also  exist a non-empty spacelike part of the boundary, at least generically.
This remains an open problem!
In fact, it would be interesting simply to exhibit a single open set in the moduli space
of initial data for the vacuum equations $(\ref{INTROvace})$
whose maximal Cauchy development is bounded at least in part by a piece
of ``singularity'' which can indeed naturally be thought of as spacelike.
(Let us note in contrast that, for the Einstein--scalar field system, such an open set has
indeed been constructed by Rodnianski--Speck~\cite{Rodnianski:2014yaa, Rodnianski:2014zaa}. The presence of
a non-trivial scalar field is, however, fundamental for the proof, as it is the scalar
field that drives the singular behaviour.)

\subsection{Previous work and a reformulation of strong cosmic censorship}
\label{INTROprework}

Though  Theorem~\ref{PRWTO} disproves the original expectations concerning
the interior structure of generic black holes and the $C^0$ formulation
of strong cosmic censorship, 
it turns out that it is still compatible with a well-motivated,
reformulated version of strong cosmic censorship originally due to Christodoulou~\cite{Chr}.
To understand this, 
we must turn back to some previous work which inspired our study
and is in fact the source for many of the ideas here. 
Indeed, the present work 
can be thought of as a
natural culmination of three distinct but related
threads.
\vskip1pc
{\bf 1.~The scalar wave equation $(\ref{linearwavePOLY})$ on Kerr.} 
The first thread  concerns the study of 
the scalar wave equation $(\ref{linearwavePOLY})$
on a fixed Kerr black hole interior
(as well as on the simpler, spherically-symmetric 
{\bf Reissner--Nordstr\"om} background~\cite{Hawking&Ellis} which
again has a Cauchy horizon).
Recall that it was precisely in this context that 
Penrose first discussed the blue-shift instability suggesting
strong cosmic censorship (cf.~Section~\ref{BSIintro}).
The mathematical study of $(\ref{linearwavePOLY})$ in these black hole interiors was initiated 
by
McNamara~\cite{McNamara1}.
Though it can indeed be rigorously shown that
the blue-shift instability  induces \emph{energy blow up}
for generic solutions of $(\ref{linearwavePOLY})$,
at the same time, one can prove $C^0$ \emph{stability} statements
all the way up to the Cauchy horizon. These  stability and instability
statements in turn depend on quantitative
upper and lower bounds, respectively, for the decay of solutions of $(\ref{linearwavePOLY})$
in the black hole \emph{exterior}.
We will review all these statements in Section~\ref{LWERNK} below.
Extrapolation of these results already suggests  {\bf a re-formulation of
strong cosmic censorship}, due to Christodoulou~\cite{Chr},
compatible
with the above behaviour. This  will be given as Conjecture~\ref{ChrSCC}.

\vskip1pc
{\bf 2.~A non-linear spherically symmetric toy model.}
The second thread, to be described 
in Section~\ref{modelprobintro}, concerns 
a toy-model~\cite{D1, D2, D3}, where $(\ref{linearwavePOLY})$
 is coupled to the Einstein--Maxwell
system \emph{in spherical symmetry}.
On the one hand, as a gravitationally coupled non-linear problem, study of this  system 
goes beyond  the  strictly
linear setting described in the previous paragraph; on the other hand, it is
more restricted 
in that it is purely spherically symmetric. This toy-model is itself inspired by even simpler 
 model problems considered by Hiscock~\cite{Hiscock}, Poisson--Israel~\cite{PI1}, 
and Ori~\cite{A-Ori} where the wave equation is replaced by  null dust.   The toy-model
completely vindicates extrapolation of the linear theory of $(\ref{linearwavePOLY})$
described above:
It has now been  proven for the spherically symmetric Einstein--Maxwell--scalar
field equations that (i) for all admissible initial data part
of the boundary is null and the spacetime extends continuously to a larger spacetime~\cite{D2, D3}
(thus, Conjecture~\ref{C0formSCC} is false for this toy model under spherical symmetry),
and that (ii) for  generic initial data in the symmetry class,
these null boundaries are singular in a sense to be described~\cite{LukOh2017one, LukOh2017two}; 
this type of singularity was given the name
``weak null singularity''. This gives further evidence that 
 Conjecture~\ref{ChrSCC}  is the proper formulation of strong cosmic censorship\footnote{though,
 as we shall see, it is difficult to infer that  the most satisfying analogue of
 Conjecture~\ref{ChrSCC} is true even in this toy model,
 in view of the geometric nature of its statement.}.
\vskip1pc
{\bf 3.~Local construction of vacuum  weak null singularities without symmetry.}
The picture of weak null singularities arising from the above toy model
 gave rise to much discussion and ongoing debate.
At the time when the spherically symmetric model was first considered,
there were no known
examples of vacuum spacetimes with null singularities
of that type.  Thus, one could still even hope that such singular fronts
\emph{never} occurred for the vacuum equations $(\ref{INTROvace})$.
Similarly, one could hope that the weak null singularities of~\cite{D2, D3, LukOh2017one} would
become spacelike when perturbed outside of symmetry.
  This brings us to the third
thread, to  be discussed in Section~\ref{ESSENTIAL!}.
Here, we consider the completely local problem of constructing 
patches of  vacuum spacetime
bounded by ``weak null singularities''. A definitive statement  was recently 
obtained in~\cite{LukWeakNull}. Imposing characteristic initial data
with a specific singular profile, one can first  show that a solution exists in a
full ``rectangular'' domain covered by a double null foliation, i.e.~no spacelike singularity
immediately arises (stability). Moreover, one can then
show \emph{a posteriori} that the initial 
singular profile propagates, naturally 
inducing a  weak null singular boundary on spacetime. 
This construction can be generalised from the vacuum equations~$(\ref{INTROvace})$
to the Einstein--Maxwell--scalar field equations, the latter now without symmetry.
In particular, one can  obtain the
stability of the weak null singularities of~\cite{D2, D3, LukOh2017one} 
in the black hole interior, 
\emph{provided that these have already formed}.  
\vskip1pc

We turn now in Sections~\ref{LWERNK},~\ref{modelprobintro}
and~\ref{ESSENTIAL!}, respectively, 
to a more detailed discussion of each of the above threads.

\subsubsection{The linear wave equation on  Kerr (and Reissner--Nordstr\"om)
 and Christodoulou's reformulation of strong cosmic censorship}
\label{LWERNK}

The most basic  setting for understanding the stability and instability properties
of the Cauchy horizon is
the study
of the linear wave equation $(\ref{linearwavePOLY})$
on a fixed Kerr  background
with initial data defined on $\Sigma$.

\begin{figure}
\centering{
\def\svgwidth{6pc}
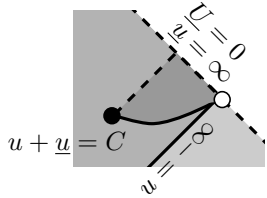}
\caption{Null coordinates in the black hole interior}\label{nullfigure}
\end{figure}

A simplified setting is the so-called Reissner--Nordstr\"om background~\cite{Hawking&Ellis}.
Recall that the Reissner--Nordstr\"om family $g_{Q,M}$ is a two parameter family
of spherically symmetric metrics which in local coordinates take the form
\begin{equation}
\label{RNlocal}
g_{Q,M} = -(1-2M/r+Q^2/r^2)dt^2+(1-2M/r+Q^2/r^2)^{-1}dr^2+r^2(d\theta^2+\sin^2\theta \, d\phi^2).
\end{equation}
The metrics are electrovacuum, i.e.~together with an appropriate
electromagnetic field $F_{\mu\nu}$ they solve  
the Einstein--Maxwell system 
\begin{equation}
\label{EM1}
{Ric}_{\mu\nu}-\frac12 g_{\mu\nu}R =8\pi T_{\mu\nu}\doteq
8\pi(   \frac1{4\pi} (F_{\mu}^{\, \, \lambda} F_{\lambda \nu} -\frac14 g_{\mu\nu} F_{\alpha\beta}F^{\alpha\beta})),
\end{equation}
\begin{equation}
\label{EM2}
\nabla^\mu F_{\mu\nu}=0, \qquad \nabla_{[\lambda} F_{\mu\nu]}=0.
\end{equation}
In the parameter range $0<|Q|<M$, the discussion in Section~\ref{KCHintro}
applies equally well to Reissner--Nordstr\"om, in particular, the form of the Penrose
diagram and existence and properties of the Cauchy horizon. 
The metric in the region between the event and Cauchy horizon can be expressed
as $(\ref{RNlocal})$, where $r$ ranges in
\begin{equation}
\label{rrangeRN}
M-\sqrt{M^2-Q^2}<r<M+\sqrt{M^2-Q^2}.
\end{equation}
Defining $r^*$  by
$\frac{dr^*}{dr}=(1-2M/r+Q^2/r^2)^{-1}$, then the region $(\ref{rrangeRN})$
is covered by the unbounded null coordinates
\begin{equation}
\label{EdFin}
u=\frac12(r^*-t), \qquad \underline{u}=\frac12(r^*+t)
\end{equation}
as $(-\infty,\infty)\times(-\infty,\infty)$, with the Cauchy horizon
formally parameterised as $\{u=\infty\}\cup \{\underline{u}=\infty\}$.
See Figure~\ref{nullfigure}.
The metric can be extended beyond the Cauchy horizon by
passing to Kruskal type coordinates:
For instance,  defining  
\begin{equation}
\label{Kruskaldefi}
\underline{U}=-e^{-2\kappa_- \underline{u}},
\end{equation}
where $\kappa_-=\frac{r_+-r_-}{2r_-^2}$ is the surface gravity of the Cauchy horizon, 
then $\underline{u}=\infty$ maps to $\underline{U}=0$, and the metric
is seen to be smooth.
We will discuss versions of the null coordinates $u$ and $\underline{u}$ appropriate
for non-spherically symmetric spacetimes including Kerr in Section~\ref{indoublenullINTRO}.
The hypersurface $\Sigma_0$ corresponds to $u+\underline{u}=C$.

As described in Section~\ref{BSIintro}, it was precisely
in the setting of $(\ref{linearwavePOLY})$
that the blue-shift effect, and its possible role
in allowing for some version of strong cosmic censorship to be true,
 was originally discussed.
Theorem~\ref{instabilitytheoremsphsym}, to follow shortly below, 
will  indeed provide a rigorous        manifestation  of  the instability associated with this 
effect. What is much less
well known, however, is that {\bf solutions of the wave equation enjoy
global
$C^0$ stability properties within the black hole interior, still compatible with the blue-shift instability.}
The proof of these \emph{interior} stability statements is in turn related to quantitative \emph{decay} estimates 
in the \emph{exterior}, which themselves have only been obtained recently.
We turn first to both of these stability statements. 

\paragraph{Stability.}
The study of the wave equation $(\ref{linearwavePOLY})$ on a Kerr background
has been the subject of intense activity in
the past years, and definitive boundedness and decay
results, both in the exterior, but, surprisingly, in the interior, have been obtained.
The following theorem summarises the statements relevant for us.

\begin{theorem}
\label{stabilityforwaveinintro}
Let $\psi$ be a solution of  $(\ref{linearwavePOLY})$ on sub-extremal Kerr or
Reissner--Nordstr\"om arising from  regular Cauchy data
on $\Sigma$
decaying sufficiently fast at spatial infinity.\footnote{For convenience, one can assume
that the data are smooth and compactly supported on $\Sigma$, but the result is true
under much weaker regularity and decay assumptions!}
Then
\begin{enumerate}
\item
The solution
$\psi$ remains globally bounded in the exterior ${J}^-(\mathcal{I}^+)$ and decays
inverse polynomially to $0$,
in particular, on the event horizon $\mathcal{H}^+$. See~\cite{partiii, Civin}.
\item
The polynomial decay on $\mathcal{H}^+$ propagates to similar
decay on the spacelike hypersurface $\Sigma_0$ in the black hole interior, with respect
to coordinates $u$, $\underline{u}$. See~\cite{propaginto, annefranzen}.
\item
The solution $\psi$ is in fact uniformly bounded in all of $\mathcal{M}$,
and extends continuously to $\mathcal{CH}^+$. See~\cite{annefranzen, Franzen2,
Hintz:2015koq, LukSbierski}.
\end{enumerate}
\end{theorem}

Let us note that the first partial result in the direction of statement 3.~was due
to McNamara~\cite{McNamara1}, who essentially showed that 3.~held for fixed spherical
harmonics $\psi_\ell$
on Reissner--Nordstr\"om, provided that the scalar field decayed
suitably fast on the event horizon, i.e.~assuming the analogue of 1.,
something which at the time had not been proven.
It is interesting that the significance of this stability does not seem to have
been explicitly noted in most subsequent papers.
As we shall see in the discussion below, statement 3.~of the above theorem
can be thought of as the first prototype of our Theorem~\ref{PRWTO}.

The proof of statement 3.~depends on  obtaining \emph{weighted energy estimates}
for $\psi$, with appropriate polynomial weights in the null coordinates $u$ and $\underline{u}$
described above.  For $(\ref{linearwavePOLY})$, such 
energy estimates naturally arise from well-chosen
vector fields.
Given a vector field $V$ one can associate currents $\mathfrak{J}^\mu[\psi]$
and $\mathfrak{K}[\psi]$ by
\[
\mathfrak{J}^\mu[\psi]=T_{\mu\nu}[\psi]V^\mu,\qquad \mathfrak{K}={}^{(V)}\pi^{\mu\nu}T_{\mu\nu}[\psi]=
\frac12(\mathcal{L}_Vg)^{\mu\nu}T_{\mu\nu}[\psi],
\]
where $T_{\mu\nu}[\psi]$ denotes the standard energy momentum tensor\footnote{This
is defined exactly as in the expression on the right hand side of $(\ref{ESF1})$.}.
The wave equation $(\ref{linearwavePOLY})$ implies the divergence identity
\begin{equation}
\label{newidentity}
\int \nabla^\mu \mathfrak{J}_\mu[\psi]=   \int \mathfrak{K}[\psi],
\end{equation}
which upon integration in an appropriate region of the form
$\mathcal{R}= J^+(\Sigma_0)\cap J^-(\Sigma_1)$ yields an energy identity
\begin{equation}
\label{integratedvfidentity}
\int_{\Sigma_1}\mathfrak{J}_\mu[\psi]n_{\Sigma_1}^\mu + \int_{\mathcal{R}} \mathfrak{K}[\psi] =\int_{\Sigma_0}\mathfrak{J}_\mu[\psi] n_{\Sigma_0}^\mu
\end{equation}
where integration is always with respect to the induced volume form
and $n^\mu_{\Sigma_i}$ denotes an appropriate normal.\footnote{We will in fact
be interested
in the case where $\Sigma_1$ is null; the choice of a volume form again then induces
a notion of normal, which is however tangential to $\Sigma_1$.}

Let us discuss first   the Reissner--Nordstr\"om case. One
considers  the vector field\footnote{Here the coordinate vector fields are defined
with respect to $(u,\underline{u},\theta,\phi)$ coordinates. The argument described
here is a streamlined version of that in~\cite{annefranzen}, where the $r$-weight and
$u$, $\underline{u}$ weights were applied in separate regions.}
\begin{equation}
\label{vectorfield}
V= r^{2N}(|u|^{1+2\delta}\partial_u+\underline{u}^{1+2\delta}\partial_{\underline{u}}),
\end{equation}
and applies $(\ref{integratedvfidentity})$
in the region bounded between $\Sigma_0$ described above and
$\Sigma_1=( \{u=c\}\cup \{\underline{u}=\underline{c}\} ) \cap J^+(\Sigma_0)$.
The  arising flux term 
$\int_{\Sigma_1}\mathfrak{J}_\mu[\psi]n_{\Sigma_1}^\mu$
on the left hand side of $(\ref{integratedvfidentity})$
is positive and 
controls the following
weighted $L^2$ energies
\begin{equation}
\label{vectorfield0}
\|\underline{u}^{\frac12+\delta}r^N\partial_{\underline{u}}\psi  \|^2_{L^2(d\underline{u} \sin\theta d\theta d\phi)}+\|u^{\frac12+\delta}\Omega r^N|\slashed\nabla\psi|\|^2_{L^2(d\underline{u} \sin\theta d\theta d\phi)}, 
\end{equation}
\begin{equation}
\label{vectorfield00}
\||u|^{\frac12+\delta} r^N\partial_{u}\psi\|^2_{L^2(du \sin\theta d\theta d\phi)}+
\|\underline{u}^{\frac12+\delta}\Omega r^N|\slashed\nabla\psi |\|^2_{L^2(du \sin\theta d\theta d\phi)},
\end{equation}
where the integrals are taken on $\{u=c\}\cap J^+(\Sigma_0)$ and
$\{\underline{u}=\underline{c}\}\cap J^+(\Sigma_0)$, respectively.
Here $|\slashed\nabla \psi|$ denotes the norm of 
the induced gradient
of $\psi$ on the constant-$(u,\underline{u})$ spheres of symmetry
and $\Omega^2=-(1-2M/r+Q^2/r^2)$.
The initial flux on $\Sigma_0$ appearing
on the right
hand side of $(\ref{integratedvfidentity})$ is 
controlled in view of statement 2.~of Theorem~\ref{stabilityforwaveinintro}
for sufficiently small
$\delta>0$. 
Thus, to infer an estimate on $(\ref{vectorfield0})$ and $(\ref{vectorfield00})$, it remains
to examine the bulk $\mathfrak{K}$ term in $(\ref{integratedvfidentity})$.

Herein lies the significance of 
the $r^N$-weight in $(\ref{vectorfield})$. This weight 
is of course globally bounded above and below to the future of $\Sigma_0$
in view of $(\ref{rrangeRN})$,
but for $N\gg 1$, this generates a favourable non-negative part of the
 ``bulk''  $\mathfrak{K}$-term on the left hand of $(\ref{integratedvfidentity})$,
\begin{equation}
\label{goodbulkoneforwave}
\int_{ \{ u+\underline{u}\ge C\} \cap \{u\ge c\} \cap \{\underline{u}\le \underline{c}\} }N r^{N-1}[ |u|^{1+2\delta}(\partial_u\psi)^2+\underline{u}^{1+2\delta}(\partial_{\underline{u}}\psi)^2 ]  \cdot r^2\Omega^2 \sin\theta\,  d\theta \, d\phi\,  du\,  d\underline{u}
\end{equation}
which can be used to absorb mixed terms  $\partial_u\psi\partial_{\underline{u}}\psi$
also appearing in $\mathfrak{K}$, 
which do not come with a sign. Moreover, quite fortuitously, the 
only   additional terms in $\mathfrak{K}$ containing angular derivatives
come with a good sign:
\begin{equation}
\label{goodbulktwoforwave}
\int_{ \{ u+\underline{u}\ge C\} \cap \{u\ge c\} \cap \{\underline{u}\le \underline{c}\} } r^N (|u|^{1+2\delta}+\underline{u}^{1+2\delta} )\|\slashed\nabla\psi|^2  \cdot r^2\Omega^2 
\sin\theta \, d\theta \, d\phi\, du\, d\underline{u}.
\end{equation}
Thus, for $N\gg 1$, the $\mathfrak{K}$ term
is nonnegative, and $(\ref{integratedvfidentity})$ indeed
implies a uniform bound for
$(\ref{vectorfield0})$ and $(\ref{vectorfield00})$
from initial data, independent of $\bar{u}\ge C$, $u\le 0$.

Note that along $u=c\le 0$, we may integrate in $\underline{u}_1\ge \underline{u}
\ge \underline{u}_0\ge C-c$ to obtain
\begin{eqnarray}
\nonumber
\psi^2(c,\underline{u}_1,\theta,\phi) &\lesssim&  \left(\int_{\underline{u}_0}^{\underline{u}_1}\partial_{\underline{u}} \psi (c,\underline{u},\theta,\phi) d\underline{u}\right)^{2} +\psi^2(c, \underline{u}_0,\theta,\phi)\\
\nonumber
&\lesssim& \int_{\underline{u}_0}^{\underline{u}_1} \underline{u}^{-1-2\delta}d\underline{u} \int_{\underline{u}_0}^{\underline{u}_1} \underline{u}^{1+2\delta} (\partial_{\underline{u}}\psi)^2 d\underline{u}
+\psi^2(c, \underline{u}_0,\theta,\phi)\\
\label{remembercontinuity}
&\lesssim&(\underline{u}_1^{-2\delta}-\underline{u}_0^{-2\delta}) \int_{\underline{u}_0}^{\underline{u}_1} \underline{u}^{1+2\delta} (\partial_{\underline{u}}\psi)^2 d\underline{u} +\psi^2(c, \underline{u}_0,\theta,\phi)\\
\label{tobecommuted}
&\lesssim& \int_{\underline{u}_0}^{\underline{u}_1} \underline{u}^{1+2\delta} (\partial_{\underline{u}}\psi)^2 d\underline{u} +\psi^2(c, \underline{u}_0,\theta,\phi),
\end{eqnarray}
where to obtain $(\ref{tobecommuted})$ from $(\ref{remembercontinuity})$
we  have used fundamentally that $\delta>0$.
Note that we may apply the above  to
$\mathscr{W}_i\mathscr{W}_j\psi$ in place of $\psi$, where $\mathscr{W}_i$ denote  
 generators of $\mathfrak{so}(3)$,
since the spherical symmetry of $(\ref{RNlocal})$ 
implies that $(\ref{linearwavePOLY})$ commutes
with $\mathscr{W}_i$.
Applying the commuted  inequality $(\ref{tobecommuted})$ with $\underline{u}_0=C-c$,
so that the second term on the right hand side is bounded by data, and integrating
further with respect to $\sin \theta d\theta d\phi$,
application of a  Sobolev inequality
\begin{equation}
\label{sobolev}
|\psi|^2 (u,\underline{u},\theta,\phi) \lesssim
\sum_{i,j=1}^3\int( |\psi|^2 + |\mathscr{W}_i\mathscr{W}_j\psi|^2 )(u,\underline{u},\theta,\phi)\sin\theta
d\theta d\phi
\end{equation}
allows 
one to immediately obtain uniform $C^0$ bounds for $\psi$ from
the boundedness of $(\ref{vectorfield0})$,
giving the first part of statement~3. 
Note that  commutation of $(\ref{linearwavePOLY})$ 
with the Killing field $\partial_t$ reveals that $\partial_t\psi$ is also uniformly bounded.
Using this latter fact, 
the continuous extendibility of $\psi$ to $\mathcal{CH}^+$ follows easily by
revisiting $(\ref{remembercontinuity})$ with $\underline{u}_0\to \infty$.

The argument in the Kerr case that statement 2.~of
Theorem~\ref{stabilityforwaveinintro} implies statement 
3.~is in fact not much more difficult than
in Reissner--Nordstr\"om. One can apply the argument in various coordinates, but
for comparison with the present work, it is most natural to work in a double null
foliation of the interior of Kerr defined by functions $u$ and $\underline{u}$, normalised
analogously to $(\ref{EdFin})$ as in Reissner--Nordstr\"om. This is in fact
the approach of~\cite{Franzen2}. (We shall review 
such a foliation in Section~\ref{indoublenullINTRO}.) 
It is convenient to replace $r$ in $(\ref{vectorfield})$ by a quantity
$\varpi$ which is related to the square root of the
area of the  constant $(u,\underline{u})$-spheres foliation. 
The essential estimate now again follows from the energy identity of the
analogue of vector field $(\ref{vectorfield})$.
One must also deal
with the difficulty that the usual commuting in the angular directions now generates error terms.
See~\cite{Hintz:2015koq, LukSbierski} for alternative approaches.

\paragraph{Instability.}
We turn now to the blue-shift instability for $(\ref{linearwavePOLY})$.

We have already noted the original heuristic considerations of 
Penrose~\cite{penrose1968battelle}. It turns out that, together with the
Gaussian beam approximation, the argument of Section~\ref{BSIintro} 
can easily be used to infer
that  generic  \emph{finite-energy} solutions on $\Sigma$ fail to be $H^1_{\rm loc}$
at the Cauchy horizon. See Sbierski~\cite{SbierskiGauss}.  
This class of solutions is too large, however.
What we are interested in are solutions which arise from \emph{localised} initial data
on $\Sigma$, as in the statement of Theorem~\ref{stabilityforwaveinintro}.  For such $\psi$,  the 
decay of waves in the exterior proven as statement 1.~of Theorem~\ref{stabilityforwaveinintro}
is a potential counterweight
to the blue-shift effect.

Let us discuss first the Reissner--Nordstr\"om case, where the most definitive
results have been obtained:

\begin{theorem}[L.--Oh~\cite{LukOhpub}]
\label{instabilitytheoremsphsym}
In the Reissner--Nordstr\"om case,
let $\psi$   arise from \underline{generic} compactly supported\footnote{One can also
consider generic data in classes allowing for specific ``tails''
at spatial infinity.} initial data on $\Sigma$.
Then 
\begin{itemize}
\item[(i)] there is a \underline{lower} bound for the rate of polynomial decay on $\mathcal{H}^+$
and
\item
[(ii)] any non-trivial extension of $\psi$ to an open 
subset of $\widetilde{\mathcal{M}}$ fails to be
$H^1_{\rm loc}$.
\end{itemize}
\end{theorem}

In the above $\widetilde{\mathcal{M}}$ can be taken to be the standard
maximal analytic extension of Reissner--Nordstr\"om.
Note that the failure of $\psi$ to be $H^1_{\rm loc}$ is compatible
with the boundedness of $(\ref{vectorfield0})$, since if ${\underline{U}}$ is 
the Kruskal type coordinate $(\ref{Kruskaldefi})$,
then the first term of $(\ref{vectorfield0})$ transforms to
\begin{equation}
\label{relationof}
\| (-\underline{U})^{\frac12} |\log (-\underline{U})|^{1/2+\delta}\partial_{\underline{U}}\psi  \|_{L^2(d\underline{U} \sin\theta d\theta d\phi)}
\end{equation}
which no longer controls
 $\| \partial_{\underline{U}}\psi \|_{L^2(d\underline{U} \sin\theta d\theta d\phi)}^2$.
We will return to the relation between $(\ref{vectorfield0})$ and $(\ref{relationof})$
in Section~\ref{baridiaeisagwgns}, 
as this will be very relevant for the methods of the present paper.

As we have remarked above,
a fundamental difficulty in obtaining  instability is that the blue-shift effect is offset
by the decay of waves in the exterior. 
The statement  (i) 
that there is a \emph{lower} bound on the rate of decay on $\mathcal{H}^+$
thus turns out to be intimately related to the proof of the  $H^1_{\rm loc}$ inextendibility (ii). 
A stronger lower bound than that of~\cite{LukOhpub}
can be obtained by using also the results of~\cite{Angelopoulos:2016wcv},
and this together with a monotonicity argument of~\cite{D1} 
allows one to show that $\psi$ at the Cauchy horizon is not in $W^{1,p}_{\rm loc}$ 
for any $p>1$. An alternative proof of this follows 
from~\cite{Gleeson:2017fqx}. Note on the other hand that solutions
are indeed in $W^{1,1}_{\rm loc}$ at the Cauchy horizon.
The special relevance of the space $H^1_{\rm loc}=W^{1,2}_{\rm loc}$ will become clear
below.

Turning to Kerr, the failure of $\psi$ to be $H^1_{\rm loc}$ at  the Cauchy horizon
has been proven
in~\cite{LukSbierski}, \emph{assuming}, however, the analogous lower bound on $\mathcal{H}^+$
given in Theorem~\ref{instabilitytheoremsphsym}. It thus remains to obtain the genericity
of these lower bounds on $\mathcal{H}^+$  to complete the picture.

An alternative approach to obtain a version of Theorem~\ref{instabilitytheoremsphsym}
in the Kerr case is to construct a suitable solution
from scattering data on past null infinity $\mathcal{I}^-$. 
This approach was pioneered by McNamara~\cite{mcnamara1978instability}
and was completed in~\cite{DafShlap}, using the results of~\cite{Dafermos:2014jwa}.
The result of~\cite{DafShlap} again yields 
that the generic localised solution on $\Sigma$ fails to be in $H^1_{\rm loc}$, but
in contrast to Theorem~\ref{instabilitytheoremsphsym},
it requires working in a class of data slightly larger than compactly supported functions.
Moreover, it does not yield a direct relation between behaviour on $\mathcal{H}^+$
and $\mathcal{CH}^+$.
%a truly satisfactory instability statement, concerning generic compactly supported initial data
%for $\psi$, has only been given in the
%Reissner--Nordstr\"om case:

Let us note finally that  the extremal case $|Q|=M$ or $|a|=M$ hides several
subtle surprises both for the stability in the black hole exterior and for the stability and instability 
questions in the interior.
See~\cite{Aretakis2, Gajic:2015csa, Gajic:2015hyu} and the discussion in our 
survey~\cite{JonathanMesurvey}.

\paragraph{Christodoulou's reformulation of strong cosmic censorship.}
Were one to naively extrapolate from the linear       wave equation  $(\ref{linearwavePOLY})$
to the non-linear Einstein vacuum equations $(\ref{INTROvace})$, under the identification
\[
g\sim \psi,\qquad
\Gamma \sim \partial\psi,
\]
then  
Theorem~\ref{stabilityforwaveinintro} 
would already suggest the picture    of $C^0$-metric stability
of our theorems of Section~\ref{MTtnseisag}.
Indeed, Theorem~\ref{stabilityforwaveinintro} can be thought of as representing
at the same time
the analogues of Conjecture~\ref{stabofkerrconj} (statement 1),
 as well as all three of the main results of this series,
 Theorems~\ref{PRWTO}--\ref{TRITO}.
 
 %We will review the proof of the  parts of Theorem~\ref{stabilityforwaveinintro} 
 %relevant
 %for Theorem~\ref{PRWTO} in Section~\ref{overviewsection}.

At the same time, naive extrapolation of
Theorem~\ref{instabilitytheoremsphsym} 
would suggest that for \emph{generic} vacuum initial data, the Christoffel symbols $\Gamma$
would in particular fail to be locally square 
integrable at the Cauchy horizon. 
The significance of the Christoffel symbols failing to be in $L^2_{\rm loc}$
is that this implies breakdown \emph{even as a weak solution} of the Einstein
vacuum equations $(\ref{INTROvace})$. Thus, though not as strong as $C^0$-inextendibility, 
this notion of inextendibility could still justify viewing the maximal Cauchy
development as the unique spacetime which one can associate with 
initial data. 
These considerations motivated  Christodoulou to propose the following reformulation
of strong cosmic censorship in his monograph~\cite{Chr}:
\begin{bigconj}[Christodoulou re-formulation of strong cosmic censorship~\cite{Chr}]
\label{ChrSCC}
For generic asymptotically flat vacuum initial data, the maximal Cauchy development
is inextendible as a Lorentzian manifold with continuous metric
and Christoffel symbols locally square integrable.
\end{bigconj}

With the hindsight 
of Theorem~\ref{PRWTO}, 
it is clear that the problem is precisely to show that the naive extrapolations described above
indeed survive in the non-linear theory governed by the Einstein vacuum equations 
$(\ref{INTROvace})$. Without, however,
one might reasonably have
thought that \emph{given the linear instability of Theorem~\ref{instabilitytheoremsphsym}}, 
non-linear effects would 
effectively invalidate the relevance of
Theorem~\ref{stabilityforwaveinintro} to the analysis of
$(\ref{INTROvace})$ in a neighbourhood
of the Cauchy horizon.
The first evidence to the contrary came from the spherically
symmetric  non-linear toy  model to be discussed immediately below.

\subsubsection{A non-linear, spherically symmetric toy-model}
\label{modelprobintro}

We have already remarked that the model $(\ref{ESF1})$ of Christodoulou
does not admit Kerr-like Cauchy horizons in spherical symmetry. One way to  allow
for such Cauchy horizons is to add charge to the model, so that Reissner--Nordstr\"om $(\ref{RNlocal})$
becomes
an explicit solution. 

The earliest attempts to probe the effects of back-reaction in the context of our problem concerned 
the Vaidya  model of a self-gravitating, spherically symmetric null dust in the presence
of charge, 
and were undertaken by Hiscock~\cite{Hiscock}.  A more realistic model
considered a spherically symmetric configuration of
\emph{two} null dusts interacting gravitationally in the presence of charge
and was first studied in a seminal work of Poisson--Israel~\cite{PI1}. 
For a detailed discussion of these early models, we refer the reader to our 
survey~\cite{JonathanMesurvey}.
The  most basic truly satisfying toy
model, however,
is the \emph{spherically symmetric Einstein--Maxwell--real scalar field system}.
Here one couples $(\ref{linearwavePOLY})$ to the Einstein--Maxwell 
equations $(\ref{EM1})$--$(\ref{EM2})$
under spherical symmetry, that is to say one considers the evolution of
\begin{equation}
\label{EMSF1}
{\rm Ric}_{\mu\nu}-\frac12 g_{\mu\nu}R =8\pi T_{\mu\nu}\doteq
8\pi(   \frac1{4\pi} (F_{\mu}^{\, \, \lambda} F_{\lambda \nu} -\frac14 g_{\mu\nu} F_{\alpha\beta}F^{\alpha\beta})+\partial_\mu\psi\partial_\nu \psi-\frac12g_{\mu\nu} \partial^\alpha\psi\partial_\alpha\psi),
\end{equation}
\begin{equation}
\label{EMSF2}
\nabla^\mu F_{\mu\nu}=0, \qquad \nabla_{[\lambda} F_{\mu\nu]}=0,\qquad \Box_g\psi =0,
\end{equation}
for asymptotically flat spherically symmetric data. If the electromagnetic tensor $F$ is non-trivial, such data necessarily
have two ends, while if $F=0$, then the model reduces to 
the self-gravitating scalar field model pioneered by Christodoulou~\cite{ChrSph1, ChrSph2, Christodoulou4},
mentioned already in Section~\ref{CI+WCC}.
We note that the model $(\ref{EMSF1})$--$(\ref{EMSF2})$
had been discussed heuristically in~\cite{BDIM} and  
studied 
numerically by~\cite{gnedin1992instability, gnedingnedin2} and~\cite{BS, Burko}, 
initially with conflicting
results.

\begin{figure}
\centering{
\def\svgwidth{12pc}
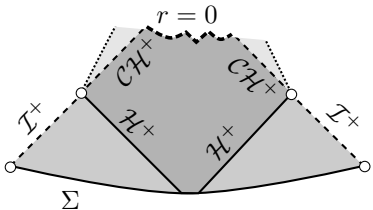}
\caption{The Penrose diagram for the spherically symmetric toy model}\label{EMSFPendiag}
\end{figure}

In the two-ended case, in view of the absence of fixed points for
the rotation symmetries, one can easily
show that, in general, the maximal future Cauchy development of spherically symmetric data will possess a complete future null infinity
$\mathcal{I}^+$ with two connected components and a non-empty black hole region, which will moreover
be bounded in the future by the union $\mathcal{CH}^+\cup \{r=0\}$,
where $\mathcal{CH}^+=\mathcal{CH}^+_1\cup\mathcal{CH}^+_2$ is itself the union of two
possibly empty null boundary components arising from the future
endpoints of the event horizon. See Figure~\ref{EMSFPendiag}.
We have already remarked that in Section~\ref{SCCintro},
for the model of Christodoulou, i.e.~$F=0$ in $(\ref{EMSF1})$--$(\ref{EMSF2})$,
one can show easily that
$\mathcal{CH}^+=\emptyset$, while one can show moreover that the boundary
$\{r=0\}$ is spacelike and singular in a suitable sense. This allows one to
easily infer a version of Conjecture~\ref{C0formSCC} for Christodoulou's model
in the $2$-ended case, without any quantitative
understanding of the global behaviour of $\psi$.\footnote{This is in contrast to the 
one-ended case, where, as mentioned already in Section~\ref{SCCintro}, there is an additional
(possibly empty) $\mathcal{CH}^+$ segment arising from the centre across
which the solution may be extendible. The deep proof~\cite{Christodoulou4}
of both cosmic censorship conjectures in the one-ended case amounts to showing that this segment
is indeed empty for generic initial data.}
Allowing charge, however, the situation changes completely:

\begin{theorem}(D.~\cite{D2, D3}, D.--Rodnianski~\cite{DRPrice})\label{ExtendibleModel}
Let $(\mathcal{M}, g, F_{\mu\nu}, \psi)$ 
be the maximal Cauchy development
arising from spherically symmetric,
admissible\footnote{The notion of ``admissibility'' is a generalisation of the no anti-trapped
surfaces condition of Christodoulou, appropriate now  for the two-ended case.}, regular
asymptotically flat two-ended initial data for the system $(\ref{EMSF1})$--$(\ref{EMSF2})$.
Then  the Penrose diagram is as in Figure~\ref{EMSFPendiag}, with $\mathcal{I}^+$ complete and
$\mathcal{H}^+$ non-empty, but with  either $\mathcal{CH}^+$ or $\{r=0\}$  possibly empty.
The scalar field $\psi$ and its derivatives
decay at least inverse polynomially along $\mathcal{H}^+$, as well as along
a suitable spacelike hypersurface $\Sigma_0$ in the black hole interior, and the geometry
towards each end approaches a subextremal\footnote{The sub-extremality
can indeed be deduced from the admissibility assumption by an argument of Kommemi.
See~\cite{LukOh2017one}.} 
Reissner--Nordstr\"om $0\le |Q|<M_i$. 
We have $Q=0$ iff $F_{\mu\nu}=0$ identically.

If $Q=0$ then $\mathcal{CH}^+=\emptyset$, while 
if $Q\ne 0$, then 
the segments $\mathcal{CH}^+$ are necessarily non-empty.
Both $\psi$ and the metric 
 $g$ are continuously extendible beyond  $\mathcal{CH}^+$.

If the initial data are globally close\footnote{Note, in this case, by Cauchy stability one can
drop the assumption of ``admissibility'' and consider Cauchy hypersurfaces $\Sigma$
as in Figure~\ref{KerrPD}.} to Reissner--Nordstr\"om with $Q\ne0$, 
then the boundary  $\{r=0\}$ is
necessarily
empty and $\mathcal{CH}^+$ is thus a bifurcate null hypersurface to which
$g$ and $\psi$ extend continuously. It follows that
the Penrose diagram of 
Reissner--Nordstr\"om of Figure~\ref{KerrPD} 
is globally stable, and $(\mathcal{M},g)$ is $C^0$-extendible
to $(\widetilde{\mathcal{M}},\tilde{g})$ 
such that all future incomplete causal geodesics $\gamma$ pass into $\widetilde{\mathcal{M}}\setminus \mathcal{M}$.
\end{theorem}

These results confirmed in particular
the numerical studies~\cite{BS, Burko}.
A corollary of the above is the following statement
\begin{corollary}
The $C^0$ formulation of strong cosmic censorship (i.e.~the analogue of 
Conjecture~\ref{C0formSCC})
is false  for the Einstein--Maxwell--scalar field system $(\ref{EMSF1})$--$(\ref{EMSF2})$  \underline{restricted to spherical symmetry}.
\end{corollary}
Here, ``restricted to spherical symmetry'' means that genericity is interpreted
within the class of spherically symmetric initial data.

The above theorem is a non-linear, but spherically symmetric, analogue of 
Theorem~\ref{stabilityforwaveinintro}. 
Thus, one can view it as containing a spherically symmetric
version of  both Conjecture~\ref{stabofkerrconj} and all Theorems~\ref{PRWTO},~\ref{DEUTERO}
and~\ref{TRITO}
of this work and our upcoming work. Proving sufficiently fast decay rates for $\psi$ along $\mathcal{H}^+$
and $\Sigma_0$ is fundamental for  showing the non-emptiness of $\mathcal{CH}^+$
and continuous extendibility beyond.
We note, however, that, in contrast to the situation in Conjecture~\ref{stabofkerrconj}, 
where one expects that  obtaining
the polynomial decay rates in the exterior is \emph{essential} in order to prove
the completeness of $\mathcal{I}^+$,
in the spherically symmetric model, the completeness of $\mathcal{I}^+$ can be obtained
\emph{a priori}~\cite{withatrapped}.
This is due to the fact that  under spherical symmetry,
 the presence of the black hole effectively breaks the super-criticality
of $(\ref{EMSF1})$--$(\ref{EMSF2})$.

The above theorem can be proven without addressing the question of whether and
in what sense the boundary $\mathcal{CH}^+$ is generically singular.
A nonlinear analogue of the instability result 
Theorem~\ref{instabilitytheoremsphsym}
has more recently been obtained in the following:

\begin{theorem}(L.--Oh~\cite{LukOh2017one, LukOh2017two})\label{LukOhte}
For generic initial data as above (open in a weighted $C^1$ norm and dense in a $C^k$ norm),
there is a lower bound for the decay rate of $\psi$ on the event horizon
$\mathcal{H}^+$ and the boundaries $\mathcal{CH}^+$ correspond 
to ``weak null singularities''.
\end{theorem}

In what sense are the boundaries $\mathcal{CH}^+$ proven singular in the above
theorem? The cleanest
statement which can be deduced is that the spacetime is inextendible as a $C^2$
Lorentzian manifold.
Recalling the $C^2$ formulation of strong cosmic censorship discussed in 
Section~\ref{SCCintro},
one has the following  Corollary
of Theorem~\ref{LukOhte}:
\begin{corollary}
The $C^2$ formulation of strong cosmic censorship is true for the
Einstein--Maxwell--scalar field system $(\ref{EMSF1})$--$(\ref{EMSF2})$
\underline{restricted to spherical symmetry}.
\end{corollary}
The analogue
of Christodoulou's formulation (Conjecture~\ref{ChrSCC}) cannot quite be deduced.
If one  restricts not only
the initial
data \emph{but also the extensions $(\widetilde{\mathcal{M}}, \widetilde{g})$ themselves}
to be spherically symmetric, then one can show that the scalar field $\psi$ does
not extend to be in $H^1_{\rm loc}$.  Restricting to such spherically symmetric
extensions, and under the additional assumption 
of a pointwise lower bound on the horizon $\mathcal{H}^+$, 
one can show also
that the Christoffel symbols are not in $L^2_{\rm loc}$ at $\mathcal{CH}^+$~\cite{D2}.
This stronger assumption on $\mathcal{H}^+$  remains however
 to be obtained for the evolution of generic initial data on $\Sigma$. Under this stronger assumption,
it is shown moreover that the Hawking mass diverges identically on $\mathcal{CH}^+$. In
terminology introduced by Poisson--Israel~\cite{PI1}, this is known as \emph{mass inflation}.

The still restrictive nature of the above results, even in the context of the spherically symmetric
toy model, reflects a difficulty in proving inextendibility statements at low regularity:
The differential structure at the Cauchy horizon
is not necessarily uniquely defined, and thus there could exist inequivalent extensions
which are more regular than those suggested naively.  This difficulty is already apparent
in the proof~\cite{SbierskiCzero} of the $C^0$-inextendibility of Schwarzschild.

Let us note finally that equations $(\ref{EMSF1})$--$(\ref{EMSF2})$ can be themselves
thought of as a simplified model of the Einstein--Maxwell--\emph{charged} scalar field
equations, where the scalar field is moreover allowed to have non-zero 
mass. In the latter system,  $\psi$ becomes complex-valued, the covariant derivatives in 
the definition of the energy momentum tensor $T_{\mu\nu}$ and in the equations of motion
are replaced by \emph{gauge} covariant derivatives $D_\mu$, 
including a gauge potential $A_\mu$
whose curvature is now $F_{\mu\nu}$, and a current density
$2\pi i e(\psi \overline{D_\mu \psi} - \overline{\psi} D_\mu\psi)$ is added to the right hand side of the first Maxwell
equation. In the massive case, a Klein--Gordon term is also added to the wave equation
and the definition of the energy momentum tensor.
For a general discussion of this system, see~\cite{kommemi}. 
The direct coupling of $F$ with $\psi$ means that the system now admits also  solutions
with a single end but non-trivial charge, which is the
physical case,
and thus allows for a setting where one can non-trivially
study both cosmic censorship conjectures and the instability of
Kerr-like Cauchy horizons. Including the mass term also allows for boson star-like solutions.
In the two-ended case, the Penrose diagram is
again as given in Figure~\ref{EMSFPendiag} (where  various boundary components are potentially empty),
whereas in the one-ended case, there is an additional (possibly empty) null component
arising from the centre of symmetry, which may or may not terminate at $\mathcal{I}^+$,
which then may or may not be complete~\cite{kommemi}. There are thus a host of open problems to understand
for this model!
Assuming, however, the non-vanishing and non-extremality of the asymptotic charge on $\mathcal{H}^+$, together with
appropriate upper and lower bounds on the decay  of $\psi$, the 
recent~\cite{VandeMoortel:2017ztd} has shown
(a) the non-emptiness of $\mathcal{CH}^+$, (b) the continuous extendibility of the metric
$g$ beyond $\mathcal{CH}^+$ and
(c) the $C^2$-inextendibility of $g$. For previous results in this direction,
see also~\cite{kommemithesis}.

\subsubsection{Local construction of vacuum weak null singularities without symmetry}
\label{ESSENTIAL!}
Are  the results of Section~\ref{modelprobintro} truly indicative of
the vacuum case, or are they due to the restrictive nature of 
 the spherically symmetric model considered?

\begin{figure}
\centering{
\def\svgwidth{6pc}
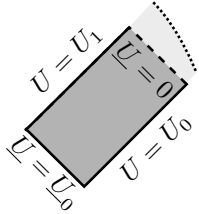}
\caption{Local existence allowing for a singular profile}\label{lukstheor}
\end{figure}

Before trying to understand this in our complicated
context of black hole interiors,
one can pose a simpler, purely local question.
Suppose one started with characteristic initial data  for $(\ref{INTROvace})$
on a null hypersurface
with the profile of a weak null singularity. Would this indeed
propagate as such, or would spacetime immediately break down forming a spacelike singularity?
A definitive answer to this question is given by the following:

\begin{theorem}[L.~\cite{LukWeakNull}]
\label{localconsLuk}
Consider vacuum initial data posed on a suitable
bifurcate null hypersurface such that the data are regular on the
ingoing part $[U_0,U_1]\times \{\underline{U}_0\}$ up to $U_1$,
while on  the outgoing part 
$\{U_0\}\times [\underline{U}_0, 0)$, 
the Christoffel symbols are assumed bounded in a norm consistent with 
the singular profile suggested by Theorem~\ref{LukOhte}
as $\underline{U}\to0$.
Then the maximal future development can still be covered by a full
rectangular domain
$[U_0,U_1]\times [\underline{U}_0,0)$
of double null coordinates $U$, $\underline{U}$,  for
which
 the metric is continuously extendible across $\underline{U}=0$.
See Figure~\ref{lukstheor}.

If the Christoffel symbols indeed have the  singular
profile  suggested by Theorem~\ref{LukOhte} on $U=U_0$, then this profile propagates
for all $U\in [U_0,U_1]$. Thus $\underline{U}=0$ can be thought
of as a weak null singularity.
\end{theorem}

See also  earlier work of Ori--Flanagan~\cite{FO}.

The proof of the above theorem relies on expressing the Einstein vacuum equations 
$(\ref{INTROvace})$ 
with respect to a double null foliation, and
moreover renormalising the resulting set of equations
so as to remove the components with the worst behaviour at $U=U_0$.
The result generalises previous work of~\cite{LR}.
We shall review the main elements of the proof in the context
of Section~\ref{firstremarksintro}, as they shall play a central role
in the proof of Theorem~\ref{PRWTO}. The precise assumption
on the initial data in Theorem~\ref{localconsLuk} 
is boundedness with respect to norms which generalise
$(\ref{relationof})$.

\begin{figure}
\centering{
\def\svgwidth{8pc}
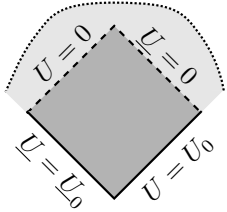}
\caption{Two singular profiles meet}\label{bifurcloc}
\end{figure}

In fact,  not only can one construct in the above way local
``weak null singularities'', but one can construct \emph{bifurcate} weak
null singularities:
\begin{theorem}[L. \cite{LukWeakNull}]
\label{BifConsLuk}
Consider initial data  such that the Christoffel symbols
have the singular profile    suggested by Theorem~\ref{LukOhte} on \underline{both}
ingoing and outgoing parts. Then, with an appropriate smallness condition, 
the maximal  future development is bounded in the future
by a bifurcate null singularity, across which the metric is globally continuously
extendible. See Figure~\ref{bifurcloc}.
\end{theorem}

As with Theorem~\ref{LukOhte}, a truly geometrical characterization of the
above weak null singularities is still lacking. 
For the continuous extensions which are constructed in~\cite{LukWeakNull}, 
indeed the Christoffel symbols
fail to be square integrable. Conjecturally this is true for \emph{any} continuous extension of
the metric, and this would provide a
geometric characterization of the singular nature of the boundary, justifying
thinking of it as ``terminal''  (see the discussion concerning Conjecture~\ref{ChrSCC}). 
At present, however,
the only geometric invariant statement that can be easily inferred is that the 
metric is inextendible with $g\in C^2$, which is clearly highly sub-optimal.

We emphasise again that the above two theorems are purely local. They show
that bifurcate, weak null singularities can occur in principle, but they
 in no way imply that weak null singularities necessarily
form in black hole interiors for $(\ref{INTROvace})$.  They do show,
however, that once they occur, they are stable to small perturbation.
We proceed to formulate this corollary below.

In view of the fact that our only examples of already-formed 
weak null singularities inside black holes concern the system
$(\ref{EMSF1})$--$(\ref{EMSF2})$, it is useful to consider this system
in place of $(\ref{INTROvace})$. 
Let us  first note then that analogues of Theorems~\ref{localconsLuk} and~\ref{BifConsLuk} 
indeed easily
generalise to the case where the vacuum equations $(\ref{INTROvace})$
are replaced by the system $(\ref{EMSF1})$--$(\ref{EMSF2})$, where the latter
are   considered without the assumption of spherical symmetry. 

\begin{figure}
\centering{
\def\svgwidth{12pc}
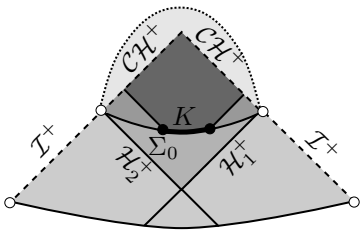}
\caption{Stability of bifurcate weak null singularities from \emph{compactly supported}
perturbations on $K\subset \Sigma_0$}\label{condstafig}
\end{figure}

Let $(\mathcal{M},g)$ now be a spacetime as in the last statement
of Theorem~\ref{ExtendibleModel}, i.e.~with a bifurcate Cauchy horizon $\mathcal{CH}^+$,
satisfying moreover the generic condition of Theorem~\ref{LukOhte}
ensuring that $\mathcal{CH}^+$ is singular.  Passing
to a Kruskal-like coordinate $\underline{U}$ with $\underline{U}=0$ on the
right segment of $\mathcal{CH}^+$, 
one can
thus think of the induced geometry on the null hypersurfaces $\{u_0\}\times [\underline{U}_0,0) $
as a special case of the initial data allowed by Theorem~\ref{localconsLuk}, the latter
generalised to apply to the system $(\ref{EMSF1})$--$(\ref{EMSF2})$.
Moreover, if $U$ is a conjugate Kruskal-like coordinate such that $U=0$ on the left segment
of $\mathcal{CH}^+$, then the induced geometry on $\{U_0\}\times [\underline{U}_0,0)\cup
[U_0, 0)\times\{\underline{U}_0\}$ satisfies the assumptions of Theorem~\ref{BifConsLuk}
(again, generalised to $(\ref{EMSF1})$--$(\ref{EMSF2})$),
if ${U}_0$ and $\underline{U}_0$ are both sufficiently late times.
Together with an appeal to Cauchy stability and the domain of dependence property, this allows
us to infer the following stability statement for the already-formed weak null singularities
of Theorem~\ref{LukOhte}:
\begin{corollary}[Stability of already-formed weak null singularities]
Let $(\mathcal{M},g, F_{\mu\nu},\psi)$ be as in the last statement
of Theorem~\ref{ExtendibleModel},
satisfying moreover the generic condition of Theorem~\ref{LukOhte}
ensuring that $\mathcal{CH}^+$ is singular,
 and let $\Sigma_0$ be
a hypersurface in the black hole interior as depicted in Figure~\ref{condstafig}. Consider a new initial data
set on $\Sigma_0$  which is a small perturbation of the induced data of $\mathcal{M}$
and which identically coincides with that data
outside a compact subset $K\subset \Sigma_0$, and consider the future development $\mathcal{M}'$
of the new
data on $\Sigma_0$. Then $\mathcal{M}'$
is again globally extendible continuously beyond a bifurcate Cauchy horizon,
which can moreover be interpreted as a weak null singularity.
\end{corollary}

\subsection{First remarks on the proof and guide to the paper}
\label{firstremarksintro}
We make some preliminary remarks concerning the proof of Theorem~\ref{PRWTO}, 
followed by a guide to the remainder of the paper. 

We begin in Section~\ref{indoublenullINTRO} where we shall introduce
the double null gauge which is central to our analysis. The logic of the proof
will rest on a bootstrap argument which is discussed briefly in Section~\ref{introbootstr}.
We then describe in Section~\ref{lukmeth} the estimates of~\cite{LukWeakNull},  and with this background
turn to the main global estimates
of the present paper in Section~\ref{baridiaeisagwgns}. We leave a brief discussion of certain auxiliary
issues of
the proof to Section~\ref{allapolla} and the issue of continuous extendibility beyond
the Cauchy horizon to Section~\ref{continuityofextenintro}.  
Finally, our guide to the rest of the paper
will be given in Section~\ref{outlineoftherest}.

%The remarks will be further fleshed
%out in Section~\ref{overviewsection} after additional notation has been introduced and a precise
%statement of Theorem~\ref{PRWTO} given.  
The reader may wish to return to this section upon reading the bulk of the paper.

\subsubsection{The Einstein equations in double null gauge}\label{indoublenullINTRO}
The proof of Theorem~\ref{PRWTO}
will employ a \emph{double null gauge} to write the Einstein
vacuum equations $(\ref{INTROvace})$ 
in what will turn out to be the entire maximal future evolution of
the data on $\Sigma_0$. The leaves of the foliation will be null hypersurfaces which
are level sets of
the ingoing null coordinate $u$ and outgoing null coordinate $\underline{u}$.
The metric then takes the form
\begin{equation}
\label{metricoftheform}
g=-2\Omega^2(du\otimes d\underline{u} +d\underline{u}\otimes du)+\gamma_{AB}(d\theta^A-b^Ad\underline{u})
\otimes (d\theta^B
-b^Bd\underline{u}),
\end{equation}
where $\Omega$ is a function, $\gamma_{AB}$ a metric on the spheres of intersection
$S_{u,\underline{u}}$ and $b^A$ a vector field tangential to the spheres.
This type of gauge has been applied most spectacularly in Christodoulou's 
formation of black holes theorem~\cite{Chr}. In the context of
the present problem, its use was in fact already suggested in~\cite{brady1995nonlinear}.
We have previously remarked that Pretorius--Israel~\cite{Pretorius} have shown
that the  Kerr solution itself can be covered by such a coordinate system in the region
of interest
as the coordinate range
\begin{equation}
\label{rangeforkerrintro}
\{-u_f+C_R<\underline{u}<\infty\}\cap \{u_f>u >-\infty\} \cap \{u+\underline{u} \ge C_R\},
\end{equation}
with $\Sigma_0$ corresponding to $u+\underline{u}=C_R$.
The normalisation should be thought of as a generalisation
of the Eddington--Finkelstein retarded and advanced coordinates $(\ref{EdFin})$
on Reissner--Nordstr\"om. 
The problem is thus to show both global existence and appropriate
stability in the range
$(\ref{rangeforkerrintro})$, allowing one to infer $C^0$-extendibility across
what will be a non-trivial Cauchy horizon corresponding to
$\underline{u}=\infty$.

Associated to such a double null foliation is a normalised null frame
$e_3=\partial_u$, and $e_4=\Omega^{-2}(\partial_{\underline{u}}+b^A\partial_{\theta^A})$, completed by a coordinate frame $e_A=\partial_{\theta^A}$ for the spheres
$S_{u,\underline{u}}$.
The content of the Einstein equations in this gauge
can be viewed as a coupled system of the so-called \emph{null structure
equations} (relating both the components of the metric itself and the so-called
Ricci coefficients (including the second fundamental
form of the foliation), decomposed into this null frame) and \emph{the Bianchi equations} satisfied
by the null-frame decomposed components of curvature.

One can compare the above system of equations with the more general
Newman--Penrose formalism which expresses the Einstein equations~$(\ref{INTROvace})$ 
in terms of a general null frame~\cite{newman1962approach}. 
In our case here, it is essential that
the null frame is tied explicitly
to  a double null foliation. This allows for the null structure equations to be estimated as \emph{transport equations} on the
constant $u$ cones and $\underline{u}$ cones or
\emph{elliptic equations} on the $S_{u,\underline{u}}$ spheres.
The Bianchi equations can be treated as 
hyperbolic systems, similar to the Maxwell system $(\ref{EM2})$.

An example of a transport equation is the well-known Raychaudhuri
equation
\begin{equation}
\label{exampleone}
\nab_4 \trch+\frac 12 (\trch)^2=-|\chih|^2
\end{equation}
relating two Ricci coefficients, the outgoing expansion $\trch$ and the shear $\chih$,
whereas an elliptic equation is given by
\begin{equation}
\label{exampletwo}
\div\chih=\frac 12 \slashed{\nabla} \trch - \eta \cdot (\chi - \trch\gamma) -\beta,
\end{equation}
relating the shear $\chih$ with the curvature component $\beta_A=\frac12R(e_A,e_4,e_3,e_4)$.
The one-form $\eta_A$ appearing above
is an additional  Ricci coefficient.
Here and in the formulas that follow,
$\slashed\nabla_3$ and $\slashed\nabla_4$ are
covariant differential operator in the direction of $e_3$ and $e_4$, respectively
while $\slashed\nabla$, $\div$ denote  differential operators
associated to the $S_{u,\underline{u}}$-spheres. 
Note that the above Ricci coefficients both also satisfy transport equations
in the $e_3$ direction:
\begin{align}
\label{newtransfortr}
\nab_3 \trch+\frac1 2 \trchb \,  \trch &=2\omegab \trch+2\rho- \chih\cdot\chibh+2\div \eta+2|\eta|^2,\\
\label{newtransforshear}
\nab_3\chih+\frac 1 2 \trchb \chih&=\nab\widehat{\otimes} \eta+2\omegab \chih-\frac 12 \trch \chibh +\eta\widehat{\otimes} \eta.
\end{align}
 In the formula above,
we see the Ricci coefficient $\underline\omega$, defined by
\begin{equation}
\label{introdefomegeb}
\omegab=-\nab_3 (\log\Omega)
\end{equation}
as well as various conjugate quantities to the ones
we have defined, where the roles of $e_3$ and $e_4$ are interchanged with various
sign conventions.

The curvature component
$\beta$ together with the couple $(\rho=\frac14R(e_4,e_3,e_4,e_3), 
\sigma=\frac14{}^*R(e_4,e_3,e_4,e_3))$ can be thought of as a ``Bianchi pair'' related
by the equations
\begin{equation}
\label{hypexample1}
\nab_3\beta+\trchb\beta=\slashed{\nabla}\rho +^*\slashed{\nabla}\sigma  + 2\omegab \beta+2\chih\cdot\betab+3(\eta\rho+^*\eta\sigma),
\end{equation}
\begin{equation}
\label{hypexample2}
\nab_4\sigma+\frac 32\trch\sigma=-\div^*\beta+\frac 12\chibh\wedge \alpha-\zeta\wedge\beta-2\etab\wedge\beta, 
\qquad
\nab_4\rho+\frac 32\trch\rho=\div\beta-\frac 12\chibh\cdot\alpha+\zeta\cdot\beta+2\etab\cdot\beta.
\end{equation}
Here $*$ denotes Hodge dual.
In estimating $(\ref{hypexample1})$--$(\ref{hypexample2})$ as a hyperbolic system 
one exploits the fact that the angular operators
$(\slashed{\nabla}, ^*\slashed{\nabla})$ and
$\div$ can be viewed as $L^2$ adjoints on the spheres $S_{u,\underline{u}}$.

As first understood in the context of the monumental proof
of the stability of Minkowski space by Christodoulou--Klainerman~\cite{CK},
use of $(\ref{exampleone})$ and $(\ref{exampletwo})$ together (as well as related pairs
of transport and elliptic equations involving  $\trchb$ and the quantities
$\mu$, $\underline{\mu}$ and $\ombs$ to be discussed later) 
allows one to eventually estimate 
all Ricci coefficients at one level of differentiability
greater than curvature. Briefly,
the point is that neither curvature nor first derivatives of Ricci coefficients appear on 
the right hand side of $(\ref{exampleone})$ and thus, integration of $(\ref{exampleone})$
does \emph{not} lose a derivative, in contrast to the generic transport equation, like
$(\ref{newtransfortr})$--$(\ref{newtransforshear})$, for which
one sees $\div \eta$ and $\nab\widehat{\otimes} \eta$. On the other
hand, the elliptic equation $(\ref{exampletwo})$ \emph{gains} a derivative with respect
its right hand side, where the curvature term $\beta$ is present.

\subsubsection{The bootstrap argument and the logic of the proof}
\label{introbootstr}

\begin{figure}
\centering{
\def\svgwidth{9pc}
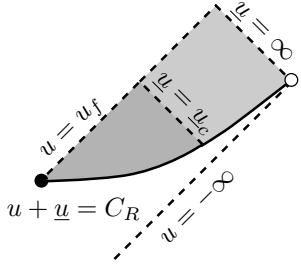}
\caption{The region defined by $(\ref{rangeforkerrintrofinite})$}\label{bootlabel}
\end{figure}

As with many non-linear analyses for the Einstein vacuum equations $(\ref{INTROvace})$,
the proof of Theorem~\ref{PRWTO} will proceed via the method of continuity (or ``bootstrap'').
One considers the set $\mathfrak{I}\subset (-u_f+C_R,\infty)$ of $\underline{u}_c\in (-u_f+C_R,\infty)$ such that there
exists a solution of the form $(\ref{metricoftheform})$ on 
\begin{equation}
\label{rangeforkerrintrofinite}
\{-u_f+C_R<\underline{u}<\underline{u}_c\}\cap \{u_f>u >-\underline{u}_c+C_R\} \cap \{u+\underline{u} \ge C_R\},
\end{equation}
satisfying a number of estimates, known in this context as ``bootstrap assumptions''
(see already $(\ref{alreadyboot})$ below). This region is depicted as the darker
shaded region in Figure~\ref{bootlabel}.

This set $\mathfrak{I}$ 
is non-empty by a suitable local existence result and it will be manifestly closed by the nature
of the inequalities. Finally, we shall show that the set is open by ``improving the 
bootstrap assumptions'', i.e.~by (a)
proving estimates on all quantities, including
stronger estimates than the assumed bootstrap assumptions, 
and then (b) applying
local existence again to obtain that the stronger estimates hold in a slightly larger
region. It will follow that $\mathfrak{I}= (-u_f+C_R,\infty)$ and thus the
solution exists on the entire domain $(\ref{rangeforkerrintro})$.

In Section~\ref{baridiaeisagwgns} below, we will focus entirely on (a), i.e.~on proving
estimates on the solution in a region $(\ref{rangeforkerrintrofinite})$ on which
the solution is already known to exist. We will return to discuss
the completion of the bootstrap argument 
in Section~\ref{allapolla}.

\subsubsection{The method of~\cite{LukWeakNull}: Renormalisation, local weights
and null structure}\label{lukmeth}

In anticipation of the fact that one expects (cf.~the discussion in
Section~\ref{INTROprework})  that the Cauchy horizon $\mathcal{CH}^+$, 
although $C^0$ stable, will generically represent
a  weak null singularity, the methods introduced in~\cite{LukWeakNull} described in Section~\ref{ESSENTIAL!}
will play an essential role in the proof. For otherwise, the estimates would be
inconsistent with  controlling the solution up to $\mathcal{CH}^+$.

Let us introduce briefly some of the main ideas of~\cite{LukWeakNull},
directly in the context of Theorem~\ref{BifConsLuk}. The proof again employs a double null foliation,
but we shall denote the coordinates by $U$ and $\underline{U}$ in place
of $u$ and $\underline{u}$
to emphasise that these coordinates  naturally have finite range 
$[U_0,0)\times[\underline{U}_0,0)$.
\begin{itemize}
\item
{\bf Renormalisation.} To estimate solutions up to potentially singular null fronts
$U=0$ and $\underline{U}=0$ in Theorem~\ref{BifConsLuk},
one must renormalise the
Bianchi equations by eliminating the ``most singular components''
$\alpha= R(e_A, e_4, e_B, e_4)$, $\underline{\alpha}=R(e_A,e_3,e_B,e_3)$ and
replacing
the curvature components $\rho$ and $\sigma$ discussed in Section~\ref{indoublenullINTRO} with 
\[
\rho\rightsquigarrow K\doteq -\rho+\frac12\chih\cdot\chibh-\frac14\trch\trchb,\qquad 
\sigma \rightsquigarrow \sigmac=\sigma+\frac12\chibh\wedge\chih.
\]
The significance of these quantities $K$ and $\sigmac$
is that $K$ represents the Gaussian curvature
of the spheres $S_{U,\underline{U}}$ and $\sigmac$ the
curvature of the normal bundle, respectively, and these quantities remain
regular up to the null boundary. These renormalisations in fact first appeared
in~\cite{LR, LR2}.
Remarkably, this renormalisation respects the structure of Section~\ref{indoublenullINTRO}, 
i.e.~the renormalised system can still be viewed as a set of coupled elliptic, transport and
hyperbolic equations. Now the elliptic estimates through $(\ref{exampletwo})$ appear
absolutely  essential, not just for the purpose of gaining regularity, but for the 
system to close.\footnote{This is in contrast, for instance,
with~\cite{vacuumscatter}, where elliptic estimates are not used. This in turn was possible
because $\chih$ could be estimated by a transport equation involving $\nab_4\chih$
with $\alpha$ appearing on the right hand side.}
\item
{\bf Local weights.}
Even though the worst behaving components are now removed,
the remaining components, e.g.~the curvature component $\beta$ and the
shear $\hat\chi$ still are not expected to have finite ``local'' energy at 
a weak null singular front. (This is in contrast to the situation in~\cite{LR, LR2}, referred
to above, where, all Ricci coefficients were locally square integrable.)
One can only hope to prove weighted estimates with respect to the local ``regular'' coordinate $\underline{U}$,  degenerating at $\underline{U}=0$, with weight which is integrable at
$\underline{U}=0$ when squared. 
An example of such a weight is
\begin{equation}
\label{localweight}
f(\underline{U})=(-\underline{U})^{\frac12}|\log(-\underline{U})|^{\frac12+\delta}
\end{equation}
 appearing
in expression $(\ref{relationof})$. 
The finiteness of
\[
\|f(\underline{U})\hat\chi\|_{L^2_{\underline{U}}L^2(S_{U, \underline{U}})}
\]
on an outgoing hypersurface $U=c$ is compatible (say in the case
of $(\ref{localweight})$) with the pointwise
behaviour 
\[
\hat\chi\sim (-\underline{U})^{-1}|\log (-\underline{U})|^{-1-2\delta}.
\]
Note that this is in turn compatible with the finiteness of the $L^1$ norm
\[
\|\hat\chi\|_{L^1_{\underline{U}}L^2(S_{U, \underline{U}})},
\]
which is to be expected if the metric is to be continuously extendible
beyond $\underline{U}=0$.

For a linear equation like $(\ref{linearwavePOLY})$,
such weighted estimates are readily seen to locally propagate.
For the non-linear Einstein equations $(\ref{INTROvace})$, however,
the question of the local propagation of 
such weighted estimates becomes highly non-trivial, dependent
on the precise \emph{null structure in the nonlinearities}.

\item 
{\bf Schematic notation and null structure.}
In order to describe the above structure, it is useful to first introduce a schematic 
notation.\footnote{The notation and conventions of~\cite{LukWeakNull} are slightly
different. Here we follow notation to be used in the present paper.}
One can group (a subset of) the Ricci coefficients into families as follows: 
\[
\psi\in\{\eta,\etab\},\quad \psi_H\in \{\trch,\chih\},\quad \psi_{\Hb}\in\{\trchb,\chibh\}
\]
according to
their expected singular behaviour.  
%component naturally appear in fluxes on outgoing, ingoing hypersurfaces, or both.
All terms $\psi_H$ have expected behaviour similar to $\hat\chi$ (and $\beta$)
and thus can be estimated either in $L^1_{\underline{U}}L^2(S)$ on outgoing hypersurfaces
$U=c$, 
or with the weight $(\ref{localweight})$ in $L^2_{\underline{U}}L^2(S)$. Analogous statements
hold for $\psi_{\underline{H}}$ with ingoing hypersurfaces  $\underline{U}=c$
replacing outgoing ones. The terms $\psi$, on the other hand, can be estimated
without degeneration.
%For components like $\psi_H$, we may
%represent its transport equation schematically as
%\[
%\slashed\nabla_3 \psi_H =   K + \slashed\nabla\psi + \psi\psi +\psi_H\psi_{\underline{H}}
%\]
Energy identities for $\beta$ lead in particular to 
having to estimate spacetime integrals of the form
\begin{equation}
\label{oftheformhere}
\| f^2(\underline{U}) (\beta \psi_{\underline{H}}\beta +\beta\psi_H\underline{\beta}
+\beta\psi K) \|_{L^1_UL^1_{\underline{U}}L^1(S)}
\end{equation}
All the terms appearing above can be seen to be admissible from the point
of view of integrability.
In contrast, terms like $\beta \psi_H\beta$ or $\beta \psi_{\underline{H}} \underline\beta$  would not have been controllable. \emph{It is thus precisely the absence of such terms 
which captures the ``null structure'' essential for this argument.}
\end{itemize}

Of course, the story is more complicated, in that the equations need
to be sufficiently commuted so that indeed trilinear terms can be controlled
by top order $L^2$ estimates and Sobolev inequalities. 
We defer discussion of these issues to the setting of our actual problem in
Section~\ref{baridiaeisagwgns} below.

\subsubsection{The main estimates}
\label{baridiaeisagwgns}
Already in the context of the
linear wave equation $(\ref{linearwavePOLY})$ discussed in Section~\ref{LWERNK}, 
it is clear that
weighted estimates with respect to the (infinite-range)
coordinates $u$ and $\underline{u}$
are necessary in view of the global aspects of the problem special to the black hole interior.

 {\bf \emph{The remarkable fact, central to the
 present work, is that  these two systems of weights, those introduced in the study of
 the wave equation as discussed in Section~\ref{LWERNK} and
 those introduced in~\cite{LukWeakNull} 
 as discussed in Section~\ref{ESSENTIAL!}, are in fact compatible, and
 the null condition is reflected in both.}}

Indeed, we have already seen that 
the polynomially growing weight in  $\underline{u}$, coupled
with the degenerate coordinate derivative $\partial_{\underline{u}}$ and degenerate
one-form $d\underline{u}$,
when 
looked at for fixed $u$, behaves precisely like $(\ref{relationof})$.

%Several of these weights are already suggested by
%the study of the wave equation on Reissner--Nordstr\"om and Kerr
%discussed already in Section~\ref{LWERNK}.
We outline below the main ideas in 
adapting the global weighted estimates of $(\ref{linearwavePOLY})$, combined
with the renormalisation of Section~\ref{lukmeth}, to the global analysis of
 the Einstein vacuum equations $(\ref{INTROvace})$ in the black hole interior.

\begin{itemize}
\item{\bf Differences, angular commutation and reduced schematic notation.}
First of all, it is now  differences which must be estimated, i.e.~one must subtract
from each quantity, e.g.~the outgoing expansion $\trch$ or the curvature component 
$\beta$, its Kerr value, considering
\begin{equation}
\label{thequantintro}
\widetilde{\trch}\doteq \trch-(\trch)_{\mathcal{K}}, \qquad \widetilde{\beta}\doteq \beta-\beta_{\mathcal{K}}.
\end{equation}
We will extend the schematic notation of Section~\ref{lukmeth} 
to differences by defining 
\[
\widetilde\psi \in  \{\widetilde{\eta}, \widetilde{\underline\eta}\} \, ,  \qquad
\widetilde{\psi}_{\underline{H}} \in \{\ttrchb, \widetilde{\underline{\hat\chi}}\}\,  , \qquad
\widetilde{\psi}_H \in  \{\widetilde{\trch}, \widetilde{\hat\chi}\} 
\]
and defining also a schematic notation for difference quantities
at the level of the metric itself:
\[
\tg \in  \{\gamma_{AB}-(\gamma_{AB})_{\mathcal{K}},\ldots, \log\Omega -\log
\Omega_{\mathcal{K}}\}.
\]
We note already that the metric component $\tb$ and the Ricci coefficient difference
$\widetilde{\underline\omega}$  defined from $(\ref{introdefomegeb})$
do not have a schematic representation and must always 
be dealt with explicitly.
Moreover, for our estimates to close, we will need to control
quantities in a higher $L^p$ norm via Sobolev inequalities (see also
Section~\ref{allapolla} below!), so we will
have to commute (in the
angular directions $\slashed\nabla$)
the null structure equations for the Ricci coefficients up
to three times, and the Bianchi equations up to two times, i.e.~we 
shall consider quantities $\slashed\nabla^3\widetilde\psi$,\ldots
$\slashed\nabla^2\widetilde\beta$,\ldots.\footnote{Compare with the role of commutation 
in obtaining the $L^\infty$ estimate $(\ref{sobolev})$. Note that in view of the
lack of exactly commuting operators $\mathscr{W}_i$, it is more natural to commute
tensorially here, just as in~\cite{LR, LR2, vacuumscatter}.
Note that $\slashed\nabla^i\psi$ is a higher rank $S_{u,\underline{u}}$-tensor
than $\psi$.}
After commuting the equations, we shall moreover adopt the convention
that lower order terms which can be controlled  in $L^\infty$
will not be written explicitly. The system for differences will retain
the structural properties of the renormalised system, but gives
rise to additional ``inhomogeneous'' terms.
We shall call this system the ``reduced schematic equations''.
They are derived in Section~\ref{sec.rse}.

For instance, 
the equation for $\slashed\nabla^3\ttrch$  arising from
taking differences and commuting $(\ref{exampleone})$ can be written
in reduced schematic form as
\begin{equation}
\label{reltoellipticstr}
\nab_4\nab^3\ttrch \eqrs \sum_{i_1+i_2+i_3\leq 3}(1+\nab^{i_1}\tg+\nab^{\min\{i_1,2\}}\tp)(1+\nab^{i_2}\tpH)(\nab^{i_3}(\tpH,\tg)+\Om_\Ke^{-2}\nab^{i_3}\tb),
\end{equation}
where $\eqrs$ denotes that the relation holds in the sense of our conventions described
above.
As we shall see, the quantity $\ttrch$ has a special role in the context of 
not losing differentiability for the second fundamental form (cf.~the discussion
at the end of Section~\ref{indoublenullINTRO}). 
We remark already that the equation $(\ref{reltoellipticstr})$ does not contain
top-order  terms (e.g.~terms like $\slashed\nabla^3\psi$ or $\slashed\nabla^2 \beta$ on its 
right hand side). 

\item{\bf Global weights in $\Omega_{\mathcal{K}}$, $\varpi^N$, $u$ and $\underline{u}$.}
In contrast to the case of the linear scalar wave equation $(\ref{linearwavePOLY})$,
there is no longer a 
standard Lagrangian structure naturally giving rise to weighted estimates via
vector fields such as $(\ref{vectorfield})$, 
so the weights must be put in by hand in  a more \emph{ad hoc} 
fashion in the context of direct integration by parts.
For instance, the curvature difference quantity $\widetilde\beta$ of
$(\ref{thequantintro})$ is estimated in the weighted norm
\begin{equation}
\label{whenestima}
%\|\underline{u}^{\frac12+\delta}\varpi^N\Omega^2\widetilde{\trch} \|_{L^\infty_uL^\infty_{\underline{u}}L^2(S_{u,v})}, \qquad
\|\underline{u}^{\frac12+\delta}\varpi^N\Omega^2_{\Ke}\widetilde{\beta} \|_{L^\infty_uL^2_{\underline{u}}L^2(S_{u,v})}.
\end{equation}
Comparing the above expression to the first expression in $(\ref{vectorfield0})$
which arose as a flux of the vector field $(\ref{vectorfield})$, we
note that $\Omega^2_{\Ke}e_4$ is in some sense equivalent to $\partial_{\underline u}$, and we
see the quantity $r^N$ is replaced by $\varpi^N$, appropriate
for Kerr, as discussed in Section~\ref{LWERNK}.

\item{\bf The controlling energies $\mathcal{N}$.}
As remarked already, the 
top-order estimates will be at the level of two angular commutations of curvature,
e.g.~$\slashed\nabla^2\widetilde\beta$.
In view of the elliptic estimates mentioned already in Section~\ref{indoublenullINTRO} 
(see also below), 
we will be able to estimate all terms $\slashed\nabla^3\widetilde\psi$,
$\slashed\nabla^3\widetilde\psi_H$, $\slashed\nabla^3\widetilde\psi_{\underline{H}}$,
from top-order curvature terms, \emph{and conversely}. 
This motivates defining a certain master energy, \emph{defined with reference only
to the metric and Ricci coefficients},
which will finally be shown to control the entire system. 

The master energy is  in
fact best separated into three parts, some of whose terms are displayed below:
\begin{eqnarray}
\label{Nhypintro}
\mathcal{N}_{hyp}&=&\|\underline{u}^{\frac12+\delta}\varpi^N\Omega_{\mathcal{K}}^2
\slashed\nabla^3\tpH\|^2_{L^\infty_u L^2_{\underline{u}} L^2(S)}
+\| |u|^{\frac12+\delta}\varpi^N\slashed\nabla^3\tpHb  \|^2
_{L^\infty_{\underline{u}} L^2_{u} L^2(S)} +\cdots\\
\nonumber
\mathcal{N}_{int}&=&\|\underline{u}^{\frac12+\delta}\varpi^N\Omega_{\mathcal{K}}^3
\slashed\nabla^3 \tpH \|^2_{L^2_uL^2_{\underline{u}}L^2(S)}+
\|{|u|}^{\frac12+\delta}\varpi^N\Omega_{\mathcal{K}}
\slashed\nabla^3 \tpHb \|^2_{L^2_uL^2_{\underline{u}}L^2(S)}\\
\label{Nintintro}
&&\qquad
+\|{\underline{u}}^{\frac12+\delta}\varpi^N\Omega_{\mathcal{K}}
\slashed\nabla^3\tp\|^2_{L^2_uL^2_{\underline{u}}L^2(S)}
\cdots\\
\label{Nsphintro}
\mathcal{N}_{sph}&=&\|\slashed\nabla^3 \tg \|^2_{L^\infty_{\underline{u}}L^\infty_u
L^2(S)} +  \cdots + \|%\slashed\nabla^2
\tpHb \|^2_{L^1_uL^\infty_{\underline{u}}L^2(S)}
+\cdots.
\end{eqnarray}
We note that the energy $\mathcal{N}_{hyp}$ of $(\ref{Nhypintro})$, 
which consists of null hypersurface integrals, controls in particular all
the top order flux terms of curvature, for instance $(\ref{whenestima})$
where $\widetilde\beta$ is replaced by $\slashed\nabla^2\widetilde\beta$.
See already Section~\ref{sec:def.energies} for the complete definitions of
these energies.

The energy $\mathcal{N}_{int}$ of $(\ref{Nintintro})$
consists of spacetime integrals. 
The integrals in the first line of $(\ref{Nintintro})$  are typically
generated by good-sign bulk terms in energy identities which arise
from the insertion of a good   weight like $\varpi^N$ above, cf.~$(\ref{goodbulkoneforwave})$
that arose in  estimates for the wave equation $(\ref{linearwavePOLY})$. 
The term
displayed in the second line of $(\ref{Nintintro})$ is analogous
to $(\ref{goodbulktwoforwave})$. As with $(\ref{linearwavePOLY})$, 
the latter  term
miraculously naturally appears in various
estimates with a good sign.
Note, however, the  different $\Omega_{\Ke}$ weights comparing $(\ref{Nhypintro})$ with $(\ref{Nintintro})$, signifying that $(\ref{Nintintro})$ is more degenerate
and thus gives less control.

Finally,  the energy $\mathcal{N}_{sph}$ 
of $(\ref{Nsphintro})$ gives higher $L^p$ control on \emph{lower} order terms.
We remark already that on lower order terms, the order of $L^\infty$ norm is inside, providing
a stronger estimate than otherwise. See for instance the second term of $(\ref{Nsphintro})$.
This will be crucial, because for the top order terms like the first two terms of
$(\ref{Nhypintro})$, the $L^\infty$ norm will have to be taken outside.

We note that 
the  bootstrap assumption (cf.~Section~\ref{introbootstr}) can in fact
be expressed entirely in terms of $(\ref{Nsphintro})$
as 
\begin{equation}
\label{alreadyboot}
\mathcal{N}_{sph}\le \epsilon.
\end{equation}
Given $(\ref{alreadyboot})$, our estimates for the null structure equations
and the Bianchi identities (see below!)~will allow us to estimate
\begin{equation}
\label{obtainingestimates}
\mathcal{N}_{hyp} \lesssim \epsilon^2 , \qquad \mathcal{N}_{int} \lesssim \epsilon^2
\end{equation}
provided that an initial data quantity $\mathcal{D}$
satisfies
\begin{equation}
\label{initialdatasmallintro}
\mathcal{D}\le\epsilon^2,
\end{equation}
for $\epsilon$ sufficiently small.
Finally, this will allow us to infer (using  the null structure
equations together with $1$-dimensional Sobolev inequalities) that
\begin{equation}
\label{hereimproveboo}
\mathcal{N}_{sph}\lesssim \epsilon^2,
\end{equation}
which, for $\epsilon$ sufficiently small, improves $(\ref{alreadyboot})$.
This will then complete part (a) of  the bootstrap argument
as outlined in Section~\ref{introbootstr}. 

We thus turn in the remainder of this section to a discussion of 
how to obtain $(\ref{obtainingestimates})$.

\item{\bf Transport estimates for Ricci coefficients and wave-type estimates for curvature.}
In general, the Ricci 
coefficients will be estimated via integrating transport equations while
the curvature will be estimated from the Bianchi identities.
Since the Bianchi identities  encode the hyperbolic aspect of the Einstein 
vacuum equations $(\ref{INTROvace})$ expressed in a double null gauge,
they are in some sense most analogous to the linear wave equation $(\ref{linearwavePOLY})$,
their tensorial structure not withstanding. 
In particular, energy estimates with weights related to those described in Section~\ref{LWERNK}
(see already Section~\ref{sec.energy})
would indeed allow one to obtain $(\ref{obtainingestimates})$ for \emph{some} 
of the top order terms
displayed in $(\ref{Nhypintro})$ and $(\ref{Nintintro})$,
provided that certain ``error terms'' can be controlled,
which can be classified into various types
$\mathcal{E}_1$, $\mathcal{E}_2$, $\mathcal{T}_1$, $\mathcal{F}_2$.
(These  are defined in Section~\ref{sec.error.def}.)

The analogous estimates for Ricci coefficients are simpler (and precede
the treatment of curvature in the paper; see already
 Section~\ref{sec.int.est}), following by weighted
estimates applied directly to transport equations. 
This again allows for control of all of the lower order estimates
for Ricci coefficients displayed in $(\ref{Nhypintro})$ and $(\ref{Nintintro})$,
conditional on being able to
recover estimates for error terms. Before turning to 
the top order estimates for the remaining Ricci coefficients let us
discuss how some of these error terms are estimated.

\item{\bf Estimating error terms.}
Let us examine briefly two 
types of error terms that arise in the process
of estimating the second term of $(\ref{whenestima})$ at highest
order (where $\widetilde\beta$ is replaced by $\slashed\nabla^2\widetilde\beta$),
via the Bianchi identities. See already Section~\ref{sec:error}.

We have already seen in Section~\ref{lukmeth}
one type of error term which arises, namely, those arising from the nonlinearities.
An example of such a term is:
\begin{equation}
\label{examplesofnlt}
\|\underline{u}^{1+2\delta}\varpi^{2N}\Omega_{\mathcal{K}}^4 \slashed
\nabla^3 \tpH \slashed\nabla^3 \tpH \tpHb \|_{L^1_u
L^1_{\underline{u}}L^1(S)}.
%\qquad
%\|\underline{u}^{1+2\delta}\varpi^{2N}\Omega_{\mathcal{K}}^4 \slashed
%\nabla^3\tilde{\psi}_H\tilde{\psi}_H\slashed\nabla^3\tilde{\psi}_{\underline{H}}\|_{L^1_u
%L^1_{\underline{u}}L^1_{S_{u,v}}}
\end{equation}
Note that this is analogous to the first term of $(\ref{oftheformhere})$.
In the notation of Section~\ref{sec.error.def}, 
this term arises from one of the error terms
contained in  $\mathcal{T}_2$.
It can be easily bounded
by the first term of $(\ref{Nhypintro})$ and the second term of $(\ref{Nsphintro})$,
noting the importance of the order of the $L^1L^\infty$ norm in $(\ref{Nsphintro})$.
Again, we see here a manifestation of the ``null condition''.
(See already the proof of Proposition~\ref{T2.bd}
of Section~\ref{sec.error.T1.T2}.)

Another type of error term which arises are linear inhomogeneous terms that are generated
from the process of taking differences. An example of such a term 
that must be estimated is  
\begin{equation}
\label{mustbeestim}
\|\underline{u}^{\frac12+\delta}\varpi^N \Omega_{\mathcal K}\slashed\nabla^3 \tpHb
\|_{L^2_uL^2_{\underline{u}}L^2(S)}.
\end{equation}
In the notation of Section~\ref{sec.error.def}, 
this term arises from one of the
error terms contained in $\mathcal{E}_1$.
(Note that this term is not directly controlled by the second term
of $\mathcal{N}_{int}$ in $(\ref{Nintintro})$ since $|u| \lesssim \underline{u}$.)
For  $(\ref{mustbeestim})$, we must divide the region of integration into
$-u>\frac12\underline{u}$ and $-u<\frac12\underline{u}$. In the first region, we may replace $\underline{u}$ by $|u|$ and estimate
this term by the second term of $(\ref{Nintintro})$. 
In the second region, we use the bound
\[
\sup_{S_{u,\underline{u}}}|\Omega_{\mathcal{K}}|(u,\underline{u})
\lesssim e^{-\frac12\kappa_-\underline{u}}, \qquad
\sup_{S_{u,\underline{u}}}\underline{u}^{\frac12+\delta}|\Omega_{\mathcal{K}}|
(u,\underline{u}) \lesssim e^{-\frac14\kappa_- \underline{u}}
\]
to bound $(\ref{mustbeestim})$ restricted to this region by
\[
\||u|^{\frac12+\delta}\varpi^N \slashed\nabla^3 \tpHb 
\|_{L_{\underline{u}}^\infty L^2_uL^2(S)}\|e^{-\frac14\kappa_-
\underline{u}}\|_{L^\infty_uL^2_{\underline{u}}},
\]
which can now be bound by $(\ref{Nhypintro})$.
Here $\kappa_-= \frac{r_+-r_-}{2(r_-^2+a^2)}$ denotes the surface gravity
of the Kerr Cauchy horizon. (See already the proof of 
Proposition~\ref{E1.bd}
of Section~\ref{sec.error.E1.E2}.)

\item{\bf Top order estimates for Ricci coefficients.}
To obtain $(\ref{obtainingestimates})$ for the remaining top-order quantities, i.e.~those
that cannot be directly
related to curvature, is tricky  because one must at the same time
use the structure described at the end of Section~\ref{indoublenullINTRO} and still be
be compatible with the global weights.  See already
Section~\ref{sec.elliptic}.

For instance, for top order estimates
we must estimate the quantity $\ttrch$
via the commuted Raychaudhuri equation $(\ref{reltoellipticstr})$.

To handle the conjugate quantity $\ttrchb$ via its analogous Raychaudhuri equation,
we  re-express that equation in terms of the geodesic vector field
to avoid the worst coupling with $\underline\omega$.  Recall that this latter
Ricci coefficient does not have a schematic representation,
and has anomalous 
decay properties. (The renormalised form of the transport
equation satisfied by $\ttrchb$ is given in Proposition~\ref{nab3trchb.eqn} and
is used in the proof of Proposition~\ref{prop.nabettrchb.1}.)
\end{itemize}

The importance of the ``null condition'' for either long term
existence or local regularity issues has been made clear many times
in previous work on the Einstein equations $(\ref{INTROvace})$.
The former role is most famously represented  by 
the monumental proof of the stability of Minkowski space~\cite{CK},
while  the latter role  can be seen in the works~\cite{LR, LukWeakNull}, 
discussed above, as well as 
the proof of the bounded $L^2$ curvature conjecture~\cite{L21}. 
In this context, it is interesting to remark how in the present work, 
our weights in $u$ and $\underline{u}$ capture \emph{both} roles of null structure,
seamlessly transitioning as $\underline{u}\to\infty$ from their role in proving sufficiently 
long-time existence to their role in ensuring the propagation of low regularity.

\subsubsection{Auxiliary issues}
\label{allapolla}
Let us briefly comment on certain auxiliary issues in the proof which 
we have not emphasised so far.

\begin{itemize}
\item{\bf Initial data and local existence.}
An initial data set for $(\ref{INTROvace})$ constitutes a triple $(\Sigma_0,\hat{g},\hat{k})$
satisfying the Einstein constraint equations (see already $(\ref{constraints})$).
The first version of our theorem (stated as
Theorem~\ref{main.thm.spacelike.formulation})
will give
a smallness condition expressed solely in terms of the geometric initial data
$\hat{g}$ and $\hat{k}$. 
Applying well posedness in the smooth category~\cite{geroch}, together
with ode arguments, we will be able to construct locally
a double null gauge $(\ref{metricoftheform})$ which will be the non-emptiness step of the bootstrap argument. 
We can also globally re-express the initial data in terms of the quantities associated to $(\ref{metricoftheform})$.
This will allow us to give a more precise version of the theorem (stated
as Theorem~\ref{main.quantitative.thm}), identifying the smallness
condition $(\ref{initialdatasmallintro})$, in terms directly related to the
quantities $(\ref{Nhypintro})$, $(\ref{Nsphintro})$ appearing in estimates.

\item{\bf Sobolev  and ellipticity constants.}
The estimates described  in Section~\ref{baridiaeisagwgns} 
require being able to apply  the
Sobolev inequality to handle lower order terms, 
and exploiting ellipticity of equations like $(\ref{exampletwo})$. This in turn requires
quantitative control of  various Sobolev constants with respect to the induced
geometry of the spheres
$S_{u,\underline{u}}$, and 
some
a priori input on the behaviour of the metric will be necessary.  The 
bootstrap assumption $(\ref{alreadyboot})$ will suffice for this, 
and thus bounding these constants
is in fact the first task of the proof, handled already in Section~\ref{secbasic}.  
Note that the conventions related to the
reduced schematic equations  hide
many terms estimated by Sobolev, for instance,
already in writing $(\ref{reltoellipticstr})$.

\item{\bf Propagation of higher regularity.}
As in Christodoulou's formation of black holes work~\cite{Chr}, 
for showing  the extendibility of our solution  (in the
context of step (b) of the bootstrap argument described in Section~\ref{introbootstr})
we will appeal to a local existence result in the $C^\infty$
category, just as we have done above to show that the evolution is non-empty.\footnote{We shall
need both local existence for the standard Cauchy problem, 
for which we quote~\cite{geroch}, but also for a characteristic initial value problem.
One could quote~\cite{Rendall}, but it will in fact be more convenient to appeal to a local
existence for the equations expressed already in the double null gauge,
as this ensures that one can obtain a whole neighbourhood of an initial ingoing 
hypersurface. See already Section~\ref{sec:continuity}.}
On the other hand, the main estimates described in Section~\ref{baridiaeisagwgns} 
close at the finite regularity defined by the quantities $(\ref{Nhypintro})$--$(\ref{Nsphintro})$.
Thus, to complete step (b) of Section~\ref{introbootstr}, we need to show
first the propagation of higher regularity. 
These estimates are typically very straightforward because the equations
are linear in the highest order quantities, and one only requires the finiteness
for given $\underline{u}$, not uniform control as $\underline{u}\to\infty$. Nonetheless,
in our problem, special care has to be made, however,
since, by having renormalised our system as in Section~\ref{lukmeth}, 
the energies $(\ref{Nhypintro})$--$(\ref{Nsphintro})$ 
no longer  control all relevant
quantities. This will be the subject of Section~\ref{sec:HR}.
\end{itemize}

\subsubsection{The continuity of the extension}
\label{continuityofextenintro}

Having shown global existence up to the boundary $\underline{u}=\infty$, the
last step in the proof of Theorem~\ref{PRWTO} is to show that the metric is indeed
continuously extendible beyond. 
This will require certain additional estimates on the quantity
${\nab_4\widetilde{\log\Om}}$ (see already 
Proposition~\ref{lemma.nab4logOm}).
In addition to defining---in complete analogy with what was already done in the proof
of Theorem~\ref{ExtendibleModel} concerning the spherically symmetric
toy model of Section~\ref{modelprobintro}---a rescaled coordinate
$\underline{U}=\underline{u}_{\mathcal{CH}^+}$, proving continuity
now also requires redefining the angular coordinates $(\theta^A_{\mathcal{CH}^+})$
which effectively serves to renormalise the metric coefficient $b$.
We will show in addition that the spacetime remains close to Kerr in the $C^0$
norm. This will be the subject of Section~\ref{sec.C0}.

\subsubsection{Outline of the paper}\label{outlineoftherest}
We end this guide to the paper with a brief section-by-section account
of its contents (referring back to our discussion above when relevant).

We begin in {\bf Section~\ref{secsetup}} with a general 
discussion of the double null gauge
$(\ref{metricoftheform})$ defined on a suitable domain $\mathcal{U}$, 
in particular, putting the Kerr metric in this framework.

We then present in {\bf Section~\ref{Eqindoublenullfolsec}}
the Einstein vacuum equations $(\ref{INTROvace})$ written in 
this gauge, moving immediately
to the renormalised form (described above in Section~\ref{lukmeth}), and introducing
the schematic notations $\psi$, $\psi_H$ and $\psi_{\underline{H}}$.
This will allow us in {\bf Section~\ref{sec.main.thm.dnf}} to 
set up the initial value problem with initial data  $(\Sigma_0, \hat{g}, \hat{K})$, 
giving the two progressively more detailed (cf.~the discussion above
in Section~\ref{allapolla})
statements 
of Theorem~\ref{PRWTO}, given as {\bf Theorem~\ref{main.thm.spacelike.formulation}}
and {\bf Theorem~\ref{aprioriestimates}},
as well as the statement of
the main estimates in the bootstrap region $\mathcal{U}_{\underline{u}_f}$, 
{\bf Theorem~\ref{main.quantitative.thm}}
(cf.~the discussion in Section~\ref{introbootstr}).

Sections~\ref{secbasic}--\ref{recover.bootstrap}
 will concern the proof of Theorem~\ref{main.quantitative.thm}.
 We shall use the bootstrap assumption
 to obtain in {\bf Section~\ref{secbasic}} preliminary bounds in 
 $\mathcal{U}_{\underline{u}_f}$
 on Sobolev constants
 and ellipticity estimates which will be useful throughout (cf.~Section~\ref{allapolla}). 
 This will allow
 us to derive a general estimate for quantities satisfying
 covariant transport equations  in {\bf Section~\ref{transportsec}}. 
 We then derive in {\bf Section~\ref{sec.rse}}
 the ``reduced schematic equations'' for differences,
 which incorporate $L^\infty$ bounds derived by Sobolev that hold in 
 $\mathcal{U}_{\underline{u}_f}$.
 {\bf Section~\ref{sec.error.def}} will define the
 error expressions $\mathcal{E}_1$, $\mathcal{E}_2$,
 $\mathcal{T}_1$, $\mathcal{T}_2$
 and state {\bf Proposition~\ref{concluding.est}},
 which corresponds to the main estimate for the quantities
 $\mathcal{N}_{int}$ and $\mathcal{N}_{hyp}$ in terms of these error expressions
(cf.~the
 discussion above in Section~\ref{baridiaeisagwgns}). 
 The proof of Proposition~\ref{concluding.est} 
 will then occupy the next three sections: {\bf Section~\ref{sec.int.est}}
will bound the Ricci coefficients via transport estimates for the null structure equations, 
{\bf Section~\ref{sec.energy}} will bound the curvature via
energy estimates for the Bianchi equations  and {\bf Section~\ref{sec.elliptic}} 
will bound the remaining
top order estimates for Ricci coefficients using elliptic estimates. Finally,
the error terms on the right hand side of the estimate of Proposition~\ref{concluding.est}
will be themselves estimated in {\bf Section~\ref{sec:error}}, thus
completing the proof of the boundedness of $\mathcal{N}_{int}$ and $\mathcal{N}_{hyp}$
in  $\mathcal{U}_{\underline{u}_f}$,
stated as {\bf Proposition~\ref{N.bd}} (cf.~$(\ref{obtainingestimates})$).
The bootstrap assumptions will be recovered
in {\bf Section~\ref{recover.bootstrap}} (cf.~$(\ref{hereimproveboo})$), 
completing the proof of Theorem~\ref{main.quantitative.thm}.

With Theorem~\ref{main.quantitative.thm}, proven,
we turn  in {\bf Section~\ref{sec:HR}} to  the propagation of higher regularity
discussed above in Section~\ref{allapolla}, which,
together with an application of local existence,
will allow us in {\bf Section~\ref{sec:continuity}} to obtain the global existence
part of Theorem~\ref{aprioriestimates}
from Theorem~\ref{main.quantitative.thm}, stated
as {\bf Theorem~\ref{I.final}}. This yields the existence of a solution 
on $\mathcal{U}_\infty$.

It  will  remain only to identify the solution $(\mathcal{U}_\infty, g)$
as the maximal Cauchy development  of data,
and show the continuous extendibility of the metric
beyond a non-trivial Cauchy horizon ({\bf Theorem~\ref{metric.cont.ext}}); this will be accomplished in
{\bf Section~\ref{sec.C0}}, completing the proof of Theorem~\ref{aprioriestimates}.

Although in this paper we do not
show that the Cauchy horizon is in any sense generically singular,
we end our paper in {\bf Section~\ref{sec:weaknull}} with a discussion of  the compatibility of our estimates with the Cauchy horizon
being a weak null singularity in the sense of~\cite{LukWeakNull}.
This is given as {\bf Theorem~\ref{thm.wns}}.

Several computations relating to the Kerr metric in double null gauge
are relegated to {\bf Appendix~\ref{sec.Kerr.geometry}}.

\subsection{Addendum: the case $\Lambda \ne 0$}
As is well known, Einstein entertained early on
a certain modification~\cite{einstein1917kosmologische} of his  equations $(\ref{INTROvace})$ 
by 
adding  a non-vanishing ``cosmological'' constant $\Lambda$:
\begin{equation}
\label{withcosmo0}
{\rm Ric}_{\mu\nu} -\frac12g_{\mu\nu}R+ \Lambda g_{\mu\nu}=8\pi T_{\mu\nu}.
\end{equation}
This constant $\Lambda$ is known observationally to be very small, effectively zero
for the scale of  astrophysical black holes which motivate the considerations
of this paper.
As of the present writing, however, the
case of a small $\Lambda>0$ is  favoured on so-called ``cosmological'' scales
where the universe is thought to be approximately isotropic and homogeneous, and can
be described with an energy momentum tensor corresponding to dust. 
In contrast, the case $\Lambda<0$ can be related to various speculations
in high energy physics, and has received much recent interest in 
that context~\cite{Gibbons:2011sg}. 

In the vacuum, equation $(\ref{withcosmo0})$
reduces to
\begin{equation}
\label{withcosmo}
{\rm Ric}_{\mu\nu}=\Lambda g_{\mu\nu}.
\end{equation}
The ``trivial'' solution of $(\ref{withcosmo})$ with $\Lambda>0$ was first studied
in~\cite{deSitref} and is known as de Sitter
space.  The analogous solution for $\Lambda<0$
is traditionally known as anti-de Sitter space~\cite{Hawking&Ellis}.
There is a generalisation of the Kerr family to $(\ref{withcosmo})$,
known as Kerr--de Sitter in the positive case and 
Kerr--anti de Sitter in the negative case which contain
analogues of ``black holes''.
These solutions are unfortunately not contained in the standard 
elementary textbooks, but one 
can find a detailed discussion of their geometry  in~\cite{kerrdesitrefcarter}.

\begin{figure}
\centering{
\def\svgwidth{14pc}
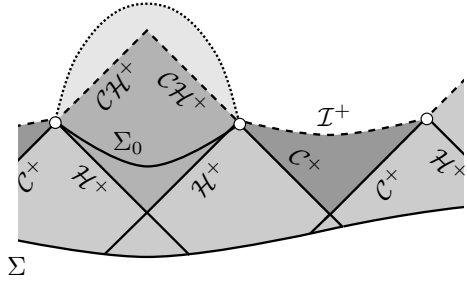}
\caption{The Kerr-de Sitter solution}\label{Kerrdesitfig}
\end{figure}

\paragraph{The cosmological case $\Lambda>0$.}
In the Kerr--de Sitter case, in addition to an event horizon $\mathcal{H}^+$ and 
Cauchy horizon $\mathcal{CH}^+$, there
is a cosmological horizon $\mathcal{C}^+$ further partitioning the exterior of the ``black hole'' into two subregions.
See Figure~\ref{Kerrdesitfig}.
The region between the event and cosmological horizons is spatially compact 
and, in remarkable recent work of Hintz--Vasy~\cite{hintz2016global}, {\bf \emph{this region
has  been shown to be nonlinearly stable}}, as a solution to $(\ref{withcosmo})$,
{\bf \emph{without symmetry assumptions}}, in the case of parameters sufficiently
close to Schwarzschild--de Sitter. This is an analogue of Conjecture~\ref{stabofkerrconj} for $\Lambda>0$, restricted to the very slowly rotating case.
In fact, their proof shows the stability of a larger
region, and in particular, this yields for all small perturbations
of Schwarzschild--de Sitter the existence of a spacelike hypersurface  $\Sigma_0$
which asymptotes 
to Kerr--de Sitter, in analogy with the
assumption of our Theorem~\ref{PRWTO}.
The rates of convergence are moreover now 
{\bf \emph{exponential}}, not inverse polynomial, and it is precisely
this which makes the global analysis of $(\ref{withcosmo})$ 
with $\Lambda>0$ so much more tractable than
the $\Lambda=0$ case of astrophysical scales (compare the difficulty of the proof of stability
of Minkowski~\cite{CK} 
with the relative ease of that of stability of de Sitter~\cite{friedrich1986existence}).

In principle, our Theorem~\ref{PRWTO} 
can be put together with~\cite{hintz2016global} to yield the $C^0$
stability of  the  region
between the event and Cauchy horizons. 
It remains to verify that our result allows for the addition of $\Lambda$
to  $(\ref{INTROvace})$, and to 
directly relate our decay assumptions on the hypersurface $\Sigma_0$ or event horizon $\mathcal{H}^+$ to the coordinate set-up of~\cite{hintz2016global}.\footnote{It is worth noting that,
even though the decay
assumptions are faster, it does not appear that this leads to a really essential
simplification in the proof of the analogue of our Theorem~\ref{PRWTO}.}
In particular,
this will {\bf \emph{disprove unconditionally}} the analogue of Conjecture~\ref{C0formSCC} for
$(\ref{withcosmo})$ with $\Lambda>0$.
For upcoming work on the stability of the ``cosmological'' region, bounded
by $\mathcal{C}^+$ and $\mathcal{I}^+$, see~\cite{schlue2016decay}.

The story for $\Lambda>0$ has an even more interesting twist, however:
We note that heuristic analysis, see~\cite{brady1998cosmic}, would indicate that 
the numerology connecting the expected exponential decay
rate on the event horizon and the strength of the blue-shift 
is such that one expects the metric to be \emph{extendible} in a $W^{1,p}$ for
a $p>1$ depending on the Kerr--de Sitter parameters, with $p\to\infty$ as
the parameters approach extremality.
In particular, there is a parameter threshold (close to extremality) beyond which
one expects extendibility in $W^{1,2}=H^1$. (This is in contrast
to the results discussed in Section~\ref{LWERNK} 
for the linear wave equation on Reissner--Nordstr\"om
where one has blowup in all $W^{1,p}$ for $p>1$, independently of the parameters $Q<M$.
Note, however, that in the exactly extremal case $|Q|=M$ considered in~\cite{Gajic:2015csa},
one again has extendibility in $W^{1,2}$!)
 While~\cite{hintz2016global} does not probe this near-extremal range in view of the restriction
to very small $a$, 
a recent result due to Hintz~\cite{hintz2016non}
extends~\cite{hintz2016global} to the Einstein--Maxwell system $(\ref{EM1})$--$(\ref{EM2})$,
proving analogous stability results for
very slowly rotating Kerr--Newmann--de Sitter, in a full neighbourhood of
subextremal Reissner--Nordstr\"om--de Sitter. Provided that sufficiently sharp decay
rates are obtained, one could then hope to show that the analogue of Christodoulou's  reformulation
of strong cosmic censorship,
i.e.~Conjecture~\ref{ChrSCC}, is false for $\Lambda>0$.
See also~\cite{Costa:2014zha, Costa2017, Costa:2017tjc}  for 
detailed work on this problem in the context of the spherically symmetric toy model
$(\ref{EMSF1})$--$(\ref{EMSF2})$, when a positive cosmological constant term
is added to the left hand side of $(\ref{EMSF1})$.

\paragraph{Instability for $\Lambda<0$?}
While the case of $\Lambda>0$ is  
more tractable than that of $\Lambda=0$ in the exterior 
of a black hole,
the case $\Lambda<0$ is yet  harder!
Anti de Sitter space itself fails to be globally hyperbolic as null infinity $\mathcal{I}$ is naturally
``timelike'', and thus, to obtain well-posed 
dynamics for $(\ref{withcosmo})$ one must add a
 ``boundary condition'' at $\mathcal{I}$~\cite{friedrich1995einstein}.  
The most interesting case is that of a reflective boundary condition at $\mathcal{I}$ 
(see however~\cite{holzegel2015asymptotic}). 
In this case, anti de Sitter space itself was conjectured to be unstable, see~\cite{eguchiha}. 
This  was first studied numerically in the seminal~\cite{bizon2011weakly}, which
moreover related the time-scale of the instability to resonant interactions understood
on the Fourier side. 
The first proof of AdS instability  has recently been given in the simplest possible
setting by Moschidis~\cite{moschidis2017proof}, via an argument
entirely physical space-based.

\begin{figure}
\centering{
\def\svgwidth{12pc}
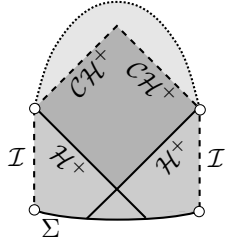}
\caption{The Kerr--AdS solution}\label{KADSfigure}
\end{figure}

What allows in principle for instability of AdS at a non-linear level is the lack of decay for the linearised
problem. The situation for     black holes is more unclear. 
The Penrose diagram of Kerr--AdS is depicted in Figure~\ref{KADSfigure}.
Holzegel--Smulevici~\cite{holzegel2013decay, HolzSmulevici} have shown that while solutions
of the linear wave equation on Kerr--AdS indeed decay, the rate of decay is
only logarithmic, and this slow decay is moreover in general sharp. 
One scenario is that at the non-linear level, backreaction builds up, and
Kerr--AdS shares the unstable fate of pure AdS, and thus
the picture of an event horizon settling down to Kerr--AdS
is generically not realised. If this is the case, then all bets are off regarding the validity
of either weak or strong cosmic censorship!
Alternatively, however, the linear picture of slow logarithmic decay could survive in the full
non-linear theory, as is suggested in~\cite{dias2012nonlinear}. The latter situation would mean
that the proper setting for understanding black hole interiors in $\Lambda<0$
requires that we replace the assumption of Theorem~\ref{PRWTO} or~\ref{DEUTERO} 
with an assumption
of data on $\Sigma_0$ or $\mathcal{H}^+$ logarithmically approaching that of
Kerr--AdS in a suitable sense.
Now, it would appear that such a 
slow logarithmic decay does not allow for our stability mechanism, for
which it seems essential that one can 
prove weighted estimates with \emph{polynomial} weights in $\underline{u}$, as in
the quantities $(\ref{Nhypintro})$--$(\ref{Nintintro})$.
See also the discussion in~\cite{bhattacharjee2016internal}.
Perversely, could this ``worse'' event horizon behaviour  be sufficient to ensure
that the original $C^0$ formulation of cosmic censorship 
(Conjecture~\ref{C0formSCC})  becomes true after all for $\Lambda<0$?

We leave such musings for another occasion!

\subsection{Acknowledgments}

M.~D.~is supported by NSF grant  DMS-1709270  as well as an EPSRC programme
grant EP/K00865X/1. J.~L.~is supported by a Sloan fellowship, a Terman fellowship, and NSF grant DMS-1709458. Much of this research
was conducted when J.~L.~was at the University of Cambridge. The authors
thank Demetrios Christodoulou for suggesting this problem many years ago 
and for many conversations over the years.

\section{Spacetimes covered by double null foliations}\label{secsetup}

In this section, we will derive the structure equations satisfied by a spacetime
$(\mathcal{U},g)$
globally foliated by a double null foliation intersecting on spheres. 
We will not require here the metric $g$ to satisfy the Einstein equations.

We begin in \textbf{Section~\ref{thegeneralclass}} with the  basic assumptions on the underlying manifold
$\mathcal{U}$
and metric in explicit form. We shall then show in \textbf{Section~\ref{secdnf}} that the geometrical
interpretation of these assumptions
is equivalent to the existence of a natural double null foliation; importantly, this
allows us to find appropriate regions of a general spacetime which allow
for parametrisation as $(\mathcal{U},g)$. In
\textbf{Section~\ref{computnewsec}}, we introduce the Ricci coefficients and curvature components
with respect to a natural null frame associated to the double null foliation,
and collect several basic facts about them.
%the equations
%for Christoffel symbols and curvature with resepct to our coordinates, which 
%geometrically represent the structure equations of the associated double null foliation.
Finally, in \textbf{Section~\ref{sec.Kerr.dbn.text}}, we describe how the work of Pretorius--Israel \cite{Pretorius} implies
that that Kerr interior can indeed be parametrised as $(\mathcal{U},g)$.

\subsection{The general class of spacetimes}\label{thegeneralclass}
We will introduce here the general class of spacetimes $(\mathcal{U},g)$ we will be considering
in this paper. The class of spacetimes is sufficiently general so as to include the Kerr interior and 
as well as the spacetimes that we construct\footnote{In fact, it will include both the full spacetimes in Theorem~\ref{PRWTO}, as well as appropriate subsets thereof that we consider in the context of the bootstrap argument.} in Theorem~\ref{PRWTO}.

In Section~\ref{coordinates}, we will define the underlying manifold $\mathcal{U}$, followed
by the metric and time orientation in Section~\ref{metandtimeo}. 
We will define an associated null frame in Section~\ref{someframes}.

\subsubsection{The differential structure}\label{coordinates}

Let us denote by $u$ and $\underline{u}$ the standard Euclidean coordinates on
$\mathbb R^2$, and let $\mathcal{W}\subset \mathbb R^2$ be a connected open subset,
and let $\mathbb S^2$ denote the standard unit sphere, with local coordinates
$\theta^A$ where $A=1,2$.
We define the smooth manifold $\mathcal{U}=\mathcal{W}\times \mathbb S^2$.

We can consider $(u, \underline{u}, \theta^1, \theta^2)$ as local coordinates on $\mathcal{U}$.

\subsubsection{The metric and time-orientation}\label{metandtimeo}
Let $\gamma:\mathcal{W}\to \Gamma(T^*\mathbb S^2\times T^*\mathbb S^2)$ denote a 
Riemannian metric on $\mathbb S^2$, let
$b:\mathcal{W}\to\Gamma(T\mathbb S^2)$ denote a vector field on $\mathbb S^2$,
and let $\Omega:\mathcal{W}\to \Gamma(\mathbb S^2 \times\mathbb R)$ denote a positive function, all depending
smoothly on $\mathcal{W}$.

We can denote the components of $b$ and $\gamma$ with respect to the local
coordinates $\theta^A$ by $b^A$ and $\gamma_{AB}$.

These choices induce a Lorentzian metric on $\mathcal{U}$ defined by:
\begin{equation}
\label{formofthemetric}
g=-2\Omega^2(du\otimes d\ub+d\ub\otimes du)+\gamma_{AB}(d\th^A-b^Ad\ub)\otimes (d\th^B-b^Bd\ub).
\end{equation}
We time-orient the above by requiring that $du$ and $d\ub$ are future directed.
With this choice, $(\mathcal{U}, g)$ is a smooth $4$-dimensional spacetime.

\subsubsection{Frames and metric components}\label{someframes}
Let $\f{\rd}{\rd u}$, $\f{\rd}{\rd\ub}$, $\f{\rd}{\rd\th^A}$ be coordinate vector fields
defined with respect to the $(u,\ub,\th^1,\th^2)$ coordinate system that we introduced above. 

Define the vector fields
\begin{equation}\label{L.Lb.def}
L=\f{\rd}{\rd\ub}+b^A\f{\rd}{\rd\th^A},\quad \Lb=\f{\rd}{\rd u}.
\end{equation}
These are easily seen to be globally defined null vector fields on $\mathcal{U}$.
We can in fact write $L=-2\Omega^{2}D u$, $\Lb=-2\Omega^{2}D \ub$,
where $D$ denotes the gradient with respect to $(\ref{formofthemetric})$.

These vector fields satisfy the properties
\begin{equation}\label{L.Lb.prop}
\Lb \,\ub=Lu=0,\quad \Lb u=L\ub=1.
\end{equation}

We have
$$g(L,\Lb)=-2\Om^2,\quad g(L,\f{\rd}{\rd\th^A})=g(\Lb,\f{\rd}{\rd\th^A})=g(L,L)=g(\Lb,\Lb)=0,$$
$$\gamma_{AB}\doteq g(\f{\rd}{\rd\th^A},\f{\rd}{\rd\th^B}).$$
As a consequence, the inverse metric $g^{-1}$ is given by
\[
g^{-1}=-\f 1{2\Om^2}(L\otimes\Lb+\Lb\otimes L)+\gamma^{-1}.
\]
The inverse metric $g^{-1}$ has the following non-vanishing components:
\[
(g^{-1})^{AB}=(\gamma^{-1})^{AB},\quad (g^{-1})^{u\ub}=-\frac{1}{2\Om^2},\quad (g^{-1})^{uA}=-\frac{1}{2\Om^2}b^A,
\]
the remaining components being zero.

We define $L'$ and $\Lb'$ as the following null vector fields
\begin{equation}\label{L'.Lb'.def.0}
\Lb'\doteq  -2D\ub = \Om^{-2}\frac{\partial}{\partial u},\quad L'\doteq -2Du = \Om^{-2}\left(\frac{\partial}{\partial \ub}+b^A\frac{\partial}{\partial \th^A}\right).
\end{equation}
It is easy to check that $L'$ and $\Lb'$ are geodesic, i.e.
\begin{equation}\label{L'.Lb'.geodesic}
D_{L'}L'=D_{\Lb'}\Lb'=0.
\end{equation}

We will define the following \emph{normalised null pair}\footnote{Notice that this is different from the usual normalisation (for example in \cite{CK} and \cite{Chr}). In this paper, we will define the null variables $u$ and $\ub$ to have Eddington-Finkelstein asymptotics and this normalisation guarantees both $e_3$ and $e_4$ to be regular at the Cauchy horizon.}:
\begin{equation}\label{e3.e4.def}
e_3\doteq \frac{\partial}{\partial u},\quad e_4\doteq\Omega^{-2}\left(\frac{\partial}{\partial \ub}+b^A\frac{\partial}{\partial \th^A}\right),
\end{equation}
where $b^A$ is as above.
We have
\[
g(e_3,e_4)=-2.
\]

\subsection{Geometric interpretation and construction of coordinates}\label{secdnf}

We give in this section a geometric interpretation of the above coordinate system
and null frame in terms of the \emph{double null foliation} that it defines.

\subsubsection{The eikonal equation and the double null foliation}\label{secdnfeik}
Considered as smooth functions 
$u:\mathcal U\to \mathbb R$ and $\ub:\mathcal U\to \mathbb R$,
the coordinates $u$ and $\ub$
each satisfy the eikonal equation
\begin{equation}\label{basic.eikonal}
g(Du, Du)=0,\quad g(D\ub, D\ub)=0.
\end{equation}
We denote the level sets of $u$ (resp. $\ub$) by $H_u$ (resp. $\Hb_{\ub}$).
The eikonal equations imply $H_u$ and $\Hb_{\ub}$ are null hypersurfaces. 
We denote
\[
S_{u,\ub} = H_u\cap \Hb_{\ub}.
\]
Note that these 
are 
diffeomorphic to $\mathbb S^2$
and $\mathcal{U}=\cup_{(u,\ub)\in\mathcal{W}}S_{u,\ub}$.

Note that the integral flows of $L$ and $\Lb$ respect the foliation $S_{u,\ub}$.
Moreover,
the coordinates $(\th^1,\th^2)$
obey the condition
\begin{equation}\label{th.transport}
\Lb \th^A=0.
\end{equation}

\subsubsection{Construction of double null coordinates}

We will need the following statement, which constructs locally a double null foliation and a system of double null coordinates:
\begin{proposition}\label{prop:u.extend}
Let $\ell:(a,b)\to \mathbb R^2$ be a simple curve, $\Sigma = \ell((a,b))\times \mathbb S^2$ and $(\mathcal{M},g_0)$ be a smooth Lorentzian $4$-manifold.  Suppose $i:\Sigma\to \mathcal M$ is an embedding such that $i(\Sigma)$ is a spacelike hypersurface. Prescribe initial data for $(u,\ub,\th^A)$ on $\Sigma$ as follows: 
\begin{itemize}
\item $(\tau,\th^A)_{A=1,2}$ is a system of local coordinates\footnote{It is to be understood that $\Sigma$ has multiple coordinate charts and $(\tau,\th^A)_{A=1,2}$ are coordinates in one of the charts.} such that $\tau = \ell^{-1}\circ \pi$ (where $\pi: \ell((a,b))\times \mathbb S^2\to \ell((a,b))$ is the natural projection map).
\item $\ub\restriction_{\Sigma}$ is a smooth, strictly increasing function of $\tau$; $u\restriction_{\Sigma}$ is a smooth strictly decreasing function of $\tau$.
\end{itemize}
Let $y \in (a,b)$. 
Then there exists an open set $\mathcal W\subset \mathbb R^2$ with $\ell(y)\in \mathcal W$ and a metric $g$ in $\mathcal U = \mathcal W\times \mathbb S^2$ taking the form \eqref{formofthemetric} such that the following holds:
\begin{itemize}
\item There exists an isometric embedding 
$$\tilde{i}: (\mathcal U,g)\to (\mathcal M,g_0).$$
\item $(u,\ub,\th^A)$ take their prescribed values on $\Sigma$.
\item $u$ and $\ub$ satisfy \eqref{basic.eikonal} and $\th^A$ (for $A=1,2$) is constant along the integral curve of $D\ub$.
\end{itemize}
\end{proposition}

\begin{proof}
According to Section~\ref{secdnfeik}, the key is to solve \eqref{basic.eikonal}, which can in turn be achieved by solving the geodesic equation. More precisely, given the initial data for $(u,\ub,\th^A)$ as above, we prescribe the initial data\footnote{Note that the coefficients in the initial data for $L'$ and $\Lb'$ are chosen to follow the notation introduced in \eqref{L'.Lb'.def.0}. Indeed, since
$$L' u \restriction_{\Sigma} = 0 \implies N u = -nu,$$
it follows that
$$Du\restriction_{\Sigma} = -(Nu) N + (nu) n = (nu) (N+n) = -\f 12 L'\restriction_{\Sigma}.$$
Similarly for $\Lb'$.}
$$L'\restriction_{\Sigma} = \f{2}{(-n u)}(N+n) ,\quad \Lb'\restriction_{\Sigma} = \f{2}{n\ub}(N-n).$$
where $n$ is the unique unit normal in $\Sigma$ to the constant-$\tau$ satisfying $n\tau>0$, and $N$ is unique future directed unit normal to $\Sigma$.

We then solve the following system of equations: $L'$ and $\Lb'$ satisfy the geodesic equation, and $(u,\ub,\th^A)$ satisfy transport equations:
$$D_{L'} L' = D_{\Lb'}\Lb' = 0,\quad L'u=\Lb\;\ub = 0,\quad \Lb\th^A=0.$$
The existence of a solution $(L',\Lb',u,\ub, \th^A)$ follows from compactness and standard local existence theory for odes.

It can then be checked that for such a solution, $u$ and $\ub$ satisfy the eikonal equation \eqref{basic.eikonal}. In view of the discussions in Section~\ref{secdnfeik}, this then implies the conclusion of the proposition.
\end{proof}

\subsection{Computations}\label{computnewsec}

\subsubsection{Derivatives and connections}\label{sec.derivatives}

Before we proceed, it is convenient to introduce the convention that lower case Greek letter $\mu$, $\nu$, etc.~are spacetime indices, while capital Latin indices $A$, $B$, $C$, etc.~are indices for the $2$-spheres $S_{u,\ub}$, i.e.~they run through $1$ and $2$. Unless otherwise stated, repeated indices will be summed over.

We denote the spacetime covariant derivative with respect to the Levi--Civita connection associated to $g$ by $D$. We compute a subset of the spacetime Christoffel symbols (which we will denote using boldface $\mathbf{\Gamma}$), which will be useful below:
\beaa
\mathbf{\Gamma}^A_{Bu}&=&\frac 12 (\gamma^{-1})^{AC}\frac{\partial}{\partial u}\gamma_{BC},\\
\mathbf{\Gamma}^A_{B\ub}&=&\frac 12 (\gamma^{-1})^{AC}(\frac{\partial}{\partial\ub}\gamma_{BC}-\frac{\partial}{\partial\th^B}(\gamma_{CD}b^D)+\frac{\partial}{\partial\th^C}(\gamma_{BD}b^D))+\frac 1{4\Omega^2}b^A(2\frac{\partial}{\partial\th^B}\Omega^2-\frac{\partial}{\partial u}(\gamma_{BC}b^C)),\\
\mathbf{\Gamma}^A_{BC}&=&\frac 12 (\gamma^{-1})^{AD}(-\frac{\partial}{\partial\th^D}\gamma_{BC}+\frac{\partial}{\partial\th^B}\gamma_{CD}+\frac{\partial}{\partial\th^C}\gamma_{BD})+\frac 1{4\Omega^2}b^A\frac{\partial}{\partial u}\gamma_{BC},\\
\mathbf{\Gamma}^{\ub}_{Au}&=&0,\\ 
\mathbf{\Gamma}^{\ub}_{A\ub}&=&-\frac 1{4\Om^2}(2\frac{\partial}{\partial \th^A}\Om^2-\frac{\partial}{\partial u}(\gamma_{AB}b^B)),\\ 
\mathbf{\Gamma}^{\ub}_{AB}&=&\frac 1{4\Om^2}\frac{\partial}{\partial u}\gamma_{AB}.
\eeaa

For the remainder of the paper, we will frequently consider tensor fields which are \emph{tangential to the spheres $S_{u,\ub}$} (or simply, \emph{$S$-tangent}), i.e.~tensorfields of the form
$$\phi = \phi_{A_1\dots A_r}^{B_1\dots B_s} \,d\th^{A_1}\otimes \dots\otimes d\th^{A_r}\otimes \f{\rd}{\rd\th^{B_1}}\otimes \dots \otimes \f{\rd}{\rd\th^{B_s}}.$$
We define the differential operators $\nab$, $\nab_3$ and $\nab_4$, which act on these $S$-tangent vector fields:
\begin{definition}\label{def.slashed}
We denote by $\nab$ the Levi--Civita connection\footnote{We note that using the above calculations for the spacetime Christoffel symbols, the induced covariant derivative operator coincides with the projection of the spacetime covariant derivatives to the tangent space of $S_{u,\ub}$.} induced by the metric $\gamma$ on $S_{u,\ub}$ and by $\nab_3$, $\nab_4$ the projections to $S_{u,\ub}$ of the covariant derivatives $D_3=D_{e_3}$, $D_4=D_{e_4}$.
\end{definition}

We now apply the above calculations for $\mathbf \Gamma$ to derive the expressions for $\nab$, $\nab_{3}$ and $\nab_4$ in coordinates. First, the formula for $\nab$ is standard: for every tensor $\phi$ of rank $r$ tangential to the spheres $S_{u,\ub}$, $\nab_B \phi_{A_1 A_2 \dots A_r}$ is a tensor of rank $r+1$ tangential to the spheres $S_{u,\ub}$ given by
\begin{equation}\label{nab.def}
\begin{split}
\nab_B \phi_{A_1 A_2 \dots A_r} = \f{\rd}{\rd\th^B} \phi_{A_1 A_2 \dots A_r} -\sum_{i=1}^r \slashed\Gamma_{A_iB}^C \phi_{A_1\dots \hat{A_i}C\dots A_r},
\end{split}
\end{equation}
where $\hat{A_i}$ denotes that the $A_i$ which was originally present is removed.
\begin{equation}\label{def.slashed.Gamma}
\slashed\Gamma^{C}_{DE}= \f 12 (\gamma^{-1})^{CF}(\f{\rd}{\rd\th^D}\gamma_{EF}+\f{\rd}{\rd\th^E}\gamma_{DF}-\f{\rd}{\rd\th^F}\gamma_{DE}).
\end{equation}
To derive the expressions for $\nab_3$ and $\nab_4$, it is convenient for later reference to introduce the null second fundamental forms as a covariant $2$-tensor tangential to the $2$-spheres $S_{u,\ub}$ by
\begin{equation}\label{chi.chib.first.def}
\chi_{AB}=g(D_{\f{\rd}{\rd\th^A}} e_4,\f{\rd}{\rd\th^B}),\quad \chib_{AB}=g(D_{\f{\rd}{\rd\th^A}} e_3,\f{\rd}{\rd\th^B}).
\end{equation}
In the double null coordinate system, for every tensor $\phi$ of rank $r$ tangential to the spheres $S_{u,\ub}$, we have
\begin{equation}\label{nab3.def}
\begin{split}
&\nab_3 \phi_{A_1 A_2 \dots A_r}\\
=&\frac{\partial}{\partial u} \phi_{A_1 A_2 \dots A_r}-\sum_{i=1}^r(\mathbf\Gamma^B_{A_i u}-b^B\mathbf\Gamma^{\ub}_{A_i u})\phi_{A_1\dots\hat{A_i}B\dots A_r}\\
=&\frac{\partial}{\partial u} \phi_{A_1 A_2 \dots A_r}-\sum_{i=1}^r(\gamma^{-1})^{BC}\chib_{A_i C}\phi_{A_1\dots\hat{A_i}B\dots A_r}.
\end{split}
\end{equation}
$\nab_{4}$ can be expressed in the double null coordinate system as
\begin{equation}\label{nab4.def}
\begin{split}
&\nab_4 \phi_{A_1 A_2\dots A_r}\\
=&\Omega^{-2}(\frac{\partial}{\partial \ub}+b^C\frac{\partial}{\partial\th^C}) \phi_{A_1 A_2 \dots A_r}-\sum_{i=1}^r\Omega^{-2}(\mathbf\Gamma^B_{A_i \ub}+b^C\mathbf\Gamma^B_{A_i C}-b^B\mathbf\Gamma^{\ub}_{A_i \ub}-b^B b^C\mathbf\Gamma^{\ub}_{A_i C})\phi_{A_1\dots\hat{A_i}B\dots A_r}\\
=&\Omega^{-2}(\frac{\partial}{\partial \ub}+b^C\frac{\partial}{\partial\th^C}) \phi_{A_1 A_2 \dots A_r}-\sum_{i=1}^r((\gamma^{-1})^{BC}\chi_{A_iC}-\Omega^{-2}\frac{\partial}{\partial\th^{A_i}}b^B)\phi_{A_1\dots\hat{A_i}B\dots A_r}.
\end{split}
\end{equation}
We can directly check by using the above formula that 
\begin{equation}\label{nab.compatibility}
\nab\gamma = 0,\quad \nab_3 \gamma = \nab_4 \gamma =0.
\end{equation}
Moreover, for $\chi$ and $\chib$ as in \eqref{chi.chib.first.def}, we have the first variation formulae
\begin{equation}\label{first.variation}
\Ls_L \gamma= 2\Omega^2\chi,\quad \Ls_{\Lb}\gamma=2\chib,
\end{equation}
where $\Ls_{\Lb}$ denotes the restriction of the Lie derivative to $TS_{u,\ub}$ (See \cite{Chr}, Chapter 1).

We end this subsection with the definitions of a few more differential operators on tensors tangential to $S_{u,\ub}$ which will be useful for later use. 
For totally symmetric tensors $\phi$ of rank $r+1$, the $\slashed{\mbox{div}}$ and $\slashed{\mbox{curl}}$ operators are defined by the formulas
\begin{equation}\label{div.def}
(\div\phi)_{A_1...A_r}\doteq (\gamma^{-1})^{BC}\slashed{\nabla}_B\phi_{CA_1...A_r},
\end{equation}
\begin{equation}\label{curl.def}
(\curl\phi)_{A_1...A_r}\doteq \in^{BC}\slashed{\nabla}_B\phi_{CA_1...A_r}.
\end{equation}
For a scalar function $\phi$, define $\slashed\Delta$ to be the Laplace--Beltrami operator associated to the metric $\gamma$, i.e.
\begin{equation}\label{LB.def}
\slashed\Delta\phi \doteq (\gamma^{-1})^{AB}\nab_A\nab_B \phi.
\end{equation}
Finally, define the operator $\nab\widehat{\otimes}$ on a $1$-form $\phi_{A}$ by
\begin{equation}\label{nab.otimes.def}
(\nab\widehat{\otimes}\phi)_{AB} \doteq   \nab_A \phi_B + \nab_B \phi_A - \gamma_{AB} \div \phi.
\end{equation}

\subsubsection{Ricci coefficients and null curvature components}\label{sec.RC.CC.def}

As before, we use capital Latin indices $A,B$ to denote $1,2$, which are indices on the spheres $S_{u,\ub}$. Letting $e_A=\f{\rd}{\rd\th^A}$ for $A=1,2$, we define the Ricci coefficients relative to $\{e_1,e_2, e_3, e_4\}$. These can be viewed as tensor fields tangent to $S_{u,\ub}$, the $2$-spheres adapted to the double null foliation. Let
 \begin{equation}\label{Ricci.coeff.def}
\begin{split}
&\chi_{AB}\doteq g(D_A e_4,e_B),\, \,\, \quad \chib_{AB}\doteq g(D_A e_3,e_B),\\
&\eta_A\doteq -\frac 12 g(D_3 e_A,e_4),\quad \etab_A\doteq -\frac 12 g(D_4 e_A,e_3),\\
&\omega\doteq -\frac 14 g(D_4 e_3,e_4),\quad\,\,\, \omegab\doteq -\frac 14 g(D_3 e_4,e_3),\\
&\zeta_A\doteq \frac 1 2 g(D_A e_4,e_3),
\end{split}
\end{equation}
where $D_A=D_{e_{A}}$, $D_3=D_{e_3}$ and $D_4=D_{e_4}$. It is clear from the definitions that these Ricci coefficients are of different rank, namely that $\chi$ and $\chib$ are symmetric $2$-tensors, $\eta$, $\etab$ and $\zeta$ are $1$-forms and $\omega$ and $\omegab$ are scalars.

We also introduce the null curvature components as follows:
 \begin{equation}\label{curv.comp.def}
\begin{split}
\a_{AB}&\doteq R(e_A, e_4, e_B, e_4),\quad \, \,\,   \ab_{AB} \doteq R(e_A, e_3, e_B, e_3),\\
\b_A&\doteq  \frac 1 2 R(e_A,  e_4, e_3, e_4) ,\quad \bb_A \doteq \frac 1 2 R(e_A,  e_3,  e_3, e_4),\\
\rho&\doteq \frac 1 4 R(e_4,e_3, e_4,  e_3),\quad \sigma \doteq \frac 1 4  \,^*R(e_4,e_3, e_4,  e_3).
\end{split}
\end{equation}
Here, $R$ is the Riemann curvature tensor of $g$, given by
$$R(X,Y,Z,W)\doteq -g(D_X D_Y Z - D_Y D_X Z, W),$$
and $\, ^*R$ denotes the Hodge dual of $R$ given by\footnote{Recall that Greek indices denote spacetime indices.}
$$^*R_{\mu\nu\lambda\sigma} \doteq  \varepsilon_{\mu\nu\alpha\beta} R^{\alpha\beta}{ }_{\lambda\sigma},$$
where $\varepsilon$ is the volume form of $(\mathcal U,g)$.
As the Ricci coefficients, the null curvature components are also tensor fields tangent to $S_{u,\ub}$. It is easy to see that $\a$ and $\ab$ are symmetric $2$-tensors, $\b_A$ and $\bb_A$ are $1$-forms and $\rho$, $\sigma$ are scalars.

For the null second fundamental forms $\chi_{AB}$ and $\chib_{AB}$, we also use $\chih_{AB}$ and $\chibh_{AB}$ to denote their respective traceless parts, i.e.
\begin{equation}\label{traceless.chi.chib.def}
\chih_{AB}\doteq \chi_{AB}-\f 12 (\trch) \gamma_{AB},\quad \chibh_{AB}\doteq \chi_{AB}-\f 12 (\trchb) \gamma_{AB},
\end{equation}
where $\trch$ and $\trchb$ denote the trace of the null second fundamental forms with respect to $\gamma$, i.e. 
\begin{equation}\label{trace.chi.chib.def}
\trch\doteq (\gamma^{-1})^{AB}\chi_{AB},\quad \trchb\doteq (\gamma^{-1})^{AB}\chib_{AB}.
\end{equation}

As we will see later, it will be convenient to also introduce two more scalars related to curvature\footnote{These are the renormalised curvature components; see Sections~\ref{lukmeth} and \ref{seceqn}.}. First, we use $K$ to denote the Gauss curvature of the $2$-surface $S_{u,\ub}$ adapted to the double null foliation, i.e.
\begin{equation}\label{Gauss.def}
\gamma_{BC} K=\f{\rd}{\rd\th^A}\slashed{\Gamma}^{A}_{BC}-\f{\rd}{\rd\th^C}\slashed{\Gamma}^A_{BA}+\slashed{\Gamma}^A_{AD}\slashed{\Gamma}^D_{BC}-\slashed{\Gamma}^A_{CD}\slashed{\Gamma}^D_{BA},
\end{equation}
where $\slashed{\Gamma}$ is as in \eqref{def.slashed.Gamma}.
We also define
\begin{equation}\label{sigmac.def}
\sigmac\doteq  \sigma+ \f12 \in^{AB}(\gamma^{-1})^{CD}\chibh_{AC}\chih_{BD}
\end{equation}
where $\in$ is the volume form associated to the metric $\gamma$. The importance of $K$ and $\sigmac$ is that together with $\beta$ and $\betab$, they form a set of \emph{renormalised null curvature components}. They behave better than general null curvature components and play a fundamental role in our argument, cf.~Section~\ref{lukmeth}.

In anticipation of the top order estimates\footnote{$\mu$ and $\mub$ satisfy better estimates than general first angular covariant derivatives of the Ricci coefficients or general curvature components (similarly, $\ombs$ satisfies better estimates than general second angular covariant derivatives of the Ricci coefficients or general first angular covariant derivatives of curvature components); see Sections~\ref{schnoteqn}, \ref{sec.rse.Ricci} and \ref{sec.rse.Ricci.add}. This observation, originating in \cite{Chr,CK}, will be crucial in Section~\ref{sec.elliptic}.}, we introduce three scalars, which are special combinations of angular covariant derivatives of the Ricci coefficients and the curvature components (or their derivatives). Define\footnote{Recall here the notations introduced in \eqref{div.def} and \eqref{LB.def}.}
\begin{equation}\label{top.quantities.def}
\mu\doteq -\div\eta +K,\quad \mub\doteq -\div\etab+K,\quad \ombs\doteq \slashed\Delta \om + \f 12\div\beta.
\end{equation}

\subsubsection{Some basic identities}
For the $S_{u,\ub}$-tangent covariant tensor fields defined in \eqref{Ricci.coeff.def} and \eqref{curv.comp.def}, when there is no risk of confusion\footnote{As we will see later, we will encounter more than one metric in the proof of our main theorem, in which case we will further specify the metric with respect to which we raise and lower indices.}, we will raise and lower indices with respect to the metric $\gamma$ on $S_{u,\ub}$ without further comments.

In the remainder of this subsection, we prove some simple identities for the Ricci coefficients, which hold because of our choice of the normalisation of $e_3$ and $e_4$ in \eqref{e3.e4.def}. In Proposition \ref{prop.gauge.con}, we prove some relations between the Ricci coefficients and in Proposition \ref{metric.der.Ricci} we will relate the Ricci coefficients and the metric components defined in Section \ref{coordinates}.
\begin{proposition}\label{prop.gauge.con}
In the gauge given by \eqref{e3.e4.def}, we have\footnote{Although $\om=0$, we will still keep it in the equations below (see Section \ref{seceqn}) for easy comparison with the formulae found in the literature where a difference normalisation of $e_3$ and $e_4$ is chosen.}
\begin{equation}\label{gauge.con}
\begin{split}
&\omega=0,\qquad \omegab=-\nab_3 (\log\Omega),\\
&\zeta_A=\eta_A,\quad \nab_A (\log\Omega)=\f12 (\eta_A+\etab_A).
\end{split}
\end{equation}
\end{proposition}
\begin{proof}
The first equality follows immediately from the fact that $e_4$ is geodesic, i.e.
$$\om=-\f 14 g(D_4 e_3, e_4)=\f 14 g( e_3, D_4 e_4)=0.$$
For the second equality, we use the fact that $\Om^{-2} e_3$ is geodesic to get
$$\omb=-\f 14g(D_3 e_4, e_3)=\f 14g( e_4, D_3(\Om^2\Om^{-2} e_3))=\f 14 (e_3\Om^2)\Om^{-2} g(e_3, e_4)=-(e_3\log\Om).$$
Since $e_3=\f{\rd}{\rd u}$ and $e_A=\f{\rd}{\rd\th^A}$, we have $g([e_3,e_A],e_4)=0$. Therefore,
$$\eta_A=-\f 12 g(D_3 e_A, e_4)=-\f 12 g(D_A e_3, e_4)=\f 12 g(D_A e_4, e_3)=\zeta_A.$$
Finally, since $e_4=\f{1}{\Om^2}(\f{\rd}{\rd\ub}+b^A\f{\rd}{\rd\th^A})$ and $e_A=\f{\rd}{\rd\th^A}$, we have 
$$g([e_4, e_A],e_3)=-(e_A \Om^{-2}) \Om^2 g(e_3,e_4)=2(e_A \Om^{-2}) \Om^2=-4(e_A\log\Om).$$
As a consequence, this implies
$$\etab_A=-\f 12 g(D_4 e_A,e_3)=-\f 12 g(D_A e_4,e_3)-\f 12g([e_4,e_A],e_3)=-\zeta_A+2(e_A\log\Om)=-\eta_A+2(e_A\log\Om).$$
After rearranging, we obtain
$$\nab_A (\log\Omega)=\f12 (\eta_A+\etab_A).$$
\end{proof}
We also have the following equations relating the metric components and the Ricci coefficients:
\begin{proposition}\label{metric.der.Ricci}
$\gamma$ satisfies the equations:
$$\f{\rd}{\rd u} \gamma_{AB}=2\chib_{AB},$$
$$(\f{\rd}{\rd\ub}+b^C\f{\rd}{\rd\th^C}) \gamma_{AB}+\gamma_{AC}\f{\rd}{\rd\th^B}b^C+\gamma_{BC}\f{\rd}{\rd\th^A}b^C=2\Om^2\chi_{AB}.$$
$\Omega$ verifies the equation:
$$\f{\rd}{\rd u}\log\Omega=-\omb.$$
$b^A$ satisfies the following equation:
$$\f{\rd b^A}{\rd u}=2\Omega^2(\eta^A-\etab^A)=4\Omega^2(\zeta^A-\nab^A\log\Om).$$
\end{proposition}
\begin{proof}
The equation for $\gamma$ follows from \eqref{first.variation} and the equation for $\Omega$ is derived in \eqref{gauge.con}. Finally, recall from \eqref{e3.e4.def} that
$$e_3=\frac{\partial}{\partial u},\quad e_4=\Omega^{-2}\left(\frac{\partial}{\partial \ub}+b^A\frac{\partial}{\partial \th^A}\right).$$
Therefore, 
$$[e_3,e_4]=(\f{\rd}{\rd u}\Om^{-2})\Om^2 e_4+\Om^{-2}\f{\rd b^A}{\rd u}\frac{\partial}{\partial \th^A}.$$
As a consequence,
\begin{equation*}
\begin{split}
\Om^{-2}\gamma_{AB}\f{\rd b^B}{\rd u}=&g([e_3,e_4],e_A)=g(D_3 e_4, e_A)-g(D_4 e_3, e_A)\\
=&-g(D_3 e_A, e_4)+g(D_4 e_A, e_3)=2\eta_A-2\etab_A.
\end{split}
\end{equation*}
The conclusion follows from combining this with \eqref{gauge.con}.
\end{proof}

\subsection{Kerr interior in double null foliation}\label{sec.Kerr.dbn.text}

As we mentioned in the beginning of Section~\ref{thegeneralclass}, the interior of the Kerr spacetime can be put into the form of the general class of spacetime that we consider. In other words, there exists a global double null foliation in the Kerr interior. This fact is a consequence of \cite{Pretorius}. Detailed discussions of the double null foliation in the Kerr interior, as well as many calculations and proofs, are deferred to Appendix~\ref{sec.Kerr.geometry}. Here, however, we gather a few facts which will be repeatedly used in the proof of the main theorem.

Fix $M,\,a\in\mathbb R$ with $0<|a|< M$. The \emph{Kerr interior} $(\mathcal M_{Kerr},g_{a,M})$ with mass $M$ and specific angular momentum $a$ is a spacetime where $\mathcal M_{Kerr} = \mathbb R^2\times \mathbb S^2$ and the metric $g_{a,M}$ in the Boyer--Lindquist coordinates $(t,r,\th,\phi)$ takes the form \eqref{BL}. Here, $t\in \mathbb R$, $r\in (r_-,r_+)$, $\th\in (0,\pi)$ and $\phi\in [0,2\pi)$, where $r_\pm = M \pm \sqrt{M^2 - a^2}$ are the roots of $\Delta=r^2+a^2-2Mr$.

The Kerr interior $(\mathcal M_{Kerr},g_{a,M})$ admits a global double null foliation so that in an appropriate double null coordinate system, the metric in $\mathcal M_{Kerr} = \mathbb R^2\times \mathbb S^2$  takes the form as in the class of spacetimes described in Section~\ref{thegeneralclass}. More precisely, $g_{a,M}$ can be written as
\begin{equation*}
g_{a,M}=-2\Omega_\Ke^2 (du\otimes d\ub+d\ub\otimes du)+(\gamma_\Ke)_{AB}(d\th^A-(b_\Ke)^A d\ub)\otimes(d\th^B-(b_{\Ke})^B d\ub).
\end{equation*}
where\footnote{Note that in Appendix~\ref{sec.Kerr.geometry}, the metric is given so that a set of spherical-like coordinates are used on the $2$-spheres. Therefore, the local coordinate system $(u,\ub,\th^1,\th^2)$ misses the two points on the rotation axis on every $\mathbb S^2$. Nevertheless, it is easy to check that one can introduce regular coordinates there.} $\Om_\Ke$, $\gamma_\Ke$ and $b_\Ke$ are given in \eqref{Kerr.metric.comp} (and expressions in \eqref{Kerr.metric.comp} are functions of $(u,\ub,\th^1,\th^2)$ and are defined implicitly in Sections~\ref{sec.PIcoord} and \ref{Kerr.dbn}).

The following are some facts regarding the Kerr metric in this coordinate system that are useful in the remaining of this paper:
\begin{itemize}
\item (Event and Cauchy horizons) Upon appropriate change of coordinates (both rescaling the null $u$ and $\ub$ coordinates and rotating the angular coordinates), one can attach appropriate boundaries to $\mathcal M_{Kerr}$ and define the \emph{event horizon} and \emph{Cauchy horizon}. The metric moreover extends smoothly up to the event and Cauchy horizons. See Section~\ref{sec.def.horizon}.
\item (Behaviour of the metric components) $\Om_\Ke$ is degenerate as $u,\ub \to \pm \infty$ and in fact obeys the following bounds in the region\footnote{We especially highlight this region since the present paper is exactly concerned with perturbations of the Kerr spacetime in this region.} $\{(u,\ub): u+\ub \geq C_R,\, u\leq -1\}\times \mathbb S^2$ for any fixed $C_R\in \mathbb R$ (see Figure~\ref{appendnewlabel} in Appendix~\ref{sec.Kerr.geometry} for a depiction of this subset)
$$e^{-\f{r_+-r_-}{r_-^2+a^2}(u+\ub)} \ls_{C_R} \Om_\Ke^2\ls_{C_R} e^{-\f{r_+-r_-}{r_-^2+a^2}(u+\ub)},$$
while $\gamma_\Ke$ and $b_\Ke$ are globally bounded and have non-vanishing limits as $u,\ub \to \pm \infty$. As before, here, $r_\pm = M \pm \sqrt{M^2 - a^2}$. Moreover, the angular coordinate partial derivatives of $\Om_\Ke^2$ and $\gamma_\Ke$ have similar behaviour, and the angular coordinate partial derivatives of $b_\Ke$ is degenerate as $u,\ub\to \pm\infty$ (see Propositions~\ref{gamma.Kerr.bounds}, \ref{b.Kerr.bounds} and \ref{Om.Kerr.bounds}).
\item (Estimates for the Ricci coefficients) For any $C_R\in \mathbb R$ and $I\in \mathbb N$, in $\{(u,\ub): u+\ub \geq C_R,\, u\leq -1\}\times \mathbb S^2$, the Ricci coefficients satisfy the following estimates (see Definition~\ref{def.Kerr.norm} and Proposition~\ref{Kerr.der.Ricci.bound})
$$\sum_{0\leq i \leq I}(|(\nab_\Ke)^i \chi_\Ke|_{\gamma_\Ke}+ |(\nab_\Ke)^i \eta_\Ke|_{\gamma_\Ke}+ |(\nab_\Ke)^i \etab_\Ke|_{\gamma_\Ke})\ls_{C_R,I} 1,$$ 
$$\sum_{0\leq i \leq I}|(\nab_\Ke)^i(\omb_\Ke-\f 12\f{r_+-r_-}{r_-^2+a^2})|_{\gamma_\Ke} + \sum_{0\leq i \leq I} |(\nab_\Ke)^i\chib_\Ke|_{\gamma_\Ke}\ls_{C_R,I} e^{-\f{r_+-r_-}{r_-^2+a^2}(u+\ub)}.$$
\item (Estimates for the renormalised null curvature components) For any $C_R\in \mathbb R$ and $I\in \mathbb N$, in $\{(u,\ub): u+\ub \geq C_R,\, u\leq -1\}\times \mathbb S^2$, the renormalised null curvature components satisfy the following estimates (see Definition~\ref{def.Kerr.norm}, Propositions~\ref{prop.Kerr.K} and \ref{Kerr.curv.est})
$$\sum_{0\leq i\leq I}|(\nab_\Ke)^i\betab_\Ke|_{\gamma_\Ke}\ls_{C_R,I} e^{-\f{r_+-r_-}{r_-^2+a^2}(u+\ub)},\quad \sum_{0\leq i\leq I}(|(\nab_\Ke)^i\sigmac_\Ke|_{\gamma_\Ke}+ |(\nab_\Ke)^i K_\Ke|_{\gamma_\Ke})\ls_{C_R,I} 1,$$
$$\sum_{0\leq i\leq I}|(\nab_\Ke)^i\beta_\Ke|_{\gamma_\Ke}\ls_{C_R,I} 1.$$
\end{itemize}

Further elaboration of these statements, as well as their proofs, can all be found Appendix~\ref{sec.Kerr.geometry}. In addition to the facts listed above, there are a number of other results regarding the Kerr interior in the double null foliation gauge that we will use in the remainder of the paper. For all those results, we will also refer the reader to Appendix~\ref{sec.Kerr.geometry}.

\section{Equations in the double null foliation and the schematic notation}\label{Eqindoublenullfolsec}
In this section, we derive consequences of the Einstein vacuum equations under the geometric setup introduced in Section \ref{secsetup}. From now on, we consider a spacetime $(\mathcal U,g)$ (with $\mathcal U=\mathcal W\times \mathbb S^2$ and $\mathcal W\subset \mathbb R^2$) where $g$ takes the form \eqref{formofthemetric}. Assume that $(\mathcal U,g)$ satisfies the Einstein vacuum equations \eqref{INTROvace}. We will recast \eqref{INTROvace} as a system for Ricci coefficients and null curvature components (recall the definition from \eqref{Ricci.coeff.def} and \eqref{curv.comp.def} in Section \ref{sec.RC.CC.def}).

In \textbf{Section~\ref{seceqn}}, we will write down the arising system for Ricci coefficients and null curvature components. As we have discussed earlier, for most of this paper, the exact form of the terms in the equations will not be too important. Instead, for most of the nonlinear terms, it suffices to understand the ``\emph{null structure}'' hidden in the equations. For this reason, we introduce a \emph{schematic notation} in \textbf{Section~\ref{schnot}} so as to simplify the exposition and to emphasise the structures of the equations that we need (cf.~Sections~\ref{lukmeth}, \ref{baridiaeisagwgns} and \cite{LukWeakNull}). We end the section by writing down the equations in schematic form in \textbf{Section~\ref{schnoteqn}}.

Let us already note here that the schematic notation introduced in this section will be further developed later in Section~\ref{sec.rse.conv}, cf.~Section~\ref{baridiaeisagwgns}.

\subsection{Equations}\label{seceqn}

Before writing down the equations, we need to introduce a few more notations. First, for $S$-tangent tensor fields, we define the following conventions for contraction with respect to the metric $\gamma$:
$$\phi^{(1)}\cdot\phi^{(2)}\doteq (\gamma^{-1})^{AC}(\gamma^{-1})^{BD}\phi^{(1)}_{AB}\phi^{(2)}_{CD} \quad\mbox{for symmetric $2$-tensors $\phi^{(1)}_{AB}$, $\phi^{(2)}_{AB}$,}$$
$$\phi^{(1)}\cdot\phi^{(2)}\doteq (\gamma^{-1})^{AB}\phi^{(1)}_{A}\phi^{(2)}_{B} \quad\mbox{for $1$-forms $\phi^{(1)}_{A}$, $\phi^{(2)}_{A}$,}$$
$$(\phi^{(1)}\cdot\phi^{(2)})_A\doteq (\gamma^{-1})^{BC}\phi^{(1)}_{AB}\phi^{(2)}_{C} \quad\mbox{for a symmetric $2$-tensor $\phi^{(1)}_{AB}$ and a $1$-form $\phi^{(2)}_{A}$,}$$
$$(\phi^{(1)}\hot\phi^{(2)})_{AB}\doteq \phi^{(1)}_A\phi^{(2)}_B+\phi^{(1)}_B\phi^{(2)}_A-\gamma_{AB}((\gamma^{-1})^{CD}\phi^{(1)}_C\phi^{(2)}_D) \quad\mbox{for one forms $\phi^{(1)}_A$, $\phi^{(2)}_A$,}$$
$$\phi^{(1)}\wedge\phi^{(2)}\doteq \in^{AB}(\gamma^{-1})^{CD}\phi^{(1)}_{AC}\phi^{(2)}_{BD}\quad\mbox{for symmetric two tensors $\phi^{(1)}_{AB}$, $\phi^{(2)}_{AB}$},$$
where $\in$ is the volume form associated to the metric $\gamma$.
Define $^*$ of $1$-forms and symmetric $2$-tensors respectively as follows (note that on $1$-forms this is the Hodge dual on $S_{u,\ub}$):
\begin{align*}
^*\phi_A \doteq  & \gamma_{AC} \in^{CB} \phi_B, \quad ^*\phi_{AB} \doteq  \gamma_{BD} \in^{DC} \phi_{AC}.
\end{align*}
Define also the trace for totally symmetric tensors of rank $r$ to be
\begin{equation}\label{tr.def}
(\tr\phi)_{A_1...A_{r-1}}\doteq (\gamma^{-1})^{BC}\phi_{BCA_1...A_{r-1}}.
\end{equation}

We now write down the equations. For this it is useful to recall the definitions \eqref{div.def}, \eqref{curl.def} and \eqref{nab.otimes.def}.
First, after separating into their trace and traceless parts (cf.~\eqref{traceless.chi.chib.def} and \eqref{trace.chi.chib.def}), $\chi$ and $\chib$ satisfy the following null structure equations:
\begin{equation}
\label{null.str1}
\begin{split}
\nab_4 \trch+\frac 12 (\trch)^2&=-|\chih|^2-2\omega \trch,\\
\nab_4\chih+\trch \chih&=-2 \omega \chih-\alpha,\\
\nab_3 \trchb+\frac 12 (\trchb)^2&=-2\omegab \trchb-|\chibh|^2,\\
\nab_3\chibh + \trchb\,  \chibh&= -2\omegab \chibh -\alphab,\\
\nab_4 \trchb+\frac1 2 \trch \trchb &=2\omega \trchb +2\rho- \chih\cdot\chibh +2\div \etab +2|\etab|^2,\\
\nab_4\chibh +\frac 1 2 \trch \chibh&=\nab\widehat{\otimes} \etab+2\omega \chibh-\frac 12 \trchb \chih +\etab\widehat{\otimes} \etab,\\
\nab_3 \trch+\frac1 2 \trchb \trch &=2\omegab \trch+2\rho- \chih\cdot\chibh+2\div \eta+2|\eta|^2,\\
\nab_3\chih+\frac 1 2 \trchb \chih&=\nab\widehat{\otimes} \eta+2\omegab \chih-\frac 12 \trch \chibh +\eta\widehat{\otimes} \eta.
\end{split}
\end{equation}
The other Ricci coefficients satisfy the following null structure equations:
\begin{equation}
\label{null.str2}
\begin{split}
\slashed{\nabla}_4\eta&=-\chi\cdot(\eta-\etab)-\b,\\
\slashed{\nabla}_3\etab &=-\chib\cdot (\etab-\eta)+\bb,\\
\nab_3\omega+\slashed{\nabla}_4\omegab&=4\omega\omegab+\zeta\cdot(\eta-\etab)-\eta\cdot\etab+ \rho.
\end{split}
\end{equation}
In our choice of gauge, since $\omega=0$ (see \eqref{gauge.con}), the last equation in \eqref{null.str2} reduces to the following equation for $\omegab$:
\begin{equation}\label{null.str.om}
\slashed{\nabla}_4\omegab =\zeta\cdot(\eta-\etab)-\eta\cdot\etab+ \rho.
\end{equation}

The Ricci coefficients also satisfy the following constraint equations:
\begin{equation}
\label{null.str3}
\begin{split}
\div\chih&=\frac 12 \slashed{\nabla} \trch - \zeta \cdot (\chi - \trch\gamma) -\beta,\\
\div\chibh&=\frac 12 \slashed{\nabla} \trchb + \zeta\cdot (\chibh-\trchb\gamma) +\betab,\\
\curl\eta &=-\curl\etab=\sigma +\frac 1 2\chibh \wedge\chih,\\
K&=-\rho+\frac 1 2 \chih\cdot\chibh-\frac 1 4 \trch \trchb.
\end{split}
\end{equation}
where we recall again that $K$ is the Gauss curvature of the spheres $S_{u,\ub}$.
The null curvature components satisfy the following null Bianchi equations:
\begin{equation}
\label{eq:null.Bianchi}
\begin{split}
&\nab_3\alpha+\frac 12 \trchb \alpha=\slashed{\nabla}\hot \beta+ 4\omegab\alpha-3(\chih\rho+^*\chih\sigma)+
(\zeta+4\eta)\hot\beta,\\
&\nab_4\beta+2\trch\beta = \div\alpha - 2\omega\beta +  (2\zeta+\etab)\cdot \alpha,\\
&\nab_3\beta+\trchb\beta=\slashed{\nabla}\rho + 2\omegab \beta +^*\slashed{\nabla}\sigma +2\chih\cdot\betab+3(\eta\rho+^*\eta\sigma),\\
&\nab_4\sigma+\frac 32\trch\sigma=-\div^*\beta+\frac 12\chibh\wedge\alpha-\zeta\wedge\beta-2\etab\wedge\beta,\\
&\nab_3\sigma+\frac 32\trchb\sigma=-\div ^*\betab-\frac 12\chih\wedge\alphab+\zeta\wedge\betab-2\eta\wedge\betab,\\
&\nab_4\rho+\frac 32\trch\rho=\div\beta-\frac 12\chibh\cdot\alpha+\zeta\cdot\beta+2\etab\cdot\beta,\\
&\nab_3\rho+\frac 32\trchb\rho=-\div\betab- \frac 12\chih\cdot\alphab+\zeta\cdot\betab-2\eta\cdot\betab,\\
&\nab_4\betab+\trch\betab=-\slashed{\nabla}\rho +^*\slashed{\nabla}\sigma+ 2\omega\betab +2\chibh\cdot\beta-3(\etab\rho-^*\etab\sigma),\\
&\nab_3\betab+2\trchb\betab=-\div\alphab-2\omegab\betab-(-2\zeta+\eta) \cdot\alphab,\\
&\nab_4\alphab+\frac 12 \trch\alphab=-\slashed{\nabla}\hot \betab+ 4\omega\alphab-3(\chibh\rho-^*\chibh\sigma)+
(\zeta-4\etab)\hot \betab.
\end{split}
\end{equation}

We now rewrite the Bianchi equations in terms of the renormalised null curvature components, i.e.~we will use the Gauss curvature $K$ of the spheres $S_{u,\ub}$ and the renormalised null curvature component $\sigmac$ defined in \eqref{sigmac.def} instead of $\rho$ and $\sigma$. Notice that with the notation that has now been introduced, we can write
$$\sigmac=\sigma+\frac 12 \chibh\wedge\chih.$$
Recall again that one of the main challenges of the problem in this paper is the potential null singularity at the Cauchy horizon. In \cite{LukWeakNull}, where the local existence problem for null singularities was treated, it was crucial that the null curvature components $\rho$ and $\sigma$ were eliminated in favor of $K$ and $\sigmac$ (cf.~Section~\ref{lukmeth}). The use of these renormalised null curvature components will also be important in this paper.

The (renormalised) Bianchi equations take the following form:
\begin{equation}
\label{eq:null.Bianchi2}
\begin{split}
\nab_3\beta+\trchb\beta=&-\slashed{\nabla} K  +^*\slashed{\nabla}\sigmac + 2\omegab \beta+2\chih\cdot\betab-3(\eta K-^*\eta\sigmac)+\frac 1 2(\slashed{\nabla}(\chih\cdot\chibh)+^*\slashed{\nabla}(\chih\wedge\chibh))\\
&+\f 32(\eta\chih\cdot\chibh+^*\eta\chih\wedge\chibh)-\frac 14 (\nab\trch \trchb+\trch\nab\trchb)-\frac 34 \eta\trch\trchb,\\
\nab_4\sigmac+\frac 32\trch\sigmac=&-\div^*\beta-\zeta\wedge\beta-2\etab\wedge
\beta-\frac 12 \chih\wedge(\nab\widehat{\otimes}\etab)-\frac 12 \chih\wedge(\etab\widehat{\otimes}\etab),\\
\nab_4 K+\trch K=&-\div\beta-\zeta\cdot\beta-2\etab\cdot\beta+\frac 12 \chih\cdot\nab\widehat{\otimes}\etab+\frac 12 \chih\cdot(\etab\widehat{\otimes}\etab)-\frac 12 \trch\div\etab-\frac 12\trch |\etab|^2,\\
\nab_3\sigmac+\frac 32\trchb\sigmac=&-\div ^*\betab+\zeta\wedge\betab-2\eta\wedge
\betab+\frac 12 \chibh\wedge(\nab\widehat{\otimes}\eta)+\frac 12 \chibh\wedge(\eta\widehat{\otimes}\eta),\\
\nab_3 K+\trchb K=&\div\betab-\zeta\cdot\betab+2\eta\cdot\betab+\frac 12 \chibh\cdot\nab\widehat{\otimes}\eta+\frac 12 \chibh\cdot(\eta\widehat{\otimes}\eta)-\frac 12 \trchb\div\eta-\frac 12 \trchb |\eta|^2,\\
\nab_4\betab+\trch\betab=&\slashed{\nabla} K +^*\slashed{\nabla}\sigmac+ 2\omega\betab +2\chibh\cdot\beta+3(\etab K+^*\etab\sigmac)-\frac 1 2(\slashed{\nabla}(\chih\cdot\chibh)-^*\slashed{\nabla}(\chih\wedge\chibh))\\
&+\frac 14 (\nab\trch \trchb+\trch\nab\trchb)-\f 32(\etab\chih\cdot\chibh-^*\etab\chih\wedge\chibh)+\frac 34 \etab\trch\trchb.
\end{split}
\end{equation}
Notice that we have obtained a system for the renormalised null curvature components in which the null curvature components $\alpha$ and $\alphab$ do not appear.\footnote{Moreover, compared to the renormalisation in \cite{LR}, this system do not contain the terms $\trch|\chibh|^2$ and $\trchb|\chih|^2$ which would be uncontrollable in the context of this paper. This observation already played an important role in \cite{LukWeakNull}.}

%From now on, we will use capital Latin letters $A\in \{1,2\}$ for indices on the spheres $S_{u,\ub}$ and Greek letters $\mu\in\{1,2,3,4\}$ for indices in the whole spacetime.

\subsection{Schematic Notation}\label{schnot}

We introduce a schematic notation for the equations that we consider. We will use the symbol $\eqs$ to emphasise that an equation is only to be understood schematically.

We first introduce schematic expressions for the Ricci coefficients according to the estimates that they will eventually be shown to obey. More precisely, let
$$\psi\in\{\eta,\etab\},\quad \psi_H\in \{\trch,\chih\},\quad \psi_{\Hb}\in\{\trchb,\chibh\}. $$
When writing an equation in the schematic notation, the following conventions will be used:
\begin{itemize}

\item (Numerical constants) We use the convention that the exact constants are kept on the left hand side of $\eqs$ while we do not keep track\footnote{Obviously, we will use this schematic notation only in situations where the exact constant in front of the term is irrelevant to the argument.} of the constants on the right hand side of $\eqs$. 

\item (Quantities on left and right hand side) When a schematic quantity, say $\psi_H$, appears on the right hand side of a schematic equation, one should understand it as summing over all possible $\psi_H\in \{\trch,\chih\}$. On the other hand, when a schematic quantity, again say $\psi_H$, appears on the left hand side, one should understand it as representing a fixed $\psi_H\in \{\trch,\chih\}$. For instance, in \eqref{tpH.S.1} below, the $\nab_3\psi_H -2\omb\psi_H$ on the left hand side is meant to represent either $\nab_3\trch -2\omb \trch$ or $\nab_3\chih - 2\omb\chih$ (but not, say, $\nab_3\chih - 2\omb\trch \gamma$).

\item (Contractions with respect to the metric) We will denote by $\psi\psi$ (or $\psi\psi_H$, etc) an arbitrary contraction with respect to the metric $\gamma$. 

\item (Derivatives and products) $\nab^i\psi^j$ will be used to denote the sum of all terms which are products of $j$ factors, such that each factor takes the form $\nab^{i_k}\psi$ and that the sum of all $i_k$'s is $i$, i.e. 
$$\nab^i\psi^j\eqs \displaystyle\sum_{i_1+i_2+...+i_j=i}\underbrace{\nab^{i_1}\psi\nab^{i_2}\psi...\nab^{i_j}\psi}_\text{j factors}.$$

\item (The bracket notation) We will use brackets to denote terms with one of the components in the brackets. For instance, the notation $\psi(\psi,\psi_H)$ denotes the sum of all terms of the form $\psi\psi$ or $\psi\psi_H$.

\item (Extra terms) We will allow in the schematic expression some terms on the right hand side of $\eqs$ which are not present in the original equation\footnote{This should already be apparent in the discussions of ``derivatives and products'' and ``the bracket notation'' since we allow the sum of all terms of a particular form.}. This creates no extra problems since in the proof of the theorem, we will be bounding all the terms on the right hand side of $\eqs$.

\item (Rank of the tensor) When using the schematic notation, we suppress the rank of the tensor involved. This will allow us to write schematically the equations for various tensors of different rank in a compact manner.
\item (Norm of a schematic quantity) We will often write $|\psi|$, $|\psi_H|$, etc.~to denote the norm of a schematic quantity, which is understood to be taken with respect to $\gamma$.
\end{itemize}

\begin{remark}[Differentiating the schematic equations]\label{rmk:diff.sch}
Note that with the conventions we introduced above, we can differentiate the schematic equation without knowing precisely the terms on the right hand side of the schematic equation. For instance, if $\phi\eqs \phi'$, then it holds that $\nab\phi \eqs \nab\phi'$. Similar statements hold for higher derivatives.
\end{remark}

We again remind the reader that we will further develop the above schematic notation in Section~\ref{sec.rse.conv}, where we introduce the \emph{reduced schematic notations}; see also Section~\ref{baridiaeisagwgns}.

\subsection{Equations in schematic notation}\label{schnoteqn}

Given the schematic notation introduced in Section~\ref{schnot}, we now rewrite some of the equations in Section~\ref{seceqn} in schematic form. These include the transport and constraint equations for the Ricci coefficients, as well as the Bianchi equations for the renormalised null curvature components. In addition, we will write down the schematic transport equations for the quantities we defined in \eqref{top.quantities.def}. (Note that while we do not explicitly write down schematic equations for the metric components, in the rest of the paper, we do need to rely on the equations in \eqref{nab.compatibility} and Proposition~\ref{metric.der.Ricci} satisfied by the metric components.)

The Ricci coefficients $\psi_H$ satisfy the schematic equation
\begin{equation}\label{tpH.S.1}
\nab_3 \psi_H -2\omb\psi_H \eqs K+ \nab\psi + \psi\psi+\psi_H \psi_{\Hb}.
\end{equation}
Here, we recall our convention that (1) all the exact constants are kept on the left hand side of the equality sign and (2) the $\psi_H$ on the left hand side refers to the same Ricci coefficient while $\psi_H$ on the right hand side can denote any of $\psi_H\in \{\trch,\chih\}$. The Ricci coefficients $\psi_{\Hb}$ similarly obey the schematic
\begin{equation}\label{tpHb.S.1}
\nab_4 \psi_{\Hb} \eqs K+ \nab\psi + \psi\psi+\psi_H \psi_{\Hb}.
\end{equation}
In fact, for later applications, it will be useful to note that $\psi_H$ and $\psi_{\Hb}$ obey slightly more precise schematic equations, which we also record here:
\begin{equation}\label{tpH.S.2}
\nab_3 \psi_H -2\omb\psi_H \eqs K+ \nab\eta + \psi\psi+\psi_H \psi_{\Hb},
\end{equation}
\begin{equation}\label{tpHb.S.2}
\nab_4 \psi_{\Hb} \eqs K+ \nab\etab + \psi\psi+\psi_H \psi_{\Hb}.
\end{equation}

The Ricci coefficients $\psi\in\{\eta,\etab\}$ obey either one of the following schematic equations:
\begin{equation}\label{etab.S.1}
\nab_3 \etab -\betab \eqs \psi\psi_{\Hb}
\end{equation}
or
\begin{equation}\label{eta.S.1}
\nab_4 \eta +\beta \eqs \psi\psi_H.
\end{equation}

The Ricci coefficient $\omb$ verifies the following schematic equation:
\begin{equation}\label{omb.S}
\nab_4\omb + K\eqs \psi\psi+\psi_H \psi_{\Hb}.
\end{equation}

The Codazzi equations, i.e.~the first two equations in \eqref{null.str3}, can be expressed schematically as follows:
\begin{equation}\label{Codazzi.S.1}
\div\chih+\beta \eqs \nab\trch+\psi\psi_H, 
\end{equation}
and
\begin{equation}\label{Codazzi.S.2}
\div\chibh-\betab \eqs \nab\trchb+\psi\psi_{\Hb}.
\end{equation}

Using the Codazzi equations, we can further simplify the schematic equations \eqref{etab.S.1} and \eqref{eta.S.1} for $\etab$ and $\eta$ to obtain
\begin{equation}\label{etab.S.2}
\nab_3 \etab \eqs \,\sum_{i_1+i_2=1}\psi^{i_1}\nab^{i_2}\psi_{\Hb}
\end{equation}
and
\begin{equation}\label{eta.S.2}
\nab_4 \eta \eqs\, \sum_{i_1+i_2=1}\psi^{i_1}\nab^{i_2}\psi_H.
\end{equation}

We also rewrite the Bianchi equations in the schematic notation:
\begin{equation}
\label{eq:null.Bianchi3}
\begin{split}
\nab_3\beta+\slashed{\nabla} K  -^*\slashed{\nabla}\sigmac -2\omb\beta \eqs\,& \sum_{i_1+i_2+i_3=1}\psi^{i_1}\nab^{i_2}\psi_{\Hb} \nab^{i_3}\psi_H+\psi K+\sum_{i_1+i_2=1} \psi^{i_1}\psi\nab^{i_2}\psi,\\
\nab_4\sigmac+\div^*\beta\eqs\,&\sum_{i_1+i_2+i_3=1}\psi^{i_1}\nab^{i_2}\psi\nab^{i_3}\psi_H,\\
\nab_4 K+\div\beta\eqs\,&\psi_H K+\sum_{i_1+i_2+i_3=1}\psi^{i_1}\nab^{i_2}\psi\nab^{i_3}\psi_H,\\
\nab_3\sigmac+\div ^*\betab \eqs\,& \sum_{i_1+i_2+i_3=1}\psi^{i_1}\nab^{i_2}\psi\nab^{i_3}\psi_{\Hb},\\
\nab_3 K-\div\betab\eqs\,&\psi_{\Hb}K+\sum_{i_1+i_2+i_3=1}\psi^{i_1}\nab^{i_2}\psi\nab^{i_3}\psi_{\Hb},\\
\nab_4\betab-\slashed{\nabla} K -^*\slashed{\nabla}\sigmac\eqs\,&\sum_{i_1+i_2+i_3=1}\psi^{i_1}\nab^{i_2}\psi_{H} \nab^{i_3}\psi_{\Hb}+\psi K+\sum_{i_1+i_2=1} \psi^{i_1}\psi\nab^{i_2}\psi.
\end{split}
\end{equation}
Notice that in order to derive the schematic equations in \eqref{eq:null.Bianchi3}, we have used the first three equations in \eqref{null.str3} to express the (renormalised) null curvature components $\beta$, $\betab$ and $\sigmac$ in terms of the Ricci coefficients and their derivatives.

Finally, we have the schematic equation for the quantities defined in \eqref{top.quantities.def}\footnote{These schematic equations can be easily derived using Proposition~\ref{repeated.com} (that we prove later), the fact $\div ^*\nab\sigmac =0$ and the schematic equations above. Specifically, \eqref{mub.S} follows from \eqref{etab.S.2} and \eqref{eq:null.Bianchi3}; \eqref{mu.S} follows from \eqref{eta.S.2} and \eqref{eq:null.Bianchi3}; \eqref{ombs.S} follows from \eqref{omb.S} and \eqref{eq:null.Bianchi3}; we omit the straightforward details.}:
\begin{equation}\label{mub.S}
\nab_3\mub \eqs \psi_{\Hb} K + \sum_{i_1+i_2+i_3=1} \psi^{i_1}\nab^{i_2}\psi \nab^{i_3}\psi_{\Hb},
\end{equation}
\begin{equation}\label{mu.S}
\nab_4\mu \eqs \psi_H K + \sum_{i_1+i_2+i_3=1} \psi^{i_1}\nab^{i_2}\psi \nab^{i_3}\psi_H,
\end{equation}
\begin{equation}\label{ombs.S}
\nab_4\ombs \eqs \sum_{i_1+i_2+i_3= 2}\psi^{i_1}\nab^{i_2}\psi\nab^{i_3}\psi + \sum_{i_1+i_2+i_3= 2} \psi^{i_1}\nab^{i_2}\psi_H\nab^{i_3}\psi_{\Hb} + \sum_{i_1+i_2+i_3= 1}\psi^{i_1}\nab^{i_2}\psi \nab^{i_3}K.
\end{equation}
Importantly, the special combinations $\mub$, $\mu$ and $\ombs$ of (derivatives of) Ricci coefficients and the curvature components (or their derivatives) remove the top order derivative of the curvature component on the right hand side\footnote{For instance, in the equation \eqref{mub.S}, if one considers $\nab\etab$ instead of $\mub$, then the schematic equation takes the form
$$\nab_3\nab\etab \eqs \nab \betab + \psi_{\Hb} K + \sum_{i_1+i_2+i_3=1} \psi^{i_1}\nab^{i_2}\psi \nab^{i_3}\psi_{\Hb}.$$
The additional term $\nab \betab$ on the right hand side is one derivative of a curvature component.
}.

\section{Statement of main theorem and the construction of the double null foliation}\label{sec.main.thm.dnf}

In this section, we formulate precisely Theorem \ref{PRWTO} of Section~\ref{sec:MT}. In fact, it will be convenient to state two versions of the theorem: the first will be given in Section \ref{precise.PRWTO} as Theorem \ref{main.thm.spacelike.formulation}, where the notion of the ``expected induced geometry of a dynamical black hole settling down to Kerr on a suitable spacelike hypersurface $\Sigma_0$ in the black hole interior'' will be given in terms of the geometric data on $\Sigma_0$. The second version will be given in Section \ref{sec.main.thm.dn} as Theorem \ref{aprioriestimates} in terms of the geometric quantities associated to a double null foliation (cf.~Section~\ref{secsetup}). The latter version has the additional advantage that it is more in line with the geometric setup of the proof of the theorem and we can therefore also describe in a quantitative manner the closeness of the maximal globally hyperbolic future development and the background Kerr solution. Moreover, we will show that the conditions on the geometric data given in Theorem~\ref{main.thm.spacelike.formulation} indeed imply the conditions in Theorem~\ref{aprioriestimates} for an appropriately defined double null foliation.

The section will be organised as follows. In \textbf{Section~\ref{precise.PRWTO}}, we state \textbf{Theorem~\ref{main.thm.spacelike.formulation}}, which is a formulation of Theorem~\ref{PRWTO} in terms of the geometric data on $\Sigma_0$. In Section~\ref{setup.double null}--Section~\ref{sec.main.thm.dn}, we discuss the formulation of the main theorem in the double null foliation gauge. More precisely, in \textbf{Section~\ref{setup.double null}}, we set up the data for a double null foliation and a system of double null coordinate system. In \textbf{Section~\ref{sec.identification}}, given an appropriate solution in a double null foliation gauge, we introduce an identification of the spacetime with the background Kerr spacetime. In \textbf{Section~\ref{sec.smallness.data}}, we derive the initial data for the geometric quantities with respect to the double null foliation gauge. In particular, we prove that given the assumptions on the geometric data in Theorem~\ref{main.thm.spacelike.formulation}, the data for the differences of the geometric quantities with respect to the double null foliation gauge are appropriately small. In \textbf{Section~\ref{sec.data}}, we define the initial data energy. In \textbf{Section~\ref{sec.norms}}, we introduce the energies that we will use to control the solution (cf.~Section~\ref{baridiaeisagwgns}). In \textbf{Section~\ref{sec.main.thm.dn}}, we state \textbf{Theorem~\ref{aprioriestimates}}, which is our second formulation\footnote{In view of Section~\ref{sec.smallness.data}, Theorem~\ref{aprioriestimates} indeed implies Theorem~\ref{main.thm.spacelike.formulation}.} of Theorem~\ref{PRWTO}, and is given in terms of the geometric quantities in the double null foliation gauge. In \textbf{Section~\ref{sec.BA}}, we discuss the main bootstrap assumption and the main bootstrap theorem (\textbf{Theorem~\ref{main.quantitative.thm}}).

\subsection{Statement of main theorem starting from geometric Cauchy data}\label{precise.PRWTO}

We begin by describing the initial data that we consider. For fixed parameters $M$ and $a$ such that $0<|a|<M$, we consider initial data $(\Sigma_0=[1,\infty)\times \mathbb S^2,\hat{g},\hat{k})$ which approach the corresponding data on an admissible spacelike hypersurface in a Kerr spacetime with parameters $M$ and $a$. We first fix a constant $C_R\in \mathbb R$. We will compare our initial data $(\Sigma_0,\hat{g},\hat{k})$ with the induced first and second fundamental forms $(\hat{g}_\Ke,\hat{k}_\Ke)$ on the hypersurface $\Sigma_\Ke\doteq \{(u,\ub): u+\ub=C_R,\, \ub\geq 1\}\times\mathbb S^2$ in Kerr spacetime (cf.~Section~\ref{sec.Kerr.spacelike}).

We require the geometric initial data $(\Sigma_0,\hat{g},\hat{k})$ to be smooth, and to solve the following constraint equations
\begin{equation}\label{constraints}
R_{\hat{g}}+(\mbox{tr}_{\hat g} \hat k)^2-|\hat k|_{\hat{g}}^2=0,\quad \mbox{div}_{\hat{g}} \hat k-\hat{\nabla} (\mbox{tr}_{\hat g} \hat k)=0,
\end{equation}
where $\hat\nabla$ is the Levi--Civita connection associated to $\hat{g}$, $R_{\hat{g}}$ is the scalar curvature of $\hat{g}$ and $\mbox{div}_{\hat{g}}$ and $\mbox{tr}_{\hat{g}}$ are both taken with respect to $\hat{g}$.

We also require $(\Sigma_0,\hat{g},\hat{k})$ to be close to $(\Sigma_\Ke,\hat{g}_\Ke,\hat{k}_\Ke)$. Since $\Sigma_0$ and $\Sigma_\Ke$ are diffeomorphic, we stipulate the existence of a diffeomorphism $\Phi:\Sigma_0 \to \Sigma_\Ke$ such that $\hat{g} -\Phi_* \hat{g}_\Ke$  and $\hat{k} - \Phi_* \hat{k}_\Ke$ are small. To precisely state the smallness condition, we first pullback the function $\tau\doteq \ub$ introduced in Section~\ref{sec.Kerr.spacelike}. Then, (from now on) slightly abusing notation and writing $\hat{g}_\Ke$, $\hat{k}_\Ke$ instead of $\Phi_* \hat{g}_\Ke$, $\Phi_* \hat{k}_\Ke$, we require $(\hat{g},\hat{k})$ to satisfy the following smallness conditions:
\begin{equation}\label{small.geometric.data}
\sum_{i=0}^{5}\|\tau^{\f 12+\delta}\hat{\nabla}_\Ke^i(\hat{g}-\hat{g}_\Ke)\|_{L^2(\Sigma_0,\hat{g}_\Ke)}+\sum_{i=0}^{4}\|\tau^{\f 12+\delta}\hat{\nabla}_\Ke^i(\hat{k}-\hat{k}_\Ke)\|_{L^2(\Sigma_0,\hat{g}_\Ke)}\leq \ep
\end{equation}
and
\begin{equation}\label{small.geometric.data.Linfty}
\sum_{i=0}^{3}\|\tau^{\f 12+\delta}\hat{\nabla}_\Ke^i(\hat{g}-\hat{g}_\Ke)\|_{L^\infty(\Sigma_0,\hat{g}_\Ke)}+\sum_{i=0}^{2}\|\tau^{\f 12+\delta}\hat{\nabla}_\Ke^i(\hat{k}-\hat{k}_\Ke)\|_{L^\infty(\Sigma_0,\hat{g}_\Ke)}\leq \ep
\end{equation}
for some $\de>0$ and $\ep>0$ to be specified below and where $\hat{\nabla}_\Ke$ is the Levi--Civita connection with respect to the induced metric $\hat{g}_\Ke$ on $\Sigma_0$. The $L^2(\Sigma_0,\hat{g}_\Ke)$ norm in \eqref{small.geometric.data} is defined for a general rank-$r$ covariant tensor $\phi_{i_1i_2\dots i_r}$ tangential to $\Sigma_0$ by
$$\|\phi\|_{L^2(\Sigma_0,\hat{g}_\Ke)}\doteq  \left(\int_{\Sigma_0} (\hat{g}_\Ke^{-1})^{i_1j_1}\cdots(\hat{g}_\Ke^{-1})^{i_rj_r}\phi_{i_1i_2\dots i_r}\phi_{j_1j_2\dots j_r}\,dV_{\hat{g}}\right)^{\f 12},$$
where $dV_{\hat{g}}$ is the volume form induced by the Riemannian metric $\hat{g}$. The $L^\infty(\Sigma_0,\hat{g}_\Ke)$ norm in \eqref{small.geometric.data.Linfty} is defined for a general rank-$r$ covariant tensor $\phi_{i_1i_2\dots i_r}$ tangential to $\Sigma_0$ by
$$\|\phi\|_{L^\infty(\Sigma_0,\hat{g})}\doteq \esssup \left((\hat{g}_\Ke^{-1})^{i_1j_1}\cdots(\hat{g}_\Ke^{-1})^{i_rj_r}\phi_{i_1i_2\dots i_r}\phi_{j_1j_2\dots j_r}\right)^{\f 12}.$$
Moreover, in both the \eqref{small.geometric.data} and \eqref{small.geometric.data.Linfty}, $\hat{\nabla}_\Ke^i(\hat{g}-\hat{g}_\Ke)$ and $\hat{\nabla}_\Ke^i(\hat{k}-\hat{k}_\Ke)$ are considered as covariant tensors tangential to $\Sigma_0$ of rank $i+2$.

\begin{remark}\label{rmk:nablatoLie.small}
It is easy to check, using the estimates for the Kerr metric in Appendix~\ref{sec.Kerr.geometry}, that \eqref{small.geometric.data} and \eqref{small.geometric.data.Linfty} imply the following smallness condition in terms of Lie derivatives
\begin{equation*}
\begin{split}
\sum_{i=0}^{5}\sum_{X_1,\dots,X_i \in\{ n, \f{\rd}{\rd \th^A}\}}&\|\tau^{\f 12+\delta}\mathcal L_{X_1}\cdots \mathcal L_{X_i}(\hat{g}-\hat{g}_\Ke)\|_{L^2(\Sigma_0,\hat{g}_\Ke)}\\
&+\sum_{i=0}^{4}\sum_{X_1,\dots,X_i \in\{ n, \f{\rd}{\rd \th^A}\}}\|\tau^{\f 12+\delta}\mathcal L_{X_1}\cdots \mathcal L_{X_i}(\hat{k}-\hat{k}_\Ke)\|_{L^2(\Sigma_0,\hat{g}_\Ke)}\leq C_L \ep,
\end{split}
\end{equation*}
and
\begin{equation*}
\begin{split}
\sum_{i=0}^{3}\sum_{X_1,\dots,X_i \in\{ n, \f{\rd}{\rd \th^A}\}}&\|\tau^{\f 12+\delta}\mathcal L_{X_1}\cdots \mathcal L_{X_i}(\hat{g}-\hat{g}_\Ke)\|_{L^\infty(\Sigma_0,\hat{g}_\Ke)}\\
&+\sum_{i=0}^{2}\sum_{X_1,\dots,X_i \in\{ n, \f{\rd}{\rd \th^A}\}}\|\tau^{\f 12+\delta}\mathcal L_{X_1}\cdots \mathcal L_{X_i}(\hat{k}-\hat{k}_\Ke)\|_{L^\infty(\Sigma_0,\hat{g}_\Ke)}\leq C_L \ep,
\end{split}
\end{equation*}
for some $C_L>0$ depending only on $C_R$, $\de$ and the background Kerr parameters $M$ and $a$, as long as $\ep$ in \eqref{small.geometric.data} and \eqref{small.geometric.data.Linfty} is sufficiently small. To see this, it suffices to use the estimates we obtained for the Kerr metric in Appendix~\ref{sec.Kerr.geometry} to control the Christoffel symbols associated to $\hat{g}_\Ke$.
\end{remark}

We can now formulate Theorem \ref{PRWTO} with the precise assumptions:
\begin{theorem}\label{main.thm.spacelike.formulation}
Let $(\Sigma_0,\hat{g},\hat{k})$ be a smooth initial data set for the Einstein vacuum equations satisfying the constraint equations \eqref{constraints}. For a fixed $C_R \in \mathbb R$, let $(\Sigma_\Ke,\hat{g},\hat{k})$ be the induced data on $\Sigma_\Ke=\{(u,\ub): u+\ub=C_R,\, \ub\geq 1\} \times \mathbb S^2$ in $(\mathcal M_{Kerr},g_{a,M})$ (see~Section~\ref{sec.Kerr.spacelike}).

Assume that $(\Sigma_0,\hat{g},\hat{k})$ is close to and approaches $(\Sigma_\Ke,\hat{g}_\Ke,\hat{k}_\Ke)$ in the sense that the smallness conditions \eqref{small.geometric.data} and \eqref{small.geometric.data.Linfty}
\begin{equation*}
\sum_{i=0}^{5}\|\tau^{\f 12+\delta}\hat{\nabla}_\Ke^i(\hat{g}-\hat{g}_\Ke)\|_{L^2(\Sigma_0,\hat{g}_\Ke)}+\sum_{i=0}^{4}\|\tau^{\f 12+\delta}\hat{\nabla}_\Ke^i(\hat{k}-\hat{k}_\Ke)\|_{L^2(\Sigma_0,\hat{g}_\Ke)}\leq \ep  \tag{\ref{small.geometric.data}}
\end{equation*}
and
\begin{equation*}
\sum_{i=0}^{3}\|\tau^{\f 12+\delta}\hat{\nabla}_\Ke^i(\hat{g}-\hat{g}_\Ke)\|_{L^\infty(\Sigma_0,\hat{g}_\Ke)}+\sum_{i=0}^{2}\|\tau^{\f 12+\delta}\hat{\nabla}_\Ke^i(\hat{k}-\hat{k}_\Ke)\|_{L^\infty(\Sigma_0,\hat{g}_\Ke)}\leq \ep \tag{\ref{small.geometric.data.Linfty}}
\end{equation*}
hold for some $\de>0$ and $\ep>0$.

Let $(\mathcal{M},g)$ be the maximal globally hyperbolic future development of the restriction of the initial data on\footnote{Note that the set $\Sigma_0\cap(\{\tau : \tau < -u_f + C_R\}\times \mathbb S^2)$ is a subset of the initial hypersurface containing the infinite end.} $\Sigma_0\cap (\{\tau : \tau < -u_f + C_R\}\times \mathbb S^2)$ as given by \cite{geroch}. Then for every $\de>0$ there exist $\ep_{Cauchy}>0$ and $u_{Cauchy}\leq C_R-1$ (depending on $\de$, $C_R$ and the Kerr parameters $a$ and $M$) such that if $\ep<\ep_{Cauchy}$ and $u_f\leq u_{Cauchy}$, $(\mathcal{M},g)$ has a non-trivial Cauchy horizon $\mathcal{CH}^+$ across which the metric is continuously extendible. Moreover, $(\mathcal{M},g)$ remains $C^0$ close to the background Kerr spacetime.
\end{theorem}

\begin{remark}
The non-trivial Cauchy horizon $\mathcal{CH}^+$ should be understood as a null boundary that one can attach to the solution in a similar manner as the Cauchy horizon in the Kerr spacetime (cf.~Section~\ref{sec.def.horizon}). This is in contrast to the ``trivial'' Cauchy horizon that one can attach to the solution simply because we restrict our initial data to have an incomplete end. We will give a more descriptive discussion in Remark~\ref{rmk:CH} in Section~\ref{sec.main.thm.dn}, after we set up the double null foliation; see also Section~\ref{sec.C0}.
\end{remark}

\begin{remark}
The sense in which $(\mathcal{M},g)$ is close to the background Kerr solution will be made precise in Theorem~\ref{aprioriestimates} in Section~\ref{sec.main.thm.dn}.
\end{remark}

\begin{remark}
It is convenient in this paper to restrict only to the maximal globally hyperbolic future development of a subset of $\Sigma_0$ as indicated in Theorem \ref{main.thm.spacelike.formulation}. As in the statement of Theorem \ref{main.thm.spacelike.formulation}, in the remainder of the paper, we will denote by $(\mathcal M,g)$ the maximal globally hyperbolic future development only of the subset $\Sigma_0\cap\{(\tau,\th^1,\th^2): \tau< -u_f+C_R \}\subset \Sigma_0$. Nevertheless, when combined with Theorem \ref{TRITO} (which is to appear in \cite{DafLuk3}), one sees that this restriction is not necessary, but instead the conclusion of Theorem \ref{main.thm.spacelike.formulation} holds for the entire maximal globally hyperbolic future development of the data on $\Sigma_0$.
\end{remark}

\begin{remark}
In principle, one can directly construct initial data sets satisfying the constraint equations that obey the bound \eqref{small.geometric.data}. Nevertheless, we will not pursue this since we will show in \cite{DafLuk2} that such an initial data set $(\Sigma_0,\hat{g},\hat{k})$ indeed arises from solving a characteristic initial value problem starting from a complete null hypersurface that approaches the event horizon of Kerr spacetimes.
\end{remark}

\subsection{Setting up a double null foliation}\label{setup.double null}

Given $(\Sigma_0,\hat{g},\hat{k})$ as in Theorem \ref{main.thm.spacelike.formulation}, our goal in this subsection is to set up the initial data for a double null foliation as well as a double null coordinate system. After setting up the initial data, we will also construct an open neighbourhood of an appropriate subset of $\Sigma_0$ to the future of $\Sigma_0$ in which a double null foliation as well as a double null coordinate system exist; see Proposition~\ref{prop:doublenull.local}.

{\bf Introduction of coordinates on $\Sigma_0$.} Let $(\tau,\th^1_\Sigma,\th^2_\Sigma)$ be a local coordinate system on the spacelike hypersurface $\Sigma_\Ke$ in Kerr spacetime as in Section~\ref{sec.Kerr.spacelike}, i.e.~$\tau\doteq \ub$ and $\th^A_\Sigma\doteq \th^A$, where $\th^A$ are spherical local coordinates for $2$-spheres $S_{u,\ub}$ in $\mathcal M_{Kerr}$ satisfying $\Lb\th^A=0$. Since $\Sigma_0$ is diffeomorphic to $\Sigma_\Ke=[1,\infty)\times \mathbb S^2$, we can (locally) pull back the coordinate functions $(\tau,\th^1_\Sigma,\th^2_\Sigma)$ such that the metric $\hat{g}$ on $\Sigma_0$ takes the form
\begin{equation}\label{metric.gh}
\hat{g}= \Phi \, d\tau\otimes d\tau+ w_A (d\tau\otimes d\th_{\Sigma}^A+d\th_{\Sigma}^A\otimes d\tau)+(\gamma_\Sigma)_{AB} d\th_{\Sigma}^A\otimes d\th_{\Sigma}^B,
\end{equation}
where $\Phi$, $w_A$ and $(\gamma_\Sigma)_{AB}$ are close to their corresponding values\footnote{Recall that the metric on $\Sigma_\Ke$ on Kerr spacetime is given by \eqref{Kerr.SigmaKe}.} on Kerr spacetime in the sense of \eqref{small.geometric.data}.

{\bf Initial data for the double null foliation and double null coordinate system.} According to Proposition~\ref{prop:u.extend}, in order to construct a double null foliation and a double null coordinate system, it suffices to prescribe appropriate initial data for $u$, $\ub$, $\th^A$. We specify the initial data for $u$, $\ub$, $\th^A$ ($A=1,2$) as follows:
\begin{equation}\label{coord.init} 
\ub\restriction_{\Sigma_0} = \tau,\quad u\restriction_{\Sigma_0} = -\tau + C_R,\quad \th^A\restriction_{\Sigma_0}=\th^A_{\Sigma}.
\end{equation}

We are now in a position to use Proposition~\ref{prop:u.extend} and show that there is always (at least) a small neighbourhood such that the double null foliation and the double null coordinate system are both well-defined:
\begin{proposition}\label{prop:doublenull.local}
There exists a solution $(\mathcal U_{\ub_{local}},g)$ to the Einstein vacuum equations, where $\mathcal U_{\ub_{local}}\doteq \mathcal W_{\ub_{local}}\times \mathbb S^2$ and $\mathcal W_{\ub_{local}} \doteq \{(u,\ub): u+\ub\geq C_R,\, u< u_f,\, \ub< \ub_{local}\}$ for some $\ub_{local}\in (-u_f+C_R,\infty)$, such that the metric $g$ takes the form \eqref{formofthemetric}. Moreover, on $\mathcal U_{\ub_{local}}\cap \{u+\ub=C_R\}$,
\begin{itemize}
\item the induced first and second fundamental forms coincide with the geometric initial data $(\hat{g},\hat{k})$, and 
\item $(u,\ub,\th^A)$ obey the initial conditions \eqref{coord.init}, $u$, $\ub$ satisfy the eikonal equation \eqref{basic.eikonal} and for $A=1,2$, $\th^A$ is constant along a integral curve of $D\ub$.
\end{itemize}
\end{proposition}
\begin{proof}
Let $(\widetilde{\mathcal M},g)$ be the maximal globally hyperbolic future development of the data $(\Sigma_0,\hat{g},\hat{k})$ as given by \cite{geroch}. (Note that $\widetilde{\mathcal M}$ is a superset of $\mathcal M$.) We then apply Proposition~\ref{prop:u.extend} (for $y=-u_f+C_R$) to obtain locally a double null foliation and a system of double null coordinates. Restricting then to $u+\ub\geq C_R$, $u< u_f$, $\ub< \ub_{local}$ yields the conclusion.
\end{proof}

\begin{remark}
We note already that in the context of the bootstrap argument (see Sections~\ref{introbootstr} and \ref{sec:continuity}), Proposition~\ref{prop:doublenull.local} plays the role of establishing the non-emptiness of $\mathfrak I$.
\end{remark}

\subsection{Identification with Kerr spacetime and the difference quantities}\label{sec.identification}

In this subsection, we introduce an identification of an appropriate spacetime arising from (appropriate subsets) of the initial data set $(\Sigma_0,\hat{g},\hat{k})$ with the background Kerr spacetime $(\mathcal M_{Kerr}, g_{a,M})$ (see~Appendix~\ref{sec.Kerr.geometry}). This identification relies on the double null coordinates and in particular applies to the solution $(\mathcal U_{\ub_{local}},g)$ that we constructed in Proposition~\ref{prop:doublenull.local}. Nevertheless, it is convenient to already consider a more general class of spacetimes.

\begin{figure}
\centering{
\def\svgwidth{9pc}
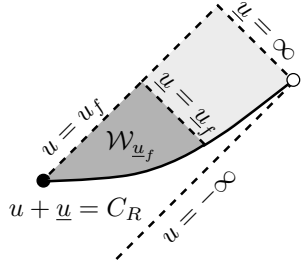}
\caption{The domain $\mathcal{W}_{\underline{u}_f}$}\label{sec4newlabel}
\end{figure}

For the rest of this subsection, assume we are given a solution $(\mathcal U_{\ub_f},g)$ (see Figure~\ref{sec4newlabel}) to the Einstein vacuum equation, where $\mathcal U_{\ub_f} = \mathcal W_{\ub_f} \times \mathbb S^2$, and $\mathcal W_{\ub_f} \subset \mathbb R^2$ is given by
$$\mathcal W_{\ub_f} = \{(u,\ub): u+\ub\geq C_R,\,u< u_f,\,\ub< \ub_f\}$$ 
for some $\ub_f\in (-u_f+C_R,\infty]$ such that both the geometric data and the data for the double null coordinates (cf.~\eqref{coord.init}) are as prescribed. On such a spacetime $(\mathcal U_{\ub_f},g)$, we can identify $\mathcal U_{\ub_f}$ with the background Kerr spacetime and compare their geometries. For this purpose, we will use the double null coordinates $(u,\ub,\th_*,\phi_*)$ on $\mathcal M_{Kerr}$ defined in Section~\ref{Kerr.dbn}; see also Section~\ref{sec.Kerr.dbn.text}.

{\bf Identification of $\mathcal U_{\ub_f}$ with Kerr spacetime.} Recall from Section~\ref{precise.PRWTO} that there exists a diffeomorphism $\Phi: \Sigma_0 \to \Sigma_\Ke$. We now extend this to a map $\widetilde{\Phi}: \mathcal U_{\ub_f} \to \mathcal M_{Kerr}$, which is a diffeomorphism onto its image. This in particular induces an identification of $\mathcal U_{\ub_f}$ and $\widetilde{\Phi}(\mathcal U_{\ub_f})\subset \mathcal M_{Kerr}$. 

Let $p\in \mathcal U_{\ub_f}$. Since $u$ and $\ub$ can be understood as functions $\mathcal U_{\ub_f} \to \mathbb R$, $u(p)$ and $\ub(p)$ are well-defined. Moreover, there exists a unique point $\tilde{p}$ on the sphere $S_{-\ub(p)+C_R,\ub(p)}\subset \Sigma_0$ such that $p$ and $\tilde{p}$ can be connected by an integral curve of $\Lb$. Define $\widetilde{\Phi}(p)$ to be the unique point $q\in S_{u(p),\ub(p)}\subset \mathcal M_{Kerr}$ such that $q$ and $\Phi(\tilde{p})$ can be connected by an integral curve of $\Lb_\Ke$. It is easy to check that $\widetilde{\Phi}$ is indeed an extension of $\Phi$, and that it is a diffeomorphism onto its image.

The above identification can also be understood after introducing local coordinates on the $2$-spheres. More precisely, consider a local coordinate system $(\tau,\th_\Sigma^1,\th_\Sigma^2)$ on $\Sigma_\Ke$, and pull back these coordinates to $\Sigma_0$ via $\Phi$. Define in each of $\mathcal U_{\ub_f}$ and $\mathcal M_{Kerr}$ a system of local double null coordinate $(u,\ub,\th^1,\th^2)$ as in Section~\ref{setup.double null}. (Note that $(u,\ub,\th^1,\th^2)$ in particular satisfies \eqref{th.transport}.) Now the identification above is equivalent to an identification of $\mathcal U_{\ub_f}$ and (the corresponding subset of) $\mathcal M_{Kerr}$ via the values of the coordinate functions $(u,\ub,\th^1,\th^2)$. (Note that we must use at least two charts in order to cover $\mathbb S^2$. Nevertheless, the identification we describe above is \emph{independent} of coordinates.)

The identification that we introduced in particular allows us to pull back tensors and functions defined $\mathcal M_{Kerr}$ to $\mathcal U_{\ub_f}$. This in particular includes the Ricci coefficients and the curvature components.

{\bf Differences of geometric quantities of $(\mathcal U_{\ub_f},g)$ and Kerr.} A particular consequence of the discussions above is that we can define the difference of the metric components $\gamma$, $\Om$, $b$; the Ricci coefficients $\chi$, $\chib$, $\eta$, $\etab$, $\om$, $\zeta$, the renormalised null curvature components $\beta$, $K$, $\sigmac$, $\betab$ and the quantities $\mu$, $\mub$, $\ombs$ of $(\mathcal U_{\ub_f},g)$ with that on the Kerr spacetime $(\mathcal M_{Kerr},g_{a,M})$. These differences will be understood as the differences of tensor fields tangential to the $2$-spheres $S_{u,\ub}$.

We now define the conventions that we will use for the difference of the geometric quantities:
\begin{itemize}
\item (The $\;{ }_\Ke\;$ convention) {\bf\emph{An unsubscripted tensor denotes a geometric quantity associated to $(\mathcal U_{\ub_f},g)$, a tensor with a subscript ${ }_{\Ke}$ denotes a geometric quantity associated to the background Kerr metric.}}

\item (The $\;\widetilde{ }\;$ convention) {\bf\emph{$\;\widetilde{ }\;$ denotes the difference of the quantities between the spacetime $(\mathcal U,g)$ and the background Kerr spacetime.}} For instance, we have
$$\widetilde{\gamma}\doteq \gamma-\gamma_{\Ke},$$
where $\gamma$ is the metric on the $2$-spheres $S_{u,\ub}$ in $(\mathcal U_{\ub_f},g)$ and $\gamma_{\Ke}$ is the corresponding metric in the background Kerr metric. 

\item (Quantities used for the $\;\widetilde{ }\;$ convention) The $\;\widetilde{ }\;$ convention that we have introduced in the previous bullet point will be used for all metric components, Ricci coefficients \eqref{Ricci.coeff.def}, curvature components \eqref{curv.comp.def}, higher order quantities associated to the Ricci coefficients \eqref{top.quantities.def}, as well as null frames, differential operators (Proposition~\ref{def.slashed}), etc. There are expressions ``involving more than one objects'' and therefore have various ways to take differences. When there is no risk of confusion, we simply distinguish them by the location that the $\;\widetilde{ }\;$ is placed. For instance,
\begin{itemize}
\item $\widetilde{\nab_4 \log\Om} = \nab_4 \log\Om - (\nab_4 \log\Om)_\Ke$, while $\nab_4\widetilde{\log\Om} = \nab_4(\log\Om-\log\Om_\Ke)$.
\item $\ttrch=(\gamma^{-1})^{AB}\chi_{AB} -(\gamma_\Ke^{-1})^{AB}\chi_\Ke$, while $(\slashed{\mbox{tr}}\widetilde{\chi})=(\gamma^{-1})^{AB}(\chi_{AB}-(\chi_\Ke)_{AB})$.
\end{itemize}

\item (Combination of the $\;{ }_\Ke$, $\;\widetilde{ }\;$ conventions with the schematic notation) We will use the same $\;{ }_\Ke$ and $\;\widetilde{ }\;$ convention for the schematic quantities $\psi$, $\psi_H$ and $\psi_{\Hb}$ (see Section \ref{schnot}). For example, $\tpH$ denotes either $\widetilde{\trch}$ or $\widetilde{\chih}$.

\item (Differences of tensorial quantities)
We use the following convention when discussing the differences of the geometric quantities on $(\mathcal U,g)$ and on Kerr spacetime. For the Ricci coefficients and the renormalised null curvature components, the $\;\widetilde{ }\;$ notation will always denote the difference of the covariant $S_{u,\ub}$-tangent tensors\footnote{Recall that the Ricci coefficients and the null curvature components are indeed defined as covariant tensors although we raise indices with respect to $\gamma$}. The same convention will be used for the difference of $\gamma$. On the other hand, for $b^A$ and $(\gamma^{-1})^{AB}$, the difference is taken for the contravariant $S_{u,\ub}$-tensors.

\item (Additional schematic notation for metric components) 
We define an additional schematic notation for the difference of the metric components. We will denote by $\tg$ one of the following four difference quantities
\begin{equation}\label{tg.possibilities}
\tg\in\{\gamma_{AB}-(\gamma_{AB})_\Ke,\, (\gamma^{-1})^{AB}-(\gamma_\Ke^{-1})^{AB},\, \frac{\Om^2-\Om_\Ke^2}{\Om^2},\, \log{\Om}-\log{\Om_\Ke}\}.
\end{equation}
We note explicitly that we do \underline{not} use the notation $\tg$ for any difference quantities associated to the metric component $b^A$. This is because we will prove slightly different estimates for them compared to that for $\tg$.

\item (Derivatives of the differences of the geometric quantities) We will also consider the derivatives of the difference quantities. Unless otherwise stated\footnote{i.e.~unless we write $(\nab_3)_\Ke$ $(\nab_4)_{\Ke}$, $\nab_{\Ke}$ explicitly.}, we differentiate these difference quantities using the connections $\nab_3$, $\nab_4$ and $\nab$ associated to the spacetime $(\mathcal U_{\ub_f},g)$ (as opposed to those associated to the background Kerr metric).

\end{itemize}

\subsection{Initial data for geometric quantities defined with respect to a double null foliation}\label{sec.smallness.data}

As we will discuss below, the geometric initial data $(\Sigma_0,\hat{g},\hat{k})$ together with the initial data for the double null coordinate system introduced in Section~\ref{setup.double null} induce the initial data for the metric components and the Ricci coefficients with respect to the double null foliation gauge. This subsection is devoted to computing and estimating these initial data.

To compute the initial data for the geometric quantities defined with respect to the double null foliation gauge, let us assume that there is a solution\footnote{It is possible to construct such a solution $(\mathcal U_{aux},g)$ using Proposition~\ref{prop:u.extend} and applying compactness arguments on a suitable exhaustion of $\Sigma_0$. However, since $(\mathcal U_{aux},g)$ will only be used to define the initial data for the geometric quantities defined with respect to the double null foliation gauge and will play no role in the rest of the paper, we will not give the details of the construction of $(\mathcal U_{aux},g)$.} $(\mathcal U_{aux},g)$ to the Einstein vacuum equations with $\mathcal U_{aux} \doteq \mathcal W_{aux}\times \mathbb S^2$, where $\mathcal W_{aux}\subset \mathbb R^2$ contains a full neighbourhood of $\{(u,\ub): \ub> 1, \, u+\ub=C_R\}$ and such that both the geometric data and the data \eqref{coord.init} are achieved on $\Sigma_0=\{(u,\ub):\ub> 1, \, u+\ub=C_R\}\times \mathbb S^2$. We can then consider the metric components and the Ricci coefficients with respect to the double null foliation gauge in $\mathcal U_{aux}$, and restrict them to $\Sigma_0$. We define this restriction to be the \emph{initial data for the geometric quantities defined with respect to the double null foliation gauge}. It can be easily checked that as long as $(\mathcal U_{aux},g)$ satisfies the above conditions, the definition of initial data for the geometric quantities defined with respect to the double null foliation gauge does not depend on the particular choice of $(\mathcal U_{aux},g)$. Moreover, as we will show below, the initial data for the geometric quantities defined with respect to the double null foliation gauge can be expressed in terms of the geometric data $(\Sigma_0,\hat{g},\hat{k})$ alone.

We will compute these initial data these initial data below. Moreover, we will show that the smallness assumptions \eqref{small.geometric.data} and \eqref{small.geometric.data.Linfty} for the geometric data translate to smallness conditions for the initial data of the geometric quantities with respect to the double null foliation gauge.

To proceed, let us compute the restriction of the geodesic vector field $L'$ and $\Lb'$ defined according to \eqref{L'.Lb'.def.0}. Let $N$ be the future-directed unit normal to $\Sigma_0$ in $\mathcal U_{aux}$. Denoting $|w|_\gamma^2 = (\gamma_\Sigma^{-1})^{AB} w_A w_B$, define
\begin{equation}\label{n.expression}
n\doteq\f{1}{\sqrt{\Phi-|w|_\gamma^2}}\left(\f{\rd}{\rd\tau}-(\gamma_\Sigma^{-1})^{CD}w_C \f{\rd}{\rd\th^D_\Sigma}\right),
\end{equation}
which is tangent to $\Sigma_0$, to be the outward-pointing\footnote{i.e.~$\tau$ increases along the integral curves of $n$.} unit normal to the constant-$\tau$ $2$-spheres. 

We claim that the vector fields $L'= -2Du$ and $\Lb'=-2D\ub$ take the following form on $\Sigma_0$:
\begin{equation}\label{L'.Lb'.def}
\begin{split}
L'\restriction_{\Sigma_0}
\doteq &\f{2}{\sqrt{\Phi-|w|_\gamma^2}}\left(N + n\right),\quad \Lb'\restriction_{\Sigma_0}\doteq \f{2}{\sqrt{\Phi-|w|_\gamma^2}}(N-n).
\end{split}
\end{equation}
To see this, first note that by \eqref{n.expression}, $n\tau = \f{1}{\sqrt{\Phi-|w|_\gamma^2}}$. Hence, using \eqref{coord.init} ,
$$\Lb'\; \ub = 0 \implies (N-n)\ub= 0 \implies  N\ub\restriction_{\Sigma_0} = n\tau = \f{1}{\sqrt{\Phi-|w|_\gamma^2}}.$$
$$L'u = 0 \implies (N+n)u= 0 \implies  Nu\restriction_{\Sigma_0} = n\tau = \f{1}{\sqrt{\Phi-|w|_\gamma^2}}.$$
Therefore,
\begin{equation}\label{Du=-f12L'}
Du\restriction_{\Sigma_0} = -(Nu)\restriction_{\Sigma_0} N + (n u) n = -\f 1{\sqrt{\Phi-|w|_\gamma^2}}(N+n) = -\f 12 L'\restriction_{\Sigma_0},
\end{equation}
\begin{equation}\label{Dub=-f12Lb'}
D \ub\restriction_{\Sigma_0} = -(N\ub)\restriction_{\Sigma_0} N + (n \ub) n = -\f 1{\sqrt{\Phi-|w|_\gamma^2}}(N-n) = -\f 12 \Lb'\restriction_{\Sigma_0}.
\end{equation}

We are now ready to compute and estimate the initial data in the double null foliation gauge. We begin with the metric components.
\begin{proposition}\label{g.initial.computation}
Given the initial metric $\hat{g}$ as in \eqref{metric.gh}, and the initial data for the double null foliation and the double null coordinates as in \eqref{coord.init} in Section~\ref{setup.double null}, the initial data for the metric components in the double null foliation gauge are given as follows:
$$ \gamma_{AB}\restriction_{\Sigma_0} = (\gamma_\Sigma)_{AB},\quad \Om\restriction_{\Sigma_0} = \f 12\sqrt{\Phi-|w|^2_\gamma},\quad b^A\restriction_{\Sigma_0} = -(\gamma^{-1})^{AB}w_B.$$
\end{proposition}
\begin{proof}
The expression for $\gamma$ is obvious since $\f{\rd}{\rd \th^A_{\Sigma}}=\f{\rd}{\rd \th^A}$. $\Om$ can be computed using $2\Om^{-2}=-g(L',\Lb')$. By \eqref{L'.Lb'.def}, we have
$$2\Om^{-2}\restriction_{\Sigma_0}=\f{8}{\Phi-|w|_\gamma^2}.$$
For $b^A$, by \eqref{L.Lb.def} and \eqref{L'.Lb'.def}, we have $b^A = L \th^A = \Om^2 L' \th^A$. Since $\Lb \th^A =0$ (by \eqref{L.Lb.def}), by \eqref{n.expression}, \eqref{L'.Lb'.def}, and the expression of $\Om \restriction_{\Sigma_0}$ derived above, we obtain
\begin{equation*}
\begin{split}
b^A\restriction_{\Sigma_0} = 2\Om\restriction_{\Sigma_0} (n \th^A) = 2\Om\restriction_{\Sigma_0} \left(\f{1}{\sqrt{\Phi-|w|_\gamma^2}}\left(\f{\rd}{\rd\tau}-(\gamma^{-1})^{CD}w_C \f{\rd}{\rd\th^D_\Sigma}\right)\right)\th^A = -(\gamma^{-1})^{AB}w_B.
\end{split}
\end{equation*}

\end{proof}

The above expressions, together with \eqref{small.geometric.data}, imply the following smallness estimates:
\begin{proposition}\label{metric.diff.initial}
For every $\de>0$, there exist $\ep_{ID,1}>0$ and $C_{ID,1}>0$  (depending on $\de$, $C_R$, $M$ and $a$) such that if the assumptions \eqref{small.geometric.data} and \eqref{small.geometric.data.Linfty} hold for some $\ep\leq \ep_{ID,1}$, then the following $L^2$ and $L^\infty$ bounds for the metric components $\gamma_{AB}$, $b^A$ and $\Om$ and their derivatives hold on $\Sigma_0$:
\begin{equation}\label{metric.diff.L2.initial}
\sum_{i\leq 5}\left\|\tau^{\f 12+\delta}\left(|\nab^i(\Om-\Om_\Ke)|_{\gamma}+|\nab^i(b-b_\Ke)|_{\gamma}+|\nab^i(\gamma-\gamma_\Ke)|_{\gamma}\right)\right\|_{L^2(\Sigma_0,\hat{g})}\leq C_{ID,1}\ep.
\end{equation}
and
\begin{equation}\label{metric.diff.Linfty.initial}
\sum_{i\leq 3}\left\|\tau^{\f 12+\delta}\left(|\nab^i(\Om-\Om_\Ke)|_{\gamma}+|\nab^i(b-b_\Ke)|_{\gamma}+|\nab^i(\gamma-\gamma_\Ke)|_{\gamma}\right)\right\|_{L^\infty(\Sigma_0,\hat{g})}\leq C_{ID,1}\ep.
\end{equation}
Here, norms are taken with respect to $\gamma$, e.g., $|b-b_\Ke|_{\gamma}\doteq (\gamma_{AB}(b^A-b_\Ke^A)(b^B-b_\Ke^B))^{\f 12}$ and $|\gamma-\gamma_\Ke|_{\gamma}\doteq ((\gamma^{-1})^{AA'}(\gamma^{-1})^{BB'}(\gamma_{AB}-(\gamma_\Ke)_{AB})(\gamma_{A'B'}-(\gamma_\Ke)_{A'B'}))^{\f12}$ and $\nab$ is the Levi--Civita connection of the induced metric $\gamma_\Sigma$ ($= \gamma$) on the $\tau=\mbox{constant}$ spheres.
\end{proposition}
\begin{proof}
This is almost an immediate consequence of \eqref{small.geometric.data}, \eqref{small.geometric.data.Linfty} and Proposition~\ref{g.initial.computation}. We only note that in \eqref{small.geometric.data} and \eqref{small.geometric.data.Linfty}, the estimates are expressed in terms of covariant derivatives in terms of the Levi--Civita connection $\hat{\nabla}$ with respect to $\hat{g}$, while according to \eqref{metric.diff.L2.initial} and \eqref{metric.diff.Linfty.initial}, we want to prove estimates so that the derivatives are covariant derivatives in terms of the Levi--Civita connection $\nab$ with respect to $\gamma$.

To handle this, first, we note that by Remark~\ref{rmk:nablatoLie.small}, the smallness assumptions \eqref{small.geometric.data} and \eqref{small.geometric.data.Linfty} imply smallness estimates with the covariant derivatives replaced by Lie derivatives. Second, by \eqref{nab.def} and \eqref{def.slashed.Gamma}, we can rewrite the left hand sides of \eqref{metric.diff.L2.initial} and \eqref{metric.diff.Linfty.initial} in terms of (potentially nonlinear) expressions that depend only on $\hat{g}$ and its Lie derivatives. Finally, in view of the fact that we have both $L^2$ and $L^\infty$ estimates in \eqref{small.geometric.data} and \eqref{small.geometric.data.Linfty}, we can control the nonlinear expressions to obtain the desired estimates.
\end{proof}

In the following proposition, we compute the initial data for the Ricci coefficients on $\Sigma_0$ in terms of $\hat{g}$ and $\hat{k}$:
\begin{proposition}\label{initial.RC.bd}
Given the geometric initial data $(\Sigma_0,\hat{g},\hat{k})$ as in Theorem~\ref{main.thm.spacelike.formulation}, and the initial data for the double null foliation and the double null coordinates as in \eqref{coord.init} in Section~\ref{setup.double null}, the initial data for the Ricci coefficients in the double null foliation gauge are given as follows:
\begin{align*}
\chib_{AB} \restriction_{\Sigma_0} = \Om \hat{k}(\f{\rd}{\rd\th^A}, \f{\rd}{\rd\th^B}) - \Om \hat{g}(\hat{\nabla}_{\f{\rd}{\rd\th^A}} n, \f{\rd}{\rd\th^B}),& \quad \chi_{AB} \restriction_{\Sigma_0} = \Om^{-1} \hat{k}(\f{\rd}{\rd\th^A}, \f{\rd}{\rd\th^B}) + \Om^{-1} \hat{g}(\hat{\nabla}_{\f{\rd}{\rd\th^A}} n, \f{\rd}{\rd\th^B}),\\
\zeta_A \restriction_{\Sigma_0}= \eta_A\restriction_{\Sigma_0} = \f{\rd}{\rd\th^A} \log\Om - k(\f{\rd}{\rd\th^A}, n),& \quad \etab_A \restriction_{\Sigma_0} = \f{\rd}{\rd\th^A} \log\Om + k(\f{\rd}{\rd\th^A}, n),\\
\omb \restriction_{\Sigma_0} = -\Om k(n, n) + \Om (n \log \Om),& \quad e_4 \log \Om \restriction_{\Sigma_0} = \Om^{-1} k(n,n) + \Om^{-1} (n \log \Om).
\end{align*}
\end{proposition}
\begin{proof}
We first note that by \eqref{L.Lb.def} and \eqref{L'.Lb'.def},
\begin{equation}\label{e3.e4.data}
e_4\restriction_{\Sigma_0} \doteq \Om^{-1}\left(N + n\right),\quad e_3 \restriction_{\Sigma_0}\doteq \Om (N-n).
\end{equation}

The identities for $\chi_{AB}$ and $\chib_{AB}$ are then immediate from the definitions. For $\zeta_A$, we compute using \eqref{e3.e4.data} and Proposition~\ref{g.initial.computation} that
\begin{equation*}
\begin{split}
\zeta_A =& \f 12 g(D_{\f{\rd}{\rd\th^A}} e_4, e_3) = \f 12 g(D_{\f{\rd}{\rd\th^A}} (\Om^{-1} (N+n )), \Om (N-n ))\\
% =& \f{\rd}{\rd\th^A} \log\Om + \f 12 g(D_{\f{\rd}{\rd\th^A}} (N+n ),  (N-n ))\\
=& \f{\rd}{\rd\th^A} \log\Om - \f 12 g(D_{\f{\rd}{\rd\th^A}} N, n )) + \f 12 g(D_{\f{\rd}{\rd\th^A}} n ,  N)\\
% =& \f{\rd}{\rd\th^A} \log\Om - g(D_{\f{\rd}{\rd\th^A}} N, n ))\\
=& \f{\rd}{\rd\th^A} \log\Om - k(\f{\rd}{\rd\th^A}, n ).
\end{split}
\end{equation*}
The formulae for $\eta_A$ and $\etab_A$ then follow from \eqref{gauge.con}.

For $\omb$ and $e_4\log\Om$, we first compute using \eqref{L'.Lb'.geodesic}, \eqref{n.expression}, \eqref{L'.Lb'.def} and Proposition~\ref{g.initial.computation} that
\begin{equation*}
\begin{split}
k(n ,n ) = & g(D_{n } N, n ) = \f 18 \Om^2 g(D_{L'-\Lb'} (\Om (L'+\Lb')), L'-\Lb')\\
= & \f 18 \Om^3 g(D_{L'}\Lb' - D_{\Lb'}L', L'-\Lb') 
= \f 12  \Om^{-1} \left(\left(\f{\rd}{\rd u} + \f{\rd}{\rd\ub} + b^A\f{\rd}{\rd\th^A}\right)\log\Om\right)
= N\log\Om,
\end{split}
\end{equation*}
where we have used
\begin{equation*}
\begin{split}
D_{L'}\Lb' - D_{\Lb'}L' = & [L', \Lb'] = \left[ \Om^{-2}\left(\f{\rd}{\rd\ub} + b^A\f{\rd}{\rd\th^A}\right), \Om^{-2} \f{\rd}{\rd u}\right]\\
 = & \left(\left(\f{\rd}{\rd\ub} + b^A\f{\rd}{\rd\th^A}\right)\Om^{-2}\right) \Lb'-\f{\rd \Om^{-2}}{\rd u} L'- \Om^{-4} \f{\rd b^A}{\rd u} \f{\rd}{\rd\th^A}.
\end{split}
\end{equation*}
As a consequence, by \eqref{gauge.con} and \eqref{e3.e4.data},
$$\omb = -e_3 \log \Om = -\Om (N\log \Om - n  \log \Om) = -\Om k(n ,n ) + \Om (n  \log \Om)$$
and
$$e_4 \log \Om = \Om^{-1} (N\log \Om + n  \log \Om) = \Om^{-1} k(n ,n ) + \Om^{-1} (n  \log \Om).$$
\end{proof}

Finally, we estimate the initial data for the Ricci coefficients:
\begin{proposition}\label{Ricci.der.diff.initial}
For every $\de>0$, there exist $\ep_{ID,2}>0$ and $C_{ID,2}>0$ (depending on $\de$, $C_R$, $M$ and $a$) such that if the assumptions \eqref{small.geometric.data} and \eqref{small.geometric.data.Linfty} hold for some $\ep\leq \ep_{ID,2}$, then the following $L^2$ bounds for the Ricci coefficients and their derivatives hold on $\Sigma_0$
\begin{equation}\label{Ricci.data.1}
\begin{split}
&\sum_{i\leq 4}\left(\|\tau^{\f 12+\delta}(|\nab^i(\chi-\chi_\Ke)|_{\gamma}+|\nab^i(\chib-\chib_\Ke)|_{\gamma})\|_{L^2(\Sigma_0,\hat{g})}\right.\\
&\qquad \left.+\|\tau^{\f 12+\delta}(|\nab^i(\eta-\eta_\Ke)|_{\gamma}+|\nab^i\nab_3^j\nab_4^k(\etab-\etab_\Ke)|_{\gamma})\|_{L^2(\Sigma_0,\hat{g})}\right.\\
&\left.+\|\tau^{\f 12+\delta}(|\nab^i(\zeta-\zeta_\Ke)|_{\gamma}+|\nab^i\nab_4(\nab\log\Om-\nab\log\Om_\Ke)|_{\gamma}+|\nab^i(\omb-\omb_\Ke)|_{\gamma})\|_{L^2(\Sigma_0,\hat{g})}\right)\leq C_{ID,2}\ep.
\end{split}
\end{equation}
and the following pointwise bounds hold on $\Sigma_0$:
\begin{equation}\label{Ricci.data.2}
\begin{split}
&\sum_{i\leq 2}\left(\|\tau^{\f 12+\delta}(|\nab^i(\chi-\chi_\Ke)|_{\gamma}+|\nab^i(\chib-\chib_\Ke)|_{\gamma})\|_{L^\infty(\Sigma_0,\hat{g})}\right.\\
&\qquad \left.+\|\tau^{\f 12+\delta}(|\nab^i(\eta-\eta_\Ke)|_{\gamma}+|\nab^i(\etab-\etab_\Ke)|_{\gamma})\|_{L^\infty(\Sigma_0,\hat{g})}\right.\\
&\left.+\|\tau^{\f 12+\delta}(|\nab^i(\zeta-\zeta_\Ke)|_{\gamma}+|\nab^i\nab_4(\nab\log\Om-\nab\log\Om_\Ke)|_{\gamma}+|\nab^i(\omb-\omb_\Ke)|_{\gamma})\|_{L^\infty(\Sigma_0,\hat{g})}\right)\leq C_{ID,2}\ep.
\end{split}
\end{equation}
\end{proposition}
\begin{proof}
Just as in the proof of Proposition~\ref{metric.diff.initial}, these estimates are almost immediate consequences of \eqref{small.geometric.data}, \eqref{small.geometric.data.Linfty} and Proposition~\ref{initial.RC.bd}. Again, the only point to note is that we need to control the Christoffel symbols associated to $\nab$.

Nevertheless, by \eqref{nab.def} and Proposition~\ref{initial.RC.bd}, we can write the left hand sides of \eqref{Ricci.data.1}, \eqref{Ricci.data.2} as nonlinear expressions in terms of Lie derivatives of $\hat{g}-\hat{g}_\Ke$ and $\hat{k}-\hat{k}_\Ke$. According to Remark~\ref{rmk:nablatoLie.small}, we have $L^2$ and $L^\infty$ bounds for these Lie derivatives of $\hat{g}-\hat{g}_\Ke$ and $\hat{k}-\hat{k}_\Ke$, and we can therefore conclude that these nonlinear expressions are small as desired.
\end{proof}

\subsection{Definition of initial data energy with respect to a double null foliation}\label{sec.data}

%From this section onwards, we specialize Theorem \ref{main.thm.spacelike.formulation} to the case where $\Sigma_\Ke$ is given in the $(u,\ub,\th_*,\phi_*)$ coordinate system by $\Sigma_\Ke=\{(u,\ub,\th_*,\phi_*): u+\ub=C_R\}$, where $C_R\in \mathbb R$ is arbitrary. As we mentioned in the beginning of Section \ref{sec.main.thm.dnf}, we make this restriction only for notational purposes and the general case of Theorem \ref{main.thm.spacelike.formulation} can be proven in a completely identical manner\footnote{One note that for this we need both \eqref{spacelike.con} and \eqref{spacelike.con.2} to hold. If \eqref{spacelike.con} is satisfied while \eqref{spacelike.con.2} \underline{fails}, then one needs an additional argument to show the existence of the solution, which will follow from ideas in \cite{DafLuk2}.}. In particular, our statement for the main theorem in terms of the double null foliation gauge (Theorem \ref{aprioriestimates}) will be stated for the class of initial data on $\Sigma_0$ which approaches the initial data on this restricted class of hypersurfaces in the Kerr spacetime.
We now define the \emph{initial data energy} $\mathcal D$, which is defined for the difference of the initial data sets $(\Sigma_0,\hat{g},\hat{k})$ and $(\Sigma_\Ke,\hat{g}_\Ke,\hat{k}_\Ke)$ in terms of the geometric quantities with respect to the double null foliation gauge. For this purpose, we will define the \emph{data in the double null gauge} by the expressions on the right hand sides of the equations in Propositions~\ref{g.initial.computation} and \ref{initial.RC.bd}. We note that this is well-defined without referring to the auxiliary spacetime $(\mathcal U_{aux},g)$.

In order that one can easily compare the quantities in the initial data energy with the quantities that we use to prove our main theorem, we define $\mathcal D$ using the schematic notation introduced in Section~\ref{schnot} and the notation for the difference quantities introduced in Section~\ref{sec.identification}, even though the data quantities are defined on $\Sigma_0$ instead of $\mathcal U_{\ub_f}$. Moreover, we will use the weights $u$ and $\ub$ (defined via \eqref{coord.init}) instead of $\tau$.
Before we proceed, let us introduce some notations
\begin{equation}\label{S.def}
\mathcal S_{\tg} \doteq \{\widetilde{\gamma},\,\widetilde{\gamma^{-1}},\,\widetilde{\log\Om},\,\Om^{-2}\widetilde{\Om^2}\}, \quad \mathcal S_{\tp} \doteq \{\widetilde{\eta},\,\widetilde{\etab}\},\quad\mathcal S_{\tpHb} \doteq \{\widetilde{\slashed{tr}\chib},\,\widetilde{\chibh}\}, \quad \mathcal S_{\tpH}\doteq \{\widetilde{\slashed{tr}\chi},\,\widetilde{\chih}\}.
\end{equation}
We define the data energy $\mathcal D$ as follows:
\begin{equation}\label{data.D.def}
\begin{split}
\mathcal D\doteq &\sum_{i\leq 3}\sum_{\substack{\tg \in \mathcal S_{\tg},\, \tp \in \mathcal S_{\tp},\\ \tpH\in \mathcal S_{\tpH},\,\tpHb\in\mathcal S_{\tpHb}}}\|\ub^{\frac 12+\de}(\nab^i(\tpHb,\widetilde{\omb},\tg,\tb,\tp,\tpH,\nab_4\widetilde{\log\Om}),\nab^{\min\{i,2\}}\tK)\|_{L^2_uL^2(S_{u,-u+C_R})}^2\\
&+\sum_{i\leq 3}\sum_{\substack{\tg \in \mathcal S_{\tg},\, \tp \in \mathcal S_{\tp},\\ \tpH\in \mathcal S_{\tpH},\,\tpHb\in\mathcal S_{\tpHb}}}\|\ub^{\frac 12+\de}(\nab^i(\tpHb,\widetilde{\omb},\tg,\tb,\tp,\tpH,\nab_4\widetilde{\log\Om}),\nab^{\min\{i,2\}}\tK)\|_{L^\infty_uL^2(S_{u,-u+C_R})}^2.
\end{split}
\end{equation}
Here (and below), we use the notation that $L^2_uL^2(S_{u,-u+C_R})$ denotes the following norm on $\Sigma_0$:
$$\|\phi\|_{L^2_u  L^2(S_{u,-u+C_R})}\doteq \left(\int_{-\infty}^{u_f}\left(\int_{S_{u,-u+C_R}}|\phi|_{\gamma}^2 \,dVol_{\gamma}\right)du\right)^{\frac 12},$$
where $dVol_{\gamma}$ is the volume form on the $2$-spheres $S_{u,-u+C+R}$ induced by the metric $\gamma$ and $du$ is the Lebesgue measure. Similarly, $L^\infty_uL^2(S_{u,-u+C_R})$ denotes
$$\|\phi\|_{L^\infty_u  L^2(S_{u,-u+C_R})}\doteq \sup_{u\in (-\infty, u_f]}\left(\int_{S_{u,-u+C_R}}|\phi|_{\gamma}^2 \, dVol_{\gamma}\right)^{\frac 12}.$$
Furthermore, the weight $\ub$ in the initial data energy $\mathcal D$ is to be understood as a function of $u$, which is defined via \eqref{coord.init}.

Propositions~\ref{metric.diff.initial} and \ref{Ricci.der.diff.initial} in Section~\ref{sec.smallness.data} show that given the assumptions on the geometric data in Theorem~\ref{main.thm.spacelike.formulation}, the $\mathcal D$ energy is small. We summarise this in the following proposition:
\begin{proposition}\label{prop.small.data.implies.small.data}
Let $(\Sigma_0,\hat{g},\hat{k})$ be an initial data set as in Theorem~\ref{main.thm.spacelike.formulation} which satisfies
\begin{equation*}
\sum_{i=0}^{5}\|\tau^{\f 12+\delta}\hat{\nabla}_\Ke^i(\hat{g}-\hat{g}_\Ke)\|_{L^2(\Sigma_0,\hat{g}_\Ke)}+\sum_{i=0}^{4}\|\tau^{\f 12+\delta}\hat{\nabla}_\Ke^i(\hat{k}-\hat{k}_\Ke)\|_{L^2(\Sigma_0,\hat{g}_\Ke)}\leq \ep  \tag{\ref{small.geometric.data}}
\end{equation*}
and
\begin{equation*}
\sum_{i=0}^{3}\|\tau^{\f 12+\delta}\hat{\nabla}_\Ke^i(\hat{g}-\hat{g}_\Ke)\|_{L^\infty(\Sigma_0,\hat{g}_\Ke)}+\sum_{i=0}^{2}\|\tau^{\f 12+\delta}\hat{\nabla}_\Ke^i(\hat{k}-\hat{k}_\Ke)\|_{L^\infty(\Sigma_0,\hat{g}_\Ke)}\leq \ep. \tag{\ref{small.geometric.data.Linfty}}
\end{equation*}
Then for every $\de>0$, there exists $\ep_{ID}>0$ sufficiently small and $C_{ID}>0$ (both depending on $\de$, $C_R$ and the background Kerr parameters $M$ and $a$) such that whenever $\ep<\ep_{ID}$, it holds that
$$\mathcal D\leq C_{ID}\ep^2.$$
\end{proposition}

In view of Proposition \ref{prop.small.data.implies.small.data}, in the rest of the paper, we will assume that $\mathcal D$ is small. In fact we will slightly abuse notation and assume
\begin{equation*}
\mathcal D\leq \ep^2,
\end{equation*}
where $\ep>0$ is small (see Theorem \ref{aprioriestimates}).

\subsection{Energies for controlling the solution}\label{sec.norms}

In this subsection, we define the spacetime energies that we use to control the difference of the geometric quantities of $(\mathcal U_{\ub_f}, g)$ and the background Kerr solution. 

We begin with some preliminaries in Sections~\ref{sec:region.int} and \ref{sec:convention.int}. After that we define the various Lebesgue norms (Section~\ref{sec:norm.intro}) and weights (Section~\ref{sec:weight.intro}) that we will use to define our energies. Finally, we define the various energies in Section~\ref{sec:def.energies}.

\subsubsection{Region of integration}\label{sec:region.int}

We will always be working in a region (see Figure~\ref{sec4newlabel}) of the form 
$$\mathcal U_{\ub_f}\doteq \mathcal W_{\ub_f}\times \mathbb S^2 \doteq \{(u,\ub): -u_f+C_R< -u+C_R\leq \ub< \ub_f\}\times \mathbb S^2.$$  
Here, 
\begin{itemize}
\item $u_f$ is a parameter quantifying the closeness of the region to timelike infinity. It will be chosen as part of the proof of Theorem~\ref{aprioriestimates} (cf.~Theorem~\ref{main.thm.spacelike.formulation}). In the proof, one can therefore view $|u_f|^{-1}$ as a small parameter.
\item $\ub_f\in (-u_f+C_R,\infty)$ is arbitrary. We will show that for \emph{any} choice of $\ub_f\in (-u_f+C_R,\infty)$, under appropriate bootstrap assumption, one can control the geometry of the region $\mathcal U_{\ub_f}$. In the context of the bootstrap argument (see Sections~\ref{introbootstr} and \ref{sec:continuity}), this allows us to conclude that the solution exists and remains regular in $\cup_{\ub\in (-u_f+C_R,\infty)} \mathcal U_{\ub}$. Therefore, it is important that (unless explicitly stated otherwise) \textbf{\emph{any estimates that we prove are \underline{independent} of $\ub_f$}}.
\end{itemize}
Without loss of generality (by choosing $u_f$ more negative depending on $C_R$ otherwise), we will assume from now on that $u\leq -1$ and $\ub\geq 1$ in $\mathcal U_f$.

\subsubsection{Conventions for integration}\label{sec:convention.int}

When there is no danger of confusion, when integrating over the $2$-spheres $S_{u,\ub}$, we will suppress in our notation the volume form $dVol_\gamma$, i.e.~we will write $\int_{S_{u,\ub}} = \int_{S_{u,\ub}}\, dVol_\gamma$.

\subsubsection{Definitions of the norms}\label{sec:norm.intro}

\begin{definition}[Spacetime norms]
We will use $L^p_{u} L^q_{\ub} L^r(S)$ or $L^q_{\ub} L^p_{u}L^r(S)$ norms to control the geometric quantities in $(\mathcal U_{\ub_f},g)$. We will use the notation that the norms on the right are taken first. Here, $L^r(S)$ is defined to be the $L^r$ norm on the $2$-sphere $S_{u,\ub}$ taken with respect to the volume form associated to the induced metric $\gamma$. On the other hand $L^p_u$ and $L^q_{\ub}$ are taken with respect to the measures $du$ and $d\ub$ respectively. Moreover, we implicitly assume that all spacetime norms are taken in the region $\mathcal U_{\ub_f}$. More precisely, for $1\leq p,q,r<\infty$ and for an arbitrary tensor $\phi$, we define
$$\|\phi\|_{L^p_u L^q_{\ub} L^r(S)}\doteq \left(\int_{-\ub_f+C_R}^{u_f}\left(\int_{-u+C_R}^{\ub_f}\left(\int_{S_{u,\ub}}|\phi|_{\gamma}^r \, dVol_{\gamma}\right)^{\frac qr}d\ub\right)^{\frac pq}du\right)^{\frac 1p}$$
and
$$\|\phi\|_{L^q_{\ub}L^p_u  L^r(S)}\doteq \left(\int_{C_R-u_f}^{\ub_f}\left(\int_{u_f}^{-\ub+C_R}\left(\int_{S_{u,\ub}}|\phi|_{\gamma}^r \, dVol_{\gamma}\right)^{\frac pr}du\right)^{\frac qp}d\ub\right)^{\frac 1q}.$$

We will also allow $p$, $q$ or $r$ to be $\infty$, where the integral norms are replaced by the esssup norms in the obvious manner.
\end{definition}

\begin{remark}[Changing the order in the $L^p_uL^q_{\ub}$ norm]\label{rmk:norm.order}
In this paper, we will frequently use the following standard facts:
$$\|\phi\|_{L^p_uL^q_{\ub} L^r(S)}\ls \|\phi\|_{L^q_{\ub}L^p_u L^r(S)},\quad\mbox{for }p\geq q$$
and 
$$\|\phi\|_{L^q_{\ub}L^p_u L^r(S)}\ls \|\phi\|_{L^p_uL^q_{\ub} L^r(S)},\quad\mbox{for }p\leq q.$$
In particular, notice that in the $\NH$ energies below, we use all of the following norms: $L^2_uL^\infty_{\ub}L^2(S)$, $L^\infty_{\ub}L^2_uL^2(S)$, $L^2_{\ub}L^\infty_uL^2(S)$ and $L^\infty_uL^2_{\ub}L^2(S)$.
\end{remark}

In addition to the above norms, we will also use norms defined on lower dimensional subsets of $\mathcal U_{\ub_f}$. More precisely, 
\begin{definition}[Norms on constant $u$ or constant $\ub$ null hypersurfaces]
For $1\leq q,r<\infty$, we define
$$\|\phi\|_{L^q_{\ub} L^r(S)}\doteq \left(\int_{-u+C_R}^{\ub_f}\left(\int_{S_{u,\ub}}|\phi|_{\gamma}^r dVol_{\gamma}\right)^{\frac qr}d\ub\right)^{\frac 1q},$$
which is to be understood as a function of $u$. For $q=\infty$ or $r=\infty$, this is defined by replacing the integral norms with the esssup norm in the obvious manner. In order to emphasise the dependence on $u$ explicitly, we will often write $\|\phi\|_{L^q_{\ub}L^r(S_{u,\ub})}=\|\phi\|_{L^q_{\ub}L^r(S)}$. We also define $\|\phi\|_{L^q_u L^r(S)}$ as a function of $\ub$ in a completely analogous manner.
\end{definition}

\begin{definition}[Norms on the $2$-spheres $S_{u,\ub}$]
For $1\leq r<\infty$, we define 
$$\|\phi\|_{L^r(S_{u,\ub})}\doteq \left(\int_{S_{u,\ub}}|\phi|_{\gamma}^r dVol_{\gamma}\right)^{\frac 1r},$$
which is to be understood as a function of $u$ and $\ub$. For $r=\infty$, we define 
$$\|\phi\|_{L^\infty(S_{u,\ub})}\doteq \esssup_{p\in S_{u,\ub}} |\phi|(p) ,$$
which is again to be understood as a function of $u$ and $\ub$. 

\end{definition}

In the remainder of this paper, we will frequently not mention the dependence of $\ub_f$ in the definition of the norms. We emphasise that all estimates that we derive will be independent of $\ub_f$.

\subsubsection{Definition of the weight functions}\label{sec:weight.intro}

We incorporate in our energies the weight functions $|u|^{\f12+\de}$, $\ub^{\f12+\de}$, $\varpi^N$ and $\Om_\Ke$, which we now describe in further details. The quantity $\varpi$ is defined as
\begin{equation}\label{varpi.def}
\varpi= 1+e^{-\f{r_+-r_-}{r_-^2+a^2}(u+\ub-C_R)}.
\end{equation}
Notice that in the region $u+\ub\geq C_R$, $\varpi$ obeys the estimate
\begin{equation}\label{varpi.bd}
1\leq \varpi\leq 2.
\end{equation}
Moreover, by Proposition \ref{Om.Kerr.bounds},
\begin{equation}\label{varpi.34}
\f{\rd}{\rd u}\varpi=\f{\rd}{\rd \ub}\varpi\ls -\Om_\Ke^2
\end{equation}
and
\begin{equation}\label{varpi.ang}
\nab \varpi=0.
\end{equation}
In \eqref{varpi.34}, we recall that $\Om_\Ke$ is defined using the identification of $(\mathcal U_{\ub_f},g)$ with the background Kerr spacetime in Section \ref{sec.identification} to be the value of the metric component $\Om$ in the background Kerr spacetime. 

Finally, $\de$ and $N$ appearing in the weights are positive constants. $\de>0$ is an arbitrary constant. $N>0$ will be a large constant (depending on $M$, $a$, $\de$ and $C_R$) to be chosen later\footnote{$N$ will moreover satisfy $\ep\leq N^{-1}$. We refer the readers to further discussions regarding the constants in Remark \ref{rmk.constant}.}.

\subsubsection{Definition of the energies for the differences of the geometric quantities in $(\protect\protect\mathcal U_{\protect\ub_f},g)$}\label{sec:def.energies}

Recalling the notations introduced in \eqref{S.def}, we now define the integrated energy $\NI$, the hypersurface energy $\NH$ and the sphere energy $\NS$. Namely\footnote{Here, we adopt the notation that $\nab^{-1}=0$},
\begin{equation}\label{NI.def}
\begin{split}
\NI\doteq &\sum_{i\leq 3}\left(\sum_{\tpHb\in \mathcal S_{\tpHb}}\||u|^{\frac 12+\de}\varpi^N\Om_\Ke\nab^i(\tpHb,\widetilde{\omb})\|_{L^2_uL^2_{\ub}L^2(S)}^2\right.\\
&\left.\qquad+\sum_{\substack{\tg \in \mathcal S_{\tg} \\ \tp \in \mathcal S_{\tp}}}\|\ub^{\frac 12+\de}\varpi^N\Om_\Ke(\nab^i(\tp,\tg,\tb),\nab^{\min\{i,2\}}\tK)\|_{L^2_uL^2_{\ub}L^2(S)}^2\right.\\
&\left.\qquad+\sum_{\tpH\in \mathcal S_{\tpH}}\|\ub^{\frac 12+\de}\varpi^N\Om_\Ke^3\nab^i\tpH\|_{L^2_uL^2_{\ub}L^2(S)}^2\right)
\end{split}
\end{equation}
and
\begin{equation}\label{NH.def}
\begin{split}
\NH\doteq &\sum_{i\leq 2}\left(\sum_{\tpHb\in \mathcal S_{\tpHb}}\||u|^{\frac 12+\de}\varpi^N\nab^i(\tpHb,\widetilde{\omb})\|_{L^2_uL^\infty_{\ub}L^2(S)}^2+\|\ub^{\frac 12+\de}\varpi^N\Om_\Ke\nab^i\widetilde{\eta}\|_{L^2_uL^\infty_{\ub}L^2(S)}^2\right.\\
&\left.+\||u|^{\frac 12+\de}\varpi^N\Om_\Ke(\nab^i\widetilde{\etab},\nab^{\min\{i,1\}}\tK)\|_{L^\infty_{\ub}L^2_uL^2(S)}^2+\|\ub^{\frac 12+\de}\varpi^N\Om_\Ke(\nab^i\widetilde{\eta},\nab^i\widetilde{\etab},\nab^{\min\{i,1\}}\tK)\|_{L^2_{\ub}L^{\infty}_uL^2(S)}^2\right.\\
&\left.+\sum_{\tpH\in \mathcal S_{\tpH}}\|\ub^{\frac 12+\de}\varpi^N\Om_\Ke^2\nab^i\tpH\|_{L^2_{\ub}L^\infty_uL^2(S)}^2\right)\\
&+\sum_{\tpHb\in \mathcal S_{\tpHb}}\||u|^{\frac 12+\de}\varpi^N\nab^3(\tpHb,\widetilde{\omb}))\|_{L^\infty_{\ub}L^2_uL^2(S)}^2+\|\ub^{\frac 12+\de}\varpi^N\Om_\Ke(\nab^3\widetilde{\eta},\nab^2 \tK)\|_{L^\infty_{\ub}L^2_uL^2(S)}^2\\
&+\||u|^{\frac 12+\de}\varpi^N\Om_\Ke(\nab^3\widetilde{\etab},\nab^2 \tK)\|_{L^{\infty}_uL^2_{\ub}L^2(S)}^2+\sum_{\tpH\in \mathcal S_{\tpH}}\|\ub^{\frac 12+\de}\varpi^N\Om_\Ke^2\nab^3\tpH\|_{L^\infty_uL^2_{\ub}L^2(S)}^2\\
&+\sum_{i\leq 3}\sum_{\tg \in \mathcal S_{\tg} }\left(\|\ub^{\frac 12+\de}\varpi^N\Om_\Ke\nab^i\tg\|_{L^2_{\ub}L^{\infty}_uL^2(S)}^2+\||u|^{\frac 12+\de}\varpi^N\Om_\Ke\nab^i\tg\|_{L^{\infty}_{\ub}L^2_uL^2(S)}^2\right)\\
&+\sum_{i\leq 3}\|\ub^{\frac 12+\de}\varpi^N\nab^i\tb\|_{L^2_{\ub}L^{\infty}_uL^2(S)}^2
\end{split}
\end{equation}
and
\begin{equation}\label{NS.def}
\begin{split}
\NS\doteq &\sum_{i\leq 3}\sum_{\tg \in \mathcal S_{\tg} }\|\nab^i\tg\|_{L^\infty_{\ub}L^\infty_uL^2(S)}^2+\sum_{i\leq 2}\sum_{\substack{ \tpHb\in \mathcal S_{\tpHb} \\ \tp \in \mathcal S_{\tp}}}\|\nab^i(\tpHb,\widetilde{\omb},\tp,\tb)\|_{L^\infty_{\ub}L^\infty_uL^2(S)}^2\\
&+\sum_{i\leq 2}\sum_{\tpH\in \mathcal S_{\tpH}}\|\Om_\Ke^2\nab^i\tpH\|_{L^\infty_{\ub}L^\infty_uL^2(S)}^2+\sum_{i\leq 1}\|\nab^{i}\tK\|_{L^\infty_{\ub}L^\infty_uL^2(S)}^2\\
&+\sum_{i\leq 2}\sum_{\tpHb\in \mathcal S_{\tpHb}}\|\nab^i\tpHb\|_{L^1_u L^\infty_{\ub}L^2(S)}^2+\sum_{i\leq 2}\sum_{\tpH\in \mathcal S_{\tpH}}\|\Om_\Ke^2\nab^i\tpH\|_{L^1_{\ub} L^\infty_uL^2(S)}^2.
\end{split}
\end{equation}
We end this subsection with some remarks on the energies defined above.

\begin{remark}[Remarks on regularity]
We now make some remarks about the number of derivatives used in the energies above.
\begin{itemize}
\item Our energies control the differences of all the Ricci coefficients up to $3$ angular derivatives. 
\item Except for the difference $\tK$ of the Gauss curvature of the $2$-spheres of the double null foliation, the curvature components do not explicitly appear in the definition of the energies. This may at the first sight seem different from the energies used in \cite{Chr, CK}. Nevertheless, in the course of the proof, we will in fact also control the differences of the renormalised curvature components $\widetilde{\beta}, \tK, \widetilde{\sigmac}, \widetilde{\betab}$ in appropriate norms up to $2$ angular derivatives.
\item Notice however that in the energies above, for the differences of the metric components $\tg$ and $\tb$, we in general only control up to $3$ angular derivatives. In other words, it is only some special combination of derivatives of the (differences of the) metric components, namely those which can be written as $3$ angular derivatives of Ricci coefficients, that we control up to $4$ derivatives.
\end{itemize}

\end{remark}

\begin{remark}[Naming of the energies]
We name our energies as the integrated energy $\NI$, hypersurface energy $\NH$ and sphere energy $\NS$ because 
\begin{itemize}
\item for $\NI$, all the geometric quantities are integrated over all spacetime in the $L^2_uL^2_{\ub}L^2(S)$ norm;
\item for $\NH$, many of the geometric quantities are integrated over a constant-$u$ or constant-$\ub$ null hypersurface (with supremum taken over all such hypersurfaces), i.e.~they are controlled in the $L^\i_u L^2_{\ub} L^2(S)$ or $L^\i_{\ub} L^2_u L^2(S)$ norms;
\item for $\NS$, many of the geometric quantities are integrated on a $2$-sphere $S_{u,\ub}$ adapted to the double null foliation (with supremum taken over all such spheres), i.e.~they are controlled in the $L^\i_u L^\i_{\ub} L^2(S)$ norm.
\end{itemize}
Notice however the following:
\begin{itemize}
\item In the $\NH$ energy, it is convenient to have a slightly stronger norm by changing the order of the $L^2$ and the $L^\infty$ norms for the lower order quantities (cf.~Remark~\ref{rmk:norm.int.order} below). 
\item In the $\NS$ energy, in addition to controlling the $L^\i_u L^\i_{\ub} L^2(S)$ norm of the geometric quantities, we also put in the $L^1_{\ub}L^\i_uL^2(S)$ control for $\nab^i\tpH$ (with $i\leq 2$) and the $L^1_u L^\i_{\ub}L^2(S)$ control for $\nab^i(\tpHb,\widetilde{\omb})$ (with $i\leq 2$). This is stronger than just using the $L^\i_u L^\i_{\ub} L^2(S)$ and directly integrating in the $u$ or $\ub$ direction, and will be crucial in closing our estimates.
\end{itemize}
\end{remark}

\begin{remark}[Order of integration in the energies]\label{rmk:norm.int.order}
Let us note that in $\NH$, while there is always one $L^2$ norm and one $L^\infty$ norm in the $u, \ub$ variable, for the lower derivatives of most of the terms (i.e.~up to two angular derivatives of the Ricci coefficients), the $L^\infty$ norm is taken \emph{before} the $L^2$ norm, while for the higher derivatives (i.e.~third derivatives of the Ricci coefficients), the $L^\infty$ norm is taken \emph{after} the $L^2$ norm. Recalling Remark~\ref{rmk:norm.order}, this means that we control a stronger norm when there are fewer derivatives.\footnote{We note also that for technical reasons, $\tg$, $\widetilde{\etab}$ and $\tK$ obey only $L^{\i}_{\ub}L^2_u$ estimates even at lower order (although they do obey $L^2_{\ub}L^{\i}_u$ estimates).} \footnote{On the other hand, although we do not explicitly indicate this in the definition of the energies, note that some of the highest order quantities, namely $\nab^3\trch$, $\nab^3\trchb$, $\nab^3\mu$, $\nab^2\mub$ and $\nab\ombs$, in fact obey stronger estimates for which the $L^\i$ norm is taken first; see Section~\ref{sec.elliptic}.}
\end{remark}

\begin{remark}[Remarks on the $\Om_\Ke$ weights]
Let us remark also the following:
\begin{itemize}
\item Recall from Proposition~\ref{Om.Kerr.bounds} that $\Om_\Ke$ vanishes on the Cauchy horizon and is moreover comparable to $e^{-\f{r_+-r_-}{r_-^2+a^2}(u+\ub)}$. Thus powers of $\Om_\Ke$ in the energies should be viewed as degenerate weights.
\item Notice that whenever $\tpH$ appears, whether it is in the $\NH$, $\NI$ or $\NS$ energies, it always comes with at least a weight of $\Om_\Ke^2$. This degeneration is associated to the expectation that $\tpH$ is potentially singular for solutions arising from a generic subset of initial data that we consider (cf.~Conjecture~\ref{ChrSCC} and related discussions in the introduction and in Section~\ref{sec:weaknull}). 
\item For $\tpHb, \widetilde{\omb}, \tpH$, note that the powers of $\Om_\Ke$ are different in $\NI$ and $\NH$. Namely, there is an additional $\Om_\Ke$ weight in the $\NI$ energy, and the $\NI$ energy can therefore be thought of as more degenerate; see also the discussions in Section~\ref{baridiaeisagwgns} in the introduction. (We remark that by re-defining $\varpi^N$, it is possible to arrange for this additional degeneration to not be a full power of $\Om_\Ke$, but be instead an additional degeneration of $\f{1}{\log^2(\f{1}{\Om_\Ke})}$; see \cite{LukSbierski} for a similar estimate for the linear wave equation. We will however not pursue this stronger estimate in order to simplify the exposition.) 
\item Importantly, however, in the energies for $\tp, \tK, \tg, \tb$, the $\Om_\Ke$ weights are the same in the $\NI$ and in $\NH$. (This is related to the fact that these weights are not generated by $\varpi^N$; see discussions in Section~\ref{baridiaeisagwgns} and Remark~\ref{rmk:bulk}.) This will be important for closing our estimates.
\end{itemize}
\end{remark}

\begin{remark}[Remarks on the $|u|^{\f 12+ \de}$ and $\ub^{\f 12+ \de}$ weights]
In $\NI$ and $\NH$, notice that sometimes there is a weight $\ub^{\f12 + \de}$ and sometimes there is a weight $|u|^{\f 12+\de}$. In the region we consider, clearly having a $\ub^{\f 12+\de}$ weight is stronger than having a $|u|^{\f 12+\de}$ weight. On the other hand, for quantities such as $(\tpHb,\widetilde{\omb})$, note that one can\underline{not} hope to put in a $\ub^{\f 12+\de}$ weight since in general, one expects $(\tpHb,\widetilde{\omb})$ to be non-vanishing at the Cauchy horizon.
\end{remark}

\begin{remark}[Remark on $\nab^i\nab_4\widetilde{\log\Om}$]
Note that the quantity $\nab^i\nab_4\widetilde{\log\Om}$ is not estimated in any of the energies $\NI$, $\NH$ or $\NS$. However, it is included in the initial data energy $\mathcal D$ in \eqref{data.D.def}. This is because we do not need any information on $\nab^i\nab_4\widetilde{\log\Om}$ when closing the bootstrap argument, but we need to ensure that $\nab^i\nab_4\widetilde{\log\Om}$ decays sufficiently fast in order to guarantee the continuous extendibility of the metric to the Cauchy horizon.
\end{remark}

\subsection{Statement of main theorem in the double null foliation gauge}\label{sec.main.thm.dn}

We are now ready to state the main theorem:
\begin{theorem}\label{aprioriestimates}
For every fixed Kerr spacetime with $0<|a|<M$ and every $\de>0$, $C_R\in(-\infty,+\infty)$, there exists $\ep_0=\ep_0(a,M,\de,C_R)>0$ and $u_f=u_f(a,M,\de,C_R)\leq -1$ such that the following holds:

Suppose that for the given geometric initial data $(\Sigma_0,\hat{g},\hat{k})$ and the initial data for the double null foliation and double null coordinates defined as in \eqref{coord.init} in Section~\ref{setup.double null}, the initial data energy $\mathcal D$ obeys
\begin{equation}\label{small.data}
\mathcal D \leq \ep^2
\end{equation}
for $\ep\leq \ep_0$, 
then the following holds:
\begin{itemize}
\item Let $(\mathcal{M},g)$ be the maximal globally hyperbolic future development of the restriction of the initial data on $\Sigma_0\cap (\{\tau : \tau < -u_f + C_R\}\times \mathbb S^2)$ as given by \cite{geroch}. Then $(\mathcal M,g)$ can be endowed with a double null foliation $(u,\ub)$ attaining the prescribed initial data for the foliation in Section~\ref{setup.double null} so that it is isometric to a smooth spacetime $(\mathcal U_\infty\doteq \mathcal W_\infty\times \mathbb S^2,g)$ with $\mathcal W_\infty\doteq \{(u,\ub):u < u_f,\,u+\ub\geq C_R\}$ and $g$ taking the form \eqref{formofthemetric}.
\item In addition, $(\mathcal U_\infty, g)$ remains close to Kerr spacetime in the sense that
\begin{equation}\label{main.apriori.est}
\NI,\,\NH,\,\NS\leq C \ep^2,
\end{equation}
for some $C>0$ which depends only on $a$, $M$, $\de$ and $C_R$. 
\item Moreover, there exists a non-trivial Cauchy horizon $\CH$, which is a future null boundary of $(\mathcal U_\infty,g)$, across which the spacetime metric extends continuously.
\end{itemize}
\end{theorem}

\begin{remark}[Theorem~\ref{aprioriestimates} implies Theorem~\ref{main.thm.spacelike.formulation}]
In view of Proposition~\ref{prop.small.data.implies.small.data}, Theorem~\ref{aprioriestimates} implies Theorem~\ref{main.thm.spacelike.formulation}.
\end{remark}

\begin{remark}[Continuous extension to the Cauchy horizon]\label{rmk:CH}
In Theorem~\ref{aprioriestimates}, we show that there is a non-trivial Cauchy horizon across which the spacetime metric extends continuously. (The non-trivial Cauchy horizon can be thought of as analogous to the Kerr Cauchy horizon. This is in contrast to the null boundary $\{u=u_f\}$, which can be thought of as a \emph{trivial} Cauchy horizon associated to the incompleteness of initial hypersurface, cf.~Section~\ref{sec.CH}.) The non-trivial Cauchy horizon can be most easily described in the double null coordinate system that we are using. We will show that by rescaling $\ub$ to $\ub_{\CH}$ and introducing a new system of coordinates $(\th^1_{\CH}, \th^2_{\CH})$ on the $2$-spheres $S_{u,\ub}$ (cf. the analogous change of variable in Kerr in Section~\ref{sec.coord.near.CH}), one can attach a null boundary $\CH$ such that the metric extends continuous to the manifold-with-boundary. Moreover, in the coordinate system $(u,\ub_{\CH},\th^1_{\CH},\th^2_{\CH})$, the metric $g$ is in fact close to the Kerr metric in $C^0$ up to and including the Cauchy horizon. See Section~\ref{sec.C0} for details.
\end{remark}

We note that the first bullet point in Theorem~\ref{aprioriestimates} will be finally obtained by combining Theorem~\ref{I.final} (proven in Section~\ref{sec:continuity}) and the fact that $(\mathcal U_\infty,g)$ is indeed the maximal globally hyperbolic future development of the restriction of the data (proven in Proposition~\ref{lem:GH} in Section~\ref{sec.C0}). The second bullet point in Theorem~\ref{aprioriestimates} will be obtained as part of Theorem~\ref{I.final} in Section~\ref{sec:continuity}. Finally, the third bullet point in Theorem~\ref{aprioriestimates} will be obtained in Theorem~\ref{metric.cont.ext} in Section~\ref{sec.C0}. We refer the reader also to Section~\ref{sec.completion.everything}, where the proof of Theorem~\ref{aprioriestimates} will be completed.

\subsection{Main bootstrap assumption}\label{sec.BA}
In order to prove the bound \eqref{main.apriori.est}, we will use a bootstrap argument. As above, we assume that we have a solution $(\mathcal U_{\ub_f},g)$ to the Einstein vacuum equations in a system of double null coordinates with the prescribed data (see precise statement in Theorem~\ref{main.quantitative.thm}). Assume that the following \emph{bootstrap assumption} holds throughout $\mathcal U_{\ub_f}$:
\begin{equation}\label{BA}
\NS\leq \ep.
\end{equation}
Under the bootstrap assumption \eqref{BA}, we will prove the bounds on $\NI$ and $\NH$ stated in \eqref{main.apriori.est} in Sections~\ref{secbasic}--\ref{sec:error}. In Section \ref{recover.bootstrap}, we will then use the estimates on $\NH$ to show that \eqref{BA} can be improved to a bound
$$\NS\leq C\ep^2$$
where $C$ is a constant depending only on $a$, $M$, $\de$ and $C_R$ and will be chosen later. Thus, for $\ep$ sufficiently small depending only on $a$, $M$, $\de$ and $C_R$, we have $C\ep^2\leq \frac{\ep}{2}$. This then improves \eqref{BA}. We summarise this discussion in the following theorem, the proof of which will be our goal in Sections~\ref{secbasic}--\ref{recover.bootstrap} (see the conclusion of the proof in Section~\ref{sec.pf.of.main.quantitative.thm}).

\begin{theorem}[Bootstrap theorem]\label{main.quantitative.thm}
For every fixed Kerr spacetime with $0<|a|<M$ and every $\de>0$, $C_R\in(-\infty,+\infty)$, there exists $\ep_0=\ep_0(a,M,\de,C_R)>0$ and $u_f=u_f(a,M,\de,C_R)\leq -1$ such that the following holds:

Let $(\mathcal U_{\ub_f}\doteq \mathcal W_{\ub_f}\times \mathbb S^2,g)$ be a smooth solution to the Einstein vacuum equation, where $\mathcal W_{\ub_f}\subset \mathbb R^2$ is given by $\mathcal W_{\ub_f}=\{(u,\ub): \ub \in [-u+C_R,\ub_f),\, u\in [-\ub+C_R,-u_f)\}$,
such that
\begin{itemize}
\item the metric $g$ takes the form \eqref{formofthemetric},
\item the metric $g$ achieves the prescribed initial data on $\{(u,\ub): u+\ub=C_R\}\times \mathbb S^2$ as in Theorem~\ref{aprioriestimates}.
\end{itemize}
Assume that \eqref{small.data} holds for some $\ep\leq \ep_0$ and that the bootstrap assumption \eqref{BA} holds in $\mathcal W_{\ub_f}\times \mathbb S^2$. Then the following estimate holds in $\mathcal W_{\ub_f}\times \mathbb S^2$:
$$\NI,\,\NH,\,\NS\leq C\ep^2$$
for some $C>0$ which depends only on $a$, $M$, $\de$ and $C_R$.
\end{theorem}

\begin{remark}[Comment regarding the smallness constants]\label{rmk.constant}
Before we proceed, we make a further comment on the constants in the proof of the theorem. In addition to the size of the initial data $\ep$, we will use another smallness parameter $N^{-1}$, which appeared\footnote{The constant $N$ appear as the exponents of the weight function $\varpi$ in the energies.} in the definition of the energies $\NI$ and $\NH$. In the proof of the theorem, the constants $a$, $M$, $\de$ and $C_R$ are taken to be given. We will then choose $N$ large and $\ep_0$ small (in that order) depending on these parameters and show that we can close the bootstrap argument as described above. In particular, we will always assume that $\ep\leq \ep_0 \leq N^{-1}$.
\end{remark}

\begin{remark}\label{rmk:setup}
Section~\ref{secbasic}--Section~\ref{recover.bootstrap} will concern the proof of the bootstrap theorem (Theorem~\ref{main.quantitative.thm}). In particular, we will work with a solution $(\mathcal U_{\ub_f},g)$ satisfying the hypotheses of Theorem~\ref{main.quantitative.thm} and assume that \eqref{BA} holds without further comments. Moreover, in the proof of the estimates, unless otherwise stated, \textbf{we will use\footnote{Note that we will use this notation both when $A,\,B>0$ and when $A,\,B<0$.} the notation $A \ls B$ to denote $A\leq CB$ for some $C>0$ depending only on $a$, $M$, $\de$ and $C_R$.} (This dependence of the constants is consistent with that in Theorem~\ref{main.quantitative.thm}.) In particular, the constants in $\ls$ are \textbf{independent of $\ub_f$, $u_f$, $\ep$ and $N$}.
\end{remark}

\section{The preliminary estimates}\label{secbasic}

Recall that we continue to work under the setting described in Remark~\ref{rmk:setup}.

\subsection{Sobolev embedding}\label{Embedding}

In this subsection, we derive bounds on the isoperimetric constant on the 2-spheres $S_{u,\ub}$ adapted to the double null foliation. This will in turn allow us to obtain the Sobolev embedding theorems that we will use the in remainder of the paper. First, we define the isoperimetric constant for any Riemannian 2-surface $(S,\gamma)$:
$${\bf I}(S,\gamma)\doteq \sup_{\substack{U\\\partial U \in C^1}} \f{\min\{\mbox{Area}(U),\mbox{Area}(U^c)\}}{(\mbox{Perimeter}(\partial U))^2}.$$
We note that in applications, $S$ will be one of the $2$-spheres $S_{u,\ub}$ adapted to the double null foliation. 

As is well-known, the constants in the Sobolev embedding theorem can be controlled by the isoperimetric constant. We will need an $L^2-L^{p}$ Sobolev embedding and also an $L^{p}-L^\i$ Sobolev embedding in this paper. To this end, we will use the following propositions quoted directly from \cite{Chr}. The first is an $L^2-L^{p}$ embedding theorem:
\begin{proposition}[\cite{Chr}, Lemma 5.1]\label{L4.prelim}
Let $2< p<\infty$ and $r\in \mathbb N\cup\{0\}$. There exists a constant $C_{p,r}>0$, depending only on $p$ and $r$, such that for any closed Riemannian $2$-manifold $(S,\gamma)$, 
$$(\mbox{Area}(S))^{-\f1p}\|\xi\|_{L^p(S)}\leq C_{p,r}\sqrt{\max\{{\bf I}(S),1\}}(\|\nab\xi\|_{L^2(S)}+(\mbox{Area}(S))^{-\f12}\|\xi\|_{L^2(S)})$$
for any covariant tensor $\xi$ of rank $r$.
\end{proposition}
The second is an $L^p-L^\i$ embedding:
\begin{proposition}[\cite{Chr}, Lemma 5.2]\label{Linfty.prelim}
Let $2< p<\infty$ and $r\in \mathbb N\cup\{0\}$. There exists a constant $C_{p,r}>0$, depending only on $p$ and $r$, such that for any closed Riemannian $2$-manifold $(S,\gamma)$,
$$\|\xi\|_{L^\infty(S)}\leq C_{p,r}\sqrt{\max\{{\bf I}(S),1\}}(\mbox{Area}(S))^{\f12-\f1p}(\|\nab\xi\|_{L^p(S)}+(\mbox{Area}(S))^{-\f12}\|\xi\|_{L^p(S)})$$
for any covariant tensor $\xi$ of rank $r$.
\end{proposition}

In order to apply Propositions~\ref{L4.prelim} and \ref{Linfty.prelim}, we need to bound the isoperimetric constants  on the 2-spheres $(S_{u,\ub},\gamma)$ in the spacetime $(\mathcal U_{\ub_f},g)$. The proof on one hand relies on the bound in Proposition \ref{isoperimetric.Kerr} for the isoperimetric constant of the background Kerr solution and on the other hand uses the fact that the metric $\gamma$ is close to the corresponding metric $\gamma_\Ke$ on the background Kerr spacetime. We first need to show that the metrics $\gamma$ and $\gamma_\Ke$ are close to each other:
\begin{proposition}\label{gamma.diff.est}
There exists a universal small constant $c>0$ such that if
$$|\gamma-\gamma_\Ke|_{\gamma}\leq c,$$
we have
$$|\gamma^{-1}-(\gamma_\Ke)^{-1}|_{\gamma}\ls |\gamma-\gamma_\Ke|_{\gamma}.$$
Moreover, let $\Lambda$ and $\lambda$ be respectively the larger and smaller eigenvalue\footnote{Notice that $(\gamma^{-1}_\Ke)\gamma$ can be viewed as a map $:T_p S_{u,\ub}\to T_p S_{u,\ub}$ and the eigenvalues are defined independently of the choice of coordinates.} of the matrix $(\gamma^{-1}_\Ke)\gamma$. Then we also have
$$|\Lambda-1|+|\lambda-1|\ls |\gamma-\gamma_\Ke|_{\gamma}.$$
\end{proposition}
\begin{proof}
By definition,
$$|\gamma-\gamma_\Ke|_{\gamma}^2=\mbox{Tr}(\gamma^{-1}(\gamma-\gamma_\Ke)(\gamma^{-1}(\gamma-\gamma_\Ke))^T)=\mbox{Tr}((\mbox{Id}-\gamma^{-1}\gamma_\Ke)(\mbox{Id}-\gamma^{-1}\gamma_\Ke)^T).$$
Here, $^T$ denotes the transpose of a matrix and $\mbox{Id}$ is the $2\times 2$ identity matrix. The assumption of the proposition thus implies that we have the following bound for every entry of the matrix\footnote{Here, we note that for a $(2\times 2)$ matrix $A=\begin{bmatrix} \begin{array}{cc}
a_{11} & a_{12} \\
a_{21} & a_{22} \\
\end{array} 
\end{bmatrix} $, $Tr(A A^T) = a_{11}^2+a_{12}^2+a_{21}^2+a_{22}^2$.}: 
$$|\mbox{Id}_{A}^B-(\gamma^{-1}\gamma_\Ke)_{A}^B|^2\leq c^2.$$
Therefore, by choosing $c$ sufficiently small, we have
$$|\det (\gamma^{-1}\gamma_\Ke) -1|\leq 3\sup_{A,\,B}|\mbox{Id}_{A}^B-(\gamma^{-1}\gamma_\Ke)_{A}^B|.$$
Using Cramer's rule, and choosing $c$ smaller if necessary, this implies
\begin{equation}\label{gamma.diff.est.1}
|\mbox{Id}_{A'}^{B'}-(\gamma_\Ke^{-1}\gamma)_{A'}^{B'}|\ls \sup_{A',\,B'}|\mbox{Id}_{A'}^{B'}-(\gamma^{-1}\gamma_\Ke)_{A'}^{B'}|.
\end{equation}
Now, notice that
\begin{equation}\label{gamma.diff.est.2}
|\gamma^{-1}-(\gamma_\Ke)^{-1}|_{\gamma}^2=\mbox{Tr}((\gamma^{-1}-(\gamma_\Ke)^{-1})\gamma((\gamma^{-1}-(\gamma_\Ke)^{-1})\gamma)^T)=\mbox{Tr}((\mbox{Id}-\gamma_\Ke^{-1}\gamma)(\mbox{Id}-\gamma_\Ke^{-1}\gamma)^T).
\end{equation}
\eqref{gamma.diff.est.1} and \eqref{gamma.diff.est.2} together imply
$$|\gamma^{-1}-(\gamma_\Ke)^{-1}|_{\gamma}\ls |\gamma-\gamma_\Ke|_{\gamma}.$$
Moreover, returning to \eqref{gamma.diff.est.1}, we obtain the desired bounds on $\Lambda$ and $\lambda$.
\end{proof}
Since our bootstrap assumption \eqref{BA} implies in particular that the metric $\gamma$ is close to the corresponding metric $\gamma_\Ke$ on the background Kerr spacetime, we can use Proposition \ref{gamma.diff.est.1} to obtain the following estimate on the isoperimetric constant:
\begin{proposition}\label{isoperimetric}
For any $S_{u,\ub}\subset \mathcal U_{\ub_f}$, the area of $S_{u,\ub}$ satisfies the following estimates for some $C>0$ depending only on $M$ and $a$:
$$Area(S_{u,\ub},\gamma_\Ke)-C\ep^{\f 12} \leq Area(S_{u,\ub},\gamma)\leq Area(S_{u,\ub},\gamma_\Ke)+C\ep^{\f 12}.$$
Moreover, the isoperimetric constant obeys the following bound:
$${\bf I}(S_{u,\ub},\gamma)\leq 2 {\bf I}_{\Ke},$$
where ${\bf I}_{\Ke}$ is the constant as in Proposition~\ref{isoperimetric.Kerr}.
\end{proposition}
\begin{proof}
By a standard continuity argument\footnote{Note that the isoperimetric constants are initially bounded on the hypersurface $\Sigma_0$ by $2 {\bf I}_{\Ke}$ due to the smallness of $\mathcal D$.}, it suffices to prove this proposition under the bootstrap assumption
\begin{equation}\label{BA.Sobolev}
\sup_{u,\ub} {\bf I}(S_{u,\ub},\gamma)\leq 4 {\bf I}_{\Ke}.
\end{equation}
Under the assumption \eqref{BA.Sobolev}, we can apply the Sobolev embedding theorems in Propositions \ref{L4.prelim} and \ref{Linfty.prelim} to obtain a bound on the $L^\i$ norm of $|\gamma-\gamma_\Ke|_\gamma$. More precisely, using the bootstrap assumption \eqref{BA} and Propositions \ref{L4.prelim} and \ref{Linfty.prelim}, we have
$$\|\gamma-\gamma_\Ke\|_{L^\i_uL^\i_{\ub}L^\i(S)}\ls \ep^{\f12}.$$
From this the area bound immediately follows. Moreover, we can apply Proposition \ref{gamma.diff.est} to obtain
\begin{equation}\label{Lambda.lambda}
\sup_{u,\ub}\sup_{S_{u,\ub}} (|\Lambda-1|+|\lambda-1|)\ls \ep^{\f12}.
\end{equation}

We now turn to the bound for the isoperimetric constant. Fix $(u,\ub)$ and a domain $U\subset S_{u,\ub}$ with $C^1$ boundary. We now compare the perimeters of $\partial U$ and the areas of $U$ and $U^c$ given by the metrics $\gamma_\Ke$ and $\gamma$. More precisely, we have
$$\f{\mbox{Perimeter}(\partial U,\gamma)}{\mbox{Perimeter}(\partial U,\gamma_\Ke)}\geq \sqrt{\inf_{S_{u,\ub}} \lambda} $$
and
$$\f{\mbox{Area}(U,\gamma)}{\mbox{Area}(U,\gamma_\Ke)}\leq \sup_{S_{u,\ub}}\sqrt{ \Lambda\lambda},\quad\f{\mbox{Area}(U^c,\gamma)}{\mbox{Area}(U^c,\gamma_\Ke)}\leq \sup_{S_{u,\ub}}\sqrt{ \Lambda\lambda}.$$
Now, the conclusion follows from the bound on the isoperimetric constant on the background Kerr spacetime in Proposition \ref{isoperimetric.Kerr} and the bounds for $\Lambda$ and $\lambda$ in \eqref{Lambda.lambda}.
\end{proof}

Finally, combining Propositions \ref{L4.prelim}, \ref{Linfty.prelim} and \ref{isoperimetric}, we obtain the following Sobolev embedding theorem in the spacetime $(\mathcal U_{\ub_f},g)$:
\begin{proposition}\label{Sobolev}
For any tensor $\xi$ on $S_{u,\ub}\subset \mathcal U_{\ub_f}$, we have the following two estimates:
$$\|\xi\|_{L^4(S_{u,\ub})}\ls \sum_{i\leq 1}\|\nab^i\xi\|_{L^2(S_{u,\ub})} $$
and 
$$\|\xi\|_{L^\infty(S_{u,\ub})}\ls \sum_{i\leq 1}\|\nab^i\xi\|_{L^4(S_{u,\ub})}.$$
Combining these two estimates, we also have the bound:
$$\|\xi\|_{L^\infty(S_{u,\ub})}\ls \sum_{i\leq 2}\|\nab^i\xi\|_{L^2(S_{u,\ub})}.$$
\end{proposition}
\begin{proof}
Proposition \ref{isoperimetric} gives an upper bound on the isoperimetric constant on every $S_{u,\ub}$ in $(\mathcal U_{\ub_f},g)$. Moreover, since the spheres $S_{u,\ub}$ in $(\mathcal M_{Kerr},g_{a,M})$ have uniformly upper and lower bounded areas, Proposition \ref{isoperimetric} gives uniform upper and lower bounds on $\mbox{Area}(S_{u,\ub})$ for $S_{u,\ub}$ in $(U_{\ub_f},g)$. The conclusion therefore follows directly from Propositions \ref{L4.prelim} and \ref{Linfty.prelim}.
\end{proof}
An immediate corollary of the bootstrap assumption \eqref{BA} and the Sobolev embedding in Proposition~\ref{Sobolev} is that we can obtain some $L^\i_uL^\i_{\ub}L^\i(S)$ bounds. Since we will repeatedly use the following estimates, we record them in the following corollary:
\begin{corollary}\label{Linfty}
The following $L^\i_uL^\i_{\ub}L^\i(S)$ estimates hold:
$$\sum_{i\leq 1}\|\nab^i\tg\|_{L^\i_uL^\i_{\ub}L^\i(S)} + \|(\tb,\tp,\tpHb,\widetilde{\omb},\Om_\Ke^2\tpH)\|_{L^\i_uL^\i_{\ub}L^\i(S)}\ls \ep^{\f 12}.$$
In particular, the estimates on $\tg$ implies
\begin{equation}\label{Om.c}
\Om_\Ke\ls \Om\ls \Om_\Ke.
\end{equation}
\end{corollary}

\subsection{General elliptic estimates for elliptic systems}\label{elliptic}

In this subsection, we will give two general elliptic estimates which hold for $2$-spheres $S_{u,\ub}\subset U_{\ub_f}$. One of them is for div-curl systems (Proposition~\ref{ellipticthm}) and the other one is for Poisson equations (Proposition~\ref{elliptic.Poisson}).

To state our first proposition, we recall the definitions of divergence, curl and trace in \eqref{div.def}, \eqref{curl.def} and \eqref{tr.def} respectively. The following elliptic estimate is standard\footnote{The estimate is restricted to the case $i\leq 3$ here because according to the bootstrap assumption \eqref{BA}, we only control up to $1$ derivative of the Gauss curvature $K$ in $L^\i_uL^\i_{\ub}L^2(S)$.} (see for example Lemmas 2.2.2, 2.2.3 in \cite{CK} or Lemmas 7.1, 7.2, 7.3 in \cite{Chr}):
\begin{proposition}\label{ellipticthm}
Let $\phi$ be a totally symmetric $r+1$ covariant tensor field on $(S_{u,\ub},\gamma)$ satisfying
$$\div\phi=f,\quad \curl\phi=g,\quad \tr\phi=h.$$
Then for $i\leq 3$, we have the estimate
$$||\slashed{\nabla}^{i}\phi||_{L^2(S_{u,\ub})}\ls \sum_{j=0}^{i-1}||\slashed{\nabla}^{j}(f,g,h,\phi)||_{L^2(S_{u,\ub})}.$$
\end{proposition}
We now also prove elliptic estimates for the Poisson's equation.
\begin{proposition}\label{elliptic.Poisson}
Let $\phi$ be a scalar function on $(S_{u,\ub},\gamma)$ satisfying 
$$\slashed\Delta\phi=f,$$
where $\slashed\Delta$ is the Laplace--Beltrami operator associated to $\gamma$ (see \eqref{LB.def}).
Then
$$\|\nab^3\phi\|_{L^2(S_{u,\ub})}\ls \sum_{i\leq 1}\|\nab^i f\|_{L^2(S_{u,\ub})}+\sum_{i\leq 2}\|\nab^i\phi\|_{L^2(S_{u,\ub})}. $$
\end{proposition}
\begin{proof}
First, notice that by the bootstrap assumption \eqref{BA}, we have the following bound on the Gauss curvature:
\begin{equation}\label{Poisson.0}
\sum_{i\leq 1}\|\nab^i K\|_{L^2(S_{u,\ub})}\ls 1.
\end{equation}
Now using Poisson's equation, we have\footnote{Note that we have the commutator estimate $|[\slashed\Delta,\nab]\phi|\ls |K||\nab\phi|$.}
\begin{equation}\label{Poisson.1}
|\slashed\Delta\nab\phi|\ls |\nab\slashed\Delta\phi| + |[\slashed\Delta,\nab]\phi| \ls |\nab f|+|K||\nab \phi|.
\end{equation}
Integrating by parts, we have the estimate
\begin{equation}\label{Poisson.2}
\left| \int_{S_{u,\ub}} |\slashed\Delta\nab\phi|^2 - \int_{S_{u,\ub}} |\nab^3\phi|^2 \right|\ls \int_{S_{u,\ub}} |K||\nab^2\phi|^2.
\end{equation}
Combining \eqref{Poisson.1} and \eqref{Poisson.2}, we obtain
$$\|\nab^3\phi\|_{L^2(S_{u,\ub})}^2\ls \|\nab f\|_{L^2(S_{u,\ub})}^2+\|K\nab\phi\|_{L^2(S_{u,\ub})}^2+\|K\nab^2\phi\nab^2\phi\|_{L^1(S_{u,\ub})}.$$
The conclusion then follows from H\"older's inequality, Sobolev embedding (Proposition \ref{Sobolev}) and the bound \eqref{Poisson.0} for the Gauss curvature.
\end{proof}

\section{Estimates for general covariant transport equations}\label{transportsec}

Recall that we continue to work under the setting described in Remark~\ref{rmk:setup}. 

In this section, we consider a general $S$-tangent tensor $\phi$ and control it in terms of either $\nab_3\phi$ or $\nab_4\phi$. In later sections, this will then be applied to estimate $\phi$ using the covariant transport equation (either in the $e_3$ or $e_4$ directions) that it satisfy. In other words, what we proved in this section can be thought of as estimates for general covariant transport equations. First, we derive the following general identities relating general $S$-tangent tensors and their $\nab_3$ or $\nab_4$ derivatives:
\begin{proposition}\label{transport}
Let $\phi$ be a tensor field of arbitrary rank tangential to the spheres $S_{u,\ub}$. The following identities hold:
\begin{equation*}
\begin{split}
 ||\phi||^2_{L^2(S_{u,\ub})}=&\:||\phi||^2_{L^2(S_{u,\ub'})}+2\int_{\ub'}^{\ub}\int_{S_{u,\ub''}} \Omega^2\left(\langle\phi,\slashed{\nabla}_4\phi\rangle_\gamma+ \frac{1}{2}\trch |\phi|^2_{\gamma}\right)d{\ub''}\\
 ||\phi||^2_{L^2(S_{u,\ub})}=&\:||\phi||^2_{L^2(S_{u',\ub})}+2\int_{u'}^{u}\int_{S_{u'',\ub}} \left(\langle\phi,\slashed{\nabla}_3\phi\rangle_\gamma+ \frac{1}{2}\trchb |\phi|^2_{\gamma}\right)d{u''}.
\end{split}
\end{equation*}
\end{proposition}

\begin{proof}
The following identities\footnote{We remark that in the first identity, $\f{\rd}{\rd \ub}$ is to be understood as the coordinate vector field in the ($2$-dimensional) $(u,\ub)$-plane. Similar remarks apply to the second identity.} hold for any scalar $f$:
\begin{equation*}
 \frac{\rd}{\rd\ub}\int_{S_{u,\ub}} f=\int_{S_{u,\ub}} \Omega^2\left(e_4(f)+ \trch f\right), \quad  \frac{\rd}{\rd u}\int_{S_{u,\ub}} f=\int_{S_{u,\ub}} \left(e_3(f)+ \trchb f\right).
\end{equation*}
Hence, taking $f=|\phi|_{\gamma}^2$ and noting $e_4(|\phi|_{\gamma}^2)=2\langle\phi,\slashed{\nabla}_4\phi\rangle_\gamma$ and $e_3(|\phi|_{\gamma}^2)=2\langle\phi,\slashed{\nabla}_3\phi\rangle_\gamma$, we obtain the conclusion.
\end{proof}
Using Proposition \ref{transport}, we show the following estimate for $\phi$ in terms of $\nab_3\phi$. In particular, we show that by choosing a $\varpi^N$ weight with $N$ sufficiently large, we have a term in $L^2_uL^2_{\ub}L^2(S)$ with a good sign and with a large coefficient $N$.
\begin{proposition}\label{transport.3.1}
Let $\phi$ be a tensor field of arbitrary rank tangential to the spheres $S_{u,\ub}$. Then $\phi$ satisfies the following estimate:
\begin{equation*}
\begin{split}
&\:\|\ub^{\frac 12+\de}\varpi^N\phi\|_{L^2_{\ub}L^\infty_uL^2(S)}^2+N\|\ub^{\frac 12+\de}\varpi^N\Om_\Ke\phi\|_{L^2_uL^2_{\ub}L^2(S)}^2\\
\ls &\:\|\ub^{\frac 12+\de}\varpi^N\phi\|_{L^2_{\ub}L^2(S_{-\ub+C_R,\ub})}^2+\| \ub^{1+2\de}\varpi^{2N}\phi\nab_3\phi\|_{L^1_{\ub}L^1_uL^1(S)},
\end{split}
\end{equation*}
where here (and below in Propositions \ref{transport.3.2} and \ref{transport.3.3}), we simply write $\phi\nab_3\phi=\langle\phi,\nab_3\phi\rangle_{\gamma}$.
\end{proposition}
\begin{proof}
We now apply the second identity in Proposition \ref{transport} to $\ub^{\frac 12+\de}\varpi^N\phi$. Noticing that $\nab_3\ub=0$, we obtain
\begin{equation}\label{id.phi}
\begin{split}
&\:\|\ub^{\frac 12+\de}\varpi^N\phi\|_{L^2(S_{u,\ub})}^2\\
=&\:\|\ub^{\frac 12+\de}\varpi^N\phi\|_{L^2(S_{-\ub+C_R,\ub})}^2+2\int_{-\ub+C_R}^{u}\int_{S_{u',\ub}}\ub^{1+2\de}\langle\varpi^N\phi,(N\varpi^{N-1}\nab_3\varpi)\phi+\varpi^N\nab_3\phi\rangle_{\gamma} \,du'\\
&\:+\int_{-\ub+C_R}^{u}\int_{S_{u',\ub}} \ub^{1+2\de}\varpi^{2N}\trchb|\phi|_{\gamma}^2 \,du'.
\end{split}
\end{equation}
Since $\varpi$ is a scalar function, $\nab_3\varpi=\frac{\partial}{\partial u}\varpi$. Recall from \eqref{varpi.bd} and \eqref{varpi.34} that
\begin{equation}\label{varpi.repeat}
1\leq \varpi\leq 2,\quad \nab_3\varpi\ls -\Om_\Ke^2 \leq 0.
\end{equation}
Now, since $\nab_3 \varpi$ is negative, we can move the second term on the right hand side of \eqref{id.phi} to the left hand side to obtain a positive term, i.e.
\begin{equation}\label{id.phi.2}
\begin{split}
&\:\|\ub^{\frac 12+\de}\varpi^N\phi\|_{L^2(S_{u,\ub})}^2+N\int_{-\ub+C_R}^{u}\|\ub^{\frac 12+\de}\varpi^N\Om_\Ke\phi\|_{L^2(S_{u',\ub})}^2 du'\\
\ls &\:\|\ub^{\frac 12+\de}\varpi^N\phi\|_{L^2(S_{-\ub+C_R,\ub})}^2+|\int_{-\ub+C_R}^{u}\int_{S_{u',\ub}}\ub^{1+2\de} \langle \varpi^N\phi,\varpi^N\nab_3\phi\rangle_{\gamma} \,du'|\\
&\:+|\int_{-\ub+C_R}^{u}\int_{S_{u',\ub}} \ub^{1+2\de}\varpi^{2N}\trchb|\phi|_{\gamma}^2 du'|.
\end{split}
\end{equation}
Now, by Proposition~\ref{Kerr.Ricci.bound} we have\footnote{We remark that the background term $\trchb_\Ke$ in fact has a good sign for this estimate, but we nonetheless simply bound its absolute value.}
\begin{equation}\label{trchb.bd}
|\trchb|\leq |\trchb_\Ke|+|\trchb-\trchb_\Ke|\ls \Om_\Ke^2+|\tpHb|.
\end{equation}
Therefore, returning to \eqref{id.phi.2}, we have
\begin{equation}\label{nab3.phi.handling.trchb}
\begin{split}
&\:\|\ub^{\frac 12+\de}\varpi^N\phi\|_{L^2(S_{u,\ub})}^2+N\int_{-\ub+C_R}^{u}\|\ub^{\frac 12+\de}\varpi^N\Om_\Ke\phi\|_{L^2(S_{u',\ub})}^2 du'\\
\ls &\:\|\ub^{\frac 12+\de}\varpi^N\phi\|_{L^2(S_{-\ub+C_R,\ub})}^2+|\int_{-\ub+C_R}^{u}\int_{S_{u',\ub}}\ub^{1+2\de} \langle\varpi^N\phi,\varpi^N\nab_3\phi\rangle_{\gamma} du'|\\
&\:+\int_{-\ub+C_R}^{u}\int_{S_{u',\ub}} \ub^{1+2\de}\varpi^{2N}\Om_\Ke^2|\phi|_{\gamma}^2 du'+\int_{-\ub+C_R}^{u}\int_{S_{u',\ub}} \ub^{1+2\de}\varpi^{2N}|\tpHb||\phi|_{\gamma}^2 du'.
\end{split}
\end{equation}
For $N$ sufficiently large, since
$$\int_{-\ub+C_R}^{u}\int_{S_{u',\ub}} \ub^{1+2\de}\varpi^{2N}\Om_\Ke^2|\phi|_{\gamma}^2 du'\ls \int_{-\ub+C_R}^{u}\|\ub^{\frac 12+\de}\varpi^N\Om_\Ke\phi\|_{L^2(S_{u',\ub})}^2 du',$$
we can absorb this term to the left hand side to obtain
\begin{equation*}
\begin{split}
&\|\ub^{\frac 12+\de}\varpi^N\phi\|_{L^2(S_{u,\ub})}^2+N\int_{-\ub+C_R}^{u}\|\ub^{\frac 12+\de}\varpi^N\Om_\Ke\phi\|_{L^2(S_{u',\ub})}^2 du'\\
\ls &\|\ub^{\frac 12+\de}\varpi^N\phi\|_{L^2(S_{-\ub+C_R,\ub})}^2+|\int_{-\ub+C_R}^{u}\int_{S_{u',\ub}}\ub^{1+2\de} \langle\varpi^N\phi,\varpi^N\nab_3\phi\rangle_{\gamma} du'|\\
&+\int_{-\ub+C_R}^{u}\int_{S_{u',\ub}} \ub^{1+2\de}\varpi^{2N}|\tpHb||\phi|_{\gamma}^2 du'.
\end{split}
\end{equation*}
Applying Gr\"onwall's inequality, we obtain
\begin{equation*}
\begin{split}
&\:\|\ub^{\frac 12+\de}\varpi^N\phi\|_{L^2(S_{u,\ub})}^2+N\int_{-\ub+C_R}^{u}\|\ub^{\frac 12+\de}\varpi^N\Om_\Ke\phi\|_{L^2(S_{u',\ub})}^2 du'\\
\ls &\:\left(\|\ub^{\frac 12+\de}\varpi^N\phi\|_{L^2(S_{-\ub+C_R,\ub})}^2+|\int_{-\ub+C_R}^{u}\int_{S_{u',\ub}}\ub^{1+2\de} \langle \varpi^N\phi,\varpi^N\nab_3\phi\rangle_{\gamma} du'|\right) \times e^{\int_{-\ub+C_R}^u \|\tpHb\|_{L^\i(S_{u',\ub})}\,du'}.
\end{split}
\end{equation*}
Finally, by Corollary~\ref{Linfty},
$$\int_{-\ub+C_R}^u \|\tpHb\|_{L^\i(S_{u',\ub})}\, du'\ls \ep^{\f 12}.$$
Therefore, we conclude by taking supremum in $u$ and then integrating in $L^2$ in $\ub$.
\end{proof}

If we put in an extra factor of $\Om_\Ke$ in the norm, then we have an additional term with a good sign. This extra term, while it does not have a factor $N$ in front, has the good property that it has the same $\Om_\Ke$ weight as the boundary term.
\begin{proposition}\label{transport.3.2}
Let $\phi$ be a tensor field of arbitrary rank tangential to the spheres $S_{u,\ub}$. Then $\phi$ satisfies the following estimate:
\begin{equation*}
\begin{split}
&\:\|\ub^{\frac 12+\de}\varpi^N\Om_\Ke\phi\|_{L^2_{\ub}L^\infty_uL^2(S)}^2+N\|\ub^{\frac 12+\de}\varpi^N\Om_\Ke^2\phi\|_{L^2_uL^2_{\ub}L^2(S)}^2+\|\ub^{\frac 12+\de}\varpi^N\Om_\Ke\phi\|_{L^2_uL^2_{\ub}L^2(S)}^2\\
\ls &\:\|\ub^{\frac 12+\de}\varpi^N\Om_\Ke\phi\|_{L^2_{\ub}L^2(S_{-\ub+C_R,\ub})}^2+\| \ub^{1+2\de}\varpi^{2N}\Om_\Ke^2\phi\nab_3\phi\|_{L^1_{\ub}L^1_uL^1(S)}.
\end{split}
\end{equation*}
\end{proposition}
\begin{proof}
We prove a slightly more general statement which will be useful later (see Proposition \ref{transport.3.3}). Let $m \in [1,3]$. We now apply the second identity in Proposition~\ref{transport} to $\ub^{\frac 12+\de}\varpi^N\Om_\Ke^m\phi$. Noting also that $\nab_3\ub^{\f 12+\de} = 0$, we obtain
\begin{equation}\label{id.t.phi}
\begin{split}
&\:\|\ub^{\frac 12+\de}\varpi^N\Om_\Ke^m\phi\|_{L^2(S_{u,\ub})}^2\\
=&\:\|\ub^{\frac 12+\de}\varpi^N\Om_\Ke^m\phi\|_{L^2(S_{-\ub+C_R,\ub})}^2 + 2\int_{-\ub+C_R}^{u}\int_{S_{u',\ub}}\ub^{1+2\de}\left(\langle\varpi^N\Om_\Ke^m\phi,(\nab_3(\varpi^N\Om_\Ke^m))\phi \rangle_{\gamma}\right) du'\\
&\:+2\int_{-\ub+C_R}^{u}\int_{S_{u',\ub}}\ub^{1+2\de}\langle\varpi^N\Om_\Ke^m\phi,\varpi^N\Om_\Ke^m\nab_3\phi \rangle_{\gamma} \,du'\\
&\:+\int_{-\ub+C_R}^{u}\int_{S_{u',\ub}} \ub^{1+2\de}\trchb \varpi^{2N}\Om_\Ke^{2m}|\phi|_{\gamma}^2 \,du'.
\end{split}
\end{equation}
The term in which $\nab_3$ acts on the weights can be expanded as follows:
$$\nab_3(\varpi^N\Om_\Ke^m) = N\varpi^{N-1}\Om_\Ke^m\nab_3\varpi+m\varpi^N\Om_\Ke^{m-1}\nab_3\Om_\Ke\doteq I+II.$$
The term $\nab_3\varpi$ has appeared also in Proposition~\ref{transport.3.1}. By \eqref{varpi.repeat}, we see that $I$ is negative and satisfies the bound 
$$I\ls -N \varpi^{N-1}\Om_\Ke^{m-2}.$$
For the term involving $\nab_3\Om_\Ke$, since $\Om_\Ke$ is a scalar function and $e_3=\frac{\partial}{\partial u}$ is a fixed vector field on the background differential structure, $\nab_3\Om_\Ke$ simply takes the value as on the background Kerr spacetime. Hence, by Proposition~\ref{Om.der.Kerr.bounds},
$$|\nab_3\Om_\Ke|\ls \Om_\Ke.$$
Moreover, we can fix a $C_{R'}$ sufficiently large so that
$$\nab_3\Om_\Ke\ls -\Om_\Ke\leq 0\quad\mbox{for }u+\ub\geq C_{R'}.$$
To proceed, we note that by Proposition~\ref{Om.Kerr.bounds},
$$1\ls \Om_\Ke \ls 1$$
in the region $-\ub+C_R\leq u\leq -\ub+C_{R'}$, where the implicit constant depends only on $M$, $a$, $C_R$ and $C_{R'}$. Therefore, by choosing $N$ large enough, the term $II$ is dominated by the term $I$. Combining these observations, we obtain the following bound which holds everywhere in $\{u+\ub\geq C_R\}$:
\begin{equation}\label{nab3.weight.good.bound}
\nab_3(\varpi^N\Om_\Ke^m)\ls -(N \Om_\Ke^{m+2}+ \Om_\Ke^m).
\end{equation}
Using \eqref{nab3.weight.good.bound} and following the proof of Proposition~\ref{transport.3.1}, we move the second term on the right hand side of \eqref{id.t.phi} to the left hand side to obtain a positive term\footnote{which is then written as two positive terms in \eqref{id.t.phi.3}.}. More precisely, we have
%\begin{equation}\label{id.t.phi.2}
%\begin{split}
%&\:\|\ub^{\frac 12+\de}\varpi^N\Om_\Ke^m\phi\|_{L^2(S_{u,\ub})}^2+N\int_{-\ub+C_R}^{u}\|\ub^{\frac 12+\de}\varpi^N\Om_\Ke^{m+1}\phi\|_{L^2(S_{u',\ub})}^2du'\\
%&\:+\int_{-\ub+C_R}^{u}\|\ub^{\frac 12+\de}\varpi^N\Om_\Ke^m\phi\|_{L^2(S_{u',\ub})}^2du'\\
%\ls &\:\|\ub^{\frac 12+\de}\varpi^N\Om_\Ke^m\phi\|_{L^2(S_{-\ub+C_R,\ub})}^2 + \int_{-\ub+C_R}^{u}\int_{S_{u',\ub}} \ub^{1+2\de}\varpi^{2N}\Om_\Ke^{2m}|\langle\phi,\nab_3\phi\rangle_{\gamma}| du'\\
%&\:+\int_{-\ub+C_R}^{u}\int_{S_{u',\ub}} \ub^{1+2\de}\varpi^{2N}\Om_\Ke^{2m}|\trchb||\phi|^2_{\gamma} du'.
%\end{split}
%\end{equation}
%To proceed, we note that we have
%$$1\ls \Om_\Ke \ls 1$$
%in the region $-\ub+C_R\leq u\leq -\ub+C_{R'}$, where the implicit constant depends only on $M$, $a$, $C_R$ and $C_{R'}$. Therefore, by choosing $N$ large enough, the second term on the right hand side of \eqref{id.t.phi.2} can be controlled by the second term on the left hand side. In fact, because of this, we can also enlarge the domain of integration in the third term on the left hand side to $-\ub+C_R\leq u'\leq u$.
%Therefore, we have
\begin{equation}\label{id.t.phi.3}
\begin{split}
&\|\ub^{\frac 12+\de}\varpi^N\Om_\Ke^m\phi\|_{L^2(S_{u,\ub})}^2+N\int_{-\ub+C_R}^{u}\|\ub^{\frac 12+\de}\varpi^N\Om_\Ke^{m+1}\phi\|_{L^2(S_{u',\ub})}^2du'\\
&+\int_{-\ub+C_{R}}^{u}\|\ub^{\frac 12+\de}\varpi^N\Om_\Ke^m\phi\|_{L^2(S_{u',\ub})}^2du'\\
\ls &\|\ub^{\frac 12+\de}\varpi^N\Om_\Ke\phi\|_{L^2(S_{-\ub+C_R,\ub})}^2+\int_{-\ub+C_R}^{u}\int_{S_{u',\ub}} \ub^{1+2\de}\varpi^{2N}\Om_\Ke^{2m}|\langle\phi,\nab_3\phi\rangle_{\gamma}| du'\\
&+\int_{-\ub+C_R}^{u}\int_{S_{u',\ub}} \ub^{1+2\de}\varpi^{2N}\Om_\Ke^{2m}|\trchb||\phi|_{\gamma}^2 du'.
\end{split}
\end{equation}
We now deal with the last term in \eqref{id.t.phi.3}, which involves $\trchb$. Using \eqref{trchb.bd}, we have
\begin{equation}\label{id.t.phi.4}
\begin{split}
&\:\int_{-\ub+C_R}^{u}\int_{S_{u',\ub}} \ub^{1+2\de}\varpi^{2N}\Om_\Ke^{2m}|\trchb||\phi|_{\gamma}^2 du'\\
\ls &\:\int_{-\ub+C_R}^{u}\int_{S_{u',\ub}} \ub^{1+2\de}\varpi^{2N}\Om_\Ke^{2m+2}|\phi|^2_{\gamma} du'+\int_{-\ub+C_R}^{u}\int_{S_{u',\ub}} \ub^{1+2\de}\varpi^{2N}\Om_\Ke^{2m}|\tpHb||\phi|^2_{\gamma} du'.
\end{split}
\end{equation}
The two terms on the right hand side in \eqref{id.t.phi.4} can be handled in exactly the same manner as the last two terms in \eqref{nab3.phi.handling.trchb} so that we obtain
%The first term on the last line of \eqref{id.t.phi.4} can be absorbed into the second term on the left hand side after choosing $N$ to be sufficiently large. On the other hand, since by the bootstrap assumption \eqref{BA} and the Sobolev embedding theorem (Proposition \ref{Sobolev}), we have
%$$\|\tpHb\|_{L^\i_uL^\infty_{\ub}L^\infty(S)}\ls \ep^{\f12},$$
%the second term on the last line of \eqref{id.t.phi.4} can be absorbed into the third term on the left hand side after choosing $\ep$ to be sufficiently small. Therefore, we have
\begin{equation}\label{id.t.phi.5}
\begin{split}
&\:\|\ub^{\frac 12+\de}\varpi^N\Om_\Ke^m\phi\|_{L^2(S_{u,\ub})}^2+N\int_{-\ub+C_R}^{u}\|\ub^{\frac 12+\de}\varpi^N\Om_\Ke^{m+1}\phi\|_{L^2(S_{u',\ub})}^2du'\\
&\:+\int_{-\ub+C_{R}}^{u}\|\ub^{\frac 12+\de}\varpi^N\Om_\Ke^m\phi\|_{L^2(S_{u',\ub})}^2du'\\
\ls &\:\|\ub^{\frac 12+\de}\varpi^N\Om_\Ke^m\phi\|_{L^2(S_{-\ub+C_R,\ub})}^2+\int_{-\ub+C_R}^{u}\int_{S_{u',\ub}} \ub^{1+2\de}\varpi^{2N}\Om_\Ke^{2m}|\langle\phi,\nab_3\phi\rangle_{\gamma}| du'.
\end{split}
\end{equation}
Let $m=1$. The proposition follows from first taking supremum in $u$ and then integrating in $L^2$ in $\ub$.
\end{proof}
It will also be useful to have the following slight variant of Proposition \ref{transport.3.2}. Notice that the following proposition is an estimate on a fixed $\ub$ hypersurface, unlike the previous bounds where we took $L^2$ over the $\ub$ variable. Moreover, the weights\footnote{Since we only work in the region $u+\ub\geq C_R$, the bounds with $|u|$ weights are weaker than those with $\ub$ weights. On the other hand, according to Proposition \ref{transport.3.3}, in this case we also only need to control the error terms in a $|u|$-weighted (instead of $\ub$-weighted) norm.} can now either be in terms of $|u|$ or in terms of $\ub$.
\begin{proposition}\label{transport.3.3}
Let $\phi$ be a tensor field of arbitrary rank tangential to the spheres $S_{u,\ub}$. Then for $m=1$, $2$ or $3$, $\phi$ satisfies the following two estimates on a fixed $\ub=\mbox{constant}$ hypersurface:
\begin{equation*}
\begin{split}
&\|\ub^{\frac 12+\de}\varpi^N\Om_\Ke^m\phi\|_{L^\infty_uL^2(S)}^2+N\|\ub^{\frac 12+\de}\varpi^N\Om_\Ke^{m+1}\phi\|_{L^2_u L^2(S)}^2+\|\ub^{\frac 12+\de}\varpi^N\Om_\Ke\phi\|_{L^2_uL^2(S)}^2\\
\ls &\|\ub^{\frac 12+\de}\varpi^N\Om_\Ke^m\phi\|_{L^2(S_{-\ub+C_R,\ub})}^2+\|\ub^{1+2\de}\varpi^{2N}\Om_\Ke^{2m}\phi\nab_3\phi\|_{L^1_uL^1(S)}
\end{split}
\end{equation*}
and
\begin{equation*}
\begin{split}
&\||u|^{\frac 12+\de}\varpi^N\Om_\Ke^m\phi\|_{L^\infty_uL^2(S)}^2+N\||u|^{\frac 12+\de}\varpi^N\Om_\Ke^{m+1}\phi\|_{L^2_u L^2(S)}^2+\||u|^{\frac 12+\de}\varpi^N\Om_\Ke^m\phi\|_{L^2_uL^2(S)}^2\\
\ls &\||u|^{\frac 12+\de}\varpi^N\Om_\Ke^m\phi\|_{L^2(S_{-\ub+C_R,\ub})}^2+\| |u|^{1+2\de}\varpi^{2N}\Om_\Ke^{2m}\phi\nab_3\phi\|_{L^1_uL^1(S)}.
\end{split}
\end{equation*}
\end{proposition}
\begin{proof}
The first claimed estimate simply follows from \eqref{id.t.phi.5} in the proof of Proposition \ref{transport.3.2}. To obtain the second claimed estimate, we now revisit the proof of Proposition \ref{transport.3.2} and show that we can put in a $|u|^{\f 12+\de}$ weight instead of the $\ub^{\f12+\de}$ weight. Applying Proposition \ref{transport} to $|u|^{\f12+\de}\varpi^N\Om_\Ke\phi$ instead of $\ub^{\f12+\de}\varpi^N\Om_\Ke\phi$, we obtain the following analogue of \eqref{id.t.phi}:
\begin{equation*}
\begin{split}
&\||u|^{\frac 12+\de}\varpi^N\Om_\Ke^m\phi\|_{L^2(S_{u,\ub})}^2\\
=&\||u|^{\frac 12+\de}\varpi^N\Om_\Ke^m\phi\|_{L^2(S_{-\ub+C_R,\ub})}^2\\
&+2\int_{-\ub+C_R}^{u}\int_{S_{u',\ub}}|u'|^{1+2\de}\left(\langle \varpi^N\Om_\Ke^m\phi,(\nab_3(|u|^{\f 12+\de}\varpi^{N}\Om_\Ke^m))\phi \rangle_{\gamma}\right) du'\\
&+2\int_{-\ub+C_R}^{u}\int_{S_{u',\ub}}|u'|^{1+2\de}\left( \langle\varpi^N\Om_\Ke^m\phi,\varpi^{N}\Om_\Ke^m\nab_3\phi \rangle_{\gamma}\right) du'\\
&+\int_{-\ub+C_R}^{u}\int_{S_{u',\ub}} |u'|^{1+2\de}\trchb \varpi^{2N}\Om_\Ke^{2m}|\phi|_{\gamma}^2 du'.
\end{split}
\end{equation*}
We now compute
$$\nab_3(|u|^{\f 12+\de}\varpi^{N}\Om_\Ke^m) = N|u|^{\f 12+\de}\varpi^{N-1}\Om_\Ke^m\nab_3\varpi+m|u|^{\f 12+\de}\varpi^N\Om_\Ke^{m-1}\nab_3\Om_\Ke + \varpi^N\Om_\Ke^m \nab_3|u|^{\f 12+\de}.$$
Notice that because we have used the $|u|$ weight instead of the $\ub$ weight, we have an additional $\nab_3|u|^{\f 12+\de}$ as compared to Proposition~\ref{transport.3.2}. Nevertheless, since $\nab_3|u|^{\f12+\de}\leq 0$, \eqref{nab3.weight.good.bound} implies that
\begin{equation}\label{nab3.weight.good.bound.2}
\nab_3(|u|^{\f 12+\de}\varpi^{N}\Om_\Ke^m)\ls -(N |u|^{\f 12+\de}\Om_\Ke^{m+2}+ |u|^{\f 12+\de}\Om_\Ke^m).
\end{equation}
Using this, we can proceed as in the proof of Proposition \ref{transport.3.2} to obtain the desired conclusion. \qedhere
\end{proof}

We now move to the corresponding estimates which control tensors $\phi$ in terms of $\nab_4\phi$. We first have the following analogue of Proposition \ref{transport.3.1}: 
\begin{proposition}\label{transport.4.1}
Let $\phi$ be a tensor field of arbitrary rank tangential to the spheres $S_{u,\ub}$. Then $\phi$ satisfies the following estimate:
\begin{equation*}
\begin{split}
&\||u|^{\frac 12+\de}\varpi^N\phi\|_{L^2_uL^\infty_{\ub}L^2(S)}^2+N\||u|^{\frac 12+\de}\varpi^N\Om_\Ke\phi\|_{L^2_uL^2_{\ub}L^2(S)}^2\\
\ls &\||u|^{\frac 12+\de}\varpi^N\phi\|_{L^2_uL^2(S_{u,-u+C_R})}^2+\| |u|^{1+2\de}\varpi^{2N}\Om_\Ke^2\phi\nab_4\phi\|_{L^1_{\ub}L^1_uL^1(S)}.
\end{split}
\end{equation*}
Here (and in Propositions \ref{transport.4.2} and \ref{transport.4.3}), we simply write $\phi\nab_4\phi=\langle\phi,\nab_4\phi\rangle_\gamma$.
\end{proposition}
\begin{proof}
Applying the first identity in Proposition \ref{transport} to $|u|^{\frac 12+\de}\varpi^N\phi$, and using $\nab_4|u|^{\frac 12+\de}=0$, we obtain
\begin{equation}\label{id.transport.4.1.1}
\begin{split}
&\:\||u|^{\frac 12+\de}\varpi^N\phi\|_{L^2(S_{u,\ub})}^2\\
=&\:\||u|^{\frac 12+\de}\varpi^N\phi\|_{L^2(S_{u,-u+C_R})}^2\\
&\:+2\int_{-u+C_R}^{\ub}\int_{S_{u,\ub'}}|u|^{1+2\de}\Om^2 \langle\varpi^N\phi,N\varpi^{N-1}(\nab_4\varpi)\phi+\varpi^{N}\nab_4\phi\rangle_{\gamma} \,d\ub'\\
&\:+\int_{-u+C_R}^{\ub}\int_{S_{u',\ub}} |u|^{1+2\de}\trch\Om^2\varpi^{2N}|\phi|_{\gamma}^2\, d\ub'.
\end{split}
\end{equation}
By \eqref{varpi.34} and \eqref{Om.c}, we have
\begin{equation}\label{nab4.varpi.neg}
\nab_4\varpi\ls -1\leq 0.
\end{equation}
Therefore, the second term on the right hand side of \eqref{id.transport.4.1.1} has a favorable sign.
Moreover, using again \eqref{Om.c},
we can move the term in \eqref{id.transport.4.1.1} with a good sign to the left hand side to obtain
\begin{equation}\label{id.transport.4.1.2}
\begin{split}
&\:\||u|^{\frac 12+\de}\varpi^N\phi\|_{L^2(S_{u,\ub})}^2+N\int_{-u+C_R}^{\ub}\||u|^{\frac 12+\de}\varpi^N\Om_\Ke\phi\|_{L^2(S_{u,\ub'})}^2d\ub'\\
\ls &\:\||u|^{\frac 12+\de}\varpi^N\Om_\Ke\phi\|_{L^2(S_{u,-u+C_R})}^2+\int_{-u+C_R}^{\ub}\int_{S_{u,\ub'}} |u|^{1+2\de}\varpi^{2N}\Om_\Ke^2|\langle\phi,\nab_4\phi\rangle_\gamma|\, d\ub'\\
&\:+\int_{-u+C_R}^{\ub}\int_{S_{u',\ub}} |u|^{1+2\de}|\trch|\Om_\Ke^2\varpi^{2N}|\phi|_{\gamma}^2\, d\ub'.
\end{split}
\end{equation}
We now control the last term in \eqref{id.transport.4.1.2}. Using Proposition~\ref{Kerr.Ricci.bound}, we bound
\begin{equation}\label{transport.4.trch}
|\trch|\leq |(\trch)_\Ke|+|\widetilde{\trch}|\ls 1+|\tpH|.
\end{equation}
Hence,
\begin{equation}\label{id.transport.4.1.3}
\begin{split}
&\int_{-u+C_R}^{\ub}\int_{S_{u',\ub}} |u|^{1+2\de}|\trch|\Om_\Ke^2\varpi^{2N}|\phi|_{\gamma}^2 d\ub'\\
\ls &\int_{-u+C_R}^{\ub}\||u|^{\frac 12+\de}\varpi^N\Om_\Ke\phi\|_{L^2(S_{u,\ub'})}^2d\ub'+\int_{-u+C_R}^{\ub}\int_{S_{u',\ub}} |u|^{1+2\de}|\tpH|\Om_\Ke^2\varpi^{2N}|\phi|_{\gamma}^2 d\ub'.
\end{split}
\end{equation}
For $N$ sufficiently large, the first term on the right hand side of \eqref{id.transport.4.1.3} can be controlled by the second term on the left hand side of \eqref{id.transport.4.1.2}. Therefore, we have
\begin{equation}\label{id.transport.4.1.4}
\begin{split}
&\||u|^{\frac 12+\de}\varpi^N\phi\|_{L^2(S_{u,\ub})}^2+N\int_{-u+C_R}^{\ub}\||u|^{\frac 12+\de}\varpi^N\Om_\Ke\phi\|_{L^2(S_{u,\ub'})}^2d\ub'\\
\ls &\||u|^{\frac 12+\de}\varpi^N\Om_\Ke\phi\|_{L^2(S_{u,-u+C_R})}^2+\int_{-u+C_R}^{\ub} \int_{S_{u,\ub'}} |u|^{1+2\de}\varpi^{2N}\Om_\Ke^2|\langle\phi,\nab_4\phi\rangle_\gamma|\, d\ub'\\
&+\int_{-u+C_R}^{\ub} \int_{S_{u,\ub'}} |u|^{1+2\de}|\Om_\Ke^2\tpH|\varpi^{2N}|\phi|_{\gamma}^2 d\ub'.
\end{split}
\end{equation}
Finally, applying Gr\"onwall's inequality to \eqref{id.transport.4.1.4},
\begin{equation*}
\begin{split}
&\||u|^{\frac 12+\de}\varpi^N\phi\|_{L^2(S_{u,\ub})}^2+N\int_{-u+C_R}^{\ub}\||u|^{\frac 12+\de}\varpi^N\Om_\Ke\phi\|_{L^2(S_{u,\ub'})}^2d\ub'\\
\ls &\left(\||u|^{\frac 12+\de}\varpi^N\Om_\Ke\phi\|_{L^2(S_{u,-u+C_R})}^2+\int_{-u+C_R}^{\ub} |u|^{1+2\de}\varpi^{2N}\|\Om_\Ke^2\phi\nab_4\phi\|_{L^1(S_{u,\ub'})} d\ub'\right)\times e^{\int_{u+C_R}^{\ub}\|\Om_\Ke^2\tpH\|_{L^\i(S_{u,\ub})}\, d\ub'}.
\end{split}
\end{equation*}
By Corollary~\ref{Linfty},
$$\|\Om_\Ke^2\tpH\|_{L^1_{\ub}L^\i_uL^\i(S)}\ls \ep^{\f12}.$$
This implies
\begin{equation*}
\begin{split}
&\||u|^{\frac 12+\de}\varpi^N\phi\|_{L^2(S_{u,\ub})}^2+N\int_{-u+C_R}^{\ub}\||u|^{\frac 12+\de}\varpi^N\Om_\Ke\phi\|_{L^2(S_{u,\ub'})}^2d\ub'\\
\ls &\||u|^{\frac 12+\de}\varpi^N\Om_\Ke\phi\|_{L^2(S_{u,-u+C_R})}^2+\int_{-u+C_R}^{\ub} |u|^{1+2\de}\varpi^{2N}\|\Om_\Ke^2\phi\nab_4\phi\|_{L^1(S_{u,\ub'})} d\ub'.
\end{split}
\end{equation*}
The conclusion follows from first taking supremum in $\ub$ and then integrating the equation \eqref{id.transport.4.1.2} in $L^2$ in $u$.
\end{proof}

We also have the following analogue of Proposition \ref{transport.3.2} where we put an extra $\Om_\Ke$ in the norms. Notice that because of the additional good term that we obtain, we have a slightly stronger\footnote{Of course, at the same time, we also need the slightly stronger $\ub$-weighted norms on the right hand side to control the error terms.} estimate where we have the weight $\ub^{\frac 12+\de}$ instead of $|u|^{\frac 12+\de}$.
\begin{proposition}\label{transport.4.2}
Let $\phi$ be a tensor field of arbitrary rank tangential to the spheres $S_{u,\ub}$. Then $\phi$ satisfies the following estimate:
\begin{equation*}
\begin{split}
&\:\|\ub^{\frac 12+\de}\varpi^N\Om_\Ke\phi\|_{L^2_uL^\infty_{\ub}L^2(S)}^2+N\|\ub^{\frac 12+\de}\varpi^N\Om_\Ke^2\phi\|_{L^2_uL^2_{\ub}L^2(S)}^2+\|\ub^{\frac 12+\de}\varpi^N\Om_\Ke\phi\|_{L^2_uL^2_{\ub}L^2(S)}^2\\
\ls &\:\|\ub^{\frac 12+\de}\varpi^N\Om_\Ke\phi\|_{L^2_uL^2(S_{u,-u+C_R})}^2+\|\ub^{1+2\de}\varpi^{2N}\Om_\Ke^4\phi\nab_4\phi\|_{L^1_{\ub}L^1_uL^1(S)}.
\end{split}
\end{equation*}
\end{proposition}
\begin{proof}
As in the proof of Proposition \ref{transport.3.2}, we will prove a slightly more general statement that will be useful later (see Proposition \ref{transport.4.3}). Let $m\in [1,3]$. Applying the identity in Proposition \ref{transport} to $\ub^{\frac 12+\de}\varpi^N\Om_\Ke^m\phi$, we obtain
\begin{equation}\label{id.transport.4.2.1}
\begin{split}
&\:\|\ub^{\frac 12+\de}\varpi^N\Om_\Ke^m\phi\|_{L^2(S_{u,\ub})}^2\\
=&\:\|\ub^{\frac 12+\de}\varpi^N\Om_\Ke^m\phi\|_{L^2(S_{u,-u+C_R})}^2 +2\int_{-u+C_R}^{\ub}\int_{S_{u,\ub'}}\ub'^{1+2\de}\Om^2 \langle\varpi^N\Om_\Ke^m\phi,(\nab_4(\ub^{\f 12+\de}\varpi^N\Om_\Ke^m))\phi\rangle_{\gamma}\,d\ub'\\
&\:+2\int_{-u+C_R}^{\ub}\int_{S_{u,\ub'}}\ub'^{1+2\de}\Om^2 \langle\varpi^N\Om_\Ke^m\phi,\varpi^{N}\Om_\Ke^m\nab_4\phi \rangle_{\gamma}\, d\ub'\\
&\:+\int_{-u+C_R}^{\ub}\int_{S_{u',\ub}} \ub'^{1+2\de}\varpi^{2N}\trch\Om^2\Om_\Ke^{2m}|\phi|_{\gamma}^2\, d\ub'.
\end{split}
\end{equation}
We first consider the term where $\nab_4$ acts on the weights. Product rule implies that
\begin{equation}\label{nab4.weight.pr}
\nab_4(\ub^{\f 12+\de}\varpi^N\Om_\Ke^m) = N\ub^{\f 12+\de}\varpi^{N-1}\Om_\Ke^m\nab_4\varpi+m\ub^{\f 12+\de}\varpi^N\Om_\Ke^{m-1}\nab_4\Om_\Ke+(\nab_4\ub^{\frac 12+\de})\varpi^N\Om_\Ke^m \doteq I+II+III.
\end{equation}
We consider each of these terms.
By \eqref{nab4.varpi.neg},
$$\nab_4\varpi\ls -1\leq 0.$$
Since $e_4=\frac{1}{\Om^2}(\frac{\partial}{\partial \ub}+b^A\frac{\partial}{\partial\th^A})$, the difference of the vector fields $e_4$ and $(e_4)_\Ke$ can be computed as follows:
\begin{equation*}
\begin{split}
e_4-(e_4)_\Ke
=&\frac{\Om_\Ke^2-\Om^2}{\Om^2\Om_\Ke^2}(\frac{\partial}{\partial \ub}+(b_{\Ke})^A\frac{\partial}{\partial\th^A})+\frac{1}{\Om^2}(b^A-b^A_\Ke)\frac{\partial}{\partial\th^A}.
\end{split}
\end{equation*}
As a consequence, using Propositions~\ref{Om.Kerr.bounds}, \ref{Om.der.Kerr.bounds} and Corollary~\ref{Linfty}, we have
$$|\nab_4\Om_\Ke|\ls \Om_\Ke^{-1}.$$
After fixing $C_{R'}$ to be sufficiently large, we moreover have
$$\nab_4\Om_\Ke\ls -\Om_\Ke^{-1}\leq 0\quad\mbox{for }u+\ub\geq C_{R'}.$$
Therefore, the term $I$ in \eqref{nab4.weight.pr} is negative everywhere (with quantitative bounds) and the term $II$ in \eqref{nab4.weight.pr} is negative (again with quantitative bounds) in the region $\{u+\ub\geq C_{R'}\}$. Moreover, as in the proof of Proposition \ref{transport.3.2}, by Proposition~\ref{Om.Kerr.bounds}, we have
$$1\ls \Om_\Ke\ls 1$$
for $-u+C_R\leq \ub'\leq -u+C_{R'}$, where the implicit constants depend only on $M$, $a$, $C_R$ and $C_{R'}$. Hence by choosing $N$ to be sufficiently large, one can absorb the term $II$ by the term $I$ in \eqref{nab4.weight.pr} in the region $\{-u+C_R\leq \ub'\leq -u+C_{R'}\}$, which then leads to a bound
$$I+II \ls - (N \ub^{\f 12+\de}\varpi^N\Om_\Ke^{m+2} + \ub^{\f 12+\de}\varpi^N\Om_\Ke^m)$$
that holds everywhere. On the other hand, while the contribution from $\nab_4\ub^{\frac 12+\de}$ has a bad sign (i.e.~it is positive), it can be controlled by
$$|\nab_4\ub^{\frac 12+\de}|\ls \Om_{\Ke}^{-2}\ub^{-\f 12+\de}.$$
Therefore, after choosing $|u_f|$ to be sufficiently large (which then guarantees $\ub^{-\f 12+\de} \ll \ub^{\frac 12+\de}$ when $u+\ub\geq C_R$), the term $III$ can be absorbed by the term $II$ in \eqref{nab4.weight.pr}. Putting together all these observations, we thus obtain the following bound everywhere in $\{u+\ub\geq C_R\}$:
\begin{equation}\label{nab4.weight.good.bound}
\nab_4(\ub^{\f 12+\de}\varpi^N\Om_\Ke^m)\ls - (N \ub^{\f 12+\de}\varpi^N\Om_\Ke^{m+2} + \ub^{\f 12+\de}\varpi^N\Om_\Ke^m).
\end{equation}
Therefore, after using \eqref{Om.c}, we can move the terms in \eqref{id.transport.4.2.1} with a good sign to the left hand side to obtain
\begin{equation}\label{id.transport.4.2.2}
\begin{split}
&\|\ub^{\frac 12+\de}\varpi^N\Om_\Ke^m\phi\|_{L^2(S_{u,\ub})}^2+N\int_{-u+C_R}^{\ub}\|\ub'^{\frac 12+\de}\varpi^N\Om_\Ke^{m+1}\phi\|_{L^2(S_{u,\ub'})}^2d\ub'\\
&+\int_{-u+C_{R}}^{\ub}\|\ub'^{\frac 12+\de}\varpi^N\Om_\Ke^m\phi\|_{L^2(S_{u,\ub'})}^2d\ub'\\
\ls &\|\ub^{\frac 12+\de}\varpi^N\Om_\Ke^m\phi\|_{L^2(S_{u,-u+C_R})}^2+\int_{-u+C_R}^{\ub} \ub'^{1+2\de}\varpi^{2N}\|\Om_\Ke^{2m+2}\phi\nab_4\phi\|_{L^1(S_{u,\ub'})} d\ub'\\
&+\int_{-u+C_R}^{\ub} \ub'^{1+2\de}\varpi^{2N}\|\Om_\Ke^{2m+2}\trch\phi\phi\|_{L^1(S_{u,\ub'})} d\ub'.
\end{split}
\end{equation}
To handle the last term on the right hand side of \eqref{id.transport.4.2.2}, we apply \eqref{transport.4.trch} in the proof of Proposition \ref{transport.4.1} to obtain
\begin{equation}\label{id.transport.4.2.2.1}
\begin{split}
&\int_{-u+C_R}^{\ub}\ub'^{1+2\de}\varpi^{2N}\|\Om_\Ke^{2m+2}\trch\phi\phi\|_{L^1(S_{u,\ub'})}  d\ub'\\
\ls &\int_{-u+C_R}^{\ub}\|\ub'^{\frac 12+\de}\varpi^N\Om_\Ke^{m+1}\phi\|_{L^2(S_{u,\ub'})}^2d\ub'+\int_{-u+C_R}^{\ub}\|\Om_\Ke^2\tpH\|_{L^{\i}(S_{u,\ub'})}\|\ub'^{\frac 12+\de}\varpi^N\Om_\Ke^{m}\phi\|_{L^2(S_{u,\ub'})}^2d\ub'.
\end{split}
\end{equation}
After choosing $N$ to be sufficiently large, the first term in \eqref{id.transport.4.2.2.1} can be controlled by the second term on the left hand side of \eqref{id.transport.4.2.2}. This term can therefore be absorbed. Moreover, since $\|\Om_\Ke^2\tpH\|_{L^{\i}_u L^{\i}_{\ub}L^{\i}(S)}\ls \ep^{\f12}$ according to Corollary~\ref{Linfty}, the second term on the right hand side of \eqref{id.transport.4.2.2.1} can be absorbed by the third term on the left hand side of \eqref{id.transport.4.2.2} for $\ep_0$ (and hence $\ep$) sufficiently small. We therefore obtain
\begin{equation}\label{id.transport.4.2.3}
\begin{split}
&\|\ub^{\frac 12+\de}\varpi^N\Om_\Ke^m\phi\|_{L^2(S_{u,\ub})}^2+N\int_{-u+C_R}^{\ub}\|\ub'^{\frac 12+\de}\varpi^N\Om_\Ke^{m+1}\phi\|_{L^2(S_{u,\ub'})}^2\,d\ub'\\
&+\int_{-u+C_{R}}^{\ub}\|\ub'^{\frac 12+\de}\varpi^N\Om_\Ke^m\phi\|_{L^2(S_{u,\ub'})}^2 \,d\ub'\\
\ls &\|\ub^{\frac 12+\de}\varpi^N\Om_\Ke^m\phi\|_{L^2(S_{u,-u+C_R})}^2+\int_{-u+C_R}^{\ub} \ub'^{1+2\de}\varpi^{2N}\|\Om_\Ke^{2m+2}\phi\nab_4\phi\|_{L^1(S_{u,\ub'})}\, d\ub'.
\end{split}
\end{equation}
Now take $m=1$. The conclusion followed from first taking supremum in $\ub$ and then integrating in $L^2$ in $u$ in equation \eqref{id.transport.4.2.3}.
\end{proof}
In the proof of the above proposition, we first prove an estimate on a fixed $u=\mbox{constant}$ hypersurface and then integrate in $u$ to obtain the desired bound. In the following proposition, we also give a version of the estimate without integrating in $u$. This can be viewed as an analogue of Proposition \ref{transport.3.3}.
\begin{proposition}\label{transport.4.3}
Let $\phi$ be a tensor field of arbitrary rank tangential to the spheres $S_{u,\ub}$. Then for $m=1$, $2$ or $3$, $\phi$ satisfies the following estimate on any fixed $u=\mbox{constant}$ hypersurface:
\begin{equation*}
\begin{split}
&\|\ub^{\frac 12+\de}\varpi^N\Om_\Ke^m\phi\|_{L^\infty_{\ub}L^2(S)}^2+N\|\ub^{\frac 12+\de}\varpi^N\Om_\Ke^{m+1}\phi\|_{L^2_{\ub}L^2(S)}^2+\|\ub^{\frac 12+\de}\varpi^N\Om_\Ke^m\phi\|_{L^2_{\ub}L^2(S)}^2\\
\ls &\|\ub^{\frac 12+\de}\varpi^N\Om_\Ke^m\phi\|_{L^2(S_{u,-u+C_R})}^2+\|\ub^{1+2\de}\varpi^{2N}\Om_\Ke^{2m+2}\phi\nab_4\phi\|_{L^1_{\ub}L^1(S)}.
\end{split}
\end{equation*}
\end{proposition}
\begin{proof}
This is a direct consequence of equation \eqref{id.transport.4.2.3} in the proof of Proposition \ref{transport.3.2}.
\end{proof}

\section{The reduced schematic equations}\label{sec.rse}

Recall that we continue to work under the setting described in Remark~\ref{rmk:setup}. 

In this section, we introduce our notion of \emph{reduced schematic equations}. These are expressions which further simplify the schematic equations introduced in Section \ref{schnoteqn} and will be used for the differences of the geometric quantities and their derivatives. The main additional simplification is that {\bf\emph{all factors bounded in $L^\infty_u L^\infty_{\ub}L^\infty(S)$ by a constant depending only on $M$ and $a$ will be suppressed}}. Reduced schematic equations will be written with the symbol $\eqrs$ and a more detailed discussion of the conventions associated to reduced schematic equations will be given in \textbf{Section~\ref{sec.rse.conv}}.

After introducing the conventions, the main goal of this section is to derive reduced schematic equations for the angular covariant derivatives (with respect to $(\mathcal U_{\ub_f},g)$) of the difference of the metric components, the Ricci coefficients and the renormalised null curvature components. 

In \textbf{Section~\ref{sec.rse.prelim}}, we will derive the commutation formula for $[\nab_4,\nab]$ and $[\nab_3,\nab]$ (Proposition~\ref{commute}) and the formulae for the differences of the connection coefficients $\nab-\nab_\Ke$ (Proposition \ref{nab.diff}), $\nab_3-(\nab_3)_\Ke$ (Proposition~\ref{3.diff.prop}) and $\nab_4-(\nab_4)_\Ke$ (Proposition~\ref{4.diff.prop}). These computations will then allow us to derive general reduced schematic equations for the difference of the geometric quantities satisfying a $\nab_3$ equation (Proposition~\ref{schematic.3}) or a $\nab_4$ equation (Proposition~\ref{schematic.4}).

To derive the reduced schematic equations, it is useful to define four types of inhomogeneous terms which arise in most of the equations for the differences of the geometric quantities. These inhomogeneous terms will be introduced in \textbf{Section~\ref{sec.rse.inho}}. We will denote these terms\footnote{In fact, we will also need the versions $\mathcal F_1'$, $\mathcal F_2'$, $\mathcal F_3'$ and $\mathcal F_4'$ which contain only a subset of the terms in $\mathcal F_1$, $\mathcal F_2$, $\mathcal F_3$ and $\mathcal F_4$. The precise structure of these terms will be used to obtain some refined estimates.} by $\mathcal F_1$, $\mathcal F_2$, $\mathcal F_3$ and $\mathcal F_4$.

In the remainder of this section, we then proceed to the derivation of the reduced schematic equations, showing that indeed only $\mathcal F_1$, $\mathcal F_2$, $\mathcal F_3$ and $\mathcal F_4$ show up as inhomogeneous terms. \textbf{Section~\ref{sec.rse.metric}} will be dedicated to the reduced schematic equations for $\widetilde{\gamma}$ and $\widetilde{\log \Om}$, $\tb$ and their angular covariant derivatives. In \textbf{Section~\ref{sec.rse.Ricci}} and \textbf{Section~\ref{sec.rse.Ricci.add}}, we derive the reduced schematic equations for the Ricci coefficients. Finally, in \textbf{Section~\ref{sec.rse.curv}}, we obtain the reduced schematic equations for the renormalised null curvature components.

\subsection{Conventions for the reduced schematic equations}\label{sec.rse.conv}

We now formally introduce our notion of reduced schematic equations, already described in Section~\ref{baridiaeisagwgns}. We will use the convention $\eqrs$ to denote that the ``equation'' holds only in the reduced schematic sense. These reduced schematic equations will be equations for the difference quantities, i.e.~the $\,\,\widetilde{ }\,\,$ quantities; see Section~\ref{sec.identification}. These include differences of metric components, Ricci coefficients, renormalised curvature components, quantities in \eqref{top.quantities.def}, and their derivatives.

The conventions that we will use are as follows:
\begin{itemize}
\item (Conventions from the schematic equations) All the conventions introduced for the schematic equations in Section~\ref{schnot} will be used for the reduced schematic equations.
\item (Conventions regarding taking differences) All the conventions introduced in Section~\ref{sec.identification} regarding taking differences will be used for the reduced schematic equations.
\item (Terms bounded in $L^\infty_u L^\infty_{\ub}L^\infty(S)$ are suppressed) On the right hand side of the reduced schematic equations, {\bf\emph{we will use the convention that all factors bounded in $L^\infty_u L^\infty_{\ub}L^\infty(S)$ by a constant depending only on $M$ and $a$ will not be written}}. These include terms associated to the geometric quantities of the background Kerr spacetime, as well as terms which are bounded by $\ep^{\f12}$ in $L^\infty$ using Corollary~\ref{Linfty}.
\item (Reduced schematic equations \underline{cannot} be differentiated) We emphasise explicitly that since we suppress the terms that are bounded in $L^\infty_u L^\infty_{\ub}L^\infty(S)$ in the reduced schematic equations, one is \underline{not} allowed to differentiate the reduced schematic equations. This stands in contrast to the schematic equations introduced in Section~\ref{schnot}; see Remark~\ref{rmk:diff.sch}.
\end{itemize}

To illustrate the conventions above, we already give an example here. Consider the schematic equation \eqref{etab.S.2}. We now take the difference of the $\nab_3\etab$ equation and the $(\nab_3)_\Ke \etab_\Ke$ equation. First, we have $\nab_3\etab-(\nab_3)_\Ke \etab_\Ke = \nab_3 \widetilde{\etab} + \widetilde{\nab_3}\etab_{\Ke}$. The first term will be the main term of our new equation, and will be put on the left hand side. The second term will be considered as an error term, and will be put on the right hand side. Now when taking the difference of $\sum_{i_1+i_2\leq 1} \psi^{i_1}\nab^{i_2}\psi_{\Hb}$ and $\sum_{i_1+i_2\leq 1} (\psi^{i_1}\nab^{i_2}\psi_{\Hb})_\Ke$, we have the differences of $\psi^{i_1}$ and $\nab^{i_2}\psi_{\Hb}$, and additionally we also have the difference $\tg$, which appears because our conventions for schematic equations (cf.~Section~\ref{schnot}) allow for arbitrary contractions with respect to $\gamma$. By Corollary~\ref{Linfty}, $\tg$ is bounded in $L^\infty$ and can be suppressed in the reduced schematic equations. We will therefore only keep track of it if it is multiplied by background Kerr quantities. In other words, we obtain the following reduced schematic equation:
$$\nab_3\widetilde{\etab} \eqrs \widetilde{\nab_3}\etab_{\Ke}+\sum_{i_1+i_2\leq 1} \tp^{i_1}(\nab^{i_2}\psi_{\Hb})_\Ke+\sum_{i_1+i_2\leq 1} \psi_\Ke^{i_1}\widetilde{\nab^{i_2}\psi_{\Hb}}+\sum_{i_1+i_2\leq 1} \tp^{i_1}\widetilde{\nab^{i_2}\psi_{\Hb}}+ \tg \sum_{i_1+i_2\leq 1} \psi_\Ke^{i_1}(\nab^{i_2}\psi_{\Hb})_\Ke.$$
Recalling \eqref{nab3.def}, and using that $\eta$ and $\chib$ can be schematically represented by $\psi$ and $\psi_{\Hb}$ respectively, we have\footnote{Note already that the reduced schematic equation for $\widetilde{\nab_4}$ is slightly more complicated than $\widetilde{\nab_3}$; see the proof of Proposition~\ref{4.diff.prop}.} $\widetilde{\nab_3}\eta_{\Ke} \eqs (\tpHb + (\psi_{\Hb})_\Ke \tg) \psi_\Ke$. Thus the reduced schematic equation above can be further simplified as follows:
$$\nab_3\widetilde{\etab} \eqrs \sum_{i_1+i_2\leq 1} \tp^{i_1}(\nab^{i_2}\psi_{\Hb})_\Ke+\sum_{i_1+i_2\leq 1} \psi_\Ke^{i_1}\widetilde{\nab^{i_2}\psi_{\Hb}} +\sum_{i_1+i_2\leq 1} \tp^{i_1}\widetilde{\nab^{i_2}\psi_{\Hb}}+ \tg \sum_{i_1+i_2\leq 1} \psi_\Ke^{i_1}(\nab^{i_2}\psi_{\Hb})_\Ke.$$
Using Corollary~\ref{Linfty} and information about the background Kerr spacetime, we know that $|\tp|,\,|\psi_\Ke|\ls 1$ and $|(\psi_{\Hb})_{\Ke}|,\,|(\nab\psi_{\Hb})_{\Ke}| \ls \Om_\Ke^2$. For the $\widetilde{\nab\psi_{\Hb}}$ term, we can write $\widetilde{\nab\psi_{\Hb}} \eqs \widetilde{\nab} (\psi_{\Hb})_\Ke + \nab \tpHb$. By \eqref{nab.def}, $\widetilde{\nab}$ can be expressed as follows:
$$\widetilde{\nab}\phi \eqrs (\slashed\Gamma - \slashed\Gamma_\Ke)\phi$$
We will show below that $(\slashed\Gamma - \slashed\Gamma_\Ke)$ is bounded by $\nab\tg$ (see Proposition~\ref{nab.diff}). In particular, together with the considerations above, this implies
$$\widetilde{\nab\psi_{\Hb}} \eqrs \nab\tg (\psi_{\Hb})_\Ke + \nab\tpHb \eqrs \Om_\Ke^2\nab\tg + \nab\tpHb.$$ 
Combining all these facts, we have the following reduced schematic equation:
$$\nab_3\widetilde{\etab} \eqrs \sum_{i\leq 1}\Om_\Ke^2\nab^i\tg + \Om_\Ke^2\tp + \sum_{i\leq 1} \nab^i\tpHb.$$
In the remainder of the section, we will derive reduced schematic equations for higher derivatives of the difference quantities. In particular, one will encounter more terms which are not in $L^\i$ and the derivation will be more involved. Nevertheless, the basic principle is as we have illustrated above.

\subsection{Preliminaries}\label{sec.rse.prelim}

As mentioned above, the main goal of this subsection is to prove Propositions~\ref{schematic.3} and \ref{schematic.4} below, which give a general recipe to obtain equations for the difference of the geometric quantities from the corresponding equations on $(\mathcal U_{\ub_f},g)$ and the background Kerr spacetime.

To begin, we need some formulae for the commutators $[\nab_4,\nab]$, $[\nab_3,\nab]$ and $[\nab_3,\nab_4]$. We have the following formula from Lemma 7.3.3 in \cite{CK}:
\begin{proposition}\label{commute}
The commutator $[\slashed{\nabla}_4,\slashed{\nabla}]$ acting on a rank-$r$ covariant S-tensor is given by
\begin{equation*}
 \begin{split}
[\slashed{\nabla}_4,\slashed{\nabla}_B]\phi_{A_1...A_r}=&\:(\etab_B+\zeta_B)\slashed{\nabla}_4\phi_{A_1...A_r}-(\gamma^{-1})^{CD}\chi_{BD}\slashed{\nabla}_C\phi_{A_1...A_r} \\
&+\sum_{i=1}^r ((\gamma^{-1})^{CD}\chi_{A_iB}\etab_{D}-(\gamma^{-1})^{CD}\chi_{BD}\etab_{A_i}+\in_{A_i}{ }^C{ }^*\beta_B)\phi_{A_1...\hat{A_i}C...A_r}.
 \end{split}
\end{equation*}
Similarly, the commutator $[\nab_3,\nab]$ acting on a rank-$r$ covariant S-tensor is given by\footnote{Notice that in our choice of gauge, we have $\eta_B-\zeta_B=0$, i.e.~the first term is in fact absent. Nevertheless, we keep this expression in the following form for easy comparison with Lemma 7.3.3 in \cite{CK}.}
\begin{equation*}
 \begin{split}
[\slashed{\nabla}_3,\slashed{\nabla}_B]\phi_{A_1...A_r}=&\:(\eta_B-\zeta_B)\slashed{\nabla}_3\phi_{A_1...A_r}-(\gamma^{-1})^{CD}\chib_{BD}\slashed{\nabla}_C\phi_{A_1...A_r} \\
&+\sum_{i=1}^r ((\gamma^{-1})^{CD}\chib_{A_iB}\eta_{D}-(\gamma^{-1})^{CD}\chib_{BD}\eta_{A_i}-\in_{A_i}{ }^C{ }^*\betab_B)\phi_{A_1...\hat{A_i}C...A_r}.
 \end{split}
\end{equation*}
Moreover, the commutator $[\nab_3,\nab_4]$ acting on a rank-$r$ covariant $S$-tensor is given by\footnote{Notice that with our choice of gauge, we have $\om=0$. Again, we keep $\om$ for easy comparison with \cite{CK}.} \footnote{Note that while $\sigma$ (which appears in this equation) is not explicitly controlled by our energies, we can nonetheless estimate it using \eqref{sigmac.def} and the bounds for $\sigmac$, $\chih$ and $\chibh$.}
\begin{equation*}
\begin{split}
[\nab_3,\nab_4]\phi_{A_1...A_r}=&\:2\om \nab_3 \phi_{A_1...A_r}+2\omb \nab_4\phi_{A_1...A_r}+(\gamma^{-1})^{BC}(\eta_B-\etab_B)\nab_C\phi_{A_1...A_r}\\
&+\sum_{i=1}^r 2(\gamma^{-1})^{BC}(\etab_{A_i}\eta_B-\eta_{A_i}\etab_B+\in_{A_iB}\sigma)\phi_{A_1...\hat{A}_iC...A_r}.
\end{split}
\end{equation*}
\end{proposition}

By induction and the schematic Codazzi equations (cf.~\eqref{Codazzi.S.1} and \eqref{Codazzi.S.2})
\begin{equation}\label{sch.Codazzi}
\beta\eqs\slashed{\nabla}\chi+\psi\chi,\quad
\betab\eqs\slashed{\nabla}\chib+\psi\chib,
\end{equation}
we obtain the following schematic formula for repeated commutations (see\footnote{While the gauge in the present paper is different from that in \cite{LR} (and as a consequence the analogue of Proposition~\ref{commute} is also different), the derivation of Proposition~\ref{repeated.com} is nonetheless completely analogous.} \cite{LR}):
\begin{proposition}\label{repeated.com}
Suppose $\slashed{\nabla}_4\phi=F_0$. Let $\slashed{\nabla}_4\slashed{\nabla}^i\phi=F_i$.
Then
\begin{equation*}
\begin{split}
F_i\eqs &\sum_{i_1+i_2+i_3=i}\slashed{\nabla}^{i_1}\psi^{i_2}\slashed{\nabla}^{i_3} F_0+\sum_{i_1+i_2+i_3+i_4=i}\slashed{\nabla}^{i_1}\psi^{i_2}\slashed{\nabla}^{i_3}\chi\slashed{\nabla}^{i_4} \phi.\\
\end{split}
\end{equation*}
Similarly, suppose $\slashed{\nabla}_3\phi=G_{0}$. Let $\slashed{\nabla}_3\slashed{\nabla}^i\phi=G_{i}$.
Then
\begin{equation*}
\begin{split}
G_{i}\eqs &\sum_{i_1+i_2+i_3=i}\slashed{\nabla}^{i_1}\psi^{i_2}\slashed{\nabla}^{i_3} G_{0}+\sum_{i_1+i_2+i_3+i_4=i}\slashed{\nabla}^{i_1}\psi^{i_2}\slashed{\nabla}^{i_3}\underline{\chi}\slashed{\nabla}^{i_4} \phi.
\end{split}
\end{equation*}
\end{proposition}

Next, we turn to the computations for the differences of the connections $\nab$, $\nab_3$ and $\nab_4$. This will then allow us to take the difference of the equations on $(\mathcal U_{\ub_f},g)$ and that on the background Kerr spacetime and thus to derive the reduced schematic equations. First, we show that the difference of the Christoffel symbols on the spheres can be expressed in terms of covariant derivatives of the difference of the metrics:
\begin{proposition}\label{nab.diff}
The following formula holds:
\begin{equation*}
\begin{split}
(\slashed{\Gamma}-\slashed{\Gamma}_\Ke)^C_{AB}=\f12 (\gamma^{-1}_\Ke)^{CD}\big(\nab_A(\gamma-\gamma_\Ke)_{BD}+\nab_B (\gamma-\gamma_\Ke)_{AD}-\nab_D(\gamma-\gamma_\Ke)_{AB}\big).
\end{split}
\end{equation*}
\end{proposition}
\begin{proof}
Using the fact that $(\nab_\Ke) \gamma_\Ke=0$, we obtain
\begin{equation*}
\begin{split}
\nab_A(\gamma_{\Ke})_{BC}=&\nab_A(\gamma_\Ke)_{BC}-(\nab_\Ke)_A (\gamma_\Ke)_{BC}\\
=&(\f{\rd}{\rd\th^A} (\gamma_\Ke)_{BC}-\slashed{\Gamma}^D_{AB}(\gamma_\Ke)_{CD}-\slashed{\Gamma}^D_{AC}(\gamma_\Ke)_{BD})\\
&-(\f{\rd}{\rd\th^A} (\gamma_\Ke)_{BC}-(\slashed{\Gamma}_\Ke)^D_{AB}(\gamma_\Ke)_{CD}-(\slashed{\Gamma}_\Ke)^D_{AC}(\gamma_\Ke)_{BD})\\
=&-(\slashed{\Gamma}-\slashed{\Gamma}_\Ke)^D_{AB}(\gamma_\Ke)_{CD}-(\slashed{\Gamma}-\slashed{\Gamma}_\Ke)^D_{AC}(\gamma_\Ke)_{BD}
\end{split}
\end{equation*}
This implies
\begin{equation*}
\begin{split}
(\slashed{\Gamma}-\slashed{\Gamma}_\Ke)^C_{AB}
=&-\f12 (\gamma^{-1}_\Ke)^{CD}(\nab_A(\gamma_\Ke)_{BD}+\nab_B(\gamma_\Ke)_{AD}-\nab_D (\gamma_K)_{AB}).
\end{split}
\end{equation*}
Using now the fact that $\nab \gamma=0$, we thus obtain
$$(\slashed{\Gamma}-\slashed{\Gamma}_\Ke)^C_{AB}=\f12 (\gamma^{-1}_\Ke)^{CD}\big(\nab_A(\gamma-\gamma_\Ke)_{BD}+\nab_B (\gamma-\gamma_\Ke)_{AD}-\nab_D(\gamma-\gamma_\Ke)_{AB}\big).$$
\end{proof}
We now turn to deriving the general formulae that will be used to obtain reduced schematic equations. First, given $\nab_3$ equations for a tensor $\phi$ on both $(\mathcal U_{\ub_f},g)$ and the background Kerr spacetime, we derive the $\nab_3$ equation for the difference $\phi-\phi_\Ke$. The following proposition is an immediate consequence of \eqref{nab3.def}:
\begin{proposition}\label{3.diff.prop}
Let $\phi$ be a rank-$r$ covariant $S$-tensor and $\phi_\Ke$ be a corresponding tensor on background Kerr spacetime. Suppose
$$\nab_3\phi=G,\quad (\nab_3)_\Ke\phi_\Ke=G_\Ke.$$
Then the difference $\phi-\phi_\Ke$ satisfies the following equation:
\begin{equation*}
\begin{split}
&\:\nab_3(\phi-\phi_\Ke)_{A_1...A_r}\\
=&\:G_{A_1...A_r}-(G_\Ke)_{A_1...A_r}-\sum_{i=1}^r ((\gamma^{-1})^{BC}\chib_{CA_i}-(\gamma_\Ke^{-1})^{BC}(\chib_\Ke)_{CA_i})(\phi_\Ke)_{A_1...\hat{A}_iB...A_r}.
\end{split}
\end{equation*}
\end{proposition}
Using the commutation formula Proposition \ref{commute} (or more specifically, its schematic version in Proposition \ref{repeated.com}), this gives the following reduced schematic $\nab_3$ equation for the angular covariant derivatives of the difference quantities:
\begin{proposition}\label{schematic.3}
Let $\phi$ be a rank-$r$ covariant $S$-tensor and $\phi_\Ke$ be a corresponding tensor on background Kerr spacetime. Suppose
$$\nab_3\phi=G,\quad (\nab_3)_\Ke\phi_\Ke=G_\Ke.$$
Then the following reduced schematic equation holds for $i\leq 3$:
\begin{equation}\label{schematic.3.main}
\begin{split}
&\:\nab_3\nab^i(\phi-\phi_\Ke)\\
\eqrs&\:\sum_{i_1+i_2\leq i}(1+\nab^{i_1}\tg+\nab^{i_1-1}\tp)\nab^{i_2}(G-G_\Ke)\\
&\:+\sum_{i_1+i_2+i_3\leq i}(1+\nab^{i_1}\tg+\nab^{i_1-1}\tp)(\Om_\Ke^2\nab^{i_2}\tg+\nab^{i_2}\tpHb)\nab^{i_3}\phi_\Ke\\
&\:+\sum_{i_1+i_2+i_3+i_4\leq i}(1+\nab^{i_1}\tg+\nab^{i_1-1}\tp)(\Om_\Ke^2+\Om_\Ke^2\nab^{i_2}\tg+\nab^{i_2}\tpHb)\nab^{i_3}(\phi-\phi_\Ke).
\end{split}
\end{equation}
Moreover, if $\phi$ is a scalar, then the second term on the right hand side is absent, i.e.
\begin{equation*}
\begin{split}
&\nab_3\nab^i(\phi-\phi_\Ke)\\
\eqrs&\sum_{i_1+i_2\leq i}(1+\nab^{i_1}\tg+\nab^{i_1-1}\tp)\nab^{i_2}(G-G_\Ke)\\
&+\sum_{i_1+i_2+i_3\leq i}(1+\nab^{i_1}\tg+\nab^{i_1-1}\tp)(\Om_\Ke^2+\Om_\Ke^2\nab^{i_2}\tg+\nab^{i_2}\tpHb)\nab^{i_3}(\phi-\phi_\Ke).
\end{split}
\end{equation*}
\end{proposition}
\begin{proof}
We combine the formula in Proposition \ref{3.diff.prop} for the derivation of the difference equation and the formula in Proposition \ref{repeated.com} for repeated commutation with $\nab$. According to Proposition \ref{repeated.com}, there are two terms in the equation for $\nab_3\nab^i(\phi-\phi_\Ke)$, namely,
\begin{equation}\label{schematic.3.1}
\sum_{i_1+i_2+i_3\leq i}\nab^{i_1}\psi^{i_2}\nab^{i_3}\nab_3(\phi-\phi_\Ke) 
\end{equation}
and 
\begin{equation}\label{schematic.3.2}
\sum_{i_1+i_2+i_3+i_4\leq i}\nab^{i_1}\psi^{i_2}\nab^{i_3}\psi_{\Hb}\nab^{i_4}(\phi-\phi_\Ke).
\end{equation}
Notice that for $i\leq 2$, we have
\begin{equation}\label{d2p.sch}
\nab^i\psi=\nab^i(\psi-\psi_\Ke)+\nab^i\psi_\Ke\eqrs\nab^i\tp+\nab^i\tg+1.
\end{equation}
The last equality (as a reduced schematic equation) in \eqref{d2p.sch} holds because $|(\nab_\Ke)^i\psi_\Ke|\ls 1$ and according to Proposition \ref{nab.diff}, the difference between the connections $\nab$ and $\nab_\Ke$ can be expressed as the derivative (with respect to the connection $\nab$) of the difference of the metrics $\gamma$ and $\gamma_\Ke$ (and hence bounded in $L^\infty$ by Corollary~\ref{Linfty}).

Therefore, for $i\leq 3$, the term \eqref{schematic.3.1} can be expressed as follows:
\begin{equation*}
\begin{split}
&\sum_{i_1+i_2+i_3\leq i}\nab^{i_1}\psi^{i_2}\nab^{i_3}\nab_3(\phi-\phi_\Ke)\\
\eqrs &\sum_{i_1\leq i}\nab^{i_1}\nab_3(\phi-\phi_\Ke)+\sum_{i_1+i_2\leq i-1}\nab^{i_1}(\tg,\tp)\nab^{i_2}\nab_3(\phi-\phi_\Ke).
\end{split}
\end{equation*}
Notice that we have used the fact that $\psi$, $\tp$, $\tg$, $\nab\tg$ are bounded in $L^\i$ here (by Corollary~\ref{Linfty}) to deal with the case $i_2>1$. Now, we combine this with Proposition~\ref{3.diff.prop} to obtain
\begin{equation*}
\begin{split}
&\sum_{i_1+i_2+i_3\leq i}\nab^{i_1}\psi^{i_2}\nab^{i_3}\nab_3(\phi-\phi_\Ke)\\
\eqrs &\sum_{i_1\leq i}\nab^{i_1}(G-G_\Ke)+\sum_{i_1+i_2\leq i-1}\nab^{i_1}(\tg,\tp)\nab^{i_2}(G-G_\Ke)+\sum_{i_1+i_2+i_3\leq i}\nab^{i_1}\tg\nab^{i_2}(\psi_{\Hb})_\Ke\nab^{i_3}\phi_\Ke\\
&+\sum_{i_1+i_2+i_3\leq i}\nab^{i_1}g_\Ke\nab^{i_2}\tpHb\nab^{i_3}\phi_\Ke+\sum_{i_1+i_2+i_3\leq i}\nab^{i_1}\tg\nab^{i_2}\tpHb\nab^{i_3}\phi_\Ke\\
&+\sum_{i_1+i_2+i_3+i_4\leq i-1}\nab^{i_1}(\tg,\tp)\nab^{i_2}\tg\nab^{i_3}(\psi_{\Hb})_\Ke\nab^{i_4}\phi_\Ke+\sum_{i_1+i_2+i_3+i_4\leq i-1}\nab^{i_1}(\tg,\tp)\nab^{i_2}g_\Ke\nab^{i_3}\tpHb\nab^{i_4}\phi_\Ke\\
&+\sum_{i_1+i_2+i_3+i_4\leq i-1}\nab^{i_1}(\tg,\tp)\nab^{i_2}\tg\nab^{i_3}\tpHb\nab^{i_4}\phi_\Ke.
\end{split}
\end{equation*}
Again, we have used the boundedness of $\psi$ in $L^\i$. This expression can be simplified using the reduced schematic expressions
\begin{equation}\label{nabgK.nabpsiHbK}
\nab^i g_\Ke\eqrs\sum_{i_1\leq i}\nab^{i_1}\tg+1,\quad \nab^i (\psi_{\Hb})_\Ke\eqrs\sum_{i_1\leq i}\Om_\Ke^2\nab^{i_1}\tg+\Om_\Ke^2,
\end{equation}
which hold as a consequence of \eqref{nab.def}, \eqref{def.slashed.Gamma}, Propositions~\ref{Kerr.der.Ricci.bound} and \ref{nab.diff}.
More precisely, we obtain
\begin{equation}\label{schematic.3.3}
\begin{split}
&\sum_{i_1+i_2+i_3\leq i}\nab^{i_1}\psi^{i_2}\nab^{i_3}\nab_3(\phi-\phi_\Ke)\\
\eqrs &\sum_{i_1+i_2\leq i}(1+\nab^{i_1}\tg+\nab^{i_1-1}\tp)\nab^{i_2}(G-G_\Ke)\\
&+\sum_{i_1+i_2+i_3\leq i}(1+\nab^{i_1}\tg+\nab^{i_1-1}\tp)(\Om_\Ke^2\nab^{i_2}\tg+\nab^{i_2}\tpHb)\nab^{i_3}\phi_\Ke,
\end{split}
\end{equation}
where we have again used the fact that $\psi$, $\tp$, $\tg$ and $\nab\tg$ are bounded in $L^{\i}$.

We now turn to the term \eqref{schematic.3.2}. The factor $\nab^{i_1}\psi^{i_2}$ can be expressed using \eqref{d2p.sch}.
The factor $\nab^{i_3}\psi_{\Hb}$ can be expressed in the following reduced schematic form:
$$\nab^i\psi_{\Hb}= \nab^i(\psi_{\Hb}-(\psi_{\Hb})_\Ke) + \nab^i(\psi_{\Hb})_\Ke\eqrs \nab^i\tpHb+\sum_{i_1\leq i}\Om_\Ke^2\nab^{i_1}\tg+\Om_{\Ke}^2,$$
where we have used
$$\nab^i(\psi_{\Hb})_\Ke \eqrs \sum_{i_1\leq i}\Om_\Ke^2\nab^{i_1}\tg+\Om_{\Ke}^2,$$
which is a consequence of \eqref{nab.def}, \eqref{def.slashed.Gamma}, Propositions~\ref{Kerr.der.Ricci.bound} and \ref{nab.diff}.
Putting these together, we have
\begin{equation}\label{schematic.3.4}
\begin{split}
&\sum_{i_1+i_2+i_3+i_4\leq i}\nab^{i_1}\psi^{i_2}\nab^{i_3}\psi_{\Hb}\nab^{i_4}(\phi-\phi_\Ke)\\
\eqrs &\sum_{i_1+i_2+i_3\leq i}(1+\nab^{i_1}\tg+\nab^{i_1-1}\tp)(\Om_\Ke^2+\Om_\Ke^2\nab^{i_2}\tg+\nab^{i_2}\tpHb)\nab^{i_3}(\phi-\phi_\Ke).
\end{split}
\end{equation}
\eqref{schematic.3.main} follows from \eqref{schematic.3.3} and \eqref{schematic.3.4}. Finally, notice that if $\phi$ is a scalar, then when applying Proposition~\ref{3.diff.prop}, the last term is absent. Tracing through the above argument, this shows that \eqref{schematic.3.3} can be replaced by 
\begin{equation*}
\begin{split}
&\sum_{i_1+i_2+i_3\leq i}\nab^{i_1}\psi^{i_2}\nab^{i_3}\nab_3(\phi-\phi_\Ke)\eqrs\sum_{i_1+i_2\leq i}(1+\nab^{i_1}\tg+\nab^{i_1-1}\tp)\nab^{i_2}(G-G_\Ke)
\end{split}
\end{equation*}
and as a consequence we obtain the improvement in that case.
\end{proof}

We now provide a general proposition which allows us to derive a $\nab_4$ transport equation for the difference quantities.
\begin{proposition}\label{4.diff.prop}
Let $\phi$ be a rank-$r$ covariant $S$-tangent tensor and $\phi_\Ke$ be a corresponding tensor on Kerr spacetime. Suppose
$$\nab_4\phi=F,\quad (\nab_4)_\Ke\phi_\Ke=F_\Ke.$$
Then the difference $\phi-\phi_\Ke$ satisfies the following equation:
\begin{equation*}
\begin{split}
&\nab_4(\phi-\phi_\Ke)_{A_1...A_r}\\
=&F_{A_1...A_r}-(F_\Ke)_{A_1...A_r}-\frac{\Om_\Ke^2-\Om^2}{\Om^2}(\nab_4)_\Ke(\phi_\Ke)_{A_1...A_r}\\
&-\sum_{i=1}^r\frac{\Om_\Ke^2-\Om^2}{\Om^2}(\gamma_\Ke^{-1})^{BC}(\chi_\Ke)_{CA_i}(\phi_\Ke)_{A_1...\hat{A}_iB...A_r}-\sum_{i=1}^r((\gamma^{-1})^{BC}\chi_{CA_i}-(\gamma_\Ke^{-1})^{BC}(\chi_\Ke)_{CA_i})(\phi_\Ke)_{A_1...\hat{A}_iB...A_r}\\
&-\f{1}{\Om^2}(b^B-(b_{\Ke})^B)\nab_B(\phi_\Ke)_{A_1...A_r}-\sum_{i=1}^r\f{1}{\Om^2}\nab_{A_i}(b^B-(b_\Ke)^B) (\phi_\Ke)_{A_1...\hat{A}_iB...A_r}.
\end{split}
\end{equation*}
\end{proposition}

\begin{proof}
Unlike for the $\nab_3$ equations, in addition to the difference of the connections, we also have to compute the difference of the vector fields $e_4$ and $(e_4)_\Ke$.
In the coordinate system $(u,\ub,\th_*,\phi_*)$, we have
$$e_4=\frac{1}{\Om^2}(\frac{\partial}{\partial \ub}+b^A\frac{\partial}{\partial\th^A}).$$
Therefore, the difference of the vector fields $e_4$ and $(e_4)_\Ke$ can be computed as follows:
\begin{equation*}
\begin{split}
e_4-(e_4)_\Ke
=&\frac{\Om_\Ke^2-\Om^2}{\Om^2\Om_\Ke^2}(\frac{\partial}{\partial \ub}+(b_{\Ke})^A\frac{\partial}{\partial\th^A})+\frac{1}{\Om^2}(b^A-b^A_\Ke)\frac{\partial}{\partial\th^A}.
\end{split}
\end{equation*}
The difference of the Christoffel symbols associated to the connection $\nab_4$ (cf.~\eqref{nab4.def}) can be computed as follows:
\begin{equation*}
\begin{split}
&(\gamma^{-1})^{BD}\chi_{DC}-\Om^{-2}\f{\partial}{\partial\th^C}b^B-((\gamma^{-1})^{BD}\chi_{DC}-\Om^{-2}\f{\partial}{\partial\th^C}b^B)_\Ke\\
=&((\gamma^{-1})^{BD}\chi_{DC}-(\gamma_\Ke^{-1})^{BD}(\chi_\Ke)_{DC})-\frac{\Om_\Ke^2-\Om^2}{\Om^2\Om_\Ke^2}\f{\partial}{\partial\th^C}b_\Ke^B-\f{1}{\Om^2}\f{\rd}{\rd \th^C}(b^B-(b_\Ke)^B)
\end{split}
\end{equation*}
Given a covariant tensor $\phi_{A_1...A_r}$ of rank $r$, we observe that by \eqref{nab.def}, \eqref{def.slashed.Gamma},
\begin{equation*}
\begin{split}
&(b^B-(b_{\Ke})^B)\f{\rd}{\rd \th^B}\phi_{A_1...A_r}+\sum_{i=1}^r\f{\rd}{\rd\th^{A_i}}(b^B-(b_\Ke)^B)\phi_{A_1...\hat{A}_iB...A_r} \\
=&(b^B-(b_{\Ke})^B)\nab_B\phi_{A_1...A_r}+\sum_{i=1}^r\nab_{A_i}(b^B-(b_\Ke)^B) \phi_{A_1...\hat{A}_iB...A_r}.
\end{split}
\end{equation*}
By \eqref{nab4.def}, this implies the following formula for $\nab_{4}-(\nab_{4})_\Ke$:
\begin{equation*}
\begin{split}
&(\nab_4-(\nab_4)_\Ke)\phi_{A_1...A_r}\\
=&\frac{\Om_\Ke^2-\Om^2}{\Om^2}(\nab_{4})_\Ke\phi_{A_1...A_r}+\sum_{i=1}^r\frac{\Om_\Ke^2-\Om^2}{\Om^2}(\gamma_\Ke^{-1})^{BC}(\chi_\Ke)_{CA_i}\phi_{A_1...\hat{A}_iB...A_r}\\
&-\sum_{i=1}^r ((\gamma^{-1})^{BC}\chi_{CA_i}-(\gamma_\Ke^{-1})^{BC}(\chi_\Ke)_{CA_i})\phi_{A_1...\hat{A}_iB...A_r}\\
&+\f{1}{\Om^2}(b^B-(b_{\Ke})^B)\nab_B\phi_{A_1...A_r}+\sum_{i=1}^r\f{1}{\Om^2}\nab_{A_i}(b^B-(b_\Ke)^B) \phi_{A_1...\hat{A}_iB...A_r}.
\end{split}
\end{equation*}
Now, since
$$\nab_4\phi-(\nab_4\phi)_\Ke=\nab_4 (\phi-\phi_\Ke)+(\nab_4-(\nab_4)_\Ke)\phi_\Ke,$$
we have
\begin{equation*}
\begin{split}
&\nab_4(\phi-\phi_\Ke)_{A_1...A_r}\\
=&\nab_4\phi_{A_1...A_r}-((\nab_4\phi)_\Ke)_{A_1...A_r}-\frac{\Om_\Ke^2-\Om^2}{\Om^2}(\nab_4)_\Ke(\phi_\Ke)_{A_1...A_r}\\
&-\sum_{i=1}^r\frac{\Om_\Ke^2-\Om^2}{\Om^2}(\gamma_\Ke^{-1})^{BC}(\chi_\Ke)_{CA_i}(\phi_\Ke)_{A_1...\hat{A}_iB...A_r}+\sum_{i=1}^r((\gamma^{-1})^{BC}\chi_{CA_i}-(\gamma_\Ke^{-1})^{BC}(\chi_\Ke)_{CA_i})(\phi_\Ke)_{A_1...\hat{A}_iB...A_r}\\
&-\f{1}{\Om^2}(b^B-(b_{\Ke})^B)\nab_B(\phi_\Ke)_{A_1...A_r}-\sum_{i=1}^r\f{1}{\Om^2}\nab_{A_i}(b^B-(b_\Ke)^B) (\phi_\Ke)_{A_1...\hat{A}_iB...A_r},
\end{split}
\end{equation*}
which is the desired formula.
\end{proof}
Using the above formula together with the commutation formula Proposition \ref{commute} (or more specifically, its schematic version in Proposition \ref{repeated.com}), we deduce the following general reduced schematic formula for difference quantities.
\begin{proposition}\label{schematic.4}
Let $\phi$ be a rank-$r$ covariant $S$-tensor and $\phi_\Ke$ be a corresponding tensor on background Kerr spacetime. Suppose
$$\nab_4\phi=F,\quad (\nab_4)_\Ke\phi_\Ke=F_\Ke.$$
Then, if $\phi$ is a scalar, the following holds for $i\leq 3$:
\begin{equation}\label{schematic.4.0.scalar}
\begin{split}
\nab_4\nab^i(\phi-\phi_\Ke)
\eqrs&\sum_{i_1+i_2\leq i}(1+\nab^{i_1}\tg+\nab^{i_1-1}\tp)\nab^{i_2}(F-F_\Ke)\\
&+\sum_{i_1+i_2+i_3\leq i}(1+\nab^{i_1}\tg+\nab^{i_1-1}\tp)\nab^{i_2}\tg\nab^{i_3}(\nab_4)_\Ke\phi_\Ke\\
&+\sum_{i_1+i_2+i_3\leq i}\Om_\Ke^{-2}(1+\nab^{i_1}\tg+\nab^{i_1-1}\tp)\nab^{i_2}\tb\nab^{i_3}(\nab_\Ke)\phi_\Ke\\
&+\sum_{i_1+i_2+i_3\leq i}(1+\nab^{i_1}\tg+\nab^{i_1-1}\tp)(1+\nab^{i_2}(\tg,\tpH))\nab^{i_3}(\phi-\phi_\Ke).
\end{split}
\end{equation}
If $\phi$ is a tensor of rank $\geq 1$, the following holds for $i\leq 2$:
\begin{equation}\label{schematic.4.0.tensor}
\begin{split}
\nab_4\nab^i(\phi-\phi_\Ke)\eqrs&\sum_{i_1+i_2\leq i}(1+\nab^{i_1}\tg+\nab^{i_1-1}\tp)\nab^{i_2}(F-F_\Ke)\\
&+\sum_{i_1+i_2+i_3\leq i}(1+\nab^{i_1}\tg+\nab^{i_1-1}\tp)\nab^{i_2}\tg\nab^{i_3}(\nab_4)_\Ke\phi_\Ke\\
&+\sum_{i_1+i_2+i_3\leq i}\Om_\Ke^{-2}(1+\nab^{i_1}\tg+\nab^{i_1-1}\tp)\nab^{i_2}\tb\nab^{i_3}(\nab_\Ke)\phi_\Ke\\
&+\sum_{i_1+i_2+i_3\leq i}(1+\nab^{i_1}\tg+\nab^{i_1-1}\tp)(1+\nab^{i_2}(\tg,\tpH))\nab^{i_3}(\phi-\phi_\Ke)\\
&+\sum_{i_1+i_2+i_3\leq i}(1+\nab^{i_1}\tg+\nab^{i_1-1}\tp)\nab^{i_2}(\tg,\tpH)\nab^{i_3}\phi_\Ke\\
&+\sum_{i_1+i_2+i_3\leq i}\Om_\Ke^{-2}(1+\nab^{i_1}\tg+\nab^{i_1-1}\tp)\nab^{i_2+1}\tb\nab^{i_3}\phi_\Ke.
\end{split}
\end{equation}
\end{proposition}
\begin{proof}
We first derive the reduced schematic equation for scalars. According to Proposition \ref{4.diff.prop}, if $\phi$ is a scalar, we have
\begin{equation}\label{schematic.4.scalar}
\nab_4(\phi-\phi_\Ke)=F-F_\Ke-\frac{\Om_\Ke^2-\Om^2}{\Om^2}(\nab_4)_\Ke(\phi_\Ke)-\f{1}{\Om^2}(b^B-(b_{\Ke})^B)\nab_B\phi_\Ke.
\end{equation}
In other words, compared to the formula for general tensors in Proposition \ref{4.diff.prop}, we do not have the terms arising from the connection coefficients associated to $\nab_4$. Most importantly from the point of view of this proposition, we do not have terms involving $\nab \tb$ on the right hand side. As we will see, this will play an important role in avoiding a loss of derivative (see Proposition \ref{est.nab3trch}).

As in the proof of Proposition \ref{schematic.3}, we now apply the schematic commutation formula in Proposition \ref{repeated.com} to obtain the reduced schematic equation for higher angular derivatives. More precisely, by Proposition \ref{repeated.com}, we have the following terms:
\begin{equation}\label{schematic.4.1}
\sum_{i_1+i_2+i_3\leq i}\nab^{i_1}\psi^{i_2}\nab^{i_3}\nab_4(\phi-\phi_\Ke) 
\end{equation}
and 
\begin{equation}\label{schematic.4.2}
\sum_{i_1+i_2+i_3+i_4\leq i}\nab^{i_1}\psi^{i_2}\nab^{i_3}\psi_H\nab^{i_4}(\phi-\phi_\Ke).
\end{equation}
We first deal with the term \eqref{schematic.4.1}. Notice that as long as $i\leq 3$, we have
\begin{equation}\label{nab.b.RS}
\nab^{i}(\Om^{-2}(b^A-(b_\Ke)^A))\eqrs\sum_{i_1+i_2\leq i}\Om_\Ke^{-2}(1+\nab^{i_1}\tg)\nab^{i_2}\tb.
\end{equation}
Therefore, using \eqref{schematic.4.scalar}, \eqref{d2p.sch} and the convention that $\tg$ can schematically represent $\frac{\Om_\Ke^2-\Om^2}{\Om^2}$ (see \eqref{tg.possibilities}), we have
\begin{equation}\label{schematic.4.3}
\begin{split}
&\sum_{i_1+i_2+i_3\leq i}\nab^{i_1}\psi^{i_2}\nab^{i_3}\nab_4(\phi-\phi_\Ke)\\
\eqrs &\sum_{i_1+i_2\leq i}(1+\nab^{i_1}\tg+\nab^{i_1-1}\tp)\nab^{i_2}\nab_4(F-F_\Ke)\\
&+\sum_{i_1+i_2+i_3\leq i}(1+\nab^{i_1}\tg+\nab^{i_1-1}\tp)\nab^{i_2}\tg\nab^{i_3}(\nab_4)_\Ke\phi_\Ke\\
&+\sum_{i_1+i_2+i_3\leq i}\Om_\Ke^{-2}(1+\nab^{i_1}\tg+\nab^{i_1-1}\tp)\nab^{i_2}\tb\nab^{i_3}(\nab_\Ke)\phi_\Ke.
\end{split}
\end{equation}
We now turn to the term \eqref{schematic.4.2}. Note that for $i\leq 3$ we have
\begin{equation}\label{nab.pH.RS}
\nab^i\psi_H\eqrs\sum_{i_1\leq i}(\nab^{i_1}\tpH+\nab^{i_1}\tg)+1.
\end{equation}
Using this together with \eqref{d2p.sch}, we thus have
\begin{equation}\label{schematic.4.4}
\begin{split}
&\sum_{i_1+i_2+i_3+i_4\leq i}\nab^{i_1}\psi^{i_2}\nab^{i_3}\psi_H\nab^{i_4}(\phi-\phi_\Ke)\\
\eqrs&\sum_{i_1+i_2+i_3\leq i}(1+\nab^{i_1}\tg+\nab^{i_1-1}\tp)(1+\nab^{i_2}(\tg,\tpH))\nab^{i_3}(\phi-\phi_\Ke).
\end{split}
\end{equation}
Combining \eqref{schematic.4.3} and \eqref{schematic.4.4}, we conclude the proof of \eqref{schematic.4.0.scalar}, i.e.~the case when $\phi$ is a scalar. 

Now, we turn to the case where $\phi$ is a tensor with rank $\geq 1$. According to Propositions \ref{repeated.com} and \ref{4.diff.prop}, we have the extra terms
$$\sum_{i_1+i_2+i_3\leq i}\nab^{i_1}\psi^{i_2}\nab^{i_3}F_e,$$
where $F_e$ is given by
\begin{equation*}
\begin{split}
F_e=&-\sum_{i=1}^r\frac{\Om_\Ke^2-\Om^2}{\Om^2}(\gamma_\Ke^{-1})^{BC}(\chi_\Ke)_{CA_i}(\phi_\Ke)_{A_1...\hat{A}_iB...A_r}\\
&-\sum_{i=1}^r((\gamma^{-1})^{BC}\chi_{CA_i}-(\gamma_\Ke^{-1})^{BC}(\chi_\Ke)_{CA_i})(\phi_\Ke)_{A_1...\hat{A}_iB...A_r}\\
&-\sum_{i=1}^r\f{1}{\Om^2}\nab_{A_i}(b^B-(b_\Ke)^B) (\phi_\Ke)_{A_1...\hat{A}_iB...A_r}.
\end{split}
\end{equation*}
These terms can be expressed as follows using \eqref{Om.c}, \eqref{d2p.sch}, \eqref{nab.b.RS} and \eqref{nab.pH.RS}:
\begin{equation*}
\begin{split}
&\sum_{i_1+i_2+i_3\leq i}\nab^{i_1}\psi^{i_2}\nab^{i_3}F_e\\
%\blue{\eqrs}&\sum_{i_1+i_2+i_3\leq i}(1+\nab^{i_1}(\tg,\tpH)+\nab^{i_1-1}\tp)\nab^{i_2}\tg\nab^{i_3}\phi_\Ke\\
%&+\sum_{i_1+i_2+i_3\leq i}(1+\nab^{i_1}\tg+\nab^{i_1-1}\tp)\nab^{i_2}\tpH\nab^{i_3}\phi_\Ke\\
%&+\sum_{i_1+i_2+i_3\leq i}\Om_\Ke^{-2}(1+\nab^{i_1}\tg+\nab^{i_1-1}\tp)\nab^{i_2+1}\tb\nab^{i_3}\phi_\Ke\\
\eqrs&\sum_{i_1+i_2+i_3\leq i}(1+\nab^{i_1}\tg+\nab^{i_1-1}\tp)\nab^{i_2}(\tg,\tpH)\nab^{i_3}\phi_\Ke\\
&+\sum_{i_1+i_2+i_3\leq i}\Om_\Ke^{-2}(1+\nab^{i_1}\tg+\nab^{i_1-1}\tp)\nab^{i_2+1}\tb\nab^{i_3}\phi_\Ke.
\end{split}
\end{equation*}
Together with the terms which were treated in \eqref{schematic.4.3} and \eqref{schematic.4.4}, we thus have that for $\phi$ being a tensor of rank $\geq 1$, 
\begin{equation}\label{schematic.4.5}
\begin{split}
&\sum_{i_1+i_2+i_3\leq i}\nab^{i_1}\psi^{i_2}\nab^{i_3}\nab_4(\phi-\phi_\Ke)\\
\eqrs&\sum_{i_1+i_2\leq i}(1+\nab^{i_1}\tg+\nab^{i_1-1}\tp)\nab^{i_2}\nab_4(F-F_\Ke)\\
&+\sum_{i_1+i_2+i_3\leq i}(1+\nab^{i_1}\tg+\nab^{i_1-1}\tp)\nab^{i_2}\tg\nab^{i_3}(\nab_4)_\Ke\phi_\Ke\\
&+\sum_{i_1+i_2+i_3\leq i}\Om_\Ke^{-2}(1+\nab^{i_1}\tg+\nab^{i_1-1}\tp)\nab^{i_2}\tb\nab^{i_3}(\nab_\Ke)\phi_\Ke\\
&+\sum_{i_1+i_2+i_3\leq i}(1+\nab^{i_1}\tg+\nab^{i_1-1}\tp)\nab^{i_2}(\tg,\tpH)\nab^{i_3}\phi_\Ke\\
&+\sum_{i_1+i_2+i_3\leq i}\Om_\Ke^{-2}(1+\nab^{i_1}\tg+\nab^{i_1-1}\tp)\nab^{i_2+1}\tb\nab^{i_3}\phi_\Ke.
\end{split}
\end{equation}
Combining \eqref{schematic.4.4} and \eqref{schematic.4.5}, we thus obtain \eqref{schematic.4.0.tensor}. This concludes the proof of the proposition.
\end{proof}

\subsection{The inhomogeneous terms}\label{sec.rse.inho}

We now turn to the derivation of the reduced schematic equations. We will show that the following four types of inhomogeneous terms will arise:
\begin{equation}\label{inho.def}
\begin{split}
\mathcal F_1\doteq &\:\Om_\Ke^2\left(\sum_{i_1+i_2\leq 3}(1+\nab^{i_1}(\tg,\tp))(\nab^{i_2}(\tg,\tp,\tpHb,\widetilde{\omb})+\Om_\Ke^2\nab^{i_2}\tpH+\nab^{i_2-1} \tK)+\nab^2\tg(\nab\tK,\nab^2\tp)\right) \\
&+\Om_\Ke^2\sum_{i_1+i_2+i_3\leq 3}(1+\nab^{i_1}\tg+\nab^{\min\{i_1,2\}}\tp)\nab^{i_2} (\tpHb,\widetilde{\omb})\nab^{i_3}\tpH,\\
\mathcal F_2\doteq &\:\sum_{i_1+i_2\leq 3}(1+\nab^{i_1}(\tg,\tp))(\Om^{-2}_\Ke\nab^{i_2}\tb+\nab^{i_2}(\tg,\tp,\tpH)+\nab^{i_2-1} \tK)+\nab^2\tg\nab^2\tpH+\nab\tK\nab\tpH, \\
\mathcal F_3\doteq &\:\sum_{i_1+i_2\leq 3}(1+\nab^{i_1}(\tg,\tp))(\Om_\Ke^2\nab^{i_2}(\tg,\tp)+\nab^{i_2}(\tpHb,\widetilde{\omb})+\nab^{i_2-1} \tK)+\nab^2\tg\nab^2\tpHb+\nab\tK\nab\tpHb, \\
\mathcal F_4\doteq &\:\sum_{i_1+i_2\leq 3}(1+\nab^{i_1}(\tg,\tp))(\nab^{i_2}(\tg,\tb,\tp,\tpHb,\widetilde{\omb})+\Om_\Ke^2\nab^{i_2}\tpH+\nab^{i_2-1}\tK) +\nab^2\tg(\nab\tK,\nab^2\tp)\\
&+\sum_{i_1+i_2+i_3\leq 3}(1+\nab^{i_1}\tg+\nab^{\min\{i_1,2\}}\tp)\nab^{i_2} (\tpHb,\widetilde{\omb})(\nab^{i_3}\tpH+\Om_\Ke^{-2}\nab^{i_3}\tb).
\end{split}
\end{equation}
Here, we have used that notation where $\nab^{-1}=0$. We immediately also define the ``primed'' versions of these inhomogeneous terms $\mathcal F_i'$. Notice that each of $\mathcal F_i'$ contains a subset of the terms that are found in $\mathcal F_i$. The difference is that some of the $\nab^3(\tpH,\tp,\tpHb)$ terms are absent in the expressions of $\mathcal F_i'$ compared to that of $\mathcal F_i$. However, at the same time, not all of the differences of Ricci coefficients with three angular derivatives are absent in these expressions -- the structure of the equations is such that the potentially most harmful terms are absent and we can use this fact to obtain the bounds for $\NS$. We remark moreover that the structure of the equations with $\mathcal F_i'$ inhomogeneous terms is useful to obtain some refined integrated estimates and hypersurface estimates. The $\mathcal F_i'$ terms are defined as follows:
\begin{equation}\label{inho.def.p}
\begin{split}
\mathcal F_1'\doteq &\:\Om_\Ke^2\nab^3(\tg,\widetilde{\eta})+\Om_\Ke^2\nab^2\tK+\Om_\Ke^2\left(\sum_{i_1+i_2\leq 2}(1+\nab^{i_1}(\tg,\tp))(\nab^{i_2}(\tg,\tp,\tpHb,\widetilde{\omb})+\Om_\Ke^2\nab^{i_2}\tpH)\right) \\
&+\Om_\Ke^2\sum_{i_1+i_2+i_3\leq 2}(1+\nab^{i_1}(\tg,\tp))\nab^{i_2} (\tpHb,\widetilde{\omb})\nab^{i_3}\tpH,\\
\mathcal F_2'\doteq & \:\sum_{i_1+i_2\leq 3}(1+\nab^{i_1}\tg+\nab^{\min\{i_1,2\}}\tp)(\Om^{-2}_\Ke\nab^{i_2}\tb+\nab^{\min\{i_2,2\}}\tp+\nab^{i_2}(\tg,\tpH)),\\
\mathcal F_3'\doteq &\:\sum_{i_1+i_2\leq 3}(1+\nab^{i_1}\tg+\nab^{\min\{i_1,2\}}\tp)(\Om_\Ke^2\nab^{i_2}\tg+\Om_\Ke^2\nab^{\min\{i_2,2\}}\tp+\nab^{i_2}(\tpHb,\widetilde{\omb})),\\
\mathcal F_4'\doteq &\:\nab^3(\tg,\widetilde{\etab})+\nab^2\tK+\sum_{i_1+i_2\leq 2}(1+\nab^{i_1}(\tg,\tp))(\nab^{i_2}(\tg,\tb,\tp,\tpHb,\widetilde{\omb})+\Om_\Ke^2\nab^{i_2}\tpH) \\
&+\sum_{i_1+i_2+i_3\leq 2}(1+\nab^{i_1}\tg+\nab^{\min\{i_1,2\}}\tp)\nab^{i_2} (\tpHb,\widetilde{\omb})(\nab^{i_3}\tpH+\Om_\Ke^{-2}\nab^{i_3}\tb).
\end{split}
\end{equation}

\subsection{Reduced schematic equations for the difference of the metric components}\label{sec.rse.metric}

In this subsection, we derive the reduced schematic equations for the difference of the metric components, i.e.~$\tg$ and $\tb$. Recall our convention (see \eqref{tg.possibilities}) that we use $\tg$ to denote any of the terms $\gamma_{AB}-(\gamma_{AB})_\Ke$, $\gamma^{AB}-(\gamma^{AB})_\Ke$, $\frac{\Om^2-\Om_\Ke^2}{\Om^2}$ and $\log{\Om}-\log{\Om_\Ke}$. On the other hand, we will show that (Propositions \ref{Omega.lemma} and \ref{gamma.lemma}) once we have the estimates for $\gamma_{AB}-(\gamma_{AB})_\Ke$ and $\log{\Om}-\log{\Om_\Ke}$ and their angular covariant derivatives, it follows that $\gamma^{AB}-(\gamma^{AB})_\Ke$ and $\frac{\Om^2-\Om_\Ke^2}{\Om^2}$ and their angular covariant derivatives are also bounded. Therefore, it suffices to obtain the reduced schematic equations for $\gamma_{AB}-(\gamma_{AB})_\Ke$ and $\log{\Om}-\log{\Om_\Ke}$. This will be achieved in Propositions \ref{gamma.eqn} and \ref{Omega.eqn} respectively. Before we end the subsection, we will then obtain the reduced schematic equation for $\tb$ and its derivatives in Proposition \ref{b.eqn}.

\begin{proposition}\label{gamma.eqn}
For $i\leq 3$, $\nab^i\widetilde\gamma$ obeys the following reduced schematic equation
$$\nab_3\nab^i\widetilde\gamma\eqrs\mathcal F'_3.$$
\end{proposition}
\begin{proof}
By \eqref{nab.compatibility},
$$\nab_3 \gamma=0,\quad (\nab_3\gamma)_\Ke=0.$$
By Proposition \ref{schematic.3}, we thus have the following for $i\leq 3$.
\begin{equation}\label{gamma.eqn.1}
\begin{split}
&\nab_3\nab^i(\gamma-\gamma_\Ke)\\
\eqrs&\sum_{i_1+i_2+i_3\leq i}(1+\nab^{i_1}\tg+\nab^{i_1-1}\tp)(\Om_\Ke^2\nab^{i_2}\tg+\nab^{i_2}\tpHb)\nab^{i_3}\gamma_\Ke\\
&+\sum_{i_1+i_2+i_3+i_4\leq i}(1+\nab^{i_1}\tg+\nab^{i_1-1}\tp)(\Om_\Ke^2+\Om_\Ke^2\nab^{i_2}\tg+\nab^{i_2}\tpHb)\nab^{i_3}(\gamma-\gamma_\Ke).%\\
%\eqrs&\sum_{i_1+i_2\leq i}(1+\nab^{i_1}\tg+\nab^{i_1-1}\tp)(\Om_\Ke^2\nab^{i_2}\tg+\nab^{i_2}\tpHb)\\
%&+\sum_{i_1+i_2+i_3+i_4\leq i}(1+\nab^{i_1}\tg+\nab^{i_1-1}\tp)(\Om_\Ke^2+\Om_\Ke^2\nab^{i_2}\tg+\nab^{i_2}\tpHb)\nab^{i_3}\tg,
\end{split}
\end{equation}
We claim that every term is acceptable, i.e.~can be expressed as one of the terms in $\mathcal F_3'$. Since this is the first such proof, we will discuss every step in detail. Note that terms on the right hand side of \eqref{gamma.eqn.1} are cubic while terms in $\mathcal F_3'$ are quadratic. To see that terms on the right hand side of \eqref{gamma.eqn.1} are acceptable, we thus use the $L^\infty$ bounds in Corollary~\ref{Linfty}.

For the first term on the right hand side of \eqref{gamma.eqn.1}, first note that by \eqref{nab.compatibility}, Corollary~\ref{Linfty} and Proposition~\ref{nab.diff},
\begin{equation}\label{nab.gammaK}
\nab^{i_3}\gamma_\Ke \eqrs 1+\sum_{i_3'\leq i_3}\nab^{i_3'}\tg. 
\end{equation}
Now since $i\leq 3$, either $i_1\leq 1$ or $i_3\leq 1$. If $i_1\leq 1$, then by Corollary~\ref{Linfty}, $1+\nab^{i_1}\tg+\nab^{i_1-1}\tp \eqrs 1$ and so using \eqref{nab.gammaK} and  relabelling the indices
$$\sum_{\substack{i_1+i_2+i_3\leq i \\ i_1\leq 1}}(1+\nab^{i_1}\tg+\nab^{i_1-1}\tp)(\Om_\Ke^2\nab^{i_2}\tg+\nab^{i_2}\tpHb)\nab^{i_3}\gamma_\Ke \eqrs \sum_{i_1+i_2\leq 3}(1+\nab^{i_1}\tg)(\Om_\Ke^2\nab^{i_2}\tg+\nab^{i_2}\tpHb).$$
If $i_3\leq 1$, by \eqref{nab.gammaK}, $\nab^{i_3}\gamma_\Ke\eqrs 1$. Hence,
$$\sum_{\substack{i_1+i_2+i_3\leq i \\ i_3\leq 1}}(1+\nab^{i_1}\tg+\nab^{i_1-1}\tp)(\Om_\Ke^2\nab^{i_2}\tg+\nab^{i_2}\tpHb)\nab^{i_3}\gamma_\Ke \eqrs \sum_{i_1+i_2\leq 3}(1+\nab^{i_1}\tg+\nab^{i_1-1}\tp)(\Om_\Ke^2\nab^{i_2}\tg+\nab^{i_2}\tpHb).$$
In both cases the terms are acceptable.

Finally, the second term on the right hand side of \eqref{gamma.eqn.1} can be handled in a similar manner.
\qedhere

\end{proof}

We now derive the reduced schematic equations for the derivatives of the difference of $\log{\Om}$:
\begin{proposition}\label{Omega.eqn}
For $i\leq 3$, $\nab^i(\log\Om-\log\Om_\Ke)$ obeys the following reduced schematic equation
$$\nab_3\nab^i(\log\Om-\log\Om_\Ke)\eqrs\mathcal F'_3.$$
\end{proposition}
\begin{proof}
Recall from \eqref{gauge.con} that
$$\nab_3\log\Om=-\omegab.$$
By Proposition \ref{schematic.3}, since $\log\Om$ is a scalar, we have the following reduced schematic equation:
\begin{equation*}
\begin{split}
&\nab_3\nab^i(\log\Om-\log\Om_\Ke)\\
\eqrs&\sum_{i_1+i_2\leq i}(1+\nab^{i_1}\tg+\nab^{i_1-1}\tp)\nab^{i_2}\widetilde{\omb}\\
&+\sum_{i_1+i_2+i_3\leq i}(1+\nab^{i_1}\tg+\nab^{i_1-1}\tp)(\Om_\Ke^2+\Om_\Ke^2\nab^{i_2}\tg+\nab^{i_2}\tpHb)\nab^{i_3}(\log\Om-\log\Om_\Ke).
\end{split}
\end{equation*}
The first term is obviously acceptable. For the second term, note that either $i_1\leq 1$ or $i_3\leq 1$ so that by Corollary~\ref{Linfty}, either $(1+\nab^{i_1}\tg+\nab^{i_1-1}\tp)$ or $\nab^{i_3}(\log\Om-\log\Om_\Ke)$ is bounded in $L^\infty$. \qedhere
\end{proof}
After deriving the equations for $\widetilde{\gamma}$ and $\widetilde{\Omega}$, we now turn to the derivation of the equation for $\tb$, thus obtaining the schematic equations for all the components of the spacetime metric. It turns out that $\tb$ obeys a $\nab_3$ equation similar to that for $\tg$. Nevertheless, there is a crucial difference that we need to exploit: None of the linear terms contain $\tpHb$. More precisely, $\tpHb$ always appears on the right hand side in combination with $\tb$. This will allow us to obtain an estimate for $\tb$ that is stronger than the bounds for $\tg$. This turns out to be necessary to close the argument.
\begin{proposition}\label{b.eqn}
For $i\leq 3$, $\nab^i\tb$ obeys the following reduced schematic equation
$$\nab_3 \nab^i\tb \eqrs\sum_{i_1+i_2\leq i}(1+\nab^{i_1}(\tg,\tp))(\Om_\Ke^2\nab^{i_2}(\tg,\tp,\tb))+\sum_{i_1+i_2+i_3\leq i}(1+\nab^{i_1}(\tg,\tp))\nab^{i_2}\tpHb\nab^{i_3}\tb.$$
\end{proposition}
\begin{proof}
To obtain the reduced schematic equation for $b^A$, we will first consider the difference equation in coordinates and then rewrite in covariant form. This is because the general formula in Proposition \ref{schematic.3} is a schematic equation which does not keep track of the exact coefficients and using it would miss a crucial cancellation.

To proceed, recall the following equation for $b^A$ in Proposition~\ref{metric.der.Ricci}:
$$\frac{\partial}{\partial u} b^A=4\Omega^2(\zeta^A-\nab^A\log\Om).$$
Taking the difference between the equations for $b^A$ and $b^A_\Ke$, we have
\begin{equation*}
\begin{split}
\frac{\partial}{\partial u} (b^A-b^A_\Ke)
=&4\Omega^2(\zeta^A-\zeta_\Ke^A-\nab^A\log\Om+(\nab_{\Ke})^A\log\Om_{\Ke})+4(\Om^2-\Om_\Ke^2)(\zeta_\Ke^A-(\nab_{\Ke})^A\log\Om_\Ke).
\end{split}
\end{equation*}
Using\footnote{With the obvious modifications for contravariant tensors.} the definition of $\nab_3$ in \eqref{nab3.def}, we have
\begin{equation}\label{b.eqn.0derivative}
\begin{split}
\nab_{3} (b^A-b^A_\Ke)
=&\frac{\partial}{\partial u} (b^A-b^A_\Ke)-\chib_B{ }^A (b^B-b^B_\Ke)\\
=&4\Omega^2(\zeta^A-\zeta_\Ke^A-\nab^A\log\f{\Om}{\Om_\Ke})+4(\Om^2-\Om_\Ke^2)(\zeta_\Ke^A-(\nab_\Ke)^A\log\Om_\Ke)\\
&+4\Om^2((\gamma_\Ke^{-1})^{AB}-(\gamma^{-1})^{AB})\nab_B\log\Om_\Ke-\chib_B{ }^A (b^B-b^B_\Ke).
\end{split}
\end{equation}
We thus obtain the desired conclusion for $i=0$. Notice in particular that $\chib$ is contracted with $b-b_\Ke$, i.e.~when expanded schematically $\chib\eqrs \chib_{\Ke}+\tpHb$, the $\tpHb$ term indeed comes in the form $\tpHb\tb$.

We now use Proposition \ref{repeated.com} to derive the reduced schematic equations for the higher derivatives. According to Proposition \ref{repeated.com}, there are two contributing terms:
\begin{equation}\label{b.eqn.1}
\sum_{i_1+i_2+i_3\leq i}\nab^{i_1}\psi^{i_2}\nab^{i_3}\nab_3(b^A-b^A_\Ke)
\end{equation}
and
\begin{equation}\label{b.eqn.2}
\sum_{i_1+i_2+i_3+i_4\leq i}\nab^{i_1}\psi^{i_2}\nab^{i_3}\psi_{\Hb}\nab^{i_4}(b^A-b^A_\Ke).
\end{equation}
Using \eqref{b.eqn.0derivative}, \eqref{nabgK.nabpsiHbK} and the fact that $\psi$, $\tp$, $\tg$, $\nab\tg$ are bounded in $L^\infty$ (by Corollary~\ref{Linfty}), the term \eqref{b.eqn.1} can be expressed as follows
\begin{equation}\label{b.eqn.3}
\begin{split}
&\sum_{i_1+i_2+i_3\leq i}\nab^{i_1}\psi^{i_2}\nab^{i_3}\nab_3(b^A-b^A_\Ke)\\
\eqrs&\sum_{i_1+i_2\leq i}\Om_\Ke^2(1+\nab^{i_1}(\tg,\tp))\nab^{i_2}(\tg,\tp)\\
&+\sum_{i_1+i_2+i_3\leq i}(1+\nab^{i_1}(\tg,\tp))(\Om_\Ke^2+\Om_\Ke^2\nab^{i_2}\tg+\nab^{i_2}\tpHb)\nab^{i_3}\tb.
\end{split}
\end{equation}
Using again \eqref{nabgK.nabpsiHbK} and the $L^\infty$ boundedness of $\psi$, $\tp$, $\tg$, $\nab\tg$, the term \eqref{b.eqn.2} can be expressed as follows
\begin{equation}\label{b.eqn.4}
\begin{split}
\sum_{i_1+i_2+i_3+i_4\leq i}\nab^{i_1}\psi^{i_2}\nab^{i_3}\psi_{\Hb}\nab^{i_4}(b^A-b^A_\Ke)
\eqrs&\sum_{i_1+i_2+i_3\leq i}(1+\nab^{i_1}(\tg,\tp))(\Om_\Ke^2+\Om_\Ke^2\nab^{i_2}\tg+\nab^{i_2}\tpHb)\nab^{i_3}\tb.
\end{split}
\end{equation}
Combining \eqref{b.eqn.3} and \eqref{b.eqn.4}, we obtain the desired conclusion of the proposition.
\end{proof}

\subsection{Reduced schematic null structure equations for the differences of the Ricci coefficients}\label{sec.rse.Ricci}

In this section, we derive the reduced schematic equations for the derivatives of the difference of the Ricci coefficients $\eta$, $\etab$, $\psi_H$, $\psi_{\Hb}$, $\omb$. Before we proceed, we first give the following preliminary result, which allows us to rewrite the difference of the Gauss curvature as combinations of angular covariant derivatives of the difference of $\gamma$. As a result, in deriving the reduced schematic equations, we only need to track the second (i.e.~the highest) derivative of $\tK$ and can rewrite the $\tK$ and $\nab\tK$ in terms of derivatives of $\tg$.
\begin{proposition}\label{K.diff.formula}
The difference of the Gauss curvatures $K-K_\Ke$ can be expressed in terms of $\gamma$, $\gamma_\Ke$, $\slashed{\Gamma}$ and $\slashed{\Gamma}_\Ke$ using the following formula:
\begin{equation*}
\begin{split}
K-K_\Ke=&\f12 ((\gamma^{-1})^{BC}-(\gamma_\Ke^{-1})^{BC})(\gamma_\Ke)_{BC} K_\Ke\\
&+\f12 \big(((\gamma^{-1})^{BC}-(\gamma_\Ke^{-1})^{BC})+(\gamma_\Ke^{-1})^{BC}\big)\times\big(\nab_A(\slashed{\Gamma}-\slashed{\Gamma}_\Ke)^{A}_{BC}-\nab_C(\slashed{\Gamma}-\slashed{\Gamma}_\Ke)^A_{BA}\\
&\quad-(\slashed{\Gamma}-\slashed{\Gamma}_\Ke)^A_{AD}(\slashed{\Gamma}-\slashed{\Gamma}_\Ke)^D_{BC}+(\slashed{\Gamma}-\slashed{\Gamma}_\Ke)^A_{CD}(\slashed{\Gamma}-\slashed{\Gamma}_\Ke)^D_{BA}\big).
\end{split}
\end{equation*}
\end{proposition}
\begin{proof}
Recall from \eqref{Gauss.def} that the Gauss curvature $K$ is given by
$$\gamma_{BC} K=\f{\rd}{\rd\th^A}\slashed{\Gamma}^{A}_{BC}-\f{\rd}{\rd\th^C}\slashed{\Gamma}^A_{BA}+\slashed{\Gamma}^A_{AD}\slashed{\Gamma}^D_{BC}-\slashed{\Gamma}^A_{CD}\slashed{\Gamma}^D_{BA}.$$
From this, we can derive the following formula:
\begin{equation*}
\begin{split}
&\gamma_{BC} K-(\gamma_\Ke)_{BC} K_\Ke\\
=&\f{\rd}{\rd\th^A}(\slashed{\Gamma}-\slashed{\Gamma}_\Ke)^{A}_{BC}-\f{\rd}{\rd\th^C}(\slashed{\Gamma}-\slashed{\Gamma}_\Ke)^A_{BA}+\slashed{\Gamma}^A_{AD}(\slashed{\Gamma}-\slashed{\Gamma}_\Ke)^D_{BC}+\slashed{\Gamma}^D_{BC}(\slashed{\Gamma}-\slashed{\Gamma}_\Ke)^A_{AD}\\
&-\slashed{\Gamma}^A_{CD}(\slashed{\Gamma}-\slashed{\Gamma}_\Ke)^D_{BA}-\slashed{\Gamma}^D_{BA}(\slashed{\Gamma}-\slashed{\Gamma}_\Ke)^A_{CD}-(\slashed{\Gamma}-\slashed{\Gamma}_\Ke)^A_{AD}(\slashed{\Gamma}-\slashed{\Gamma}_\Ke)^D_{BC}+(\slashed{\Gamma}-\slashed{\Gamma}_\Ke)^A_{CD}(\slashed{\Gamma}-\slashed{\Gamma}_\Ke)^D_{BA}\\
=&\nab_A(\slashed{\Gamma}-\slashed{\Gamma}_\Ke)^{A}_{BC}-\nab_C(\slashed{\Gamma}-\slashed{\Gamma}_\Ke)^A_{BA}-(\slashed{\Gamma}-\slashed{\Gamma}_\Ke)^A_{AD}(\slashed{\Gamma}-\slashed{\Gamma}_\Ke)^D_{BC}+(\slashed{\Gamma}-\slashed{\Gamma}_\Ke)^A_{CD}(\slashed{\Gamma}-\slashed{\Gamma}_\Ke)^D_{BA}.
\end{split}
\end{equation*}
Therefore,
\begin{equation*}
\begin{split}
&K-K_\Ke=\f12 \big((\gamma^{-1})^{BC}\gamma_{BC} K-(\gamma_{\Ke}^{-1})^{BC}(\gamma_\Ke)_{BC} K_\Ke\big)\\
=&\f12 ((\gamma^{-1})^{BC}-(\gamma_\Ke^{-1})^{BC})(\gamma_\Ke)_{BC} K_\Ke+\f12 ((\gamma^{-1})^{BC}-(\gamma_\Ke^{-1})^{BC})(\gamma_{BC} K-(\gamma_\Ke)_{BC} K_\Ke)\\
&+\f12 (\gamma_\Ke^{-1})^{BC}(\gamma_{BC} K-(\gamma_\Ke)_{BC} K_\Ke)\\
=&\f12 ((\gamma^{-1})^{BC}-(\gamma_\Ke^{-1})^{BC})(\gamma_\Ke)_{BC} K_\Ke\\
&+\f12 \big(((\gamma^{-1})^{BC}-(\gamma_\Ke^{-1})^{BC})+(\gamma_\Ke^{-1})^{BC}\big)\times\big(\nab_A(\slashed{\Gamma}-\slashed{\Gamma}_\Ke)^{A}_{BC}-\nab_C(\slashed{\Gamma}-\slashed{\Gamma}_\Ke)^A_{BA}\\
&\quad-(\slashed{\Gamma}-\slashed{\Gamma}_\Ke)^A_{AD}(\slashed{\Gamma}-\slashed{\Gamma}_\Ke)^D_{BC}+(\slashed{\Gamma}-\slashed{\Gamma}_\Ke)^A_{CD}(\slashed{\Gamma}-\slashed{\Gamma}_\Ke)^D_{BA}\big).
\end{split}
\end{equation*}
\end{proof}
As mentioned earlier, an easy consequence of the formula above is that $\tK$ and $\nab \tK$ can be controlled by $\tg$ and its first two angular covariant derivatives:
\begin{proposition}\label{K.diff.covariant}
For $i\leq 1$, $\tK$ obeys the following reduced schematic equation
$$\nab^i\tK\eqrs\sum_{i_1\leq i+2}\nab^{i_1}\tg.$$
\end{proposition}
\begin{proof}
First, apply Proposition \ref{K.diff.formula}. Next, we apply Proposition \ref{nab.diff} to express $\slashed{\Gamma}-\slashed{\Gamma}_\Ke$ in terms of the covariant derivative of $\tg$. Recall now that $\tg$ and $\nab\tg$ are in $L^\i$ and thus suppressed in the reduced schematic equations. The conclusion follows.
\end{proof}
We now derive the reduced schematic equations for the Ricci coefficients and their first and second angular covariant derivatives. We begin with $\etab$:
\begin{proposition}\label{etab.eqn}
For $i\leq 2$, $\nab^i\widetilde{\etab}$ obeys the following reduced schematic equation:
$$\nab_3\nab^i\widetilde{\etab}\eqrs\mathcal F_3'.$$
\end{proposition}
\begin{proof}
We first recall the schematic equation \eqref{etab.S.2}
$$\nab_3\etab \eqs\sum_{i_1+i_2=1}\psi^{i_1}\nab^{i_2} \psi_{\Hb}.$$
Applying Proposition \ref{schematic.3}, we obtain
\begin{equation}\label{etab.eqn.1}
\begin{split}
&\nab_3\nab^i(\etab-\etab_\Ke)\\
\eqrs &\sum_{i_1+i_2\leq i}(1+\nab^{i_1}\tg+\nab^{i_1-1}\tp)\nab^{i_2}(G-G_\Ke)\\
&+\sum_{i_1+i_2+i_3\leq i}(1+\nab^{i_1}\tg+\nab^{i_1-1}\tp)(\Om_\Ke^2\nab^{i_2}\tg+\nab^{i_2}\tpHb)\nab^{i_3}\etab_\Ke\\
&+\sum_{i_1+i_2+i_3\leq i}(1+\nab^{i_1}\tg+\nab^{i_1-1}\tp)(\Om_\Ke^2+\Om_\Ke^2\nab^{i_2}\tg+\nab^{i_2}\tpHb)\nab^{i_3}(\etab-\etab_\Ke),
\end{split}
\end{equation}
where 
$$G\eqs\sum_{i_1+i_2=1}\psi^{i_1}\nab^{i_2} \psi_{\Hb}.$$
First, notice that for $j\leq 2$
\begin{equation}\label{etab.eqn.G}
\begin{split}
\nab^j(G-G_\Ke)\eqrs\sum_{j_1+j_2\leq j+1}(1+\nab^{j_1}\tg+\nab^{j_1-1}\tp)(\Om_\Ke^2\nab^{j_2-1}\tp+\Om_\Ke^2\nab^{j_2}\tg+\nab^{j_2}\tpHb).
\end{split}
\end{equation}
We have used here that if $j\leq 2$, in the terms
$$\sum_{j_1+j_2+j_3\leq j+1}\nab^{j_1}\tg\nab^{j_2-1}\tp\nab^{j_3}\tpHb,$$
we must either have $j_1\leq 1$ or $j_2-1\leq 0$. Since $\tg$, $\nab\tg$, $\tp$ are in $L^\i$ (by Corollary~\ref{Linfty}), according to our convention for the reduced schematic equations, we can write these terms as
$$\sum_{j_1+j_2\leq j+1}(\nab^{j_1}\tg+\nab^{j_1-1}\tp)\nab^{j_2}\tpHb.$$
Now \eqref{etab.eqn.G} implies that for $i\leq 2$
\begin{equation*}
\begin{split}
&\sum_{i_1+i_2\leq i}(1+\nab^{i_1}\tg+\nab^{i_1-1}\tp)\nab^{i_2}(G-G_\Ke)\\
\eqrs &\sum_{i_1+i_2\leq 3}(1+\nab^{i_1}\tg+\nab^{i_1-1}\tp)(\Om_\Ke^2\nab^{i_2-1}\tp+\Om_\Ke^2\nab^{i_2}\tg+\nab^{i_2}\tpHb).
\end{split}
\end{equation*}
Returning to \eqref{etab.eqn.1}, we thus have
\begin{equation*}
\begin{split}
&\nab_3\nab^i(\etab-\etab_\Ke)\\
\eqrs &\sum_{i_1+i_2\leq 3}(1+\nab^{i_1}\tg+\nab^{i_1-1}\tp)(\Om_\Ke^2\nab^{i_2-1}\tp+\Om_\Ke^2\nab^{i_2}\tg+\nab^{i_2}\tpHb)\\
&+\sum_{i_1+i_2+i_3\leq i}(1+\nab^{i_1}\tg+\nab^{i_1-1}\tp)(\Om_\Ke^2\nab^{i_2}\tg+\nab^{i_2}\tpHb)(1+\nab^{i_3}\tg)\\
&+\sum_{i_1+i_2+i_3\leq i}(1+\nab^{i_1}\tg+\nab^{i_1-1}\tp)(\Om_\Ke^2+\Om_\Ke^2\nab^{i_2}\tg+\nab^{i_2}\tpHb)\nab^{i_3}\tp.
\end{split}
\end{equation*}
Finally, using the $L^\infty$ estimates in Corollary~\ref{Linfty}, it can be checked that all these terms are contained in $\mathcal F_3'$ (cf.~the argument in the proof of Proposition~\ref{gamma.eqn}).
\end{proof}

The reduced schematic equation for $\widetilde{\mub}$ (recall \eqref{top.quantities.def}) can be derived in a similar way as Proposition~\ref{etab.eqn}, using \eqref{mub.S} instead of \eqref{etab.S.2}. It is important to note that
\begin{itemize}
\item $\mub$ is defined so that there are no terms containing first derivatives of curvature in the equation $\nab_3\mub$;
\item there are some extra terms so that we need $\mathcal F_3$ instead of $\mathcal F_3'$ on the right hand side.
\end{itemize}
We omit the details of the proof.
\begin{proposition}\label{mub.eqn}
For $i\leq 2$, $\nab^i\widetilde{\mub}$ satisfies the following reduced schematic equation:
$$\nab_3\nab^i\widetilde{\mub}=\mathcal F_3.$$
\end{proposition}

We now consider the $\nab_3$ equation for $\tpH$, i.e.~the terms $\ttrch$ and $\widetilde{\chih}$.
\begin{proposition}\label{tpH.eqn}
For $i\leq 2$, $\nab^i\tpH$ obeys the following reduced schematic equation:
$$\nab_3(\nab^i(\Om^2_\Ke\tpH))\eqrs\mathcal F_1'.$$
\end{proposition}
\begin{proof}
We recall the schematic equation \eqref{tpH.S.2}
$$\nab_3\psi_H-2\omegab\psi_H \eqs K+\nab\eta+\psi\psi+\psi_{\Hb}\psi_H.$$
Using (cf.~\eqref{gauge.con})
$$\nab_3\log\Om=-\omegab,$$
we derive the following equation for $\Om^2_\Ke\psi_H$:
$$\nab_3(\Om^2_\Ke\psi_H) \eqs\Om^2_\Ke(K+\nab\eta+\psi_{\Hb}\psi_H+\psi\psi+\widetilde{\omb}\psi_H).$$
We now apply Proposition~\ref{schematic.3} to obtain
\begin{equation}\label{tpH.eqn.1}
\begin{split}
&\nab_3\nab^i(\Om_\Ke^2\tpH)\\
\eqrs &\sum_{i_1+i_2\leq i}(1+\nab^{i_1}\tg+\nab^{i_1-1}\tp)\nab^{i_2}(G-G_{\Ke})\\
&+\sum_{i_1+i_2+i_3\leq i}(1+\nab^{i_1}\tg+\nab^{i_1-1}\tp)(\Om_\Ke^2\nab^{i_2}\tg+\nab^{i_2}\tpHb)\nab^{i_3}(\Om_\Ke^2(\psi_H)_\Ke)\\
&+\sum_{i_1+i_2+i_3\leq i}(1+\nab^{i_1}\tg+\nab^{i_1-1}\tp)(\Om_\Ke^2+\Om_\Ke^2\nab^{i_2}\tg+\nab^{i_2}\tpHb)\nab^{i_3}(\Om^2_\Ke\tpH),
\end{split}
\end{equation}
where in this context $G-G_\Ke$ is given schematically by
$$\Om_\Ke^2((K+\nab\eta+\psi_{\Hb}\psi_H+\psi\psi)-(K+\nab\eta+\psi_{\Hb}\psi_H+\psi\psi)_\Ke+\widetilde{\omb}\psi_H).$$
Therefore, $\sum_{i_1+i_2\leq i}(1+\nab^{i_1}\tg+\nab^{i_1-1}\tp)\nab^{i_2}(G-G_{\Ke})$ can be expressed as
\begin{equation*}
\begin{split}
&\sum_{i_1+i_2\leq i}(1+\nab^{i_1}\tg+\nab^{i_1-1}\tp)\nab^{i_2}(G-G_{\Ke})\\
\eqrs&\sum_{i_1+i_2\leq i}\Om_\Ke^2(1+\nab^{i_1}\tg+\nab^{i_1-1}\tp)(\nab^{i_2}(\tK,\tp,\tpHb,\widetilde{\omb})+\nab^{i_2+1}(\tg,\widetilde{\eta})+\Om_\Ke^2\nab^{i_2}\tpH)\\
&+\sum_{i_1+i_2\leq i}\Om_\Ke^2(1+\nab^{i_1}\tg+\nab^{i_1-1}\tp)\nab^{i_2}\tpH\nab^{i_3}(\tpHb,\widetilde{\omb}).
\end{split}
\end{equation*}
Substituting this into \eqref{tpH.eqn.1} and using the fact that
$$\nab^j(\Om_\Ke^2(\psi_H)_\Ke) \eqrs \Om_\Ke^2(1+\sum_{j_1\leq j}(\nab^{j_1}\tpH+\nab^{j_1}\tg)),\quad \nab^j(\Om^2_\Ke\tpH)\eqrs\Om_\Ke^2(\sum_{j_1\leq j}(\nab^{j_1}\tpH+\nab^{j_1}\tg)),$$
we obtain
\begin{equation*}
\begin{split}
&\nab_3\nab^i(\Om_\Ke^2\tpH)\\
\eqrs&\sum_{i_1+i_2\leq i}\Om_\Ke^2(1+\nab^{i_1}\tg+\nab^{i_1-1}\tp)(\nab^{i_2}(\nab^{i_2}(\tK,\tp,\tpHb,\widetilde{\omb})+\nab^{i_2+1}(\tg,\widetilde{\eta})+\Om_\Ke^2\nab^{i_2}\tpH)\\
&+\sum_{i_1+i_2\leq i}\Om_\Ke^2(1+\nab^{i_1}\tg+\nab^{i_1-1}\tp)\nab^{i_2}\tpH\nab^{i_3}(\tpHb,\widetilde{\omb})\\
&+\sum_{i_1+i_2+i_3\leq i}\Om_\Ke^2(1+\nab^{i_1}\tg+\nab^{i_1-1}\tp)(\Om_\Ke^2\nab^{i_2}\tg+\nab^{i_2}\tpHb)(1+\nab^{i_3}\tpH+\nab^{i_3}\tg)\\
&+\sum_{i_1+i_2+i_3\leq i}\Om_\Ke^2(1+\nab^{i_1}\tg+\nab^{i_1-1}\tp)(\Om_\Ke^2+\Om_\Ke^2\nab^{i_2}\tg+\nab^{i_2}\tpHb)(\nab^{i_3}\tpH+\nab^{i_3}\tg).
\end{split}
\end{equation*}
One finally checks that all these terms are acceptable using Corollary~\ref{Linfty}. \qedhere
\end{proof}

We now turn to the Ricci coefficients satisfying a $\nab_4$ equation. We begin with the reduced schematic equations for $\widetilde{\eta}$:
\begin{proposition}\label{eta.eqn}
For $i\leq 2$, $\nab^i\widetilde{\eta}$ satisfies the following reduced schematic equation:
$$\nab_4\nab^i\widetilde{\eta}\eqrs\mathcal F_2'.$$
\end{proposition}
\begin{proof}
Since $\eta$ is a rank $1$ tensor, we apply \eqref{schematic.4.0.tensor} in Proposition \ref{schematic.4} to obtain
\begin{equation}\label{eqn.eta.diff.1}
\begin{split}
\nab_4\nab^i\widetilde{\eta}\eqrs&\sum_{i_1+i_2\leq i}(1+\nab^{i_1}\tg+\nab^{i_1-1}\tp)\nab^{i_2}\nab_4(F-F_\Ke)\\
&+\sum_{i_1+i_2+i_3\leq i}(1+\nab^{i_1}\tg+\nab^{i_1-1}\tp)\nab^{i_2}\tg\nab^{i_3}(\nab_4)_\Ke\eta_\Ke\\
&+\sum_{i_1+i_2+i_3\leq i}\Om_\Ke^{-2}(1+\nab^{i_1}\tg+\nab^{i_1-1}\tp)\nab^{i_2}\tb\nab^{i_3}(\nab_\Ke)\eta_\Ke\\
&+\sum_{i_1+i_2+i_3\leq i}(1+\nab^{i_1}\tg+\nab^{i_1-1}\tp)(1+\nab^{i_2}(\tg,\tpH))\nab^{i_3}\widetilde{\eta}\\
&+\sum_{i_1+i_2+i_3\leq i}(1+\nab^{i_1}\tg+\nab^{i_1-1}\tp)\nab^{i_2}(\tg,\tpH)\nab^{i_3}\eta_\Ke\\
&+\sum_{i_1+i_2+i_3\leq i}\Om_\Ke^{-2}(1+\nab^{i_1}\tg+\nab^{i_1-1}\tp)\nab^{i_2+1}\tb\nab^{i_3}\eta_\Ke,
\end{split}
\end{equation}
where according to \eqref{eta.S.2},
$$F\eqs \sum_{i_1+i_2=1}\psi^{i_1}\nab^{i_2}\psi_H.$$
Note that by Propositions~\ref{nab.diff} and \ref{Kerr.der.Ricci.bound}, for $i\leq 2$,
$$\nab^i\eta_\Ke \eqrs 1+\sum_{i'\leq i}\nab^{i'}\tg$$
Using also Corollary~\ref{Linfty}, one then checks that indeed all terms on the right hand side of \eqref{eqn.eta.diff.1} can be written as $\mathcal F_2'$.
\end{proof}
A similar equation is obeyed by $\widetilde{\mu}$ (recall \eqref{top.quantities.def}), except that we use \eqref{mu.S} instead of \eqref{eta.S.2}. We will omit the proof, except for noting that
\begin{itemize}
\item $\mu$ is defined such that in the equation $\nab_4\mu$, we do not have terms which are derivatives of null curvature components;
\item there are some extra terms too so that we need to use $\mathcal F_2$ instead of $\mathcal F_2'$ on the right hand side.
\end{itemize}
\begin{proposition}\label{mu.eqn}
For $i\leq 2$, $\nab^i\widetilde{\mu}$ satisfies the following reduced schematic equation:
$$\nab_4\nab^i\widetilde{\mu}\eqrs\mathcal F_2.$$
\end{proposition}
We use Proposition~\ref{4.diff.prop} to also derive the reduced schematic equations for $\nab^i\tpHb$, $\nab^i\widetilde{\omb}$ and $\nab\tombs$:
\begin{proposition}\label{tpHb.eqn}
For $i\leq 2$, $\nab^i\tpHb$ and $\nab^i\widetilde{\omb}$ satisfy the following reduced schematic equations:
$$\nab_4\nab^i\tpHb\eqrs\mathcal F_4',\quad\nab_4\nab^i\widetilde{\omb}\eqrs\mathcal F_4'.$$
Moreover, $\nab\tombs$ (recall \eqref{top.quantities.def}) obeys the following reduced schematic equation:
$$\nab_4\nab\tombs\eqrs\mathcal F_4.$$
\end{proposition}
\begin{proof}
We begin by considering $\tpHb$. Recall \eqref{tpHb.S.2}
$$\nab_4 \psi_{\Hb} \eqs K+ \nab\etab + \psi\psi+\psi_H \psi_{\Hb}.$$ 
By Proposition~\ref{4.diff.prop}, this implies
\begin{equation*}
\begin{split}
\nab_4\nab^i\tpHb \eqrs &\sum_{i_1+i_2\leq i}(1+\nab^{i_1}\tg+\nab^{i_1-1}\tp)\nab^{i_2}\nab_4(F-F_\Ke)\\
&+\sum_{i_1+i_2+i_3\leq i}(1+\nab^{i_1}\tg+\nab^{i_1-1}\tp)\nab^{i_2}\tg\nab^{i_3}(\nab_4)_\Ke(\psi_{\Hb})_\Ke\\
&+\sum_{i_1+i_2+i_3\leq i}\Om_\Ke^{-2}(1+\nab^{i_1}\tg+\nab^{i_1-1}\tp)\nab^{i_2}\tb\nab^{i_3}(\nab_\Ke)(\psi_{\Hb})_\Ke\\
&+\sum_{i_1+i_2+i_3\leq i}(1+\nab^{i_1}\tg+\nab^{i_1-1}\tp)(1+\nab^{i_2}(\tg,\tpH))\nab^{i_3}\tpHb\\
&+\sum_{i_1+i_2+i_3\leq i}(1+\nab^{i_1}\tg+\nab^{i_1-1}\tp)\nab^{i_2}(\tg,\tpH)\nab^{i_3}(\psi_{\Hb})_\Ke\\
&+\sum_{i_1+i_2+i_3\leq i}\Om_\Ke^{-2}(1+\nab^{i_1}\tg+\nab^{i_1-1}\tp)\nab^{i_2+1}\tb\nab^{i_3}(\psi_{\Hb})_\Ke,
\end{split}
\end{equation*}
where 
$$F \eqs K+ \nab\etab + \psi\psi+\psi_H \psi_{\Hb}.$$
By \eqref{nabgK.nabpsiHbK}, 
$$\nab^i(\psi_{\Hb})_\Ke \eqrs \sum_{i'\leq i} \Om_\Ke^2 \nab^{i'}\tg+ \Om_\Ke^2.$$
In a similar manner, we have
$$\nab^i(\nab_\Ke)(\psi_{\Hb})_\Ke \eqrs \sum_{i'\leq i} \Om_\Ke^2 \nab^{i'}\tg+ \Om_\Ke^2.$$
Moreover, using \eqref{tpHb.S.2} applied to the Kerr background, and applying Propositions~\ref{nab.diff} and \ref{Kerr.der.Ricci.bound} to the right hand side, one obtains
$$\nab^i(\nab_4)_\Ke(\psi_{\Hb})_\Ke \eqrs \sum_{i'\leq i} \nab^{i'}\tg + 1.$$
Therefore, using also Corollary~\ref{Linfty}, one checks that all terms are acceptable.

The reduced schematic equations for $\nab^i\widetilde{\omb}$ can be derived in a completely identical manner since according to \eqref{omb.S} and \eqref{tpHb.S.2}, $\nab_4\omb$ (schematically) has a only subset of the terms in $\nab_4\psi_{\Hb}$.

Finally, to obtain the reduced schematic equation for $\nab\tombs$, we first recall \eqref{ombs.S} and then apply Proposition~\ref{4.diff.prop} and argue similarly as above, except noticing that because of the additional higher order terms, we need to use $\mathcal F_4$ instead of $\mathcal F_4'$ on the right hand side. \qedhere
\end{proof}

\subsection{Additional reduced schematic equations for the highest order derivatives of the Ricci coefficients}\label{sec.rse.Ricci.add}

We now consider the highest order derivatives for $\widetilde{\trch}$ and $\widetilde{\trchb}$. These equations will be used together with the Codazzi equations in \eqref{null.str3} to obtain the highest order estimates for $\tpH$ and $\tpHb$. Recall that in the previous subsection, we have derived the reduced schematic equations for $\nab_3\nab^i\tpH$ and $\nab_4\nab^i\tpHb$. On the other hand, as in the case of the local weak null singularity \cite{LukWeakNull}, at the level of three derivatives, the use of these equations leads to a loss of derivatives. Instead, we will derive the reduced schematic equations for $\nab_4\nab^3\widetilde{\trch}$ and $\nab_3\nab^3\widetilde{\trchb}$. Notice that the inhomogeneous terms in these equations cannot be written in terms of $\mathcal F_1$, $\mathcal F_2$, $\mathcal F_3$ and $\mathcal F_4$ and are therefore treated independently.

First, we consider $\nab_3\nab^3\widetilde{\trchb}$:
\begin{proposition}\label{nab3trchb.eqn}
$\nab^3\ttrchb$ obeys the following reduced schematic equation:
\begin{equation*}
\begin{split}
\nab_3(\Om_\Ke^{-2}\nab^3\ttrchb)
\eqrs \Om_{\Ke}^{-2}\sum_{i_1+i_2+i_3\leq 3}(1+\nab^{i_1}\tg+\nab^{i_1-1}\tp)(\nab^{i_2}\tpHb+\Om_\Ke^2)(\nab^{i_3}(\widetilde{\omb},\tpHb)+\Om_\Ke^2\nab^{i_3}\tg).
\end{split}
\end{equation*}
\end{proposition}
\begin{proof}
We begin with the following equation in \eqref{null.str1}:
$$\nab_3 \trchb+\frac 12 (\trchb)^2=-2\omegab \trchb-|\chibh|^2.$$
We now apply Proposition~\ref{schematic.3} to the above equation except we keep the term\footnote{Notice that we have crucially used the fact that according to \eqref{gauge.con}, $\eta-\zeta=0$. Hence, according to Proposition~\ref{commute}, there are no other terms which involve $\omb$ or $\omb_\Ke$.} $2\nab^3(\omegab \trchb-\omb_\Ke\trchb_\Ke)$ on the left hand side. Using moreover the fact that $\trchb$ is a scalar, we obtain the following reduced schematic equation:
\begin{equation}\label{nab3trchb.eqn.1}
\begin{split}
&\nab_3\nab^3_{ABC}\widetilde{\trchb}+2\nab^3_{ABC}(\omb \trchb-\omb_\Ke(\trchb)_\Ke)\\
\eqrs &\sum_{i_1+i_2\leq 3}(1+\nab^{i_1}\tg+\nab^{i_1-1}\tp)\nab^{i_2}(G-G_\Ke)\\
&+\sum_{i_1+i_2+i_3\leq 3}(1+\nab^{i_1}\tg+\nab^{i_1-1}\tp)(\Om_\Ke^2+\Om_\Ke^2\nab^{i_2}\tg+\nab^{i_2}\tpHb)\nab^{i_3}\widetilde{\trchb},
\end{split}
\end{equation}
where $G$ is given by
$$G=-\f 12(\trchb)^2-|\chibh|^2.$$
It can now be checked using Corollary~\ref{Linfty} that the right hand side of \eqref{nab3trchb.eqn.1} can be expressed as $\Om_\Ke^2$ multiplied by the expression on the right hand side of the reduced schematic equation in the statement of the proposition.

Next, we handle the terms on the left hand side of \eqref{nab3trchb.eqn.1}.  We now make the following simple observation. According to Propositions \ref{Kerr.Ricci.bound} and \ref{Kerr.der.Ricci.bound}, while $\omb_\Ke$ approaches a constant near the Cauchy horizon, the tensor $(\nab^i\omb)_\Ke$ obeys the bound 
$$\|(\nab^i\omb)_\Ke\|_{L^\infty(S_{u,\ub})}\ls \|\Om_\Ke^2\|_{L^\infty(S_{u,\ub})}$$
as long as $i\geq 1$. This in turn implies the reduced schematic equation that
$$\nab^i(\omb_\Ke)\eqrs \Om_\Ke^2 (1+\sum_{i_1\leq 2}\nab^{i_1}\tg)$$
whenever $1\leq i\leq 3$. Therefore, we can write the following reduced schematic equation
\begin{equation*}
\begin{split}
&\nab^i(\omb \trchb-\omb_\Ke\trchb_\Ke)-\omb_{\Ke}\nab^i\ttrchb\\
\eqrs &\sum_{\substack{i_1+i_2\leq i\\ i_1\geq 1}}\nab^{i_1}\omb_{\Ke}\nab^{i_2}\ttrchb+\sum_{i_1+i_2\leq i} \nab^{i_1}\widetilde{\omb}\nab^{i_2}\trchb_\Ke+\sum_{i_1+i_2\leq i}\nab^{i_1}\widetilde{\omb}\nab^{i_2}\ttrchb\\
\eqrs &\sum_{i_1+i_2\leq i}(\Om_\Ke^2+\Om_\Ke^2\nab^{i_1}\tg+\nab^{i_1}\tpHb)\nab^{i_2}(\widetilde{\omb},\tpHb)
\end{split}
\end{equation*}
for $i\leq 3$. Now, we substitute this into \eqref{nab3trchb.eqn.1} to obtain the following reduced schematic equation:
\begin{equation*}
\begin{split}
\nab_3\nab^3\ttrchb+2\omb_\Ke\nab^3\ttrchb
\eqrs \sum_{i_1+i_2+i_3\leq 3}(1+\nab^{i_1}\tg+\nab^{i_1-1}\tp)(\nab^{i_2}\tpHb+\Om_\Ke^2)(\nab^{i_3}(\widetilde{\omb},\tpHb)+\Om_\Ke^2\nab^{i_3}\tg).
\end{split}
\end{equation*}
Finally, we recall that $e_3=\f{\rd}{\rd u}$ represents the same vector field in $(\mathcal U_{\ub_f},g)$ as in the background Kerr solution. Therefore, we can use that fact (cf.~\eqref{gauge.con})
$$\omb_{\Ke}=-\nab_3(\log\Om_\Ke)$$
to obtain
$$\nab_3\nab^3\ttrchb+2\omb_\Ke\nab^3\ttrchb=\Om_\Ke^2\nab_3(\Om_\Ke^{-2}\nab^3\ttrchb)$$ 
and derive the final conclusion.
\end{proof}

We now consider the equation for $\nab_4\nab^3\widetilde{\trch}$. Unlike for $\nab_3\nab^3\widetilde{\trchb}$, we now do not need to treat the coefficients containing $\omb_\Ke$ and the derivation is more straightforward using Proposition \ref{schematic.4}.
\begin{proposition}\label{nab3trch.eqn}
$\nab^3\widetilde{\trch}$ obeys the following reduced schematic equation:
\begin{equation*}
\nab_4\nab^3\ttrch\eqrs\sum_{i_1+i_2+i_3\leq 3}(1+\nab^{i_1}\tg+\nab^{i_1-1}\tp)(1+\nab^{i_2}\tpH)(\nab^{i_3}(\tpH,\tg)+\Om_\Ke^{-2}\nab^{i_3}\tb).
\end{equation*}
\end{proposition}
\begin{proof}
Recall that we have the following equation from \eqref{null.str1}
$$\nab_4 \trch+\frac 12 (\trch)^2=-|\chih|^2.$$
We now apply Proposition \ref{schematic.4} to the above equation, noting that $\trch$ is a scalar\footnote{We remark that it is important that we do not have $\nab^4\tb$ in this equation. This can potentially happen since we are taking the third angular derivatives (for the first time!)~of a difference quantity that satisfies a $\nab_4$ equation. The crucial point here is that $\trch$ is a \emph{scalar} and therefore we do not see the Christoffel symbols associated to $\nab_4$ in the equation for the difference quantity $\ttrch$.}:
\begin{equation}\label{schematic.nab4.trch}
\begin{split}
\nab_4\nab^3\ttrch
\eqrs &\sum_{i_1+i_2\leq 3}(1+\nab^{i_1}\tg+\nab^{i_1-1}\tp)\nab^{i_2}(F-F_\Ke)\\
&+\sum_{i_1+i_2+i_3\leq 3}(1+\nab^{i_1}\tg+\nab^{i_1-1}\tp)\nab^{i_2}\tg\nab^{i_3}(\nab_4)_\Ke\trch_\Ke\\
&+\sum_{i_1+i_2+i_3\leq 3}\Om_\Ke^{-2}(1+\nab^{i_1}\tg+\nab^{i_1-1}\tp)\nab^{i_2}\tb\nab^{i_3}(\nab_\Ke)\trch_\Ke\\
&+\sum_{i_1+i_2+i_3\leq 3}(1+\nab^{i_1}\tg+\nab^{i_1-1}\tp)(1+\nab^{i_2}(\tg,\tpH))\nab^{i_3}\ttrch,
\end{split}
\end{equation}
where 
$$F=-\f12 (\trch)^2-|\chih|^2.$$
Now, one can check using Corollary~\ref{Linfty} that all the terms in \eqref{schematic.nab4.trch} are acceptable.
\end{proof}

\subsection{Reduced schematic Bianchi equations for the difference of the renormalised null curvature components}\label{sec.rse.curv}
Finally, we derive the reduced schematic Bianchi equations for the second derivatives of the difference of renormalised null curvature components. They will be used in Section~\ref{sec.energy} to obtain energy estimates. First, we have the following reduced schematic equations for $\nab^2\widetilde{\beta}$, $\nab^2\tK$ and $\nab^2\widetilde{\sigmac}$:
\begin{proposition}\label{curv.red.sch.1}
$\widetilde{\beta}$, $\tK$ and $\widetilde{\sigmac}$ obey the following reduced schematic equations:
\begin{equation*}
\begin{split}
\nab_3\nab^2_{AB}(\Om\Om_\Ke\widetilde{\beta}_C)+\slashed{\nabla}_C(\Om\Om_\Ke \nab^2_{AB}\tK)  -\in_{CD}\slashed{\nabla}^D(\Om\Om_\Ke\nab^2_{AB}\widetilde{\sigmac})
 \eqrs& \:\mathcal F_1,\\
\nab_4\nab^2_{AB}\widetilde{\sigmac}+\nab^C\nab^2_{AB}(\in_{CD}\widetilde{\beta}^D)\eqrs &\:\mathcal F_2,\\
\nab_4\nab^2_{AB} \tK+\nab^C\nab^2_{AB}\widetilde{\beta}_C\eqrs &\:\mathcal F_2.
\end{split}
\end{equation*}
\end{proposition}
\begin{proof}
The main difference in deriving these reduced schematic equations compared to those in the previous subsections (for the metric components and Ricci coefficients) is that we need to keep the terms with the highest order angular derivatives with the precise structure and coefficient. This is necessary because these highest order terms cannot be treated as error terms, for otherwise this will result in a loss of derivatives.

We begin with the following schematic equations from \eqref{eq:null.Bianchi3}:
\begin{equation}\label{curv.1.eqn.0}
\begin{split}
\nab_3\beta+\slashed{\nabla} K  -^*\slashed{\nabla}\sigmac -2\omb\beta\eqs& \:\sum_{i_1+i_2+i_3=1}\psi^{i_1}\nab^{i_2}\psi_{\Hb} \nab^{i_3}\psi_H+\psi K+\sum_{i_1+i_2=1} \psi^{i_1}\psi\nab^{i_2}\psi,\\
\nab_4\sigmac+\div^*\beta\eqs&\:\sum_{i_1+i_2+i_3=1}\psi^{i_1}\nab^{i_2}\psi\nab^{i_3}\psi_H,\\
\nab_4 K+\div\beta \eqs&\:\psi_H K+\sum_{i_1+i_2+i_3=1}\psi^{i_1}\nab^{i_2}\psi\nab^{i_3}\psi_H.
\end{split}
\end{equation}
For the first equation, we treat the term $-2\omb\beta$ in a similar way as the $-2\omb\psi_H$ in Proposition \ref{tpH.eqn}, except that for technical reasons that we become clear later (see\footnote{More precisely, this is done so that the cancellation that is needed in Proposition \ref{EE.1} can be manifestly seen.} Proposition \ref{EE.1}), we use the weight $\Om\Om_\Ke$ instead of $\Om_\Ke^2$. More precisely, we can use the relations (cf.~\eqref{gauge.con})
$$\nab_3\log\Om=-\omb,\quad (\nab_3\log\Om)_\Ke=-\omb_\Ke$$
together with the schematic Codazzi equation \eqref{Codazzi.S.1} (see also \eqref{sch.Codazzi}) to rewrite the equation schematically as
$$\nab_3(\Om\Om_\Ke\beta)+\Om\Om_\Ke\slashed{\nabla} K  -\Om\Om_\Ke{ }^*\slashed{\nabla}\sigmac \eqs \Om\Om_\Ke\sum_{i_1+i_2+i_3=1} \left(\psi^{i_1} \nab^{i_2}(\psi_{\Hb},\widetilde{\omb})\nab^{i_3}\psi_H+ \psi^{i_1}\psi\nab^{i_2}\psi+\psi K\right).$$
We now apply Proposition \ref{schematic.3} with $i=2$, except we also keep the highest order terms in the renormalised null curvature components $K$ and $\sigmac$ on the left hand side. More precisely, we have
\begin{equation}\label{curv.1.eqn.1}
\begin{split}
&\nab_3\nab^2_{AB}(\Om\Om_\Ke\widetilde{\beta}_C)+\nab^2_{AB}(\Om\Om_\Ke(\slashed{\nabla}_C K-(\slashed{\nabla}_C K)_\Ke))  -\nab^2_{AB}(\Om\Om_\Ke({ }^*\slashed{\nabla}_C\sigmac-({ }^*\slashed{\nabla}_C\sigmac)_\Ke))\\
\eqrs &\sum_{\substack{i_1+i_2\leq 2\\i_2\leq 1}}\Om_\Ke^2(1+\nab^{i_1}\tg+\nab^{i_1-1}\tp)\nab^{i_2+1}(\tK,\widetilde{\sigmac})\\
&+\sum_{i_1+i_2\leq 2}\Om_\Ke^2(1+\nab^{i_1}\tg+\nab^{i_1-1}\tp)\nab^{i_2}(G-G_{\Ke})\\
&+\sum_{i_1+i_2+i_3\leq 2}\Om_\Ke^2(1+\nab^{i_1}\tg+\nab^{i_1-1}\tp)(\Om_\Ke^2\nab^{i_2}\tg+\nab^{i_2}\tpHb)\nab^{i_3}\beta_\Ke\\
&+\sum_{i_1+i_2+i_3\leq 2}\Om_\Ke^2(1+\nab^{i_1}\tg+\nab^{i_1-1}\tp)(\Om_\Ke^2+\Om_\Ke^2\nab^{i_2}\tg+\nab^{i_2}\tpHb)\nab^{i_3}\widetilde{\beta},
\end{split}
\end{equation}
where $$G \eqs\sum_{i_1+i_2+i_3=1}\psi^{i_1}\nab^{i_2}(\psi_{\Hb},\widetilde{\omb}) \nab^{i_3}\psi_H+\psi K+\sum_{i_1+i_2=1} \psi^{i_1}\psi\nab^{i_2}\psi.$$

We now consider the terms involving $K$ and $\sigmac$ on the left hand side of \eqref{curv.1.eqn.1}. Using that $K$ is a scalar and expressing the commutators $[\nab_A,\nab_B]$, $[\nab_B,\nab_C]$ in terms of $K$, we obtain
\begin{equation*}
\begin{split}
&\nab^2_{AB}(\Om\Om_\Ke(\slashed{\nabla}_C K-(\slashed{\nabla}_C K)_\Ke))-\nab_C(\Om\Om_\Ke\nab^2_{AB} \tK)\\
=&\nab^2_{AB}(\Om\Om_\Ke\nab_C \tK)-\nab_C(\Om\Om_\Ke\nab^2_{AB} \tK)\\
=&\nab^2_{AB}(\Om\Om_\Ke)\nab_C\tK+2\nab_{(A}(\Om\Om_\Ke)\nab^2_{B)C}\tK-\nab_C(\Om\Om_\Ke)\nab^2_{AB}\tK+\Om\Om_\Ke(\nab_{ABC}^3\tK-\nab_{CAB}^3\tK)\\
=&\nab^2_{AB}(\Om\Om_\Ke)\nab_C\tK+2\nab_{(A}(\Om\Om_\Ke)\nab^2_{B)C}\tK-\nab_C(\Om\Om_\Ke)\nab^2_{AB}\tK+\Om\Om_\Ke K(\gamma_{BC}\nab_A\tK-\gamma_{AB}\nab_C\tK).
\end{split}
\end{equation*}
Using Propositions~\ref{nab.diff} and \ref{K.diff.formula}, we obtain the following reduced schematic equation
\begin{equation}\label{curv.1.eqn.2}
\begin{split}
\nab^2_{AB}(\Om\Om_\Ke(\slashed{\nabla}_C K-(\slashed{\nabla}_C K)_\Ke))-\nab_C(\Om\Om_\Ke\nab^2_{AB} \widetilde K)
\eqrs &\sum_{i_1+i_2\leq 3}\Om_\Ke^2(1+\nab^{i_1}\tg)\nab^{\min\{i_2,2\}}\widetilde K.
\end{split}
\end{equation}
For the highest order term in $\sigmac$, we have an additional contribution arising from the difference of ${ }^*\nab$. Hence, we have
\begin{equation*}
\begin{split}
&\nab^2_{AB}(\Om\Om_\Ke({\in}_{C}{}^{D}\slashed{\nabla}_{D} \sigmac-({\in}_{C}{}^{D}\slashed{\nabla}_{D} \sigmac)_\Ke))-{\in}_{C}{}^{D}\slashed{\nabla}_{D}(\Om\Om_\Ke\nab^2_{AB}\widetilde{\sigmac})\\
=&\nab^2_{AB}(\Om\Om_\Ke){\in}_{C}{}^{D}\slashed{\nabla}_{D} \widetilde{\sigmac}+\nab^2_{AB}(\Om\Om_\Ke)({\in}-{\in_\Ke})_{C}{}^{D}(\slashed{\nabla}_{D} \sigmac)_\Ke\\
&+2\nab_{(A}(\Om\Om_\Ke)\nab_{B)}({\in}_{C}{}^{D}\slashed{\nabla}_{D} \widetilde{\sigmac})+2\nab_{(A}(\Om\Om_\Ke)\nab_{B)}(({\in}-{\in_\Ke})_{C}{}^{D}(\slashed{\nabla}_{D} \sigmac)_\Ke)-{\in}_{C}{}^{D}\slashed{\nabla}_{D}(\Om\Om_\Ke)\nab^2_{AB}\widetilde{\sigmac}\\
&+\Om\Om_\Ke\nab^2_{AB}(({\in}-{\in_\Ke})_{C}{}^{D}(\slashed{\nabla}_{D} \widetilde\sigmac)_\Ke)+\Om\Om_\Ke{\in}_{C}{}^{D}(\nab^2_{AB}\slashed{\nabla}_{D} \widetilde\sigmac-\slashed{\nabla}_{D}\nab^2_{AB}\widetilde{\sigmac}).
\end{split}
\end{equation*}
Notice that the last term can be rewritten using
\begin{equation*}
\begin{split}
\nab^2_{AB}\slashed{\nabla}_D \widetilde\sigmac-\slashed{\nabla}_D\nab^2_{AB}\widetilde{\sigmac}
=\nab_{A}\slashed{\nabla}_D\nab_B \widetilde\sigmac-\slashed{\nabla}_D\nab^2_{AB}\widetilde{\sigmac}
%=&R_{DAB}{ }^M\nab_M \widetilde\sigmac=K(\gamma_{DB}\gamma_A{ }^M-\gamma_{D}{ }^M}\gamma_{AB})\nab_M \widetilde\sigmac\\
=K(\gamma_{DB}\nab_A\widetilde\sigmac-\gamma_{AB}\nab_D\widetilde\sigmac).
\end{split}
\end{equation*}
Therefore, using the constraint equation 
$$\curl \eta=\sigmac$$
from \eqref{null.str3} and also expressing $\tK$ in terms of $\tg$ and its derivatives using Proposition~\ref{K.diff.covariant}, we have the following reduced schematic equation
\begin{equation}\label{curv.1.eqn.3}
\begin{split}
&\nab^2_{AB}(\Om\Om_\Ke({\in}_{C}{}^{D}\slashed{\nabla}_{D} \sigmac-({\in}_{C}{}^{D}\slashed{\nabla}_{D} \sigmac)_\Ke))-{\in}_{C}{}^{D}\slashed{\nabla}_{D}(\Om\Om_\Ke\nab^2_{AB}\widetilde{\sigmac})\\
\eqrs &\sum_{i_1+i_2\leq 3}\Om_\Ke^2(1+\nab^{i_1}\tg)(\nab^{i_2}\tg+\nab^{i_2}\tp)+\Om_\Ke^2\nab^2\tg\nab^2\tp.
\end{split}
\end{equation}
We then note that all the terms on the right hand sides of \eqref{curv.1.eqn.1}, \eqref{curv.1.eqn.2} and \eqref{curv.1.eqn.3} are terms that are contained in $\mathcal F_1$. Therefore, we have
$$\nab_3\nab^2_{AB}(\Om\Om_\Ke\widetilde{\beta}_C)+\slashed{\nabla}_C(\Om\Om_\Ke \nab^2_{AB}\widetilde{K})  -\in_{CD}\slashed{\nabla}^D(\Om\Om_\Ke\nab^2_{AB}\widetilde{\sigmac}) \eqrs \mathcal F_1.$$
We now turn to the second equation in \eqref{curv.1.eqn.0}. Applying Proposition \ref{schematic.4} to the second equation in \eqref{curv.1.eqn.0} while keeping the highest order terms in the null curvature components $\beta$ on the left hand side, we obtain
\begin{equation}\label{nab4.nab2.sigmac}
\begin{split}
&\nab_4\nab^2_{AB}\widetilde{\sigmac}+\nab^2_{AB}\nab^C({\in}_{C}{ }^D\beta_D)-\nab^2_{AB}(\nab^C({\in}_{C}{ }^D\beta_D))_\Ke\\
\eqrs &\sum_{i_1+i_2\leq i}(1+\nab^{i_1}\tg+\nab^{i_1-1}\tp)\nab^{i_2}(F-F_\Ke)\\
&+\sum_{i_1+i_2+i_3\leq i}(1+\nab^{i_1}\tg+\nab^{i_1-1}\tp)\nab^{i_2}\tg\nab^{i_3}(\nab_4)_\Ke\sigmac_\Ke\\
&+\sum_{i_1+i_2+i_3\leq i}\Om_\Ke^{-2}(1+\nab^{i_1}\tg+\nab^{i_1-1}\tp)\nab^{i_2}\tb\nab^{i_3}(\nab_\Ke)\sigmac_\Ke\\
&+\sum_{i_1+i_2+i_3\leq i}(1+\nab^{i_1}\tg+\nab^{i_1-1}\tp)(1+\nab^{i_2}(\tg,\tpH))\nab^{i_3}\widetilde{\sigmac}\\
&+\sum_{i_1+i_2+i_3\leq i}(1+\nab^{i_1}\tg+\nab^{i_1-1}\tp)\nab^{i_2}(\tg,\tpH)\nab^{i_3}\sigmac_\Ke\\
&+\sum_{i_1+i_2+i_3\leq i}\Om_\Ke^{-2}(1+\nab^{i_1}\tg+\nab^{i_1-1}\tp)\nab^{i_2+1}\tb\nab^{i_3}\sigmac_\Ke,
\end{split}
\end{equation}
where 
$$F\eqs \sum_{i_1+i_2+i_3=1}\psi^{i_1}\nab^{i_2}\psi\nab^{i_3}\psi_H.$$
To handle the terms on the left hand side of \eqref{nab4.nab2.sigmac}, we commute the angular derivatives to show that
\begin{equation}\label{beta.angular.commute}
\begin{split}
&\nab^2_{AB}\nab_C({\in}^{CD}\beta_D)-\nab^2_{AB}(\nab_C(\in^{CD}\beta_D))_\Ke - \nab^3_{CAB}(\in^{CD}\widetilde{\beta}_D) \\
%= & \nab^2_{AB} \nab_C({\in}^{CD}\beta_D-({\in}^{CD}\beta_D)_\Ke)+\nab^2_{AB} (\nab-\nab_\Ke)_C (\in^{CD}\beta_D)_\Ke - \nab^3_{CAB}(\in^{CD}\widetilde{\beta}_D)\\
= & \nab^2_{AB} \nab_C({\in}^{CD}\widetilde{\beta}_D+(\in-{\in}_\Ke)^{CD}(\beta_\Ke)_D)+\nab^2_{AB} (\nab-\nab_\Ke)_C (\in^{CD}\beta_D)_\Ke- \nab^3_{CAB}(\in^{CD}\widetilde{\beta}_D).
\end{split}
\end{equation}
We gather the first term and the last term and compute as follows
\begin{equation*}
\begin{split}
& \nab^3_{ABC} ({\in}^{CD}\widetilde{\beta}_D) - \nab^3_{CAB}(\in^{CD}\widetilde{\beta}_D)
=  (\nab^3_{ABC} -\nab^3_{ACB}+ \nab^3_{ACB} - \nab^3_{CAB})(\in^{CD}\widetilde{\beta}_D) \\
%= & \nab_A ((-K (\gamma_{B}^C\gamma_{CE}-\gamma_{BE}\gamma_{C}^C)(\in^{ED}\widetilde{\beta}_D))-K(\gamma_{AB}\gamma_{CE}-\gamma_{AE}\gamma_{CB})\nab^E(\in^{CD}\widetilde{\beta}_D)\\
%& -K (\gamma_A^C\gamma_{CE}-\gamma_{AE}\gamma_C^C)\nab_B ({\in}^{ED}\widetilde{\beta}_D) \\
= & \nab_A (K {\in}_B{}^{D}\widetilde{\beta}_D)- K\gamma_{AB}\nab_C(\in^{CD}\widetilde{\beta}_D)+K\nab_A({\in}_B{}^{D}\widetilde{\beta}_D)+ K \nab_B ({\in}_A{}^{D}\widetilde{\beta}_D) \eqrs \mathcal F_2,
\end{split}
\end{equation*}
where at the very last step we have used Proposition~\ref{K.diff.covariant}. Plugging this back into \eqref{beta.angular.commute} and estimating the remaining terms in \eqref{beta.angular.commute}, we obtain 
$$\nab^2_{AB}\nab_C({\in}^{CD}\beta_D)-\nab^2_{AB}(\nab_C(\in^{CD}\beta_D))_\Ke - \nab^3_{CAB}(\in^{CD}\widetilde{\beta}_D) \eqrs \mathcal F_2.$$
Plugging this back into \eqref{nab4.nab2.sigmac}, and noting\footnote{To see this, we use in particular $\curl\eta=\sigmac$ in \eqref{null.str3}.} that all terms on the right hand side take the general form $\mathcal F_2$, we obtain
$$\nab_4\nab^2_{AB}\widetilde{\sigmac}+\nab^C\nab^2_{AB}(\in_{CD}\widetilde{\beta}^D)\eqrs\mathcal F_2$$
as desired.

Finally, we can obtain the third reduced schematic equation stated in the proposition from the third equation in \eqref{curv.1.eqn.0} in a completely analogous manner as for the second equation; we omit the details.
\end{proof}

We now turn to the remaining reduced schematic Bianchi equations that we will use in this paper. The proof is similar to the previous proposition\footnote{except that in fact it is now easier because we do not have to introduce the extra $\Om$ or $\Om_\Ke$ weights.} and we only sketch the proof.
\begin{proposition}\label{curv.red.sch.2}
$\widetilde{\betab}$, $\widetilde{K}$ and $\widetilde{\sigmac}$ obey the following reduced schematic equations: 
\begin{equation*}
\begin{split}
\nab_4\nab^2_{AB}\widetilde{\betab}_C- \nab^3_{CAB}\widetilde{K}  -\in_{CD}\slashed{\nabla}^D\nab^2_{AB}\widetilde{\sigmac}\eqrs&\:\mathcal F_4,\\
\nab_3\nab^2_{AB}\widetilde{\sigmac}+\nab^C\nab^2_{AB}(\in_{CD}\widetilde{\betab}^D)\eqrs &\: \mathcal F_3,\\
\nab_3\nab^2_{AB} \widetilde{K}-\nab^C\nab^2_{AB}\widetilde{\betab}_C \eqrs &\:\mathcal F_3.
\end{split}
\end{equation*}
\end{proposition}
\begin{proof}[Sketch of the proof]
We begin with the schematic equations in \eqref{eq:null.Bianchi3}.
\begin{equation*}
\begin{split}
\nab_4\betab-\slashed{\nabla} K -^*\slashed{\nabla}\sigmac\eqs\,&\sum_{i_1+i_2+i_3=1}\psi^{i_1}\nab^{i_2}\psi_{H} \nab^{i_3}\psi_{\Hb}+\psi K+\sum_{i_1+i_2=1} \psi^{i_1}\psi\nab^{i_2}\psi,\\
\nab_3\sigmac+\div ^*\betab \eqs\,& \sum_{i_1+i_2+i_3=1}\psi^{i_1}\nab^{i_2}\psi\nab^{i_3}\psi_{\Hb},\\
\nab_3 K-\div\betab\eqs\,&\psi_{\Hb}K+\sum_{i_1+i_2+i_3=1}\psi^{i_1}\nab^{i_2}\psi\nab^{i_3}\psi_{\Hb}.
\end{split}
\end{equation*}
Applying Propositions~\ref{schematic.3} and \ref{schematic.4} for $i=2$, but keeping the highest order derivatives of the renormalised null curvature components on the left hand side, we obtain
\begin{equation}\label{curv.red.sch.2.almost}
\begin{split}
\nab_4\nab^2_{AB} \betab_C-\nab^2_{AB}\nab_C K -\nab^2_{AB}(\in_{CD}\slashed{\nabla}^D\sigmac)\eqrs\,&\mathcal F_4,\\
\nab_3\nab^2_{AB}\sigmac+\nab^2_{AB} \nab_C (\in^{CD}\betab_D) \eqrs\,& \mathcal F_3,\\
\nab_3 \nab^2_{AB} K-\nab^2_{AB}\nab^C\betab_C\eqrs\,&\mathcal F_3,
\end{split}
\end{equation}
where the right hand sides indeed take the forms $\mathcal F_4$ and $\mathcal F_3$ can be easily checked as in Propositions~\ref{eta.eqn} and \ref{tpHb.eqn}. Finally, to get from \eqref{curv.red.sch.2.almost} to the conclusion of the proposition, we commute the angular derivatives as in the proof of Proposition~\ref{curv.red.sch.1} and check that the commutators indeed give rise to acceptable terms; we omit the details.
\end{proof}

\section{The error terms and the main estimate for $\NI$ and $\NH$}\label{sec.error.def}

Recall that we continue to work under the setting described in Remark~\ref{rmk:setup}. 

In Sections~\ref{sec.int.est} to \ref{sec.elliptic}, our goal is to bound the integrated energy $\NI$ and the hypersurface energy $\NH$ for the metric components, Ricci coefficients and the renormalised null curvature components. For this purpose, it will be convenient to introduce some general error terms. Define $\Er_1$, $\Er_2$, $\mathcal T_1$ and $\mathcal T_2$ as follows (recall the notation in \eqref{S.def}):
\begin{equation}\label{Er.def}
\begin{split}
\Er_1\doteq &\:\sum_{i_1+i_2\leq 3}(1+\sum_{\tg\in\mathcal S_{\tg},\,\tp\in \mathcal S_{\tp} }|\nab^{i_1}(\tg,\tp)|+|\nab^{i_1-1}\widetilde{K}|)\\
&\:\quad \times (\sum_{\tpHb \in\mathcal S_{\tpHb} }\Om_\Ke|\nab^{i_2}(\tpHb,\widetilde{\omb},\tb)|+\sum_{\substack{\tg\in\mathcal S_{\tg},\,\tp\in \mathcal S_{\tp} \\ \tpH\in \mathcal S_{\tpH}}}\Om_\Ke^3|\nab^{i_2}(\tp,\tg,\tpH)|+\Om_\Ke^3|\nab^{i_2-1}\widetilde{K}|)\\
&\:+\sum_{\tg\in\mathcal S_{\tg},\,\tpHb\in \mathcal S_{\tpHb} }|\Om_\Ke\nab^2\tg\nab^2\tpHb|+\sum_{\tg\in\mathcal S_{\tg},\,\tpH\in \mathcal S_{\tpH} }|\Om_\Ke^3\nab^2\tg\nab^2\tpH|,\\
\Er_2\doteq &\:\sum_{i_1+i_2\leq 3}\Om_\Ke(1+\sum_{\tg\in\mathcal S_{\tg},\,\tp\in \mathcal S_{\tp} }|\nab^{i_1}(\tg,\tp)|+|\nab^{i_1-1}\widetilde{K}|)(\sum_{\tg\in\mathcal S_{\tg},\,\tp\in \mathcal S_{\tp}}|\nab^{i_2}(\tp,\tg)|+|\nab^{i_2-1}\widetilde{K}|)\\
&\:+\sum_{\tg\in\mathcal S_{\tg},\,\tp\in \mathcal S_{\tp}}\Om_\Ke|\nab^2\tg(\nab\widetilde{K},\nab^2\tp)|,\\
\mathcal T_1\doteq &\:(\sum_{\tpHb\in\mathcal S_{\tpHb}}\sum_{i_1+i_2\leq 3}(1+|\slashed{\nabla}^{i_1}\tg|+|\nab^{\min\{i_1,2\}}\tp|)|\slashed{\nabla}^{i_2}(\tpHb,\omb)|)\\
&\:\quad \times(\sum_{\substack{\tg\in\mathcal S_{\tg},\,\tp\in \mathcal S_{\tp} \\ \tpH\in \mathcal S_{\tpH},\, \tpHb\in\mathcal S_{\tpHb}}}\sum_{i_3+i_4+i_5\leq 3}(1+|\slashed{\nabla}^{i_3}\tg|+|\nab^{\min\{i_3,2\}}\tp|)(|\nab^{i_4}\tpH|,\Om_\Ke^{-2}|\nab^{i_4}\tb|)|\slashed{\nabla}^{i_5}(\tpHb,\widetilde{\omb})|),\\
\mathcal T_2\doteq &\:(\sum_{\tpH\in \mathcal S_{\tpH}} \sum_{i_1+i_2\leq 3}(1+|\slashed{\nabla}^{i_1}\tg|+|\nab^{\min\{i_1,2\}}\tp|)(|\slashed{\nabla}^{i_2}\tpH|,\Om_\Ke^{-2}|\slashed{\nabla}^{i_2}\tb|))\\
&\:\quad \times (\sum_{\substack{\tg\in\mathcal S_{\tg},\,\tp\in \mathcal S_{\tp} \\ \tpH\in \mathcal S_{\tpH},\, \tpHb\in\mathcal S_{\tpHb}}}\sum_{i_3+i_4+i_5\leq 3}(1+|\slashed{\nabla}^{i_3}\tg|+|\nab^{\min\{i_3,2\}}\tp|)(|\nab^{i_4}\tpH|,\Om_\Ke^{-2}|\nab^{i_4}\tb|)|\slashed{\nabla}^{i_5}(\tpHb,\widetilde{\omb})|).
\end{split}
\end{equation}
We briefly comment on these terms. $\Er_1$ and $\Er_2$ are error terms which, after bounding appropriate quantities using the bootstrap assumption \eqref{BA} on $\NS$, can be treated essentially by linear estimates. Notice that $\Er_1$ contains all of the quantities $\tpH$, $\tpHb$, $\tp$, $\tg$, $\tb$, $\tK$ while $\Er_2$ only contains the subset $\tp$, $\tg$, $\tK$. Moreover, for the quantities that appear in both $\Er_1$ and $\Er_2$, they show up in $\Er_2$ with a worse weight. On the other hand, the terms $\mathcal T_1$ and $\mathcal T_2$ cannot be treated by linear estimates alone. It will be important that we can capture the full trilinear structure of these terms.

The following proposition will be the goal of Sections~\ref{sec.int.est} to \ref{sec.elliptic}:
\begin{proposition}\label{concluding.est}
The following estimate holds:
\begin{equation*}
\begin{split}
\mathcal N_{int}+\mathcal N_{hyp}\ls &\: N2^{2N}\mathcal D+\|\ub^{\frac 12+\de}\varpi^N \Er_1\|_{L^2_uL^2_{\ub}L^2(S)}^2+N^{-1}\|\ub^{\frac 12+\de}\varpi^N \Er_2\|_{L^2_uL^2_{\ub}L^2(S)}^2\\
&\:+\||u|^{1+2\de}\varpi^{2N}\Om_\Ke^2\mathcal T_1\|_{L^1_uL^1_{\ub}L^1(S)}+\|\ub^{1+2\de}\varpi^{2N}\Om_\Ke^4\mathcal T_2\|_{L^1_uL^1_{\ub}L^1(S)}.
\end{split}
\end{equation*}
Moreover, some terms in $\mathcal N_{int}$ can be controlled with an additional factor of $N$. More precisely, we have
\begin{equation}\label{extra.good.terms}
\begin{split}
&\:N\sum_{i\leq 3}\left(\||u|^{\frac 12+\de}\varpi^N\Om_\Ke\nab^i(\tpHb,\widetilde{\omb})\|_{L^2_uL^2_{\ub}L^2(S)}^2+\|\ub^{\frac 12+\de}\varpi^N\Om_\Ke^2\nab^i(\tp,\tg)\|_{L^2_uL^2_{\ub}L^2(S)}^2\right.\\
&\:\left.\qquad+\|\ub^{\frac 12+\de}\varpi^N\Om_\Ke^2\nab^{\min\{i,2\}}\tK\|_{L^2_uL^2_{\ub}L^2(S)}^2+\|\ub^{\frac 12+\de}\varpi^N\Om_\Ke^3\nab^i\tpH\|_{L^2_uL^2_{\ub}L^2(S)}^2\right.\\
&\left.\qquad+\|\ub^{\frac 12+\de}\varpi^N\Om_\Ke\nab^i\tb\|_{L^2_uL^2_{\ub}L^2(S)}^2\right)\\
\ls &\:N2^{2N}\mathcal D+\|\ub^{\frac 12+\de}\varpi^N \Er_1\|_{L^2_uL^2_{\ub}L^2(S)}^2+N^{-1}\|\ub^{\frac 12+\de}\varpi^N \Er_2\|_{L^2_uL^2_{\ub}L^2(S)}^2\\
&\:+\||u|^{1+2\de}\varpi^{2N}\Om_\Ke^2\mathcal T_1\|_{L^1_uL^1_{\ub}L^1(S)}+\|\ub^{1+2\de}\varpi^{2N}\Om_\Ke^4\mathcal T_2\|_{L^1_uL^1_{\ub}L^1(S)}.
\end{split}
\end{equation}
\end{proposition}

\begin{remark}[Bulk terms with a good sign and the estimates for $\NI$]\label{rmk:bulk}
As we already discussed in Section~\ref{baridiaeisagwgns}, an important aspect of our proof is that while proving the transport and energy estimates, some bulks terms (i.e.~those that are integrated over spacetime) with a good sign can be generated. These terms give us the desired bounds for $\NI$.

There are two ways that these terms are generated. Some of these good bulk terms are generated because of the $\varpi^N$ weights that we have put in the energies. Note that these terms are moreover proportional to $N$ (cf.~\eqref{extra.good.terms} in the statement of Proposition~\ref{concluding.est}). Hence, by choosing $N$ to be large, these terms are useful in handling many error terms.

However, not all of the good bulk terms are generated in this way by the $\varpi$ weight. In fact, the bulk terms for $\tg$, $\tp$ and $\tK$ with only $\Om_\Ke$ weights in $\NI$ estimates\footnote{To further illustrate this, notice that the term
$$\|\ub^{\frac 12+\de}\varpi^N\Om_\Ke\nab^i(\tp,\tg)\|_{L^2_uL^2_{\ub}L^2(S)}^2$$
 in $\NI$ does \underline{not} come with a factor of $N$ in the estimates, and it is only the weaker (in terms of $\Om_\Ke$) term 
$$\|\ub^{\frac 12+\de}\varpi^N\Om_\Ke^2\nab^i(\tp,\tg)\|_{L^2_uL^2_{\ub}L^2(S)}^2$$
that can be controlled with a factor $N$ according to Proposition~\ref{concluding.est}. The fact that the former term has a good sign (and so that we can control it in the $\NI$ energy) is due the geometry of the Kerr interior, or more precisely the behaviour of $\Om_\Ke^2$.
} are not generated by the $\varpi^N$ weights. Instead, the signs of these terms, which fortunately work in our favour in this case, are dictated by the geometry of the interior of Kerr solution.

Finally, note that this type of bulk terms with a good sign is already useful in establishing energy estimates for the linear scalar wave equation \eqref{linearwavePOLY}; see Section~\ref{LWERNK}.
\end{remark}

\section{Estimates for the integrated energies via transport equations}\label{sec.int.est}
Recall that we continue to work under the setting described in Remark~\ref{rmk:setup}. 

From this section to Section~\ref{sec.elliptic}, our goal is to bound the integrated energy $\NI$ and the hypersurface energy $\NH$ for the metric components, Ricci coefficients and the renormalised null curvature components and to prove Proposition~\ref{concluding.est}. In this section, we begin with estimating a subset of the terms in $\NI$ and $\NH$ which have the property that their bounds can be proved using transport equations alone. This section is organised as follows. We begin with some preliminary pointwise estimates for the difference of the metric components and their derivatives (\textbf{Section~\ref{metric.prelim}}).We then prove the main estimates of this section, including the bounds for the difference of the metric components up to their third angular covariant derivatives (\textbf{Section~\ref{transport.est.metric}}) and the difference of the Ricci coefficients up to their second angular covariant derivatives\footnote{However, notice that even for these components, we do not obtain at this point all the necessary $\NH$ bounds. In particular, the bounds for $\nab^i\tg$ and $\nab^i\widetilde{\eta}$ in $L^\infty_{\ub}L^2_u L^2(S)$ are coupled to the highest order estimates for $\nab^3\tpHb$ and $\nab^3\widetilde{\omb}$ and their proof will therefore be postponed to Section~\ref{sec.elliptic}.} (\textbf{Section~\ref{transport.est.RC}}). In Section~\ref{transport.est.RC}, we also prove some estimates for the quantities \eqref{top.quantities.def} and their angular covariant derivatives. (These estimates will be used later in Section \ref{sec.elliptic}.)

As we will see below, given the reduced schematic equations in Section \ref{sec.rse} and the general propositions for deriving estimates from transport equations in Section \ref{transportsec}, the estimate that we derive in this section simply follows from comparing the inhomogeneous terms $\mathcal F_1$, $\mathcal F_2$, $\mathcal F_3$, $\mathcal F_4$ with the error terms $\Er_1$, $\Er_2$, $\mathcal T_1$, $\mathcal T_2$.

\subsection{Preliminary pointwise estimates for the metric components}\label{metric.prelim}

Before we prove the estimates for the difference of the metric components and their angular covariant derivatives, recall from \eqref{tg.possibilities} that we use $\tg$ to denote any of the following: $\gamma_{AB}-(\gamma_{AB})_\Ke$, $\gamma^{AB}-(\gamma^{AB})_\Ke$, $\frac{\Om^2-\Om_\Ke^2}{\Om_\Ke^2}$ and $\log{\Om}-\log{\Om_\Ke}$. In this subsection, we prove the following two propositions, which shows that in order to control $\tg$, it suffices to obtain estimates for $\gamma_{AB}-(\gamma_{AB})_\Ke$ and $\log{\Om}-\log{\Om_\Ke}$.

We first show that the derivatives of the difference of $\log{\Om}$ controls the difference of the derivatives of $\Om$ in a weighted space. More precisely,
\begin{proposition}\label{Omega.lemma}
We have the following pointwise bound:
$$\sum_{i\leq 2}\left|\frac{\nab^i(\Om^2-\Om_\Ke^2)}{\Om_\Ke^2}\right|_{\gamma}\ls \sum_{i\leq 2}|\nab^i(\log\Om-\log\Om_\Ke)|_{\gamma}$$
and the following schematic pointwise bound:
$$\left|\frac{\nab^3(\Om^2-\Om_\Ke^2)}{\Om_\Ke^2}\right|_{\gamma}\ls \sum_{i_1+i_2\leq 3}(1+|\nab^{i_1}\tg|_{\gamma})|\nab^{i_2}(\log\Om-\log\Om_\Ke)|_{\gamma}.$$
\end{proposition}
\begin{proof}
If $i=0$, the desired bound is equivalent to the statement 
$$\left|\frac{(\Om^2-\Om_\Ke^2)}{\Om_\Ke^2}\right|\ls |\log\Om-\log\Om_\Ke|,$$
which, by the calculus inequality $|e^{\vartheta}-1|\leq |\vartheta|e^{|\vartheta|}$, is true as long as, say, $|\log\Om-\log\Om_\Ke|\leq \f12$. This latter estimate holds in view of Corollary~\ref{Linfty}.

For $i=1$, we have
$$\nab(\log\Om-\log\Om_\Ke)=\f 12(\f{\nab\Om^2}{\Om^2}-\f{\nab\Om_\Ke^2}{\Om_\Ke^2})=\f12 \f{\nab(\Om^2-\Om_\Ke^2)}{\Om^2}+\f12 (\nab\Om_\Ke^2)(\f{1}{\Om^2}-\f{1}{\Om_\Ke^2}).$$
By \eqref{Om.c}, we have $\f{1}{\Om^2}\ls \f{1}{\Om_\Ke^2}$. Also, we have $|\nab\Om_\Ke^2|\ls 1$. Finally, using the result for $i=0$ to control $\f{1}{\Om^2}-\f{1}{\Om_\Ke^2}$, we obtain the desired result for $i=1$.

For $i=2$, we have
\begin{equation*}
\begin{split}
&\nab^2_{AB}(\log\Om-\log\Om_\Ke)\\
=&\f12 \f{\nab^2_{AB}(\Om^2-\Om_\Ke^2)}{\Om^2}+\f12(\nab^2_{AB}\Om_\Ke^2)(\f{1}{\Om^2}-\f{1}{\Om_\Ke^2})-\f12\f{\nab_A\Om^2\nab_B\Om^2}{\Om^4}+\f12\f{\nab_A\Om_\Ke^2\nab_B\Om_\Ke^2}{\Om_\Ke^4}\\
=&\f12 \f{\nab^2_{AB}(\Om^2-\Om_\Ke^2)}{\Om^2}+\f12(\nab^2_{AB}\Om_\Ke^2+\nab_A\Om_\Ke\nab_B\Om_\Ke)(\f{1}{\Om^2}-\f{1}{\Om_\Ke^2})\\
&+\f1{2\Om^4}(\nab_A\Om_\Ke^2-\nab_A\Om^2)(\nab_B\Om_\Ke^2+\nab_B\Om^2).
\end{split}
\end{equation*}
We now apply Corollary~\ref{Linfty}, use the result for the $i=0,1$ cases and argue as before to obtain the desired bound for $i=2$.

%Detailed computation
%\begin{equation*}
%\begin{split}
%&-\f12\f{\nab\Om^2\nab\Om^2}{\Om^4}+\f12\f{\nab\Om_\Ke^2\nab\Om_\Ke^2}{\Om_\Ke^4}\\
%=&-\f12\f{\nab\Om^2\nab\Om^2}{\Om^4}+\f12\f{\nab\Om_\Ke^2\nab\Om_\Ke^2}{\Om^4}+\f12\nab\Om_\Ke^2\nab\Om_\Ke^2(\f{1}{\Om_\Ke^2}-\f{1}{\Om^2})\\
%=&\f1{2\Om^4}(\nab\Om_\Ke^2-\nab\Om^2)(\nab\Om_\Ke^2+\nab\Om^2)+\f12\nab\Om_\Ke^2\nab\Om_\Ke^2(\f{1}{\Om_\Ke^2}-\f{1}{\Om^2})
%\end{split}
%\end{equation*}

We finally turn to the case of $i=3$:
\begin{equation*}
\begin{split}
&\nab^3_{CAB}(\log\Om-\log\Om_\Ke)\\
=&\f12 \f{\nab^2_{CAB}(\Om^2-\Om_\Ke^2)}{\Om^2}-\f 12\f{\nab^2_{AB}(\Om^2-\Om_\Ke^2)\nab_C\Om}{\Om^4}+\f12(\nab^3_{CAB}\Om_\Ke^2+2\nab^2_{C(A}\Om_\Ke\nab_{B)}\Om_\Ke)(\f{1}{\Om^2}-\f{1}{\Om_\Ke^2})\\
&+\f12(\nab^2_{AB}\Om_\Ke^2+\nab_A\Om_\Ke\nab_B\Om_\Ke)(-\f{\nab_C\Om^2}{\Om^4}+\f{\nab_C\Om_\Ke^2}{\Om_\Ke^4})-\f{\nab_C\Om^2}{\Om^6}\nab_A(\Om_\Ke^2-\Om^2)\nab_B(\Om_\Ke^2+\Om^2)\\
&+\f1{\Om^4}\nab^2_{CA}(\Om_\Ke^2-\Om^2)\nab_{B}(\Om_\Ke^2+\Om^2)+\f1{\Om^4}\nab_{A}(\Om_\Ke^2-\Om^2)\nab^2_{CB}(\Om_\Ke^2+\Om^2).
\end{split}
\end{equation*}
The main difference between the $i=3$ case and the $i=0,1,2$ cases is that we now have the terms $\f{1}{\Om_{\Ke}}\nab^3\Om_\Ke^2$ and $\f{1}{\Om^2}\nab^2\Om^2$, which are not bounded pointwise by Corollary~\ref{Linfty}, unlike the corresponding lower order terms in the previous cases. Nevertheless, we can still control these terms by $(1+\sum_{i\leq 2}|\nab^i\tg|_{\gamma})$. We can then conclude by arguing as before: using Corollary~\ref{Linfty} as well as the $i=0,1,2$ cases.
\end{proof}
We now show that the differences of the inverses of $\gamma$ and their derivatives can be controlled by the differences of the derivatives of $\gamma$. This is an extension of Proposition \ref{gamma.diff.est} to higher derivatives. More precisely, we have
\begin{proposition}\label{gamma.lemma}
For $i\leq 3$, we have the following pointwise bound:
$$|\nab^i(\gamma^{-1}-(\gamma_\Ke)^{-1})|_\gamma\ls \sum_{i_1\leq i}|\nab^{i_1}(\gamma-\gamma_\Ke)|_{\gamma}.$$
\end{proposition}
\begin{proof}
The $i=0$ case is proved in Proposition \ref{gamma.diff.est}. To deal with the $i\geq 1$, we apply Proposition \ref{nab.diff}. More precisely, using the fact that $\nab\gamma^{-1}=(\nab_\Ke)(\gamma_\Ke^{-1})=0$, we have
$$\nab(\gamma^{-1}-(\gamma_\Ke)^{-1})=-\nab(\gamma_\Ke)^{-1}=(\nab_\Ke-\nab)(\gamma_\Ke)^{-1}.$$
Therefore, using Proposition \ref{nab.diff}, we have
\begin{equation}\label{gamma.inverse.1}
\begin{split}
&\nab_A(\gamma^{-1}-(\gamma_\Ke)^{-1})^{BC}\\
=&-(\slashed{\Gamma}-\slashed{\Gamma}_\Ke)_{AD}^C(\gamma_\Ke^{-1})^{BD}-(\slashed{\Gamma}-\slashed{\Gamma}_\Ke)_{AD}^B(\gamma_\Ke^{-1})^{CD}\\
=&-\f 12(\gamma_\Ke^{-1})^{BD}(\gamma_\Ke^{-1})^{CE}(\nab_A(\gamma-\gamma_\Ke)_{DE}+\nab_D(\gamma-\gamma_\Ke)_{AE}-\nab_E(\gamma-\gamma_\Ke)_{AD})\\
&-\f 12(\gamma_\Ke^{-1})^{BE}(\gamma_\Ke^{-1})^{CD}(\nab_A(\gamma-\gamma_\Ke)_{DE}+\nab_D(\gamma-\gamma_\Ke)_{AE}-\nab_E(\gamma-\gamma_\Ke)_{AD}).
\end{split}
\end{equation}
This implies the desired conclusion for $i=1$ as long as $\gamma^{-1}(\gamma_\Ke)$ and its inverse are both bounded, which is the case by Proposition \ref{gamma.diff.est}. We now turn to the $i=2$ case. Using again $\nab\gamma^{-1}=(\nab_\Ke)(\gamma_\Ke^{-1})=0$, we obtain
\begin{equation*}
\begin{split}
\nab^2_{FA}(\gamma^{-1}-(\gamma_\Ke)^{-1})^{BC}=&((\nab^2_\Ke)_{FA}-\nab^2_{FA})(\gamma_\Ke^{-1})^{BC}\\
=&(\nab_\Ke)_F((\nab_\Ke)_A-\nab_A)(\gamma_\Ke^{-1})^{BC}+((\nab_\Ke)_F-\nab_F)(\nab_A-(\nab_{\Ke})_A)(\gamma_\Ke^{-1})^{BC}\\
=&\nab_F((\nab_\Ke)_A-\nab_A)(\gamma_\Ke^{-1})^{BC}.
\end{split}
\end{equation*}
Combining this with \eqref{gamma.inverse.1}, we thus obtain
\begin{equation}\label{gamma.inverse.2}
\begin{split}
&\nab^2_{FA}(\gamma^{-1}-(\gamma_\Ke)^{-1})^{BC}\\
=&-\f 12\nab_F\left((\gamma_\Ke^{-1})^{BD}(\gamma_\Ke^{-1})^{CE}(\nab_A(\gamma-\gamma_\Ke)_{DE}+\nab_D(\gamma-\gamma_\Ke)_{AE}-\nab_E(\gamma-\gamma_\Ke)_{AD})\right)\\
&-\f 12\nab_F\left((\gamma_\Ke^{-1})^{BE}(\gamma_\Ke^{-1})^{CD}(\nab_A(\gamma-\gamma_\Ke)_{DE}+\nab_D(\gamma-\gamma_\Ke)_{AE}-\nab_E(\gamma-\gamma_\Ke)_{AD})\right).
\end{split}
\end{equation}
Consider the first term in \eqref{gamma.inverse.2} since all the other terms can be treated similarly.
\begin{equation*}
\begin{split}
&\f 12\nab_F\left((\gamma_\Ke^{-1})^{BD}(\gamma_\Ke^{-1})^{CE}\nab_A(\gamma-\gamma_\Ke)_{DE}\right)\\
=&\f 12(\gamma_\Ke^{-1})^{BD}(\gamma_\Ke^{-1})^{CE}\nab_F\nab_A(\gamma-\gamma_\Ke)_{DE}+\f 12\nab_A(\gamma-\gamma_\Ke)_{DE}(\nab_F-(\nab_\Ke)_F)((\gamma_\Ke^{-1})^{BD}(\gamma_\Ke^{-1})^{CE}).
\end{split}
\end{equation*}
By Propositions \ref{gamma.diff.est} and \ref{nab.diff}, this can be controlled (in the norm induced by $\gamma$) by 
\begin{equation}\label{thistermingammainverse}
\sum_{i_1+i_2\leq 2}(1+|\nab^{i_1}(\gamma-\gamma_\Ke)|_{\gamma})|\nab^{i_2}(\gamma-\gamma_\Ke)|_{\gamma}.
\end{equation}
On the other hand, by Corollary~\ref{Linfty}, 
\begin{equation}\label{gamma.inverse.BA}
\sum_{i\leq 1}\|\nab^i (\gamma-\gamma_\Ke)\|_{L^\i_uL^\i_{\ub}L^\i(S)}\ls \ep^{\f12}.
\end{equation}
Therefore, \eqref{thistermingammainverse} is bounded by
$$\sum_{i_1\leq 2}|\nab^{i_1}(\gamma-\gamma_\Ke)|_{\gamma}.$$
Finally, the $i=3$ case can be treated in a completely similar manner. We use the fact that $\nab\gamma^{-1}=(\nab_\Ke)(\gamma_\Ke^{-1})=0$ to obtain
\begin{equation}\label{gamma.inverse.3}
\begin{split}
&\nab^3_{GFA}(\gamma^{-1}-(\gamma_\Ke)^{-1})^{BC}\\
=&((\nab^3_\Ke)_{GFA}-\nab^3_{GFA})(\gamma_\Ke^{-1})^{BC}\\
=&(\nab_\Ke)_G(\nab_\Ke)_F((\nab_\Ke)_A-\nab_A)(\gamma_\Ke^{-1})^{BC}+(\nab_\Ke)_G((\nab_\Ke)_F-\nab_F)\nab_A(\gamma_\Ke^{-1})^{BC}\\
&+((\nab_\Ke)_G-\nab_G)\nab_F\nab_A(\gamma_\Ke^{-1})^{BC}.
\end{split}
\end{equation}
It is easy to see that after expanding this using \eqref{gamma.inverse.1}, the term is bounded by
$$\sum_{i_1+i_2+i_3\leq 3}(1+|\nab^{i_1}(\gamma-\gamma_\Ke)|_{\gamma})(1+|\nab^{i_2}(\gamma-\gamma_\Ke)|_{\gamma})|\nab^{i_3}(\gamma-\gamma_\Ke)|_{\gamma}.$$
Using \eqref{gamma.inverse.BA}, we can therefore control \eqref{gamma.inverse.3} by
$$\sum_{i_1\leq 3}|\nab^{i_1}(\gamma-\gamma_\Ke)|_{\gamma}.$$
This concludes the proof of the proposition.
\end{proof}

\subsection{Estimates for the metric components}\label{transport.est.metric}
By Propositions \ref{Omega.lemma} and \ref{gamma.lemma}, in order to bound $\nab^i\tg$, it suffices to control $\widetilde{\log\Om}$ and $\widetilde{\gamma}$ and their derivatives. We now show the following bounds for $\tg$ and their angular covariant derivatives using this fact:
\begin{proposition}\label{prop.g}
For $i\leq 3$, $\nab^i\tg$ obeys the following $L^2_{\ub}L^\infty_uL^2(S)$ and $L^2_{\ub}L^2_uL^2(S)$ estimates:
\begin{equation*}
\begin{split}
&\sum_{i\leq 3}(\|\ub^{\frac 12+\de}\varpi^N\Om_\Ke\nab^i\tg\|^2_{L^2_{\ub}L^\infty_u L^2(S)}+\|\ub^{\frac 12+\de}\varpi^N\Om_\Ke\nab^i\tg\|^2_{L^2_{\ub} L^2_{u}L^2(S)}+N\|\ub^{\frac 12+\de}\varpi^N\Om_\Ke^2\nab^i\tg\|^2_{L^2_{\ub} L^2_{u}L^2(S)})  \\
\ls &2^{2N}\mathcal D+\|\ub^{\frac 12+\de}\varpi^{N} \Er_1\|_{L^2_uL^2_{\ub}L^2(S)}^2.
\end{split}
\end{equation*}
\end{proposition}
\begin{proof}
Recall from Propositions~\ref{gamma.eqn} and \ref{Omega.eqn} that for $i\leq 3$, we have the reduced schematic equations\footnote{According to Propositions~\ref{gamma.eqn} and \ref{Omega.eqn}, we in fact have the stronger statement that the right hand sides take the form $\mathcal F_3'$. Nevertheless, we only use the weaker bounds for the proof of this proposition so that the proof can be directly adapted for Proposition~\ref{prop.etab}.}
$$\nab_3 \nab^i\widetilde{\gamma}\eqrs\mathcal F_3,\quad \nab_3 \nab^i(\log\Om-\log\Om_\Ke)\eqrs\mathcal F_3.$$
By Proposition~\ref{transport.3.2}, \eqref{data.D.def} and \eqref{varpi.bd}, we have
\begin{equation*}
\begin{split}
&\sum_{i\leq 3}\|\ub^{\frac 12+\de}\varpi^N\Om_\Ke\nab^i(\widetilde{\gamma},\log\Om-\log\Om_\Ke)\|^2_{L^2_{\ub}L^\infty_u L^2(S)}
+\sum_{i\leq 3}\|\ub^{\frac 12+\de}\varpi^N\Om_\Ke\nab^i(\widetilde{\gamma},\log\Om-\log\Om_\Ke)\|^2_{L^2_{\ub} L^2_{u}L^2(S)}\\
&+\sum_{i\leq 3}N\|\ub^{\frac 12+\de}\varpi^N\Om_\Ke^2\nab^i(\widetilde{\gamma},\log\Om-\log\Om_\Ke)\|^2_{L^2_{\ub} L^2_{u}L^2(S)}  \\
\ls &2^{2N}\mathcal D+\|\ub^{1+2\de}\varpi^{2N}\Om_\Ke^2 (\sum_{i\leq 3}\nab^i\tg)\mathcal F_3\|_{L^1_uL^1_{\ub}L^1(S)}.
\end{split}
\end{equation*}
By Propositions \ref{Omega.lemma} and \ref{gamma.lemma}, the terms $\nab^i(\widetilde{\gamma},\log\Om-\log\Om_\Ke)$ control all the $\nab^i\tg$ terms pointwise. Therefore,
\begin{equation}\label{prop.g.1}
\begin{split}
&\sum_{i\leq 3}\|\ub^{\frac 12+\de}\varpi^N\Om_\Ke\nab^i\tg\|^2_{L^2_{\ub}L^\infty_u L^2(S)}\\
&+\sum_{i\leq 3}(\|\ub^{\frac 12+\de}\varpi^N\Om_\Ke\nab^i\tg\|^2_{L^2_{\ub} L^2_{u}L^2(S)}+N\|\ub^{\frac 12+\de}\varpi^N\Om_\Ke^2\nab^i\tg\|^2_{L^2_{\ub} L^2_{u}L^2(S)})  \\
\ls &2^{2N}\mathcal D+\|\ub^{1+2\de}\varpi^{2N}\Om_\Ke^2 (\sum_{i\leq 3}\nab^i\tg)\mathcal F_3\|_{L^1_uL^1_{\ub}L^1(S)}\\
\ls &2^{2N}\mathcal D+\|\ub^{\f 12+\de}\varpi^{N}\Om_\Ke\mathcal F_3\|_{L^2_uL^2_{\ub}L^2(S)}^2,
\end{split}
\end{equation}
where in the last line we have used the Cauchy--Schwarz inequality and absorbed the quantity 
$$\sum_{i\leq 3}\|\ub^{\frac 12+\de}\varpi^N\Om_\Ke\nab^i\tg\|^2_{L^2_{\ub} L^2_{u}L^2(S)}$$ 
to the left hand side.
Comparing the definitions of $\mathcal F_3$ in \eqref{inho.def} and $\Er_1$ in \eqref{Er.def}, we see that
\begin{equation}\label{prop.g.2}
\Om_\Ke|\mathcal F_3|\ls \Er_1.
\end{equation}
The conclusion thus follows from combining \eqref{prop.g.1} and \eqref{prop.g.2}.
\end{proof}

\begin{remark}\label{rmk:tg.needs.improvement}
We recall here that in the $\NH$ energy, we also need the estimates for $\nab^i\tg$ in a weighted $L^\i_{\ub}L^2_uL^2(S)$ space. However, this will be postponed to Proposition~\ref{improved.tg} in Section~\ref{sec.elliptic}. This is because the estimate for $\nab^i\tg$ in $L^\i_{\ub}L^2_uL^2(S)$ will be achieved by integrating the equations for on the fixed $\{\ub=\mbox{constant}\}$ hypersurface without integrating in $\ub$. As we will see, the estimates will therefore be coupled with that for $\nab^i\tpHb$, which we will derive later.
\end{remark}

We now turn to the terms $\nab^i\tb$. It is easy to see from the schematic equation that $\nab^i\tb$ obeys that we can also obtain the same estimates for $\tb$ and its angular covariant derivative as those for $\tg$ and its angular covariant derivatives. Nevertheless, this will be insufficient to close the estimates. Instead, we need to take advantage of the fact that on the right hand side of the equation $\nab_3\nab^i\tb$, there are no linear terms of the form $\nab^i\tpHb$. Indeed, all the appearances of $\nab^i\tpHb$ arise from the commutator $[\nab_4,\nab]$ and contain the factor $\nab^{i_1}\tpHb\nab^{i_2}\tb$. Using this observation, we obtain the following bounds for $\nab^i\tb$, which ``gain'' an $\Omega_\Ke$ weight compared to the corresponding estimates for $\nab^i\tg$:

\begin{proposition}\label{prop.b}
For $i\leq 3$, $\nab^i\tb$ obeys the following $L^2_{\ub}L^\infty_uL^2(S)$ and $L^2_{\ub}L^2_uL^2(S)$ estimates:
\begin{equation*}
\begin{split}
&\sum_{i\leq 3}(\|\ub^{\frac 12+\de}\varpi^N\nab^i\tb\|^2_{L^2_{\ub}L^\infty_u L^2(S)} +N\|\ub^{\frac 12+\de}\varpi^N\Om_\Ke\nab^i\tb\|^2_{L^2_{\ub} L^2_{u}L^2(S)})  \\
\ls &2^{2N}\mathcal D+N^{-1}\|\ub^{\frac 12+\de}\varpi^{N}\Er_1\|_{L^2_uL^2_{\ub}L^2(S)}^2+N^{-1}\|\ub^{\frac 12+\de}\varpi^{N}\Er_2\|_{L^2_uL^2_{\ub}L^2(S)}^2+\|\ub^{1+2\de}\varpi^{2N}\Om_\Ke^4\mathcal T_2\|_{L^1_uL^1_{\ub}L^1(S)}.
\end{split}
\end{equation*}
\end{proposition}
\begin{proof}
Recall from Proposition \ref{b.eqn} that for $i\leq 3$, the reduced schematic equation for $\nab^i\tb$ takes the form
$$\nab_3\nab^i\tb\eqrs\sum_{i_1+i_2\leq 3}(1+\nab^{i_1}(\tg,\tp))(\Om_\Ke^2\nab^{i_2}(\tg,\tp,\tb))+\sum_{i_1+i_2+i_3\leq 3}(1+\nab^{i_1}(\tg,\tp))\nab^{i_2}\tpHb\nab^{i_3}\tb.$$
By Proposition \ref{transport.3.1}, \eqref{data.D.def} and \eqref{varpi.bd}, we have
\begin{equation}\label{prop.b.main}
\begin{split}
&\sum_{i\leq 3}(\|\ub^{\frac 12+\de}\varpi^N\nab^i\tb\|^2_{L^2_{\ub}L^\infty_u L^2(S)} +N\|\ub^{\frac 12+\de}\varpi^N\Om_\Ke\nab^i\tb\|^2_{L^2_{\ub} L^2_{u}L^2(S)})  \\
\ls &2^{2N}\mathcal D+\|\ub^{1+2\de}\varpi^{2N}(\sum_{i\leq 3} \nab^i\tb)(\sum_{i_1+i_2\leq 3}(1+\nab^{i_1}(\tg,\tp))(\Om_\Ke^2\nab^{i_2}(\tg,\tp,\tb)))\|_{L^1_uL^1_{\ub}L^1(S)}\\
&+\|\ub^{1+2\de}\varpi^{2N}(\sum_{i\leq 3} \nab^i\tb)(\sum_{i_1+i_2+i_3\leq 3}(1+\nab^{i_1}(\tg,\tp))\nab^{i_2}\tpHb\nab^{i_3}\tb)\|_{L^1_uL^1_{\ub}L^1(S)}.
\end{split}
\end{equation}
For the first error term, we can use the Cauchy--Schwarz inequality and absorb 
$$\f{N}{2}\sum_{i\leq 3}\|\ub^{\frac 12+\de}\varpi^N\Om_\Ke\nab^i\tb\|^2_{L^2_{\ub}L^2_u L^2(S)}$$ to the left hand side. It thus suffices to control the term
$$N^{-1}\sum_{i_1+i_2\leq 3}\|\ub^{\f12+\de}\varpi^N\Om_\Ke(1+\nab^{i_1}(\tg,\tp))\nab^{i_2}(\tg,\tp,\tb)\|_{L^2_uL^2_{\ub}L^2(S)}^2,$$
which by inspection (cf.~\eqref{Er.def}) can be controlled by 
$$N^{-1}\|\ub^{\f12+\de}\varpi^N\mathcal E_1\|_{L^2_uL^2_{\ub}L^2(S)}^2+N^{-1}\|\ub^{\f12+\de}\varpi^N\mathcal E_2\|_{L^2_uL^2_{\ub}L^2(S)}^2.$$
Finally, we notice that by the definition of $\mathcal T_2$ in \eqref{Er.def}, the second error term in \eqref{prop.b.main} can be controlled by 
$$\|\ub^{1+2\de}\varpi^{2N}\Om_\Ke^4 \mathcal T_2\|_{L^1_uL^1_{\ub}L^1(S)}.$$
\end{proof}

This concludes the estimates for the differences of the metric components and their derivatives that we prove in this section. 

\subsection{Estimates for the Ricci coefficients}\label{transport.est.RC}

We now move on to the differences of the Ricci coefficients and their derivatives. We begin by proving the bounds for $\nab^i\widetilde{\etab}$ ($i\leq 2$). We will also simultaneously control $\nab^i\widetilde{\mub}$ ($i\leq 2$). (This will be later used to control the top derivative of $\widetilde{\etab}$ via elliptic estimates; see Section~\ref{sec.elliptic.tp}.)

According to Propositions \ref{etab.eqn} and \ref{mub.eqn}, for $i\leq 2$, $\nab^i\widetilde{\etab}$ and $\nab^i\widetilde{\mub}$ both satisfy a $\nab_3$ reduced schematic equation which is similar to that for $\nab^i\widetilde{\gamma}$ and $\nab^i\widetilde{\log\Om}$, i.e.~the right hand sides have the terms $\mathcal F_3$. As a consequence, $\nab^i\widetilde{\etab}$ and $\nab^i\widetilde{\mub}$ obey the same estimates as the ones for $\nab^i\tg$ that we proved in Proposition \ref{prop.g}:
\begin{proposition}\label{prop.etab}
For $i\leq 2$, $\nab^i\widetilde{\etab}$ and $\nab^i\widetilde{\mub}$ obey the following $L^2_{\ub}L^\infty_uL^2(S)$ and $L^2_{\ub}L^2_uL^2(S)$ estimates:
\begin{equation*}
\begin{split}
&\sum_{i\leq 2}\|\ub^{\frac 12+\de}\varpi^N\Om_\Ke\nab^i(\widetilde{\etab},\widetilde{\mub})\|^2_{L^2_{\ub}L^\infty_u L^2(S)}\\
&+\sum_{i\leq 2}(\|\ub^{\frac 12+\de}\varpi^N\Om_\Ke\nab^i(\widetilde{\etab},\widetilde{\mub})\|^2_{L^2_{\ub} L^2_{u}L^2(S)}+N\|\ub^{\frac 12+\de}\varpi^N\Om_\Ke^2\nab^i(\widetilde{\etab},\widetilde{\mub})\|^2_{L^2_{\ub} L^2_{u}L^2(S)})  \\
\ls &2^{2N}\mathcal D+\|\ub^{\frac 12+\de}\varpi^{N}\Er_1\|_{L^2_uL^2_{\ub}L^2(S)}^2.
\end{split}
\end{equation*}
\end{proposition}
We now prove the estimates for $\nab^i\widetilde{\eta}$ and $\nab^i\mu$ for $i\leq 2$, using the $\nab_4$ equation that they satisfy.
\begin{proposition}\label{prop.eta}
For $i\leq 2$, $\nab^i\widetilde{\eta}$ and $\nab^i\widetilde{\mu}$ obey the following $L^2_uL^\infty_{\ub}L^2(S)$ and $L^2_{\ub}L^2_uL^2(S)$ estimates:
\begin{equation*}
\begin{split}
&\:\sum_{i\leq 2}\|\ub^{\frac 12+\de}\varpi^N\Om_\Ke\nab^i(\widetilde{\eta},\widetilde{\mu})\|^2_{L^2_uL^\infty_{\ub} L^2(S)}\\
&\:+\sum_{i\leq 2}(\|\ub^{\frac 12+\de}\varpi^N\Om_\Ke\nab^i(\widetilde{\eta},\widetilde{\mu})\|^2_{L^2_{\ub} L^2_{u}L^2(S)}+N\|\ub^{\frac 12+\de}\varpi^N\Om_\Ke^2\nab^i(\widetilde{\eta},\widetilde{\mu})\|^2_{L^2_{\ub} L^2_{u}L^2(S)})  \\
\ls &\:2^{2N}\mathcal D+\|\ub^{\frac 12+\de}\varpi^{N}\Er_1\|_{L^2_uL^2_{\ub}L^2(S)}^2.
\end{split}
\end{equation*}
\end{proposition}
\begin{proof}
Applying Propositions~\ref{transport.4.2}, \ref{eta.eqn} and \ref{mu.eqn}, we have
\begin{equation*}
\begin{split}
&\sum_{i\leq 2}\|\ub^{\frac 12+\de}\varpi^N\Om_\Ke\nab^i(\widetilde{\eta},\widetilde{\mu})\|^2_{L^2_uL^\infty_{\ub} L^2(S)}\\
& +\sum_{i\leq 2}(\|\ub^{\frac 12+\de}\varpi^N\Om_\Ke\nab^i(\widetilde{\eta},\widetilde{\mu})\|^2_{L^2_{\ub} L^2_{u}L^2(S)}+N\|\ub^{\frac 12+\de}\varpi^N\Om_\Ke^2\nab^i(\widetilde{\eta},\widetilde{\mu})\|^2_{L^2_{\ub} L^2_{u}L^2(S)})  \\
\ls &\sum_{i\leq 2}\|\ub^{\frac 12+\de}\varpi^{N}\nab^i(\widetilde{\eta},\widetilde{\mu})\|^2_{L^2_{u}L^2(S_{u,-u+C_R})}+\|\ub^{1+2\de}\varpi^{2N}\Om_\Ke^4(\sum_{i\leq 2}\nab^i(\widetilde{\eta},\widetilde{\mu}))\mathcal F_2\|_{L^1_uL^1_{\ub}L^1(S)}\\
\ls &2^{2N}\mathcal D+\|\ub^{\f12+\de}\varpi^{N}\Om_\Ke^3\mathcal F_2\|_{L^2_uL^2_{\ub}L^2(S)}^2,
\end{split}
\end{equation*}
where in the last line we have used the Cauchy--Schwarz inequality and absorbed the term
$$\sum_{i\leq 2}\|\ub^{\frac 12+\de}\varpi^N\Om_\Ke\nab^i(\widetilde{\eta},\widetilde{\mu})\|_{L^2_{\ub} L^2_{u}L^2(S)}^2$$
to the left hand side. Finally, the conclusion follows after noting that 
\begin{equation}\label{F2.est}
\Om_\Ke^3|\mathcal F_2|\ls \mathcal E_1.
\end{equation}
\end{proof}
Notice that in Proposition~\ref{prop.eta}, $\nab^i\widetilde{\eta}$ is only bounded in (appropriately weighted) $L^2_{u}L^\infty_{\ub}L^2(S)$ norm but not in the $L^2_{\ub}L^{\i}_uL^2(S)$ norm (as is also stated in the $\NH$ energy). Nevertheless, if we combine the bounds for $\nab^i\widetilde{\log\Om}$ in Proposition~\ref{prop.g} and that for $\nab^i\widetilde{\etab}$ in Proposition~\ref{prop.etab} and use the fact (cf.~\eqref{gauge.con}) that
$$\eta=2\nab(\log\Om)-\etab,$$
we immediately obtain
\begin{proposition}\label{improved.eta}
For $i\leq 2$, $\nab^i\widetilde{\eta}$ satisfies the following $L^2_uL^\i_{\ub}L^2(S)$ estimates:
\begin{equation*}
\sum_{i\leq 2}\|\ub^{\frac 12+\de}\varpi^N\Om_\Ke\nab^i\widetilde{\eta}\|^2_{L^2_{\ub}L^\infty_u L^2(S)}
\ls 2^{2N}\mathcal D+\|\ub^{\frac 12+\de}\varpi^{N}\Er_1\|_{L^2_uL^2_{\ub}L^2(S)}^2.
\end{equation*}
\end{proposition}

\begin{remark}\label{rmk:tetab.needs.improvement}
Eventually, we will also need to bound $\nab^i\widetilde{\etab}$ in (appropriately weighted) $L^\i_{\ub}L^2_{u}L^2(S)$. (Notice that Proposition~\ref{prop.etab} gives an estimate for $\nab^i\widetilde{\etab}$ in $L^2_{\ub}L^\i_uL^2(S)$ instead.) This will be postponed to Proposition~\ref{improved.etab} in Section~\ref{sec.elliptic}, after we obtain the bounds for $\nab^i\widetilde{\log\Om}$ in $L^\i_{\ub}L^2_uL^2(S)$.
\end{remark}

We now derive the estimates for $\nab^i\tpHb$.

\begin{proposition}\label{prop.tpHb}
For $i\leq 2$, $\nab^i\tpHb$ obeys the following $L^2_uL^\infty_{\ub}L^2(S)$ and $L^2_{\ub}L^2_uL^2(S)$ estimates:
\begin{equation*}
\begin{split}
&\sum_{i\leq 2}(\| |u|^{\frac 12+\delta}\varpi^N\nab^i\tpHb\|^2_{L^2_uL^\infty_{\ub} L^2(S)}  +N\| |u|^{\frac 12+\delta}\varpi^N\Om_\Ke\nab^i\tpHb\|^2_{L^2_{\ub} L^2_{u}L^2(S)})  \\
\ls &2^{2N}\mathcal D+\| |u|^{\frac 12+\delta}\varpi^{N} \Er_1\|_{L^2_uL^2_{\ub}L^2(S)}^2+N^{-1}\| |u|^{\frac 12+\delta}\varpi^{N} \Er_2\|_{L^2_uL^2_{\ub}L^2(S)}^2+\| |u|^{1+2\delta}\varpi^{2N}\Om_\Ke^2 \mathcal T_1\|_{L^1_uL^1_{\ub}L^1(S)}.
\end{split}
\end{equation*}
\end{proposition}
\begin{proof}
Recall from Proposition \ref{tpHb.eqn} that we have
$$\nab_4\nab^i\tpHb \eqrs\mathcal F_4.$$
By Proposition \ref{transport.4.1}, and using \eqref{data.D.def} and \eqref{varpi.bd} for the data term, we thus have 
\begin{equation}\label{tpHb.4.0}
\begin{split}
&\sum_{i\leq 2}(\| |u|^{\frac 12+\delta}\varpi^N\nab^i\tpHb\|^2_{L^2_uL^\infty_{\ub} L^2(S)}  +N \| |u|^{\frac 12+\delta}\varpi^N\Om_\Ke\nab^i\tpHb\|^2_{L^2_{\ub} L^2_{u}L^2(S)})  \\
\ls &2^{2N}\mathcal D+\| |u|^{1+2\delta}\varpi^{2N}\Om_\Ke^2 (\sum_{i\leq 2}\nab^i\tpHb)\mathcal F_4\|_{L^1_uL^1_{\ub}L^1(S)}.
\end{split}
\end{equation}
Recall from the definition of $\mathcal F_4$ in \eqref{inho.def} that there are two contributions, i.e.
$$\mathcal F_4=\mathcal F_{4,1}+\mathcal F_{4,2},$$
where 
$$\mathcal F_{4,1}\doteq \sum_{i_1+i_2\leq 3}(1+\nab^{i_1}(\tg,\tp))(\nab^{i_2}(\tg,\tb,\tp,\tpHb,\widetilde{\omb})+\Om_\Ke^2\nab^{i_2}\tpH+\nab^{i_2-1}\tK)+\nab^2\tg(\nab\tK,\nab^2\tp)$$
and
\begin{equation}\label{F42.def}
\mathcal F_{4,2}\doteq \sum_{i_1+i_2+i_3\leq 3}(1+\nab^{i_1}\tg+\nab^{\min\{i_1,2\}}\tp)\nab^{i_2} (\tpHb,\widetilde{\omb})(\nab^{i_3}\tpH+\Om_\Ke^{-2}\nab^{i_3}\tb).
\end{equation}
Now, it is easy to check that
\begin{equation}\label{F41.first.bound}
\Om_\Ke|\mathcal F_{4,1}|\ls  \Er_1+\Er_2
\end{equation} 
by comparing $\Om_\Ke\mathcal F_{4,1}$ with $\Er_1$ and $\Er_2$ in \eqref{Er.def}. By the Cauchy--Schwarz inequality, this implies 
\begin{equation}\label{tpHb.4.1}
\begin{split}
&\||u|^{1+2\delta}\varpi^{2N}\Om_\Ke^2 (\sum_{i\leq 2}\nab^i\tpHb)\mathcal F_{4,1}\|_{L^1_uL^1_{\ub}L^1(S)}\\
\ls &\left(\sum_{i\leq 2}N^{\f12}\||u|^{1+2\delta}\varpi^{2N}\Om_\Ke \nab^i\tpHb\|_{L^2_uL^2_{\ub}L^2(S)}\right)N^{-\f12}\||u|^{1+2\delta}\varpi^{2N}\Om_\Ke \mathcal F_{4,1}\|_{L^2_uL^2_{\ub}L^2(S)}\\
\ls &\left(\sum_{i\leq 2}N^{\f12}\||u|^{1+2\delta}\varpi^{2N}\Om_\Ke \nab^i\tpHb\|_{L^2_uL^2_{\ub}L^2(S)}\right)\\
&\times\left(N^{-\f12}\||u|^{1+2\delta}\varpi^{2N}\mathcal E_1\|_{L^2_uL^2_{\ub}L^2(S)}+N^{-\f12}\||u|^{1+2\delta}\varpi^{2N}\mathcal E_2\|_{L^2_uL^2_{\ub}L^2(S)}\right).
\end{split}
\end{equation}

The term $\mathcal F_{4,2}$ can be controlled by
\begin{equation}\label{F42.first.bound}
\sum_{i\leq 3}|\nab^i\tpHb\mathcal F_{4,2}|\ls \mathcal T_1
\end{equation}
so that
\begin{equation}\label{tpHb.4.2}
\begin{split}
\||u|^{1+2\delta}\varpi^{2N}\Om_\Ke^2 (\sum_{i\leq 2}\nab^i\tpHb)\mathcal F_{4,2}\|_{L^1_uL^1_{\ub}L^1(S)}
\ls &\||u|^{1+2\delta}\varpi^{2N}\Om_\Ke^2 \mathcal T_1\|_{L^1_uL^1_{\ub}L^1(S)}.
\end{split}
\end{equation}
The conclusion then follows from \eqref{tpHb.4.0}, \eqref{tpHb.4.1} and \eqref{tpHb.4.2} after applying Young's inequality to \eqref{tpHb.4.1} and absorbing the term
$$\f{N}{2}\sum_{i\leq 2}\||u|^{1+2\delta}\varpi^{2N}\Om_\Ke \nab^i\tpHb\|_{L^2_uL^2_{\ub}L^2(S)}^2$$
to the left hand side.
\end{proof}
By Proposition~\ref{tpHb.eqn}, $\nab^i\widetilde{\omb}$ satisfies a similar schematic equation as $\nab^i\tpHb$ for $i\leq 2$. Therefore, we also have the following estimates for $\nab^i\widetilde{\omb}$:
\begin{proposition}\label{prop.omb}
For $i\leq 2$, $\nab^i\widetilde{\omb}$ obeys the following $L^2_uL^\infty_{\ub}L^2(S)$ and $L^2_{\ub}L^2_uL^2(S)$ estimates:
\begin{equation*}
\begin{split}
&\sum_{i\leq 2}(\| |u|^{\frac 12+\delta}\varpi^N\nab^i\widetilde{\omb}\|^2_{L^2_uL^\infty_{\ub} L^2(S)}  +N\||u|^{\frac 12+\delta}\varpi^N\Om_\Ke\nab^i\widetilde{\omb}\|^2_{L^2_{\ub} L^2_{u}L^2(S)})  \\
\ls &2^{2N}\mathcal D+\||u|^{\frac 12+\delta}\varpi^{N} \Er_1\|_{L^2_uL^2_{\ub}L^2(S)}^2+N^{-1}\||u|^{\frac 12+\delta}\varpi^{N} \Er_2\|_{L^2_uL^2_{\ub}L^2(S)}^2+\||u|^{1+2\delta}\varpi^{2N}\Om_\Ke^2 \mathcal T_1\|_{L^1_uL^1_{\ub}L^1(S)}.
\end{split}
\end{equation*}
\end{proposition}
In general, the third derivative of $\widetilde{\omb}$ cannot be controlled using transport equations alone. Nevertheless, by Proposition~\ref{tpHb.eqn}, $\nab\tombs$ (cf.~\eqref{top.quantities.def}), which is a special linear combination of third derivatives of $\widetilde{\omb}$ and second derivatives of the null curvature component $\widetilde{\betab}$, satisfies a similar equation
$$\nab_4\nab\tombs \eqrs\mathcal F_4$$
 and therefore we have
\begin{proposition}\label{prop.kappab}
$\nab\tombs$ obeys the following $L^2_uL^\infty_{\ub}L^2(S)$ and $L^2_{\ub}L^2_uL^2(S)$ estimates:
\begin{equation*}
\begin{split}
&\sum_{i\leq 2}(\||u|^{\frac 12+\delta}\varpi^N\nab\tombs\|^2_{L^2_uL^\infty_{\ub} L^2(S)}  +N\||u|^{\frac 12+\delta}\varpi^N\Om_\Ke\nab\tombs\|^2_{L^2_{\ub} L^2_{u}L^2(S)})  \\
\ls &2^{2N}\mathcal D+\||u|^{\frac 12+\delta}\varpi^{N} \Er_1\|_{L^2_uL^2_{\ub}L^2(S)}^2+N^{-1}\||u|^{\frac 12+\delta}\varpi^{N} \Er_2\|_{L^2_uL^2_{\ub}L^2(S)}^2+\||u|^{1+2\delta}\varpi^{2N}\Om_\Ke^2 \mathcal T_1\|_{L^1_uL^1_{\ub}L^1(S)}.
\end{split}
\end{equation*}
\end{proposition}
Finally, we derive the estimates for $\tpH$ and its derivatives:
\begin{proposition}\label{prop.tpH}
For $i\leq 2$, $\nab^i\tpH$ obeys the following $L^2_{\ub}L^\infty_{u}L^2(S)$ and $L^2_{\ub}L^2_uL^2(S)$ estimates:
\begin{equation*}
\begin{split}
&\sum_{i\leq 2}(\|\ub^{\frac 12+\delta}\varpi^N\Om_\Ke^2\nab^i\tpH\|^2_{L^2_{\ub}L^\infty_u L^2(S)}  +N\|\ub^{\frac 12+\delta}\varpi^N\Om_\Ke^3\nab^i\tpH\|^2_{L^2_{\ub} L^2_{u}L^2(S)})  \\
\ls &2^{2N}\mathcal D+\|\ub^{\frac 12+\delta}\varpi^{N} \Er_1\|_{L^2_uL^2_{\ub}L^2(S)}^2+N^{-1}\|\ub^{\frac 12+\delta}\varpi^{N} \Er_2\|_{L^2_uL^2_{\ub}L^2(S)}^2+\|\ub^{1+2\delta}\varpi^{2N}\Om_\Ke^4 \mathcal T_2\|_{L^1_uL^1_{\ub}L^1(S)}.
\end{split}
\end{equation*}
\end{proposition}
\begin{proof}
Recall from Proposition \ref{tpH.eqn} that for $i\leq 2$, $\nab^i\tpH$ obeys the following reduced schematic equation\footnote{Here, we recall that the terms in $\mathcal F_1'$ is of course by definition a subset of those in $\mathcal F_1$.}:
$$\nab_3\nab^i(\Om_\Ke^2\tpH)\eqrs\mathcal F_1.$$
Combining this with Proposition \ref{transport.3.1}, and using \eqref{data.D.def} and \eqref{varpi.bd} for the data term, we have
\begin{equation}\label{tpH.1.0}
\begin{split}
&\:\sum_{i\leq 2}(\|\ub^{\frac 12+\delta}\varpi^N\nab^i(\Om_\Ke^2\tpH)\|^2_{L^2_{\ub}L^\infty_u L^2(S)}  +N\|\ub^{\frac 12+\delta}\varpi^N\Om_\Ke\nab^i(\Om_\Ke^2\tpH)\|^2_{L^2_{\ub} L^2_{u}L^2(S)})  \\
\ls &\:2^{2N}\mathcal D+\| \ub^{1+2\de}\varpi^{2N}(\sum_{i\leq 2}\nab^i(\Om_\Ke^2\tpH))\mathcal F_1\|_{L^1_{\ub}L^1_uL^1(S)}.
\end{split}
\end{equation}
Similar to the proof of Proposition \ref{prop.tpHb}, we consider each contribution to $\mathcal F_1$ separately. Recall that we can write
$$\mathcal F_1=\mathcal F_{1,1}+\mathcal F_{1,2},$$
where
\begin{equation}\label{F11.def}
\mathcal F_{1,1}\doteq \Om_\Ke^2\left(\sum_{i_1+i_2\leq 3}(1+\nab^{i_1}(\tg,\tp))(\nab^{i_2}(\tg,\tp,\tpHb,\widetilde{\omb})+\Om_\Ke^2\nab^{i_2}\tpH+\nab^{i_2-1} \tK)+\nab^2\tg(\nab \tK,\nab^2\tp)\right)
\end{equation}
and
\begin{equation}\label{F12.def}
\mathcal F_{1,2}\doteq \Om_\Ke^2\sum_{i_1+i_2+i_3\leq 3}(1+\nab^{i_1}\tg+\nab^{\min\{i_1,2\}}\tp)\nab^{i_2} (\tpHb,\widetilde{\omb})\nab^{i_3}\tpH.
\end{equation}
Observe now (by comparing directly with $\Er_1$ and $\Er_2$ in \eqref{Er.def}) that $\mathcal F_{1,1}$ obeys
\begin{equation}\label{F11.first.bound}
\Om_\Ke^{-1}|\mathcal F_{1,1}| \ls  \Er_1+\Er_2.
\end{equation}
Hence,
\begin{equation}\label{tpH.1.1}
\begin{split}
&\| \ub^{1+2\de}\varpi^{2N}(\sum_{i\leq 2}(\Om_\Ke^2\nab^i\tpH))\mathcal F_{1,1}\|_{L^1_{\ub}L^1_uL^1(S)}\\
\ls &(\sum_{i\leq 2}N^{\f12}\| \ub^{1+2\de}\varpi^{2N}\Om_\Ke\nab^i(\Om_\Ke^2\tpH)\|_{L^2_uL^2_{\ub}L^2(S)})N^{-\f12}\|\ub^{1+2\de}\varpi^{2N}\Om_\Ke^{-1}\mathcal F_{1,1}\|_{L^2_{\ub}L^2_uL^2(S)}\\
\ls &(\sum_{i\leq 2}N^{\f12}\| \ub^{1+2\de}\varpi^{2N}\Om_\Ke\nab^i(\Om_\Ke^2\tpH)\|_{L^2_uL^2_{\ub}L^2(S)})\\
&\times N^{-\f12}(\|\ub^{1+2\de}\varpi^{2N}\Er_1\|_{L^2_{\ub}L^2_uL^2(S)}+\|\ub^{1+2\de}\varpi^{2N}\Er_2\|_{L^2_{\ub}L^2_uL^2(S)}).
\end{split}
\end{equation}
For $\mathcal F_{1,2}$, notice that 
\begin{equation}\label{F12.first.bound}
\sum_{i\leq 3}|\nab^i\tpH\mathcal F_{1,2}| \ls \Om_\Ke^2\mathcal T_2.
\end{equation}
Since
\begin{equation}\label{tpH.OmKe}
\sum_{1\leq i\leq 2}\|\nab^i(\log\Om_\Ke)\|_{L^\i_uL^\i_{\ub}L^\i(S)}\ls 1
\end{equation}
using\footnote{Clearly $\sum_{1\leq i\leq 2}\|(\nab^i(\log\Om))_\Ke)\|_{L^\i_uL^\i_{\ub}L^\i(S)}\ls 1$. Hence, we only need to use Corollary~\ref{Linfty} (and Proposition~\ref{nab.diff}) to control the difference of the connection $\nab-\nab_\Ke$.} Corollary~\ref{Linfty}, \eqref{F12.first.bound} implies
\begin{equation}\label{tpH.1.2}
\begin{split}
\| \ub^{1+2\de}\varpi^{2N}(\sum_{i\leq 2}\nab^i(\Om_\Ke^2\tpH))\mathcal F_{1,2}\|_{L^1_{\ub}L^1_uL^1(S)} \ls &\:\| \ub^{1+2\de}\varpi^{2N}\Om_\Ke^2(\sum_{i\leq 2}\nab^i\tpH)\mathcal F_{1,2}\|_{L^1_{\ub}L^1_uL^1(S)}\\
\ls &\:\|\ub^{1+2\delta}\varpi^{2N}\Om_\Ke^4 \mathcal T_2\|_{L^1_uL^1_{\ub}L^1(S)}.
\end{split}
\end{equation}
Combining \eqref{tpH.1.0}, \eqref{tpH.1.1} and \eqref{tpH.1.2}, using Young's inequality on \eqref{tpH.1.1} and absorbing
$$\f N2 \sum_{i\leq 2}\| \ub^{1+2\de}\varpi^{2N}\Om_\Ke^3\nab^i\tpH\|_{L^2_uL^2_{\ub}L^2(S)}^2$$ 
to the left hand side, we obtain 
\begin{equation}\label{tpH.1.3}
\begin{split}
&\sum_{i\leq 2}(\|\ub^{\frac 12+\delta}\varpi^N\nab^i(\Om_\Ke^2\tpH)\|^2_{L^2_{\ub}L^\infty_u L^2(S)}  +N\|\ub^{\frac 12+\delta}\varpi^N\Om_\Ke\nab^i(\Om_\Ke^2\tpH)\|^2_{L^2_{\ub} L^2_{u}L^2(S)})  \\
\ls &2^{2N}\mathcal D+\|\ub^{\frac 12+\delta}\varpi^{N} \Er_1\|_{L^2_uL^2_{\ub}L^2(S)}^2+N^{-1}\|\ub^{\frac 12+\delta}\varpi^{N} \Er_2\|_{L^2_uL^2_{\ub}L^2(S)}^2+\|\ub^{1+2\delta}\varpi^{2N}\Om_\Ke^4 \mathcal T_2\|_{L^1_uL^1_{\ub}L^1(S)}.
\end{split}
\end{equation}
Finally, we apply \eqref{tpH.OmKe} again to obtain
\begin{equation*}
\begin{split}
&\sum_{i\leq 2}(\|\ub^{\frac 12+\delta}\varpi^N\Om_\Ke^2\nab^i\tpH\|^2_{L^2_{\ub}L^\infty_u L^2(S)}  +N\|\ub^{\frac 12+\delta}\varpi^N\Om_\Ke^3\nab^i\tpH\|^2_{L^2_{\ub} L^2_{u}L^2(S)})  \\
\ls &\sum_{i\leq 2}(\|\ub^{\frac 12+\delta}\varpi^N\nab^i(\Om_\Ke^2\tpH)\|^2_{L^2_{\ub}L^\infty_u L^2(S)}  +N\|\ub^{\frac 12+\delta}\varpi^N\Om_\Ke\nab^i(\Om_\Ke^2\tpH)\|^2_{L^2_{\ub} L^2_{u}L^2(S)}),
\end{split}
\end{equation*}
which together with \eqref{tpH.1.3} give the desired conclusion.
\end{proof}

\section{Energy estimates for the null curvature components}\label{sec.energy}
Recall that we continue to work under the setting described in Remark~\ref{rmk:setup}. 

In this section, we derive the energy estimates for the second angular covariant derivatives of the difference of the renormalised null curvature components $\beta$, $K$, $\sigmac$, $\betab$. 

The derivation of the energy estimates in this section is similar to that for the estimates in the previous section, except that the equations for $\beta$, $K$, $\sigmac$, $\betab$ form a system of hyperbolic equations instead of transport equations. For this reason, we need an extra integration by parts on the 2-spheres $S_{u,\ub}$ to avoid the loss of derivatives. To this end, we recall the following standard integration by parts formula on $S_{u,\ub}$, which can be proved by a direct computation:
\begin{proposition}\label{intbypartssph}
Given $S$-tangent tensor fields $^{(1)}\phi$ and $^{(2)}\phi$ with ranks $r$ and $r-1$ respectively,
$$\int_{S_{u,\ub}}\left({ }^{(1)}\phi^{A_1A_2...A_r}\slashed{\nabla}_{A_r}{ }^{(2)}\phi_{A_1...A_{r-1}}\right)+\int_{S_{u,\ub}}\left(\slashed{\nabla}^{A_r}{ }^{(1)}\phi_{A_1A_2...A_r}{ }^{(2)}\phi^{A_1...A_{r-1}}\right)= 0.$$
\end{proposition}
There are two main energy estimates that will be proven in this section. Before we prove these energy estimates, we first make some general comments on how we apply \eqref{BA} and Sobolev embedding to control the nonlinear terms in \textbf{Section~\ref{sec:rmk.BA.S}}. This discussion will also be relevant in later sections. Then, the estimates for $\nab^2\widetilde{\beta}$, $\nab^2\tK$ and $\nab^2\widetilde{\sigmac}$ will be proven in \textbf{Section~\ref{sec:EE.1}}; the estimates for $\nab^2\widetilde{\betab}$, $\nab^2\tK$ and $\nab^2\widetilde{\sigmac}$ will be proven in \textbf{Section~\ref{sec:EE.2}}.

\subsection{First remarks on using \eqref{BA} and Sobolev embedding to control the nonlinear terms}\label{sec:rmk.BA.S}

In Section~\ref{sec.int.est}, since we mostly control the inhomogeneous $\mathcal F$ terms by directly comparing them with $\Er_1$, $\Er_2$, $\mathcal T_1$ and $\mathcal T_2$, we did not need to handle many nonlinear terms. However, starting from this section (see in particular the terms $III$ in Proposition~\ref{EE.1} and \ref{EE.2}), we will have to control more nonlinear error terms. In particular, we will often need to bound terms with a factor of $(1+\nab^{i_1}\tg+\nab^{\min\{i_1,2\}}\tp)$. In the case of $\tg$, $\nab\tg$ and $\tp$, we can simply bound these terms in $L^\i$ using Corollary~\ref{Linfty} (as we have done in the derivation of the reduced schematic equations in Section~\ref{sec.rse}). In general, while these terms are not in $L^{\infty}$, they can still be handled using a combination of \eqref{BA} and Proposition~\ref{Sobolev}.

Since we will repeatedly use this type of estimate, we include here a detailed discussion. For this purpose, it is instructive to consider the following example\footnote{Note that we often have 
$$\sum_{i_1+i_2\leq 3} (1+\nab^{i_1}\tg+\nab^{i_1-1}\tp)\nab^{i_2} F,$$
which is a strictly better expression.}:
$$\sum_{i_1+i_2\leq 3} (1+\nab^{i_1}\tg+\nab^{\min\{i_1,2\}}\tp)\nab^{i_2} F.$$
To control this we consider various possibilities of $i_1$ and $i_2$ which are allowed by the numerology. It is easy to check that while the following are not mutually exclusive, they exhaust all possibilities.
\begin{enumerate}
\item $i_1=0$. This is the simplest case since by Corollary~\ref{Linfty}, we can simply control $(\tg,\tp)$ in $L^\i_uL^\i_{\ub}L^\i(S)$. This yields
$$\|\sum_{i\leq 3} (1+\tg+\tp)\nab^{i_2} F \|_{L^2(S_{u,\ub})}\ls \sum_{i_2\leq 3}\|\nab^{i_2} F\|_{L^2(S_{u,\ub})}.$$
\item $i_2\leq 1$. In this case, we first control the $(\nab^{i_1}\tg,\nab^{\min\{i_1,2\}}\tp)$ term in $L^2(S)$ and the $\nab^{i_2}F$ term in $L^\i(S)$ and then use Sobolev embedding (Proposition~\ref{Sobolev}) and the bootstrap assumption \eqref{BA}. More precisely,
\begin{equation*}
\begin{split}
&\:\|\sum_{i_1\leq 3,\,i_2\leq 1} (1+\nab^{i_1}\tg+\nab^{\min\{i_1,2\}}\tp)\nab^{i_2} F\|_{L^2(S_{u,\ub})} \\
\ls &\:(1+\sum_{i_1\leq 3}\|(\nab^{i_1}\tg,\nab^{\{\min\{i_1,2\}}\tp)\|_{L^2(S_{u,\ub})}) (\sum_{i_2\leq 1}\|\nab^{i_2} F\|_{L^\i(S_{u,\ub})})\\
\ls &\:(1+\sum_{i_1\leq 3}\|(\nab^{i_1}\tg,\nab^{\{\min\{i_1,2\}}\tp)\|_{L^2(S_{u,\ub})}) (\sum_{i_2\leq 3}\|\nab^{i_2} F\|_{L^2(S_{u,\ub})}) \ls \sum_{i_2\leq 3}\|\nab^{i_2} F\|_{L^2(S_{u,\ub})}.
\end{split}
\end{equation*}
\item $i_1=1,\,i_2\leq 2$. In this case, we first use H\"older's inequality to put the terms in $L^4$, and then use Sobolev embedding (Proposition~\ref{Sobolev}) and finally the bootstrap assumption \eqref{BA}.
\begin{equation*}
\begin{split}
&\:\|\sum_{i\leq 2} (1+\nab\tg+\nab\tp)\nab^{i} F\|_{L^2(S_{u,\ub})} \ls (1+\|\nab(\tg,\tp)\|_{L^4(S_{u,\ub})}) (\sum_{i\leq 2}\|\nab^i F\|_{L^4(S_{u,\ub})})\\
\ls &\:(1+\sum_{i_1\leq 2}\|\nab^{i_1}(\tg,\tp)\|_{L^2(S_{u,\ub})}) (\sum_{i_2\leq 3}\|\nab^{i_2} F\|_{L^2(S_{u,\ub})}) \ls \sum_{i_2\leq 3}\|\nab^{i_2} F\|_{L^2(S_{u,\ub})}.
\end{split}
\end{equation*}
\end{enumerate}
Combining all these observations, we have
\begin{equation}\label{BA.S.ii2.nw}
\begin{split}
\|\sum_{i_1+i_2\leq 3} (1+\nab^{i_1}\tg+\nab^{\min\{i_1,2\}}\tp)\nab^{i_2} F\|_{L^2(S_{u,\ub})}
\ls  \sum_{i\leq 3}\|\nab^i F\|_{L^2(S_{u,\ub})}.
\end{split}
\end{equation}
The estimate \eqref{BA.S.ii2.nw} obviously also holds with weights depending only on $u$ and $\ub$. Moreover, since (by Proposition~\ref{Om.Kerr.bounds}) $e^{-\f{r_+-r_-}{r_-^2+a^2}(u+\ub)} \ls \Om_\Ke^2\ls e^{-\f{r_+-r_-}{r_-^2+a^2}(u+\ub)}$, this estimates can also be implemented with $\Om_\Ke$ weights. More precisely, for $w$ a positive smooth function of $u$ and $\ub$ (for $u<u_f$, $u+\ub\geq C_R$) and for $m\in \mathbb Z$, the following holds (with implicit constants depending also on $w$ and $m$):
\begin{equation}\label{BA.S.ii2}
\begin{split}
\|w(u,\ub) \Om_\Ke^m \sum_{i_1+i_2\leq 3} (1+\nab^{i_1}\tg+\nab^{\min\{i_1,2\}}\tp)\nab^{i_2} F\|_{L^2(S_{u,\ub})}
\ls  \sum_{i\leq 3}\|w(u,\ub) \Om_\Ke^m \nab^i F\|_{L^2(S_{u,\ub})}.
\end{split}
\end{equation}
Clearly we can also apply this when there are additional integrations in $u$ and $\ub$. In particular, for any $p,q\in [1,\infty]$,
\begin{equation}\label{BA.S.222}
\begin{split}
&\:\|w(u,\ub) \Om_\Ke^m\sum_{i_1+i_2\leq 3} (1+\nab^{i_1}\tg+\nab^{\min\{i_1,2\}}\tp)\nab^{i_2} F\|_{L^p_uL^q_{\ub}L^2(S)}\\
\ls &\: (1+\sum_{i_1\leq 2}\|(\nab^{i_1}\tg,\nab^{\min\{i_1,2\}}\tp)\|_{L^\i_uL^\i_{\ub}L^2(S)})\sum_{i_2\leq 3}\|w(u,\ub) \Om_\Ke^m\nab^{i_2} F\|_{L^p_uL^q_{\ub}L^2(S)} \\
\ls &\:\sum_{i\leq 3}\|w(u,\ub) \Om_\Ke^m\nab^i F\|_{L^p_uL^q_{\ub}L^2(S)}.
\end{split}
\end{equation}
There are other variations of these estimates. For instance, we have
\begin{equation}\label{BA.S.lower}
\begin{split}
\|w(u,\ub) \Om_\Ke^m\sum_{i_1+i_2\leq 2} (1+\nab^{i_1}\tp)\nab^{i_2} F\|_{L^2(S_{u,\ub})}
\ls  \sum_{i\leq 2}\|w(u,\ub) \Om_\Ke^m\nab^i F\|_{L^2(S_{u,\ub})}, 
\end{split}
\end{equation}
and also
\begin{equation}\label{BA.S.222.lower}
\begin{split}
\|w(u,\ub) \Om_\Ke^m\sum_{i_1+i_2\leq 2} (1+\nab^{i_1}\tp)\nab^{i_2} F\|_{L^p_uL^q_{\ub}L^2(S)}
\ls  \sum_{i\leq 2}\|w(u,\ub) \Om_\Ke^m\nab^i F\|_{L^p_uL^q_{\ub}L^2(S)},
\end{split}
\end{equation}
where the latter estimate holds for any $p,q\in [1,\infty]$. Since these are broadly similar to the estimates above, we omit the details except to note that when $i_1=i_2=1$, we first use H\"older's inequality to put each term in $L^4(S_{u,\ub})$ as in case 3. above.

In what follows, when estimates in the style of \eqref{BA.S.ii2.nw}, \eqref{BA.S.ii2}, \eqref{BA.S.222}, \eqref{BA.S.lower} and \eqref{BA.S.222.lower} are used, we will often not include all the details of considering the various cases that the numerology allows. The reader will be referred here for details.

\subsection{Energy estimates for $\nab^2\widetilde{\beta}$, $\nab^2\tK$ and $\nab^2\widetilde{\sigmac}$}\label{sec:EE.1}
We have the following energy estimates for $\nab^2\widetilde{\beta}$, $\nab^2\tK$ and $\nab^2\widetilde{\sigmac}$:

\begin{proposition}\label{EE.1}
$\nab^2(\widetilde{\beta},\tK,\widetilde{\sigmac})$ satisfy the following bound:
\begin{equation*}
\begin{split}
&\|\ub^{\frac 12+\delta}\varpi^N\Om_\Ke^2\nab^2\widetilde{\beta}\|^2_{L^\infty_u L^2_{\ub}L^2(S)}  +N\|\ub^{\frac 12+\delta}\varpi^N\Om_\Ke^3\nab^2\widetilde{\beta}\|^2_{L^2_{\ub} L^2_{u}L^2(S)}  \\
&+\|\ub^{\frac 12+\delta}\varpi^N\Om_\Ke\nab^2(\tK,\widetilde{\sigmac})\|^2_{L^\infty_{\ub} L^2_uL^2(S)} +N\|\ub^{\frac 12+\delta}\varpi^N\Om_\Ke^2\nab^2(\tK,\widetilde{\sigmac})\|^2_{L^2_{\ub} L^2_{u}L^2(S)}  \\
&+\|\ub^{\frac 12+\delta}\varpi^N\Om_\Ke\nab^2(\tK,\widetilde{\sigmac})\|^2_{L^2_{\ub} L^2_{u}L^2(S)} \\
\ls &2^{2N}\mathcal D+\|\ub^{\frac 12+\delta}\varpi^{N} \Er_1\|_{L^2_uL^2_{\ub}L^2(S)}^2+N^{-1}\|\ub^{\frac 12+\delta}\varpi^{N} \Er_2\|_{L^2_uL^2_{\ub}L^2(S)}^2+\|\ub^{1+2\delta}\varpi^{2N}\Om_\Ke^4 \mathcal T_2\|_{L^1_uL^1_{\ub}L^1(S)}.
\end{split}
\end{equation*}
\end{proposition}
\begin{proof}
\textbf{Application of Propositions~\ref{transport} and \ref{curv.red.sch.1}.} To prove the desired estimates, we will use the reduced schematic equations in Proposition~\ref{curv.red.sch.1}. First, we apply the identities in Proposition~\ref{transport}. Fix $u$ and $\ub$. Applying Proposition~\ref{transport} to $\ub^{\frac 12+\de}\varpi^N\nab^2(\Om\Om_\Ke\widetilde{\beta})$, we obtain
\begin{equation}\label{eqn.beta}
\begin{split}
&\|\ub^{\frac 12+\de}\varpi^N\nab^2(\Om\Om_\Ke\widetilde{\beta})\|^2_{L^2(S_{u,\ub})}\\
=&\|\ub^{\frac 12+\de}\varpi^N\nab^2(\Om\Om_\Ke\widetilde{\beta})\|^2_{L^2(S_{-\ub+C_R,\ub})}\\
&+2\int_{-\ub+C_R}^{u}\int_{S_{u',\ub}} \left(\ub^{1+2\de}\varpi^{N}\nab^2_{AB}(\Om\Om_\Ke\widetilde{\beta}_C)\slashed{\nabla}_3\left(\varpi^N(\nab^2)^{AB}(\Om\Om_\Ke\widetilde{\beta}^C)\right)\right)d{u'}\\
&+ \int_{-\ub+C_R}^{u}\int_{S_{u',\ub}}\left(\ub^{1+2\de}\varpi^{2N}\trchb|\nab^2(\Om\Om_\Ke\widetilde{\beta})|^2_{\gamma}\right)d{u'}.
\end{split}
\end{equation}
Applying Proposition \ref{transport} to $\ub^{\frac 12+\de}\varpi^N\Om_\Ke\nab^2\widetilde{\sigmac}$ and $\ub^{\frac 12+\de}\varpi^N\Om_\Ke\nab^2\tK$, we obtain
\begin{equation}\label{eqn.sigmac}
\begin{split}
&\|\ub^{\frac 12+\de}\varpi^N\Om_\Ke\nab^2\widetilde{\sigmac}\|^2_{L^2(S_{u,\ub})}\\
=&\|\ub^{\frac 12+\de}\varpi^N\Om_\Ke\nab^2\widetilde{\sigmac}\|^2_{L^2(S_{u,-u+C_R})}\\
&+2\int_{-u+C_R}^{\ub}\int_{S_{u,\ub'}} \Omega^2\left(\ub'^{\frac 12+\de}\varpi^N\Om_\Ke\nab^2_{AB}\widetilde{\sigmac}\slashed{\nabla}_4(\ub^{\frac 12+\de}\varpi^N\Om_\Ke(\nab^2)^{AB}\widetilde{\sigmac})\right) d\ub'\\
&+ \int_{-u+C_R}^{\ub}\int_{S_{u,\ub'}} \Omega^2\left(\trch |\ub'^{\frac 12+\de}\varpi^N\Om_\Ke\nab^2\widetilde{\sigmac}|^2_{\gamma}\right)d{\ub'}
\end{split}
\end{equation}
and
\begin{equation}\label{eqn.K}
\begin{split}
&\|\ub^{\frac 12+\de}\varpi^N\Om_\Ke\nab^2\tK\|^2_{L^2(S_{u,\ub})}\\
=&\|\ub^{\frac 12+\de}\varpi^N\Om_\Ke\nab^2\tK\|^2_{L^2(S_{u,-u+C_R})}\\
&+2\int_{-u+C_R}^{\ub}\int_{S_{u,\ub'}}  \Omega^2\left(\ub'^{\frac 12+\de}\varpi^N\Om_\Ke\nab^2_{AB}\tK\slashed{\nabla}_4(\ub^{\frac 12+\de}\varpi^N\Om_\Ke(\nab^2)^{AB}\tK)\right)d\ub'\\
&+ \int_{-u+C_R}^{\ub}\int_{S_{u,\ub'}}  \Omega^2\left(\trch |\ub'^{\frac 12+\de}\varpi^N\Om_\Ke\nab^2\tK|^2_{\gamma}\right)d{\ub'}
\end{split}
\end{equation}
respectively.

We now integrate \eqref{eqn.beta} in $\ub$ and integrate \eqref{eqn.sigmac} and \eqref{eqn.K} in $u$. We then add up these three equations to obtain an equation of the form
\begin{equation}\label{EE.1.1}
\begin{split}
&\int_{-u+C_R}^{\ub}\|\ub^{\frac 12+\de}\varpi^N\nab^2(\Om\Om_\Ke\widetilde{\beta})\|^2_{L^2(S_{u,\ub'})}d\ub'\\
&+\int_{-\ub+C_R}^{u}\left(\|\ub^{\frac 12+\de}\varpi^N\Om_\Ke\nab^2\widetilde{\sigmac}\|^2_{L^2(S_{u',\ub})}+\|\ub^{\frac 12+\de}\varpi^N\Om_\Ke\nab^2\tK\|^2_{L^2(S_{u',\ub})}\right)du'\\
=& \int_{-u+C_R}^{\ub}\|\ub^{\frac 12+\de}\varpi^N\nab^2(\Om\Om_\Ke\widetilde{\beta})\|^2_{L^2(S_{-\ub'+C_R,\ub'})}d\ub'\\
&+\int_{-\ub+C_R}^{u}\left(\|\ub^{\frac 12+\de}\varpi^N\Om_\Ke\nab^2\widetilde{\sigmac}\|^2_{L^2(S_{u',-u'+C_R})}+\|\ub^{\frac 12+\de}\varpi^N\Om_\Ke\nab^2\tK\|^2_{L^2(S_{u',-u'+C_R})}\right)du'\\
&+I+II+III+IV,
\end{split}
\end{equation}
where $I$, $II$, $III$ and $IV$ are terms arising from combining the right hand sides of \eqref{eqn.beta}, \eqref{eqn.sigmac} and \eqref{eqn.K} and using the reduced schematic equations in Proposition~\ref{curv.red.sch.1}. We now define these terms. First, we have the terms where one of the curvature components has three angular derivatives:
\begin{equation*}
\begin{split}
I=&\:-2\int_{-u+C_R}^{\ub}\int_{-\ub'+C_R}^{u}\int_{S_{u',\ub'}} \ub'^{1+2\de} \varpi^{2N} \nab^2_{AB}(\Om\Om_\Ke\widetilde{\beta}_C)\slashed{\nabla}^C(\Om\Om_\Ke (\nab^2)^{AB}\tK) du' d\ub'\\
&\: +2\int_{-u+C_R}^{\ub}\int_{-\ub'+C_R}^{u}\int_{S_{u',\ub'}} \ub'^{1+2\de} \varpi^{2N} \nab^2_{AB}(\Om\Om_\Ke\widetilde{\beta}_C)\in^C{ }_{D}\slashed{\nabla}^D(\Om\Om_\Ke(\nab^2)^{AB}\widetilde{\sigmac}) d{u'}d\ub'\\
&\:-2\int_{-u+C_R}^{\ub}\int_{-\ub'+C_R}^{u}\int_{S_{u',\ub'}} \ub'^{1+2\de}\varpi^{2N}\Omega^2\Om_\Ke^2 \nab^2_{AB}\widetilde{\sigmac}\nab^C(\nab^2)^{AB}(\in_{CD}\widetilde{\beta}^D) du'd{\ub'}\\
&\:-2\int_{-u+C_R}^{\ub}\int_{-\ub'+C_R}^{u}\int_{S_{u',\ub'}} \ub'^{1+2\de}\varpi^{2N}\Omega^2\Om_\Ke^2\nab^2_{AB}\tK\nab^C(\nab^2)^{AB}\widetilde{\beta}_C du'd{\ub'}.
\end{split}
\end{equation*}
Second, we have terms where $\nab_3$ and $\nab_4$ act on the weight functions. These terms have a good sign and can be used to control the error terms as long as $N$ and $|u_f|$ are both chosen to be sufficiently large (cf.~\eqref{est.intbyparts.2}):
\begin{equation*}
\begin{split}
II=&\:2\int_{-u+C_R}^{\ub}\int_{-\ub'+C_R}^{u}\int_{S_{u',\ub'}} \ub'^{1+2\de}\varpi^N(\slashed{\nabla}_3 \varpi^N)|\nab^2(\Omega\Om_\Ke\widetilde{\beta})|^2_{\gamma}du'd{\ub'}\\
&\:+2\int_{-u+C_R}^{\ub}\int_{-\ub'+C_R}^{u}\int_{S_{u',\ub'}} \ub'^{\frac 12+\de}\varpi^N\Omega^2\Om_\Ke\slashed{\nabla}_4(\ub^{\frac 12+\de}\varpi^N\Om_\Ke)(|\nab^2\widetilde{\sigmac}|^2_{\gamma}+|\nab^2\tK|^2_{\gamma})du'd{\ub'}.
\end{split}
\end{equation*}
Third, we have the terms which arise from the error terms in the reduced schematic equations in Proposition~\ref{curv.red.sch.1}. For these terms, we do not need their precise form and we will therefore only give their reduced schematic form:
\begin{equation*}
\begin{split}
III\eqrs &\:2\int_{-u+C_R}^{\ub}\int_{-\ub'+C_R}^{u}\int_{S_{u',\ub'}} \ub'^{1+2\de}\varpi^{2N} \langle\nab^2(\Om\Om_\Ke\widetilde{\beta}),\mathcal F_1\rangle_\gamma d{u'}d\ub'\\
&\:+2\int_{-u+C_R}^{\ub}\int_{-\ub'+C_R}^{u}\int_{S_{u',\ub'}} \ub'^{1+2\de}\varpi^{2N}\Omega^2\Om_\Ke^2\langle\nab^2\widetilde{\sigmac}+\nab^2\tK,\mathcal F_2\rangle_\gamma d{\ub'}du'.
\end{split}
\end{equation*}
Finally, we have the term
\begin{equation*}
\begin{split}
IV=&\:\int_{-u+C_R}^{\ub}\int_{-\ub'+C_R}^{u}\int_{S_{u',\ub'}} \ub'^{1+2\de}\varpi^{2N}\left(\trchb|\nab^2(\Omega\Om_\Ke\widetilde{\beta})|^2_{\gamma}+\trch\Omega^2\Om_\Ke^2\left( |\nab^2\widetilde{\sigmac}|^2_{\gamma}+|\nab^2\tK|^2_{\gamma}\right)\right)du' d{\ub'}.
\end{split}
\end{equation*}

\textbf{Estimates for the term $I$.} We now estimate each of these terms. To handle $I$, we perform integration by parts on the $2$-spheres $S_{u,\ub}$ using Proposition \ref{intbypartssph}. Applying Proposition \ref{intbypartssph} to the first line, it is easy to see that the term where $\widetilde{\beta}$ has three angular derivatives cancel with the term on the fourth line. Similarly, applying Proposition \ref{intbypartssph} to the second line, the term with $\nab^3\widetilde{\beta}$ cancel with the term on the third line. Therefore, the only contributions come from when $\nab$ hits on $\Om$ or $\Om_\Ke$ (notice that $\nab \ub = \nab \varpi=0$). Also, by the bootstrap assumption \eqref{BA}, we have the bound (when $i\leq 3$)
\begin{equation}\label{EE.1.Om.bd}
|\nab^i\log\Om_\Ke |_\gamma+|\nab^i\log\Om |_\gamma\ls 1+\sum_{j\leq i} |\nab^j\tg|.
\end{equation}
Therefore, using \eqref{EE.1.Om.bd}, the Cauchy--Schwarz inequality and then \eqref{BA.S.222.lower}, we obtain
\begin{equation}\label{est.intbyparts.1}
\begin{split}
|I| \ls &\:\|\ub^{1+2\de}\varpi^{2N}\Om_\Ke^4(\sum_{i_1+i_2\leq 2}(1+\nab^{i_1}\tg)\nab^{i_2}\widetilde{\beta})\nab^2(\tK,\widetilde{\sigmac}) \|_{L^1_uL^1_{\ub}L^1(S)}\\
%\ls &\:(1+\sum_{i_1\leq 2}\|\nab^{i_1}\tg\|_{L^\i_uL^\i_{\ub} L^2(S)})(\sum_{i_2\leq 2}\|\ub^{\frac 12+\de}\varpi^{N}\Om_\Ke^3\nab^{i_2}\widetilde{\beta} \|_{L^2_uL^2_{\ub}L^2(S)})\|\ub^{\frac 12+\de}\varpi^{N}\Om_\Ke\nab^2(\tK,\widetilde{\sigmac}) \|_{L^2_uL^2_{\ub}L^2(S)}\\
\ls &\:(\sum_{i\leq 2}\|\ub^{\frac 12+\de}\varpi^{N}\Om_\Ke^3\nab^{i}\widetilde{\beta} \|_{L^2_uL^2_{\ub}L^2(S)})\|\ub^{\frac 12+\de}\varpi^{N}\Om_\Ke\nab^2(\tK,\widetilde{\sigmac}) \|_{L^2_uL^2_{\ub}L^2(S)}\\
\ls &\:\|\ub^{\frac 12+\de}\varpi^{N}\Om_\Ke\nab^2(\tK,\widetilde{\sigmac}) \|_{L^2_uL^2_{\ub}L^2(S)}\|\ub^{\frac 12+\de}\varpi^{N}\Er_1 \|_{L^2_uL^2_{\ub}L^2(S)}.
\end{split}
\end{equation}
Here, in the final inequality, we used the following reduced schematic equation (with $j=2$), which can be deduced from the schematic Codazzi equation \eqref{Codazzi.S.1} (or \eqref{sch.Codazzi}),
\begin{equation}\label{EE.1.Codazzi.2}
\sum_{i\leq j}\nab^i\widetilde{\beta} \eqrs \sum_{i_1+i_2\leq j+1}(1+\nab^{i_1-1}\tp+\nab^{i_1}\tg)(\nab^{i_2}(\tg,\tpH),\nab^{\min\{i_2,2\}}\tp)
\end{equation}
to bound the $\nab^i\widetilde{\beta}$ term by $\Er_1$.

\textbf{Estimates for the term $II$.} The term $II$ will be handled similarly as in the proof of Propositions \ref{transport.3.1} and \ref{transport.4.2}, i.e.~we will use the fact that $\slashed{\nabla}_3 \varpi^N$ and $\slashed{\nabla}_4(\ub^{\frac 12+\de}\varpi^N\Om_\Ke)$ both have good signs (after choosing $|u_f|$ to be sufficiently large). More precisely, by \eqref{varpi.repeat} and \eqref{nab4.weight.good.bound}, there exists some constant $C>0$ depending only on $M$, $a$ and $C_R$, we have
\begin{equation}\label{est.intbyparts.2}
\begin{split}
II \leq & -C^{-1}\int_{-u+C_R}^{\ub}\int_{-\ub'+C_R}^{u}\|\ub^{\frac 12+\delta}\varpi^N\Om_\Ke\nab^2(\tK,\widetilde{\sigmac})\|^2_{L^2(S_{u',\ub'})} du'd\ub'\\
&-C^{-1}N\int_{-u+C_R}^{\ub}\int_{-\ub'+C_R}^{u}\|\ub^{\frac 12+\delta}\varpi^N\Om_\Ke\nab^2(\Omega\Om_\Ke\widetilde{\beta})\|^2_{L^2(S_{u',\ub'})} du'd\ub'\\
&-C^{-1}N\int_{-u+C_R}^{\ub}\int_{-\ub'+C_R}^{u}\|\ub^{\frac 12+\delta}\varpi^N\Om_\Ke^2\nab^2(\tK,\widetilde{\sigmac})\|^2_{L^2(S_{u'\ub'})} du'd\ub'.
\end{split}
\end{equation}

\textbf{Estimates for the term $III$.} The term $III$ is an error term. As in the proof of Proposition~\ref{prop.tpH}, we split the term $\mathcal F_1$ into
$$\mathcal F_1=\mathcal F_{1,1}+\mathcal F_{1,2},$$
where, the following bounds hold:
\begin{equation}\label{EE.1.F1.1}
\Om_\Ke^{-1}|\mathcal F_{1,1}| \ls  \Er_1+\Er_2, \quad |(\sum_{i_1+i_2\leq 3}(1+\nab^{i_1}\tg+\nab^{i_1-1}\tp)\nab^{i_2}\tpH)\mathcal F_{1,2}| \ls \Om_\Ke^2\mathcal T_2.
\end{equation}
The first bound in \eqref{EE.1.F1.1} is exactly \eqref{F11.first.bound} in the proof of Proposition~\ref{prop.tpH}, while the second bound is an easy variation of \eqref{F12.first.bound}, which again follows from comparing the terms with $\mathcal T_2$ in \eqref{Er.def}.
The first relation in \eqref{EE.1.F1.1} immediately implies the following bound for the $\mathcal F_{1,1}$ term using the Cauchy--Schwarz inequality:
\begin{equation}\label{EE.1.F11}
\begin{split}
&\:\|\ub^{1+2\de} \varpi^{2N} \nab^2(\Om\Om_\Ke\widetilde{\beta})\mathcal F_{1,1} \|_{L^1_uL^1_{\ub}L^1(S)} \\
\ls &\:\|\ub^{\frac 12+\delta}\varpi^N\Om_\Ke\nab^2(\Om\Om_\Ke\widetilde{\beta}) \|_{L^2_uL^2_{\ub}L^2(S)}(\|\ub^{\frac 12+\de}\varpi^N\Er_1 \|_{L^2_uL^2_{\ub}L^2(S)}+\|\ub^{\frac 12+\de}\varpi^N\Er_2 \|_{L^2_uL^2_{\ub}L^2(S)}).
\end{split}
\end{equation}
To handle the $\mathcal F_{2,2}$ term, we want to use the second relation in \eqref{EE.1.F1.1}. For this purpose, we need to first control $\sum_{i\leq 2}\nab^i(\Om\Om_\Ke\widetilde{\beta})$ by $\sum_{i\leq 3}\Om_\Ke^2\nab^i\tpH$ (plus some extra error terms) using the Codazzi equation for $\beta$. This process introduces some error terms, which we will control using $\Er_1$. We now turn to the details. Using the reduced schematic Codazzi equation \eqref{EE.1.Codazzi.2}, and appropriately controlling the derivatives of $\Om$ and $\Om_\Ke$ using \eqref{EE.1.Om.bd}, we deduce that
\begin{equation}\label{EE.1.F12.express}
\begin{split}
(\sum_{i\leq 2}\nab^i(\Om\Om_\Ke\widetilde{\beta}))\mathcal F_{1,2}
\eqrs &\:\Om_\Ke^2 \sum_{i_1+i_2\leq 3}(1+\nab^{i_1}\tg+\nab^{i_1-1}\tp)(\nab^{i_2}(\tg,\tpH),\nab^{\min\{i_2,2\}}\tp)\mathcal F_{1,2}.
\end{split}
\end{equation}
The terms in \eqref{EE.1.F12.express} that contain a factor of $\nab^{i_2}\tpH$ are the main terms. Using the second bound in \eqref{EE.1.F1.1}, we obtain
\begin{equation}\label{EE.1.F12.1}
\begin{split}
&\:\|\ub^{1+2\de}\varpi^{2N}\Om_\Ke^2\sum_{i_1+i_2\leq 3}(1+\nab^{i_1}\tg+\nab^{i_1-1}\tp)\nab^{i_2}\tpH\mathcal F_{1,2} \|_{L^1_uL^1_{\ub}L^1(S)} 
\ls  \|\ub^{1+2\de}\varpi^{2N}\Om_\Ke^4\mathcal T_2 \|_{L^1_uL^1_{\ub}L^1(S)}.
\end{split}
\end{equation}
For the remaining terms in \eqref{EE.1.F12.express}, we use \eqref{F12.def} to further estimate as follows
\begin{equation}\label{EE.1.F1.3}
\begin{split}
&\:\Om_\Ke^2\sum_{i_1+i_2\leq 3}(1+\nab^{i_1}\tg+\nab^{i_1-1}\tp)(\nab^{i_2}\tg,\nab^{\min\{i_2,2\}}\tp)\mathcal F_{1,2} \\
\eqrs &\:\Om_\Ke^4 (\sum_{i_1+i_2\leq 3}(1+\nab^{i_1}\tg+\nab^{i_1-1}\tp)(\nab^{i_2}\tg,\nab^{\min\{i_2,2\}}\tp)) \\
&\quad \times(\sum_{i_3+i_4+i_5\leq 3}(1+\nab^{i_3}\tg+\nab^{\min\{i_3,2\}}\tp)\nab^{i_4}\tpH \nab^{i_5}(\tpHb,\widetilde{\omb})).
\end{split}
\end{equation}
Using the Cauchy--Schwarz inequality and applying \eqref{BA.S.222} twice, we obtain
\begin{equation}\label{EE.1.F1.4}
\begin{split}
&\:\|\ub^{1+2\de} \varpi^{2N} \Om_\Ke^2(\sum_{i_1+i_2\leq 3}(1+\nab^{i_1}\tg+\nab^{i_1-1}\tp)(\nab^{i_2}\tg,\nab^{\min\{i_2,2\}}\tp))\mathcal F_{1,2}\|_{L^1_uL^1_{\ub}L^1(S)}\\
\ls &\: \|\ub^{\f 12+\de} \varpi^{N} \Om_\Ke(\sum_{i_1+i_2\leq 3}(1+\nab^{i_1}\tg+\nab^{i_1-1}\tp)(\nab^{i_2}\tg,\nab^{\min\{i_2,2\}}\tp))\|_{L^2_uL^2_{\ub}L^2(S)}\\
&\:\quad \times \|\ub^{\f 12+\de} \varpi^{N} \Om_\Ke(\sum_{i_3+i_4+i_5\leq 3}(1+\nab^{i_3}\tg+\nab^{\min\{i_3,2\}}\tp)\nab^{i_4}\tpH \nab^{i_5}(\tpHb,\widetilde{\omb}))\|_{L^2_uL^2_{\ub}L^2(S)}\\
\ls &\: (\sum_{i\leq 3}\|\ub^{\f 12+\de} \varpi^{N} \Om_\Ke(\nab^{i}\tg,\nab^{\min\{i,2\}}\tp)\|_{L^2_uL^2_{\ub}L^2(S)}) (\sum_{i_1+i_2\leq 3}\|\ub^{\f 12+\de} \varpi^{N} \Om_\Ke\nab^{i_1}\tpH \nab^{i_2}(\tpHb,\widetilde{\omb})\|_{L^2_uL^2_{\ub}L^2(S)})\\
\ls &\: 2^{2N}\mathcal D +\|\ub^{\f 12+\de} \varpi^N \Er_1\|_{L^2_uL^2_{\ub}L^2(S)}^2 + \sum_{i_1+i_2\leq 3}\|\ub^{\f 12+\de} \varpi^{N} \Om_\Ke\nab^{i_1}\tpH \nab^{i_2}(\tpHb,\widetilde{\omb})\|_{L^2_uL^2_{\ub}L^2(S)}^2,
\end{split}
\end{equation}
where in the last line, we used Young's inequality and the estimates for $\nab^i\tp$ ($i\leq 2$) and $\nab^i\tg$ ($i\leq 3$) derived in Propositions~\ref{prop.g}, \ref{prop.etab} and \ref{prop.eta}.

To estimate the last term in \eqref{EE.1.F1.4}, we separate the cases $i_1\leq i_2$ and $i_1>i_2$. Using H\"older's inequality and Sobolev embedding (Proposition~\ref{Sobolev}) in the style of Section~\ref{sec:rmk.BA.S}, we obtain
\begin{equation}\label{EE.1.F1.5}
\begin{split}
&\:\sum_{i_1+i_2\leq 3}\|\ub^{\f 12+\de} \varpi^{N} \Om_\Ke\nab^{i_1}\tpH \nab^{i_2}(\tpHb,\widetilde{\omb})\|_{L^2_uL^2_{\ub}L^2(S)}^2\\
\ls & \: \sum_{i_1\leq 2,\,i_2\leq 3}\|\nab^{i_1}(\tpHb,\widetilde{\omb})\|_{L^\i_uL^\i_{\ub}L^2(S)}\|\ub^{\f 12+\de}\varpi^N\Om_\Ke^3\nab^{i_2}\tpH\|_{L^2_uL^2_{\ub}L^2(S)} \\
&\:+\sum_{i_1\leq 2,\,i_2\leq 3}\|\Om_\Ke^2\nab^{i_1}\tpH\|_{L^\i_uL^\i_{\ub}L^2(S)}^2\|\ub^{\f 12+\de}\varpi^N\Om_\Ke\nab^{i_2}(\tpHb,\widetilde{\omb})\|_{L^2_uL^2_{\ub}L^2(S)}^2\\
\ls &\: \|\ub^{\f 12+\de} \varpi^N \Er_1\|_{L^2_uL^2_{\ub}L^2(S)}^2,
\end{split}
\end{equation}
where in the last line we have controlled the weighted $L^\i_uL^\i_{\ub}L^2(S)$ norms of $\nab^i\tpH$ and $\nab^{i}(\tpHb,\widetilde{\omb})$ ($i\leq 2$) by the bootstrap assumption \eqref{BA}, and controlled the weighted $L^2_uL^2_{\ub}L^2(S)$ norms of $\nab^i\tpH$ and $\nab^i(\tpHb,\widetilde{\omb})$ ($i\leq 3$) by directly comparing with $\Er_1$ in \eqref{Er.def}.
Plugging \eqref{EE.1.F1.5} back into \eqref{EE.1.F1.4}, we then obtain
\begin{equation}\label{EE.1.F12.2}
\begin{split}
&\:\|\ub^{1+2\de} \varpi^{2N} \Om_\Ke^2\sum_{i_1+i_2\leq 3}(1+\nab^{i_1}\tg+\nab^{i_1-1}\tp)(\nab^{i_2}\tg,\nab^{\min\{i_2,2\}}\tp)\mathcal F_{1,2}\|_{L^1_uL^1_{\ub}L^1(S)}\\
\ls &\:2^{2N}\mathcal D +\|\ub^{\f 12+\de} \varpi^N \Er_1\|_{L^2_uL^2_{\ub}L^2(S)}^2.
\end{split}
\end{equation}
Combining \eqref{EE.1.F11}, \eqref{EE.1.F12.express}, \eqref{EE.1.F12.1} and \eqref{EE.1.F12.2}, and applying the Young's inequality, we thus obtain
\begin{equation}\label{EE.1.F1.final}
\begin{split}
&\:\|\ub^{1+2\de} \varpi^{2N} \nab^2(\Om\Om_\Ke\widetilde{\beta})\mathcal F_{1} \|_{L^1_uL^1_{\ub}L^1(S)}\\
\ls &\: 2^{2N}\mathcal D+\|\ub^{\frac 12+\delta}\varpi^N\Om_\Ke\nab^2(\Om\Om_\Ke\widetilde{\beta}) \|_{L^2_uL^2_{\ub}L^2(S)} \|\ub^{\frac 12+\de}\varpi^N\Er_2 \|_{L^2_uL^2_{\ub}L^2(S)}\\
&\:+\|\ub^{\frac 12+\delta}\varpi^N\Om_\Ke\nab^2(\Om\Om_\Ke\widetilde{\beta}) \|_{L^2_uL^2_{\ub}L^2(S)}^2+\|\ub^{1+2\de}\varpi^{2N}  \Om_\Ke^4\mathcal T_2\|_{L^1_uL^1_{\ub}L^1(S)}+\|\ub^{\f 12+\de} \varpi^N \Er_1\|_{L^2_uL^2_{\ub}L^2(S)}^2.
\end{split}
\end{equation}
Moreover, we have, as in the proof of Proposition~\ref{prop.eta} (see \eqref{F2.est}),
$$\Om_\Ke^3|\mathcal F_2|\ls \Er_1,$$
which implies (after using also \eqref{Om.c})
\begin{equation}\label{EE.1.F2.final}
\|\ub^{1+2\de} \varpi^{2N} \Om\Om_\Ke\nab^2(\tK,\widetilde{\sigmac})\mathcal F_{2} \|_{L^1_uL^1_{\ub}L^1(S)}\ls \|\ub^{\frac 12+\delta}\varpi^N\Om_\Ke\nab^2(\tK,\widetilde{\sigmac}) \|_{L^2_uL^2_{\ub}L^2(S)}\|\ub^{\frac 12+\de}\varpi^N\Er_1 \|_{L^2_uL^2_{\ub}L^2(S)}.
\end{equation}
Therefore, recalling the definition of $III$ and using \eqref{EE.1.F1.final} and \eqref{EE.1.F2.final}, we obtain
\begin{equation}\label{est.intbyparts.3}
\begin{split}
|III| \ls &\: 2^{2N}\mathcal D + \|\ub^{\frac 12+\delta}\varpi^N\Om_\Ke\nab^2(\Om\Om_\Ke\widetilde{\beta}) \|_{L^2_uL^2_{\ub}L^2(S)}\|\ub^{\frac 12+\de}\varpi^N\Er_2 \|_{L^2_uL^2_{\ub}L^2(S)}\\
&\:+\|\ub^{\frac 12+\delta}\varpi^N\Om_\Ke\nab^2(\Om\Om_\Ke\widetilde{\beta}) \|_{L^2_uL^2_{\ub}L^2(S)}^2+\|\ub^{1+2\de}\varpi^{2N}\Om_\Ke^4\mathcal T_2 \|_{L^1_uL^1_{\ub}L^1(S)}+\|\ub^{\frac 12+\de}\varpi^N\Er_1 \|_{L^2_uL^2_{\ub}L^2(S)}^2\\
&\:+\|\ub^{\frac 12+\delta}\varpi^N\Om_\Ke\nab^2(\tK,\widetilde{\sigmac}) \|_{L^2_uL^2_{\ub}L^2(S)}\|\ub^{\frac 12+\de}\varpi^N\Er_1 \|_{L^2_uL^2_{\ub}L^2(S)}.
\end{split}
\end{equation}

\textbf{Estimates for the term $IV$.} Using the following estimates for $\trchb$ and $\trch$ (which follow from the bounds for the background $(\trchb)_\Ke$ and $(\trch)_\Ke$ in Proposition~\ref{Kerr.Ricci.bound})
$$|\trchb|\ls \Om_\Ke^2+|\tpHb|,\quad |\trch| \ls 1+|\tpH|,$$
and \eqref{Om.c}, the term $IV$ obeys the bound
\begin{equation}\label{est.intbyparts.4.0}
\begin{split}
|IV| \ls &\: \|\ub^{\frac 12+\delta}\varpi^N\Om_\Ke\nab^2(\Om\Om_\Ke\widetilde{\beta}) \|_{L^2_uL^2_{\ub}L^2(S)}^2+\|\ub^{1+2\delta}\varpi^{2N}\tpHb\nab^2(\Om\Om_\Ke\widetilde{\beta})\nab^2(\Om\Om_\Ke\widetilde{\beta}) \|_{L^1_uL^1_{\ub}L^1(S)}\\
&\:+\|\ub^{\frac 12+\delta}\varpi^N\Om_\Ke^2\nab^2(\tK,\widetilde{\sigmac}) \|_{L^2_uL^2_{\ub}L^2(S)}^2+\|\ub^{1+2\delta}\varpi^{2N}\Om_\Ke^4\tpH\nab^2(\tK,\widetilde{\sigmac})\nab^2(\tK,\widetilde{\sigmac}) \|_{L^1_uL^1_{\ub}L^1(S)}.
\end{split}
\end{equation}
We now deal with the second term on the right hand side of \eqref{est.intbyparts.4.0}. Using H\"older's inequality and then using Sobolev embedding (Proposition~\ref{Sobolev}) and the bootstrap assumption \eqref{BA} to control the $L^1_{\ub}L^\i_uL^\i(S)$ norm of $\tpHb$, we obtain
\begin{equation*}
\begin{split}
&\|\ub^{1+2\delta}\varpi^{2N}\tpHb\nab^2(\Om\Om_\Ke\widetilde{\beta})\nab^2(\Om\Om_\Ke\widetilde{\beta}) \|_{L^1_uL^1_{\ub}L^1(S)}\\
\ls &\:\| \tpHb\|_{L^1_uL^\i_{\ub}L^\i(S)} \|\ub^{\f 12+\delta}\varpi^{N}\nab^2(\Om\Om_\Ke\widetilde{\beta})\|_{L^\i_uL^2_{\ub}L^2(S)}^2\\ 
\ls &\:\ep^{\f 12} \|\ub^{\f 12+\delta}\varpi^{N}\nab^2(\Om\Om_\Ke\widetilde{\beta})\|_{L^\i_uL^2_{\ub}L^2(S)}^2.
\end{split}
\end{equation*}
Similarly, for the last term on the right hand side of \eqref{est.intbyparts.4.0}, we use H\"older's inequality and then use Proposition~\ref{Sobolev} and \eqref{BA} to control the $L^1_{\ub}L^\i_uL^\i(S)$ norm of $\Om_\Ke^2\tpH$ to obtain
\begin{equation*}
\begin{split}
&\:\|\ub^{1+2\delta}\varpi^{2N}\Om_\Ke^4\tpH\nab^2(\tK,\widetilde{\sigmac})\nab^2(\tK,\widetilde{\sigmac}) \|_{L^1_uL^1_{\ub}L^1(S)}\\
\ls &\:\|\ub^{\f 12+ \delta}\varpi^{N}\Om_\Ke\nab^2(\tK,\widetilde{\sigmac})\|_{L^\i_{\ub} L^2_{u}L^2(S)}^2 \|\Om_\Ke^2 \tpH\|_{L^1_{\ub} L^\i_u L^\i(S)}\\
\ls &\: \ep^{\f 12}\|\ub^{\f 12+ \delta}\varpi^{N}\Om_\Ke\nab^2(\tK,\widetilde{\sigmac})\|_{L^\i_{\ub} L^2_{u}L^2(S)}^2.
\end{split}
\end{equation*}
Therefore, returning to \eqref{est.intbyparts.4.0}, we obtain
\begin{equation}\label{est.intbyparts.4}
\begin{split}
|IV| \ls & \:\|\ub^{\frac 12+\delta}\varpi^N\Om_\Ke\nab^2(\Om\Om_\Ke\widetilde{\beta}) \|_{L^2_uL^2_{\ub}L^2(S)}^2+\|\ub^{\frac 12+\delta}\varpi^N\Om_\Ke^2\nab^2(\tK,\widetilde{\sigmac}) \|_{L^2_uL^2_{\ub}L^2(S)}^2\\
&\: +\ep^{\f 12} \|\ub^{\f 12+\delta}\varpi^{N}\nab^2(\Om\Om_\Ke\widetilde{\beta})\|_{L^\i_uL^2_{\ub}L^2(S)}^2+\ep^{\f 12}\|\ub^{\f 12+ \delta}\varpi^{N}\Om_\Ke\nab^2(\tK,\widetilde{\sigmac})\|_{L^\i_{\ub} L^2_{u}L^2(S)}^2.
\end{split}
\end{equation}

\textbf{Putting together the terms $I$, $II$, $III$ and $IV$.} Gathering the bounds \eqref{EE.1.1}, \eqref{est.intbyparts.1}, \eqref{est.intbyparts.2}, \eqref{est.intbyparts.3} and \eqref{est.intbyparts.4}, putting the terms in \eqref{est.intbyparts.2} with good signs to the left hand side, taking supremums in $u$ and $\ub$, and controlling the data term by $2^{2N}\mathcal D$ using \eqref{data.D.def} and \eqref{varpi.bd}, we obtain
\begin{equation*}
\begin{split}
&\:\|\ub^{\frac 12+\delta}\varpi^N\nab^2(\Om\Om_\Ke\widetilde{\beta})\|^2_{L^\infty_u L^2_{\ub}L^2(S)}  +N\|\ub^{\frac 12+\delta}\varpi^N\Om_\Ke\nab^2(\Om\Om_\Ke\widetilde{\beta})\|^2_{L^2_{\ub} L^2_{u}L^2(S)}  \\
&\:+\|\ub^{\frac 12+\delta}\varpi^N\Om_\Ke\nab^2(\tK,\widetilde{\sigmac})\|^2_{L^\infty_{\ub} L^2_uL^2(S)} +N\|\ub^{\frac 12+\delta}\varpi^N\Om_\Ke^2\nab^2(\tK,\widetilde{\sigmac})\|^2_{L^2_{\ub} L^2_{u}L^2(S)}  \\
&\:+\|\ub^{\frac 12+\delta}\varpi^N\Om_\Ke\nab^2(\tK,\widetilde{\sigmac})\|^2_{L^2_{\ub} L^2_{u}L^2(S)} \\
\ls &\:2^{2N}\mathcal D+\|\ub^{\frac 12+\delta}\varpi^{N} \Er_1\|_{L^2_uL^2_{\ub}L^2(S)}^2+\|\ub^{1+2\delta}\varpi^{2N}\Om_\Ke^4 \mathcal T_2\|_{L^1_uL^1_{\ub}L^1(S)}\\
&\:+\ep^{\f 12}\|\ub^{\f 12+\delta}\varpi^{N}\nab^2(\Om\Om_\Ke\widetilde{\beta})\|_{L^\i_uL^2_{\ub}L^2(S)}^2+\ep^{\f 12}\|\ub^{\f 12+ \delta}\varpi^{N}\Om_\Ke\nab^2(\tK,\widetilde{\sigmac})\|_{L^\i_{\ub} L^2_{u}L^2(S)}^2\\
&\:+\|\ub^{\frac 12+\delta}\varpi^N\Om_\Ke\nab^2(\Om\Om_\Ke\widetilde{\beta}) \|_{L^2_uL^2_{\ub}L^2(S)}^2+\|\ub^{\frac 12+\delta}\varpi^N\Om_\Ke^2\nab^2(\tK,\widetilde{\sigmac}) \|_{L^2_uL^2_{\ub}L^2(S)}^2\\
&\:+\|\ub^{\frac 12+\delta}\varpi^N\Om_\Ke\nab^2(\Om\Om_\Ke\widetilde{\beta}) \|_{L^2_uL^2_{\ub}L^2(S)}\|\ub^{\frac 12+\de}\varpi^N\Er_2 \|_{L^2_uL^2_{\ub}L^2(S)}\\
&\:+\|\ub^{\frac 12+\delta}\varpi^N\Om_\Ke\nab^2(\tK,\widetilde{\sigmac}) \|_{L^2_uL^2_{\ub}L^2(S)}\|\ub^{\frac 12+\de}\varpi^N\Er_1 \|_{L^2_uL^2_{\ub}L^2(S)}.
\end{split}
\end{equation*}
For the last two terms on the right hand side, we can apply Young's inequality and absorb the terms with second derivatives of curvature to the left hand side. For the remaining terms, notice that for $\ep_0$ (and hence $\ep$) sufficiently small, the terms $\ep^{\f 12}\|\ub^{\frac 12+\de}\varpi^N\nab^2(\Om\Om_\Ke\widetilde{\beta})\|^2_{L^\infty_uL^2_{\ub}L^2(S)}$ and $\ep^{\f 12}\|\ub^{\f 12+ \delta}\varpi^{N}\Om_\Ke\nab^2(\tK,\widetilde{\sigmac})\|_{L^\i_{\ub} L^2_{u}L^2(S)}^2$ can be absorbed to the left hand side. Similarly, for $N$ sufficiently large, the term $\|\ub^{\frac 12+\delta}\varpi^N\Om_\Ke\nab^2(\Om\Om_\Ke\widetilde{\beta}) \|_{L^2_uL^2_{\ub}L^2(S)}^2$ and the term $\|\ub^{\frac 12+\delta}\varpi^N\Om_\Ke^2\nab^2(\tK,\widetilde{\sigmac}) \|_{L^2_uL^2_{\ub}L^2(S)}^2$ can be absorbed to the left hand side. Therefore, we have
\begin{equation}\label{EE1.almost}
\begin{split}
&\|\ub^{\frac 12+\delta}\varpi^N\nab^2(\Om\Om_\Ke\widetilde{\beta})\|^2_{L^\infty_u L^2_{\ub}L^2(S)}  +N\|\ub^{\frac 12+\delta}\varpi^N\Om_\Ke\nab^2(\Om\Om_\Ke\widetilde{\beta})\|^2_{L^2_{\ub} L^2_{u}L^2(S)}  \\
&+\|\ub^{\frac 12+\delta}\varpi^N\Om_\Ke\nab^2(\tK,\widetilde{\sigmac})\|^2_{L^\infty_{\ub} L^2_uL^2(S)} +N\|\ub^{\frac 12+\delta}\varpi^N\Om_\Ke^2\nab^2(\tK,\widetilde{\sigmac})\|^2_{L^2_{\ub} L^2_{u}L^2(S)}  \\
&+\|\ub^{\frac 12+\delta}\varpi^N\Om_\Ke\nab^2(\tK,\widetilde{\sigmac})\|^2_{L^2_{\ub} L^2_{u}L^2(S)} \\
\ls &2^{2N}\mathcal D+\|\ub^{\frac 12+\delta}\varpi^{N} \Er_1\|_{L^2_uL^2_{\ub}L^2(S)}^2+N^{-1}\|\ub^{\frac 12+\delta}\varpi^{N} \Er_2\|_{L^2_uL^2_{\ub}L^2(S)}^2+\|\ub^{1+2\delta}\varpi^{2N}\Om_\Ke^4 \mathcal T_2\|_{L^1_uL^1_{\ub}L^1(S)}.
\end{split}
\end{equation}

\textbf{Estimating the $\widetilde{\beta}$ terms on the left hand side.} It now remains to show that the two terms $\|\ub^{\frac 12+\delta}\varpi^N\nab^2(\Om\Om_\Ke\widetilde{\beta})\|^2_{L^\infty_u L^2_{\ub}L^2(S)}$ and $N\|\ub^{\frac 12+\delta}\varpi^N\Om_\Ke\nab^2(\Om\Om_\Ke\widetilde{\beta})\|^2_{L^2_{\ub} L^2_{u}L^2(S)}$ on the left hand side of \eqref{EE1.almost} indeed control $\|\ub^{\frac 12+\delta}\varpi^N\Om_\Ke^2\nab^2\widetilde{\beta}\|^2_{L^\infty_u L^2_{\ub}L^2(S)}$ and $N\|\ub^{\frac 12+\delta}\varpi^N\Om_\Ke^3\nab^2\widetilde{\beta}\|^2_{L^2_{\ub} L^2_{u}L^2(S)}$ up to some acceptable error terms. 

We first consider $\|\ub^{\frac 12+\delta}\varpi^N\nab^2(\Om\Om_\Ke\widetilde{\beta})\|^2_{L^\infty_u L^2_{\ub}L^2(S)}$ and commute the functions $\Om\Om_\Ke$ out of the angular covariant derivatives. More precisely, using \eqref{EE.1.Om.bd}, we obtain
\begin{equation}\label{EE.1.bdry}
\begin{split}
&\left|\int_{-u+C_R}^{\ub}\|\ub^{\frac 12+\de}\varpi^N\Om_\Ke^2\nab^2\widetilde{\beta}\|^2_{L^2(S_{u,\ub'})}d\ub'
-\int_{-u+C_R}^{\ub}\|\ub^{\frac 12+\de}\varpi^N\nab^2(\Om\Om_\Ke\widetilde{\beta})\|^2_{L^2(S_{u,\ub'})}d\ub'\right|\\
\ls &\sum_{\substack{i_1+i_2\leq 2}}\|\ub^{\frac 12+\de}\varpi^N\Om_\Ke^2\nab^{i_1}\tg\nab^{i_2}\widetilde{\beta}\|^2_{L^\infty_uL^2_{\ub}L^2(S)}+\sum_{\substack{i\leq 1}}\|\ub^{\frac 12+\de}\varpi^N\Om_\Ke^2\nab^{i}\widetilde{\beta}\|^2_{L^\infty_uL^2_{\ub}L^2(S)}.
\end{split}
\end{equation}
We control each of the terms in \eqref{EE.1.bdry}. For the first term, using the H\"older's inequality, Sobolev embedding theorem in Proposition \ref{Sobolev} and the bootstrap assumption \eqref{BA} (in the style of Section~\ref{sec:rmk.BA.S}), we have
\begin{equation}\label{firsttermin.EE.1.bdry}
\begin{split}
&\:\sum_{\substack{i_1+i_2\leq 2}}\|\ub^{\frac 12+\de}\varpi^N\Om_\Ke^2\nab^{i_1}\tg\nab^{i_2}\widetilde{\beta}\|^2_{L^\infty_uL^2_{\ub}L^2(S)}\\
\ls &\:(\sum_{i_1\leq 2}\|\nab^{i_1}\tg \|_{L^\i_uL^\i_{\ub}L^2(S)})(\sum_{i_2\leq 2}\|\ub^{\frac 12+\de}\varpi^N\Om_\Ke^2\nab^{i_2}\widetilde{\beta}\|^2_{L^\infty_uL^2_{\ub}L^2(S)})\ls \ep^{\f 12}\sum_{i\leq 2}\|\ub^{\frac 12+\de}\varpi^N\Om_\Ke^2\nab^i\widetilde{\beta}\|^2_{L^\infty_uL^2_{\ub}L^2(S)}.
\end{split}
\end{equation}
In the case $\widetilde{\beta}$ has at most $1$ derivative, we first use the reduced schematic Codazzi equation in \eqref{EE.1.Codazzi.2} (in the $j=1$ case), and then use \eqref{BA.S.222.lower} to obtain the following estimate
\begin{equation}\label{beta.commute.1.0}
\begin{split}
&\:\sum_{i\leq 1}\|\ub^{\frac 12+\de}\varpi^N\Om_\Ke^2\nab^{i}\widetilde{\beta}\|^2_{L^\infty_uL^2_{\ub}L^2(S)}\\
\ls &\:\sum_{i_1+i_2\leq 2}\|\ub^{\frac 12+\de}\varpi^N\Om_\Ke^2(1+\nab^{i_1-1}\tp+\nab^{i_1}\tg)(\nab^{i_2}(\tg,\tpH),\nab^{\min\{i_2,2\}}\tp)\|^2_{L^\infty_uL^2_{\ub}L^2(S)}\\
\ls &\:\sum_{i \leq 2}\|\ub^{\frac 12+\de}\varpi^N\Om_\Ke^2\nab^{i}(\tg,\tp,\tpH)\|^2_{L^\infty_uL^2_{\ub}L^2(S)}\\
\ls &\:2^{2N}\mathcal D+\|\ub^{\frac 12+\delta}\varpi^{N} \Er_1\|_{L^2_uL^2_{\ub}L^2(S)}^2+N^{-1}\|\ub^{\frac 12+\delta}\varpi^{N} \Er_2\|_{L^2_uL^2_{\ub}L^2(S)}^2+\|\ub^{1+2\delta}\varpi^{2N}\Om_\Ke^4 \mathcal T_2\|_{L^1_uL^1_{\ub}L^1(S)},
\end{split}
\end{equation}
where in the last step we have used the estimates derived in Propositions \ref{prop.g}, \ref{prop.etab}, \ref{improved.eta} and \ref{prop.tpH}. Plugging this back into \eqref{firsttermin.EE.1.bdry}, we obtain the following bound for the first term in \eqref{EE.1.bdry}:
\begin{equation}\label{beta.commute.1}
\begin{split}
&\:\sum_{i_1+i_2\leq 2}\|\ub^{\frac 12+\de}\varpi^N\Om_\Ke^2\nab^{i_1}\tg\nab^{i_2}\widetilde{\beta}\|^2_{L^\infty_uL^2_{\ub}L^2(S)}\\
\ls &\:\ep^{\f12}(2^{2N}\mathcal D+\|\ub^{\frac 12+\delta}\varpi^{N} \Er_1\|_{L^2_uL^2_{\ub}L^2(S)}^2+N^{-1}\|\ub^{\frac 12+\delta}\varpi^{N} \Er_2\|_{L^2_uL^2_{\ub}L^2(S)}^2+\|\ub^{1+2\delta}\varpi^{2N}\Om_\Ke^4 \mathcal T_2\|_{L^1_uL^1_{\ub}L^1(S)})\\
&\:+\ep^{\f 12}\|\ub^{\frac 12+\de}\varpi^N\Om_\Ke^2\nab^2\widetilde{\beta}\|^2_{L^\infty_uL^2_{\ub}L^2(S)}.
\end{split}
\end{equation}
Now note that when proving the above bound, we have also establish the estimate \eqref{beta.commute.1.0} for the second term in \eqref{EE.1.bdry}. Therefore, combining \eqref{beta.commute.1} and \eqref{beta.commute.1.0}, and returning to \eqref{EE.1.bdry}, we obtain
\begin{equation}\label{beta.commute.1.final}
\begin{split}
&\: \|\ub^{\frac 12+\de}\varpi^N\Om_\Ke^2\nab^2\widetilde{\beta}\|^2_{L^\infty_u L^2_{\ub}L^2(S)}\\
\ls &\: \|\ub^{\frac 12+\de}\varpi^N\nab^2(\Om\Om_\Ke\widetilde{\beta})\|^2_{L^\infty_u L^2_{\ub}L^2(S)} + 2^{2N}\mathcal D+\|\ub^{\frac 12+\delta}\varpi^{N} \Er_1\|_{L^2_uL^2_{\ub}L^2(S)}^2\\
&\: +N^{-1}\|\ub^{\frac 12+\delta}\varpi^{N} \Er_2\|_{L^2_uL^2_{\ub}L^2(S)}^2 +\|\ub^{1+2\delta}\varpi^{2N}\Om_\Ke^4 \mathcal T_2\|_{L^1_uL^1_{\ub}L^1(S)}) + \ep^{\f 12}\|\ub^{\frac 12+\de}\varpi^N\Om_\Ke^2\nab^2\widetilde{\beta}\|^2_{L^\infty_uL^2_{\ub}L^2(S)}\\
\ls &\: \|\ub^{\frac 12+\de}\varpi^N\nab^2(\Om\Om_\Ke\widetilde{\beta})\|^2_{L^\infty_u L^2_{\ub}L^2(S)} + 2^{2N}\mathcal D+\|\ub^{\frac 12+\delta}\varpi^{N} \Er_1\|_{L^2_uL^2_{\ub}L^2(S)}^2 +N^{-1}\|\ub^{\frac 12+\delta}\varpi^{N} \Er_2\|_{L^2_uL^2_{\ub}L^2(S)}^2 \\
&\:+\|\ub^{1+2\delta}\varpi^{2N}\Om_\Ke^4 \mathcal T_2\|_{L^1_uL^1_{\ub}L^1(S)}),
\end{split}
\end{equation}
where the last line is achieved by absorbing $\ep^{\f 12}\|\ub^{\frac 12+\de}\varpi^N\Om_\Ke^2\nab^2\widetilde{\beta}\|^2_{L^\infty_uL^2_{\ub}L^2(S)}$ to the left hand side (for $\ep_0$, and hence $\ep$, sufficiently small).

The term $N\|\ub^{\frac 12+\delta}\varpi^N\Om_\Ke\nab^2(\Om\Om_\Ke\widetilde{\beta})\|^2_{L^2_{\ub} L^2_{u}L^2(S)}$ can be treated similarly using \eqref{EE.1.Om.bd}, \eqref{EE.1.Codazzi.2} and \eqref{BA.S.222.lower} so that we obtain
\begin{equation}\label{beta.commute.2}
\begin{split}
&N\|\ub^{\frac 12+\delta}\varpi^N\Om_\Ke^3\nab^2\widetilde{\beta}\|^2_{L^2_{\ub} L^2_{u}L^2(S)}\\
\ls &N\|\ub^{\frac 12+\delta}\varpi^N\Om_\Ke\nab^2(\Om\Om_\Ke\widetilde{\beta})\|^2_{L^2_{\ub} L^2_{u}L^2(S)}+N\ep^{\f 12}\|\ub^{\frac 12+\delta}\varpi^N\Om_\Ke^3\nab^2\widetilde{\beta}\|^2_{L^2_{\ub} L^2_{u}L^2(S)}\\
&+N\ep^{\f 12}(\sum_{i\leq 2}\|\ub^{\frac 12+\delta}\varpi^N\Om_\Ke^3\nab^{i}(\tg,\tp,\tpH)\|^2_{L^2_{\ub} L^2_{u}L^2(S)})\\
\ls &N\|\ub^{\frac 12+\delta}\varpi^N\Om_\Ke\nab^2(\Om\Om_\Ke\widetilde{\beta})\|^2_{L^2_{\ub} L^2_{u}L^2(S)}+2^{2N}\mathcal D+\|\ub^{\frac 12+\delta}\varpi^{N} \Er_1\|_{L^2_uL^2_{\ub}L^2(S)}^2+N^{-1}\|\ub^{\frac 12+\delta}\varpi^{N} \Er_2\|_{L^2_uL^2_{\ub}L^2(S)}^2\\
&+\|\ub^{1+2\delta}\varpi^{2N}\Om_\Ke^4 \mathcal T_2\|_{L^1_uL^1_{\ub}L^1(S)},
\end{split}
\end{equation}
where in the last line, we have 
\begin{itemize}
\item used Propositions \ref{prop.g}, \ref{prop.etab}, \ref{improved.eta} and \ref{prop.tpH} to estimate $\sum_{i_2\leq 2}\|\ub^{\frac 12+\delta}\varpi^N\Om_\Ke^3\nab^{i_2}(\tg,\tp,\tpH)\|^2_{L^2_{\ub} L^2_{u}L^2(S)}$;
\item absorbed the term $N\ep^{\f 12}\|\ub^{\frac 12+\delta}\varpi^N\Om_\Ke^3\nab^2\widetilde{\beta}\|^2_{L^2_{\ub} L^2_{u}L^2(S)}$ to the left hand side.
\end{itemize}
Finally, combining \eqref{EE1.almost}, \eqref{beta.commute.1.final} and \eqref{beta.commute.2}, we conclude the proof of the proposition.
\end{proof}

\subsection{Energy estimates for $\nab^2\tK$, $\nab^2\widetilde{\sigmac}$ and $\protect\nab^2\protect\widetilde{\protect\betab}$}\label{sec:EE.2}
Next, we prove the energy estimates for the second derivatives of the remaining differences of curvature components, i.e.~$\tK$, $\widetilde{\sigmac}$ and $\widetilde{\betab}$. Here, we need to use $|u|$ weights instead of $\ub$ weights. As a consequence, the spacetime term that we obtain for $\nab^i(\tK,\widetilde{\sigmac})$ in the following proposition are weaker than those we have already derived in Proposition \ref{EE.1}. Therefore, we will simply drop these terms in the following proposition.

\begin{proposition}\label{EE.2}
$\nab^2(\tK,\widetilde{\sigmac},\widetilde{\betab})$ obey the following estimate:
\begin{equation*}
\begin{split}
&\||u|^{\frac 12+\delta}\varpi^N\nab^2\widetilde{\betab}\|^2_{L^\infty_{\ub} L^2_{u}L^2(S)}  +N\||u|^{\frac 12+\delta}\varpi^N\Om_\Ke\nab^2\widetilde{\betab}\|^2_{L^2_{\ub} L^2_{u}L^2(S)} +\||u|^{\frac 12+\delta}\varpi^N\Om_\Ke\nab^2(\tK,\widetilde{\sigmac})\|^2_{L^\infty_{u} L^2_{\ub}L^2(S)} \\
\ls &2^{2N}\mathcal D+\|\ub^{\frac 12+\delta}\varpi^{N} \Er_1\|_{L^2_uL^2_{\ub}L^2(S)}^2+N^{-1}\||u|^{\frac 12+\delta}\varpi^{N} \Er_2\|_{L^2_uL^2_{\ub}L^2(S)}^2+\||u|^{1+2\delta}\varpi^{2N}\Om_\Ke^2 \mathcal T_1\|_{L^1_uL^1_{\ub}L^1(S)}.
\end{split}
\end{equation*}
\end{proposition}
\begin{proof}
The proof will be very similar to that of Proposition~\ref{EE.1}, except that we will use the reduced schematic equations for the curvature components from Proposition~\ref{curv.red.sch.2} instead.

\textbf{Application of Propositions~\ref{transport} and \ref{curv.red.sch.2}.} We now apply the identities in Proposition \ref{transport}. Fix $u$ and $\ub$. Applying Proposition \ref{transport} to $|u|^{\frac 12+\de}\varpi^N\nab^2\widetilde{\betab}$, we obtain
\begin{equation}\label{eqn.betab}
\begin{split}
&\:\||u|^{\frac 12+\de}\varpi^N\nab^2\widetilde{\betab}\|^2_{L^2(S_{u,\ub})}\\
=&\:\||u|^{\frac 12+\de}\varpi^N\nab^2\widetilde{\betab}\|^2_{L^2(S_{u,-u+C_R})}\\
&\:+2\int_{-u+C_R}^{\ub}\int_{S_{u,\ub'}} |u'|^{1+2\de}\Om^2\left(\varpi^N\nab^2_{AB}\widetilde{\betab}_C\slashed{\nabla}_4(\varpi^N(\nab^2)^{AB}\widetilde{\betab}^C)\right) d{\ub'}\\
&\:+ \int_{-u+C_R}^{\ub}\int_{S_{u,\ub'}} |u'|^{1+2\de}\Om^2\left(\trch \varpi^{2N}|\nab^2\widetilde{\betab}|^2_{\gamma}\right)d{\ub'}.
\end{split}
\end{equation}
Applying Proposition \ref{transport} to $|u|^{\frac 12+\de}\varpi^N\Om\nab^2\widetilde{\sigmac}$ and $|u|^{\frac 12+\de}\varpi^N\Om\nab^2\tK$, we obtain, respectively,
\begin{equation}\label{eqn.sigmac.2}
\begin{split}
&\:\||u|^{\frac 12+\de}\varpi^N\Om\nab^2\widetilde{\sigmac}\|^2_{L^2(S_{u,\ub})}\\
=&\:\||u|^{\frac 12+\de}\varpi^N\Om\nab^2\widetilde{\sigmac}\|^2_{L^2(S_{-\ub+C_R,\ub})}\\
&\:+2\int_{-\ub+C_R}^{u}\int_{S_{u',\ub}} \left(|u'|^{\frac 12+\de}\varpi^N\Om\nab^2_{AB}\widetilde{\sigmac}\slashed{\nabla}_3(|u|^{\frac 12+\de}\varpi^N\Om(\nab^2)^{AB}\widetilde{\sigmac})\right) du'\\
&\:+ \int_{-\ub+C_R}^{u}\int_{S_{u',\ub}} \left(\trchb ||u'|^{\frac 12+\de}\varpi^N\Om\nab^2\widetilde{\sigmac}|^2_{\gamma}\right)du'
\end{split}
\end{equation}
and
\begin{equation}\label{eqn.K.2}
\begin{split}
&\:\||u|^{\frac 12+\de}\varpi^N\Om\nab^2\tK\|^2_{L^2(S_{u,\ub})}\\
=&\:\||u|^{\frac 12+\de}\varpi^N\Om\nab^2\tK\|^2_{L^2(S_{-\ub+C_R,\ub})}\\
&\:+2\int_{-\ub+C_R}^{u}\int_{S_{u',\ub}}  \left(|u'|^{\frac 12+\de}\varpi^N\Om\nab^2_{AB}\tK\slashed{\nabla}_3(|u|^{\frac 12+\de}\varpi^N\Om(\nab^2)^{AB}\tK)\right) du'\\
&\:+ \int_{-\ub+C_R}^{u}\int_{S_{u',\ub}}  \left(\trchb ||u'|^{\frac 12+\de}\varpi^N\Om\nab^2\tK|^2_{\gamma}\right)du'.
\end{split}
\end{equation}
We now integrate \eqref{eqn.sigmac.2} and \eqref{eqn.K.2} in $\ub$ and integrate \eqref{eqn.betab} in $u$. Then add up these three equations and substituting in the reduced schematic equations in Proposition~\ref{curv.red.sch.2}, we obtain the following equation
\begin{equation}\label{EE2.main}
\begin{split}
&\:\int_{-\ub+C_R}^u \||u'|^{\frac 12+\de}\varpi^N\nab^2\widetilde{\betab}\|^2_{L^2(S_{u',\ub})}\, du'\\
&\:+\int_{-u+C_R}^{\ub} \left(\||u|^{\frac 12+\de}\varpi^N\Om\nab^2\widetilde{\sigmac}\|^2_{L^2(S_{u,\ub'})}+\||u|^{\frac 12+\de}\varpi^N\Om\nab^2\tK\|^2_{L^2(S_{u,\ub'})}\right)\, d\ub'\\
= & \:\int_{-\ub+C_R}^u \||u'|^{\frac 12+\de}\varpi^N\nab^2\widetilde{\betab}\|^2_{L^2(S_{u',-u'+C_R})}\, du'\\
&\:\int_{-u+C_R}^{\ub} \left(\||u|^{\frac 12+\de}\varpi^N\Om\nab^2\widetilde{\sigmac}\|^2_{L^2(S_{-\ub'+C_R,\ub'})}+\||u|^{\frac 12+\de}\varpi^N\Om\nab^2\tK\|^2_{L^2(S_{-\ub'+C_R,\ub'})}\right)\, d\ub'+ I+II+III+IV,
\end{split}
\end{equation}
where $I$, $II$, $III$ and $IV$ are to be defined now. First, we have the terms where one of the curvature components has three angular derivatives:
\begin{equation*}
\begin{split}
I=&\:2\int_{-u+C_R}^{\ub}\int_{-\ub'+C_R}^{u}\int_{S_{u',\ub'}} |u'|^{1+2\de}\varpi^{2N}\Om^2 \nab^2_{AB}\widetilde{\betab}_C(\nab^3_{CAB} \tK  +\in^C{ }_{D}(\slashed{\nabla}^3)^{DAB}\widetilde{\sigmac}) d{u'}d\ub'\\
&\:-2\int_{-u+C_R}^{\ub}\int_{-\ub'+C_R}^{u}\int_{S_{u',\ub'}} |u'|^{1+2\de}\varpi^{2N}\Omega^2\nab^2_{AB}\widetilde{\sigmac}\nab^C\nab^2_{AB}\in_{C}{ }^D\widetilde{\betab}_D du'd{\ub'}\\
&\:+2\int_{-u+C_R}^{\ub}\int_{-\ub'+C_R}^{u}\int_{S_{u',\ub'}} |u'|^{1+2\de}\varpi^{2N}\Omega^2\nab^2_{AB}\tK\nab^C(\nab^2)^{AB}\widetilde{\betab}_C du'd{\ub'}.
\end{split}
\end{equation*}
Second, we have terms where $\nab_3$ and $\nab_4$ act on the weight functions:
\begin{equation*}
\begin{split}
II=&\:2\int_{-u+C_R}^{\ub}\int_{-\ub'+C_R}^{u}\int_{S_{u',\ub'}} |u'|^{1+2\de}\varpi^N\Omega^2(\slashed{\nabla}_4 \varpi^N)|\nab^2\widetilde{\betab}|^2_{\gamma}du'd{\ub'}\\
&\:+2\int_{-u+C_R}^{\ub}\int_{-\ub'+C_R}^{u}\int_{S_{u',\ub'}} |u'|^{\frac 12+\de}\varpi^N\Omega\slashed{\nabla}_3(|u|^{\frac 12+\de}\varpi^N\Om)(|\nab^2\widetilde{\sigmac}|^2_{\gamma}+|\nab^2\tK|^2_{\gamma})du'd{\ub'}.
\end{split}
\end{equation*}
Third, we have the terms which arise from the error terms in the reduced schematic equations in Proposition~\ref{curv.red.sch.2}. Similarly as the proof of Proposition~\ref{EE.1}, it is sufficient for our purpose to give $III$ in its reduced schematic form.
\begin{equation*}
\begin{split}
III \eqrs &\:2\int_{-u+C_R}^{\ub}\int_{-\ub'+C_R}^{u}\int_{S_{u',\ub'}} |u'|^{1+2\de}\varpi^{2N}\Om^2 \langle\nab^2\widetilde{\betab},\mathcal F_4\rangle_\gamma d{u'}d\ub'\\
&\:+2\int_{-u+C_R}^{\ub}\int_{-\ub'+C_R}^{u}\int_{S_{u',\ub'}} |u'|^{1+2\de}\varpi^{2N}\Omega^2\langle\nab^2\widetilde{\sigmac}+\nab^2\tK,\mathcal F_3\rangle_\gamma du'd{\ub'}.
\end{split}
\end{equation*}
Finally, we have the term
\begin{equation*}
\begin{split}
IV=&\int_{-u+C_R}^{\ub}\int_{-\ub'+C_R}^{u}\int_{S_{u',\ub'}} |u'|^{1+2\de}\varpi^{2N}\Omega^2\left(\trch|\nab^2\widetilde{\betab}|^2_{\gamma}+\trchb |\nab^2\widetilde{\sigmac}|^2_{\gamma}+\trchb|\nab^2\tK|^2_{\gamma}\right)du' d{\ub'}.
\end{split}
\end{equation*}

\textbf{Estimates for the term $I$.} We now estimate each of these terms. $I$ will be handled as in the proof of Proposition~\ref{EE.1}, where we perform an integration by parts on the $2$-spheres $S_{u,\ub}$ using Proposition~\ref{intbypartssph} to avoid a loss of derivatives. More precisely, we apply Proposition \ref{intbypartssph} to the first line and note that all the terms where one of the curvature components has three angular derivatives cancel. Notice moreover that we have $\nab \varpi=0$ (by \eqref{varpi.ang}) and $\nab\Om^2\ls \Om_\Ke^2$ (by Corollary~\ref{Linfty}). Therefore, using Young's inequality, we have
\begin{equation}\label{est.intbyparts.1.1}
\begin{split}
|I| \ls &\:\||u|^{1+2\de}\varpi^{2N}\Om_\Ke^2\nab^2\widetilde{\betab}\nab^2(\tK,\widetilde{\sigmac}) \|_{L^1_uL^1_{\ub}L^1(S)}\\
\ls &\:\||u|^{\frac 12+\de}\varpi^{N}\Om_\Ke\nab^2\widetilde{\betab} \|_{L^2_uL^2_{\ub}L^2(S)}\||u|^{\frac 12+\de}\varpi^{N}\Om_\Ke\nab^2(\tK,\widetilde{\sigmac}) \|_{L^2_uL^2_{\ub}L^2(S)}\\
\ls &\: N^{\f 12}\||u|^{\frac 12+\de}\varpi^{N}\Om_\Ke\nab^2\widetilde{\betab} \|_{L^2_uL^2_{\ub}L^2(S)}^2+ N^{-\f 12}\||u|^{\frac 12+\de}\varpi^{N}\Om_\Ke\nab^2(\tK,\widetilde{\sigmac}) \|_{L^2_uL^2_{\ub}L^2(S)}^2.
\end{split}
\end{equation}

\textbf{Estimates for the term $II$.} The term $II$ will be handled similarly as in the proof of Propositions~\ref{transport.3.3} and \ref{transport.4.1}, i.e.~we use the estimates \eqref{nab3.weight.good.bound.2} and \eqref{nab4.varpi.neg} for $\slashed{\nabla}_3(|u|^{\frac 12+\de}\varpi^N\Om_\Ke)$ and $\slashed{\nabla}_4 \varpi^N$ respectively to show that $II$ has a good sign (after choosing $|u_f|$ to be sufficiently large), except we also need to take into account the difference between $\slashed\nab_3(|u|^{\frac 12+\de}\varpi^N\Om)$ and $\slashed\nab_3(|u|^{\frac 12+\de}\varpi^N\Om_\Ke)$. This difference can be controlled by (cf.~\eqref{gauge.con}, \eqref{varpi.34} and \eqref{Om.c})
\begin{equation*}
\begin{split}
&\slashed\nab_3(|u|^{\frac 12+\de}\varpi^N\Om) -\slashed \nab_3(|u|^{\frac 12+\de}\varpi^N\Om_\Ke)\\
= &\nab_3(|u|^{\f 12+\de}\varpi^N) (\Om-\Om_\Ke)- |u|^{\f 12+\de} \varpi^N(\Om \omb - \Om_\Ke \omb_\Ke) \eqrs N |u|^{\f 12+\de} \varpi^N\Om_\Ke^3 \tg+ |u|^{\f 12+\de} \varpi^N\Om_\Ke(\tg,\widetilde{\omb}).
\end{split}
\end{equation*}
Therefore, using Corollary~\ref{Linfty} for $\tg$ and $\widetilde{\omb}$, and choosing $\ep_0$ (and hence $\ep$) to be sufficiently small, these terms can be controlled by the main terms in \eqref{nab3.weight.good.bound.2} and \eqref{nab4.varpi.neg}. In other words, for some constant $C>0$ depending only on $M$, $a$ and $C_R$,
\begin{equation}\label{est.intbyparts.1.2}
\begin{split}
II \leq & -C^{-1}\||u|^{\frac 12+\delta}\varpi^N\Om_\Ke\nab^2(\tK,\widetilde{\sigmac})\|^2_{L^2_{\ub} L^2_uL^2(S)}  \\
&-C^{-1}N\||u|^{\frac 12+\delta}\varpi^N\Om_\Ke\nab^2\widetilde{\betab}\|^2_{L^2_{\ub} L^2_{u}L^2(S)}  -C^{-1}N\||u|^{\frac 12+\delta}\varpi^N\Om_\Ke^2\nab^2(\tK,\widetilde{\sigmac})\|^2_{L^2_{\ub} L^2_{u}L^2(S)}.
\end{split}
\end{equation}

\textbf{Estimates for the term $III$.} The term $III$ is an error term. As in the proof of Proposition~\ref{prop.tpHb}, we split the term $\mathcal F_4$ into
$$\mathcal F_4=\mathcal F_{4,1}+\mathcal F_{4,2},$$
where, $\mathcal F_{4,1}$ and $\mathcal F_{4,2}$ obey
\begin{equation}\label{F4.basic.est}
\Om_\Ke|\mathcal F_{4,1}|\ls  \Er_1+\Er_2, \quad |(\sum_{i_1+i_2\leq 3}(1+\nab^{i_1}\tg+\nab^{i_1-1}\tp)\nab^{i_2}\tpHb)\mathcal F_{4,2}|\ls \mathcal T_1.
\end{equation}
Here, the first bound is \eqref{F41.first.bound} in the proof of Proposition~\ref{prop.tpHb} and the second bound is a variation of \eqref{F42.first.bound}, which again follows from comparing the terms with those in \eqref{Er.def}. Using the Cauchy--Schwarz inequality and \eqref{Om.c}, the first estimate in \eqref{F4.basic.est} implies
\begin{equation}\label{EE.2.III.F41.final}
\begin{split}
&\:\||u|^{1+2\de} \varpi^{2N} \Om^2 \nab^2\widetilde{\betab} \mathcal F_{4,1} \|_{L^1_uL^1_{\ub}L^1(S)}\\
\ls &\: \||u|^{\frac 12+\delta}\varpi^N\Om_\Ke\nab^2\widetilde{\betab} \|_{L^2_uL^2_{\ub}L^2(S)}(\||u|^{\frac 12+\de}\varpi^N\Er_1 \|_{L^2_uL^2_{\ub}L^2(S)}+\||u|^{\frac 12+\de}\varpi^N\Er_2 \|_{L^2_uL^2_{\ub}L^2(S)}).
\end{split}
\end{equation}
The term $\nab^2\widetilde{\betab}\mathcal F_{4,2}$ will be handled in a similar manner as the corresponding term in Proposition~\ref{EE.1}, namely, we control $\nab^2\widetilde{\betab}$ by $\sum_{i\leq 3}\nab^i\tpH$ plus some error terms using the Codazzi equation for $\betab$. More precisely, using the schematic Codazzi equation \eqref{Codazzi.S.2} (cf.~\eqref{sch.Codazzi}), we have the following reduced schematic equation for $\nab^2\widetilde{\betab}$:
\begin{equation*}
\nab^2\widetilde{\betab} \eqrs \sum_{i_1+i_2\leq 3}(1+\nab^{i_1}\tg+\nab^{i_1-1}\tp)(\nab^{i_2}\tpHb,\Om_\Ke^2\nab^{i_2}\tg,\Om_\Ke^2\nab^{\min\{i_2,2\}}\tp),
\end{equation*}
which implies
\begin{equation*}
\begin{split}
 \nab^2\widetilde{\betab} \mathcal F_{4,2} 
\eqrs &\sum_{i_1+i_2\leq 3}(1+\nab^{i_1}\tg+\nab^{i_1-1}\tp)(\nab^{i_2}\tpHb,\Om_\Ke^2\nab^{i_2}\tg,\Om_\Ke^2\nab^{\min\{i_2,2\}}\tp)\mathcal F_{4,2}.
\end{split}
\end{equation*}
The terms containing $\nab^{i_2}\tpHb$ are the main terms, for which we use the second estimate in \eqref{F4.basic.est} and \eqref{Om.c} to obtain
\begin{equation}\label{EE.2.III.F42.1.final}
\begin{split}
&\:\||u|^{1+2\de} \varpi^{2N} \Om^2 \sum_{i_1+i_2\leq 3}(1+\nab^{i_1}\tg+\nab^{i_1-1}\tp)\nab^{i_2}\tpHb \mathcal F_{4,2} \|_{L^1_uL^1_{\ub}L^1(S)}
\ls \||u|^{1+2\de} \varpi^{2N} \Om_\Ke^2 \mathcal T_1\|_{L^1_uL^1_{\ub}L^1(S)}.
\end{split}
\end{equation}
For the remaining terms in $\nab^2\widetilde{\betab} \mathcal F_{4,2}$, we note that by \eqref{F42.def},
\begin{equation*}
\begin{split}
&\: \sum_{i_1+i_2\leq 3}(1+\nab^{i_1}\tg+\nab^{i_1-1}\tp)(\Om_\Ke^2\nab^{i_2}\tg,\Om_\Ke^2\nab^{\min\{i_2,2\}}\tp) \mathcal F_{4,2}\\
\eqrs &\: \sum_{i_1+i_2\leq 3}(1+\nab^{i_1}\tg+\nab^{i_1-1}\tp)(\Om_\Ke^2\nab^{i_2}\tg,\Om_\Ke^2\nab^{\min\{i_2,2\}}\tp) \\
&\:\quad \times(\sum_{i_3+i_4+i_5\leq 3} (1+\nab^{i_3}\tg+\nab^{\min\{i_3,2\}}\tp)\nab^{i_4}(\tpHb,\widetilde{\omb})(\nab^{i_5}\tpH+\Om_\Ke^{-2}\nab^{i_5}\tb)).
\end{split}
\end{equation*}
This term is almost the same as the term \eqref{EE.1.F1.3} in the proof of Proposition~\ref{EE.1} (except for having an extra $\Om_\Ke^{-2}\nab^{i_5}\tb$ term, which can be dealt with in a similar manner). We therefore proceed as in \eqref{EE.1.F1.4} and \eqref{EE.1.F1.5} (and use \eqref{Om.c}) to obtain
\begin{equation}\label{EE.2.III.F42.2.final}
\begin{split}
&\:\||u|^{1+2\de} \varpi^{2N} \Om^2 \sum_{i_1+i_2\leq 3}(1+\nab^{i_1}\tg+\nab^{i_1-1}\tp)(\Om_\Ke^2\nab^{i_2}\tg,\Om_\Ke^2\nab^{\min\{i_2,2\}}\tp) \mathcal F_{4,2} \|_{L^1_uL^1_{\ub}L^1(S)}\\
\ls &\: (\sum_{i\leq 3}\||u|^{\f 12+\de} \varpi^N \Om_{\Ke} (\nab^{i}\tg,\nab^{\min\{i,2\}}\tp)\|_{L^2_uL^2_{\ub}L^2(S)})\\
&\:\quad\times(\sum_{i_1\leq 2,\,i_2\leq 3} \|\nab^{i_1}(\tpHb,\widetilde{\omb})\|_{L^\i_uL^\i_{\ub}L^2(S)} \||u|^{\f 12+\de} \varpi^N (\Om_\Ke^3 \nab^{i_2} \tpH, \Om_\Ke \nab^{i_2}\tb) \|_{L^2_uL^2_{\ub}L^2(S)}\\
&\: \qquad + \sum_{i_1\leq 3\,i_2\leq 2} \||u|^{\f 12+\de} \varpi^N \Om_\Ke\nab^{i_1}(\tpHb,\widetilde{\omb})\|_{L^2_uL^2_{\ub}L^2(S)} \|(\Om_\Ke^2 \nab^{i_2} \tpH,\nab^{i_2}\tb) \|_{L^\i_uL^\i_{\ub}L^2(S)})\\
\ls &\:(\sum_{i\leq 3}\||u|^{\f 12+\de} \varpi^N \Om_{\Ke} (\nab^{i}\tg,\nab^{\min\{i,2\}}\tp)\|_{L^2_uL^2_{\ub}L^2(S)})\||u|^{\f 12+\de}\varpi^N \Er_1\|_{L^2_uL^2_{\ub}L^2(S)}\\
\ls &\: 2^{2N}\mathcal D+ \|\ub^{\f 12+\de}\varpi^N \Er_1\|_{L^2_uL^2_{\ub}L^2(S)}^2.
\end{split}
\end{equation}
Moreover, by \eqref{prop.g.2}, $\mathcal F_3$ satisfies
$$\Om_\Ke|\mathcal F_3|\ls \Er_1,$$
which, after using the Cauchy--Schwarz inequality and \eqref{Om.c}, implies
\begin{equation}\label{EE.2.III.F3.final}
\begin{split}
&\:\||u|^{1+2\de} \varpi^{2N} \Om^2 \nab^2(\tK, \widetilde{\sigmac}) \mathcal F_3 \|_{L^1_uL^1_{\ub}L^1(S)}\\
\ls &\: \||u|^{\frac 12+\delta}\varpi^N\Om_\Ke\nab^2(\tK,\widetilde{\sigmac}) \|_{L^2_uL^2_{\ub}L^2(S)}\||u|^{\frac 12+\de}\varpi^N\Er_1 \|_{L^2_uL^2_{\ub}L^2(S)}.
\end{split}
\end{equation}
Combining \eqref{EE.2.III.F41.final}, \eqref{EE.2.III.F42.1.final}, \eqref{EE.2.III.F42.2.final} and \eqref{EE.2.III.F3.final}, we obtain
\begin{equation}\label{est.intbyparts.1.3}
\begin{split}
|III| \ls &\: 2^{2N} \mathcal D+\||u|^{\frac 12+\delta}\varpi^N\Om_\Ke\nab^2\widetilde{\betab} \|_{L^2_uL^2_{\ub}L^2(S)}(\||u|^{\frac 12+\de}\varpi^N\Er_1 \|_{L^2_uL^2_{\ub}L^2(S)}+\||u|^{\frac 12+\de}\varpi^N\Er_2 \|_{L^2_uL^2_{\ub}L^2(S)})\\
&\:+\|\ub^{\frac 12+\de}\varpi^N\Er_1 \|_{L^2_uL^2_{\ub}L^2(S)}^2+\||u|^{1+2\de}\varpi^{2N}\Om_\Ke^2\mathcal T_1 \|_{L^1_uL^1_{\ub}L^1(S)}\\
&\:+\||u|^{\frac 12+\delta}\varpi^N\Om_\Ke\nab^2(\tK,\widetilde{\sigmac}) \|_{L^2_uL^2_{\ub}L^2(S)}\||u|^{\frac 12+\de}\varpi^N\Er_1 \|_{L^2_uL^2_{\ub}L^2(S)}\\
\ls &\: 2^{2N} \mathcal D+ \||u|^{\frac 12+\delta}\varpi^N\Om_\Ke\nab^2\widetilde{\betab} \|_{L^2_uL^2_{\ub}L^2(S)}\||u|^{\frac 12+\de}\varpi^N\Er_2 \|_{L^2_uL^2_{\ub}L^2(S)}\\
&\:+\||u|^{\frac 12+\delta}\varpi^N\Om_\Ke\nab^2(\tK,\widetilde{\sigmac}) \|_{L^2_uL^2_{\ub}L^2(S)}\||u|^{\frac 12+\de}\varpi^N\Er_1 \|_{L^2_uL^2_{\ub}L^2(S)}\\
&\:+\||u|^{\frac 12+\delta}\varpi^N\Om_\Ke\nab^2\widetilde{\betab} \|_{L^2_uL^2_{\ub}L^2(S)}^2 +\|\ub^{\frac 12+\de}\varpi^N\Er_1 \|_{L^2_uL^2_{\ub}L^2(S)}^2+\||u|^{1+2\de}\varpi^{2N}\Om_\Ke^2\mathcal T_1 \|_{L^1_uL^1_{\ub}L^1(S)},
\end{split}
\end{equation}
where in the last line we have additionally used the Young's inequality and noted that $|u|\ls \ub$.

\textbf{Estimates for the term $IV$.} Finally, using the bounds for the background $(\trchb)_\Ke$ and $(\trch)_\Ke$ in Proposition~\ref{Kerr.Ricci.bound}, we have the following estimates for $\trchb$ and $\trch$:
$$|\trchb|\ls \Om_\Ke^2+|\tpHb|\ls \Om_\Ke^2+\ep^{\f12},\quad |\trch| \ls 1+|\tpH|.$$
Hence, the term $IV$ obeys the bound
\begin{equation}\label{est.intbyparts.1.4.prelim}
\begin{split}
|IV| \ls & \:\||u|^{\frac 12+\delta}\varpi^N\Om_\Ke\nab^2\widetilde{\betab} \|_{L^2_uL^2_{\ub}L^2(S)}^2+\||u|^{1+2\delta}\varpi^{2N}\Om_\Ke^2\tpH\nab^2\widetilde{\betab}\nab^2\widetilde{\betab} \|_{L^1_uL^1_{\ub}L^1(S)}\\
&\:+\||u|^{\frac 12+\delta}\varpi^N\Om_\Ke^2\nab^2(\tK,\widetilde{\sigmac}) \|_{L^2_uL^2_{\ub}L^2(S)}^2+\ep^{\f12}\||u|^{\frac 12+\delta}\varpi^N\Om_\Ke\nab^2(\tK,\widetilde{\sigmac}) \|_{L^2_uL^2_{\ub}L^2(S)}^2.
\end{split}
\end{equation}
For the second term on the right hand side, we use H\"older's inequality and then apply Sobolev embedding (Proposition~\ref{Sobolev}) and \eqref{BA} to bound the $L^1_{\ub}L^\i_uL^\i(S)$ norm of $\Om_\Ke^2\tpH$ to obtain
\begin{equation*}
\begin{split}
& \: \||u|^{1+2\delta}\varpi^{2N}\Om_\Ke^2\tpH\nab^2\widetilde{\betab}\nab^2\widetilde{\betab} \|_{L^1_uL^1_{\ub}L^1(S)}\\
\ls &\: \|\Om_\Ke^2\tpH\|_{L^1_{\ub}L^\i_uL^\i(S)} \||u|^{\f 12+\de}\varpi^N\nab^2\widetilde{\betab}\|^2_{L^\i_{\ub}L^2_uL^2(S)}\\
\ls &\: \ep^{\f 12} \||u|^{\f 12+\de}\varpi^N\nab^2\widetilde{\betab}\|^2_{L^\i_{\ub}L^2_uL^2(S)}.
\end{split}
\end{equation*}
Therefore, plugging this back into \eqref{est.intbyparts.1.4.prelim}, we obtain
\begin{equation}\label{est.intbyparts.1.4}
\begin{split}
|IV| \ls & \:\||u|^{\frac 12+\delta}\varpi^N\Om_\Ke\nab^2\widetilde{\betab} \|_{L^2_uL^2_{\ub}L^2(S)}^2+\ep^{\f 12} \||u|^{\f 12+\de}\varpi^N\nab^2\widetilde{\betab}\|^2_{L^\i_{\ub}L^2_uL^2(S)}\\
&\:+\||u|^{\frac 12+\delta}\varpi^N\Om_\Ke^2\nab^2(\tK,\widetilde{\sigmac}) \|_{L^2_uL^2_{\ub}L^2(S)}^2+\ep^{\f12}\||u|^{\frac 12+\delta}\varpi^N\Om_\Ke\nab^2(\tK,\widetilde{\sigmac}) \|_{L^2_uL^2_{\ub}L^2(S)}^2.
\end{split}
\end{equation}

\textbf{Putting everything together.} Gathering the bounds \eqref{est.intbyparts.1.1}, \eqref{est.intbyparts.1.2}, \eqref{est.intbyparts.1.3} and \eqref{est.intbyparts.1.4}, plugging into \eqref{EE2.main}, putting the terms in \eqref{est.intbyparts.1.2} with good signs to the left hand side of \eqref{EE2.main}, taking supremum in $u$ and $\ub$, and bounding the initial data term using \eqref{data.D.def} and \eqref{varpi.bd}, we obtain
\begin{equation*}
\begin{split}
&\||u|^{\frac 12+\delta}\varpi^N\nab^2\widetilde{\betab}\|^2_{L^\infty_{\ub} L^2_uL^2(S)}  +N\||u|^{\frac 12+\delta}\varpi^N\Om_\Ke\nab^2\widetilde{\betab}\|^2_{L^2_{\ub} L^2_{u}L^2(S)}  \\
&+\||u|^{\frac 12+\delta}\varpi^N\Om_\Ke\nab^2(\tK,\widetilde{\sigmac})\|^2_{L^\infty_u L^2_{\ub}L^2(S)} +N\||u|^{\frac 12+\delta}\varpi^N\Om_\Ke^2\nab^2(\tK,\widetilde{\sigmac})\|^2_{L^2_{\ub} L^2_{u}L^2(S)}  \\
&+\||u|^{\frac 12+\delta}\varpi^N\Om_\Ke\nab^2(\tK,\widetilde{\sigmac})\|^2_{L^2_{\ub} L^2_{u}L^2(S)} \\
\ls &2^{2N}\mathcal D+\|\ub^{\frac 12+\de}\varpi^N\Er_1 \|_{L^2_uL^2_{\ub}L^2(S)}^2+|||u|^{1+2\delta}\varpi^{2N}\Om_\Ke^2 \mathcal T_1||_{L^1_uL^1_{\ub}L^1(S)}\\
&+\||u|^{\frac 12+\delta}\varpi^N\Om_\Ke\nab^2\widetilde{\betab} \|_{L^2_uL^2_{\ub}L^2(S)}\||u|^{\frac 12+\de}\varpi^N\Er_2 \|_{L^2_uL^2_{\ub}L^2(S)}\\
&+\||u|^{\frac 12+\delta}\varpi^N\Om_\Ke\nab^2(\tK,\widetilde{\sigmac}) \|_{L^2_uL^2_{\ub}L^2(S)}\||u|^{\frac 12+\de}\varpi^N\Er_1 \|_{L^2_uL^2_{\ub}L^2(S)}\\
&+(1+N^{\f 12}) \||u|^{\frac 12+\delta}\varpi^N\Om_\Ke\nab^2\widetilde{\betab} \|_{L^2_uL^2_{\ub}L^2(S)}^2+\ep^{\f 12} \||u|^{\f 12+\de}\varpi^N\nab^2\widetilde{\betab}\|^2_{L^\i_{\ub}L^2_uL^2(S)}\\
&+\||u|^{\frac 12+\delta}\varpi^N\Om_\Ke^2\nab^2(\tK,\widetilde{\sigmac}) \|_{L^2_uL^2_{\ub}L^2(S)}^2+(N^{-\f 12}+\ep^{\f12})\||u|^{\frac 12+\delta}\varpi^N\Om_\Ke\nab^2(\tK,\widetilde{\sigmac}) \|_{L^2_uL^2_{\ub}L^2(S)}^2.
\end{split}
\end{equation*}
For the second and third lines on the right hand side, we can apply Young's inequality and absorb the terms with second derivatives of curvature to the left hand side. Notice that for $N$ sufficiently large and $\ep_0$ (and hence $\ep$) sufficiently small, the terms $(1+N^{\f 12})\||u|^{\frac 12+\delta}\varpi^N\Om_\Ke\nab^2\widetilde{\betab} \|_{L^2_uL^2_{\ub}L^2(S)}^2$, $\ep^{\f 12} \||u|^{\f 12+\de}\varpi^N\nab^2\widetilde{\betab}\|^2_{L^\i_{\ub}L^2_uL^2(S)}$, $\||u|^{\frac 12+\delta}\varpi^N\Om_\Ke^2\nab^2(\tK,\widetilde{\sigmac}) \|_{L^2_uL^2_{\ub}L^2(S)}^2$ and $(N^{-\f 12}+\ep^{\f12})\||u|^{\frac 12+\delta}\varpi^N\Om_\Ke\nab^2(\tK,\widetilde{\sigmac}) \|_{L^2_uL^2_{\ub}L^2(S)}^2$ can be completely absorbed to the left hand side. Therefore, we have
\begin{equation*}
\begin{split}
&\||u|^{\frac 12+\delta}\varpi^N\nab^2\widetilde{\betab}\|^2_{L^\infty_{\ub} L^2_uL^2(S)}  +N\||u|^{\frac 12+\delta}\varpi^N\Om_\Ke\nab^2\widetilde{\betab}\|^2_{L^2_{\ub} L^2_{u}L^2(S)}  \\
&+\||u|^{\frac 12+\delta}\varpi^N\Om_\Ke\nab^2(\tK,\widetilde{\sigmac})\|^2_{L^\infty_u L^2_{\ub}L^2(S)} +N\||u|^{\frac 12+\delta}\varpi^N\Om_\Ke^2\nab^2(\tK,\widetilde{\sigmac})\|^2_{L^2_{\ub} L^2_{u}L^2(S)}  \\
&+\||u|^{\frac 12+\delta}\varpi^N\Om_\Ke\nab^2(\tK,\widetilde{\sigmac})\|^2_{L^2_{\ub} L^2_{u}L^2(S)} \\
\ls &2^{2N}\mathcal D+\|\ub^{\frac 12+\delta}\varpi^{N} \Er_1\|_{L^2_uL^2_{\ub}L^2(S)}^2+N^{-1}\||u|^{\frac 12+\delta}\varpi^{N} \Er_2\|_{L^2_uL^2_{\ub}L^2(S)}^2+\||u|^{1+2\delta}\varpi^{2N}\Om_\Ke^2 \mathcal T_1\|_{L^1_uL^1_{\ub}L^1(S)}.
\end{split}
\end{equation*}
This concludes the proof of the proposition.
\end{proof}

\section{Elliptic estimates}\label{sec.elliptic}

Recall that we continue to work under the setting described in Remark~\ref{rmk:setup}. 

In this section, we obtain the bounds for the third angular derivatives of the difference of the Ricci coefficients, i.e.~we will prove $\NH$ and $\NI$ estimates for $\nab^3\widetilde{\eta}$, $\nab^3\widetilde{\etab}$, $\nab^3\ttrch$, $\nab^3\widetilde{\chih}$, $\nab^3\widetilde{\trchb}$, $\nab^3\widetilde{\chibh}$ and $\nab^3\widetilde{\omb}$. These estimates will be derived using a combination of transport estimates and elliptic estimates. More precisely, we first use transport equations to derive estimates for $\nab^2\widetilde{\mu}$, $\nab^2\widetilde{\mub}$, $\nab^3\ttrch$, $\nab^3\widetilde{\trchb}$ and $\nab\widetilde{\ombs}$. (Some of these transport estimates were already derived in Section~\ref{transport.est.RC}.) These quantities, while being top order in terms of derivatives, are chosen so that the transport equations they obey do not contain terms with three derivatives of the curvature components\footnote{Terms with third derivatives of the curvature components are uncontrollable and would show up in the transport equation for a general third angular derivative of a Ricci coefficient.}. We will then use the elliptic structure that was understood in \cite{Chr,CK} hidden in the equations for the Ricci coefficients, and apply the elliptic estimates in Propositions~\ref{ellipticthm} and \ref{elliptic.Poisson} to prove all the top order estimates for the Ricci coefficients.

This section will be organised as follows: The estimates for $\nab^3\widetilde{\eta}$ and $\nab^3\widetilde{\etab}$ will be proved in \textbf{Section~\ref{sec.elliptic.tp}}; the estimates for $\nab^3\ttrch$ and $\nab^3\widetilde{\chih}$ will be proved in \textbf{Section~\ref{sec.elliptic.tpH}}; the estimates for $\nab^3\widetilde{\trchb}$, $\nab^3\widetilde{\chibh}$ and $\nab^3\widetilde{\omb}$ will be proved in \textbf{Section~\ref{sec.elliptic.tpHb}}.

In order to close our elliptic estimates in Section~\ref{sec.elliptic.tpHb}, we will also need to obtain some additional estimates for $\tg$ and their angular covariant derivatives (Proposition~\ref{improved.tg.prop.1}). These will in turn also give the bounds $\nab^i\widetilde{\etab}$ for $i\leq 2$ in $L^\i_{\ub}L^2_uL^2(S)$ (Proposition~\ref{improved.etab}). With all of these estimates, we will thus conclude the proof of Proposition~\ref{concluding.est} in \textbf{Section~\ref{sec:proof.of.concluding.est}}.

\subsection{Estimates for $\protect\nab^3\widetilde{\eta}$ and $\protect\nab^3\widetilde{\protect\underline{\eta}}$}\label{sec.elliptic.tp}

We begin with the bounds for $\nab^3\widetilde{\eta}$ and $\nab^3\widetilde{\etab}$.  First, we apply the elliptic estimates in Proposition \ref{ellipticthm} to each fixed 2-sphere $S_{u,\ub}$ to control $\nab^3\widetilde{\eta}$ and $\nab^3\widetilde{\etab}$. 
\begin{proposition}\label{elliptic.psi}
\begin{equation*}
\begin{split}
\|\nab^3\widetilde{\eta}\|_{L^2(S_{u,\ub})}
\ls &\sum_{i\leq 2}\|\nab^i(\tp,\widetilde{\mu},\tK,\widetilde{\sigmac})\|_{L^2(S_{u,\ub})}+\sum_{i\leq 3}\|\nab^i\tg\|_{L^2(S_{u,\ub})}.
\end{split}
\end{equation*}
and
\begin{equation*}
\begin{split}
\|\nab^3\widetilde{\etab}\|_{L^2(S_{u,\ub})}
\ls &\sum_{i\leq 2}\|\nab^i(\tp,\widetilde{\mub},\tK,\widetilde{\sigmac})\|_{L^2(S_{u,\ub})}+\sum_{i\leq 3}\|\nab^i\tg\|_{L^2(S_{u,\ub})}.
\end{split}
\end{equation*}
\end{proposition}
\begin{proof}
We will only prove the bounds for $\nab^3\widetilde{\eta}$ as the estimates for $\nab^3\widetilde{\etab}$ can be derived in a completely analogous manner. Recall that
\begin{equation*}
\begin{split}
\div\eta=-\mu+K,\quad \curl\eta=\sigmac.
\end{split}
\end{equation*}
Taking the difference of these equations on $(\mathcal U_{\ub_f},g)$ and the background Kerr metric, we deduce that for $i\leq 2$, $\nab^i\div\widetilde{\eta}$ and $\nab^i\curl\widetilde{\eta}$ take the following reduced schematic form
$$\nab^i (\div\widetilde{\eta},\curl{\widetilde{\eta}})\eqrs \tg\nab^3\widetilde{\eta}+\sum_{i_1+i_2\leq 3} (1+\nab^{i_1}\tg)\nab^{\min\{i_2,2\}}(\tg,\widetilde{\eta})+\sum_{i\leq 2}\nab^{i}(\widetilde{\mu},\tK,\widetilde{\sigmac}).$$
We now apply Proposition~\ref{ellipticthm} to obtain the bound
\begin{equation*}
\begin{split}
&\|\nab^3\widetilde{\eta}\|_{L^2(S_{u,\ub})}\\
\ls &\|\tg\nab^3\widetilde{\eta}\|_{L^2(S_{u,\ub})}+\sum_{i_1+i_2\leq 3} \|(1+\nab^{i_1}\tg)\nab^{\min\{i_2,2\}}(\tg,\widetilde{\eta})\|_{L^2(S_{u,\ub})}+\sum_{i\leq 2}\|\nab^{i}(\widetilde{\mu},\tK,\widetilde{\sigmac})\|_{L^2(S_{u,\ub})}\\
\ls &\ep^{\f12}\|\nab^3\widetilde{\eta}\|_{L^2(S_{u,\ub})}+\sum_{i\leq 3} \|\nab^i\tg\|_{L^2(S_{u,\ub})}+\sum_{i\leq 2}\|\nab^{i}(\widetilde{\eta},\widetilde{\mu},\tK,\widetilde{\sigmac})\|_{L^2(S_{u,\ub})},
\end{split}
\end{equation*}
where in the last step we have used Corollary~\ref{Linfty} to control $\tg$ in the first term and used \eqref{BA.S.ii2.nw} for the second term.

For $\ep_0$ (and hence $\ep$) sufficiently small, we can subtract $\f12 \|\nab^3\widetilde{\eta}\|_{L^2(S_{u,\ub})}$ on both sides to obtain the desired conclusion for $\nab^3\widetilde{\eta}$. The estimate for $\nab^3\widetilde{\etab}$ follows similarly using instead the equations
\begin{equation*}
\div\etab=-\mub+K,\quad \curl\etab=-\sigmac.
\end{equation*}

\end{proof}
Given the elliptic estimates for $\nab^3\tp$ in Proposition \ref{elliptic.psi}, we obtain the $\NH$ and $\NI$ bounds for $\nab^3\tp$ using the estimates proved in the previous sections:
\begin{proposition}\label{est.nab3psi}
$\nab^3\tp$ obeys the following $L^\infty_uL^2_{\ub}L^2(S)$, $L^\infty_{\ub}L^2_uL^2(S)$ and $L^2_{\ub}L^2_uL^2(S)$ estimates:
\begin{equation*}
\begin{split}
&\|\ub^{\frac 12+\de}\varpi^N\Om_\Ke\nab^3\widetilde{\etab}\|^2_{L^\infty_uL^2_{\ub} L^2(S)}+\|\ub^{\frac 12+\de}\varpi^N\Om_\Ke\nab^3\widetilde{\eta}\|^2_{L^\infty_{\ub}L^2_u L^2(S)}  \\
&+\|\ub^{\frac 12+\de}\varpi^N\Om_\Ke\nab^3\tp\|^2_{L^2_{\ub} L^2_{u}L^2(S)}+N\|\ub^{\frac 12+\de}\varpi^N\Om_\Ke^2\nab^3\tp\|^2_{L^2_{\ub} L^2_{u}L^2(S)}  \\
\ls &2^{2N}\mathcal D+\|\ub^{\frac 12+\de}\varpi^{N}\Er_1\|_{L^2_uL^2_{\ub}L^2(S)}^2+N^{-1}\|\ub^{\frac 12+\de}\varpi^{N}\Er_2\|_{L^2_uL^2_{\ub}L^2(S)}^2\\
&+\|\ub^{1+2\de}\varpi^{2N}\Om_\Ke^4\mathcal T_2\|_{L^1_uL^1_{\ub}L^1(S)}+\||u|^{1+2\de}\varpi^{2N}\Om_\Ke^2\mathcal T_1\|_{L^1_uL^1_{\ub}L^1(S)}.
\end{split}
\end{equation*}
\end{proposition}
\begin{proof}
This is a direct consequence of Propositions \ref{prop.etab}, \ref{prop.eta}, \ref{EE.1}, \ref{EE.2} and \ref{elliptic.psi}.
\end{proof}

This concludes the estimates for $\nab^3\widetilde{\eta}$ and $\nab^3\widetilde{\etab}$. 

\subsection{Estimates for $\protect\nab^3\protect\tpH$}\label{sec.elliptic.tpH}
We now turn to the second group of terms and control $\nab^3\ttrch$ and $\nab^3\widetilde{\chih}$. We will first prove estimates for $\nab^3\ttrch$ using the transport equation (see Proposition \ref{nab3trch.eqn}) for $\nab_4\nab^3\ttrch$. This will be achieved in Propositions \ref{est.nab3trch} and \ref{est.nab3trch.2}. We then derive the elliptic estimates in Proposition \ref{elliptic.chih} which allow us to control $\nab^3\widetilde{\chih}$ in terms of the bounds $\nab^3\ttrch$ and the estimates for lower order terms that we have already obtained in the previous sections. Finally, in Proposition \ref{elliptic.tpH}, we combine these estimates and conclude that $\nab^3\tpH$ can be controlled by the terms in Proposition~\ref{concluding.est}.

We begin with deriving the estimates for $\nab^3\ttrch$ via transport estimates. Notice that unlike for the lower order derivatives of $\ttrch$ (see Proposition \ref{prop.tpH}), we now need to prove an estimate for $\nab^3\ttrch$ which is $L^2$ in the $\ub$ direction using a transport equation which is also in the $\nab_4$ direction. Therefore, instead of integrating in the spacetime region and obtain the $L^2_{\ub}L^\infty_u$ estimate as a boundary term, we integrate on a fix constant $u$ hypersurface and after choosing appropriate weight functions, we obtain the $L^\i_uL^2_{\ub}$ estimate as a bulk term. More precisely, we have the following bounds:
\begin{proposition}\label{est.nab3trch}
$\nab^3\widetilde{\trch}$ obeys the following $L^\infty_uL^\infty_{\ub} L^2(S)$ and $L^\infty_uL^2_{\ub} L^2(S)$ estimates:
\begin{equation*}
\begin{split}
&\:\|\ub^{\frac 12+\delta}\varpi^N\Om_\Ke^2\nab^3\widetilde{\trch}\|^2_{L^\infty_uL^\infty_{\ub} L^2(S)}+\|\ub^{\frac 12+\delta}\varpi^N\Om_\Ke^2\nab^3\widetilde{\trch}\|^2_{L^\infty_uL^2_{\ub} L^2(S)}\\
&\:+N\|\ub^{\frac 12+\delta}\varpi^N\Om_\Ke^3\nab^3\widetilde{\trch}\|^2_{L^\infty_uL^2_{\ub} L^2(S)}  \\
\ls &\:2^{2N}\mathcal D+\|\ub^{\frac 12+\delta}\varpi^{N}\Er_1\|_{L^2_uL^2_{\ub}L^2(S)}^2+N^{-1}\|\ub^{\frac 12+\delta}\varpi^{N}\Er_2\|_{L^2_uL^2_{\ub}L^2(S)}^2\\
&\:+\|\ub^{1+2\delta}\varpi^{2N}\Om_\Ke^4 \mathcal T_2\|_{L^1_uL^1_{\ub}L^1(S)}+N^{-1}\|\ub^{\f12+\de}\varpi^N\Om_\Ke^2\nab^3\tpH\|_{L^\i_uL^2_{\ub}L^2(S)}.
\end{split}
\end{equation*}
\end{proposition}
\begin{proof}
By Proposition \ref{nab3trch.eqn}, the reduced schematic equation for $\nab_4\nab^3\ttrch$ takes the form
\begin{equation}\label{eq.nab3ttch}
\nab_4\nab^3\ttrch \eqrs\sum_{i_1+i_2+i_3\leq 3}(1+\nab^{i_1}\tg+\nab^{i_1-1}\tp)(1+\nab^{i_2}\tpH)(\nab^{i_3}(\tpH,\tg)+\Om_\Ke^{-2}\nab^{i_3}\tb).
\end{equation}
Applying Proposition \ref{transport.4.3} with $m=2$, and using \eqref{data.D.def} and \eqref{varpi.bd}, we obtain that for every fixed $u$, the following holds:
\begin{equation}\label{id.nab3trch}
\begin{split}
&\:\|\ub^{\frac 12+\delta}\varpi^N\Om_\Ke^2\nab^3\widetilde{\trch}\|^2_{L^\infty_{\ub} L^2(S)}+\|\ub^{\frac 12+\delta}\varpi^N\Om_\Ke^2\nab^3\widetilde{\trch}\|^2_{L^2_{\ub} L^2(S)}+N\|\ub^{\frac 12+\delta}\varpi^N\Om_\Ke^3\nab^3\widetilde{\trch}\|^2_{L^2_{\ub} L^2(S)}  \\
\ls &\: 2^{2N}\mathcal D+\|\ub^{1+2\delta}\varpi^{2N}\Om_\Ke^6 \nab^3\ttrch\nab_4\nab^3\ttrch\|_{L^1_{\ub}L^1(S)}.
\end{split}
\end{equation}
We now control the term
$$\|\ub^{1+2\delta}\varpi^{2N}\Om_\Ke^6 \nab^3\ttrch\nab_4\nab^3\ttrch\|_{L^1_{\ub}L^1(S)}.$$
We split the terms in equation \eqref{eq.nab3ttch} for $\nab_4\nab^3\ttrch$ into the following two subsets:
\begin{equation}\label{nab3trch.0.1}
\sum_{i_1+i_2\leq 3}(1+\nab^{i_1}\tg+\nab^{i_1-1}\tp)(\nab^{i_2}(\tpH,\tg)+\Om_\Ke^{-2}\nab^{i_2}\tb)
\end{equation}
and
\begin{equation}\label{nab3trch.0.2.1}
\sum_{i_1+i_2+i_3\leq 3}(1+\nab^{i_1}\tg+\nab^{i_1-1}\tp)\nab^{i_2}\tpH(\nab^{i_3}(\tpH,\tg)+\Om_\Ke^{-2}\nab^{i_3}\tb).
\end{equation}
Note that since $\nab^i\tg \eqrs 1$ for $i\leq 1$ and $\tp \eqrs 1$ (by Corollary~\ref{Linfty}), it is easy to see that \eqref{nab3trch.0.2.1} can be expressed as a sum of the term \eqref{nab3trch.0.1} and the following term
\begin{equation}\label{nab3trch.0.2}
\sum_{i_1+i_2+i_3\leq 3}(1+\nab^{i_1}\tg+\nab^{i_1-1}\tp)\nab^{i_2}\tpH(\nab^{i_3}\tpH+\Om_\Ke^{-2}\nab^{i_3}\tb).
\end{equation}
We therefore control \eqref{nab3trch.0.1} and \eqref{nab3trch.0.2} below. Notice in particular that \eqref{nab3trch.0.1} is at most linear in $(\tpH,\tb)$ and \eqref{nab3trch.0.2} is quadratic in $(\tpH,\tb)$. For the term \eqref{nab3trch.0.1}, we apply the Cauchy--Schwarz inequality and \eqref{BA.S.222} to obtain
\begin{equation}\label{nab3trch.1}
\begin{split}
&\:\|\ub^{1+2\de}\varpi^{2N}\Om_\Ke^6\nab^3\ttrch(\sum_{i_1+i_2\leq 3}(1+\nab^{i_1}\tg+\nab^{i_1-1}\tp)(\nab^{i_2}(\tpH,\tg)+\Om_\Ke^{-2}\nab^{i_2}\tb))\|_{L^1_{\ub}L^{1}(S)}\\
\ls &\:\|\ub^{\f12+\de}\varpi^{N}\Om_\Ke^3\nab^3\ttrch\|_{L^2_{\ub}L^2(S)}\\
&\qquad\times\|\ub^{\frac 12+\de}\varpi^N\Om_\Ke^3\sum_{i_1+i_2\leq 3}(1+\nab^{i_1}\tg+\nab^{i_1-1}\tp)(\nab^{i_2}(\tpH,\tg)+\Om_\Ke^{-2}\nab^{i_2}\tb)\|_{L^2_{\ub}L^2(S)}\\
\ls &\|\ub^{\f12+\de}\varpi^{N}\Om_\Ke^3\nab^3\ttrch\|_{L^2_{\ub}L^2(S)}(\sum_{i\leq 3}(\|\ub^{\frac 12+\de}\varpi^N\Om_\Ke^3\nab^{i}(\tpH,\tg)\|_{L^2_{\ub}L^2(S)}+\|\ub^{\frac 12+\de}\varpi^N\Om_\Ke\nab^{i}\tb\|_{L^2_{\ub}L^2(S)})).
\end{split}
\end{equation}
For the term \eqref{nab3trch.0.2}, we apply the Cauchy--Schwarz inequality, \eqref{BA.S.222} and then H\"older's inequality and the Sobolev embedding in Proposition~\ref{Sobolev} (in the style of Section~\ref{sec:rmk.BA.S}) to obtain
\begin{equation}\label{nab3trch.2}
\begin{split}
&\:\|\ub^{1+2\de}\varpi^{2N}\Om_\Ke^6\nab^3\ttrch(\sum_{i_1+i_2+i_3\leq 3}(1+\nab^{i_1}\tg+\nab^{i_1-1}\tp)\nab^{i_2}\tpH(\nab^{i_3}\tpH+\Om_\Ke^{-2}\nab^{i_3}\tb)\|_{L^{1}_{\ub}L^{1}(S)}\\
\ls &\: \|\ub^{\frac 12+\de}\varpi^N\Om_\Ke^2\nab^3\ttrch\|_{L^2_{\ub}L^2(S)}(\sum_{i_2+i_3\leq 3} \| \nab^{i_2}\tpH(\nab^{i_3}\tpH+\Om_\Ke^{-2}\nab^{i_3}\tb)\|_{L^2_{\ub}L^2(S)})\\
\ls &\:\|\ub^{\frac 12+\de}\varpi^N\Om_\Ke^2\nab^3\ttrch\|_{L^2_{\ub}L^2(S)}(\sum_{i_2\leq 2}\|\Om_\Ke^2\nab^{i_2}\tpH\|_{L^\infty_{\ub}L^2(S)})\\
&\:\quad\times(\sum_{i_3\leq 3}(\|\ub^{\frac 12+\de}\varpi^N\Om_\Ke^2\nab^{i_3}\tpH\|_{L^2_{\ub}L^2(S)}+\|\ub^{\frac 12+\de}\varpi^N\nab^{i_3}\tb\|_{L^2_{\ub}L^2(S)}))\\
&\:+\|\ub^{\frac 12+\de}\varpi^N\Om_\Ke^2\nab^3\ttrch\|_{L^2_{\ub}L^2(S)}(\sum_{i_2\leq 3}\|\ub^{\frac 12+\de}\varpi^N\Om_\Ke^2\nab^{i_2}\tpH\|_{L^2_{\ub}L^2(S)})(\sum_{i_3\leq 2}\|\nab^{i_3}\tb\|_{L^\i_{\ub}L^2(S)})\\
\ls &\:\ep^{\f12}\|\ub^{\frac 12+\de}\varpi^N\Om_\Ke^2\nab^3\ttrch\|_{L^2_{\ub}L^2(S)}(\sum_{i\leq 3}(\|\ub^{\frac 12+\de}\varpi^N\Om_\Ke^2\nab^{i}\tpH\|_{L^2_{\ub}L^2(S)}+\|\ub^{\frac 12+\de}\varpi^N\nab^{i}\tb\|_{L^2_{\ub}L^2(S)})),
\end{split}
\end{equation}
where we have again used Sobolev embedding (Proposition~\ref{Sobolev}) and also the bootstrap assumption \eqref{BA} for $\tb$ and $\tpH$. Combining \eqref{id.nab3trch}, \eqref{nab3trch.1} and \eqref{nab3trch.2}, we obtain
\begin{equation*}
\begin{split}
&\:\|\ub^{\frac 12+\delta}\varpi^N\Om_\Ke^2\nab^3\widetilde{\trch}\|^2_{L^\infty_{\ub} L^2(S)}+\|\ub^{\frac 12+\delta}\varpi^N\Om_\Ke^2\nab^3\widetilde{\trch}\|^2_{L^2_{\ub} L^2(S)}+N\|\ub^{\frac 12+\delta}\varpi^N\Om_\Ke^3\nab^3\widetilde{\trch}\|^2_{L^2_{\ub} L^2(S)}  \\
\ls &\:2^{2N}\mathcal D\\
&\:+\|\ub^{\f12+\de}\varpi^{N}\Om_\Ke^3\nab^3\ttrch\|_{L^2_{\ub}L^2(S)}(\sum_{i\leq 3}(\|\ub^{\frac 12+\de}\varpi^N\Om_\Ke^3\nab^{i}(\tpH,\tg)\|_{L^2_{\ub}L^2(S)}+\|\ub^{\frac 12+\de}\varpi^N\Om_\Ke\nab^{i}\tb\|_{L^2_{\ub}L^2(S)}))\\
&\:+\ep^{\f12}\|\ub^{\frac 12+\de}\varpi^N\Om_\Ke^2\nab^3\ttrch\|_{L^2_{\ub}L^2(S)}(\sum_{i\leq 3}(\|\ub^{\frac 12+\de}\varpi^N\Om_\Ke^2\nab^{i}\tpH\|_{L^2_{\ub}L^2(S)}+\|\ub^{\frac 12+\de}\varpi^N\nab^{i}\tb\|_{L^2_{\ub}L^2(S)})).
\end{split}
\end{equation*}
After applying Young's inequality and absorbing the $\nab^3\widetilde{\trch}$ terms to the left hand side, we obtain
\begin{equation*}
\begin{split}
&\:\|\ub^{\frac 12+\delta}\varpi^N\Om_\Ke^2\nab^3\widetilde{\trch}\|^2_{L^\infty_{\ub} L^2(S)}+\|\ub^{\frac 12+\delta}\varpi^N\Om_\Ke^2\nab^3\widetilde{\trch}\|^2_{L^2_{\ub} L^2(S)}+N\|\ub^{\frac 12+\delta}\varpi^N\Om_\Ke^3\nab^3\widetilde{\trch}\|^2_{L^2_{\ub} L^2(S)}  \\
\ls &\:2^{2N}\mathcal D+\sum_{i\leq 3}(N^{-1}\|\ub^{\frac 12+\de}\varpi^N\Om_\Ke^3\nab^{i}\tg\|_{L^2_{\ub}L^2(S)}^2+(\ep+N^{-1})\|\ub^{\frac 12+\de}\varpi^N\nab^{i}\tb\|_{L^2_{\ub}L^2(S)}^2)\\
&\:+(\ep+N^{-1})\sum_{i\leq 3}\|\ub^{\frac 12+\de}\varpi^N\Om_\Ke^2\nab^{i}\tpH\|_{L^2_{\ub}L^2(S)}^2.
\end{split}
\end{equation*}
The desired conclusion then follows after applying the estimates in Propositions \ref{prop.g}, \ref{prop.b} and \ref{prop.tpH} for $\nab^i\tg$ ($i\leq 3$), $\nab^i\tb$ ($i\leq 3$) and $\nab^i\tpH$ ($i\leq 2$) respectively, recalling that $\ep\leq N^{-1}$ (see Remark~\ref{rmk.constant}), and dropping extra favourable $\ep$ and $N^{-1}$.
\end{proof}
In addition to the above bounds for $\nab^3\widetilde{\trch}$, we also have the following $L^2_uL^2_{\ub}L^2(S)$ estimates for $\nab^3\widetilde{\trch}$:
\begin{proposition}\label{est.nab3trch.2}
$\nab^3\ttrch$ satisfies the following $L^2_{\ub}L^2_uL^{2}(S)$ estimate:
\begin{equation*}
\begin{split}
&\:N\|\ub^{\frac 12+\delta}\varpi^N\Om_\Ke^3\nab^3\widetilde{\trch}\|^2_{L^2_{\ub} L^2_{u}L^2(S)}  \\
\ls &\:N 2^{2N}\mathcal D+\|\ub^{\frac 12+\delta}\varpi^{N}\Er_1\|_{L^2_uL^2_{\ub}L^2(S)}^2+N^{-1}\|\ub^{\frac 12+\delta}\varpi^{N}\Er_2\|_{L^2_uL^2_{\ub}L^2(S)}^2\\
&\:+\|\ub^{1+2\delta}\varpi^{2N}\Om_\Ke^4 \mathcal T_2\|_{L^1_uL^1_{\ub}L^1(S)}+\ep N\|\ub^{\frac 12+\delta}\varpi^N\Om_\Ke^3\nab^3\tpH\|^2_{L^2_{\ub} L^2_{u}L^2(S)}.
\end{split}
\end{equation*}
\end{proposition}
\begin{proof}
Applying Proposition \ref{transport.4.3} with $m=3$, we obtain, for every fixed $u$,
\begin{equation*}
\begin{split}
&\:\|\ub^{\frac 12+\delta}\varpi^N\Om_\Ke^3\nab^3\widetilde{\trch}\|^2_{L^\infty_{\ub} L^2(S)}+\|\ub^{\frac 12+\delta}\varpi^N\Om_\Ke^3\nab^3\widetilde{\trch}\|^2_{L^2_{\ub} L^2(S)}+N\|\ub^{\frac 12+\delta}\varpi^N\Om_\Ke^4\nab^3\widetilde{\trch}\|^2_{L^2_{\ub} L^2(S)}  \\
\ls &\:2^{2N}\|\ub^{\frac 12+\delta}\Om_\Ke^3\nab^3\widetilde{\trch}\|^2_{L^2(S_{u,-u+C_R})}+\|\ub^{1+2\delta}\varpi^{2N}\Om_\Ke^8 \nab^3\ttrch\nab_4\nab^3\ttrch\|_{L^1_{\ub}L^1(S)}.
\end{split}
\end{equation*}
Integrating in $u$, and using \eqref{data.D.def} and \eqref{varpi.bd} for the data term, we obtain
\begin{equation}\label{nab3ttrch.int.1}
\begin{split}
&\:\|\ub^{\frac 12+\delta}\varpi^N\Om_\Ke^3\nab^3\widetilde{\trch}\|^2_{L^2_uL^\infty_{\ub} L^2(S)}+\|\ub^{\frac 12+\delta}\varpi^N\Om_\Ke^3\nab^3\widetilde{\trch}\|^2_{L^2_uL^2_{\ub} L^2(S)}\\
&\:+N\|\ub^{\frac 12+\delta}\varpi^N\Om_\Ke^4\nab^3\widetilde{\trch}\|^2_{L^2_uL^2_{\ub} L^2(S)}  \\
\ls &\:2^{2N}\mathcal D+\|\ub^{1+2\delta}\varpi^{2N}\Om_\Ke^8 \nab^3\ttrch\nab_4\nab^3\ttrch\|_{L^1_u L^1_{\ub}L^1(S)}.
\end{split}
\end{equation}
We now use the equation \eqref{eq.nab3ttch} for $\nab_4\nab^3\ttrch$ and the bootstrap assumption \eqref{BA} to control the term $\|\ub^{1+2\delta}\varpi^{N}\Om_\Ke^6 \nab^3\ttrch\nab_4\nab^3\ttrch\|_{L^1_u L^1_{\ub}L^1(S)}$. As in the proof of Proposition \ref{est.nab3trch}, we control the two terms \eqref{nab3trch.0.1} and \eqref{nab3trch.0.2} separately. For the term \eqref{nab3trch.0.1}, dropping a good $\Om_\Ke$ factor, using the Cauchy--Schwarz inequality and comparing the terms with $\Er_1$ in \eqref{Er.def}, we have
\begin{equation}\label{nab3ttrch.int.2}
\begin{split}
&\:\|\ub^{1+2\delta}\varpi^{2N}\Om_\Ke^8 \nab^3\ttrch(\sum_{i_1+i_2\leq 3}(1+\nab^{i_1}\tg+\nab^{i_1-1}\tp)(\nab^{i_2}(\tpH,\tg)+\Om_\Ke^{-2}\nab^{i_2}\tb))\|_{L^1_u L^1_{\ub}L^1(S)}\\
\ls &\:\|\ub^{\frac 12+\de}\varpi^N\Om_\Ke^4\nab^3\ttrch\|_{L^2_uL^2_{\ub}L^2(S)}\|\ub^{\frac 12+\de}\varpi^N\Er_1\|_{L^2_uL^2_{\ub}L^2(S)}.
\end{split}
\end{equation}
For the term \eqref{nab3trch.0.2}, we apply the Cauchy--Schwarz inequality, \eqref{BA.S.222}, H\"older's inequality, the bootstrap assumption \eqref{BA} and Sobolev embedding (Proposition~\ref{Sobolev}) to obtain
\begin{equation}\label{nab3ttrch.int.3}
\begin{split}
&\|\ub^{1+2\delta}\varpi^{2N}\Om_\Ke^8 \nab^3\ttrch(\sum_{i_1+i_2+i_3\leq 3}(1+\nab^{i_1}\tg+\nab^{i_1-1}\tp)\nab^{i_2}\tpH(\nab^{i_3}\tpH+\Om_\Ke^{-2}\nab^{i_3}\tb))\|_{L^1_u L^1_{\ub}L^1(S)}\\
\ls &\|\ub^{\frac 12+\de}\varpi^N\Om_\Ke^3\nab^3\ttrch\|_{L^2_uL^2_{\ub}L^2(S)}(\sum_{i_2\leq 2}\|\Om_\Ke^2\nab^{i_2}\tpH\|_{L^\infty_uL^\infty_{\ub}L^2(S)})\\
&\quad\times(\sum_{i_3\leq 3}(\|\ub^{\frac 12+\de}\varpi^N\Om_\Ke^3\nab^{i_3}\tpH\|_{L^2_uL^2_{\ub}L^2(S)}+\|\ub^{\frac 12+\de}\varpi^N\Om_\Ke\nab^{i_3}\tb\|_{L^2_uL^2_{\ub}L^2(S)}))\\
&+\|\ub^{\frac 12+\de}\varpi^N\Om_\Ke^3\nab^3\ttrch\|_{L^2_uL^2_{\ub}L^2(S)}(\sum_{i_2\leq 3}\|\ub^{\frac 12+\de}\varpi^N\Om_\Ke^3\nab^{i_2}\tpH\|_{L^2_uL^2_{\ub}L^2(S)})(\sum_{i_3\leq 2}\|\nab^{i_3}\tb\|_{L^{\i}_uL^{\i}_{\ub}L^2(S)})\\
\ls &\ep^{\f12}\|\ub^{\frac 12+\de}\varpi^N\Om_\Ke^3\nab^3\ttrch\|_{L^2_uL^2_{\ub}L^2(S)}\times(\sum_{i\leq 3}(\|\ub^{\frac 12+\de}\varpi^N\Om_\Ke^3\nab^{i}\tpH\|_{L^2_uL^2_{\ub}L^2(S)}+\|\ub^{\frac 12+\de}\varpi^N\Om_\Ke\nab^{i}\tb\|_{L^2_uL^2_{\ub}L^2(S)})).
\end{split}
\end{equation}
We now return to \eqref{nab3ttrch.int.1}, substituting in the estimates \eqref{nab3ttrch.int.2} and \eqref{nab3ttrch.int.3} to obtain
\begin{equation*}
\begin{split}
&\:\|\ub^{\frac 12+\delta}\varpi^N\Om_\Ke^3\nab^3\widetilde{\trch}\|^2_{L^2_uL^\infty_{\ub} L^2(S)}+\|\ub^{\frac 12+\delta}\varpi^N\Om_\Ke^3\nab^3\widetilde{\trch}\|^2_{L^2_uL^2_{\ub} L^2(S)}+N\|\ub^{\frac 12+\delta}\varpi^N\Om_\Ke^4\nab^3\widetilde{\trch}\|^2_{L^2_uL^2_{\ub} L^2(S)}  \\
\ls &\:2^{2N}\mathcal D+\|\ub^{\frac 12+\de}\varpi^N\Om_\Ke^4\nab^3\ttrch\|_{L^2_uL^2_{\ub}L^2(S)}\|\ub^{\frac 12+\de}\varpi^N\Er_1\|_{L^2_uL^2_{\ub}L^2(S)}\\
&\:+\ep^{\f12}\|\ub^{\frac 12+\de}\varpi^N\Om_\Ke^3\nab^3\ttrch\|_{L^2_uL^2_{\ub}L^2(S)}\times(\sum_{i\leq 3}(\|\ub^{\frac 12+\de}\varpi^N\Om_\Ke^3\nab^{i}\tpH\|_{L^2_uL^2_{\ub}L^2(S)}+\|\ub^{\frac 12+\de}\varpi^N\Om_\Ke\nab^{i}\tb\|_{L^2_uL^2_{\ub}L^2(S)})).
\end{split}
\end{equation*}
Applying Young's inequality and absorbing the terms involving $\nab^3\ttrch$ to the left hand side, we obtain
\begin{equation*}
\begin{split}
&\|\ub^{\frac 12+\delta}\varpi^N\Om_\Ke^3\nab^3\widetilde{\trch}\|^2_{L^2_uL^\infty_{\ub} L^2(S)}+\|\ub^{\frac 12+\delta}\varpi^N\Om_\Ke^3\nab^3\widetilde{\trch}\|^2_{L^2_uL^2_{\ub} L^2(S)}+N\|\ub^{\frac 12+\delta}\varpi^N\Om_\Ke^4\nab^3\widetilde{\trch}\|^2_{L^2_uL^2_{\ub} L^2(S)}  \\
\ls &2^{2N}\mathcal D+N^{-1}\|\ub^{\frac 12+\de}\varpi^N\Er_1\|_{L^2_uL^2_{\ub}L^2(S)}^2+\ep(\sum_{i\leq 3}(\|\ub^{\frac 12+\de}\varpi^N\Om_\Ke^3\nab^{i}\tpH\|_{L^2_uL^2_{\ub}L^2(S)}^2+\|\ub^{\frac 12+\de}\varpi^N\Om_\Ke\nab^{i}\tb\|_{L^2_uL^2_{\ub}L^2(S)}^2)).
\end{split}
\end{equation*}
We now multiply by $N$,\footnote{noting that $\ep N\leq 1$ by Remark~\ref{rmk.constant}.} apply Proposition~\ref{prop.b} to $\sum_{i\leq 3}\|\ub^{\frac 12+\de}\varpi^N\Om_\Ke\nab^{i}\tb\|_{L^2_uL^2_{\ub}L^2(S)}^2$ and apply Proposition~\ref{prop.tpH} to $\sum_{i\leq 2}\|\ub^{\frac 12+\de}\varpi^N\Om_\Ke^3\nab^{i}\tpH\|_{L^2_uL^2_{\ub}L^2(S)}^2$
to obtain the desired conclusion.
\end{proof}
We now use the above estimates for $\nab^3\ttrch$ to obtain the desired bound for $\nab^3\widetilde{\chih}$ via elliptic estimates. Notice that 
in Propositions \ref{est.nab3trch} and \ref{est.nab3trch.2}, the estimates have an error term involving $\nab^3\tpH$. Nevertheless, since this term comes with an appropriate smallness constant, the term can be absorbed after combining the bounds for both $\nab^3\ttrch$ and $\nab^3\widetilde{\chih}$. We begin with elliptic estimates for $\nab^3\widetilde{\chih}$ on each $2$-sphere $S_{u,\ub}$:
\begin{proposition}\label{elliptic.chih}
The following estimate holds on any sphere $S_{u,\ub}$:
\begin{equation*}
\begin{split}
\|\nab^3\widetilde{\chih}\|_{L^2(S_{u,\ub})}
\ls &\sum_{i\leq 3}\|\nab^i(\widetilde{\trch},\tg)\|_{L^2(S_{u,\ub})}+\sum_{i\leq 2}\|\nab^i(\widetilde{\beta},\tp,\tpH)\|_{L^2(S_{u,\ub})}.
\end{split}
\end{equation*}
\end{proposition}
\begin{proof}
Recall the Codazzi equation in \eqref{null.str3}:
$$\div\chih=\frac 12\nab\trch-\zeta\cdot(\chih-\frac 12\trch)-\beta.$$
Since $\chih$ is traceless, we compute the curl of $\chih$ to be
\begin{equation}\label{curl.of.traceless}
\curl\chih=^*\div\chih.
\end{equation}
From these equations, we can derive the following system, written in schematic form
\beaa
\div\widetilde{\chih}-\frac 12\nab\widetilde{\trch}+\widetilde{\beta}&\eqs &I,\\
\curl\widetilde{\chih}-\frac 12{^*\nab}\widetilde{\trch}+^*\widetilde{\beta}&\eqs &I,\\
\tr\widetilde{\chih}&\eqs & II,
\eeaa
where for $i\leq 2$, $\nab^i I$ takes the following reduced schematic form
$$\nab^i I \eqrs \tg\nab^3\tpH+\nab^3\tg+\sum_{i_1+i_2\leq 3}\nab^{\min\{i_1,2\}}\tpH\nab^{i_2}\tg+\sum_{i_1+i_2\leq 2}(1+\nab^{i_1}(\tg,\tp))\nab^{i_2}(\tg,\tp,\tpH)$$
and $\nab^i II$ takes the following reduced schematic form
$$\nab^i II \eqrs \sum_{i_1+i_2\leq 2}(1+\nab^{i_1}(\tg,\tp))\nab^{i_2}(\tg,\tp,\tpH).$$
(Recall that $\slashed{\mbox{div}}$, $\slashed{\mbox{curl}}$, $\slashed{\mbox{tr}}$ and ${ }^*$ above are defined with respect to $\gamma$, and hence while $\chih$ is by definition traceless, $\widetilde{\chih}$ is not necessarily traceless.) Therefore, controlling $\tg$ in $L^\infty$ by Corollary~\ref{Linfty} and using \eqref{BA.S.ii2.nw} and \eqref{BA.S.lower}, $\nab^i I$ and $\nab^i II$ are bounded as follows:
\begin{equation*}
\begin{split}
&\:\sum_{i\leq 2}\|\nab^i(I,II)\|_{L^2(S_{u,\ub})}\\
\ls &\: \|\tg\nab^3\tpH\|_{L^2(S_{u,\ub})} + \|\nab^3\tg\|_{L^2(S_{u,\ub})} + \|\sum_{i_1+i_2\leq 3}\nab^{\min\{i_1,2\}}\tpH\nab^{i_2}\tg\|_{L^2(S_{u,\ub})}\\
&\:+\|\sum_{i_1+i_2\leq 2}(1+\nab^{i_1}(\tg,\tp))\nab^{i_2}(\tg,\tp,\tpH)\|_{L^2(S_{u,\ub})}\\
\ls &\: \ep^{\f 12}\|\nab^3\tpH\|_{L^2(S_{u,\ub})} + \|\nab^3\tg\|_{L^2(S_{u,\ub})}+\sum_{i\leq 2}\|\nab^{i}(\tg,\tp,\tpH)\|_{L^2(S_{u,\ub})}.
\end{split}
\end{equation*}
Applying Proposition \ref{ellipticthm}, and using the above estimate, we obtain that for any $u$, $\ub$, we have
\begin{equation*}
\begin{split}
\|\nab^3\widetilde{\chih}\|_{L^2(S_{u,\ub})}
\ls &\sum_{i\leq 3}\|\nab^i\widetilde{\trch}\|_{L^2(S_{u,\ub})}+\sum_{i\leq 2}\|\nab^i(\widetilde{\beta},\widetilde{\chih})\|_{L^2(S_{u,\ub})}+\sum_{i\leq 2}\|\nab^i(I,II)\|_{L^2(S_{u,\ub})}\\
\ls &\sum_{i\leq 3}\|\nab^i(\widetilde{\trch},\tg)\|_{L^2(S_{u,\ub})}+\sum_{i\leq 2}\|\nab^i(\widetilde{\beta},\tp,\tpH)\|_{L^2(S_{u,\ub})}+\ep^{\f12}\|\nab^3\widetilde{\chih}\|_{L^2(S_{u,\ub})}.
\end{split}
\end{equation*}
Finally, notice that for $\ep_0$ (and hence $\ep$) sufficiently small, we can absorb the term 
$$\ep^{\f12}\|\nab^3\widetilde{\chih}\|_{L^2(S_{u,\ub})}$$ 
to the left hand side to obtain the desired conclusion. \qedhere
\end{proof}
Applying the estimates proved above for $\nab^3\widetilde{\chih}$, together with the bounds for $\nab^3\ttrch$ in Propositions \ref{est.nab3trch} and \ref{est.nab3trch.2}, as well as the estimates from the previous sections, we obtain
\begin{proposition}\label{elliptic.tpH}
$\nab^3\tpH$ obeys the following $L^\infty_uL^\infty_{\ub} L^2(S)$ and $L^\infty_uL^2_{\ub} L^2(S)$ estimates:
\begin{equation*}
\begin{split}
&\|\ub^{\frac 12+\delta}\varpi^N\Om_\Ke^2\nab^3\tpH\|^2_{L^\infty_uL^2_{\ub} L^2(S)}+N\|\ub^{\frac 12+\delta}\varpi^N\Om_\Ke^3\nab^3\tpH\|^2_{L^\infty_uL^2_{\ub} L^2(S)}  \\
\ls &N2^{2N}\mathcal D+\|\ub^{\frac 12+\delta}\varpi^{N}\Er_1\|_{L^2_uL^2_{\ub}L^2(S)}^2+N^{-1}\|\ub^{\frac 12+\delta}\varpi^{N}\Er_2\|_{L^2_uL^2_{\ub}L^2(S)}^2\\
&+\|\ub^{1+2\delta}\varpi^{2N}\Om_\Ke^4 \mathcal T_2\|_{L^1_uL^1_{\ub}L^1(S)}.
\end{split}
\end{equation*}
\end{proposition}
\begin{proof}
To prove the desired result, we apply Proposition \ref{elliptic.chih} and integrate in appropriate weighted spaces. First, by Propositions \ref{prop.g}, \ref{prop.etab}, \ref{prop.eta}, \ref{improved.eta}, \ref{prop.tpH}, \ref{EE.1}, we note that except for the term involving $\nab^3\ttrch$, all of the terms on the right hand side of the estimate in Proposition \ref{elliptic.chih} can be controlled after integrating in weighted $L^\i_uL^2_{\ub}L^2(S)$ and $L^2_uL^2_{\ub}L^2(S)$ spaces, i.e.~we have
\begin{equation*}
\begin{split}
&\:\sum_{i\leq 3}\|\ub^{\f12+\de}\varpi^N\Om_\Ke^2\nab^i\tg\|_{L^\i_uL^2_{\ub}L^2(S)}^2+\sum_{i\leq 2}\|\ub^{\f12+\de}\varpi^N\Om_\Ke^2\nab^i(\widetilde{\beta},\tp,\tpH)\|_{L^\i_uL^2_{\ub}L^2(S)}^2\\
&\:+\sum_{i\leq 3}N\|\ub^{\f12+\de}\varpi^N\Om_\Ke^3\nab^i\tg\|_{L^2_uL^2_{\ub}L^2(S)}^2+\sum_{i\leq 2}N\|\ub^{\f12+\de}\varpi^N\Om_\Ke^2\nab^i(\widetilde{\beta},\tp,\tpH)\|_{L^2_uL^2_{\ub}L^2(S)}^2\\
\ls &\:2^{2N}\mathcal D+\|\ub^{\frac 12+\delta}\varpi^{N}\Er_1\|_{L^2_uL^2_{\ub}L^2(S)}^2+N^{-1}\|\ub^{\frac 12+\delta}\varpi^{N}\Er_2\|_{L^2_uL^2_{\ub}L^2(S)}^2+\|\ub^{1+2\delta}\varpi^{2N}\Om_\Ke^4 \mathcal T_2\|_{L^1_uL^1_{\ub}L^1(S)}.
\end{split}
\end{equation*}
Therefore, Proposition \ref{elliptic.chih} implies that
\begin{equation*}
\begin{split}
&\|\ub^{\f12+\de}\varpi^N\Om_\Ke^2\nab^3\widetilde{\chih}\|_{L^\i_uL^2_{\ub}L^2(S)}^2+N\|\ub^{\f12+\de}\varpi^N\Om_\Ke^3\nab^3\widetilde{\chih}\|_{L^2_uL^2_{\ub}L^2(S)}^2\\
\ls &2^{2N}\mathcal D+\|\ub^{\frac 12+\delta}\varpi^{N}\Er_1\|_{L^2_uL^2_{\ub}L^2(S)}^2+N^{-1}\|\ub^{\frac 12+\delta}\varpi^{N}\Er_2\|_{L^2_uL^2_{\ub}L^2(S)}^2+\|\ub^{1+2\delta}\varpi^{2N}\Om_\Ke^4 \mathcal T_2\|_{L^1_uL^1_{\ub}L^1(S)}\\
&+\|\ub^{\f12+\de}\varpi^N\Om_\Ke^2\nab^3\ttrch\|_{L^\i_uL^2_{\ub}L^2(S)}^2+N\|\ub^{\f12+\de}\varpi^N\Om_\Ke^3\nab^3\ttrch\|_{L^2_uL^2_{\ub}L^2(S)}^2.
\end{split}
\end{equation*}
We then apply Propositions \ref{est.nab3trch} and \ref{est.nab3trch.2} to control the $\nab^3\ttrch$ terms and to obtain
\begin{equation*}
\begin{split}
&\|\ub^{\f12+\de}\varpi^N\Om_\Ke^2\nab^3\widetilde{\chih}\|_{L^\i_uL^2_{\ub}L^2(S)}^2+N\|\ub^{\f12+\de}\varpi^N\Om_\Ke^3\nab^3\widetilde{\chih}\|_{L^2_uL^2_{\ub}L^2(S)}^2\\
&+\|\ub^{\f12+\de}\varpi^N\Om_\Ke^2\nab^3\ttrch\|_{L^\i_uL^2_{\ub}L^2(S)}^2+N\|\ub^{\f12+\de}\varpi^N\Om_\Ke^3\nab^3\ttrch\|_{L^2_uL^2_{\ub}L^2(S)}^2\\
\ls & N 2^{2N}\mathcal D+\|\ub^{\frac 12+\delta}\varpi^{N}\Er_1\|_{L^2_uL^2_{\ub}L^2(S)}^2+N^{-1}\|\ub^{\frac 12+\delta}\varpi^{N}\Er_2\|_{L^2_uL^2_{\ub}L^2(S)}^2+\|\ub^{1+2\delta}\varpi^{2N}\Om_\Ke^4 \mathcal T_2\|_{L^1_uL^1_{\ub}L^1(S)}\\
&+N^{-1}\|\ub^{\f12+\de}\varpi^N\Om_\Ke^2\nab^3\tpH\|_{L^\i_uL^2_{\ub}L^2(S)}^2+\ep N\|\ub^{\f12+\de}\varpi^N\Om_\Ke^3\nab^3\tpH\|_{L^2_uL^2_{\ub}L^2(S)}^2.
\end{split}
\end{equation*}
Finally, since $N^{-1}\ll 1$ and $\ep N\ll N$, we can absorb the last two terms to the left hand side and obtain the desired conclusion.
\end{proof}
We have thus obtained all the estimates for $\nab^3\tpH$ and this concludes the subsection.

\subsection{Estimates for $\protect\nab^3\protect\tpHb$ and $\protect\nab^3\widetilde{\protect\omb}$}\label{sec.elliptic.tpHb}

In this subsection, we prove the estimates for $\nab^3\widetilde{\omb}$ (Proposition~\ref{prop.omb3}) and $\nab^3\tpHb$ (Proposition~\ref{elliptic.tpHb}). The elliptic estimates for $\nab^3\widetilde{\omb}$ and $\nab^3\tpHb$ are coupled to each other and we have to prove them simultaneously. Moreover, these bounds are also coupled with that for $\nab^{i}\tg$ in $L^{\i}_{\ub}L^2_uL^2(S)$ for $i\leq 3$. (Recall that in Proposition~\ref{prop.g}, we have only obtain $L^{2}_{\ub}L^{\i}_uL^2(S)$ and $L^2_{\ub}L^2_uL^2(S)$ estimates for $\nab^i\tg$; see Remark~\ref{rmk:tg.needs.improvement}.)

In view of the coupling, we organise the estimates as follows. In \textbf{Section~\ref{sec:elliptic.prelim.tg}} and \textbf{Section~\ref{sec:elliptic.prelim.tomb}}, we begin with some preliminary estimates for $\nab^i\tg$ and $\nab^3\widetilde{\omb}$ respectively. To handle the coupling, we allow on the right hand side of the ($L^{\i}_{\ub}L^2_uL^2(S)$) estimates of $\nab^i\tg$ the term $\||u|^{\frac 12+\de}\varpi^N\Om_\Ke\nab^3(\tpHb,\widetilde{\omb})\|_{L^\infty_{\ub}L^2_uL^2(S)}^2$. Similarly, we allow on the right hand side of the estimates for $\nab^3\widetilde{\omb}$ the term $\||u|^{\frac 12+\de}\varpi^N\Om_\Ke\nab^3\tpHb\|_{L^\infty_{\ub}L^2_uL^2(S)}^2$. Then in \textbf{Section~\ref{sec:elliptic.chibh.elliptic}}, we prove an elliptic estimate controlling $\nab^3\chibh$. In \textbf{Section~\ref{sec:elliptic.tpHb.i22}}, we use the estimates obtained so far to obtain $L^\infty_{\ub}L^2_uL^2(S)$ estimates for $\nab^3\tpHb$. This in particular allows us to prove in \textbf{Section~\ref{sec:elliptic.improved}} estimates for $\nab^i\tg$ and $\nab^3\widetilde{\omb}$, which improve those in Sections~\ref{sec:elliptic.prelim.tg} and \ref{sec:elliptic.prelim.tomb}. We will also at this point prove appropriate $L^\i_{\ub}L^2_uL^2(S)$ bounds for $\nab^i\widetilde{\etab}$ (cf.~Remark~\ref{rmk:tetab.needs.improvement}). Finally, in \textbf{Section~\ref{sec:elliptic.tpHb.222}}, we prove the $L^2_{\ub}L^2_uL^2(S)$ estimates for $\nab^3\tpHb$ and conclude this subsection.

\subsubsection{Preliminary estimate for $\protect\nab^i\protect\tg$}\label{sec:elliptic.prelim.tg}

We begin with the following estimate for $\nab^i\tg$ ($i\leq 3$). Recall from our earlier discussion that on the right hand side of this estimate, we allow the term $\||u|^{\frac 12+\de}\varpi^N\Om_\Ke\nab^3(\tpHb,\widetilde{\omb})\|_{L^\infty_{\ub}L^2_uL^2(S)}^2$, which will be controlled later; see Propositions~\ref{prop.omb3} and \ref{prop.nabettrchb.2} and \ref{improved.tg.prop.1}.
\begin{proposition}\label{improved.tg}
For $i\leq 3$, $\nab^i\tg$ obeys the following $L^\infty_{\ub}L^\infty_uL^2(S)$ and $L^\infty_{\ub}L^2_uL^2(S)$ estimates:
\begin{equation*}
\begin{split}
&\:\sum_{i\leq 3}(\||u|^{\frac 12+\de}\varpi^N\Om_\Ke\nab^i\tg\|^2_{L^\infty_{\ub}L^\infty_u L^2(S)}+\||u|^{\frac 12+\de}\varpi^N\Om_\Ke\nab^i\tg\|^2_{L^\infty_{\ub} L^2_{u}L^2(S)}+N\||u|^{\frac 12+\de}\varpi^N\Om_\Ke^2\nab^i\tg\|^2_{L^\infty_{\ub} L^2_{u}L^2(S)})  \\
\ls &\:2^{2N}\mathcal D+\||u|^{\frac 12+\de}\varpi^N\Om_\Ke\nab^3(\tpHb,\widetilde{\omb})\|_{L^\infty_{\ub}L^2_uL^2(S)}^2\\
&\:+\|\ub^{\frac 12+\de}\varpi^{N}\Er_1\|_{L^2_uL^2_{\ub}L^2(S)}^2+N^{-1}\||u|^{\frac 12+\de}\varpi^{N}\Er_2\|_{L^2_uL^2_{\ub}L^2(S)}^2+\||u|^{1+2\de}\varpi^{2N}\Om_\Ke^2\mathcal T_1\|_{L^1_uL^1_{\ub}L^1(S)}.
\end{split}
\end{equation*}
\end{proposition}
\begin{proof}
Recall from Propositions \ref{Omega.lemma} and \ref{gamma.lemma} that it suffices to control $\widetilde{\gamma}$ and $\widetilde{\log\Om}$. Applying the second estimate in Proposition~\ref{transport.3.3} (with $m=1$) to the equation for $\nab^i\widetilde{\gamma}$ in Proposition \ref{gamma.eqn} and the equation for $\nab^i(\log\Om-\log\Om_\Ke)$ in Proposition~\ref{Omega.eqn}, taking supremum in $\ub$, and estimating the data term using \eqref{data.D.def} and \eqref{varpi.bd}, we obtain
\begin{equation}\label{improved.tg.1}
\begin{split}
&\:\sum_{i\leq 3}\||u|^{\frac 12+\de}\varpi^N\Om_\Ke\nab^i\tg\|^2_{L^\infty_{\ub}L^\infty_u L^2(S)}\\
&\:+\sum_{i\leq 3}(\||u|^{\frac 12+\de}\varpi^N\Om_\Ke\nab^i\tg\|^2_{L^\infty_{\ub} L^2_{u}L^2(S)}+N\||u|^{\frac 12+\de}\varpi^N\Om_\Ke^2\nab^i\tg\|^2_{L^\infty_{\ub} L^2_{u}L^2(S)})  \\
\ls &\:2^{2N}\mathcal D+\||u|^{1+2\de}\varpi^{2N}\Om_\Ke^2 (\sum_{i\leq 3}\nab^i\tg)\mathcal F'_3\|_{L^\infty_{\ub}L^1_uL^1(S)}.
\end{split}
\end{equation}
Here, we recall from \eqref{inho.def.p} that
$$\mathcal F'_3=\sum_{i_1+i_2\leq 3}(1+\nab^{i_1}\tg+\nab^{\min\{i_1,2\}}\tp)(\Om_\Ke^2\nab^{i_2}\tg+\Om_\Ke^2\nab^{\min\{i_2,2\}}\tp+\nab^{i_2}(\tpHb,\widetilde{\omb})).$$
Notice that we can in fact assume that $\nab^{\min\{i_2,2\}}\tp$ in the last pair of brackets to be $\nab^{\min\{i_2,2\}}\widetilde{\eta}$, since we have $\etab=2\nab\log\Om-\eta$ by \eqref{gauge.con}.  %In other words, we have
%$$\mathcal F'_3=\sum_{i_1+i_2\leq 3}(1+\nab^{i_1}\tg+\nab^{\min\{i_1,2\}}\tp)(\Om_\Ke^2\nab^{i_2}\tg+\Om_\Ke^2\nab^{\min\{i_2,2\}}\widetilde{\eta}+\nab^{i_2}(\tpHb,\widetilde{\omb})).$$

We now control the error term in \eqref{improved.tg.1}. Using the Cauchy--Schwarz inequality and \eqref{BA.S.222}, we obtain
\begin{equation}\label{improved.tg.2}
\begin{split}
&\:\||u|^{1+2\de}\varpi^{2N}\Om_\Ke^2 (\sum_{i\leq 3}\nab^i\tg)\mathcal F'_3\|_{L^\infty_{\ub}L^1_uL^1(S)}\\
%\ls &(\sum_{i\leq 3}\||u|^{\frac 12+\de}\varpi^{N}\Om_\Ke \nab^i\tg\|_{L^\infty_{\ub}L^2_uL^2(S)})(1+\sum_{i_1\leq 3}\|\nab^{i_1}\tg\|_{L^\infty_{\ub}L^\infty_uL^2(S)}+\sum_{i_1\leq 2}\|\nab^{i_1}\tp\|_{L^\infty_{\ub}L^\infty_uL^2(S)})\\
%&\times(\sum_{i_2\leq 3}\||u|^{\frac 12+\de}\varpi^N\Om_\Ke^2\nab^{i_2}\tg\|_{L^\infty_{\ub}L^2_uL^2(S)}+\sum_{i_2\leq 3}\||u|^{\frac 12+\de}\varpi^N\Om_\Ke\nab^{i_2}(\tpHb,\widetilde{\omb})\|_{L^\infty_{\ub}L^2_uL^2(S)}\\
%&\quad+\sum_{i_2\leq 2}\||u|^{\frac 12+\de}\varpi^N\Om_\Ke\nab^{i_2}\widetilde{\eta}\|_{L^\infty_{\ub}L^2_uL^2(S)})\\
\ls &\:(\sum_{i\leq 3}\||u|^{\frac 12+\de}\varpi^{N}\Om_\Ke \nab^i\tg\|_{L^\infty_{\ub}L^2_uL^2(S)})\\
&\:\quad\times(\sum_{i_2\leq 3}\||u|^{\frac 12+\de}\varpi^N\Om_\Ke^2\nab^{i_2}\tg\|_{L^\infty_{\ub}L^2_uL^2(S)}+\sum_{i_2\leq 3}\||u|^{\frac 12+\de}\varpi^N\Om_\Ke\nab^{i_2}(\tpHb,\widetilde{\omb})\|_{L^\infty_{\ub}L^2_uL^2(S)}\\
&\:\qquad+\sum_{i_2\leq 2}\||u|^{\frac 12+\de}\varpi^N\Om_\Ke\nab^{i_2}\widetilde{\eta}\|_{L^\infty_{\ub}L^2_uL^2(S)}).
\end{split}
\end{equation}
By Proposition \ref{prop.eta}, and using that $|u|\ls \ub$, we have
\begin{equation}\label{improved.tg.3}
\begin{split}
&\sum_{i_2\leq 2}\||u|^{\frac 12+\de}\varpi^N\Om_\Ke\nab^{i_2}\widetilde{\eta}\|_{L^\infty_{\ub}L^2_uL^2(S)}\ls \sum_{i_2\leq 2}\|\ub^{\frac 12+\de}\varpi^N\Om_\Ke\nab^{i_2}\widetilde{\eta}\|_{L^\infty_{\ub}L^2_uL^2(S)}\ls 2^N\mathcal D^{\f 12}+\|\ub^{\frac 12+\de}\varpi^{N}\Er_1\|_{L^2_uL^2_{\ub}L^2(S)}.
\end{split}
\end{equation}
Returning to \eqref{improved.tg.1} and substituting in \eqref{improved.tg.2} and \eqref{improved.tg.3}, we obtain
\begin{equation*}
\begin{split}
&\:\sum_{i\leq 3}(\||u|^{\frac 12+\de}\varpi^N\Om_\Ke\nab^i\tg\|^2_{L^\infty_{\ub}L^\infty_u L^2(S)}+\||u|^{\frac 12+\de}\varpi^N\Om_\Ke\nab^i\tg\|^2_{L^\infty_{\ub} L^2_{u}L^2(S)}+N\||u|^{\frac 12+\de}\varpi^N\Om_\Ke^2\nab^i\tg\|^2_{L^\infty_{\ub} L^2_{u}L^2(S)})  \\
\ls &\:2^{2N}\mathcal D+(\sum_{i\leq 3}\||u|^{\frac 12+\de}\varpi^{N}\Om_\Ke \nab^i\tg\|_{L^\infty_{\ub}L^2_uL^2(S)}) (2^N\mathcal D^{\f 12}+\|\ub^{\frac 12+\de}\varpi^{N}\Er_1\|_{L^2_uL^2_{\ub}L^2(S)})\\
&\:+(\sum_{i\leq 3}\||u|^{\frac 12+\de}\varpi^{N}\Om_\Ke \nab^i\tg\|_{L^\infty_{\ub}L^2_uL^2(S)})(\sum_{i_2\leq 3}\||u|^{\frac 12+\de}\varpi^N\Om_\Ke^2\nab^{i_2}\tg\|_{L^\infty_{\ub}L^2_uL^2(S)})\\
&\:+(\sum_{i\leq 3}\||u|^{\frac 12+\de}\varpi^{N}\Om_\Ke \nab^i\tg\|_{L^\infty_{\ub}L^2_uL^2(S)})(\sum_{i_2\leq 3}\||u|^{\frac 12+\de}\varpi^N\Om_\Ke\nab^{i_2}(\tpHb,\widetilde{\omb})\|_{L^\infty_{\ub}L^2_uL^2(S)}).
\end{split}
\end{equation*}
Apply Young's inequality to all these terms and absorb $\sum_{i\leq 3}\||u|^{\frac 12+\de}\varpi^{N}\Om_\Ke \nab^i\tg\|_{L^\infty_{\ub}L^2_uL^2(S)}^2$ to the left hand side. Notice moreover that for the second line, the remaining term $\sum_{i\leq 3}\||u|^{\frac 12+\de}\varpi^N\Om_\Ke^2\nab^{i}\tg\|_{L^\infty_{\ub}L^2_uL^2(S)}$ can also be absorbed to the left hand side after choosing $N$ to be sufficiently large. Hence, we obtain
\begin{equation}\label{improved.tg.4}
\begin{split}
&\sum_{i\leq 3}(\||u|^{\frac 12+\de}\varpi^N\Om_\Ke\nab^i\tg\|^2_{L^\infty_{\ub}L^\infty_u L^2(S)}\\
&+\sum_{i\leq 3}(\||u|^{\frac 12+\de}\varpi^N\Om_\Ke\nab^i\tg\|^2_{L^\infty_{\ub} L^2_{u}L^2(S)}+N\|\ub^{\frac 12+\de}\varpi^N\Om_\Ke^2\nab^i\tg\|^2_{L^\infty_{\ub} L^2_{u}L^2(S)})  \\
\ls &2^{2N}\mathcal D+\sum_{i\leq 3}\||u|^{\frac 12+\de}\varpi^N\Om_\Ke\nab^{i}(\tpHb,\widetilde{\omb})\|_{L^\infty_{\ub}L^2_uL^2(S)}^2+\|\ub^{\frac 12+\de}\varpi^{N}\Er_1\|_{L^2_uL^2_{\ub}L^2(S)}^2.
\end{split}
\end{equation}
Finally, notice that in the second term on the right hand side, we can bound the contributions from $i\leq 2$ using Propositions~\ref{prop.tpHb} and \ref{prop.omb}, i.e.
\begin{equation*}
\begin{split}
&\sum_{i\leq 2}\||u|^{\frac 12+\de}\varpi^N\Om_\Ke\nab^{i}(\tpHb,\widetilde{\omb})\|_{L^\infty_{\ub}L^2_uL^2(S)}^2\\
\ls & 2^{2N}\mathcal D+\||u|^{\frac 12+\de}\varpi^{N}\Er_1\|_{L^2_uL^2_{\ub}L^2(S)}^2+N^{-1}\||u|^{\frac 12+\de}\varpi^{N}\Er_2\|_{L^2_uL^2_{\ub}L^2(S)}^2+\||u|^{1+2\de}\varpi^{2N}\Om_\Ke^2\mathcal T_1\|_{L^1_uL^1_{\ub}L^1(S)}.
\end{split}
\end{equation*}
Plugging this back into \eqref{improved.tg.4} concludes the proof of the proposition.
\end{proof}

\subsubsection{Preliminary estimate for $\protect\nab^3\protect\widetilde{\protect\omb}$}\label{sec:elliptic.prelim.tomb}
Next, we prove the following preliminary estimate for $\nab^3\widetilde{\omb}$. For this, we need the elliptic estimates in Proposition~\ref{elliptic.Poisson}. Notice that on the right hand side of this bound, there is a term $N^{-1}\||u|^{\frac 12+\de}\varpi^N\Om_\Ke\nab^3\tpHb\|_{L^\infty_{\ub}L^2_uL^2(S)}^2$, which will be controlled later; see Propositions~\ref{prop.nabettrchb.2} and \ref{improved.tg.prop.1}.
\begin{proposition}\label{prop.omb3}
$\nab^3\widetilde{\omb}$ obeys that following $L^\infty_{\ub} L^2_{u}L^2(S)$ and $L^2_{\ub}L^2_uL^2(S)$ estimates:
\begin{equation*}
\begin{split}
&\||u|^{\frac 12+\delta}\varpi^N\nab^3\widetilde{\omb}\|^2_{L^\infty_{\ub}L^2_u L^2(S)}+ N\||u|^{\frac 12+\delta}\varpi^N\Om_\Ke\nab^3\widetilde{\omb}\|^2_{L^2_{\ub} L^2_{u}L^2(S)}  \\
\ls &2^{2N}\mathcal D+\|\ub^{\frac 12+\delta}\varpi^{N}\Er_1\|_{L^2_{\ub}L^2_uL^2(S)}^2+N^{-1}\||u|^{\frac 12+\delta}\varpi^{N}\Er_2\|_{L^2_{\ub}L^2_uL^2(S)}^2\\
&+\||u|^{1+2\de}\varpi^{2N}\Om_\Ke^2\mathcal T_1\|_{L^1_uL^1_{\ub}L^1(S)}+N^{-1}\||u|^{\frac 12+\de}\varpi^N\Om_\Ke\nab^3\tpHb\|_{L^\infty_{\ub}L^2_uL^2(S)}^2.
\end{split}
\end{equation*}
\end{proposition}
\begin{proof}
Recall that $\ombs$ is defined by
$$\ombs\doteq \slashed\Delta\omb+\f12 \div\betab.$$
In other words, $\omb$ satisfies a Poisson equation with $\ombs$ and $\div\betab$ on the right hand side. Taking appropriate differences and derivatives, we obtain the following reduced schematic equations\footnote{Note that we have in particular used Propositions~\ref{Kerr.der.Ricci.bound} and \ref{Kerr.curv.est} to control the background Kerr quantities.}:
$$\slashed\Delta\widetilde{\omb}\eqrs \sum_{i\leq 1}\nab^i\widetilde{\ombs}+\sum_{i\leq 2}\nab^i(\widetilde{\betab},\widetilde{\omb})+\sum_{i\leq 3}\Om_\Ke^2\nab^i\tg$$
and 
$$\nab\slashed\Delta\widetilde{\omb} \eqrs \sum_{i\leq 1}\nab^i\widetilde{\ombs}+\sum_{i\leq 2}\nab^i(\widetilde{\betab},\widetilde{\omb})+\sum_{i\leq 3}\Om_\Ke^2\nab^i\tg.$$
Therefore, elliptic estimates from Proposition~\ref{elliptic.Poisson} imply that
\begin{equation}\label{omb.elliptic}
\begin{split}
\||u|^{\frac 12+\delta}\varpi^N\nab^3\widetilde{\omb}\|^2_{L^\infty_{\ub}L^2_{u} L^2(S)}
\ls &\:\sum_{i\leq 1}\||u|^{\frac 12+\de}\varpi^N\nab^i\widetilde{\ombs}\|^2_{L^\infty_{\ub}L^2_{u} L^2(S)}+\sum_{i\leq 2}\||u|^{\frac 12+\de}\varpi^N\nab^i(\widetilde{\betab},\widetilde{\omb})\|^2_{L^\infty_{\ub}L^2_u L^2(S)}\\
&\:+\sum_{i\leq 3}\||u|^{\frac 12+\de}\varpi^N\Om_\Ke^2\nab^i\tg\|^2_{L^\infty_{\ub}L^2_u L^2(S)}.
\end{split}
\end{equation}

We can now use the estimates in Propositions~\ref{prop.omb}, \ref{prop.kappab}, \ref{EE.2} and \ref{improved.tg} to control the right hand side. Notice in particular that the bounds in Proposition \ref{improved.tg} for $\nab^i\tg$ in the $L^\infty_{\ub}L^2_u L^2(S)$ norm involve $\tpHb$ and $\widetilde{\omb}$ on the right hand side (in addition to the terms involving $\Er_1$, $\Er_2$ and $\mathcal T_1$). Nevertheless, since the bound for $\sum_{i\leq 3}\||u|^{\frac 12+\de}\varpi^N\Om_\Ke^2\nab^i\tg\|^2_{L^\infty_{\ub}L^2_u L^2(S)}$ in Proposition \ref{improved.tg} comes with a factor of $N$, the terms involving $\tpHb$ and $\widetilde{\omb}$ in the estimate below come with an extra factor of $N^{-1}$. More precisely, we obtain
\begin{equation*}
\begin{split}
\||u|^{\frac 12+\delta}\varpi^N\nab^3\widetilde{\omb}\|^2_{L^\i_{\ub}L^2_u L^2(S)}
\ls &\:2^{2N}\mathcal D+\|\ub^{\frac 12+\de}\varpi^{N}\Er_1\|_{L^2_uL^2_{\ub}L^2(S)}^2+N^{-1}\||u|^{\frac 12+\de}\varpi^{N}\Er_2\|_{L^2_uL^2_{\ub}L^2(S)}^2\\
&\:+\||u|^{1+2\de}\varpi^{2N}\Om_\Ke^2\mathcal T_1\|_{L^1_uL^1_{\ub}L^1(S)}+N^{-1}\||u|^{\frac 12+\de}\varpi^N\Om_\Ke\nab^3(\tpHb,\widetilde{\omb})\|_{L^\infty_{\ub}L^2_uL^2(S)}^2.
\end{split}
\end{equation*}
For $N$ sufficiently large, the last term with $\nab^3\widetilde{\omb}$ can be absorbed to the left hand side to obtain
\begin{equation}\label{omb.elliptic.2}
\begin{split}
&\||u|^{\frac 12+\delta}\varpi^N\nab^3\widetilde{\omb}\|^2_{L^\i_{\ub} L^2_u L^2(S)}\\
\ls &2^{2N}\mathcal D+\|\ub^{\frac 12+\de}\varpi^{N}\Er_1\|_{L^2_uL^2_{\ub}L^2(S)}^2+N^{-1}\||u|^{\frac 12+\de}\varpi^{N}\Er_2\|_{L^2_uL^2_{\ub}L^2(S)}^2\\
&+\||u|^{1+2\de}\varpi^{2N}\Om_\Ke^2\mathcal T_1\|_{L^1_uL^1_{\ub}L^1(S)}+N^{-1}\||u|^{\frac 12+\de}\varpi^N\Om_\Ke\nab^3\tpHb\|_{L^\infty_{\ub}L^2_uL^2(S)}^2.
\end{split}
\end{equation}
In a manner analogous to \eqref{omb.elliptic}, we also have the bounds for $\nab^3\widetilde{\omb}$ in $L^2_{\ub}L^2_u L^2(S)$ using elliptic estimate given by Proposition \ref{elliptic.Poisson}:
\begin{equation*}
\begin{split}
&N\||u|^{\frac 12+\delta}\varpi^N\Om_\Ke\nab^3\widetilde{\omb}\|^2_{L^2_{\ub} L^2_{u}L^2(S)}\\
\ls &N\sum_{i\leq 1}\||u|^{\frac 12+\de}\varpi^N\Om_\Ke\nab^i\widetilde{\ombs}\|^2_{L^2_{\ub}L^2_u L^2(S)}+N\sum_{i\leq 2}\||u|^{\frac 12+\de}\varpi^N\Om_\Ke\nab^i(\widetilde{\betab},\widetilde{\omb})\|^2_{L^2_{\ub}L^2_u L^2(S)}\\
&+N\sum_{i\leq 3}\||u|^{\frac 12+\de}\varpi^N\Om_\Ke^2\nab^i\tg\|^2_{L^2_{\ub}L^2_u L^2(S)}.
\end{split}
\end{equation*}
We now bound the right hand side\footnote{Although we do not need this fact for later estimates, the reader may note that since we used Proposition \ref{prop.g} instead of Proposition \ref{improved.tg}, the right hand side of \eqref{omb.elliptic.3} does not contain terms involving $\nab^3\tpHb$ or $\nab^3\widetilde{\omb}$.} using Propositions \ref{prop.g}, \ref{prop.omb}, \ref{prop.kappab} and \ref{EE.2} to obtain
\begin{equation}\label{omb.elliptic.3}
\begin{split}
&N\||u|^{\frac 12+\delta}\varpi^N\Om_\Ke\nab^3\widetilde{\omb}\|^2_{L^\infty_{\ub} L^2_{u}L^2(S)}\\
\ls &2^{2N}\mathcal D+\|\ub^{\frac 12+\de}\varpi^{N}\Er_1\|_{L^2_uL^2_{\ub}L^2(S)}^2+N^{-1}\||u|^{\frac 12+\de}\varpi^{N}\Er_2\|_{L^2_uL^2_{\ub}L^2(S)}^2+\||u|^{1+2\de}\varpi^{2N}\Om_\Ke^2\mathcal T_1\|_{L^1_uL^1_{\ub}L^1(S)}.
\end{split}
\end{equation}
The conclusion of the proposition follows from \eqref{omb.elliptic.2} and \eqref{omb.elliptic.3}.
\end{proof}

\subsubsection{Elliptic estimates for $\protect\nab^3\protect\chibh$}\label{sec:elliptic.chibh.elliptic}
We now begin to derive estimates for $\nab^3\tpHb$. The general strategy for bounding $\nab^3\tpHb$ is similar to that for $\nab^3\tpH$ in Propositions~\ref{elliptic.chih} and \ref{elliptic.tpH}: we will control $\nab^3\ttrchb$ by the transport equation and then bound $\nab^3\widetilde{\chibh}$ by elliptic estimates. The main result of this subsubsection is the following proposition, which is an analogue of Proposition \ref{elliptic.chih}, in which we obtain the bounds for the third angular covariant derivative of the difference of $\chibh$. The proof of the following proposition is very similar to Proposition \ref{elliptic.chih}, except for the fact that we have some extra factors of $\Om_\Ke^2$ in the estimates below, which arise from the bounds for $(\trchb)_\Ke$ and $\chibh_{\Ke}$.
\begin{proposition}\label{elliptic.chibh}
The following estimate holds on any sphere $S_{u,\ub}$:
\begin{equation*}
\begin{split}
\|\nab^3\widetilde{\chibh}\|_{L^2(S_{u,\ub})}
\ls \sum_{i\leq 3}(\|\nab^i\widetilde{\trchb}\|_{L^2(S_{u,\ub})}+\|\Om_\Ke^2\nab^i\tg\|_{L^2(S_{u,\ub})})+\sum_{i\leq 2}(\|\nab^i(\widetilde{\betab},\tpHb)\|_{L^2(S_{u,\ub})}+\|\Om_\Ke^2\nab^i\tp\|_{L^2(S_{u,\ub})}).
\end{split}
\end{equation*}
\end{proposition}
\begin{proof}
As in the proof of Proposition \ref{elliptic.chih}, we begin with the Codazzi equation, which is this case is
$$\div\chibh =\frac 12 \slashed{\nabla} \trchb + \zeta\cdot (\chibh-\trchb\gamma) +\betab.$$
Since $\chibh$ is traceless, we again have (cf.~\eqref{curl.of.traceless})
$$\curl\chibh=^*\div\chibh.$$
We now derive schematically the system of equations that $\widetilde{\chibh}$ satisfies:
\beaa
\div\widetilde{\chibh}-\frac 12\nab\widetilde{\trchb}+\widetilde{\betab}&\eqs&I,\\
\curl\widetilde{\chibh}-\frac 12{^*\nab}\widetilde{\trchb}+^*\widetilde{\betab}&\eqs&I,\\
\tr\widetilde{\chibh}&\eqs& II,
\eeaa
where $I$ is a schematic expression whose $\nab^i$ derivatives take the following reduced schematic form for $i\leq 2$
\begin{equation*}
\begin{split}
\nab^i I \eqrs &\:\tg\nab^3\tpHb+\Om_\Ke^2\nab^3\tg+\sum_{i_1+i_2\leq 3}\nab^{\min\{i_1,2\}}\tpHb\nab^{i_2}\tg\\
&\:+\sum_{i_1+i_2\leq 2}\Om_\Ke^2(1+\nab^{i_1}(\tg,\tp))\nab^{i_2}(\tg,\tp)+\sum_{i_1+i_2\leq 2}(1+\nab^{i_1}(\tg,\tp))\nab^{i_2}\tpHb
\end{split}
\end{equation*}
and the $\nab^i$ derivatives of $II$ verifies, for $i\leq 2$, the following reduced schematic equations
$$\nab^i II \eqrs \sum_{i_1+i_2\leq 2}\Om_\Ke^2(1+\nab^{i_1}(\tg,\tp))\nab^{i_2}(\tg,\tp)+\sum_{i_1+i_2\leq 2}(1+\nab^{i_1}(\tg,\tp))\nab^{i_2}\tpHb.$$
We bound $\nab^i I$ and $\nab^i II$ by controlling $\tg$ in $L^\infty$ by Corollary~\ref{Linfty} and using \eqref{BA.S.ii2} and \eqref{BA.S.lower}:
\begin{equation*}
\begin{split}
&\:\sum_{i\leq 2}\|\nab^i(I,II)\|_{L^2(S_{u,\ub})}\\
\ls &\: \ep^{\f 12} \|\nab^3\tpHb\|_{L^2(S_{u,\ub})}+ \|\Om_\Ke^2\nab^3\tg\|_{L^2(S_{u,\ub})}+\sum_{i\leq 2}\|\nab^i\tpHb\|_{L^2(S_{u,\ub})}+\sum_{i\leq 2}\|\Om_\Ke^2\nab^{i}(\tg,\tp)\|_{L^2(S_{u,\ub})}.
\end{split}
\end{equation*}
Applying Proposition~\ref{ellipticthm}, we therefore obtain that for any $u$, $\ub$,
\begin{equation*}
\begin{split}
&\|\nab^3\widetilde{\chibh}\|_{L^2(S_{u,\ub})}\\
\ls & \sum_{i\leq 3}\|\nab^i\widetilde{\trch}\|_{L^2(S_{u,\ub})}+\sum_{i\leq 2}\|\nab^i(\widetilde{\beta},\widetilde{\chih})\|_{L^2(S_{u,\ub})}+\sum_{i\leq 2}\|\nab^i(I,II)\|_{L^2(S_{u,\ub})}\\
\ls &\sum_{i\leq 3}(\|\nab^i\widetilde{\trchb}\|_{L^2(S_{u,\ub})}+\|\Om_\Ke^2\nab^i\tg\|_{L^2(S_{u,\ub})})+\sum_{i\leq 2}(\|\nab^i(\widetilde{\betab},\tpHb)\|_{L^2(S_{u,\ub})}+\|\Om_\Ke^2\nab^i\tp\|_{L^2(S_{u,\ub})})\\
&+\ep^{\f12}\|\nab^3\widetilde{\chibh}\|_{L^2(S_{u,\ub})}.
\end{split}
\end{equation*}
Finally, notice that for $\ep_0$ (and hence $\ep$) sufficiently small, we can absorb the term 
$$\ep^{\f12}\|\nab^3\widetilde{\chibh}\|_{L^2(S_{u,\ub})}$$ 
to the left hand side to obtain the desired conclusion.
\end{proof}

\subsubsection{$\protect L^{\protect\infty}_{\protect\ub} \protect L^2_{u}L^2(S)$ estimate for $\protect\nab^3\protect\tpHb$}\label{sec:elliptic.tpHb.i22}
We can now prove the bounds for $\nab^3\ttrchb$ in $L^\i_{\ub}L^2_uL^2(S)$ using the elliptic estimates derived in Proposition~\ref{elliptic.chibh} above. Using Proposition~\ref{elliptic.chibh} again, this then implies the bounds for $\nab^3\tpHb$ in $L^\i_{\ub}L^2_uL^2(S)$; see Proposition~\ref{prop.nabettrchb.2}.

Before we proceed, let us note that it is convenient to have already expressed the reduced schematic equation (cf.~Proposition~\ref{nab3trchb.eqn}) in terms of $\Om_{\Ke}^{-2}\nab^3\ttrchb$ instead of $\nab^3\ttrchb$. What this essentially does is to re-express $\nab^3\ttrchb$ in terms of a geodesic vector field, and the new equation captures the linear behaviour of $\nab^3\ttrchb$. When combined with Proposition~\ref{transport.3.3}, it is easy to see that this gives an $L^\infty_{\ub} L^2_{u}L^2(S)$ term with a good sign with no $\Om_\Ke$ weights (see the second term on the left hand side in the estimate in Proposition~\ref{prop.nabettrchb.1}), which is what we need to control in the $\NH$ energy (see \eqref{NH.def}).

\begin{proposition}\label{prop.nabettrchb.1}
$\nab^3\widetilde{\trchb}$ obeys the following $L^\infty_{\ub}L^\infty_u L^2(S)$ and $L^\infty_{\ub} L^2_{u}L^2(S)$ estimates:
\begin{equation*}
\begin{split}
&\:\||u|^{\frac 12+\delta}\varpi^N\nab^3\widetilde{\trchb}\|^2_{L^\infty_{\ub}L^\infty_u L^2(S)}+ \||u|^{\frac 12+\delta}\varpi^N\nab^3\widetilde{\trchb}\|^2_{L^\infty_{\ub} L^2_{u}L^2(S)}  \\
&\:+N\||u|^{\frac 12+\delta}\varpi^N\Om_\Ke\nab^3\widetilde{\trchb}\|^2_{L^\infty_{\ub} L^2_{u}L^2(S)}  \\
\ls &\:2^{2N}\mathcal D+\|\ub^{\frac 12+\de}\varpi^{N}\Er_1\|_{L^2_uL^2_{\ub}L^2(S)}^2+N^{-1}\||u|^{\frac 12+\de}\varpi^{N}\Er_2\|_{L^2_uL^2_{\ub}L^2(S)}^2+\||u|^{1+2\de}\varpi^{2N}\Om_\Ke^2\mathcal T_1\|_{L^1_uL^1_{\ub}L^1(S)}.
\end{split}
\end{equation*}
\end{proposition}

\begin{proof}
Recall the reduced schematic equation for $\nab_3\nab^3\ttrchb$ from Proposition~\ref{nab3trchb.eqn}:
\begin{equation}\label{eq.nab3ttrchb}
\begin{split}
\nab_3(\Om_{\Ke}^{-2}\nab^3\ttrchb)
\eqrs &\:\Om_{\Ke}^{-2}\sum_{i_1+i_2+i_3\leq 3}(1+\nab^{i_1}\tg+\nab^{i_1-1}\tp)(\nab^{i_2}\tpHb+\Om_\Ke^2)(\nab^{i_3}(\tpHb,\widetilde{\omb})+\Om_\Ke^2\nab^{i_3}\tg).
\end{split}
\end{equation}
Applying the second estimate in Proposition \ref{transport.3.3} with $\phi=\Om_{\Ke}^{-2}\nab^3\ttrchb$ and with $m=2$, and using \eqref{data.D.def} and \eqref{varpi.bd} for the data term, we obtain
\begin{equation}\label{nab3ttrchb.1}
\begin{split}
&\||u|^{\frac 12+\delta}\varpi^N\nab^3\widetilde{\trchb}\|^2_{L^\infty_u L^2(S)}+ \||u|^{\frac 12+\delta}\varpi^N\nab^3\widetilde{\trchb}\|^2_{ L^2_{u}L^2(S)}  \\
&+N\||u|^{\frac 12+\delta}\varpi^N\Om_\Ke\nab^3\widetilde{\trchb}\|^2_{ L^2_{u}L^2(S)}  \\
\ls &2^{2N}\mathcal D+\||u|^{1+2\delta}\varpi^{2N} \Om_\Ke^2\nab^3\ttrchb(\nab_3(\Om_\Ke^{-2}\nab^3\ttrchb))\|_{L^1_uL^1(S)},
\end{split}
\end{equation}
for every fixed $\ub$. According to \eqref{eq.nab3ttrchb}, there are two terms that contribute to the error term in \eqref{nab3ttrchb.1},
which are
\begin{equation}\label{nab3ttrchb.er.2}
\||u|^{1+2\delta}\varpi^{2N}\Om_\Ke^2 \nab^3\ttrchb(\sum_{i_1+i_2\leq 3}(1+\nab^{i_1}\tg+\nab^{\min\{i_1,2\}}\tp)(\nab^{i_2}(\tpHb,\widetilde{\omb})+\Om_\Ke^2\nab^{i_2}\tg))\|_{L^1_uL^1(S)}
\end{equation}
and
\begin{equation}\label{nab3ttrchb.er.3}
\||u|^{1+2\delta}\varpi^{2N} \nab^3\ttrchb(\sum_{i_1+i_2+i_3\leq 3}(1+\nab^{i_1}\tg+\nab^{\min\{i_1,2\}}\tp)\nab^{i_2}\tpHb(\nab^{i_3}(\tpHb,\widetilde{\omb})+\Om_\Ke^2\nab^{i_3}\tg))\|_{L^1_uL^1(S)}.
\end{equation}
The term \eqref{nab3ttrchb.er.2} can be estimated using the Cauchy--Schwarz inequality and \eqref{BA.S.222} as follows 
\begin{equation}\label{nab3ttrchb.2}
\begin{split}
&\:\||u|^{1+2\delta}\varpi^{2N}\Om_\Ke^2 \nab^3\ttrchb(\sum_{i_1+i_2\leq 3}(1+\nab^{i_1}\tg+\nab^{\min\{i_1,2\}}\tp)(\nab^{i_2}(\tpHb,\widetilde{\omb})+\Om_\Ke^2\nab^{i_2}\tg))\|_{L^1_uL^1(S)}\\
\ls &\:\||u|^{\frac 12+\delta}\varpi^{N}\Om_\Ke \nab^3\ttrchb\|_{L^2_uL^2(S)}\\
&\:\times(\sum_{i_2\leq 3}(\||u|^{\frac 12+\delta}\varpi^{N}\Om_\Ke\nab^{i_2}(\tpHb,\widetilde{\omb})\|_{L^2_uL^2(S)}+\||u|^{\frac 12+\delta}\varpi^{N}\Om_\Ke^3\nab^{i_2}\tg\|_{L^2_uL^2(S)})).
\end{split}
\end{equation}
The term \eqref{nab3ttrchb.er.3} can be handled similarly, with an additional application of H\"older's inequality and Sobolev embedding (Proposition~\ref{Sobolev})
\begin{equation}\label{nab3ttrchb.3}
\begin{split}
&\:\||u|^{1+2\delta}\varpi^{2N} \nab^3\ttrchb(\sum_{i_1+i_2+i_3\leq 3}(1+\nab^{i_1}\tg+\nab^{i_1-1}\tp)\nab^{i_2}\tpHb(\nab^{i_3}(\tpHb,\widetilde{\omb})+\Om_\Ke^2\nab^{i_3}\tg))\|_{L^1_uL^1(S)}\\
\ls &\:\||u|^{\frac 12+\delta}\varpi^{N} \nab^3\ttrchb\|_{L^2_uL^2(S)}(\sum_{i_2\leq 2}\|\nab^{i_2}\tpHb\|_{L^\infty_uL^2(S)})(\sum_{i_3\leq 3}\||u|^{\frac 12+\delta}\varpi^{N}(\nab^{i_3}(\tpHb,\widetilde{\omb})+\Om_\Ke^2\nab^{i_3}\tg))\|_{L^2_uL^2(S)})\\
\ls &\:\ep^{\f12}\||u|^{\frac 12+\delta}\varpi^{N} \nab^3\ttrchb\|_{L^2_uL^2(S)}(\sum_{i_2\leq 3}(\||u|^{\frac 12+\delta}\varpi^{N}\nab^{i_2}(\tpHb,\widetilde{\omb})\|_{L^2_uL^2(S)}+\||u|^{\frac 12+\delta}\varpi^{N}\Om_\Ke^2\nab^{i_2}\tg\|_{L^2_uL^2(S)})),
\end{split}
\end{equation}
where in the last line we used \eqref{BA}.
Combining \eqref{nab3ttrchb.1}, \eqref{nab3ttrchb.2} and \eqref{nab3ttrchb.3}, we obtain
\begin{equation*}
\begin{split}
&\:\||u|^{\frac 12+\delta}\varpi^N\nab^3\widetilde{\trchb}\|^2_{L^\infty_u L^2(S)}+ \||u|^{\frac 12+\delta}\varpi^N\nab^3\widetilde{\trchb}\|^2_{ L^2_{u}L^2(S)}  \\
&\:+N\||u|^{\frac 12+\delta}\varpi^N\Om_\Ke\nab^3\widetilde{\trchb}\|^2_{ L^2_{u}L^2(S)}  \\
\ls &\: 2^{2N}\mathcal D\\
&\:+\||u|^{\frac 12+\delta}\varpi^{N}\Om_\Ke \nab^3\ttrchb\|_{L^2_uL^2(S)}(\sum_{i\leq 3}(\||u|^{\frac 12+\delta}\varpi^{N}\Om_\Ke\nab^{i}(\tpHb,\widetilde{\omb})\|_{L^2_uL^2(S)}+\||u|^{\frac 12+\delta}\varpi^{N}\Om_\Ke^3\nab^{i}\tg\|_{L^2_uL^2(S)}))\\
&\:+\ep^{\f12}\||u|^{\frac 12+\delta}\varpi^{N} \nab^3\ttrchb\|_{L^2_uL^2(S)}(\sum_{i\leq 3}(\||u|^{\frac 12+\delta}\varpi^{N}\nab^{i}(\tpHb,\widetilde{\omb})\|_{L^2_uL^2(S)}+\||u|^{\frac 12+\delta}\varpi^{N}\Om_\Ke^2\nab^{i}\tg\|_{L^2_uL^2(S)})).
\end{split}
\end{equation*}
Applying Young's inequality, we can absorb the $\nab^3\ttrchb$ terms to the left hand side to obtain
\begin{equation}\label{nab3ttrchb.4}
\begin{split}
&\:\||u|^{\frac 12+\delta}\varpi^N\nab^3\widetilde{\trchb}\|^2_{L^\infty_u L^2(S)}+ \||u|^{\frac 12+\delta}\varpi^N\nab^3\widetilde{\trchb}\|^2_{ L^2_{u}L^2(S)}  +N\||u|^{\frac 12+\delta}\varpi^N\Om_\Ke\nab^3\widetilde{\trchb}\|^2_{ L^2_{u}L^2(S)}  \\
\ls &\: 2^{2N}\mathcal D+N^{-1}\sum_{i\leq 3}(\||u|^{\frac 12+\delta}\varpi^{N}\Om_\Ke\nab^{i}(\tpHb,\widetilde{\omb})\|_{L^2_uL^2(S)}^2+\||u|^{\frac 12+\delta}\varpi^{N}\Om_\Ke^3\nab^{i}\tg\|_{L^2_uL^2(S)}^2)\\
&\:+\ep \sum_{i\leq 3}(\||u|^{\frac 12+\delta}\varpi^{N}\nab^{i}(\tpHb,\widetilde{\omb})\|_{L^2_uL^2(S)}^2+\||u|^{\frac 12+\delta}\varpi^{N}\Om_\Ke^2\nab^{i}\tg\|_{L^2_uL^2(S)}^2).
\end{split}
\end{equation}
Comparing the terms in the two lines on the right hand side, we see that the terms on the last line have worse weights. We therefore only have to consider those terms. By Propositions~\ref{prop.tpHb}, \ref{prop.omb}, \ref{improved.tg} and \ref{prop.omb3}, we have\footnote{Notice that there is an additional term $\||u|^{\frac 12+\de}\varpi^N\Om_\Ke\nab^3\widetilde{\omb}\|_{L^\infty_{\ub}L^2_uL^2(S)}^2$ on the right hand side in Proposition~\ref{improved.tg}, which can be controlled by the terms in \eqref{nab3ttrchb.5} using Proposition~\ref{prop.omb3}.}
\begin{equation}\label{nab3ttrchb.5}
\begin{split}
&\:\sum_{i\leq 3}(\||u|^{\frac 12+\delta}\varpi^{N}\nab^{i}\widetilde{\omb}\|_{L^2_uL^2(S)}+\||u|^{\frac 12+\delta}\varpi^{N}\Om_\Ke^2\nab^{i}\tg\|_{L^2_uL^2(S)}^2)+\sum_{i\leq 2}\||u|^{\frac 12+\delta}\varpi^{N}\nab^{i}\tpHb\|_{L^2_uL^2(S)}\\
\ls &\:2^{2N}\mathcal D+\|\ub^{\frac 12+\de}\varpi^{N}\Er_1\|_{L^2_uL^2_{\ub}L^2(S)}^2+N^{-1}\||u|^{\frac 12+\de}\varpi^{N}\Er_2\|_{L^2_uL^2_{\ub}L^2(S)}^2\\
&\:+\||u|^{1+2\de}\varpi^{2N}\Om_\Ke^2\mathcal T_1\|_{L^1_uL^1_{\ub}L^1(S)}+N^{-1}\||u|^{\frac 12+\de}\varpi^N\Om_\Ke\nab^3\tpHb\|_{L^\infty_{\ub}L^2_uL^2(S)}^2.
\end{split}
\end{equation}
In order to control the right hand side of \eqref{nab3ttrchb.4}, we also need to estimate $\nab^3\tpHb$. By the elliptic estimates in Proposition \ref{elliptic.chibh}, we can bound $\nab^3\widetilde{\chibh}$ by $\nab^3\ttrchb$, $\nab^{2}\widetilde{\betab}$ and lower order terms. More precisely, we have
\begin{equation*}
\begin{split}
&\||u|^{\frac 12+\de}\varpi^N\nab^3\tpHb\|_{L^\infty_{\ub}L^2_uL^2(S)}^2\\
\ls &\||u|^{\frac 12+\de}\varpi^N\nab^3\ttrchb\|_{L^\infty_{\ub}L^2_uL^2(S)}^2+\||u|^{\frac 12+\de}\varpi^N\nab^2\widetilde{\betab}\|_{L^\infty_{\ub}L^2_uL^2(S)}^2\\
&+\sum_{i\leq 3}\||u|^{\frac 12+\de}\varpi^N\Om_\Ke(\nab^i\tg,\nab^{\min\{i,2\}}\tp)\|_{L^\infty_{\ub}L^2_uL^2(S)}^2+\sum_{i\leq 2}\||u|^{\frac 12+\de}\varpi^N\nab^i\tpHb\|_{L^\infty_{\ub}L^2_uL^2(S)}^2.
\end{split}
\end{equation*}
Using \eqref{gauge.con}, for $i\leq 2$, we have $\nab^i\tp \eqrs \sum_{i_1\leq 3}(\nab^{i_1}\tg,\nab^{\min\{i_1,2\}}\widetilde{\eta})$. Therefore, applying Propositions~\ref{prop.eta}, \ref{prop.tpHb}, \ref{EE.2} and \ref{improved.tg}, the previous estimate implies
\begin{equation*}
\begin{split}
&\:\||u|^{\frac 12+\de}\varpi^N\nab^3\tpHb\|_{L^\infty_{\ub}L^2_uL^2(S)}^2\\
\ls &\:2^{2N}\mathcal D+\||u|^{\frac 12+\de}\varpi^N\nab^3\ttrchb\|_{L^\infty_{\ub}L^2_uL^2(S)}^2+\|\ub^{\frac 12+\de}\varpi^{N}\Er_1\|_{L^2_uL^2_{\ub}L^2(S)}^2\\
&\:+N^{-1}\||u|^{\frac 12+\de}\varpi^{N}\Er_2\|_{L^2_uL^2_{\ub}L^2(S)}^2+\||u|^{1+2\de}\varpi^{2N}\Om_\Ke^2\mathcal T_1\|_{L^1_uL^1_{\ub}L^1(S)}+N^{-1}\||u|^{\frac 12+\de}\varpi^N\Om_\Ke\nab^3\tpHb\|_{L^\infty_{\ub}L^2_uL^2(S)}^2.
\end{split}
\end{equation*}
For $N$ sufficiently large, since $\Om_\Ke\ls 1$, the last term can be absorbed to the left hand side so that
\begin{equation}\label{nab3ttrchb.6}
\begin{split}
\||u|^{\frac 12+\de}\varpi^N\nab^3\tpHb\|_{L^\infty_{\ub}L^2_uL^2(S)}^2
\ls &\:2^{2N}\mathcal D+\||u|^{\frac 12+\de}\varpi^N\nab^3\ttrchb\|_{L^\infty_{\ub}L^2_uL^2(S)}^2+\|\ub^{\frac 12+\de}\varpi^{N}\Er_1\|_{L^2_uL^2_{\ub}L^2(S)}^2\\
&+N^{-1}\||u|^{\frac 12+\de}\varpi^{N}\Er_2\|_{L^2_uL^2_{\ub}L^2(S)}^2+\||u|^{1+2\de}\varpi^{2N}\Om_\Ke^2\mathcal T_1\|_{L^1_uL^1_{\ub}L^1(S)}.
\end{split}
\end{equation}
Substituting the bounds in \eqref{nab3ttrchb.5} and \eqref{nab3ttrchb.6} into \eqref{nab3ttrchb.4}, we obtain
\begin{equation*}
\begin{split}
&\:\||u|^{\frac 12+\delta}\varpi^N\nab^3\widetilde{\trchb}\|^2_{L^\infty_u L^2(S)}+ \||u|^{\frac 12+\delta}\varpi^N\nab^3\widetilde{\trchb}\|^2_{ L^2_{u}L^2(S)}  +N\||u|^{\frac 12+\delta}\varpi^N\Om_\Ke\nab^3\widetilde{\trchb}\|^2_{ L^2_{u}L^2(S)}  \\
\ls &\:2^{2N}\mathcal D+\|\ub^{\frac 12+\de}\varpi^{N}\Er_1\|_{L^2_uL^2_{\ub}L^2(S)}^2+N^{-1}\||u|^{\frac 12+\de}\varpi^{N}\Er_2\|_{L^2_uL^2_{\ub}L^2(S)}^2\\
&\:+\||u|^{1+2\de}\varpi^{2N}\Om_\Ke^2\mathcal T_1\|_{L^1_uL^1_{\ub}L^1(S)}+(\ep+ N^{-1})\||u|^{\frac 12+\de}\varpi^N\Om_\Ke\nab^3\tpHb\|_{L^\infty_{\ub}L^2_uL^2(S)}^2\\
\ls &\:2^{2N}\mathcal D+\|\ub^{\frac 12+\de}\varpi^{N}\Er_1\|_{L^2_uL^2_{\ub}L^2(S)}^2+N^{-1}\||u|^{\frac 12+\de}\varpi^{N}\Er_2\|_{L^2_uL^2_{\ub}L^2(S)}^2\\
&\:+\||u|^{1+2\de}\varpi^{2N}\Om_\Ke^2\mathcal T_1\|_{L^1_uL^1_{\ub}L^1(S)}+(\ep+ N^{-1})\||u|^{\frac 12+\de}\varpi^N\nab^3\ttrchb\|_{L^\infty_{\ub}L^2_uL^2(S)}^2.
\end{split}
\end{equation*}
For $\ep_0$ (and hence $\ep$) sufficiently small and $N$ sufficiently large, the last term can be absorbed to the left hand side to yield the desired conclusion.
\end{proof}
Proposition \ref{prop.nabettrchb.1} and the estimate \eqref{nab3ttrchb.6} together imply a bound on $\nab^3\tpHb$, which we record in the following proposition:
\begin{proposition}\label{prop.nabettrchb.2}
$\nab^3\tpHb$ obeys the following $L^\infty_{\ub}L^2_uL^2(S)$ estimate:
\begin{equation*}
\begin{split}
&\||u|^{\frac 12+\de}\varpi^N\nab^3\tpHb\|_{L^\infty_{\ub}L^2_uL^2(S)}^2\\
\ls &2^{2N}\mathcal D+\|\ub^{\frac 12+\de}\varpi^{N}\Er_1\|_{L^2_uL^2_{\ub}L^2(S)}^2+N^{-1}\||u|^{\frac 12+\de}\varpi^{N}\Er_2\|_{L^2_uL^2_{\ub}L^2(S)}^2+\||u|^{1+2\de}\varpi^{2N}\Om_\Ke^2\mathcal T_1\|_{L^1_uL^1_{\ub}L^1(S)}.
\end{split}
\end{equation*}
\end{proposition}

\subsubsection{Estimates for $\protect\nab^i\protect\tg$, $\protect\nab^3\protect\widetilde{\protect\omb}$ and $\protect\nab^i\protect\widetilde{\protect\etab}$}\label{sec:elliptic.improved}
In this subsubsection, we derive some immediate corollaries of Proposition~\ref{prop.nabettrchb.2}. First, substituting in the estimates from Proposition~\ref{prop.nabettrchb.2} to Propositions~\ref{improved.tg} and \ref{prop.omb3}, we obtain the following estimates:
\begin{proposition}\label{improved.tg.prop.1}
$\nab^i\tg$ for $i\leq 3$ and $\nab^3\widetilde{\omb}$ obey the following estimates:
\begin{equation*}
\begin{split}
&\sum_{i\leq 3}\||u|^{\frac 12+\de}\varpi^N\Om_\Ke\nab^i\tg\|^2_{L^\infty_{\ub}L^\infty_u L^2(S)}\\
&+\sum_{i\leq 3}(\||u|^{\frac 12+\de}\varpi^N\Om_\Ke\nab^i\tg\|^2_{L^\infty_{\ub} L^2_{u}L^2(S)}+N\||u|^{\frac 12+\de}\varpi^N\Om_\Ke^2\nab^i\tg\|^2_{L^\infty_{\ub} L^2_{u}L^2(S)})  \\
&+\||u|^{\frac 12+\delta}\varpi^N\nab^3\widetilde{\omb}\|^2_{L^\infty_{\ub}L^2_u L^2(S)}+ N\||u|^{\frac 12+\delta}\varpi^N\Om_\Ke\nab^3\widetilde{\omb}\|^2_{L^2_{\ub} L^2_{u}L^2(S)}  \\
\ls &2^{2N}\mathcal D+\|\ub^{\frac 12+\de}\varpi^{N}\Er_1\|_{L^2_uL^2_{\ub}L^2(S)}^2+N^{-1}\||u|^{\frac 12+\de}\varpi^{N}\Er_2\|_{L^2_uL^2_{\ub}L^2(S)}^2+\||u|^{1+2\de}\varpi^{2N}\Om_\Ke^2\mathcal T_1\|_{L^1_uL^1_{\ub}L^1(S)}.
\end{split}
\end{equation*}
\end{proposition}
\begin{proof}
Apply Propositions \ref{improved.tg} and \ref{prop.omb3} and substitute in the bound from Proposition \ref{prop.nabettrchb.2}.
\end{proof}
An immediate consequence of Proposition~\ref{improved.tg.prop.1} is the following additional bound\footnote{Recall that previously in Proposition \ref{prop.etab}, we only have $L^2_{\ub}L^{\i}_uL^2(S)$ bounds (and $L^2_uL^2_{\ub}L^2(S)$ bounds) for $\nab^i\etab$.} for $\nab^i\etab$:
\begin{proposition}\label{improved.etab}
For $i\leq 2$, $\nab^i\widetilde{\etab}$ obeys the following $L^\infty_{\ub}L^2_uL^2(S)$ estimate:
\begin{equation*}
\begin{split}
&\sum_{i\leq 2}\||u|^{\frac 12+\de}\varpi^N\Om_\Ke\nab^i\widetilde{\etab}\|^2_{L^\infty_{\ub}L^2_uL^2(S)}\\
\ls &2^{2N}\mathcal D+\|\ub^{\frac 12+\de}\varpi^{N}\Er_1\|_{L^2_uL^2_{\ub}L^2(S)}^2+N^{-1}\||u|^{\frac 12+\de}\varpi^{N}\Er_2\|_{L^2_uL^2_{\ub}L^2(S)}^2+\||u|^{1+2\de}\varpi^{2N}\Om_\Ke^2\mathcal T_1\|_{L^1_uL^1_{\ub}L^1(S)}.
\end{split}
\end{equation*}
\end{proposition}
\begin{proof}
Use the identity (cf.~\eqref{gauge.con}) $\etab=2\nab(\log\Om)-\eta$ and apply the bounds in Propositions~\ref{prop.eta} and \ref{improved.tg.prop.1}.
\end{proof}

\subsubsection{$\protect L^2_{\protect\ub} L^2_{u}L^2(S)$ estimate for $\protect\nab^3\protect\tpHb$}\label{sec:elliptic.tpHb.222}
In addition to the $L^\infty_{\ub}L^\infty_u L^2(S)$ and $L^\infty_{\ub} L^2_{u}L^2(S)$ estimates for $\nab^3\ttrchb$ in Proposition \ref{prop.nabettrchb.1} above, we also need the following $L^2_uL^2_{\ub}L^2(S)$ estimates. Note that using Proposition~\ref{elliptic.chibh}, this will then imply $L^2_{\ub} L^2_{u}L^2(S)$ estimates for $\nab^3\tpH$; see Proposition~\ref{elliptic.tpHb}.
\begin{proposition}\label{prop.nabettrchb.3}
$\nab^3\widetilde{\trchb}$ obeys the following $L^2_{\ub} L^2_{u}L^2(S)$ estimates:
\begin{equation*}
\begin{split}
N\||u|^{\frac 12+\delta}\varpi^N\Om_\Ke\nab^3\widetilde{\trchb}\|^2_{L^2_{\ub} L^2_{u}L^2(S)}
\ls N2^{2N}\mathcal D+\||u|^{\frac 12+\de}\varpi^{N}\Er_1\|_{L^2_uL^2_{\ub}L^2(S)}^2.
\end{split}
\end{equation*}
\end{proposition}
\begin{proof}
Using the second estimate in Proposition \ref{transport.3.3} with $\phi=\Om_{\Ke}^{-2}\nab^3\ttrchb$ and $m=3$, we have
\begin{equation*}
\begin{split}
&\||u|^{\frac 12+\delta}\varpi^N\Om_\Ke\nab^3\widetilde{\trchb}\|^2_{L^\infty_u L^2(S)}+ \||u|^{\frac 12+\delta}\varpi^N\Om_\Ke\nab^3\widetilde{\trchb}\|^2_{ L^2_{u}L^2(S)}  \\
&+N\||u|^{\frac 12+\delta}\varpi^N\Om_\Ke^2\nab^3\widetilde{\trchb}\|^2_{ L^2_{u}L^2(S)}  \\
\ls &2^{2N}\||u|^{\frac 12+\delta}\Om_\Ke\nab^3\widetilde{\trchb}\|^2_{L^2(S_{-\ub+C_R,\ub})}+\||u|^{1+2\delta}\varpi^{2N}\Om_\Ke^4 \nab^3\ttrchb(\nab_3(\Om_\Ke^{-2}\nab^3\ttrchb))\|_{L^1_uL^1(S)}
\end{split}
\end{equation*}
for every fixed $\ub$. Integrating in $\ub$, we obtain
\begin{equation}\label{nab3ttrchb.2.1}
\begin{split}
&\||u|^{\frac 12+\delta}\varpi^N\Om_\Ke\nab^3\widetilde{\trchb}\|^2_{L^2_{\ub}L^\infty_u L^2(S)}+ \||u|^{\frac 12+\delta}\varpi^N\Om_\Ke\nab^3\widetilde{\trchb}\|^2_{ L^2_{\ub}L^2_{u}L^2(S)}  \\
&+N\||u|^{\frac 12+\delta}\varpi^N\Om_\Ke^2\nab^3\widetilde{\trchb}\|^2_{L^2_{\ub} L^2_{u}L^2(S)}  \\
\ls &2^{2N}\||u|^{\frac 12+\delta}\Om_\Ke\nab^3\widetilde{\trchb}\|^2_{L^2_{\ub}L^2(S_{-\ub+C_R,\ub})}+\||u|^{1+2\delta}\varpi^{2N}\Om_\Ke^4 \nab^3\ttrchb(\nab_3(\Om_\Ke^{-2}\nab^3\ttrchb))\|_{L^1_{\ub}L^1_uL^1(S)}.
\end{split}
\end{equation}
We now control the term 
$$\||u|^{1+2\delta}\varpi^{2N}\Om_\Ke^4 \nab^3\ttrchb(\nab_3(\Om_\Ke^{-2}\nab^3\ttrchb))\|_{L^1_{\ub}L^1_uL^1(S)}.$$
As in the proof of Proposition \ref{prop.nabettrchb.1}, we use Proposition~\ref{nab3trchb.eqn} and divide into two cases, i.e.~we consider separately
\begin{equation}\label{nab3ttrchb.2.Er.2}
\||u|^{1+2\delta}\varpi^{2N}\Om_\Ke^4 \nab^3\ttrchb(\sum_{i_1+i_2\leq 3}(1+\nab^{i_1}\tg+\nab^{\min\{i_1,2\}}\tp)(\nab^{i_2}(\tpHb,\widetilde{\omb})+\Om_\Ke^2\nab^{i_2}\tg))\|_{L^1_{\ub}L^1_uL^1(S)}
\end{equation}
and
\begin{equation}\label{nab3ttrchb.2.Er.3}
\||u|^{1+2\delta}\varpi^{2N} \Om_\Ke^2\nab^3\ttrchb(\sum_{i_1+i_2+i_3\leq 3}(1+\nab^{i_1}\tg+\nab^{\min\{i_1,2\}}\tp)\nab^{i_2}\tpHb(\nab^{i_3}(\tpHb,\widetilde{\omb})+\Om_\Ke^2\nab^{i_3}\tg))\|_{L^1_{\ub}L^1_uL^1(S)}.
\end{equation}
For the term \eqref{nab3ttrchb.2.Er.2}, after using the Cauchy--Schwarz inequality and comparing the terms with $\Er_1$ in \eqref{Er.def}, we have the estimate
\begin{equation}\label{nab3ttrchb.2.2}
\begin{split}
&\||u|^{1+2\delta}\varpi^{2N}\Om_\Ke^4 \nab^3\ttrchb(\sum_{i_1+i_2\leq 3}(1+\nab^{i_1}\tg+\nab^{\min\{i_1,2\}}\tp)(\nab^{i_2}(\tpHb,\widetilde{\omb})+\Om_\Ke^2\nab^{i_2}\tg))\|_{L^1_{\ub}L^1_uL^1(S)}\\
\ls &\||u|^{\frac 12+\delta}\varpi^{N}\Om_\Ke^2 \nab^3\ttrchb\|_{L^2_{\ub}L^2_uL^2(S)}\||u|^{\frac 12+\delta}\varpi^{N}\Er_1\|_{L^2_{\ub}L^2_uL^2(S)}.
\end{split}
\end{equation}
Using the Cauchy--Schwarz inequality, \eqref{BA.S.222}, H\"older's inequality, Proposition~\ref{Sobolev} and \eqref{BA}, and comparing the terms with $\Er_1$ in \eqref{Er.def}, we bound the term \eqref{nab3ttrchb.2.Er.3} by
\begin{equation}\label{nab3ttrchb.2.3}
\begin{split}
&\:\||u|^{1+2\delta}\varpi^{2N} \Om_\Ke^2\nab^3\ttrchb(\sum_{i_1+i_2+i_3\leq 3}(1+\nab^{i_1}\tg+\nab^{\min\{i_1,2\}}\tp)\nab^{i_2}\tpHb(\nab^{i_3}(\tpHb,\widetilde{\omb})+\Om_\Ke^2\nab^{i_3}\tg))\|_{L^1_{\ub}L^1_uL^1(S)}\\
\ls &\:\||u|^{\frac 12+\delta}\varpi^{N} \Om_\Ke\nab^3\ttrchb\|_{L^2_{\ub}L^2_uL^2(S)}\\
&\:\quad\times(\sum_{i_2\leq 2}\|\nab^{i_2}\tpHb\|_{L^\infty_{\ub}L^\infty_uL^2(S)})(\sum_{i_3\leq 3}\||u|^{\frac 12+\delta}\varpi^{N}\Om_\Ke(\nab^{i_3}(\tpHb,\widetilde{\omb})+\Om_\Ke^2\nab^{i_3}\tg))\|_{L^2_{\ub}L^2_uL^2(S)})\\
&\:+\||u|^{\frac 12+\delta}\varpi^{N} \Om_\Ke\nab^3\ttrchb\|_{L^2_{\ub}L^2_uL^2(S)}\\
&\:\quad\times(\sum_{i_2\leq 3}\|(\nab^{i_2}\tg,\nab^{\min\{i_2,2\}}(\tpHb,\widetilde{\omb}))\|_{L^\infty_{\ub}L^\infty_uL^2(S)})(\sum_{i_3\leq 3}\||u|^{\frac 12+\delta}\varpi^{N}\Om_\Ke\nab^{i_3}\tpHb\|_{L^2_{\ub}L^2_uL^2(S)})\\
\ls &\:\ep^{\f12}\||u|^{\frac 12+\delta}\varpi^{N}\Om_\Ke \nab^3\ttrchb\|_{L^2_{\ub}L^2_uL^2(S)}\\
&\:\quad\times(\sum_{i_2\leq 3}(\||u|^{\frac 12+\delta}\varpi^{N}\Om_\Ke\nab^{i_2}(\tpHb,\widetilde{\omb})\|_{L^2_{\ub}L^2_uL^2(S)}+\||u|^{\frac 12+\delta}\varpi^{N}\Om_\Ke^3\nab^{i_2}\tg\|_{L^2_{\ub}L^2_uL^2(S)}))\\
\ls &\:\ep^{\f12}\||u|^{\frac 12+\delta}\varpi^{N}\Om_\Ke \nab^3\ttrchb\|_{L^2_{\ub}L^2_uL^2(S)}\||u|^{\frac 12+\delta}\varpi^{N}\Er_1\|_{L^2_{\ub}L^2_uL^2(S)}.
\end{split}
\end{equation}
Substituting \eqref{nab3ttrchb.2.2} and \eqref{nab3ttrchb.2.3} into \eqref{nab3ttrchb.2.1}, and using \eqref{data.D.def} and \eqref{varpi.bd} for the data term, we obtain
\begin{equation*}
\begin{split}
&\||u|^{\frac 12+\delta}\varpi^N\Om_\Ke\nab^3\widetilde{\trchb}\|^2_{L^2_{\ub}L^\infty_u L^2(S)}+ \||u|^{\frac 12+\delta}\varpi^N\Om_\Ke\nab^3\widetilde{\trchb}\|^2_{ L^2_{\ub}L^2_{u}L^2(S)}  \\
&+N\||u|^{\frac 12+\delta}\varpi^N\Om_\Ke^2\nab^3\widetilde{\trchb}\|^2_{L^2_{\ub} L^2_{u}L^2(S)}  \\
\ls &2^{2N}\mathcal D+\||u|^{\frac 12+\delta}\varpi^{N}\Om_\Ke^2 \nab^3\ttrchb\|_{L^2_{\ub}L^2_uL^2(S)}\||u|^{\frac 12+\delta}\varpi^{N}\Er_1\|_{L^2_{\ub}L^2_uL^2(S)}\\
&+\ep^{\f12}\||u|^{\frac 12+\delta}\varpi^{N}\Om_\Ke \nab^3\ttrchb\|_{L^2_{\ub}L^2_uL^2(S)}\||u|^{\frac 12+\delta}\varpi^{N}\Er_1\|_{L^2_{\ub}L^2_uL^2(S)}.
\end{split}
\end{equation*}
Using Young's inequality and absorbing the $\nab^3\ttrchb$ terms to the left hand side, we obtain
\begin{equation}\label{nab3ttrchb.2.4}
\begin{split}
&\:\||u|^{\frac 12+\delta}\varpi^N\Om_\Ke\nab^3\widetilde{\trchb}\|^2_{L^2_{\ub}L^\infty_u L^2(S)}+ \||u|^{\frac 12+\delta}\varpi^N\Om_\Ke\nab^3\widetilde{\trchb}\|^2_{ L^2_{\ub}L^2_{u}L^2(S)}  \\
&\:+N\||u|^{\frac 12+\delta}\varpi^N\Om_\Ke^2\nab^3\widetilde{\trchb}\|^2_{L^2_{\ub} L^2_{u}L^2(S)}  \\
\ls &\:2^{2N}\mathcal D+(N^{-1}+\ep)\||u|^{\frac 12+\de}\varpi^{N}\Er_1\|_{L^2_uL^2_{\ub}L^2(S)}^2.
\end{split}
\end{equation}
Multiplying \eqref{nab3ttrchb.2.4} by $N$ and taking $\ep_0$ (and hence $\ep$) to be sufficiently small\footnote{We recall again that the constants $\ep_0$ and $N$ satisfy $\ep_0\leq N^{-1}$ (see Remark \ref{rmk.constant}).}, we obtain the desired estimate.
\end{proof}
We now apply the elliptic estimates for $\nab^3\widetilde{\chibh}$ thus obtaining control for all the terms $\nab^3\tpHb$:
\begin{proposition}\label{elliptic.tpHb}
$\nab^3\tpHb$ obeys the following $L^{\i}_{\ub}L^2_uL^2(S)$ and $L^2_{\ub}L^2_uL^2(S)$ estimates:
\begin{equation*}
\begin{split}
&\||u|^{\frac 12+\delta}\varpi^N\nab^3\tpHb\|^2_{L^\infty_{\ub} L^2_{u}L^2(S)}  +N\||u|^{\frac 12+\delta}\varpi^N\Om_\Ke\nab^3\tpHb\|^2_{L^2_{\ub} L^2_{u}L^2(S)}  \\
\ls &N2^{2N}\mathcal D+\|\ub^{\frac 12+\delta}\varpi^{N}\Er_1\|_{L^2_{\ub}L^2_uL^2(S)}^2+N^{-1}\||u|^{\frac 12+\delta}\varpi^{N}\Er_2\|_{L^2_{\ub}L^2_uL^2(S)}^2+\||u|^{1+2\de}\varpi^{2N}\Om_\Ke^2\mathcal T_1\|_{L^1_uL^1_{\ub}L^1(S)}.
\end{split}
\end{equation*}
\end{proposition}
\begin{proof}
The estimates for $\nab^3\ttrchb$ and the $L^\i_{\ub}L^2_uL^2(S)$ estimate for $\nab^3\widetilde{\chibh}$ follow directly from Propositions~\ref{prop.nabettrchb.2} and \ref{prop.nabettrchb.3}. To obtain the bound for $\nab^3\widetilde{\chibh}$ in $L^2_{\ub}L^2_uL^2(S)$, we simply combine the elliptic estimates in Proposition~\ref{elliptic.chibh} with the bounds in Propositions~\ref{prop.g}, \ref{prop.etab}, \ref{prop.eta}, \ref{prop.tpHb}, \ref{EE.2} and \ref{prop.nabettrchb.3}.
\end{proof}

\subsection{Proof of Proposition~\ref{concluding.est}}\label{sec:proof.of.concluding.est}

We have now controlled the third angular covariant derivatives of all the differences of Ricci coefficients. Together with the estimates obtained in the previous sections, we have therefore controlled $\NI$ and $\NH$ (see \eqref{NI.def} and \eqref{NH.def}). In particular, we conclude the proof of Proposition~\ref{concluding.est}:
\begin{proof}[Proof of Proposition~\ref{concluding.est}]
Proposition~\ref{concluding.est} is a consequence of the combination of Propositions \ref{prop.g}, \ref{prop.b}, \ref{prop.etab}, \ref{prop.eta}, \ref{improved.eta}, \ref{prop.tpHb}, \ref{prop.omb}, \ref{prop.tpH}, \ref{EE.1}, \ref{EE.2}, \ref{est.nab3psi}, \ref{elliptic.tpH}, \ref{improved.tg.prop.1}, \ref{improved.etab} and \ref{elliptic.tpHb}.
\end{proof}

\section{Controlling the error terms}\label{sec:error}
Recall that we continue to work under the setting described in Remark~\ref{rmk:setup}. 

By Proposition~\ref{concluding.est}, we have obtained estimates for the metric components, the Ricci coefficients and the curvature components in the hypersurface energy $\NH$ and the integrated energy $\NI$ in terms of the error terms $\Er_1$, $\Er_2$, $\mathcal T_1$ and $\mathcal T_2$. In this section, we will prove the opposite bounds: we will show that the error terms $\Er_1$, $\Er_2$, $\mathcal T_1$ and $\mathcal T_2$ can be controlled by the $\NH$ energy and the $\NI$ energy. Moreover, by choosing $N$ sufficiently large and $\ep_0$ (and hence $\ep$) sufficiently small, we can show that there is a smallness constant in the estimates such that we can indeed derive the boundedness of $\NH$ and $\NI$. The following proposition is the main goal of this section, the proof of which will be achieved in Section~\ref{sec.end.of.error}:
\begin{proposition}\label{N.bd}
For $N$ sufficiently large and $\ep_0$ sufficiently small depending on $M$, $a$, $\de$ and $C_R$, we have
$$\NH+\NI\ls \ep^2,$$
whenever $\ep\leq \ep_0$ where the implicit constant depends only on $M$, $a$, $\de$ and $C_R$.
\end{proposition}
After we prove this proposition in Section~\ref{sec.end.of.error}, we will then fix the parameter $N$.

The section is organised as follows. In \textbf{Section~\ref{sec:add.energy}}, we define the energies $\protect\mathcal N_{int,1}$, $\protect\mathcal N_{int,2}$ and $\protect\mathcal N_{hyp,1}$, which contain subsets of terms in $\NI$ and $\NH$, and will be used to estimate the error terms. We then briefly explain the strategy used to control the error terms. In \textbf{Section~\ref{sec.error.E1.E2}}, we bound the $\Er_1$, $\Er_2$ error terms; in \textbf{Section~\ref{sec.error.T1.T2}}, we bound the $\mathcal T_1$, $\mathcal T_2$ error terms. Finally, in \textbf{Section~\ref{sec.end.of.error}}, we put together these estimates to conclude the proof of Proposition~\ref{N.bd}.

\subsection{The $\protect\mathcal N_{int,1}$, $\protect\mathcal N_{int,2}$ and $\protect\mathcal N_{hyp,1}$ energies}\label{sec:add.energy}
In order to control the error terms, we in fact do not need the full strength of $\NH$ and $\NI$ and only need a subset of the terms.\footnote{Notice that while we have dropped some of the terms here, they will nevertheless be useful when we control the $\NS$ energy in the next section.} To facilitate the exposition, we introduce some additional notations. Recalling the notations in \eqref{S.def}, define
\begin{equation*}
\begin{split}
\mathcal N_{int,1}\doteq &\sum_{i\leq 3}\left(\sum_{\tpHb\in \mathcal S_{\tpHb}}\||u|^{\frac 12+\de}\varpi^N\Om_\Ke\nab^i(\tpHb,\widetilde{\omb})\|_{L^2_uL^2_{\ub}L^2(S)}^2+\sum_{\tpH\in \mathcal S_{\tpH}}\|\ub^{\frac 12+\de}\varpi^N\Om_\Ke^3\nab^i\tpH\|_{L^2_uL^2_{\ub}L^2(S)}^2\right.\\
&\left.\qquad +\|\ub^{\frac 12+\de}\varpi^N\Om_\Ke^2\nab^{\min\{i,2\}}\tK)\|_{L^2_uL^2_{\ub}L^2(S)}^2+\|\ub^{\frac 12+\de}\varpi^N\Om_\Ke\nab^i\tb\|_{L^2_uL^2_{\ub}L^2(S)}^2\right.\\
&\left.\qquad+\sum_{\tg \in \mathcal S_{\tg},\, \tp \in \mathcal S_{\tp}}\|\ub^{\frac 12+\de}\varpi^N\Om_\Ke^2\nab^i(\tp,\tg)\|_{L^2_uL^2_{\ub}L^2(S)}^2\right),
\end{split}
\end{equation*}
\begin{equation*}
\begin{split}
\mathcal N_{int,2}\doteq &\sum_{i\leq 3}\left(\sum_{\tg \in \mathcal S_{\tg},\, \tp \in \mathcal S_{\tp}}\|\ub^{\frac 12+\de}\varpi^N\Om_\Ke\nab^i(\tp,\tg)\|_{L^2_uL^2_{\ub}L^2(S)}^2+\|\ub^{\frac 12+\de}\varpi^N\Om_\Ke\nab^{\min\{i,2\}}\tK)\|_{L^2_uL^2_{\ub}L^2(S)}^2\right)
\end{split}
\end{equation*}
and
\begin{equation*}
\begin{split}
\mathcal N_{hyp,1}\doteq &\sum_{i\leq 2}\left(\sum_{\tpHb\in \{\widetilde{\slashed{tr}\chib},\,\widetilde{\chibh}\}}\||u|^{\frac 12+\de}\varpi^N\nab^i\tpHb\|_{L^2_uL^\infty_{\ub}L^2(S)}^2+\sum_{\tpH\in \{\widetilde{\slashed{tr}\chi},\,\widetilde{\chih}\}}\|\ub^{\frac 12+\de}\varpi^N\Om_\Ke^2\nab^i\tpH\|_{L^2_{\ub}L^\infty_uL^2(S)}^2\right)\\
&+\sum_{\tpHb\in \{\widetilde{\slashed{tr}\chib},\,\widetilde{\chibh}\}}\||u|^{\frac 12+\de}\varpi^N\nab^3\tpHb\|_{L^\infty_{\ub}L^2_uL^2(S)}^2+\sum_{\tpH\in \{\widetilde{\slashed{tr}\chi},\,\widetilde{\chih}\}}\|\ub^{\frac 12+\de}\varpi^N\Om_\Ke^2\nab^3\tpH\|_{L^\infty_uL^2_{\ub}L^2(S)}^2\\
&+\sum_{i\leq 3}\|\ub^{\f12+\de}\varpi^N\nab^i\tb\|_{L^2_{\ub}L^\i_uL^2(S)}^2.
\end{split}
\end{equation*}
Combining the estimates in the previous sections (see Proposition~\ref{concluding.est}), we have in particular that
\begin{equation}\label{main.est.2}
\begin{split}
&\:N\mathcal N_{int,1}+\mathcal N_{int,2}+\mathcal N_{hyp,1}\\
\ls &\:N2^{2N}\mathcal D+\|\ub^{\frac 12+\de}\varpi^N \Er_1\|_{L^2_uL^2_{\ub}L^2(S)}^2+N^{-1}\|\ub^{\frac 12+\de}\varpi^N \Er_2\|_{L^2_uL^2_{\ub}L^2(S)}^2\\
&\:+\||u|^{1+2\de}\varpi^{2N}\Om_\Ke^2\mathcal T_1\|_{L^1_uL^1_{\ub}L^1(S)}+\|\ub^{1+2\de}\varpi^{2N}\Om_\Ke^4\mathcal T_2\|_{L^1_uL^1_{\ub}L^1(S)}.
\end{split}
\end{equation}
In the remainder of this section, we will control each of the four error terms in \eqref{main.est.2}. Our goal in Sections~\ref{sec.error.E1.E2} and \ref{sec.error.T1.T2} is to control the right hand side of \eqref{main.est.2} so as to establish an estimate of the form
$$N\mathcal N_{int,1}+\mathcal N_{int,2}+\mathcal N_{hyp,1}\ls N2^{2N}\ep^2+\mathcal N_{int,1}+(N^{-1}+\ep)\mathcal N_{int,2}+N^{-1}\mathcal N_{hyp,1}+\mathcal N_{hyp,1}^{\frac 32},$$
which will then allow us to conclude (after choosing $N$ to be sufficiently large and $\ep_0$ to be sufficiently small\footnote{Recall that $\ep\leq \ep_0$.}) that
$$N\mathcal N_{int,1}+\mathcal N_{int,2}+\mathcal N_{hyp,1}\ls N2^{2N}\ep^2.$$

\subsection{Estimates for $\Er_1$ and $\Er_2$}\label{sec.error.E1.E2}
We will first begin with the $\Er_1$ term. Most of the terms can be directly controlled by $\mathcal N_{int,1}$ or $(N^{-1}+\ep)\mathcal N_{int,2}$. The main difficulty in dealing with the $\Er_1$ term is that it has a $\ub^{\frac 12+\de}$ weight, while the control we have for $\nab^i(\tpHb,\widetilde{\omb})$ in the $\mathcal N_{int,1}$ energy is only weighted in $|u|^{\frac 12+\de}$. To overcome this challenge, we divide the spacetime in two regions: in one region $|u|$ and $\ub$ are comparable, and in the other region $\Om_\Ke\ls e^{-\frac 12\kappa_-\ub}$. As a consequence, in the second region, we can dominate any polynomial powers of $\ub$ by $\Om_\Ke$ after choosing $|u_f|$ to be sufficiently large. We now turn to the estimates.
\begin{proposition}\label{E1.bd}
$\Er_1$ obeys the following bound:
$$\|\ub^{\frac 12+\de}\varpi^N \Er_1\|_{L^2_uL^2_{\ub}L^2(S)}^2\ls \mathcal N_{int,1}+\ep\mathcal N_{int,2}+N^{-1}\mathcal N_{hyp,1}.$$
\end{proposition}
\begin{proof}
We recall from \eqref{Er.def} that $\Er_1$ denotes the terms
\begin{equation}\label{E1.def}
\begin{split}
\Er_1\doteq &\:\sum_{i_1+i_2\leq 3}(1+\sum_{\tg\in\mathcal S_{\tg},\,\tp\in \mathcal S_{\tp} }|\nab^{i_1}(\tg,\tp)|+|\nab^{i_1-1}\widetilde{K}|)\\
&\:\quad \times (\sum_{\tpHb \in\mathcal S_{\tpHb} }\Om_\Ke|\nab^{i_2}(\tpHb,\widetilde{\omb},\tb)|+\sum_{\substack{\tg\in\mathcal S_{\tg},\,\tp\in \mathcal S_{\tp} \\ \tpH\in \mathcal S_{\tpH}}}\Om_\Ke^3|\nab^{i_2}(\tp,\tg,\tpH)|+\Om_\Ke^3|\nab^{i_2-1}\widetilde{K}|)\\
&\:+\sum_{\tg\in\mathcal S_{\tg},\,\tpHb\in \mathcal S_{\tpHb} }|\Om_\Ke\nab^2\tg\nab^2\tpHb|+\sum_{\tg\in\mathcal S_{\tg},\,\tpH\in \mathcal S_{\tpH} }|\Om_\Ke^3\nab^2\tg\nab^2\tpH|.
\end{split}
\end{equation}

We consider the following three contributions in the first two line. The first is the sum of all terms where there are no $\nab^i(\tpHb,\widetilde{\omb},\tb)$ terms in the last pair of brackets and that in the first pair of brackets, there are at most two derivatives on $\tp$ and at most one derivative on $\tK$, i.e.
\begin{equation}\label{E1.bd.I}
\begin{split}
I\doteq \|\ub^{\frac 12+\de}\varpi^N\sum_{i_1+i_2\leq 3}&(1+\nab^{i_1}\tg+\nab^{\min\{i_1,2\}}\tp+\nab^{\min\{i_1-1,1\}}\tK)\\
&\times(\Om_\Ke^3\nab^{i_2}(\tp,\tg,\tpH)+\Om_\Ke^3\nab^{i_2-1}\tK)\|_{L^2_uL^2_{\ub}L^2(S)}^2.
\end{split}
\end{equation}
For II, we consider the terms that contain the factor $\nab^{i_2}(\tpHb,\widetilde{\omb},\tb)$ and that moreover in the first pair of brackets, there are at most two derivatives on $\tp$ and at most one derivative on $\tK$, i.e.
\begin{equation}\label{E1.bd.II}
II\doteq \|\ub^{\frac 12+\delta}\varpi^N\sum_{i_1+i_2\leq 3}(1+\nab^{i_1}\tg+\nab^{\min\{i_1,2\}}\tp+\nab^{\min\{i_1-1,1\}}\tK)\Om_\Ke\nab^{i_2}(\tpHb,\widetilde{\omb},\tb)\|_{L^2_uL^2_{\ub}L^2(S)}^2.
\end{equation}
Next, the third contribution contains the terms where there are three derivatives on $\tp$ or two derivatives on $\tK$ in the first pair of brackets. In this case, suppose we have $\tg$, $\tp$ in the last pair of brackets. Then notice that there cannot be any derivatives on $\tg$ and $\tp$. In particular, these factors can be controlled in $L^\i$ and after relabelling the indices, we can consider this as part of term $I$ above. Thus we only need to bound the term:
\begin{equation}\label{E1.bd.III}
III\doteq \|\ub^{\frac 12+\delta}\varpi^N(\nab^3\tp,\nab^2\tK)(\Om_\Ke^3\tpH,\Om_\Ke\tpHb,\Om_\Ke\widetilde{\omb},\Om_\Ke\tb)\|_{L^2_uL^2_{\ub}L^2(S)}^2.
\end{equation}
Finally, we have the last two terms in \eqref{E1.def}, which have a larger total number of derivatives than the other terms. We will call these terms $IV$.
\begin{equation}\label{E1.bd.IV}
IV\doteq \|\ub^{\frac 12+\delta}\varpi^N(\Om_\Ke\nab^2\tg\nab^2\tpHb+\Om_\Ke^3\nab^2\tg\nab^2\tpH)\|_{L^2_uL^2_{\ub}L^2(S)}^2.
\end{equation}
It is easy to check that controlling $I$, $II$, $III$ and $IV$ by $\mathcal N_{int,1}+\ep\mathcal N_{int,2}+N^{-1}\mathcal N_{hyp,1}$ indeed implies the desired estimate for $\Er_1$.

We now control the term $I$ in \eqref{E1.bd.I}. First, by Proposition~\ref{K.diff.covariant}, $\sum_{i_1\leq 3}\nab^{\min\{i_1-1,1\}}\tK \eqrs \sum_{i_1\leq 3}\nab^{i_1}\tg$. We then apply \eqref{BA.S.222} in Section~\ref{sec:rmk.BA.S} and recall the definition of $\mathcal N_{int,1}$ (and use $\Om_\Ke\ls 1$) to obtain
\begin{equation}\label{E1.0}
\begin{split}
I \ls &\:\sum_{i\leq 3}(\|\ub^{\frac 12+\de}\varpi^N\Om_\Ke^3\nab^{i}(\tp,\tg,\tpH)\|_{L^2_uL^2_{\ub}L^2(S)}^2+\|\ub^{\frac 12+\de}\varpi^N\Om_\Ke^3\nab^{i-1}\tK \|_{L^2_uL^2_{\ub}L^2(S)}^2)\ls \mathcal N_{int,1}.
\end{split}
\end{equation}

We now bound the term $II$ in \eqref{E1.bd.II}. As in the estimates for $I$, we note $\sum_{i_1\leq 3}\nab^{\min\{i_1-1,1\}}\tK \eqrs \sum_{i_1\leq 3}\nab^{i_1}\tg$ and use \eqref{BA.S.222} to obtain
\begin{equation}\label{E1.1}
\begin{split}
II \ls &\:\sum_{i_2\leq 3}\|\ub^{\frac 12+\de}\varpi^N\Om_\Ke\nab^{i_2}(\tpHb,\widetilde{\omb},\tb)\|_{L^2_uL^2_{\ub}L^2(S)}^2\\
\ls &\:\mathcal N_{int,1}+\sum_{i\leq 3}\|\ub^{\frac 12+\de}\varpi^N\Om_\Ke\nab^{i}(\tpHb,\widetilde{\omb})\|_{L^2_uL^2_{\ub}L^2(S)}^2,
\end{split}
\end{equation}
since we observe that the $\tb$ term can be bounded by the $\mathcal N_{int,1}$ energy.
We now have to control
$$\sum_{i\leq 3}\|\ub^{\frac 12+\de}\varpi^N\Om_\Ke\nab^{i}(\tpHb,\widetilde{\omb})\|_{L^2_uL^2_{\ub}L^2(S)}^2.$$
Note that this is a difficult term that we have already discussed in the introduction; see \eqref{mustbeestim}. The difficulty in estimating this term is due to the fact that the energies for $\tpHb$ have weights in $|u|$ while we need to control a term with weights in $\ub$. To overcome this problem, we divide the region of integration into $\{-u> \frac 12\ub\}$ and $\{-u\leq \frac 12\ub\}$. In the first region, since $\ub\leq 2|u|$, we have
\begin{equation}\label{E1.1.1}
\begin{split}
&\:\sum_{i\leq 3}\|\ub^{\frac 12+\de}\varpi^N\Om_\Ke\nab^i(\tpHb,\widetilde{\omb})\|_{L^2_{u,\ub}(\{(u,\ub):-u> \frac 12\ub\})L^2(S)}\\
\leq &\:2^{\f 12+\de}\sum_{i\leq 3}\||u|^{\frac 12+\de}\varpi^N\Om_\Ke\nab^i(\tpHb,\widetilde{\omb})\|_{L^2_uL^2_{\ub}L^2(S)}\ls \mathcal N_{int,1}^{\f 12}.
\end{split}
\end{equation}
In the second region, by Proposition~\ref{Om.Kerr.bounds}, $\Om_\Ke$ satisfies the pointwise bound (recall $r_\pm = M \pm \sqrt{M^2 - a^2}$)
$$\sup_{\{(u,\ub)\times \mathbb S^2:-u\leq \frac 12\ub\}}|\Om_\Ke |\ls \sup_{\{(u,\ub)\times \mathbb S^2:-u\leq \frac 12\ub\}} e^{-\f{r_+-r_-}{2(r_-^2+a^2)}(\ub+u)}\ls e^{-\f{r_+-r_-}{2(r_-^2+a^2)} \ub+\f{r_+-r_-}{4(r_-^2+a^2)}\ub}\ls e^{-\f{r_+-r_-}{4(r_-^2+a^2)} \ub}.$$
This then implies
$$\sup_{\{(u,\ub)\times \mathbb S^2:-u\leq \frac 12\ub\}}\ub^{\frac 12+\de}|\Om_\Ke |\ls e^{-\f{r_+-r_-}{8(r_-^2+a^2)} \ub}.$$
Therefore, in the second region, we can control $\tpHb$ and $\widetilde{\omb}$ as follows:
\begin{equation}\label{E1.1.2}
\begin{split}
&\:\sum_{i\leq 3}\|\ub^{\frac 12+\de}\varpi^N\Om_\Ke\nab^i(\tpHb,\widetilde{\omb})\|_{L^2_{u,\ub}(\{(u,\ub):-u\leq \frac 12\ub\})L^2(S)}\\
\ls &\: \sum_{i\leq 3}\|\varpi^N\nab^i(\tpHb,\widetilde{\omb})\|_{L^\infty_{\ub}L^2_uL^2(S)}\|\ub^{\f 12+\de}\Om_\Ke\|_{(L^2_{\ub}L^\i_u)(\{(u,\ub):-u\leq \frac 12\ub\})L^\i(S)}\\
\ls &\:\sum_{i\leq 3}\||u|^{\frac 12+\delta}\varpi^{N}\slashed{\nabla}^{i}(\tpHb,\widetilde{\omb})\|_{L^\infty_{\ub}L^2_uL^2(S)}\|e^{-\f{r_+-r_-}{8(r_-^2+a^2)} \ub} \|_{L^2_{\ub}}\\
\ls &\:e^{-\f{r_+-r_-}{8(r_-^2+a^2)} |u_f|}\mathcal N_{hyp,1}^{\f12}.
\end{split}
\end{equation}
Clearly, we can choose $|u_f|$ to be sufficiently large to that $e^{-\f{r_+-r_-}{8(r_-^2+a^2)} |u_f|}\leq N^{-\f12}$.
Therefore, by \eqref{E1.1.1} and \eqref{E1.1.2}, we have
\begin{equation}\label{E1.2}
\begin{split}
&\sum_{i\leq 3}\|\ub^{\frac 12+\de}\varpi^N\Om_\Ke\nab^i(\tpHb,\widetilde{\omb})\|_{L^2_uL^2_{\ub}L^2(S)}^2\ls \mathcal N_{int,1}+N^{-1}\mathcal N_{hyp,1}.
\end{split}
\end{equation}
Substituting \eqref{E1.2} into \eqref{E1.1}, we obtain
\begin{equation}\label{E1.2.final}
II\ls \mathcal N_{int,1}+N^{-1}\mathcal N_{hyp,1}.
\end{equation}

We now control $III$ in \eqref{E1.bd.III}, i.e.~the term where the highest derivatives fall on $\tp$ and $\tK$.
For this term, we use Corollary~\ref{Linfty} to bound $(\tpHb,\widetilde{\omb},\tb,\Om_\Ke^2\tpH)$ in $L^\i_uL^\i_{\ub}L^\i(S)$ to obtain
\begin{equation}\label{E1.3}
\begin{split}
III \ls &\:\|\ub^{\frac 12+\delta}\varpi^{N}\Om_\Ke(\slashed{\nabla}^3\tp,\nab^2\tK)\|_{L^2_uL^2_{\ub}L^2(S)}^2(\|(\tpHb,\widetilde{\omb},\tb)\|_{L^\infty_uL^\infty_{\ub}L^\infty(S)}^2+\|\Om_\Ke^2\tpH\|_{L^\infty_uL^\infty_{\ub}L^\infty(S)}^2)\\
\ls &\:\ep\mathcal N_{int,2}.
\end{split}
\end{equation}

Finally, we estimate the term $IV$ from \eqref{E1.bd.IV}. For this term, we apply H\"older's inequality and the Sobolev embedding theorem in Proposition \ref{Sobolev} to obtain
\begin{equation}\label{E1.4}
\begin{split}
IV\ls &\:\|\nab^2\tg\|_{L^\i_uL^\i_{\ub}L^4(S)}^2\|\ub^{\frac 12+\delta}\varpi^N(\Om_\Ke\nab^2\tpHb,\Om_\Ke^3\nab^2\tpH)\|_{L^2_uL^2_{\ub}L^4(S)}^2\\
\ls &\:(\sum_{i_1\leq 3}\|\nab^{i_1}\tg\|_{L^\i_uL^\i_{\ub}L^2(S)}^2)(\sum_{i_2\leq 3}\|\ub^{\frac 12+\delta}\varpi^N(\Om_\Ke\nab^{i_2}\tpHb,\Om_\Ke^3\nab^{i_2}\tpH)\|_{L^2_uL^2_{\ub}L^2(S)}^2\\
\ls &\:\ep \mathcal N_{int,1} +\ep N^{-1}\mathcal N_{hyp,1},
\end{split}
\end{equation}
where we have again used the bootstrap assumption \eqref{BA} and \eqref{E1.2} in the last line.

Combining the estimates for the terms $I$, $II$, $III$ and $IV$ in \eqref{E1.0}, \eqref{E1.2.final}, \eqref{E1.3} and \eqref{E1.4} respectively, this concludes the proof of the proposition.
\end{proof}
We now move on to the second term in \eqref{main.est.2}, i.e.~the $\Er_2$ term. Notice that since this term has a prefactor of $N^{-1}$, it suffices to bound it by $\mathcal N_{int,2}$. This is what we prove in the following proposition:
\begin{proposition}\label{E2.bd}
$\Er_2$ obeys the following bound:
$$\|\ub^{\frac 12+\de}\varpi^N \Er_2\|_{L^2_uL^2_{\ub}L^2(S)}^2\ls \mathcal N_{int,2}.$$
\end{proposition}
\begin{proof}
We recall from \eqref{Er.def} that $\Er_2$ denotes the terms
\begin{equation*}
\begin{split}
\Er_2\doteq &\:\sum_{i_1+i_2\leq 3}\Om_\Ke(1+\sum_{\tg\in\mathcal S_{\tg},\,\tp\in \mathcal S_{\tp} }|\nab^{i_1}(\tg,\tp)|+|\nab^{i_1-1}\widetilde{K}|)(\sum_{\tg\in\mathcal S_{\tg},\,\tp\in \mathcal S_{\tp}}|\nab^{i_2}(\tp,\tg)|+|\nab^{i_2-1}\widetilde{K}|)\\
&\:+\sum_{\tg\in\mathcal S_{\tg},\,\tp\in \mathcal S_{\tp}}\Om_\Ke|\nab^2\tg(\nab\widetilde{K},\nab^2\tp)|.
\end{split}
\end{equation*}
We first handle the term with the summation over $i_1$ and $i_2$. Without loss of generality, we can assume $i_1\leq i_2$. Moreover, by Proposition~\ref{K.diff.covariant}, since $i_1\leq 1$, $\nab^{i_1}\tK \eqrs \sum_{i_1'\leq 3}\nab^{i_1'}\tg$. We then apply \eqref{BA.S.222} in Section~\ref{sec:rmk.BA.S} and recall the definition of $\mathcal N_{int,2}$ to obtain
\begin{equation*}
\begin{split}
&\sum_{\substack{i_1+i_2\leq 3\\i_1\leq i_2}}\|\ub^{\frac 12+\delta}\varpi^N \Om_\Ke(1+\nab^{i_1}(\tg,\tp)+\nab^{i_1-1}\tK)(\nab^{i_2}(\tp,\tg)+\nab^{i_2-1}\tK)\|_{L^2_uL^2_{\ub}L^2(S)}^2\\
\ls &\sum_{i\leq 3}\|\ub^{\frac 12+\delta}\varpi^{N}\Om_\Ke\nab^{i}(\tp,\tg)\|_{L^2_uL^2_{\ub}L^2(S)}^2+\sum_{i\leq 2}\|\ub^{\frac 12+\delta}\varpi^{N}\Om_\Ke\nab^{i}\tK\|_{L^2_uL^2_{\ub}L^2(S)}^2 \ls \mathcal N_{int,2}.
\end{split}
\end{equation*}
For the remaining term, we control it using H\"older's inequality, the bootstrap assumption \eqref{BA} and Sobolev embedding (Proposition \ref{Sobolev}) in the style of Section~\ref{sec:rmk.BA.S}
\begin{equation*}
\begin{split}
&\:\|\ub^{\frac 12+\delta}\varpi^N \Om_\Ke\nab^2\tg(\nab\tK,\nab^2\tp)\|_{L^2_uL^2_{\ub}L^2(S)}^2\\
\ls &\: \|\nab^2\tg\|_{L^\infty_uL^\infty_{\ub}L^4(S)}^2(\|\ub^{\frac 12+\delta}\varpi^{N}\Om_\Ke \nab \tK\|_{L^2_uL^2_{\ub}L^4(S)}^2+\|\ub^{\frac 12+\delta}\varpi^{N}\Om_\Ke\nab^2\tp\|_{L^2_uL^2_{\ub}L^4(S)}^2)\\
\ls &\:(\sum_{i_1\leq 3}\|\nab^{i_1}\tg\|_{L^\infty_uL^\infty_{\ub}L^2(S)}^2)(\sum_{i_2\leq 2}\|\ub^{\frac 12+\delta}\varpi^{N}\Om_\Ke\nab^{i_2}\tK\|_{L^2_uL^2_{\ub}L^2(S)}^2+\sum_{i_2\leq 3}\|\ub^{\frac 12+\delta}\varpi^{N}\Om_\Ke\nab^{i_2}\tp\|_{L^2_uL^2_{\ub}L^2(S)}^2)\\
\ls &\:\ep\mathcal N_{int,2}.
\end{split}
\end{equation*}
This concludes the proof of the proposition.
\end{proof}

\subsection{Estimates for $\mathcal T_1$ and $\mathcal T_2$}\label{sec.error.T1.T2}
We now turn to the terms $\mathcal T_1$ and $\mathcal T_2$. In order to control these error terms, we need to exploit their trilinear structure. In particular, we will bound these terms using the $\mathcal N_{hyp,1}$ energy. 

Before we proceed, let us make one remark. For the estimates in this subsection, the order of $u$ and $\ub$ in the norms is crucial, and we heavily exploit the fact that according to the definition of $\NH$ in \eqref{NH.def}, for many of the lower order terms, we can take $L^\infty$ before taking $L^2$. To exploit this, we will often use Remark~\ref{rmk:norm.order} without explicit comment.

We begin with the $\mathcal T_1$ term.
\begin{proposition}\label{T1.bd}
The $\mathcal T_1$ term obeys the following bound:
$$\||u|^{1+2\de}\varpi^{2N}\Om_\Ke^2\mathcal T_1\|_{L^1_uL^1_{\ub}L^1(S)}\ls \mathcal N_{hyp,1}^{\frac 32}.$$
\end{proposition}
\begin{proof}
Recall from \eqref{Er.def} that $\mathcal T_1$ is defined as
\begin{equation*}
\begin{split}
\mathcal T_1\doteq &\:(\sum_{\tpHb\in\mathcal S_{\tpHb}}\sum_{i_1+i_2\leq 3}(1+|\slashed{\nabla}^{i_1}\tg|+|\nab^{\min\{i_1,2\}}\tp|)|\slashed{\nabla}^{i_2}(\tpHb,\omb)|)\\
&\:\quad \times(\sum_{\substack{\tg\in\mathcal S_{\tg},\,\tp\in \mathcal S_{\tp} \\ \tpH\in \mathcal S_{\tpH},\, \tpHb\in\mathcal S_{\tpHb}}}\sum_{i_3+i_4+i_5\leq 3}(1+|\slashed{\nabla}^{i_3}\tg|+|\nab^{\min\{i_3,2\}}\tp|)(|\nab^{i_4}\tpH|,\Om_\Ke^{-2}|\nab^{i_4}\tb|)|\slashed{\nabla}^{i_5}(\tpHb,\widetilde{\omb})|).
\end{split}
\end{equation*}
Using the H\"older's inequality and applying the estimate \eqref{BA.S.222} twice, we obtain\footnote{Note that the indices have been relabelled.}
\begin{equation*}
\begin{split}
&\:\||u|^{1+2\delta}\varpi^{2N}\Om_\Ke^2\mathcal T_1\|_{L^1_uL^1_{\ub}L^1(S)}\\
\ls &(\sum_{i_1\leq 3}\||u|^{\frac 12+\delta}\varpi^{N}\slashed{\nabla}^{i_1}(\tpHb,\widetilde{\omb})\|_{L^\infty_{\ub}L^2_{u}L^2(S)})(\sum_{i_2+i_3\leq 3}\|\ub^{\frac 12+\delta}\varpi^N\Om_\Ke^2(\slashed{\nabla}^{i_2}\tpH,\Om_\Ke^{-2}\slashed{\nabla}^{i_2}\tb)\slashed{\nabla}^{i_3}(\tpHb,\widetilde{\omb})\|_{L^1_{\ub}L^2_uL^2(S)})\\
\ls &\:\mathcal N_{hyp,1}^{\f 12} (\sum_{i_2+i_3\leq 3}\||u|^{\frac 12+\delta}\varpi^N\Om_\Ke^2(\slashed{\nabla}^{i_2}\tpH,\Om_\Ke^{-2}\slashed{\nabla}^{i_2}\tb)\slashed{\nabla}^{i_3}(\tpHb,\widetilde{\omb})\|_{L^1_{\ub}L^2_uL^2(S)}).
\end{split}
\end{equation*}
It remains to estimate the term in the last line. Notice that since $i_2+i_3\leq 3$, we have either $i_2\leq 1$ or $i_3\leq 1$. In the case $i_2\leq 1$, we have
\begin{equation*}
\begin{split}
&\:\sum_{\substack{i_2+i_3\leq 3 \\ i_2\leq 1}}\||u|^{\frac 12+\delta}\varpi^N\Om_\Ke^2(\slashed{\nabla}^{i_2}\tpH,\Om_\Ke^{-2}\slashed{\nabla}^{i_2}\tb)\slashed{\nabla}^{i_3}(\tpHb,\widetilde{\omb})\|_{L^1_{\ub}L^2_uL^2(S)}\\
\ls &\:(\|\ub^{\frac 12+\delta}\varpi^N\Om_\Ke^2(\tpH,\Om_\Ke^{-2}\tb)\|_{L^2_{\ub}L^\infty_uL^{\i}(S)})(\sum_{i_3\leq 3}\||u|^{\frac 12+\delta}\slashed{\nabla}^{i_3}(\tpHb,\widetilde{\omb})\|_{L^\infty_{\ub}L^2_uL^2(S)})\|\ub^{-\frac 12-\delta}\|_{L^2_{\ub}}\\
&\:+(\sum_{i_2\leq 1}\|\ub^{\frac 12+\delta}\varpi^N\Om_\Ke^2(\slashed{\nabla}^{i_2}\tpH,\Om_\Ke^{-2}\slashed{\nabla}^{i_2}\tb)\|_{L^2_{\ub}L^\infty_uL^4(S)})(\sum_{i_3\leq 2}\||u|^{\frac 12+\delta}\slashed{\nabla}^{i_3}(\tpHb,\widetilde{\omb})\|_{L^\infty_{\ub}L^2_uL^4(S)})\|\ub^{-\frac 12-\delta}\|_{L^2_{\ub}}\\
\ls &\:(\sum_{i_2\leq 2}\|\ub^{\frac 12+\delta}\varpi^N\Om_\Ke^2(\slashed{\nabla}^{i_2}\tpH,\Om_\Ke^{-2}\slashed{\nabla}^{i_2}\tb)\|_{L^2_{\ub}L^\infty_uL^2(S)})(\sum_{i_3\leq 3}\||u|^{\frac 12+\delta}\slashed{\nabla}^{i_3}(\tpHb,\widetilde{\omb})\|_{L^\infty_{\ub}L^2_uL^2(S)})
\ls \mathcal N_{hyp,1},
\end{split}
\end{equation*}
where we have used H\"older's inequality, \eqref{varpi.bd} and Sobolev embedding (Proposition~\ref{Sobolev}).

The case $i_3\leq 1$ can be estimated in the similar manner:
\begin{equation*}
\begin{split}
&\:\sum_{\substack{i_2+i_3\leq 3 \\ i_3\leq 1}}\||u|^{\frac 12+\delta}\varpi^N\Om_\Ke^2(\slashed{\nabla}^{i_2}\tpH,\Om_\Ke^{-2}\slashed{\nabla}^{i_2}\tb)\slashed{\nabla}^{i_3}(\tpHb,\widetilde{\omb})\|_{L^1_{\ub}L^2_uL^2(S)}\\
\ls &\: (\sum_{i_2\leq 3}\|\ub^{\frac 12+\delta}\varpi^N\Om_\Ke^2(\slashed{\nabla}^{i_2}\tpH,\Om_\Ke^{-2}\nab^{i_2}\tb)\|_{L^\infty_uL^2_{\ub}L^2(S)})(\||u|^{\frac 12+\delta}(\tpHb,\widetilde{\omb})\|_{L^2_uL^\infty_{\ub}L^{\i}(S)})\|\ub^{-\frac 12-\delta}\|_{L^2_{\ub}}\\
&\:+(\sum_{i_2\leq 2}\|\ub^{\frac 12+\delta}\varpi^N\Om_\Ke^2(\slashed{\nabla}^{i_2}\tpH,\Om_\Ke^{-2}\nab^{i_2}\tb)\|_{L^\infty_uL^2_{\ub}L^4(S)})(\sum_{i_3\leq 1}\||u|^{\frac 12+\delta}\slashed{\nabla}^{i_3}(\tpHb,\widetilde{\omb})\|_{L^2_uL^\infty_{\ub}L^4(S)})\|\ub^{-\frac 12-\delta}\|_{L^2_{\ub}}\\
\ls &\:(\sum_{i_2\leq 3}\|\ub^{\frac 12+\delta}\varpi^N\Om_\Ke^2(\slashed{\nabla}^{i_2}\tpH,\Om_\Ke^{-2}\nab^{i_2}\tb)\|_{L^\infty_uL^2_{\ub}L^2(S)})(\sum_{i_3\leq 2}\||u|^{\frac 12+\delta}\slashed{\nabla}^{i_3}(\tpHb,\widetilde{\omb})\|_{L^2_uL^\infty_{\ub}L^2(S)})\ls \mathcal N_{hyp,1}.
\end{split}
\end{equation*}
As before, we used \eqref{varpi.bd}, \eqref{BA} and Proposition~\ref{Sobolev}. Combining all the above estimates yields the conclusion of the proposition.
\end{proof}
Finally, we turn to the last error term $\mathcal T_2$. 
\begin{proposition}\label{T2.bd}
The $\mathcal T_2$ term obeys the following bound:
$$\|\ub^{1+2\de}\varpi^{2N}\Om_\Ke^4\mathcal T_2\|_{L^1_uL^1_{\ub}L^1(S)}\ls \mathcal N_{hyp,1}^{\frac 32}.$$
\end{proposition}
\begin{proof}
Recall from \eqref{Er.def} that $\mathcal T_2$ is defined as
\begin{equation*}
\begin{split}
\mathcal T_2\doteq &\:(\sum_{\tpH\in \mathcal S_{\tpH}} \sum_{i_1+i_2\leq 3}(1+|\slashed{\nabla}^{i_1}\tg|+|\nab^{\min\{i_1,2\}}\tp|)(|\slashed{\nabla}^{i_2}\tpH|,\Om_\Ke^{-2}|\slashed{\nabla}^{i_2}\tb|))\\
&\:\quad \times (\sum_{\substack{\tg\in\mathcal S_{\tg},\,\tp\in \mathcal S_{\tp} \\ \tpH\in \mathcal S_{\tpH},\, \tpHb\in\mathcal S_{\tpHb}}}\sum_{i_3+i_4+i_5\leq 3}(1+|\slashed{\nabla}^{i_3}\tg|+|\nab^{\min\{i_3,2\}}\tp|)(|\nab^{i_4}\tpH|,\Om_\Ke^{-2}|\nab^{i_4}\tb|)|\slashed{\nabla}^{i_5}(\tpHb,\widetilde{\omb})|).
\end{split}
\end{equation*}
Using the H\"older's inequality and applying the estimate \eqref{BA.S.222} twice, we obtain\footnote{Note that the indices have been relabelled.}
\begin{equation*}
\begin{split}
&\:\|\ub^{1+2\delta}\varpi^{2N}\Om_\Ke^4\mathcal T_{2}\|_{L^1_uL^1_{\ub}L^1(S)}\\
\ls &\:(\sum_{i_1\leq 3}\|\ub^{\frac 12+\delta}\varpi^{N}(\Om_\Ke^2\slashed{\nabla}^{i_1}\tpH,\nab^{i_1}\tb)\|_{L^\infty_{u}L^2_{\ub}L^2(S)})(\sum_{i_2+i_3\leq 3}\|\ub^{\frac 12+\delta}\varpi^N(\Om_\Ke^2\slashed{\nabla}^{i_2}\tpH,\nab^{i_2}\tb)\slashed{\nabla}^{i_3}(\tpHb,\widetilde{\omb})\|_{L^1_u L^2_{\ub}L^2(S)})\\
\ls &\: \mathcal N_{hyp,1}^{\f12}(\sum_{i_2+i_3\leq 3}\|\ub^{\frac 12+\delta}\varpi^N(\Om_\Ke^2\slashed{\nabla}^{i_2}\tpH,\nab^{i_2}\tb)\slashed{\nabla}^{i_3}(\tpHb,\widetilde{\omb})\|_{L^1_u L^2_{\ub}L^2(S)}).
\end{split}
\end{equation*}
We now control the last term on the right hand side. As in the proof of Proposition \ref{T1.bd}, we note that since $i_2+i_3\leq 3$, we have either $i_2\leq 1$ or $i_3\leq 1$. In the case $i_2\leq 1$, we use H\"older's inequality, \eqref{varpi.bd} and Proposition~\ref{Sobolev} to obtain
\begin{equation*}
\begin{split}
&\:\sum_{\substack{i_2+i_3\leq 3 \\ i_2\leq 1}}\|\ub^{\frac 12+\delta}\varpi^N(\Om_\Ke^2\slashed{\nabla}^{i_2}\tpH,\nab^{i_2}\tb)\slashed{\nabla}^{i_3}(\tpHb,\widetilde{\omb})\|_{L^1_u L^2_{\ub}L^2(S)}\\
\ls &\:(\|\ub^{\frac 12+\delta}\varpi^N(\Om_\Ke^2\tpH,\tb)\|_{L^2_{\ub}L^\infty_uL^\i(S)})(\sum_{i_3\leq 3}\||u|^{\frac 12+\delta}\slashed{\nabla}^{i_3}(\tpHb,\widetilde{\omb})\|_{L^\infty_{\ub}L^2_uL^2(S)})\||u|^{-\frac 12-\delta}\|_{L^2_u}\\
&\:+(\sum_{i_2\leq 1}\|\ub^{\frac 12+\delta}\varpi^N(\Om_\Ke^2\slashed{\nabla}^{i_2}\tpH,\nab^{i_2}\tb)\|_{L^2_{\ub}L^\infty_uL^4(S)})(\sum_{i_3\leq 2}\||u|^{\frac 12+\delta}\slashed{\nabla}^{i_3}(\tpHb,\widetilde{\omb})\|_{L^\infty_{\ub}L^2_uL^4(S)})\||u|^{-\frac 12-\delta}\|_{L^2_u}\\
\ls &\:(\sum_{i_2\leq 2}\|\ub^{\frac 12+\delta}\varpi^N(\Om_\Ke^2\slashed{\nabla}^{i_2}\tpH,\nab^{i_2}\tb)\|_{L^2_{\ub}L^\infty_uL^2(S)})(\sum_{i_3\leq 3}\||u|^{\frac 12+\delta}\slashed{\nabla}^{i_3}(\tpHb,\widetilde{\omb})\|_{L^\infty_{\ub}L^2_uL^2(S)})\ls \mathcal N_{hyp,1}.
\end{split}
\end{equation*}
When $i_3\leq 1$, we similarly use H\"older's inequality, \eqref{varpi.bd} and Proposition~\ref{Sobolev} to obtain
\begin{equation*}
\begin{split}
&\:\sum_{\substack{i_2+i_3\leq 3 \\ i_3\leq 1}}\|\ub^{\frac 12+\delta}\varpi^N(\Om_\Ke^2\slashed{\nabla}^{i_2}\tpH,\nab^{i_2}\tb)\slashed{\nabla}^{i_3}(\tpHb,\widetilde{\omb})\|_{L^1_u L^2_{\ub}L^2(S)}\\
\ls &\:(\sum_{i_2\leq 3}\|\ub^{\frac 12+\delta}\varpi^N(\Om_\Ke^2\slashed{\nabla}^{i_2}\tpH,\nab^{i_2}\tb)\|_{L^\infty_uL^2_{\ub}L^2(S)}(\||u|^{\frac 12+\delta}(\tpHb,\widetilde{\omb})\|_{L^2_uL^\infty_{\ub}L^\i(S)})\||u|^{-\frac 12-\delta}\|_{L^2_u}\\
&\:+ (\sum_{i_2\leq 2}\|\ub^{\frac 12+\delta}\varpi^N(\Om_\Ke^2\slashed{\nabla}^{i_2}\tpH,\nab^{i_2}\tb)\|_{L^\infty_uL^2_{\ub}L^4(S)}(\sum_{i_3\leq 1}\||u|^{\frac 12+\delta}\slashed{\nabla}^{i_3}(\tpHb,\widetilde{\omb})\|_{L^2_uL^\infty_{\ub}L^4(S)})\||u|^{-\frac 12-\delta}\|_{L^2_u}\\
\ls &\:(\sum_{i_2\leq 3}\|\ub^{\frac 12+\delta}\varpi^N(\Om_\Ke^2\slashed{\nabla}^{i_2}\tpH,\nab^{i_2}\tb)\|_{L^\infty_uL^2_{\ub}L^2(S)}(\sum_{i_3\leq 2}\||u|^{\frac 12+\delta}\slashed{\nabla}^{i_3}(\tpHb,\widetilde{\omb})\|_{L^2_uL^\infty_{\ub}L^2(S)})\ls \mathcal N_{hyp,1}.
\end{split}
\end{equation*}
Combining all the above estimates yields the conclusion of the proposition. \qedhere
\end{proof}

\subsection{Proof of Proposition~\ref{N.bd}}\label{sec.end.of.error}

We have thus controlled all the error terms involving $\Er_1$, $\Er_2$, $\mathcal T_1$ and $\mathcal T_2$. Together with the estimates from the previous sections, we can show that the energies $\NH$ and $\NI$ are bounded. In other words, we can prove Proposition~\ref{N.bd}. \textbf{\emph{Note that in the following proof, we will also fix the (large) auxiliary parameter $N$.}}
\begin{proof}[Proof of Proposition~\ref{N.bd}]
Recall from \eqref{main.est.2} that we have
\begin{equation*}
\begin{split}
&N\mathcal N_{int,1}+\mathcal N_{int,2}+\mathcal N_{hyp,1}\\
\ls &N2^{2N}\mathcal D+\|\ub^{\frac 12+\de}\varpi^N \Er_1\|_{L^2_uL^2_{\ub}L^2(S)}^2+N^{-1}\|\ub^{\frac 12+\de}\varpi^N \Er_2\|_{L^2_uL^2_{\ub}L^2(S)}^2\\
&+\||u|^{1+2\de}\varpi^{2N}\Om_\Ke^2\mathcal T_1\|_{L^1_uL^1_{\ub}L^1(S)}+\|\ub^{1+2\de}\varpi^{2N}\Om_\Ke^4\mathcal T_2\|_{L^1_uL^1_{\ub}L^1(S)}.
\end{split}
\end{equation*}
Combining this with the estimates for $\Er_1$, $\Er_2$, $\mathcal T_1$ and $\mathcal T_2$ in Propositions \ref{E1.bd}, \ref{E2.bd}, \ref{T1.bd} and \ref{T2.bd}, and using \eqref{small.data} to control $\mathcal D$, we have thus obtained
$$N\mathcal N_{int,1}+\mathcal N_{int,2}+\mathcal N_{hyp,1}\ls N2^{2N}\ep^2+\mathcal N_{int,1}+(N^{-1}+\ep)\mathcal N_{int,2}+N^{-1}\mathcal N_{hyp,1}+\mathcal N_{hyp,1}^{\frac 32}.$$
Since $\ep\leq \ep_0$, by choosing $N$ sufficiently large and $\ep_0$ sufficiently small depending\footnote{Here, the choice of $N$ and $\ep_0$ is only dependent on the implicit constant in $\ls$, which in turn only depends on $M$, $a$, $\de$ and $C_R$.} only $M$, $a$, $\de$ and $C_R$, this implies
$$N\mathcal N_{int,1}+\mathcal N_{int,2}+\mathcal N_{hyp,1}\ls N2^{2N}\ep^2+\mathcal N_{hyp,1}^{\frac 32}.$$
A simple continuity argument then implies that 
$$N\mathcal N_{int,1}+\mathcal N_{int,2}+\mathcal N_{hyp,1}\ls N2^{2N}\ep^2.$$
Recall now that the choice of $N$ is only dependent on $M$, $a$, $\de$ and $C_R$, {\bf\emph{from this point onwards we consider $N$ to be fixed and absorb it into the constants in our inequality}}. Hence, we have
$$\mathcal N_{int,1}+\mathcal N_{int,2}+\mathcal N_{hyp,1}\ls \ep^2.$$
Returning to Propositions \ref{E1.bd}, \ref{E2.bd}, \ref{T1.bd} and \ref{T2.bd} again, we can thus conclude that 
\begin{equation*}
\begin{split}
&\|\ub^{\frac 12+\de}\varpi^N \Er_1\|_{L^2_uL^2_{\ub}L^2(S)}^2+\|\ub^{\frac 12+\de}\varpi^N \Er_2\|_{L^2_uL^2_{\ub}L^2(S)}^2\\
&+\||u|^{1+2\de}\varpi^{2N}\Om_\Ke^2\mathcal T_1\|_{L^1_uL^1_{\ub}L^1(S)}+\|\ub^{1+2\de}\varpi^{2N}\Om_\Ke^4\mathcal T_2\|_{L^1_uL^1_{\ub}L^1(S)}\ls \ep^2.
\end{split}
\end{equation*}
Finally, we apply Proposition~\ref{concluding.est} to obtain the desired conclusion.
\end{proof}

\section{Recovering the bootstrap assumptions}\label{recover.bootstrap}

Recall that we continue to work under the setting described in Remark~\ref{rmk:setup}. 

Our goal in this section is to improve the bounds in the bootstrap assumptions \eqref{BA}, hence concluding the proof of the a priori estimates \eqref{main.apriori.est} (and also Theorem~\ref{main.quantitative.thm}). Recall that our bootstrap assumption \eqref{BA} reads $\NS\leq \ep$,
where $\NS$ is defined in \eqref{NS.def}. Our goal in this section is to obtain an improvement of this and to prove
$$\NS\ls \ep^2.$$
The implicit constant here may depend on $M$, $a$, $\de$ and $C_R$. We note that in the course of the proof, the implicit constant (as do all the constants in this section) here may depend on the parameter $N$ in the previous section, which is now fixed (see Section~\ref{sec.end.of.error}).
 
This section is organised as follows. First, in \textbf{Section~\ref{sec.L1.Linfty}}, we estimate the terms
\begin{equation}\label{recover.bootstrap.1}
\sum_{i\leq 2}\|\nab^i\tpHb\|_{L^1_u L^\infty_{\ub}L^2(S)}^2,\quad\sum_{i\leq 2}\|\Om_\Ke^2\nab^i\tpH\|_{L^1_{\ub} L^\infty_uL^2(S)}^2
\end{equation}
in the definition $\NS$, which are the most straightforward in view of the bounds on $\NH$ and the Cauchy--Schwarz inequality. For the rest of the terms, we prove in \textbf{Section~\ref{sec.gen.Li.Li.L2}} a general proposition (Proposition \ref{transport.2}) which will be used to obtain $L^{\i}_uL^{\i}_{\ub}L^2(S)$ estimates from transport equations. In \textbf{Section~\ref{sec.main.Li.Li.L2.est}}, we then control the remaining terms in $\NS$. The key to achieving these estimates is to use the transport equations and noting now that because of the weights in the $\NH$ energies, the error terms are integrable in either $u$ and $\ub$. Finally, in \textbf{Section~\ref{sec.pf.of.main.quantitative.thm}}, we put together all the estimates for $\NS$ and combine them with Proposition~\ref{N.bd} to conclude the proof of Theorem~\ref{main.quantitative.thm}.

\subsection{$\protect L^1_{\protect\ub}\protect L^\infty_u\protect L^2(S)$ estimate for $\protect\nab^{\protect i}\protect\tpH$ and $\protect L^1_u\protect L^\infty_{\protect\ub}\protect L^2(S)$ estimate for $\protect\nab^{\protect i}\protect\tpHb$}\label{sec.L1.Linfty}
We first estimate the terms \eqref{recover.bootstrap.1}:
\begin{proposition}\label{integral.recover}
The following estimates hold:
$$\sum_{i\leq 2}(\|\nab^i\tpHb\|_{L^1_u L^\infty_{\ub}L^2(S)}^2+\|\Om_\Ke^2\nab^i\tpH\|_{L^1_{\ub} L^\infty_uL^2(S)}^2)\ls \ep^2.$$
\end{proposition}
\begin{proof}
By the bound on $\NH$ in Proposition~\ref{N.bd}, we have
$$\sum_{i\leq 2}(\||u|^{\f12+\de}\varpi^N\nab^i\tpHb\|_{L^2_u L^\infty_{\ub}L^2(S)}^2+\|\ub^{\f12+\de}\varpi^N\Om_\Ke^2\nab^i\tpH\|_{L^2_{\ub} L^\infty_uL^2(S)}^2)\ls \ep^2.$$
By the Cauchy--Schwarz inequality in $u$ and $\ub$ respectively and the observation that
$$\||u|^{-\f12-\de}\|_{L^2_u}\ls 1,\quad \|\ub^{-\f12-\de}\|_{L^2_{\ub}}\ls 1,$$
we have
$$\sum_{i\leq 2}(\|\nab^i\tpHb\|_{L^1_u L^\infty_{\ub}L^2(S)}^2+\|\Om_\Ke^2\nab^i\tpH\|_{L^1_{\ub} L^\infty_uL^2(S)}^2)\ls \ep^2.$$
\end{proof}

\subsection{General transport estimates for obtaining $\protect L^{\i}_u\protect L^{\i}_{\protect\ub}L^2(S)$ bounds}\label{sec.gen.Li.Li.L2}

For the remaining terms in $\NS$, we will apply a general proposition which gives $\protect L^{\i}_u\protect L^{\i}_{\protect\ub}L^2(S)$ bounds using the transport equations. This is an easy consequence of Proposition~\ref{transport}.
\begin{proposition}\label{transport.2}
Let $\phi$ be a tensor of arbitrary rank tangential to the spheres $S_{u,\ub}$. Then the following hold:
\begin{equation*}
\begin{split}
\|\phi\|_{L^2(S_{u,\ub})}&\ls \|\phi\|_{L^2(S_{u,-u+C_R})}+\int_{-u+C_R}^{\ub} \|\Om_\Ke^2\nab_4\phi\|_{L^2(S_{u,\ub'})} d\ub'\\
\|\phi\|_{L^2(S_{u,\ub})}&\ls \|\phi\|_{L^2(S_{-\ub+C_R,\ub})}+\int_{-\ub+C_R}^{u} \|\nab_3\phi\|_{L^2(S_{u',\ub})} du'.
\end{split}
\end{equation*}
\end{proposition}
\begin{proof}
Recall the following identities in Proposition \ref{transport}:
\begin{equation*}
\begin{split}
 \|\phi\|^2_{L^2(S_{u,\ub})}=&\|\phi\|^2_{L^2(S_{u,-u+C_R})}+2\int_{-u+C_R}^{\ub}\int_{S_{u,\ub'}} \Omega^2\left(\langle\phi,\slashed{\nabla}_4\phi\rangle_\gamma+ \frac{1}{2}\trch |\phi|^2_{\gamma}\right)d{\ub'}\\
 \|\phi\|^2_{L^2(S_{u,\ub})}=&\|\phi\|^2_{L^2(S_{-\ub+C_R,\ub})}+2\int_{-\ub+C_R}^{u}\int_{S_{u',\ub}} \left(\langle\phi,\slashed{\nabla}_3\phi\rangle_\gamma+ \frac{1}{2}\trchb |\phi|^2_{\gamma}\right)d{u'}.
\end{split}
\end{equation*}
Taking supremum over $u$ and $\ub$ respectively and using Young's inequality, we obtain
\begin{equation*}
\begin{split}
&\sup_{\ub'\in [-u+C_R,\ub]}\|\phi\|^2_{L^2(S_{u,\ub'})} \\
\ls &\|\phi\|^2_{L^2(S_{u,-u+C_R})} +  \left(\int_{-u+C_R}^{\ub} \|\Om^2\nab_4\phi\|_{L^2(S_{u,\ub'})} d\ub'\right)^2+ \int_{-u+C_R}^{\ub}\int_{S_{u,\ub'}} \Omega^2 |\trch| |\phi|^2_{\gamma}\,d\ub'
\end{split}
\end{equation*}
and 
\begin{equation*}
\begin{split}
&\sup_{u'\in [-\ub+C_R,u]} \|\phi\|_{L^2(S_{u',\ub})}\\
\ls &\|\phi\|_{L^2(S_{-\ub+C_R,\ub})}^2+\left(\int_{-\ub+C_R}^{u} \|\nab_3\phi\|_{L^2(S_{u',\ub})} du'\right)^2+ \int_{-\ub+C_R}^{u}\int_{S_{u',\ub}} |\trchb| |\phi|^2_{\gamma}\,d{u'}.
\end{split}
\end{equation*}
By Gr\"onwall's inequality, we have
\begin{equation*}
\begin{split}
\|\phi\|_{L^2(S_{u,\ub})}
\ls &\left(\|\phi\|_{L^2(S_{u,-u+C_R})}+\int_{-u+C_R}^{\ub} \|\Om^2\nab_4\phi\|_{L^2(S_{u,\ub'})} d\ub'\right)e^{C\|\Om^2 tr\chi\|_{L^1_{\ub}L^\infty(S)}}
\end{split}
\end{equation*}
and
\begin{equation*}
\begin{split}
\|\phi\|_{L^2(S_{u,\ub})}
\ls &\left(\|\phi\|_{L^2(S_{-\ub+C_R,\ub})}+\int_{-\ub+C_R}^{u} \|\nab_3\phi\|_{L^2(S_{u',\ub})} du'\right)e^{C\|tr\chib\|_{L^1_{u}L^\infty(S)}}.
\end{split}
\end{equation*}
The conclusion follows by noting that using \eqref{BA} and Proposition~\ref{Kerr.Ricci.bound}, we have
$$\Om^2\leq 2\Om_\Ke^2,\quad \|\Om^2\trch\|_{L^1_{\ub}L^\infty(S)},\|\trchb\|_{L^1_{u}L^\infty(S)}\ls 1.$$
\end{proof}

\subsection{Main $\protect L^{\i}_u \protect L^{\i}_{\protect\ub} L^2(S)$ estimates}\label{sec.main.Li.Li.L2.est}

We now apply Proposition~\ref{transport.2} to obtain the bounds for the remaining terms in $\NS$. First, we estimate all those terms without an $\Om_\Ke^2$ weight which verify a transport equation in the $\nab_3$ equation. More precisely, we have 

\begin{proposition}\label{g.recover}
$\nab^i\tg$ (for $i\leq 3$), $\nab^i(\widetilde{\etab},\tb)$ (for $i\leq 2$) and $\nab^i\tK$ (for $i\leq 1$) satisfy the following $L^\infty_uL^\infty_{\ub}L^2(S)$ bounds:
$$\sum_{i\leq 3}\|\nab^i\tg\|_{L^\infty_uL^\infty_{\ub}L^2(S)}^2+\sum_{i\leq 2}\|\nab^i(\widetilde{\etab},\tb)\|_{L^\infty_uL^\infty_{\ub}L^2(S)}^2+\sum_{i\leq 1}\|\nab^i\tK\|_{L^\infty_uL^\infty_{\ub}L^2(S)}^2\ls \ep^2.$$
\end{proposition}
\begin{proof}
Fix $\ub$. Our strategy is to apply Proposition~\ref{transport.2}. By \eqref{data.D.def} and \eqref{small.data}, the initial data contribution for these terms is appropriately bounded. Recall from Propositions~\ref{Omega.lemma} and \ref{gamma.lemma} that in order to estimate $\nab^i\tg$, it suffices to control $\nab^i(\widetilde{\gamma},\widetilde{\log\Om})$. Therefore, by Proposition~\ref{transport.2}, we need to bound
$$\sup_u\int_{-\ub+C_R}^{u} \|\nab_3\phi\|_{L^2(S_{u',\ub})} \,du',$$
by $\ep$ for $\phi$ being $\sum_{i\leq 3}\nab^i(\widetilde{\gamma},\widetilde{\log\Om})$, $\sum_{i\leq 2}\nab^i(\widetilde{\etab},\tb)$. (Notice that by Proposition~\ref{nab.diff} and \ref{K.diff.formula}, $\sum_{i\leq 1}\nab^i\tK$ can be controlled by $\sum_{i\leq 3}\nab^i \tg$.)
By Propositions~\ref{gamma.eqn}, \ref{Omega.eqn}, \ref{b.eqn}, \ref{etab.eqn}, the reduced schematic equations for $\nab_3\phi$ have the following terms:
\begin{equation}\label{g.recover.1}
\begin{split}
&\sum_{i_1+i_2\leq 3}(1+\nab^{i_1}\tg+\nab^{\min\{i_1,2\}}\tp)(\Om_\Ke^2\nab^{i_2}\tg+\Om_\Ke^2\nab^{\min\{i_2,2\}}\tp+\nab^{i_2}(\tpHb,\widetilde{\omb}))\\
&+\sum_{i_1+i_2+i_3\leq 2}(1+\nab^{i_1}(\tg,\tp))(\Om_\Ke^2\nab^{i_2}\tb+\nab^{i_2}\tpHb\nab^{i_3}\tb).
\end{split}
\end{equation}
Notice that these include the terms in $\mathcal F_3'$ (see \eqref{inho.def.p}) and also the terms in the equation for $\nab_3 \nab^i\tb$ (for $i\leq 2$) (see Proposition~\ref{b.eqn}) which contain $\tb$.
We now control each of these terms. We will show below that all of the terms above can be controlled in the $L^1_uL^2(S)$ norm uniformly in $\ub$ using the $\NH$ energy, which is already bounded in Proposition~\ref{N.bd}. 

First, using \eqref{BA.S.222}, the Cauchy--Schwarz inequality and Proposition~\ref{N.bd}, we have,
\begin{equation}\label{g.recover.2}
\begin{split}
&\:\sum_{i_1+i_2\leq 3}\|(1+\nab^{i_1}\tg+\nab^{\min\{i_1,2\}}\tp)\Om_\Ke^2\nab^{i_2}\tg\|_{L^1_uL^2(S)}
\ls \sum_{i\leq 3}\|\Om_\Ke^2\nab^{i}\tg\|_{L^1_uL^2(S)}\\
\ls &\:(\sum_{i\leq 3}\||u|^{\f12+\de}\varpi^N\Om_\Ke^2\nab^{i}\tg\|_{L^2_uL^2(S)}\||u|^{-\f12-\de}\|_{L^2_u})\ls \NH^{\f12}\ls \ep.
\end{split}
\end{equation}
The other terms in the first line of \eqref{g.recover.1} can be estimated in a completely analogous manner, i.e.~we have
\begin{equation}\label{g.recover.3}
\begin{split}
&\sum_{i_1+i_2\leq 3}\|(1+\nab^{i_1}\tg+\nab^{\min\{i_1,2\}}\tp)\Om_\Ke^2\nab^{\min\{i_2,2\}}\tp\|_{L^1_uL^2(S)}\ls  \ep
\end{split}
\end{equation}
and
\begin{equation}\label{g.recover.4}
\begin{split}
&\sum_{i_1+i_2\leq 3}\|(1+\nab^{i_1}\tg+\nab^{\min\{i_1,2\}}\tp)\nab^{i_2}(\tpHb,\widetilde{\omb})\|_{L^1_uL^2(S)}\ls  \ep.
\end{split}
\end{equation}
Notice that we have used the fact that there are no more than two derivatives on $\tp$. This is important because $\NH$ only controls $\nab^3\widetilde{\eta}$ in $L^1_uL^2(S)$ but does not control $\nab^3\widetilde{\etab}$ in the same space. (Instead, $\nab^3\widetilde{\etab}$ can only be bounded in $L^1_{\ub}L^2(S)$.)

We now consider the terms involving $\tb$, i.e.~terms on the second line of \eqref{g.recover.1}. They can also be bounded similarly by first applying \eqref{BA.S.222.lower} and then putting $\nab^{i_2}\tb$ in $L^\i_uL^2(S)$ using H\"older's inequality and Sobolev embedding (Proposition~\ref{Sobolev}):
\begin{equation*}
\begin{split}
&\:\sum_{i_1+i_2+i_3\leq 2}\|(1+\nab^{i_1}(\tg,\tp))(\Om_\Ke^2,\nab^{i_2}\tpHb) \nab^{i_3}\tb\|_{L^1_uL^2(S)}\\
\ls &\:\sum_{i_1+i_2\leq 2}\|(\Om_\Ke^2,\nab^{i_1}\tpHb) \nab^{i_2}\tb\|_{L^1_uL^2(S)}\\
\ls &\:(\|\Om_\Ke^2\|_{L^1_uL^\infty(S)}+\sum_{i_1 \leq 2}\|\nab^{i_1}\tpHb\|_{L^1_uL^2(S)})(\sum_{i_2\leq 2}\|\nab^{i_2}\tb\|_{L^\infty_u L^2(S)})\ls  \ep,
\end{split}
\end{equation*}
where we have used Proposition~\ref{Om.Kerr.bounds} to estimate the integral of $\Om_\Ke^2$. We now conclude the proof of the proposition by taking supremum in $\ub$.
\end{proof}
We then retrieve the bootstrap assumption for $\tpH$ and its angular covariant derivatives. These components also obey a $\nab_3$ equation. Unlike for $\nab^i\tp$, the $L^\i_uL^\i_{\ub}L^2(S)$ bounds for $\nab^i\tpH$ have a weight of $\Om_\Ke^2$.
\begin{proposition}\label{tpH.recover}
For $i\leq 2$, $\nab^i\tpH$ satisfies the following $L^\infty_uL^\infty_{\ub}L^2(S)$ estimate:
$$\sum_{i\leq 2}\|\Om_\Ke^2\nab^i\tpH\|_{L^\infty_uL^\infty_{\ub}L^2(S)}^2\ls \ep^2.$$
\end{proposition}
\begin{proof}
In order to apply Proposition~\ref{transport.2}, we first need to control $\nab_3(\Om_\Ke^2\nab^i\tpH)$ for $i\leq 2$. By Proposition~\ref{tpH.eqn}, for $i\leq 2$, $\nab^i\tpH$ obeys the following reduced schematic equation:
\begin{equation*}
\begin{split}
\nab_3(\Om^2_\Ke\nab^i\tpH) \eqrs\mathcal F_1',
\end{split}
\end{equation*}
where by \eqref{inho.def.p}, $\mathcal F_1'$ contains the terms
\begin{equation}\label{tpH.recover.1}
\begin{split}
&\Om_\Ke^2\nab^3(\tg,\widetilde{\eta})+\Om_\Ke^2\nab^2\tK+\Om_\Ke^2\left(\sum_{i_1+i_2\leq 2}(1+\nab^{i_1}(\tg,\tp))(\nab^{i_2}(\tg,\tp,\tpHb,\widetilde{\omb})+\Om_\Ke^2\nab^{i_2}\tpH)\right) \\
&+\Om_\Ke^2\sum_{i_1+i_2+i_3\leq 2}(1+\nab^{i_1}(\tg,\tp))\nab^{i_2} (\tpHb,\widetilde{\omb})\nab^{i_3}\tpH.
\end{split}
\end{equation}
Here, we need to take advantage of the fact that we have $\mathcal F_1'$ instead of $\mathcal F_1$, i.e.~firstly the only $\nab^3\tp$ term is the $\nab^3\widetilde{\eta}$ term and secondly that there are no $\nab^3\tpH$ terms:
\begin{itemize}
\item If a $\nab^3\widetilde{\etab}$ term is present, it would be problematic since it cannot be controlled in the $L^1_uL^2(S)$ norm. 
\item Since the only $\nab^i\tpH$ terms have $i\leq 2$, we can use Gr\"onwall's inequality to control them. 
\end{itemize}
We now turn to the estimates for each of the term in $\mathcal F_1'$. First, we focus on the terms that do not contain $\tpH$. Notice that most of these terms were already bounded in the proof of Proposition \ref{g.recover}: more precisely in \eqref{g.recover.2}, \eqref{g.recover.3} and \eqref{g.recover.4}\footnote{The only difference here is that the $(\tpHb,\widetilde{\omb})$ term has an additional favourable weight $\Om_\Ke^2$ compared to \eqref{g.recover.4}, which of course only makes the estimate easier.}. The only terms that where $\tpH$ is absent which are not yet controlled in Proposition \ref{g.recover} can also be bounded similarly by
\begin{equation*}
\begin{split}
\|\Om_\Ke^2\nab^3\widetilde{\eta}\|_{L^1_uL^2(S)}\ls  \ep,\quad \|\Om_\Ke^2\nab^2\tK\|_{L^1_uL^2(S)}\ls  \ep.
\end{split}
\end{equation*}
To achieve this, we simply use the $\NH$ energy together with the Cauchy--Schwarz inequality as in the proof of Proposition \ref{g.recover}.

Now we move on to the two terms where $\tpH$ is present. Notice that the $\NH$ energy does not\footnote{Indeed, $\NH$ only controls $\Om_\Ke^2\nab^i\tpH$ in $L^1_{\ub}L^\i_uL^2(S)$ (for $i\leq 2$).} control $\Om_\Ke^2\nab^i\tpH$ in $L^1_uL^2(S)$. Instead, we will used Gr\"onwall's inequality. More precisely, using \eqref{BA.S.222.lower}, for the term on the first line of \eqref{tpH.recover.1} with $\tpH$, we have
\begin{equation*}
\begin{split}
\sum_{i_1+i_2\leq 2}\int_{-\ub+C_R}^{u}\|\Om_\Ke^4(1+\nab^{i_1}(\tg,\tp))\nab^{i_2}\tpH\|_{L^2(S_{u',\ub})}\,du'
\ls \sum_{i\leq 2}\int_{-\ub+C_R}^{u}\|\Om_\Ke^4\nab^{i}\tpH\|_{L^2(S_{u',\ub})}\,du'
\end{split}
\end{equation*}
and for the term on the second line of \eqref{tpH.recover.1} with $\tpH$, we have
\begin{equation*}
\begin{split}
&\sum_{i_1+i_2+i_3\leq 2}\|\Om_\Ke^2(1+\nab^{i_1}(\tg,\tp))\nab^{i_2} (\tpHb,\widetilde{\omb})\nab^{i_3}\tpH\|_{L^1_uL^2(S)}\\
\ls &\int_{-\ub+C_R}^{u}(\sum_{i_2\leq 2}\|\nab^{i_2}(\tpHb,\widetilde{\omb})\|_{L^2(S_{u',\ub})})(\sum_{i_3\leq 2}\|\Om_\Ke^2\nab^{i_3}\tpH\|_{L^2(S_{u',\ub})})\,du'.
\end{split}
\end{equation*}
Combining the above estimates, using Proposition~\ref{transport.2}, and estimating the initial data contribution by \eqref{data.D.def} and \eqref{small.data}, we have, for every fixed $u$, $\ub$,
\begin{equation}\label{tpH.recover.2}
\begin{split}
&\:\sum_{i\leq 2}\|\Om_\Ke^2\nab^i\tpH\|_{L^2(S_{u,\ub})}\\
\ls & \:\ep+\int_{-\ub+C_R}^{u}(\|\Om_\Ke^2\|_{L^2(S_{u',\ub})}+\sum_{i_2\leq 2}\|\nab^{i_2}(\tpHb,\widetilde{\omb})\|_{L^2(S_{u',\ub})})(\sum_{i_3\leq 2}\|\Om_\Ke^2\nab^{i_3}\tpH\|_{L^2(S_{u',\ub})})\,du'.
\end{split}
\end{equation}
By Proposition~\ref{Om.Kerr.bounds},
$$\int_{-\ub+C_R}^{u}\|\Om_\Ke^2\|_{L^2(S_{u',\ub})}du'\ls 1$$
uniformly in $\ub$, and by the Cauchy--Schwarz inequality and the estimates of $\NH$ in Proposition~\ref{N.bd},
$$\sum_{i\leq 2}\|\nab^{i}(\tpHb,\widetilde{\omb})\|_{L^1_uL^2(S)}\ls \sum_{i\leq 2}\||u|^{\f12+\de}\varpi^N\nab^i(\tpHb,\widetilde{\omb})\|_{L^2_uL^2(S)}\||u|^{-\f 12-\de}\|_{L^2_u}\ls \ep,$$
also uniformly in $\ub$. Therefore, we can conclude by applying Gr\"onwall's inequality to \eqref{tpH.recover.2}.
\end{proof}

The bootstrap assumption for $\widetilde{\eta}$ and its angular covariant derivatives can also be retrieved in a similar manner, using the $\nab_4$ equation instead of a $\nab_3$ equation. 
\begin{proposition}\label{eta.recover}
For $i\leq 2$, $\nab^i\widetilde{\eta}$ satisfies the following $L^\infty_uL^\infty_{\ub}L^2(S)$ estimates:
$$\sum_{i\leq 2}\|\nab^i\widetilde{\eta}\|_{L^\infty_uL^\infty_{\ub}L^2(S)}^2\ls \ep^2.$$
\end{proposition}
\begin{proof}
Fix $u$. In order to use Proposition \ref{transport.2}, we need to control
$$\int_{-u+C_R}^{\ub} \|\Om_\Ke^2\nab_4\nab^i\widetilde{\eta}\|_{L^2(S_{u,\ub'})} d\ub'$$
for $i\leq 2$. By Proposition~\ref{eta.eqn}, for $i\leq 2$, we have 
$$\nab_4\nab^i\widetilde{\eta} \eqrs\mathcal F_2'$$ 
where by \eqref{inho.def.p}, $\mathcal F_2'$ has the following terms:
$$\sum_{i_1+i_2\leq 3}(1+\nab^{i_1}\tg+\nab^{\min\{i_1,2\}}\tp)(\Om^{-2}_\Ke\nab^{i_2}\tb+\nab^{\min\{i_2,2\}}\tp+\nab^{i_2}(\tg,\tpH)).$$
The bounds for these terms are similar to the estimates in Proposition \ref{g.recover} except for having to control the terms in $L^1_{\ub}L^2(S)$ instead of $L^1_uL^2(S)$. Indeed, we can show that every term in $\Om_\Ke^2\mathcal F_2'$ can be bounded by the $\NH$ energy. We treat the first term in detail as an example. Using \eqref{BA.S.222}, the Cauchy--Schwarz inequality, \eqref{varpi.bd} and the estimates for $\NH$ established in Proposition~\ref{N.bd}, we have
\begin{equation}\label{eta.recover.1}
\begin{split}
&\:\sum_{i_1+i_2\leq 3}\|(1+\nab^{i_1}\tg+\nab^{\min\{i_1,2\}}\tp)\nab^{i_2}\tb\|_{L^1_{\ub}L^2(S)}\\
\ls &\: \sum_{i\leq 3}\|\nab^{i}\tb\|_{L^1_{\ub}L^2(S)} \ls \sum_{i\leq 3}\|\ub^{\f12+\de}\varpi^N\nab^{i}\tb\|_{L^2_{\ub}L^2(S)}\|\ub^{-\f12-\de}\|_{L^2_{\ub}}\ls \NH^{\f12}\ls \ep.
\end{split}
\end{equation}
The other terms can be controlled similarly as above:
\begin{equation}\label{eta.recover.2}
\begin{split}
\sum_{i_1+i_2\leq 3}\|(1+\nab^{i_1}\tg+\nab^{\min\{i_1,2\}}\tp)\Om_\Ke^2\nab^{\min\{i_2,2\}}\tp\|_{L^1_{\ub}L^2(S)}
\ls  \ep
\end{split}
\end{equation}
and
\begin{equation}\label{eta.recover.3}
\begin{split}
&\sum_{i_1+i_2\leq 3}\|(1+\nab^{i_1}\tg+\nab^{\min\{i_1,2\}}\tp)\Om_\Ke^2\nab^{i_2}(\tg,\tpH)\|_{L^1_{\ub}L^2(S)}\ls \ep.
\end{split}
\end{equation}
Combining \eqref{eta.recover.1}, \eqref{eta.recover.2} and \eqref{eta.recover.3}, using Proposition \ref{transport.2}, estimating the initial data contribution by \eqref{data.D.def} and \eqref{small.data}, and taking supremum in $u$, we obtain the desired conclusion.
\end{proof}

Finally, we show the estimates for $\nab^i(\tpHb,\widetilde{\omb})$ in $L^\i_uL^\i_{\ub}L^2(S)$ for $i\leq 2$ to obtain all the necessary bounds for $\NS$, which will allow us to recover the bootstrap assumptions.
\begin{proposition}\label{tpHb.recover}
For $i\leq 2$, $\nab^i\tpHb$ and $\nab^i\widetilde{\omb}$ satisfy the following $L^\i_uL^\i_{\ub}L^2(S)$ bounds
$$\sum_{i\leq 2}\|\nab^i(\tpHb,\widetilde{\omb})\|_{L^\i_uL^\i_{\ub}L^2(S)}^2\ls \ep^2.$$
\end{proposition}
\begin{proof}
In order to apply Proposition~\ref{transport.2}, we first need to control
$$\int_{-u+C_R}^{\ub} \|\Om_\Ke^2\nab_4\nab^i(\tpHb,\widetilde{\omb})\|_{L^2(S_{u,\ub'})} d\ub'$$
for $i\leq 2$. By Proposition~\ref{tpHb.eqn}, for $i\leq 2$, we have 
$$\nab_4\nab^i(\tpHb,\widetilde{\omb})\eqrs\mathcal F_4'$$ 
where by \eqref{inho.def.p}, $\mathcal F_4'$ has the following terms:
\begin{equation*}
\begin{split}
&\nab^3(\tg,\widetilde{\etab})+\nab^2\tK+\sum_{i_1+i_2\leq 2}(1+\nab^{i_1}(\tg,\tp))(\nab^{i_2}(\tg,\tb,\tp,\tpHb,\widetilde{\omb})+\Om_\Ke^2\nab^{i_2}\tpH) \\
&+\sum_{i_1+i_2+i_3\leq 2}(1+\nab^{i_1}\tg+\nab^{\min\{i_1,2\}}\tp)\nab^{i_2} (\tpHb,\widetilde{\omb})(\nab^{i_3}\tpH+\Om_\Ke^{-2}\nab^{i_3}\tb).
\end{split}
\end{equation*}
Now, notice that all of the terms that do not contain $\nab^i(\tpHb,\widetilde{\omb})$, $\nab^3\widetilde{\etab}$ and $\nab^2\tK$ have already been controlled in \eqref{eta.recover.1}, \eqref{eta.recover.2} and \eqref{eta.recover.3} in the proof of Proposition \ref{eta.recover}. The terms $\nab^3\widetilde{\etab}$ and $\nab^2\tK$ can be controlled similarly as in Proposition~\ref{eta.recover} using the bound on $\NH$ in Proposition~\ref{N.bd}. More precisely, we obtain
$$\|\Om_\Ke^2\nab^3\widetilde{\etab}\|_{L^1_{\ub}L^2(S)}\ls \ep,\quad \|\Om_\Ke^2\nab^2\tK\|_{L^1_{\ub}L^2(S)}\ls \ep.$$
We now turn to those terms containing $(\tpHb,\widetilde{\omb})$. Like the $\tpH$ terms in the proof of Proposition~\ref{tpH.recover}, these terms are controlled using Gr\"onwall's inequality. More precisely, we use \eqref{BA.S.lower} and H\"older's inequality to bound these terms by
\begin{equation*}
\begin{split}
&\sum_{i_1+i_2\leq 2}\int_{-u+C_R}^{\ub} \|\Om_\Ke^2(1+\nab^{i_1}(\tg,\tp))\nab^{i_2}(\tpHb,\widetilde{\omb})\|_{L^2(S_{u,\ub'})} \,d\ub'\\
\ls &\sum_{i\leq 2}\int_{-u+C_R}^{\ub} \|\Om_\Ke^2\|_{L^\i(S_{u,\ub'})}\|\nab^{i}(\tpHb,\widetilde{\omb})\|_{L^2(S_{u,\ub'})} \,d\ub'
\end{split}
\end{equation*}
and
\begin{equation*}
\begin{split}
&\sum_{i_1+i_2+i_3\leq 2}\int_{-u+C_R}^{\ub} \|\Om_\Ke^2(1+\nab^{i_1}\tg+\nab^{\min\{i_1,2\}}\tp)\nab^{i_2} (\tpHb,\widetilde{\omb})(\nab^{i_3}\tpH+\Om_\Ke^{-2}\nab^{i_3}\tb)\|_{L^2(S_{u,\ub'})} \,d\ub'\\
\ls &\int_{-u+C_R}^{\ub} (\sum_{i_1\leq 2}(\|\Om_\Ke^2\nab^{i_1}\tpH\|_{L^2(S_{u,\ub'})}+\|\nab^{i_1}\tb\|_{L^2(S_{u,\ub'})}))(\sum_{i_2\leq 2}\|\nab^{i_2}(\tpHb,\widetilde{\omb})\|_{L^2(S_{u,\ub'})}) \,d\ub'.
\end{split}
\end{equation*}
Therefore, using the above estimates and Proposition~\ref{transport.2}, and estimating the initial data contribution by \eqref{data.D.def} and \eqref{small.data}, we have, for every fixed $u$, $\ub$,
\begin{equation}\label{tpHb.recover.2}
\begin{split}
&\sum_{i\leq 2}\|\nab^i(\tpHb,\widetilde{\omb})\|_{L^2(S_{u,\ub})}\\
\ls &\:\ep+\sum_{i\leq 2}\int_{-u+C_R}^{\ub} \|\Om_\Ke^2\|_{L^\i(S_{u,\ub'})}\|\nab^{i}(\tpHb,\widetilde{\omb})\|_{L^2(S_{u,\ub'})}\, d\ub'\\
&+\int_{-u+C_R}^{\ub} (\sum_{i_1\leq 2}(\|\Om_\Ke^2\nab^{i_1}\tpH\|_{L^2(S_{u,\ub'})}+\|\nab^{i_1}\tb\|_{L^2(S_{u,\ub'})}))(\sum_{i_2\leq 2}\|\nab^{i_2}(\tpHb,\widetilde{\omb})\|_{L^2(S_{u,\ub'})}) \,d\ub'.
\end{split}
\end{equation}
Notice now that by Proposition~\ref{Om.Kerr.bounds},
$$\int_{-u+C_R}^{\ub} \|\Om_\Ke^2\|_{L^\i(S_{u,\ub'})} \,d\ub'\ls 1,$$
and by the estimates of $\NH$ in Proposition~\ref{N.bd} and the Cauchy--Schwarz inequality,
$$\int_{-u+C_R}^{\ub} (\sum_{i_1\leq 2}(\|\Om_\Ke^2\nab^{i_1}\tpH\|_{L^2(S_{u,\ub'})}+\|\nab^{i_1}\tb\|_{L^2(S_{u,\ub'})})) \,d\ub'\ls \ep,$$
where both bounds hold uniformly in $u$ and $\ub$. The conclusion of the proposition follows after applying Gr\"onwall's inequality to \eqref{tpHb.recover.2} and taking supremum in $u$ and $\ub$.
\end{proof}

\subsection{Proof of Theorem~\ref{main.quantitative.thm}}\label{sec.pf.of.main.quantitative.thm}

\begin{proof}[Proof of Theorem~\ref{main.quantitative.thm}]
The estimates for $\NH$ and $\NI$ were already obtained in Proposition~\ref{N.bd} in Section~\ref{sec.end.of.error}. It thus remains to prove the estimates for $\NS$. Recalling \eqref{NS.def} and combining the estimates in Propositions~\ref{integral.recover},\ref{g.recover}, \ref{tpH.recover}, \ref{eta.recover} and \ref{tpHb.recover}, we have
$$\NS\ls \ep^2.$$
Here, the constant may depend on $N$ but recall that it has been fixed and depends only on $a$, $M$, $\de$ and $C_R$ (see Section~\ref{sec.end.of.error}).
Therefore, by our convention of $\ls$, the implicit constant depends only on $a$, $M$, $\de$ and $C_R$. This concludes the proof of Theorem~\ref{main.quantitative.thm}.\qedhere
\end{proof}

\section{Propagation of higher regularity}\label{sec:HR}

After having proven Theorem~\ref{main.quantitative.thm} in the last section, we return to the proof of Theorem~\ref{aprioriestimates}. Our next goal will be to prove Theorem~\ref{I.final} in which we establish the existence of a solution $(\mathcal U_\infty,g)$. Meanwhile, we first have to propagate higher regularity (cf.~Section~\ref{allapolla}). This is achieved in the present section.

More precisely, we use the estimates derived in Theorem~\ref{main.quantitative.thm} to propagate estimates for any \emph{higher} derivatives of the Ricci coefficients and renormalised curvature components. As is standard, since we have already closed the lower order estimates, the higher order estimates are essentially linear in the unknown. These higher order terms can therefore be controlled using a Gr\"onwall-type argument. The only slightly subtle issue is that because we are only controlling the renormalised curvature components $(\beta,K,\sigmac,\betab)$, it is not a priori clear that this is sufficient to bound, say, all the spacetime curvature components. Nonetheless, we will show that restricting to in a finite $\ub$ region, precisely because the Ricci coefficients and the renormalised curvature components form a closed system of equations, we can differentiate that system to obtain all higher order estimates. A similar propagation of higher regularity argument can be found in \cite{LR}.

The setting of this section is as follows. We will consider the domain $\mathcal U_{\ub_f} = \mathcal W_{\ub_f}\times \mathbb S^2$, where, as before, $\mathcal W_{\ub_f}$ is the closed-to-the-past, open-to-the-future triangular domain in the $(u,\ub)$-plane given by
\begin{equation}\label{def.W.ub.f}
\mathcal W_{\ub_f} = \{(u,\ub): \ub \in [-u+C_R,\ub_f),\, u\in [-\ub+C_R,u_f)\}\subset \mathbb R^2.
\end{equation}
From now on until the end of this section, assume that we have a smooth solution to the Einstein vacuum equations \eqref{INTROvace} in $\mathcal U_{\ub_f}$ in the double null coordinate system, which achieves the given data in Theorem~\ref{aprioriestimates} and assume that the estimate \eqref{BA} holds. By Theorem~\ref{main.quantitative.thm}, this implies the estimate $\NI,\,\NH,\,\NS\ls \ep^2$.

Our goal in this section is to prove that all derivatives of $\psi$, $\psi_H$, $\psi_{\Hb}$, $\omb$ and $K$ are bounded in $\mathcal U_{\ub_f}$. Of course all the higher derivative estimates may degenerate as $\ub_f\to \infty$, which is why we control the renormalised quantities instead in the first place. Therefore, unlike in the previous sections, all estimates in this section \textbf{\emph{are allowed to depend on $\ub_f$, in addition to $M$, $a$, $\de$ and $C_R$}}. We will emphasise this by adding a subscript ${ }_{\ub_f}$ in all the constants. (These constants are allowed to differ from line to line.) These estimates, though they degenerate as $\ub_f\to \infty$, will be sufficient for showing that the metric remains smooth in $\mathcal U_{\ub_f} = W_{\ub_f}\times \mathbb S^2$ and extends to a smooth metric on the manifold-with-corner $\overline{W_{\ub_f}}\times \mathbb S^2$ (cf.~Proposition~\ref{prop:smooth.up.to.bdry}). We will then use this fact to conclude the bootstrap argument in the next section.

We begin with the following simple calculus lemma, which can be thought of as a $2$-dimensional analogue of the Gr\"onwall's inequality:
\begin{lemma}\label{2D.Gronwall}
Let $f_1,f_2:\mathfrak D\doteq  \{(x,y): 0\leq -x\leq y\leq 1\}\to \mathbb R_{\geq 0}$ be continuous functions. Suppose there exist constants $A>0$ and $B>0$ such that for every $(x,y)\in \mathfrak D$, the following holds:
\begin{equation}\label{2D.Gronwall.main}
\int_{-y}^x f_1^2(x',y)\, dx'+\int_{-x}^y f_2^2(x,y')\, dy'\leq A + B \int_{-x}^y \int_{-y'}^{x} f_1^2(x',y')\,dx'\,dy'+ B\int_{-y}^x \int_{-x'}^{y} f_2^2(x',y')\,dy'\,dx'.
\end{equation}
Then, the following estimate holds for every $(x,y)\in \mathfrak D$:
$$\int_{-y}^x f_1^2(x',y)\, dx'+\int_{-x}^y f_2^2(x,y')\, dy'\leq 4Ae^{8B}.$$
\end{lemma}
\begin{proof}
Consider the subset $\mathfrak S\subset \mathfrak D$ which consists of all points $(x,y)\in \mathfrak D$ such that the following estimate holds for all $(x',y')$ such that $0\leq -x'\leq y'\leq 1$, $y'\leq y$ and $x'\leq x$:
\begin{equation}\label{2D.Gronwall.BA}
\int_{-y'}^{x'} f_1^2(x'',y')\, dx''\leq 2Ae^{8B(y'+x')},\quad \int_{-x'}^{y'} f_2^2(x',y'')\, dy''\leq 2Ae^{8B(y'+x')}.
\end{equation}
It is obvious that $\mathfrak S$ is non-empty and closed. We will proceed to show that $\mathfrak S$ is open. First, take a point $(x,y)\in \mathfrak S$ and estimate
$$\int_{-x}^y \int_{-y'}^{x} f_1^2(x',y')\,dx'\,dy'\leq 2A\int_{-x}^y e^{8B(y'+x)} dy' = \f{A}{4B} (e^{8B(y+x)} - 1)\leq \f {A}{4B} e^{8B(y+x)},$$
and 
$$\int_{-y}^x \int_{-x'}^{y} f_2^2(x',y')\,dy'\,dx'\leq 2A \int_{-y}^x e^{8B(y+x')} dx' = \f {A}{4B} (e^{8B(y+x)} - 1)\leq \f {A}{4B} e^{8B(y+x)}.$$
By \eqref{2D.Gronwall.main}, we thus obtain 
$$\int_{-y}^x f_1^2(x',y)\, dx'+\int_{-x}^y f_2^2(x,y')\, dy'\leq A + \f {AB}{2B} e^{B(y+x)}\leq \f {3A}2e^{B(y+x)}.$$
It therefore follows from continuity that for every $(x,y)\in \mathfrak S$, there is an open neighbourhood $U_{x,y}$ of $(x,y)$ such that for every $(x',y')\in U_{x,y}\cap \mathfrak D$,
$$\int_{-y'}^{x'} f_1^2(x'',y')\, dx''\leq 2Ae^{8B(y'+x')},\quad \int_{-x'}^{y'} f_2^2(x',y'')\, dy''\leq 2Ae^{8B(y'+x')}.$$
Now take $(x_*,y_*)\in \mathfrak S$. By definition of $\mathfrak S$, the set $\mathfrak S'\doteq \{(x,y): 0\leq -x\leq y\leq 1,\, x\leq x_*,\, y\leq y_*\}\subset \mathfrak S$. By the above discussions and the compactness of $\mathfrak S'$, it follows that there is an open neighbourhood $U$ of $\mathfrak S'$ such that for every
$(x',y')\in U\cap \mathfrak D$,
$$\int_{-y'}^{x'} f_1^2(x'',y')\, dx''\leq 2Ae^{8B(y'+x')},\quad \int_{-x'}^{y'} f_2^2(x',y'')\, dy''\leq 2Ae^{8B(y'+x')}.$$
In particular, this implies that there is an open neighbourhood $U_*$ of $(x_*,y_*)$ such that $U_*\subset \mathfrak S$. This implies that $\mathfrak S$ is open. Since $\mathfrak S$ is a non-empty, open and closed subset of $\mathfrak D$, by the connectedness of $\mathfrak D$, it follows that $\mathfrak S = \mathfrak D$. In particular, the desired estimate holds.
\end{proof}
We now prove higher order derivative bounds for $\psi_H$, $\psi$, $\psi_{\Hb}$, $\omb$ and $K$. We first bound the higher $\nab$ derivatives of these quantities:
\begin{proposition}\label{prop:HO.renormalised}
All higher angular derivatives of the Ricci coefficients $\psi_H$, $\psi$, $\psi_{\Hb}$, $\omb$ and the Gauss curvature $K$ are bounded as follows: For every $I\geq 0$,
\begin{equation}\label{main.higher.order}
\begin{split}
\sup_{(u,\ub)\in \mathcal W_{\ub_f}}\sum_{i\leq I}\|(\nab^i(\psi,\psi_H,\psi_{\Hb},\omb),\nab^{i-1} K)\|_{L^2(S_{u,\ub})}\leq C_{I,\ub_f},
\end{split}
\end{equation}
where $C_{I,\ub_f} > 0$ are constants depending on $I$ and $\ub_f$ (in addition to the profile of the initial data).
\end{proposition}
\begin{proof}
The key point is that we have a closed system of equations (cf.~discussions in Section~\ref{seceqn}) on the metric components, the Ricci coefficients, and the renormalised curvature components. In particular, in this system of equations, the spacetime curvature components $\a$ and $\ab$ do \underline{not} appear. We can therefore repeat the argument in Sections~\ref{sec.int.est}-\ref{recover.bootstrap} for the higher derivative estimates.

We will inductively prove both \eqref{main.higher.order} together with an $L^2$ energy type estimate. The base case, which consists of the following estimates, has been handled by Theorem~\ref{main.quantitative.thm}. 
\begin{equation*}
\begin{split}
&\sup_{(u,\ub)\in \mathcal W_{\ub_f}}\sum_{i\leq 3}\int_{-u+C_R}^{\ub}\|(\nab^i (\psi_H,\psi), \nab^{i-1}K)\|_{L^2(S_{u,\ub'})}^2\, d\ub' \\
&\qquad + \sup_{(u,\ub)\in \mathcal W_{\ub_f}}\sum_{i\leq 3}\int_{-\ub+C_R}^{u}\|(\nab^i (\psi_{\Hb},\omb,\psi),\nab^{i-1}K)\|_{L^2(S_{u',\ub})}^2\, du' \leq C_{\ub_f}
\end{split}
\end{equation*}
and
\begin{equation*}
\sup_{(u,\ub)\in \mathcal W_{\ub_f}}\sum_{i\leq 2}\|(\nab^i (\psi,\psi_H,\psi_{\Hb},\omb),\nab^{i-1}K)\|_{L^2(S_{u,\ub})}\leq C_{\ub_f}. 
\end{equation*}
By induction, we assume that there exists $I\geq 4$ such that the following two estimates both hold for all $I'<I$:
\begin{equation}\label{prop.reg.induct.1}
\begin{split}
&\sup_{(u,\ub)\in \mathcal W_{\ub_f}}\sum_{i\leq I'}\int_{-u+C_R}^{\ub}\|(\nab^i (\psi_H,\psi), \nab^{i-1}K)\|_{L^2(S_{u,\ub'})}^2\, d\ub' \\
&\qquad + \sup_{(u,\ub)\in \mathcal W_{\ub_f}}\sum_{i\leq I'}\int_{-\ub+C_R}^{u}\|(\nab^i (\psi_{\Hb},\omb,\psi),\nab^{i-1}K)\|_{L^2(S_{u',\ub})}^2\, du' \leq C_{I',\ub_f}
\end{split}
\end{equation}
and
\begin{equation}\label{prop.reg.induct.2}
\sup_{(u,\ub)\in \mathcal W_{\ub_f}}\sum_{i\leq I'-1}\|(\nab^i (\psi,\psi_H,\psi_{\Hb},\omb),\nab^{i-1}K)\|_{L^2(S_{u,\ub})}\leq C_{I',\ub_f}. 
\end{equation}
Below, we will prove \eqref{prop.reg.induct.1} and \eqref{prop.reg.induct.2} for $I'=I$.

\textbf{Energy and elliptic estimates.} We apply energy estimates as in Section~\ref{sec.energy} and elliptic estimates as in Section~\ref{sec.elliptic}. Note that the ``initial data'' term is bounded since the initial data are smooth and the corresponding term is an integral over a compact region. Moreover, using the induction hypothesis \eqref{prop.reg.induct.2} and Sobolev embedding (Proposition~\ref{Sobolev}), all the lower order terms can be controlled. The estimate is therefore linear in the highest order term. Hence, for any $(u,\ub)\in \mathcal W_{\ub_f}$, the following estimate holds:
\begin{equation}\label{prop.reg.main}
\begin{split}
&\sum_{i\leq I}\int_{-u+C_R}^{\ub}\|(\nab^i (\psi_H,\psi), \nab^{i-1}K)\|_{L^2(S_{u,\ub'})}^2\, d\ub' + \sum_{i\leq I}\int_{-\ub+C_R}^{u}\|(\nab^i (\psi_{\Hb},\omb,\psi),\nab^{i-1}K)\|_{L^2(S_{u',\ub})}^2\, du'\\
\leq & C_{I,\ub_f} + C_{I,\ub_f} \int_{-\ub+C_R}^{u_f} \int_{-u'+C_R}^{\ub} (\sum_{i\leq I}\|(\nab^{i_2} (\psi_{H},\psi), \nab^{i-1} K)\|_{L^2(S_{u',\ub'})})^2\, d\ub' \,du' \\
&+ C_{I,\ub_f} \int_{-u_f+C_R}^{\ub} \int_{-\ub'+C_R}^{u_f} (\sum_{i\leq I}\|(\nab^{i} (\psi_{\Hb},\omb,\psi),\nab^{i-1}K)\|_{L^2(S_{u',\ub'})})^2\, du'\, d\ub'.
\end{split}
\end{equation}
Given \eqref{prop.reg.main}, an application of Lemma~\ref{2D.Gronwall} then gives the estimate:
\begin{equation}\label{prop.reg.induct.improve}
\begin{split}
&\sup_{(u,\ub)\in \mathcal W_{\ub_f}}\sum_{i\leq I}\int_{-u+C_R}^{\ub}\|\nab^i (\psi_H,\psi), \nab^{i-1}K\|_{L^2(S_{u,\ub'})}^2\, d\ub' \\
&\qquad+ \sup_{(u,\ub)\in \mathcal W_{\ub_f}}\sum_{i\leq I}\int_{-\ub+C_R}^{u}\|(\nab^i (\psi_{\Hb},\omb,\psi),\nab^{i-1}K)\|_{L^2(S_{u',\ub})}^2\, du' \leq C_{I,\ub_f},
\end{split}
\end{equation}
which is the estimate \eqref{prop.reg.induct.1} for $I' = I$.

\textbf{Estimates using transport equations.} We first consider $\eta$ and $K$. For each $i\leq I-1$, commute the $\nab_4\eta$ equation in \eqref{null.str2} and the $\nab_4 K$ equation in \eqref{eq:null.Bianchi2} with $\nab^i$ and $\nab^{i-1}$ respectively. Since we are on a compact region of $u$ and $\ub$, and we have obtained $L^1_{\ub}L^{\i}_uL^{\i}(S)$ estimates for $\trch$, we can use the identity in Proposition~\ref{transport} and Gr\"onwall's inequality to obtain that for $(u,\ub)\in \mathcal W_{\ub_f}$,
\begin{equation*}
\begin{split}
&\sum_{i\leq I-1}\|(\nab^i \eta,\nab^{i-1}K)\|_{L^2(S_{u,\ub})}\\
\ls  &\sum_{i\leq I-1}\|(\nab^i \eta,\nab^{i-1}K)\|_{L^2(S_{u,-u+C_R})} + \sum_{i\leq I-1}\int_{-u+C_R}^{\ub} \|(\nab_4\nab^i \eta,\nab_4\nab^{i-1}K)\|_{L^2(S_{u,\ub'})} \,d\ub'.
\end{split}
\end{equation*}
The initial data term, i.e.~the term on $S_{u,-u+C_R}$ is bounded (uniformly in $u$), since the initial data are smooth and $u$ varies over a compact set. To control $\nab_4\nab^i \eta$ and $\nab_4\nab^{i-1}K$, we simply note the following features of these equations:
\begin{itemize}
\item On the right hand side of the equations $\nab_4\nab^i \eta$ and $\nab_4\nab^{i-1}K$, there are only $\psi_H$, $\psi$ and $K$ terms (and their derivatives), i.e.~there is no appearance of $\psi_{\Hb}$.
\item In these equations, there are at most $I$ $\nab$ derivatives on $\psi_H$ and $\psi$, and at most $I-1$ $\nab$ derivatives on $K$.
\item Notice that for any nonlinear terms, using the induction hypothesis \eqref{prop.reg.induct.2} and Sobolev embedding (Proposition~\ref{Sobolev}), we can control up to $I-2$ $\nab$ derivatives on $\psi_H$ and $\psi$ and up to $I-3$ $\nab$ derivatives on $K$.
\end{itemize}
Combining these observations, we have
$$\sum_{i\leq I-1}\|(\nab^i \eta,\nab^{i-1}K)\|_{L^2(S_{u,\ub})}\leq C_{I,\ub_f} + C_{I,\ub_f} \int_{-u+C_R}^{\ub} (\sum_{i\leq I}\|\nab^i (\psi_H,\psi), \nab^{i-1}K\|_{L^2(S_{u,\ub'})})\, d\ub'.$$
Since $\ub$ only varies over a finite interval, we can control the last term by Cauchy--Schwarz inequality and the estimate \eqref{prop.reg.induct.improve} that we just proved above. This gives the following uniform estimate:
\begin{equation}\label{prop.reg.psi.1}
\begin{split}
\sup_{(u,\ub)\in \mathcal W_{\ub_f}}\sum_{i\leq I-1}\|(\nab^i \eta,\nab^{i-1}K)\|_{L^2(S_{u,\ub})}\leq &C_{I,\ub_f}.
\end{split}
\end{equation}
In a completely analogous manner, expect for using the $\nab_3$ equation (instead of $\nab_4$ equation) for $\nab^i\etab$, we obtain
\begin{equation}\label{prop.reg.psi.2}
\sup_{(u,\ub)\in \mathcal W_{\ub_f}}\sum_{i\leq I-1}\|\nab^i \etab\|_{L^2(S_{u,\ub})}\leq C_{I,\ub_f}.
\end{equation}
Combining \eqref{prop.reg.psi.1} and \eqref{prop.reg.psi.2}, we thus obtain
\begin{equation}\label{prop.reg.psi}
\sup_{(u,\ub)\in \mathcal W_{\ub_f}}\sum_{i\leq I-1}\|(\nab^i \psi,\nab^{i-1}K)\|_{L^2(S_{u,\ub})}\leq C_{I,\ub_f}.
\end{equation}
For $\psi_H$, $\psi_{\Hb}$ and $\omb$, we need to be slightly more careful. This is because in the $\nab_3\psi_H$ equation, we have terms that have a $\psi_H$ (for instance a $\psi_H\psi_{\Hb}$ term). In particular, we need to control $\int_{-\ub+C_R}^u \|\nab^{I-1} \psi_H\|_{L^2(S_{u',\ub})}\, du'$. Our estimate \eqref{prop.reg.induct.improve} does \underline{not} give a bound of the $L^{\i}_{\ub}L^2_{u}L^2(S)$ norm for $\nab^{I-1} \psi_H$ (since \eqref{prop.reg.induct.improve} only bounds its integral in $\ub$) and our induction hypothesis \eqref{prop.reg.induct.2} also does not control this (since \eqref{prop.reg.induct.2} only bounds up to $I-2$ derivatives of $\psi_H$). Instead, we will control this using Gr\"onwall's inequality. More precisely, consider the equation for $\nab_3\nab^i\psi_H$ for $i\leq I-1$, we see that all of $\psi_H$, $\psi_{\Hb}$, $\psi$, $\omb$ and $K$. $\psi$ has up to $I$ $\nab$ derivatives, $\psi_H$, $\psi_{\Hb}$, $\omb$ and $K$ have up to $I-1$ $\nab$ derivatives. Hence, using \eqref{prop.reg.induct.2} and Sobolev embedding (Proposition~\ref{Sobolev}), we have, for $(u,\ub)\in \mathcal W_{\ub_f}$,
\begin{equation*}
\begin{split}
\sum_{i\leq I-1}\|\nab^i \psi_H\|_{L^2(S_{u,\ub})}\leq &C_{I,\ub_f} +C_{I,\ub_f}\int_{-\ub+C_R}^u (\sum_{i\leq I}\|(\nab^{i}(\psi_{\Hb},\omb,\psi),\nab^{i-1}K)\|_{L^2(S_{u',\ub})})\, du'\\
&+C_{I,\ub_f} \int_{-\ub+C_R}^u (\sum_{i\leq I-1}\|\nab^{i}\psi_H\|_{L^2(S_{u',\ub})})\, du'\\
\leq & C_{I,\ub_f} (u+\ub-C_R)^{\f 12}e^{C_{I,\ub_f} (u+\ub-C_R)}\sum_{i\leq I}\|(\nab^{i}(\psi_{\Hb},\omb,\psi),\nab^{i-1}K)\|_{L^{\i}_{\ub}L^2_uL^2(S)},
\end{split}
\end{equation*}
where in the last line we have used Gr\"onwall's inequality and Cauchy--Schwarz inequality. Using \eqref{prop.reg.induct.improve}, this implies
\begin{equation}\label{prop.reg.psiH}
\sup_{(u,\ub)\in \mathcal W_{\ub_f}} \sum_{i\leq I-1}\|\nab^i \psi_H\|_{L^2(S_{u,\ub})}\leq C_{I,\ub_f}.
\end{equation}
Moreover, using the $\nab_4\trchb$, $\nab_4\chibh$ equations in \eqref{null.str1} and the $\nab_4\omb$ equation in \eqref{null.str2}, and arguing similarly as for $\psi_H$, we also have the following estimate for $(u,\ub)\in \mathcal W_{\ub_f}$:
\begin{equation*}
\begin{split}
 \sum_{i\leq I-1}\|\nab^i (\psi_{\Hb},\omb)\|_{L^2(S_{u,\ub})}\leq &C_{I,\ub_f} +C_{I,\ub_f}\int_{-u+C_R}^{\ub} (\sum_{i\leq I}\|(\nab^{i}(\psi_H,\psi),\nab^{i-1}K)\|_{L^2(S_{u,\ub'})})\, d\ub'\\
&+C_{I,\ub_f} \int_{-u+C_R}^{\ub} (\sum_{i\leq I-1}\|\nab^{i}(\psi_{\Hb},\omb)\|_{L^2(S_{u,\ub'})})\, d\ub'\\
\leq & C_{I,\ub_f} (u+\ub-C_R)^{\f 12}e^{C_{I,\ub_f} (u+\ub-C_R)}\sum_{i\leq I}\|(\nab^{i}(\psi_H,\psi),\nab^{i-1}K)\|_{L^{\i}_uL^2_{\ub}L^2(S)},
\end{split}
\end{equation*}
where again we used Gr\"onwall's inequality and Cauchy--Schwarz inequality. Using \eqref{prop.reg.induct.improve}, we thus obtain
\begin{equation}\label{prop.reg.psiHb}
\sup_{(u,\ub)\in \mathcal W_{\ub_f}} \sum_{i\leq I-1}\|\nab^i (\psi_{\Hb},\omb)\|_{L^2(S_{u,\ub})}\leq C_{I,\ub_f}.
\end{equation}
Combining \eqref{prop.reg.psi}, \eqref{prop.reg.psiH} and \eqref{prop.reg.psiHb}, we have thus proven the analogue of \eqref{prop.reg.induct.2} for $I'=I$.
In particular, this concludes the induction argument and the proof of the proposition.
\end{proof}

A similar argument proves higher order estimates for \emph{all} derivatives of $\psi_H$, $\psi$, $\psi_{\Hb}$, $\omb$ and $K$
\begin{proposition}\label{higher.order.34}
For every $I, J, L\geq 0$
\begin{equation}\label{main.higher.order.34}
\begin{split}
&\sup_{(u,\ub)\in \mathcal W_{\ub_f}}\sum_{i\leq I,\,j\leq J, \, \ell\leq L} \|\nab^i \nab_3^j \nab_4^{\ell} (\psi, \psi_H,\psi_{\Hb},\omb), \nab^{i-1}\nab_3^j \nab_4^{\ell} K \|_{L^2(S_{u,\ub})} \leq C_{I, J, L,\ub_f},
\end{split}
\end{equation}
where $C_{I, J, L,\ub_f}>0$ are constants depending on $I$, $J$, $L$ and $\ub_f$ (in addition to the profile of the initial data).
\end{proposition}
\begin{proof}
Proposition~\ref{prop:HO.renormalised} implies the desired estimates in the case $J=L=0$. We prove the proposition by induction. First, we show how to induct in $I$ for $J$ and $L$ fixed. Then, we show how to induct in $J$ or $L$.

\textbf{Inducting on $\nab$.} First, we induct on the number of $\nab$ derivatives for fixed $J,L\geq 0$ (but not both $=0$, as that case has been handled in Proposition~\ref{prop:HO.renormalised}). More precisely, let $J,L\geq 0$ be fixed (with $J>0$ or $L>0$). We will make four inductions hypotheses. The first two concern the case where $j<J$ or $\ell<L$. We assume that for \underline{all} $I'\geq 0$,
\begin{equation}\label{prop.reg.34.lower.induct.1}
\sup_{(u,\ub)\in \mathcal W_{\ub_f}}\sum_{i\leq I',\,j< J, \, \ell\leq L}\|(\nab^i\nab_3^j \nab_4^{\ell} (\psi,\psi_H,\psi_{\Hb},\omb),\nab^{i-1}\nab_3^j \nab_4^{\ell} K)\|_{L^2(S_{u,\ub})}\leq C_{I',J,L,\ub_f}
\end{equation}
and
\begin{equation}\label{prop.reg.34.lower.induct.2}
\sup_{(u,\ub)\in \mathcal W_{\ub_f}}\sum_{i\leq I',\,j\leq J, \, \ell< L}\|(\nab^i\nab_3^j \nab_4^{\ell} (\psi,\psi_H,\psi_{\Hb},\omb),\nab^{i-1}\nab_3^j \nab_4^{\ell} K)\|_{L^2(S_{u,\ub})}\leq C_{I',J,L,\ub_f}.
\end{equation}
Moreover, we assume that the following two estimates hold for some $I\geq 1$:
\begin{equation}\label{prop.reg.34.induct.1}
\begin{split}
&\sup_{(u,\ub)\in \mathcal W_{\ub_f}}\sum_{i\leq I-1,\,j= J, \, \ell= L}\int_{-u+C_R}^{\ub}\|\nab^i \nab_3^j \nab_4^{\ell} (\psi_H,\psi), \nab^{i-1}\nab_3^j \nab_4^{\ell} K\|_{L^2(S_{u,\ub'})}^2\, d\ub' \\
&\qquad + \sup_{(u,\ub)\in \mathcal W_{\ub_f}}\sum_{i\leq I-1,\,j= J, \, \ell= L}\int_{-\ub+C_R}^{u}\|(\nab^i \nab_3^j \nab_4^{\ell}(\psi_{\Hb},\omb,\psi),\nab^{i-1}\nab_3^j \nab_4^{\ell} K)\|_{L^2(S_{u',\ub})}^2\, du' \leq C_{I-1, J, L,\ub_f}
\end{split}
\end{equation}
and
\begin{equation}\label{prop.reg.34.induct.2}
\sup_{(u,\ub)\in \mathcal W_{\ub_f}}\sum_{i\leq I-2,\,j= J, \, \ell= L}\|(\nab^i\nab_3^j \nab_4^{\ell} (\psi,\psi_H,\psi_{\Hb},\omb),\nab^{i-1}\nab_3^j \nab_4^{\ell} K)\|_{L^2(S_{u,\ub})}\leq C_{I-1,J,L,\ub_f}. 
\end{equation}
Here, we use the convention that whenever $i<0$, then the corresponding term is dropped. Note that if it were the case that $J=L=0$, then the induction hypotheses \eqref{prop.reg.34.induct.1} and \eqref{prop.reg.34.induct.2} reduce exactly to \eqref{prop.reg.induct.1} and \eqref{prop.reg.induct.2}.

Our goal is to prove the analogue of \eqref{prop.reg.34.induct.1} with $I-1$ replaced by $I$, and the analogue of \eqref{prop.reg.34.induct.2} with $I-2$ replaced by $I-1$ (in both cases, with $J$ and $L$ unchanged).

To achieve this, we argue as in the proof of Proposition~\ref{prop:HO.renormalised}. Namely, we first carry out energy and elliptic estimates (to prove the analogue of \eqref{prop.reg.34.induct.1}) and then carry out transport estimates (to prove the analogue of \eqref{prop.reg.34.induct.2}).

Note that by the induction hypotheses \eqref{prop.reg.34.lower.induct.1} and \eqref{prop.reg.34.lower.induct.2} and Sobolev embedding (Proposition~\ref{Sobolev}), as long as there are strictly fewer than $J$ $\nab_3$'s or strictly fewer than $L$ $\nab_4$'s (even if there are more $\nab$'s), then the corresponding term can be controlled in $L^\i$. Once we notice this, it is straightforward to show that all the estimates are in fact linear\footnote{Note that in the proof of Proposition~\ref{prop:HO.renormalised}, there are quadratic terms of $\psi_H$, $\psi_{\Hb}$, $\psi$, $\omb$ and $K$, each of which has no $\nab_3$ or $\nab_4$ derivatives. We argued, by counting the number of $\nab$'s, that all but one factor can be controlled in $L^\i$. In contrast, in the present case, since $J>0$ or $K>0$, it is in fact easier to deduce that the estimates are linear.}, as in the proof of Proposition~\ref{prop:HO.renormalised}. The argument therefore proceeds in a completely analogous manner as in the proof of Proposition~\ref{prop:HO.renormalised}; we omit the details.

\textbf{Inducting on $\nab_3$ or $\nab_4$.}
Since the induction on $\nab_3$ and on $\nab_4$ proceed in a similar manner, we only consider the induction on the number of $\nab_3$ derivatives. More precisely, assume that for some $J\geq 1$, $L\geq 0$ such that for \underline{every} $I\geq 0$, the following holds:
\begin{equation*}
\begin{split}
&\sup_{(u,\ub)\in \mathcal W_{\ub_f}}\sum_{i\leq I,\,j\leq J-1, \, \ell\leq L}\|\nab^i \nab_3^j \nab_4^{\ell} (\psi,\psi_H,\psi_{\Hb},\omb), \nab^{i-1}\nab_3^j \nab_4^{\ell} K\|_{L^2(S_{u,\ub})}\leq C_{I,J,L,\ub_f}.
\end{split}
\end{equation*}
Our goal is to show that
\begin{equation*}
\begin{split}
&\sup_{(u,\ub)\in \mathcal W_{\ub_f}}\sum_{i\leq 1,\,j\leq J, \, \ell\leq L}\int_{-u+C_R}^{\ub}\|\nab^i \nab_3^j \nab_4^{\ell} (\psi_H,\psi), \nab_3^j \nab_4^{\ell} K\|_{L^2(S_{u,\ub'})}^2\, d\ub' \\
&\qquad + \sup_{(u,\ub)\in \mathcal W_{\ub_f}}\sum_{i\leq 1,\,j\leq J, \, \ell\leq L}\int_{-\ub+C_R}^{u}\|(\nab^i \nab_3^j \nab_4^{\ell}(\psi_{\Hb},\omb,\psi),\nab_3^j \nab_4^{\ell} K)\|_{L^2(S_{u',\ub})}^2\, du' \leq C_{J, L,\ub_f}
\end{split}
\end{equation*}
and
\begin{equation*}
\sup_{(u,\ub)\in \mathcal W_{\ub_f}}\sum_{j\leq J, \, \ell\leq L}\|\nab_3^j \nab_4^{\ell} (\psi,\psi_H,\psi_{\Hb},\omb)\|_{L^2(S_{u,\ub})}\leq C_{J,L,\ub_f}.
\end{equation*}
Again, it is straightforward to check that the estimates are linear and we proceed in a similar manner as in the proof of Proposition~\ref{prop:HO.renormalised}. Combining all these discussions, we conclude the proof of the proposition.
\end{proof}

\begin{remark}[Changing the order of the derivatives]
Given Proposition~\ref{higher.order.34}, it follows easily that any number of mixed derivatives $\nab$, $\nab_3$, $\nab_4$ of any of $\psi$, $\psi_H$, $\psi_{\Hb}$, $\omb$ and $K$ \underline{in any order} are be bounded uniformly in the region under consideration. This can be proven using an induction argument and estimating the commutators using Proposition~\ref{commute}; we omit the details.
\end{remark}

\begin{remark}[Bounds for the spacetime curvature components]
Let us note that the estimates for the derivatives of the Ricci coefficients in particular imply bounds for the spacetime curvature components $\a$ and $\ab$ (and their derivatives). That this is the case follows from the equations for $\nab_4\chih$ and $\nab_3\chibh$ in \eqref{null.str1}.
\end{remark}

The estimates in Proposition~\ref{higher.order.34} imply that in an appropriate coordinate system, the metric itself obeys higher regularity estimates. We now introduce the coordinates that we use. (Notice that up to this point, even though for any given coordinates $(\th^1,\th^2)$ on the $2$-spheres in the initial hypersurface $\Sigma_0$, we have constructed double null coordinates (see Section~\ref{setup.double null}), we have not used any specific systems of spherical coordinates. Indeed, all of our estimates so far can be obtained tensorially.)

To proceed, we first introduce the spherical coordinates $(\th_*,\phi_*)$ on $\mathcal U_{\ub_f}$ by taking the $(\th_*,\phi_*)$ coordinates on $\mathcal M_{Kerr}$ as in Section~\ref{Kerr.dbn} (note that indeed $\Lb_\Ke \th_* = \Lb_\Ke \phi_* = 0$), and pull them back to $\mathcal U_{\ub_f}$ using the identification introduced in Section~\ref{sec.identification}. $(u,\ub,\th_*,\phi_*)$ is thus a system of local coordinates on $\mathcal U_{\ub_f}$.

Next, we introduce new systems of coordinates on $\mathbb S^2$ using the stereographic projection. Note that we do not use spherical-type coordinates for the estimates of the metric components since the inverse metric components become infinite at the axis.

Given $(\th_*,\phi_*)$ as above, consider the usual coordinates on the $2$-sphere defined by the stereographic projection. Namely, let $\mathcal V_1$ and $\mathcal V_2$ be open sets on $\mathbb S^2$ defined as $\mathcal V_1\doteq \mathbb S^2\setminus\{\th_*=0\}$ and $\mathcal V_2\doteq \mathbb S^2\setminus\{\th_*=\pi\}$ respectively. Consider in $\mathcal V_1$ the coordinate transformation $(\th^1_{(1)},\th^2_{(1)}) = (\cot \f{\th_*}{2}\cos\phi_*, \cot \f{\th_*}{2}\sin\phi_*)$; and in $\mathcal V_2$ the coordinate transformation $(\th_{(2)}^1,\th_{(2)}^2) = (\tan \f{\th_*}{2}\cos\phi_*, \tan \f{\th_*}{2}\sin\phi_*)$.

Given $(\th^1_{(1)},\th^2_{(1)})$ and $(\th^1_{(2)},\th^2_{(2)})$ defined as above, we then define $(u,\ub,\th^1_{(1)},\th^2_{(1)})$ and $(u,\ub,\th^1_{(2)},\th^2_{(2)})$ as systems of coordinates on $\mathcal W_{\ub_f}\times \mathcal V_1$ and $\mathcal W_{\ub_f}\times \mathcal V_2$ respectively. We remark that since $(\th_*,\phi_*)$ are defined to be constant along the integral curves of $\Lb$, the new coordinate functions $\th^1_{(i)}$ and $\th^2_{(i)}$ (with $i=1,2$) that we define here still satisfy the property that they are constant along integral curves of $\Lb$.

Fix two precompact open subsets $\mathcal V_1'\subset \mathcal V_1$ and $\mathcal V_2'\subset \mathcal V_2$ such that $\mathcal V_1'\cup\mathcal V_2'=\mathbb S^2$. In the following proposition, the metric will be estimated in the coordinate charts $\mathcal V_1'$ and $\mathcal V_2'$, with the coordinates $(\th^1_{(1)},\th^2_{(1)})$ and $(\th^1_{(2)},\th^2_{(2)})$ defined above.

\begin{proposition}\label{prop:metric.higher}
Consider either $\mathcal V_i'=\mathcal V_1'$ or $\mathcal V_i'=\mathcal V_2'$ as above. For every $I\geq 0$, with respect to the coordinates defined above (from now on we suppress the subscripts ${ }_{(i)}$ in the $\th$ coordinates), the coordinates derivatives of $\gamma_{AB}$, $b^A$ and $\log\Om$ are uniformly bounded in $\mathcal W_{\ub_f}\times \mathcal V_i'$ (recall $\mathcal W_{\ub_f}$ in \eqref{def.W.ub.f}) as follows:
\begin{equation*}
\begin{split}
&\sup_{(u,\ub,\vartheta)\in \mathcal W_{\ub_f}\times \mathcal V_i'}|(\gamma^{-1})^{AB}|(u,\ub,\vartheta)+\sup_{(u,\ub,\vartheta)\in \mathcal W_{\ub_f}\times \mathcal V_i'}\sum_{i_1+i_2+i_3+i_4=I}\sup_{A,B} \left|(\f{\rd}{\rd\th^1})^{i_1}(\f{\rd}{\rd\th^2})^{i_2}(\f{\rd}{\rd u})^{i_3}(\f{\rd}{\rd\ub})^{i_4}\gamma_{AB}\right|(u,\ub,\vartheta)\\
&\qquad\qquad + \sup_{(u,\ub,\vartheta)\in \mathcal W_{\ub_f}\times \mathcal V_i'}\sum_{i_1+i_2+i_3+i_4=I} \sup_A \left|(\f{\rd}{\rd\th^1})^{i_1}(\f{\rd}{\rd\th^2})^{i_2}(\f{\rd}{\rd u})^{i_3}(\f{\rd}{\rd\ub})^{i_4}b^A\right|(u,\ub,\vartheta) \\
&\qquad\qquad + \sup_{(u,\ub,\vartheta)\in \mathcal W_{\ub_f}\times \mathcal V_i'}\sum_{i_1+i_2+i_3+i_4=I} \left|(\f{\rd}{\rd\th^1})^{i_1}(\f{\rd}{\rd\th^2})^{i_2}(\f{\rd}{\rd u})^{i_3}(\f{\rd}{\rd\ub})^{i_4}\log\Om\right|(u,\ub,\vartheta) \leq C_{I,\ub_f},
\end{split}
\end{equation*}
where $C_{I, \ub_f}>0$ is a constant depending on $I$ and $\ub_f$ (in addition to the profile of the initial data and the choice of $\mathcal V_1'$ and $\mathcal V_2'$).
\end{proposition}
\begin{proof}
\textbf{Estimates for the metric in $L^\infty$.} First, the $L^\infty$ bound for $\log\Om$ follows immediately from the bounds for $\NS$ proven in Theorem~\ref{main.quantitative.thm}. For $\gamma$, note that by taking $\ep$ smaller if necessary, we have pointwise bounds for $|\gamma-(\gamma_\Ke)|_{\gamma}$ and $|\gamma^{-1}-(\gamma_\Ke)^{-1}|_{\gamma}$ by Theorem~\ref{main.quantitative.thm}. By the estimates on Kerr geometry in Section~\ref{sec.Kerr.geometry}, this therefore implies that in $\mathcal W_{\ub_f}\times \mathcal V_i'$, $\gamma_{AB}$ is a matrix with uniformly upper bounded entries and uniformly lower bounded determinant. This in particular also implies the bounds for $(\gamma^{-1})^{AB}$. As a consequence, this shows that a bound on the norm of an $S$-tangent tensor $\phi$ with respect to the norm induced by $\gamma$ implies a bound on each component of $\phi$. Finally, using the fact that we just mentioned, the estimate for $b^A$ follows from the bounds for $\NS$ proven in Theorem~\ref{main.quantitative.thm}.

\textbf{Estimates for higher derivatives.} We now estimate the higher derivatives by induction. Suppose there exists $I> 0$ such that if $i_1+i_2+i_3+i_4<I$, then the following three estimates hold:
\begin{equation}\label{metric.higher.induct.1}
\sup_{(u,\ub,\vartheta)\in \mathcal W_{\ub_f}\times \mathcal V_i'} \left|(\f{\rd}{\rd\th^1})^{i_1}(\f{\rd}{\rd\th^2})^{i_2}(\f{\rd}{\rd u})^{i_3}(\f{\rd}{\rd\ub})^{i_4}\log\Om \right|(u,\ub,\vartheta) \leq C_{I-1,\ub_f},
\end{equation}
\begin{equation}\label{metric.higher.induct.2}
\sup_{(u,\ub,\vartheta)\in \mathcal W_{\ub_f}\times \mathcal V_i'} \sup_{A,B}\left|(\f{\rd}{\rd\th^1})^{i_1}(\f{\rd}{\rd\th^2})^{i_2}(\f{\rd}{\rd u})^{i_3}(\f{\rd}{\rd\ub})^{i_4}\gamma_{AB}\right|(u,\ub,\vartheta) \leq C_{I-1,\ub_f}
\end{equation}
and
\begin{equation}\label{metric.higher.induct.3}
\sup_{(u,\ub,\vartheta)\in \mathcal W_{\ub_f}\times \mathcal V_i'} \sup_{A}\left|(\f{\rd}{\rd\th^1})^{i_1}(\f{\rd}{\rd\th^2})^{i_2}(\f{\rd}{\rd u})^{i_3}(\f{\rd}{\rd\ub})^{i_4}b^A\right|(u,\ub,\vartheta) \leq C_{I-1,\ub_f}.
\end{equation}
Our goal is to show that these estimates also hold in the case $i_1+i_2+i_3+i_4=I$. Before proceeding, let us give an overview of the strategy. We will use the transport equations from Proposition~\ref{metric.der.Ricci}:
\begin{equation}\label{metric.transport.again}
\f{\rd}{\rd u} \gamma_{AB}=2\chib_{AB},\quad 
\f{\rd}{\rd u}\log\Omega=-\omb,\quad 
\f{\rd b^A}{\rd u}=2\Omega^2(\eta^A-\etab^A).
\end{equation}
These equations will be differentiated (up to $I$-th (mixed) derivatives) and we will bound their right hand sides. With the estimates in Proposition~\ref{higher.order.34} and the estimates for $\gamma_{AB}$ and $(\gamma^{-1})^{AB}$ that we just derived above, the only thing that remains to be controlled is the Christoffel symbols associated to the connections $\nab$, $\nab_3$ and $\nab_4$. By \eqref{nab.def}, \eqref{def.slashed.Gamma}, \eqref{nab3.def} and \eqref{nab4.def}, we in particular need to bound the derivatives of $\f{\rd\gamma_{AB}}{\rd\th^C}$ and $\f{\rd b^A}{\rd\th^B}$. We will therefore derive estimates a coupled system of transport equations.

We start with $\log\Om$. Commute the corresponding equation in \eqref{metric.transport.again} with $(\f{\rd}{\rd\th^1})^{i_1}(\f{\rd}{\rd\th^2})^{i_2}(\f{\rd}{\rd u})^{i_3}(\f{\rd}{\rd\ub})^{i_4}$ for $i_1+i_2+i_3+i_4=I$. Since $\omb$ is a scalar function, when we use the estimates in Proposition~\ref{higher.order.34} and bound the Christoffel symbols associated to $\nab$, $\nab_3$ and $\nab_4$, it can be checked using \eqref{nab.def}, \eqref{def.slashed.Gamma}, \eqref{nab3.def} and \eqref{nab4.def} that we only need to control up to $I-1$ coordinate derivatives of $\gamma$ and $b$. These terms can be bounded using the induction hypotheses \eqref{metric.higher.induct.1}, \eqref{metric.higher.induct.2} and \eqref{metric.higher.induct.3}. It therefore follows that
\begin{equation}\label{metric.higher.Om.induction}
\sup_{(u,\ub,\vartheta)\in \mathcal W_{\ub_f}\times \mathcal V_i'} \sum_{i_1+i_2+i_3+i_4\leq I} \left|(\f{\rd}{\rd\th^1})^{i_1}(\f{\rd}{\rd\th^2})^{i_2}(\f{\rd}{\rd u})^{i_3}(\f{\rd}{\rd\ub})^{i_4}\log\Om \right|(u,\ub,\vartheta) \leq C_{I,\ub_f}.
\end{equation}
To control the derivatives of $\gamma_{AB}$, we commute the equation by $(\f{\rd}{\rd\th^1})^{i_1}(\f{\rd}{\rd\th^2})^{i_2}(\f{\rd}{\rd u})^{i_3}(\f{\rd}{\rd\ub})^{i_4}$ for $i_1+i_2+i_3+i_4=I$. Now notice that has the right hand side contains coordinate derivatives of $\chib$, which a priori we do not know are bounded by the norms of the covariant derivatives of $\chib$ (which are bounded in Proposition~\ref{higher.order.34}). Nevertheless, by we can control the Christoffel symbols by derivatives of the metric as follows. First, note that terms that do not have top order derivatives on the metric can be controlled using induction hypotheses \eqref{metric.higher.induct.1}, \eqref{metric.higher.induct.2} and \eqref{metric.higher.induct.3}. Then note that according to \eqref{nab.def}, \eqref{def.slashed.Gamma}, \eqref{nab3.def} and \eqref{nab4.def}, there are two possible contributions of highest derivatives of $\gamma$ and $b$. Namely,
\begin{itemize}
\item when one bounds a coordinate $\f{\rd}{\rd\th^A}$ derivative by a $\nab$ derivative, one needs to bound $\slashed\Gamma$ and hence $\f{\rd\gamma_{AB}}{\rd\th^C}$ (and higher derivatives thereof);
\item when controlling a $\f{\rd}{\rd\ub}$ derivative by a $\nab_4$ derivative, one needs to bound $\f{\rd b^A}{\rd\th^B}$ (and higher derivatives thereof).
\end{itemize}
Therefore, for every $(u,\ub)\in \mathcal W_{\ub_f}$, $\vartheta\in \mathcal V_i'$ and $i_1+i_2+i_3+i_4=I$,
\begin{equation}\label{gamma.higher.order}
\begin{split}
&\sup_{A,B}\left|(\f{\rd}{\rd\th^1})^{i_1}(\f{\rd}{\rd\th^2})^{i_2}(\f{\rd}{\rd u})^{i_3}(\f{\rd}{\rd\ub})^{i_4}\gamma_{AB}\right|(u,\ub,\vartheta)\\
\leq &C_{I,\ub_f} + C_{I,\ub_f}\int_{-\ub+C_R}^u  \sup_{A,B} \left|(\f{\rd}{\rd\th^1})^{i_1}(\f{\rd}{\rd\th^2})^{i_2}(\f{\rd}{\rd u})^{i_3}(\f{\rd}{\rd\ub})^{i_4-1}(\f{\rd b^A}{\rd \th^B})\right|(u',\ub,\vartheta)\,du'\\
&+C_{I,\ub_f} \sup_{A,B}\int_{-\ub+C_R}^u \left|(\f{\rd}{\rd\th^1})^{i_1}(\f{\rd}{\rd\th^2})^{i_2}(\f{\rd}{\rd u})^{i_3}(\f{\rd}{\rd\ub})^{i_4}\gamma_{AB} \right|(u',\ub,\vartheta)\, du',
\end{split}
\end{equation}
where it is understood that if $i_4=0$, then $(\f{\rd}{\rd\ub})^{i_4-1}=0$.

Next, we estimate the derivatives of $b^A$. For this we commute derivatives with the corresponding equation in \eqref{metric.transport.again}. Note that contributions of highest derivatives of $\gamma$ and $b$ would arise in exactly the same situations as above. (Note that since there is a factor $\Om^2$ on the right hand side of the equation for $b^A$, whose derivatives could not all be bounded using Proposition~\ref{higher.order.34}, but they can nonetheless be controlled using the bound \eqref{metric.higher.Om.induction} that we just proved.) Therefore, for every $(u,\ub)\in \mathcal W_{\ub_f}$, $\vartheta\in \mathcal V_i'$ and $i_1+i_2+i_3+i_4=I$,
\begin{equation}\label{b.higher.order}
\begin{split}
&\sup_A \left|(\f{\rd}{\rd\th^1})^{i_1}(\f{\rd}{\rd\th^2})^{i_2}(\f{\rd}{\rd u})^{i_3}(\f{\rd}{\rd\ub})^{i_4}b^A\right|(u,\ub,\vartheta)\\
\leq &C_{I,\ub_f} + C_{I,\ub_f}\int_{-\ub+C_R}^u  \sup_{A,B} \left|(\f{\rd}{\rd\th^1})^{i_1}(\f{\rd}{\rd\th^2})^{i_2}(\f{\rd}{\rd u})^{i_3}(\f{\rd}{\rd\ub})^{i_4-1}(\f{\rd b^A}{\rd \th^B})\right|(u,\ub,\vartheta)\,du'\\
&+C_{I,\ub_f} \sup_{A,B}\int_{-\ub+C_R}^u \left|(\f{\rd}{\rd\th^1})^{i_1}(\f{\rd}{\rd\th^2})^{i_2}(\f{\rd}{\rd u})^{i_3}(\f{\rd}{\rd\ub})^{i_4}\gamma_{AB} \right|(u,\ub,\vartheta)\, du',
\end{split}
\end{equation}
where, again, if $i_4=0$, then we take $(\f{\rd}{\rd\ub})^{i_4-1}=0$.

Now, sum \eqref{gamma.higher.order} and \eqref{b.higher.order} and sum over all $i_1+i_2+i_3+i_4 = I$. Applying Gr\"onwall's inequality yields
\begin{equation*}
\begin{split}
&\sup_{(u,\ub,\vartheta)\in \mathcal W_{\ub_f}\times \mathcal V_i'}\sum_{i_1+i_2+i_3+i_4=I}\sup_{A,B} \left|(\f{\rd}{\rd\th^1})^{i_1}(\f{\rd}{\rd\th^2})^{i_2}(\f{\rd}{\rd u})^{i_3}(\f{\rd}{\rd\ub})^{i_4}\gamma_{AB}\right|(u,\ub,\vartheta)\\
&\qquad+ \sup_{(u,\ub,\vartheta)\in \mathcal W_{\ub_f}\times \mathcal V_i'}\sum_{i_1+i_2+i_3+i_4=I}\sup_A \left|(\f{\rd}{\rd\th^1})^{i_1}(\f{\rd}{\rd\th^2})^{i_2}(\f{\rd}{\rd u})^{i_3}(\f{\rd}{\rd\ub})^{i_4}b^A\right|(u,\ub,\vartheta) \leq C_{I,\ub_f}.
\end{split}
\end{equation*}
Combining this with \eqref{metric.higher.Om.induction}, we have completed the induction argument. \qedhere

\end{proof}

\begin{proposition}\label{prop:smooth.up.to.bdry}
In the coordinate system defined above, the metric extends smoothly to the manifold-with-corner $\overline{\mathcal W_{\ub_f}}\times \mathbb S^2$, where the closure is to be understood as the closure in $\mathbb R^2$.
\end{proposition}
\begin{proof}
Define the extension in the natural manner: For any $p\in (\rd \mathcal W_{\ub_f}\setminus \mathcal W_{\ub_f})\times \mathbb S^2$, take a sequence $p_n\to p$ with $p_n\in \mathcal W_{\ub_f}\times \mathbb S^2$ and define $\gamma_{AB}(p)= \lim_{n\to \infty}\gamma_{AB}(p_n)$, $b^A(p)= \lim_{n\to \infty}b^A(p_n)$, $\log\Om(p)= \lim_{n\to \infty}\log\Om(p_n)$.

That this is well-defined, independent of the choice of the sequence $p_n$ and that the extension is a smooth metric on $\overline{\mathcal W_{\ub_f}}\times \mathbb S^2$ all follow from the uniform bounds for $\gamma_{AB}$, $b^A$ and $\log\Om$ and their derivatives proven in Proposition~\ref{prop:metric.higher}.
\end{proof}

\section{Completion of the bootstrap argument}\label{sec:continuity}

We now carry out the bootstrap argument and complete the proof (cf.~\textbf{Theorem~\ref{I.final}}) of 
\begin{itemize}
\item the existence of a smooth solution to the Einstein vacuum equations \eqref{INTROvace} in the double null foliation gauge with prescribed initial data (i.e.~both the geometric initial data and initial data for the double null gauge as in Section~\ref{setup.double null}) in $\mathcal U_\infty=\{(u,\ub): u\in [-\ub+C_R,u_f),\,\ub\in [-u+C_R,\infty)\}\times \mathbb S^2$, and
\item the estimate \eqref{main.apriori.est} holds everywhere in $\mathcal U_\infty$
\end{itemize}
as stated in Theorem~\ref{aprioriestimates}. (The statement on the continuous extendibility of the metric to the Cauchy horizon will be proven in Section~\ref{sec.C0}; see Theorem~\ref{metric.cont.ext}.)

Define
$\mathfrak I \subset (-u_f+C_R,\infty)$ to be the set of all $\ub_c \in (-u_f+C_R,\infty)\subset \mathbb R$ such that the following holds for $\mathcal W = \mathcal W_{\ub_c} = \{(u,\ub): \ub \in [-u+C_R,\ub_c),\, u\in [-\ub+C_R,u_f)\} \subset \mathbb R^2$:
\begin{enumerate}[(A)] 
\item There is a smooth solution $g$ on $\mathcal U= \mathcal W\times \mathbb S^2$ satisfying $Ric(g)=0$ in a system of double null coordinates (cf.~Sections~\ref{secsetup} and \ref{seceqn}) achieving the geometric initial data $(\hat{g},\hat{k})$ and the initial data for the double null gauge as in Section~\ref{setup.double null}.
\item The estimate \eqref{main.apriori.est} holds on $\mathcal U = \mathcal W \times \mathbb S^2$.
\end{enumerate}
Our goal is to show that $\mathfrak I = (-u_f+C_R,\infty)$. Using connectedness of $(-u_f+C_R,\infty)$, this will be achieved in Propositions~\ref{I.nonempty}, \ref{I.open}, \ref{I.closed} and Theorem~\ref{I.final} below.

\begin{proposition}\label{I.nonempty}
$\mathfrak I$ is non-empty.
\end{proposition}
\begin{proof}
By Proposition~\ref{prop:doublenull.local}, there exists $\ub_{local}> -u_f+C_R$ such that for $\mathcal W_{\ub_{local}} \doteq \{(u,\ub): u+\ub\geq C_R,\, u< u_f,\, \ub< \ub_{local}\}$ and $\mathcal U_{\ub_{local}}\doteq \mathcal W_{\ub_{local}}\times \mathbb S^2$, there is a solution $(\mathcal U_{\ub_{local}},g)$ to the Einstein vacuum equations in the double null foliation gauge \eqref{formofthemetric} attaining the geometric initial data and the prescribed data for the double null foliation and double null coordinates. 

Moreover, by \eqref{small.data}, $\mathcal D\leq \ep^2$. Hence, by continuity, and taking $\ub_{local}$ closer to $-u_f+C_R$ if necessary, the estimate \eqref{main.apriori.est} holds for $(\mathcal U_{local},g)$

It then follows that $\ub_{local} \in \mathfrak I$. In particular, $\mathfrak I$ is non-empty.
\end{proof}

\begin{proposition}\label{I.closed}
$\mathfrak I$ is closed.
\end{proposition}
\begin{proof}
Let $\{\ub_n\}_{n=1}^\infty$ be an increasing sequence of real numbers with $\ub_n\to \ub_c$ such that (A) and (B) above both hold on $\mathcal U_{\ub_n} = \mathcal W_{\ub_n} \times \mathbb S^2$ for all $n\in \mathbb N$, where $\mathcal W_{\ub_n}= \{(u,\ub): \ub \in [-u+C_R,\ub_n),\, u\in [-\ub+C_R,-u_f)\} \subset \mathbb R^2$. We can assume that $\ub_n$ is an increasing sequence (since if there exists $\ub_n\geq \ub_c$ for some $n$, then the conclusion is trivial). It now follows that (A) and (B) both hold on $\mathcal W_{\ub_c}$:
\begin{enumerate}[(A)]
\item This is immediate from the fact $\mathcal W_{\ub_c} = \displaystyle \cup_{\ub_n} \mathcal W_{\ub_n}$.
\item Again, we use the fact that $\mathcal W_{\ub_c} = \displaystyle \cup_{\ub_n} \mathcal W_{\ub_n}$: Since the estimate \eqref{main.apriori.est} holds on $\mathcal W_{\ub_n}$ for all $n\in \mathbb N$ and the constant in \eqref{main.apriori.est} is manifestly \emph{independent} of $n$, \eqref{main.apriori.est} also holds on $\mathcal W_{\ub_c}$.
\end{enumerate}
\qedhere
\end{proof}

We need a couple of preliminary results before proving the openness of $\mathfrak I$. The following proposition can be viewed as a local existence theorem for the characteristic initial value problem for the Einstein vacuum equations\footnote{Strictly speaking, \cite{L} assumes that $\Om$ equals a constant on both initial characteristic hypersurface. Nevertheless, it is easy to check that the same proof works when $\Om$ is non-constant and satisfies the bounds as stated in Proposition~\ref{char.local}.}:
\begin{proposition}[\cite{L}]\label{char.local}
Consider the characteristic initial value problem for the Einstein vacuum equations \eqref{INTROvace} with smooth characteristic initial data on two null hypersurfaces $H\doteq \{0\}\times [0,I_{H}]\times \mathbb S^2$ and $\underline{H}\doteq [0,I_{\underline{H}}]\times \{0\}\times \mathbb S^2$ transversely intersecting at a $2$-surface $S_{0,0} \doteq \{0\}\times\{0\}\times \mathbb S^2$. Suppose the characteristic initial data are given in terms of the double null foliation gauge and satisfy the following properties:
\begin{itemize}
\item $H$ is foliated by $2$-spheres $\{S_{0,\ub}\}_{\ub \in [0,I_{H}]}$. $\gamma$, $\Om$, $\chi$ are prescribed to be smooth such that
\begin{itemize}
\item the following constraint equations hold:
$$\mathcal L_{L} \gamma = 2\Om^2\chi,\quad \nab_4\trch + \f 12(\trch)^2 = - |\chih|^2. $$
\item the following estimates\footnote{The assumptions in \cite{L} are stated for some other Ricci coefficients and curvature components in addition to the ones below. Nevertheless, it is easy to see that the bounds below imply those estimates. Moreover, much fewer derivatives are required in \cite{L}, but we simply state the version below since the number of derivatives will be irrelevant for our application.} for $\Om$, $\chi$ and $K$ hold for some $C>0$:
$$\sup_{\ub \in [0,I_H]}\sum_{i+j\leq 10}( \|\nab_4^j \nab^i \log\Om\|_{L^\i(S_{0,\ub})}+\|\nab_4^j \nab^i 
\chi \|_{L^\i(S_{0,\ub})}+\|\nab_4^j \nab^i K\|_{L^\i(S_{0,\ub})})\leq C.$$
\end{itemize}
\item $\underline{H}$ is foliated by $2$-spheres $\{S_{u,0}\}_{u\in [0,I_{\underline{H}}]}$. $\gamma$, $\Om$, $\chib$ are prescribed to be smooth such that
\begin{itemize}
\item the following constraint equations hold:
$$\mathcal L_{\Lb} \gamma = 2\chib,\quad \nab_3\trchb + \f 12(\trchb)^2 = 2(\nab_3\log\Om)\trchb - |\chibh|^2, $$
\item the following estimates for $\Om$, $\chib$ and $K$ hold for some $C>0$:
$$\sup_{u \in [0,I_{\underline{H}}]}\sum_{i+j\leq 10}( \|\nab_3^j \nab^i \log\Om\|_{L^\i(S_{u,0})}+\|\nab_3^j \nab^i 
\chib \|_{L^\i(S_{u,0})}+\|\nab_3^j \nab^i K\|_{L^\i(S_{u,0})})\leq C.$$
\end{itemize}
\item $\gamma$ and $\Om$ are continuous up to $S_{0,0}$.
\item $\zeta$ is prescribed on $S_{0,0}$ and is smooth.
\end{itemize}
Then there exists $\ub_\ep\in (0, I_H)$ such that there is a smooth solution to the Einstein vacuum equations with the prescribed data in the double null foliation gauge in $\mathcal U = \mathcal W\times \mathbb S^2$, where $\mathcal W\doteq \{(u,\ub): 0\leq u\leq I_{\underline{H}},\, 0\leq \ub\leq \ub_{\ep}\}$. 
\end{proposition}

Using also the result of Choquet-Bruhat--Geroch \cite{geroch}, this immediately implies a local existence result for a mixed characteristic Cauchy initial value problem. We will only state this result in the particular setting of our problem, where the Cauchy initial data are given in terms of a double null foliation as in Theorem~\ref{aprioriestimates} and the characteristic initial data are given on a constant $\{\ub=\ub_c\}$ null hypersurface:
\begin{lemma}\label{mixed.local}
Given the following mixed characteristic Cauchy initial value problem:
\begin{itemize}
\item smooth Cauchy data $(\Sigma_0,\hat{g},\hat{k})$ are given as in Theorem~\ref{aprioriestimates};
\item smooth characteristic initial data are given on the null hypersurface $\Hb$ with $\ub=\ub_c$ (i.e.~the set $\{(u,\ub_c): u\in [-\ub_c+C_R,u_f)\}\times \mathbb S^2$) such that
\begin{itemize}
\item the following constraint equations hold:
$$\mathcal L_{\Lb} \gamma = 2\chib,\quad \nab_3\trchb + \f 12(\trchb)^2 = 2(\nab_3\log\Om)\trchb - |\chibh|^2, $$
\item the following estimates for $\Om$, $\chib$ and $K$ hold for some $C>0$:
$$\sup_{u \in [-\ub_c+C_R,u_f]}\sum_{i+j\leq 10}( \|\nab_3^j \nab^i \log\Om\|_{L^\i(S_{u,0})}+\|\nab_3^j \nab^i 
\chib \|_{L^\i(S_{u,0})}+\|\nab_3^j \nab^i K\|_{L^\i(S_{u,0})})\leq C.$$
\end{itemize}
\item $\gamma$ and $\Om$ are continuous up to the $2$-sphere $S_{-\ub_c+C_R,\ub_c}$.
\end{itemize}
Then there exists $\ub_\ep>0$ such that there is a smooth solution to the Einstein vacuum equations \eqref{INTROvace} with the prescribed data in the double null foliation gauge in $\mathcal U = \mathcal W\times \mathbb S^2$, where $\mathcal W\doteq \{(u,\ub): -\ub+C_R\leq u < u_f,\, \ub_c \leq \ub < \ub_c+\ub_{\ep}\}$
\end{lemma}
\begin{proof}
Refer to Figure~\ref{mixedfigure}.
\begin{figure}
\centering{
\def\svgwidth{9pc}
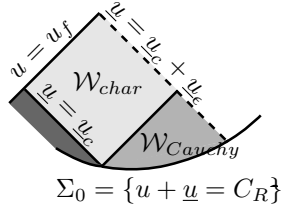}
\caption{Solving the mixed characteristic Cauchy initial value problem}\label{mixedfigure}
\end{figure}
As in the proof of Proposition~\ref{prop:doublenull.local}, let $(\widetilde{\mathcal M},g)$ be the maximal globally hyperbolic future development of the data $(\Sigma_0,\hat{g},\hat{k})$ as given by \cite{geroch}. Applying Proposition~\ref{prop:u.extend} to $(\widetilde{\mathcal M},g)$, there exists $\ub_{\ep}$ sufficiently small such that there exists a solution with the prescribed initial data to the Einstein vacuum equations in the double null foliation gauge in $\mathcal W_{Cauchy}\times \mathbb S^2$, where $\mathcal W_{Cauchy}$ is given by
$$\mathcal W_{Cauchy} \doteq \{(u,\ub): u\in [-\ub+C_R,-\ub_c+C_R),\,\ub \in [\ub_c,\ub_c+\ub_\ep)\}.$$
Moreover, the solution extends smoothly up to the null hypersurface 
\begin{equation}\label{C0.H.def}
H \doteq \{(u,\ub): u=-\ub_c+C_R,\,\ub \in [\ub_c,\ub_c+\ub_\ep)\}\times \mathbb S^2.
\end{equation}
Next, we apply Proposition~\ref{char.local} to solve the characteristic initial value problem, where the initial data are given on $\Hb$ (as in the statement of the Lemma) and $H$ (as in \eqref{C0.H.def}). Proposition~\ref{char.local} then implies that by taking $\ub_{\ep}$ smaller if necessary, there exists a solution with the prescribed initial data to the Einstein vacuum equations in the double null foliation gauge in $\mathcal W_{char}\times \mathbb S^2$, where $\mathcal W_{char}$ is given by
$$\mathcal W_{char}\doteq \{(u,\ub): u\in [-\ub_c+C_R,u_f),\,\ub\in [\ub_c,\ub_c+\ub_\ep)\}$$
We now take $\mathcal W \doteq \mathcal W_{Cauchy}\cup \mathcal W_{char}$ and combine the solution in $\mathcal W_{Cauchy}\times \mathbb S^2$ with the solution in $\mathcal W_{char}\times \mathbb S^2$. Since the solution is smooth in $\mathcal W_{Cauchy}\times \mathbb S^2$ and $\mathcal W_{char}\times S^2$, the solution in $\mathcal W\times \mathbb S^2$ is piecewise smooth. Moreover, the solution in $\mathcal W\times \mathbb S^2$ is everywhere continuous and is smooth up to $H$ on each side. Finally, it remains to note that in the double null foliation gauge, the derivatives of the metric to transversal\footnote{Note that this is true for all higher order transversal derivatives.} to $H$ are determined by the Einstein vacuum equation \eqref{INTROvace}, and the limits from each side of $H$ must agree. Therefore, the solution in $\mathcal W\times \mathbb S^2$ is smooth, and we have obtained the desired conclusion.
\end{proof}

\begin{proposition}\label{I.open}
Choosing $\ep_0$ smaller if necessary, $\mathfrak I$ is open.
\end{proposition}
\begin{proof}
Suppose $\ub_c\in \mathfrak I$.

By Proposition~\ref{prop:smooth.up.to.bdry}, the metric can be extended smoothly to $\overline{\mathcal W_{\ub_c}}\times \mathbb S^2$. Therefore, by Lemma~\ref{mixed.local}, we can extend the solution by solving the mixed characteristic Cauchy initial value problem to obtain a smooth solution in the double null foliation gauge in $\mathcal W_{\ub_c+\ub_{\ep}}\times \mathbb S^2$ for some $\ub_{\ep}>0$.

Finally, by definition of $\mathfrak I$, the estimate \eqref{main.apriori.est} holds in $\mathcal W_{\ub_c}\times \mathbb S^2$. By continuity, this implies that the estimate \eqref{main.apriori.est} holds in $\overline{\mathcal W_{\ub_c}}\times \mathbb S^2$. In particular, since \eqref{main.apriori.est} is a strictly stronger estimate than \eqref{BA} for $\ep_0$ (and hence $\ep$) sufficiently small, by continuity, every point in $\overline{\mathcal W_{\ub_c}}\times \mathbb S^2$ has an open neighbourhood such that the bootstrap assumption \eqref{BA} holds. By compactness of $\overline{\mathcal W_{\ub_c}}\times \mathbb S^2$, this implies that there is some $\ub>\ub_c$ such that $\mathcal W_{\ub}$ satisfies (A) and \eqref{BA}. Applying Theorem~\ref{main.quantitative.thm}, it follows that \eqref{main.apriori.est}, and therefore (B), holds in $\mathcal W_{\ub}$.
\end{proof}

By the connectedness of $(-u_f+C_R,\infty)$, Propositions~\ref{I.nonempty}, \ref{I.closed} and \ref{I.open} imply that $\mathfrak I = (-u_f+C_R,\infty)$. This then implies that the conditions (A) and (B) stated in the beginning of the section hold for $\mathcal W_{\infty} \doteq \{(u,\ub): \ub \in [-u+C_R,\infty),\, u\in [-\ub+C_R,-u_f)\} \subset \mathbb R^2$. In other words,
\begin{theorem}\label{I.final}
Let $\mathcal U_{\infty} \doteq \mathcal W_{\infty}\times \mathbb S^2$, where
$$\mathcal W_{\infty} \doteq \{(u,\ub):u+\ub\geq C_R ,\, u< u_f \},$$
Then there exists a solution $(\mathcal U_{\infty},g)$ to the Einstein vacuum equations \eqref{INTROvace} where $g$ takes the form \eqref{formofthemetric} and 
\begin{itemize}
\item attains the prescribed initial geometric data $(\hat{g},\hat{k})$ on $\Sigma\cap \{\tau:\tau<-u_f+C_R\}$ (so that $\Sigma\cap \{\tau:\tau<-u_f+C_R\}$ is a Cauchy hypersurface in $(\mathcal U_{\infty},g)$), and
\item attains the prescribed data for the double null foliation as in \eqref{coord.init} in Section~\ref{setup.double null}.
\end{itemize}
Moreover, the estimate \eqref{main.apriori.est} holds in $\mathcal U_\infty$.
\end{theorem}

\section{Continuity of the metric up to the Cauchy horizon}\label{sec.C0}

In this section, we show that a non-trivial Cauchy horizon $\CH$ can be attached to $(\mathcal U_\infty,g)$ (cf.~Theorem~\ref{I.final}) and moreover that the metric can be extended continuously to the Cauchy horizon. This will be achieved in \textbf{Theorem~\ref{metric.cont.ext}} in Section~\ref{sec.metric.cont.ext}. This will be the main remaining ingredient which we use to complete the proof of Theorem~\ref{aprioriestimates} in Section~\ref{sec.completion.everything}.

For the purpose of showing the continuous extendibility of the metric, it is important to have precise estimates for various geometric quantities $\ub\to \infty$. For this reason, in this section, \textbf{\emph{all the implicit constants in $\ls$ are only allowed to depend only on $M$, $a$, $\de$ and $C_R$, and all the estimates are required to hold uniformly in $\mathcal U_{\infty}$.}}

This section is organised as follows. In \textbf{Section~\ref{sec.CH}}, we define the Cauchy horizon $\CH$ by attaching an appropriate boundary in the $(u,\ub_{\CH},\th^1,\th^2)$ coordinate system as in the Kerr case (see Section~\ref{sec.def.horizon}). In \textbf{Section~\ref{sec.nab4logOm.est}}, we prove some auxiliary estimates for $\nab_4\widetilde{\log\Om}$. Already at this point we show in \textbf{Section~\ref{sec.prelim.cont}} that some geometric quantities can be extended continuously to $\CH$. In \textbf{Section~\ref{sec.CH.coord}}, we introduce a different coordinate system $(u,\ub_{\CH},\th^1_{\CH},\th^2_{\CH})$ and prove estimates related to the change of variable. We then prove in \textbf{Section~\ref{sec.metric.cont.ext}} that in this new coordinate system, the metric extends continuously to $\CH$ (Theorem \ref{metric.cont.ext}). In \textbf{Section~\ref{sec.C0.closeness}}, we prove that in this new coordinate system, the metric is moreover close to the Kerr metric (in an appropriate coordinate system) in $C^0$. We end this section by concluding the proof of Theorem~\ref{aprioriestimates} in \textbf{Section~\ref{sec.completion.everything}}.

\subsection{The Cauchy horizon $\protect\CH$}\label{sec.CH}

\subsubsection{Definition of $\protect\ub_{\protect\CH}$ and the Cauchy horizon}

Let $(\mathcal U_{\infty},g)$ be as in Theorem~\ref{I.final}. Notice that according to Proposition~\ref{prop:smooth.up.to.bdry}, the metric extends smoothly up to $\{u=u_f\}$ (or more precisely, up to $\overline{\mathcal W_{\infty}}\times \mathbb S^2$, where the closure is to be understood with respect to the natural topology of the $(u,\ub)$-plane.) The boundary $\{u=u_f\}$ can be viewed as a ``trivial'' Cauchy horizon, whose existence is simply associated to the incompleteness of the initial hypersurface $\Sigma_0\cap \{\tau>-u_f+C_R\}$ (as $\tau \to -u_f+C_R$). On the other hand, we will define a \emph{non-trivial} Cauchy horizon, which can be thought of as analogous to the Kerr Cauchy horizon. From now on, when we refer to the \emph{Cauchy horizon}, we will be referring to this non-trivial Cauchy horizon.

We define the \emph{Cauchy horizon} $\CH$ as follows: Introduce the function $\ub_{\CH}$ as in the Kerr case in Section~\ref{sec.def.horizon}, i.e.~$\ub_{\CH}(\ub)$ is a smooth and strictly increasing function of $\ub$, which agrees with $\ub$ for $\ub\leq -1$, satisfies $\ub_{\CH}\to 0$ as $\ub\to \infty$ and moreover there exists $\ub_-\geq -1$ such that 
\begin{equation}\label{CH.def.last}
\f{d \ub_{\CH}(\ub)}{d\ub}=e^{-\f{r_+-r_-}{r_-^2+a^2}\ub}\mbox{ for }\ub\geq \ub_-.
\end{equation}
Given $u$ as before, $\ub_{\CH}$ as above, and $(\th^1,\th^2)$ a system of local coordinates on the spheres with initial data \eqref{coord.init} and satisfying $\Lb \th^A=0$, attach the boundary $\CH$ to the set $\{(u,\ub,\th^1,\th^2):u < u_f,\, \ub_{\CH}< 0\}$ where $\CH=(-\infty,u_f) \times \{\ub_{\CH}=0\}\times \mathbb S^2$.

Given the above, we then consider $\mathcal U_{\infty} \cup \CH$ as a manifold-with-boundary. Defining moreover $S_{u_f,\CH} = \{(u,\ub,\th^1,\th^2):u = u_f,\, \ub_{\CH} = 0\}$ in the coordinate system, we can consider $\mathcal U_{\infty}\cup\{u=u_f\}\cup\CH\cup S_{u_f,\CH}$ as a manifold-with-corner.

\subsubsection{Differential structures on $\mathcal U_\infty\cup \CH$}\label{sec:two.diff.str}

We define two differential structures on $\mathcal U_\infty\cup\CH$. The first one is defined as the natural differential structure associated to the $(u,\ub_{\CH},\th^1,\th^2)$ coordinate system. The second one is the natural differential structure associated to $(u,\ub_{\CH},\th^1_{\CH}, \th^2_{\CH})$, which is a new coordinate system that we will define (see Section~\ref{sec.CH.coord}). We note already that we will only show the continuous extendibility of the metric up to the Cauchy horizon with respect to the latter differential structure.\footnote{Indeed, in the $(u,\ub_{\CH},\th^1,\th^2)$ coordinate system, the metric takes the form
$$g=-2e^{\f{r_+-r_-}{r_-^2+a^2}\ub}\Omega^2(du\otimes d\ub_{\CH}+d\ub_{\CH}\otimes du)+\gamma_{AB}(d\th^A-e^{\f{r_+-r_-}{r_-^2+a^2}\ub}b^A d\ub_{\CH})\otimes (d\th^B-e^{\f{r_+-r_-}{r_-^2+a^2}\ub}b^B d\ub_{\CH})$$
and $e^{\f{r_+-r_-}{r_-^2+a^2}\ub}b^A$ may not be bounded as $\ub\to \infty$ (or equivalently $\ub_{\CH}\to 0$).
} We make two important remarks regarding the difference between the $(\th^1,\th^2)$ coordinates and the $(\th^1_{\CH},\th^2_{\CH})$ coordinates:
\begin{itemize}
\item First, according to Section~\ref{sec.def.horizon}, even in the Kerr case, one needs a change of coordinates to obtain a regular system of double null coordinates near the horizon. To compare with the Kerr case, one should think that $(\th^1,\th^2)$ is analogous to $(\th_*,\phi_*)$ on Kerr, while $(\th^1_{\CH},\th^2_{\CH})$ is analogous to $(\th_*,\phi_{*,\CH})$.
\item Second, while the $(\th^1,\th^2)$ coordinates are defined starting from the initial data, the $(\th^1_{\CH},\th^2_{\CH})$ coordinates are defined starting from the hypersurface $\{u=u_f\}$ \emph{in the future}.
\end{itemize} 
As we will show later, both these coordinate systems give the same $C^0$ structure on $\mathcal U_\infty\cup\CH$, but in general give different $C^1$ structures for $\mathcal U_\infty\cup \CH$. In particular, we note that $\f{\rd\th_{\CH}}{\rd \ub_{\CH}}$ is \emph{not} necessarily bounded up to $\CH$. Nevertheless, we will have the following important property regarding the change of spherical coordinates $(\th^1, \th^2) \leftrightarrow (\th^1_{\CH}, \th^2_{\CH})$, namely that in appropriate coordinate charts,
\begin{equation}\label{dthdth.ce}
\begin{split}
\mbox{$((\f{\rd \th_{\CH}}{\rd \th})^{-1})_A{ }^B$ is}&\mbox{ continuously extendible to $\CH$}\\
\mbox{with respect to the differential}&\mbox{ structure corresponding to $(u,\ub_{\CH},\th^1,\th^2)$.}
\end{split}
\end{equation}
By virtue of \eqref{dthdth.ce}, if an $S$-tangent tensor $\phi$ is continuous up to the boundary $\CH$ in the $(u,\ub_{\CH},\th^1,\th^2)$ coordinate system, then $\phi$ is also continuous up to the boundary $\CH$ in the $(u,\ub_{\CH},\th^1_{\CH},\th^2_{\CH})$ coordinate system; see already Corollary~\ref{prop:sphere.eq}.

\subsection{Additional estimates for $\protect\nab_4\protect\widetilde{\protect\log\Om}$}\label{sec.nab4logOm.est}

In this subsection, we give some technical estimates for $\nab_4\widetilde{\log\Om}\doteq \nab_4(\log\Om-\log\Om_\Ke)$. This will be used in the next subsection to show the continuity of various geometric quantities up to $\CH$.

The following simple lemma will be useful for deriving the estimates in this section:
\begin{lemma}\label{lem:34.commute}
Let $\phi$ be a scalar function. $\nab_3(\nab^i(\Om^2\nab_4\phi))$ obeys the following schematic equation for $i\leq 3$,
\begin{equation*}
\begin{split}
\nab_3(\nab^i(\Om^2\nab_4\phi)) 
\eqs& \sum_{i_1+i_2+i_3 = i} \nab^{i_1}\psi^{i_2}\nab^{i_3} (\Om^2 \nab_4\nab_3\phi)+ \sum_{i_1+i_2+i_3+i_4 = i} \Om^2 \nab^{i_1}\psi^{i_2}\nab^{i_3}\psi \nab^{i_4+1}\phi\\
& + \sum_{i_1+i_2+i_3+i_4 = i} \nab^{i_1}\psi^{i_2}\nab^{i_3}\psi_{\Hb} \nab^{i_4} (\Om^2\nab_4\phi).
\end{split}
\end{equation*}

\end{lemma}
\begin{proof}
We prove this lemma by using
$$\nab_3(\nab^i(\Om^2\nab_4\phi)) = [\nab_3,\nab^i\Om^2\nab_4]\phi + \nab^i(\Om^2\nab_4\nab_3\phi),$$
where we note that the last term is an acceptable term, i.e.~it is of the form of one of terms on the right hand side. For the commutator term, we have
$$[\nab_3,\nab^i\Om^2\nab_4] = [\nab_3,\nab^i]\Om^2 \nab_4 + \nab^i [\nab_3,\Om^2\nab_4 ].$$
The desired schematic equation hence follows from repeated applications of the formulae in Propositions~\ref{commute} and \ref{repeated.com}. 
Let us note the following two key points:
\begin{itemize}
\item Since $\nab_3\log\Om = -\omb$ (by \eqref{gauge.con}), if we look at $[\nab_3,\Om^2\nab_4]$ instead of $\Om^2[\nab_3,\nab_4]$, we cancel off the term $2\Om^2\omb\nab_4 \phi$.
\item When applying the formula for $[\nab_3,\nab_4]$ in Propositions~\ref{commute}, note that since $\phi$ is a scalar function, the term $\sum_{i=1}^r 2(\gamma^{-1})^{BC}(\etab_{A_i}\eta_B-\eta_{A_i}\etab_B+\in_{A_iB}\sigma)\phi_{A_1...\hat{A}_iC...A_r}$ is absent.
\end{itemize}
\qedhere
\end{proof}

\begin{proposition}\label{lemma.nab4logOm}
Choosing $\ep_0>0$ smaller if necessary\footnote{Since we have already closed our bootstrap argument in Section~\ref{recover.bootstrap}, we could in principle carry out this argument without further choosing $\ep_0$ to be small. This is only done for expositional simplicity.}, the following estimate for $\nab_4\widetilde{\log\Om}$ holds in $\mathcal U_\infty$:
\begin{equation}\label{prop.nab4logOm.main}
\sum_{i\leq 3}\|\ub^{\f 12+\de}\nab^i(\Om^2\nab_4\widetilde{\log\Om})\|_{L^{\i}_uL^2_{\ub}L^2(S)}\ls \ep.
\end{equation}
Moreover, the following bounds hold:
\begin{equation}\label{prop.nab4logOm}
\|\Om_\Ke^2\nab_4\widetilde{\log\Om}\|_{L^{\i}_uL^1_{\ub}L^{\i}(S)}\ls \ep
\end{equation}
and
\begin{equation}\label{prop.nab4nablogOm}
\|\Om_\Ke^2\nab_4\nab\widetilde{\log\Om}\|_{L^{\i}_uL^1_{\ub}L^{\i}(S)}\ls \ep.
\end{equation}

\end{proposition}
\begin{proof}
We will first estimate the weighted $L^\i_uL^2_{\ub}L^2(S)$ norm of $\nab^i\nab_4\widetilde{\log\Om}$ for $i\leq 3$ as in \eqref{prop.nab4logOm.main}. Applying Propositions \ref{Sobolev} and \ref{commute} and using the estimates for $\NS$ proved in Theorem~\ref{main.quantitative.thm}, this then gives \eqref{prop.nab4logOm} and \eqref{prop.nab4nablogOm} (see \eqref{prop.nab4logOm.prelim}, \eqref{prop.nab4logOm.1} and \eqref{prop.nab4nablogOm.1}). Note that since we do not directly have a relation between $\nab_4\log\Om$ and the Ricci coefficients, we will use the equation $\omb=-\nab_3\log\Om$ from \eqref{gauge.con} to control $\nab_4\nab\log\Om$.

\textbf{Proof of \eqref{prop.nab4logOm.main} for $i\leq 2$.} Applying Lemma~\ref{lem:34.commute} with $\phi= \widetilde{\log\Om}$, we have, for $i\leq 2$, the following reduced schematic equation:
\begin{equation}\label{nab3nab4logOm}
\nab_3(\nab^i\Om^2\nab_4\widetilde{\log\Om})\eqrs\mathcal F_1+\sum_{i_1+i_2+i_3\leq i}(1+\nab^{i_1}(\tg,\tp)) (\Om^2_\Ke+\Om^2_\Ke\nab^{i_2}\tg+\nab^{i_2}\tpHb)(\nab^{i_3}(\Om^2\nab_4\widetilde{\log\Om})),
\end{equation}
where $\mathcal F_1$ is as in \eqref{inho.def}. Indeed, for the term $\sum_{i_1+i_2+i_3 = i} \nab^{i_1}\psi^{i_2}\nab^{i_3} (\Om^2 \nab_4\nab_3\widetilde{\log\Om})$ (i.e.~the first term on the right hand side in Lemma~\ref{lem:34.commute}), we use the fact $\nab_3 \widetilde{\log\Om} = -\widetilde{\omb}$ (which follows from \eqref{gauge.con} and the fact that when acting on a scalar function, $\nab_3 = (\nab_3)_\Ke$) and the equation for $\nab_4\widetilde{\omb}$ (cf.~Proposition~\ref{tpHb.eqn}) to deduce that this term has the reduced schematic form of $\mathcal F_1$. The second term in Lemma~\ref{lem:34.commute} can easily be seen to be of the form $\mathcal F_1$ after using the estimates in Theorem~\ref{main.quantitative.thm}. (Note in particular that $\nab\widetilde{\log\Om}$ can be written in terms of $\widetilde{\eta}$ and $\widetilde{\etab}$ by \eqref{gauge.con}.) Finally, for the term $\sum_{i_1+i_2+i_3+i_4 = i} \nab^{i_1}\psi^{i_2}\nab^{i_3}\psi_{\Hb} \nab^{i_4} (\Om^2\nab_4\widetilde{\log\Om})$, we use the estimates in Theorem~\ref{main.quantitative.thm} to control $\psi$, $\psi_{\Hb}$ and their derivatives and show that they take the form of the second term in \eqref{nab3nab4logOm}.

Let $\bar{N}$ be a large parameter to be chosen later. By Proposition~\ref{transport.3.1}, we obtain\footnote{Here, we actually use a slight variant of Proposition~\ref{transport.3.1}. Namely, we have introduced a new parameter $\bar{N}$ in place of $N$, which will be chosen to be large. Here, we recall that since $N$ has been fixed earlier, we suppress the explicit dependence of $N$ in our notations and absorb them into the implicit constant in $\ls$.}
\begin{equation}\label{main.est.for.nab4logOm.ileq2}
\begin{split}
&\|\ub^{\f 12+\de}\varpi^{\bar{N}}\nab^i(\Om^2\nab_4\widetilde{\log\Om})\|_{L^2_{\ub}L^{\i}_uL^2(S)}^2+\bar{N}\|\ub^{\f 12+\de}\varpi^{\bar{N}}\Om_\Ke\nab^i(\Om^2\nab_4\widetilde{\log\Om})\|_{L^2_uL^2_{\ub}L^2(S)}^2\\
\ls &2^{2\bar{N}}\ep+\|\ub^{1+2\delta} \varpi^{2\bar{N}}\nab^i(\Om^2\nab_4\widetilde{\log\Om}) \mathcal F_1\|_{L^1_uL^1_{\ub}L^1(S)}\\
&+\sum_{i_1+i_2+i_3\leq i}\|\ub^{1+2\de} \varpi^{2\bar{N}} \nab^i(\Om^2\nab_4\widetilde{\log\Om})(1+\nab^{i_1}(\tg,\tp)) (\Om^2_\Ke+\Om^2_\Ke\nab^{i_2}\tg+\nab^{i_2}\tpHb)(\nab^{i_3}(\Om^2 \nab_4\widetilde{\log\Om}))\|_{L^1_uL^1_{\ub}L^1(S)},
\end{split}
\end{equation}
where we have used \eqref{data.D.def}, \eqref{varpi.bd} and \eqref{small.data} to control the initial data term.

Recall now that $\mathcal F_1$ can be decomposed into $\mathcal F_{1,1}$ and $\mathcal F_{1,2}$ as in \eqref{F11.def} and \eqref{F12.def} in Proposition~\ref{prop.tpH}. Using the Cauchy--Schwarz inequality, the estimates for $\mathcal F_{1,1}$ in \eqref{F11.first.bound} and the bounds for $\Er_1$ and $\Er_2$ in Propositions~\ref{E1.bd}, \ref{E2.bd} and \ref{N.bd}, we have
\begin{equation}\label{nab4logOm.ileq2.main.0}
\begin{split}
&\:\|\ub^{1+2\delta} \varpi^{2\bar{N}}\nab^i(\Om^2\nab_4\widetilde{\log\Om}) \mathcal F_{1,1}\|_{L^1_uL^1_{\ub}L^1(S)}\\
\ls &\: \|\ub^{\f 12+\de}\varpi^{\bar{N}}\Om_\Ke\nab^i(\Om^2\nab_4\widetilde{\log\Om})\|_{L^2_uL^2_{\ub}L^2(S)}^2 + \|\ub^{\frac 12+\delta}\varpi^{\bar{N}} \Er_1\|_{L^2_uL^2_{\ub}L^2(S)}^2+\bar{N}^{-1}\|\ub^{\frac 12+\delta} \varpi^{\bar{N}} \Er_2\|_{L^2_uL^2_{\ub}L^2(S)}^2\\
\ls &\: \|\ub^{\f 12+\de}\varpi^{\bar{N}}\Om_\Ke\nab^i(\Om^2\nab_4\widetilde{\log\Om})\|_{L^2_uL^2_{\ub}L^2(S)}^2 + 2^{2\bar{N}}\ep^2.
\end{split}
\end{equation}
The $\mathcal F_{1,2}$ term can be controlled as follows using the estimates for $\NH$ and $\NS$ in Theorem~\ref{main.quantitative.thm}, \eqref{varpi.bd}, H\"older's inequality, Remark~\ref{rmk:norm.order} and Proposition~\ref{Sobolev} (cf.~Section~\ref{sec:rmk.BA.S}):
\begin{equation}\label{nab4logOm.ileq2.main.1}
\begin{split}
& \|\ub^{1+2\de} \varpi^{2\bar{N}} \nab^i(\Om^2\nab_4\widetilde{\log\Om})\mathcal F_{1,2}\|_{L^1_uL^1_{\ub}L^1(S)}\\
\ls & 2^{\bar{N}}\|\ub^{\f 12+\de} \varpi^{\bar{N}} \nab^i(\Om^2\nab_4\widetilde{\log\Om})\|_{L^2_{\ub}L^{\i}_uL^2(S)}(1+\sum_{i_1\leq 3}\|(\nab^{i_1}\tg,\nab^{\min\{i_1,2\}}\tp)\|_{L^\i_uL^\i_{\ub}L^2(S)})\\
&\quad\times(\sum_{i_2\leq 2} \|\nab^{i_2}\tpHb\|_{L^1_uL^\i_{\ub}L^2(S)})(\sum_{i_3\leq 3}\|\ub^{\f 12+ \de} \Om_\Ke^2\nab^{i_3}\tpH \|_{L^{\i}_uL^2_{\ub}L^2(S)})\\
&+ 2^{\bar{N}} \|\ub^{\f 12+\de} \varpi^{\bar{N}} \nab^i(\Om^2\nab_4\widetilde{\log\Om})\|_{L^2_{\ub}L^{\i}_uL^2(S)}(1+\sum_{i_1\leq 3}\|(\nab^{i_1}\tg,\nab^{\min\{i_1,2\}}\tp)\|_{L^\i_uL^\i_{\ub}L^2(S)})\||u|^{-\f 12-\de}\|_{L^2_u}\\
&\quad\times(\sum_{i_2\leq 3} \||u|^{\f 12+\de}\nab^{i_2}\tpHb\|_{L^\i_{\ub}L^2_uL^2(S)})(\sum_{i_3\leq 2}\|\ub^{\f 12+ \de} \Om_\Ke^2\nab^{i_3}\tpH \|_{L^2_{\ub}L^{\i}_uL^2(S)})\\
\ls & 2^{\bar{N}}\ep^2\|\ub^{\f 12+\de} \varpi^{\bar{N}} \nab^i(\Om^2\nab_4\widetilde{\log\Om})\|_{L^\i_uL^2_{\ub}L^2(S)}.
\end{split}
\end{equation}
The last term in \eqref{main.est.for.nab4logOm.ileq2} can be estimated using H\"older's inequality, Sobolev embedding (Proposition~\ref{Sobolev}) and the estimates in Theorem~\ref{main.quantitative.thm} as follows (cf.~Section~\ref{sec:rmk.BA.S}):
\begin{equation}\label{nab4logOm.ileq2.main.2}
\begin{split}
&\sum_{i_1+i_2+i_3\leq 2}\|\ub^{1+2\de} \varpi^{2\bar{N}} \nab^i(\Om^2\nab_4\widetilde{\log\Om})(1+\nab^{i_1}(\tg,\tp)) (\Om^2_\Ke+\Om^2_\Ke\nab^{i_2}\tg+\nab^{i_2}\tpHb)(\nab^{i_3}(\Om^2 \nab_4\widetilde{\log\Om}))\|_{L^1_uL^1_{\ub}L^1(S)}\\
\ls & (1+\sum_{i_1\leq 2}\|\nab^{i_1}(\tg,\tp)\|_{L^\i_uL^\i_{\ub} L^2(S)})(\sum_{i_2\leq 2}\|\ub^{1+2\de} \varpi^{\bar{N}}\Om_\Ke(\nab^{i_2}(\Om^2 \nab_4\widetilde{\log\Om}))\|_{L^2_uL^2_{\ub}L^2(S)}^2)\\
&+(1+\sum_{i_1\leq 2}\|\nab^{i_1}(\tg,\tp)\|_{L^\i_uL^\i_{\ub} L^2(S)}) (\sum_{i_2\leq 2}\|\nab^{i_2}\tpHb\|_{L^1_uL^\i_{\ub}L^2(S)})(\sum_{i_3\leq 2}\|\ub^{\f 12+ \de} \varpi^{\bar{N}}(\nab^{i_3}(\Om^2 \nab_4\widetilde{\log\Om}))\|_{L^{\i}_uL^2_{\ub}L^2(S)}^2)\\
\ls &\sum_{i\leq 2}\|\ub^{1+2\de} \varpi^{\bar{N}}\Om_\Ke(\nab^{i}(\Om^2 \nab_4\widetilde{\log\Om}))\|_{L^2_uL^2_{\ub}L^2(S)}^2+\ep\sum_{i\leq 2}\|\ub^{\f 12+\de} \varpi^{\bar{N}}(\nab^{i}(\Om^2 \nab_4\widetilde{\log\Om}))\|_{L^2_{\ub}L^{\i}_uL^2(S)}^2.
\end{split}
\end{equation}
Plugging the estimate \eqref{nab4logOm.ileq2.main.0}, \eqref{nab4logOm.ileq2.main.1} and \eqref{nab4logOm.ileq2.main.2} into \eqref{main.est.for.nab4logOm.ileq2}, choosing $\bar{N}^{-1}$, $\ep_0$ (and hence $\ep$) sufficiently small (in that order), and absorbing terms to the left hand side, this implies
\begin{equation}\label{nab4logOm.main.2}
\begin{split}
&\sum_{i\leq 2}\|\ub^{\f 12+\de}\varpi^{\bar{N}}\nab^i(\Om^2\nab_4\widetilde{\log\Om})\|_{L^2_{\ub}L^{\i}_uL^2(S)}^2+\sum_{i\leq 2} \|\ub^{\f 12+\de}\varpi^{\bar{N}}\Om_\Ke\nab^i(\Om^2\nab_4\widetilde{\log\Om})\|_{L^2_uL^2_{\ub}L^2(S)}^2\ls 2^{2\bar{N}}\ep^2.
\end{split}
\end{equation}

\textbf{Proof of \eqref{prop.nab4logOm.main} for $i= 3$.} In the $i=3$ case, the equation for $\nab_3\left(\nab^3\left(\Om^2\nab_4\widetilde{\log\Om}\right)\right)$ has a term $\Om^2\nab^3\tK$, which has more derivatives than is allowed by $\mathcal F_1$. We therefore need to introduce a renormalisation that we now explain. Instead of estimating general third derivatives of $\Om^2\nab_4\widetilde{\log\Om}$, we consider the reduced schematic equation only for a special combination of three angular derivatives\footnote{Recall here our convention that terms on the left hand side of the reduced schematic equation carry precise signs and constant.}, namely when the three angular derivatives take the form $\nab\slashed\Delta$. Using Proposition~\ref{lem:34.commute}, we have\footnote{Notice that according to Proposition~\ref{lem:34.commute}, we have a term with $\nab^4\widetilde{\log\Om}$. While we do not allow four derivatives of $\tg$ in $\mathcal F_1$, this term is nonetheless acceptable since by \eqref{gauge.con}, $\nab\log\Om = \f 12(\eta+\etab)$ and the $\nab^4\widetilde{\log\Om}$ can be considered as a $\nab^3\tp$ term (plus acceptable error terms).}
\begin{equation}\label{nab4logOm.renorm.1}
\begin{split}
&\:\nab_3( \nab\slashed\Delta(\Om^2\nab_4\widetilde{\log\Om}))- \nab\slashed\Delta (\Om^2 \tK)\\
\eqrs &\: \mathcal F_{1}+\sum_{i_1+i_2+i_3\leq 3}(1+(\nab^{i_1}\tg,\nab^{\min\{i_1,2\}}\tp)) (\Om^2_\Ke+\Om^2_\Ke\nab^{i_2}\tg+\nab^{i_2}\tpHb)\nab^{i_3}(\Om^2\nab_4\widetilde{\log\Om}).
\end{split}
\end{equation}
This can be derived in a similar manner as \eqref{nab3nab4logOm}, except that we put the term $\nab\slashed\Delta (\Om^2 \tK)$, which has three derivatives on $\tK$, on the left hand side. Consider now the reduced schematic equation
$$\nab_3\nab^2_{AB}(\Om^2\widetilde{\beta}_C)+\slashed{\nabla}_C(\Om^2 \nab^2_{AB}\tK)  -\in_{CD}\slashed{\nabla}^D(\Om^2\nab^2_{AB}\widetilde{\sigmac})
 \eqrs \mathcal F_1,$$
which is a variation from that in Proposition \ref{curv.red.sch.1} (with an identical proof). Contracting this with $(\gamma^{-1})^{BC}$, we obtain
\begin{equation}\label{nab4logOm.renorm.2}
(\gamma^{-1})^{BC}\nab_3\nab^2_{AB}(\Om^2\widetilde{\beta}_C)+(\gamma^{-1})^{BC}\slashed{\nabla}_C(\Om^2 \nab^2_{AB}\tK)  
 \eqrs \mathcal F_1,
\end{equation}
where the term $(\gamma^{-1})^{BC}\in_{CD}\slashed{\nabla}^D(\Om^2\nab^2_{AB}\widetilde{\sigmac})$ drops off since by the symmetry properties, we do not have the term with $3$ derivatives on $\widetilde{\sigmac}$ and hence this term can be expressed as $(\gamma^{-1})^{BC}\in_{CD}\slashed{\nabla}^D(\Om^2\nab^2_{AB}\widetilde{\sigmac})\eqrs\mathcal F_1$. Moreover, we have
\begin{equation}\label{nab4logOm.renorm.3}
(\gamma^{-1})^{BC}\slashed{\nabla}_C(\Om^2 \nab^2_{AB}\tK)-\Om^2\nab_A\slashed\Delta\tK\eqrs\mathcal F_1.
\end{equation}
Combining \eqref{nab4logOm.renorm.1}, \eqref{nab4logOm.renorm.2} and \eqref{nab4logOm.renorm.3}, we thus obtain
\begin{equation}\label{nab4logOm.renorm.4}
\begin{split}
&\:\nab_3\left( \nab_A\slashed\Delta(\Om^2\nab_4\widetilde{\log\Om})+(\gamma^{-1})^{BC}\nab^2_{AB}(\Om^2\widetilde{\beta}_C)\right)\\
\eqrs &\:\mathcal F_1+\sum_{i_1+i_2+i_3\leq 3}(1+\nab^{i_1}(\tg,\tp)) (\Om^2_\Ke+\Om^2_\Ke\nab^{i_2}\tg+\nab^{i_2}\tpHb)\nab^{i_3}(\Om^2\nab_4\widetilde{\log\Om}).
\end{split}
\end{equation}
Denote by $F$ the $S$-tangent $1$-form
\begin{equation}\label{F.nab4logOm.renorm}
F_A\doteq \nab_A\slashed\Delta(\Om^2\nab_4\widetilde{\log\Om})+(\gamma^{-1})^{BC}\nab^2_{AB}(\Om\Om_\Ke\widetilde{\beta}_C).
\end{equation}
\eqref{nab4logOm.renorm.4} is then a transport equation for $F_A$. We then apply Proposition~\ref{transport.3.1} as in the derivation of \eqref{main.est.for.nab4logOm.ileq2}. Estimating also $\mathcal F_1$ as in \eqref{nab4logOm.ileq2.main.0} and \eqref{nab4logOm.ileq2.main.1}, and absorbing terms to the left hand side appropriately, we have
\begin{equation}\label{nab4logOm.3}
\begin{split}
&\|\ub^{\f 12+\de}\varpi^{\bar{N}} F\|_{L^2_{\ub} L^{\i}_uL^2(S)}^2+\bar{N}\|\ub^{\f 12+\de}\varpi^{\bar{N}}\Om_\Ke F\|_{L^2_uL^2_{\ub}L^2(S)}^2\\
\ls &2^{2\bar{N}}\ep^2\\
&+\sum_{i_1+i_2+i_3\leq 3}\|\ub^{1+2\de} \varpi^{2\bar{N}} F(1+(\nab^{i_1}\tg,\nab^{\min\{i_1,2\}}\tp)) (\Om^2_\Ke+\Om^2_\Ke\nab^{i_2}\tg+\nab^{i_2}\tpHb)(\nab^{i_3}(\Om^2 \nab_4\widetilde{\log\Om}))\|_{L^1_uL^1_{\ub}L^1(S)}.
\end{split}
\end{equation}
We apply the elliptic estimate in Proposition~\ref{elliptic.Poisson} and use \eqref{F.nab4logOm.renorm} to obtain
\begin{equation}\label{nab4logOm.3.1}
\begin{split}
& \|\ub^{\f 12+\de} \varpi^{\bar{N}}\nab^3 (\Om^2  \nab_4\widetilde{\log\Om})\|_{L^{\i}_uL^2_{\ub}L^2(S)}\\
\ls & \sum_{i\leq 1}\|\ub^{\f 12+\de} \varpi^{\bar{N}} \nab^i\slashed\Delta(\Om^2  \nab_4\widetilde{\log\Om})\|_{L^{\i}_uL^2_{\ub}L^2(S)}+\sum_{i\leq 2}\|\ub^{\f 12+\de} \varpi^{\bar{N}}\nab^i (\Om^2  \nab_4\widetilde{\log\Om})\|_{L^{\i}_uL^2_{\ub}L^2(S)} \\
\ls & \|\ub^{\f 12+\de}\varpi^{\bar{N}} F\|_{L^{\i}_uL^2_{\ub}L^2(S)} +\sum_{i\leq 2}\|\ub^{\f 12+\de} \varpi^{\bar{N}}\nab^i (\Om^2  \nab_4\widetilde{\log\Om})\|_{L^{\i}_uL^2_{\ub}L^2(S)} +\|\ub^{\f 12+\de}\varpi^{\bar{N}} \nab^2(\Om\Om_\Ke \widetilde{\beta})\|_{L^{\i}_uL^2_{\ub}L^2(S)}\\
\ls & \|\ub^{\f 12+\de}\varpi^{\bar{N}} F\|_{L^{\i}_uL^2_{\ub}L^2(S)} +2^{\bar{N}}\ep,
\end{split}
\end{equation}
where in the last line we have used the estimates in \eqref{nab4logOm.main.2}, the $\NH$ bound for $\widetilde{\beta}$ and the $\NS$ bound for $\widetilde{\log\Om}$ proven in Theorem~\ref{main.quantitative.thm}.

Similarly, using Proposition~\ref{elliptic.Poisson}, \eqref{F.nab4logOm.renorm}, \eqref{nab4logOm.main.2} and Theorem~\ref{main.quantitative.thm}, we have
\begin{equation}\label{nab4logOm.3.3}
\begin{split}
&\|\ub^{\f 12+\de} \varpi^{\bar{N}}\Om_\Ke\nab^3 (\Om^2  \nab_4\widetilde{\log\Om})\|_{L^{2}_uL^2_{\ub}L^2(S)}\\
\ls & \|\ub^{\f 12+\de} \varpi^{\bar{N}}\Om_\Ke F\|_{L^{2}_uL^2_{\ub}L^2(S)} +\sum_{i\leq 2}\|\ub^{\f 12+\de} \varpi^{\bar{N}}\nab^i (\Om^2  \nab_4\widetilde{\log\Om})\|_{L^2_uL^2_{\ub}L^2(S)} + \|\ub^{\f 12+\de}\Om_\Ke \varpi^{\bar{N}}\nab^2(\Om\Om_\Ke \widetilde{\beta})\|_{L^2_uL^2_{\ub}L^2(S)}\\
\ls & \|\ub^{\f 12+\de} \varpi^{\bar{N}}\Om_\Ke F\|_{L^{2}_uL^2_{\ub}L^2(S)} + 2^{\bar{N}}\ep.
\end{split}
\end{equation}
We now estimate the right hand side of \eqref{nab4logOm.3} using H\"older's inequality together with Sobolev embedding (Proposition~\ref{Sobolev}); cf.~Section~\ref{sec:rmk.BA.S}:
\begin{equation*}
\begin{split}
&\: \sum_{i_1+i_2+i_3\leq 3}\|\ub^{1+2\de} \varpi^{2\bar{N}} F(1+(\nab^{i_1}\tg,\nab^{\min\{i_1,2\}}\tp)) (\Om^2_\Ke+\Om^2_\Ke\nab^{i_2}\tg+\nab^{i_2}\tpHb)(\nab^{i_3}(\Om^2 \nab_4\widetilde{\log\Om}))\|_{L^1_uL^1_{\ub}L^1(S)}\\
\ls &\: (1+\sum_{i_1\leq 3}\|(\nab^{i_1}\tg,\nab^{\min\{i_1,2\}}\tp)\|_{L^\i_uL^\i_{\ub}L^2(S)})\|\ub^{\f 12+\de}\varpi^{\bar{N}} \Om_\Ke F\|_{L^2_uL^2_{\ub}L^2(S)}\\
&\:\quad\times(\sum_{i_2\leq 3}\|\ub^{\f 12+\de} \varpi^{\bar{N}}\Om_\Ke \nab^{i_2}(\Om^2\nab_4\widetilde{\log\Om})\|_{L^2_uL^2_{\ub}L^2(S)})\\
&\:+(1+\sum_{i_1\leq 3}\|(\nab^{i_1}\tg,\nab^{\min\{i_1,2\}}\tp)\|_{L^{\i}_uL^{\i}_{\ub}L^2(S)})(\sum_{i_2\leq 2} \|\nab^{i_2}\tpHb\|_{L^{1}_{u} L^{\i}_{\ub} L^2(S)})\|\ub^{\f 12+\de}\varpi^{\bar{N}} F\|_{L^\i_uL^2_{\ub}L^2(S)}\\
&\:\quad\times(\sum_{i_3\leq 3}\|\ub^{\f 12+\de}\varpi^{\bar{N}} \nab^{i_3}(\Om^2\nab_4\widetilde{\log\Om})\|_{L^{\i}_u L^2_{\ub}L^2(S)})\\
&\:+(1+\sum_{i_1\leq 3}\|(\nab^{i_1}\tg,\nab^{\min\{i_1,2\}}\tp)\|_{L^{\i}_uL^{\i}_{\ub}L^2(S)})(\sum_{i_2\leq 3} \||u|^{\f 12+\de}\nab^{i_2}\tpHb\|_{L^{\i}_{\ub} L^2_{u}  L^2(S)})\|\ub^{\f 12+\de}\varpi^{\bar{N}} F\|_{L^2_{\ub}L^\i_uL^2(S)}\\
&\:\quad\times(\sum_{i_3\leq 2}\|\ub^{\f 12+\de}\varpi^{\bar{N}} \nab^{i_3}(\Om^2\nab_4\widetilde{\log\Om})\|_{L^2_{\ub} L^{\i}_uL^2(S)})\||u|^{-\f12-\de}\|_{L^2_u}\\
\ls & \:2^{2\bar{N}}\ep^2 + \ep^2\|\ub^{\f 12+\de}\varpi^{\bar{N}} F\|_{L^2_{\ub} L^{\i}_uL^2(S)}^2+\|\ub^{\f 12+\de}\varpi^{\bar{N}}\Om_\Ke F\|_{L^2_uL^2_{\ub}L^2(S)}^2,
\end{split}
\end{equation*}
where in the last line we have used \eqref{nab4logOm.main.2}, \eqref{nab4logOm.3.1} and \eqref{nab4logOm.3.3}, the estimates for $\NS$ and $\NH$ proved in Theorem~\ref{main.quantitative.thm}, and Young's inequality. We then plug this into the right hand side of \eqref{nab4logOm.3}. Choosing $\ep_0$ (and hence $\ep$) sufficiently small and $\bar{N}$ sufficiently large, the last two terms on the right hand side can be absorbed to the left hand side. This implies
$$\|\ub^{\f 12+\de}\varpi^{\bar{N}} F\|_{L^2_{\ub} L^{\i}_uL^2(S)}^2+\bar{N}\|\ub^{\f 12+\de}\varpi^{\bar{N}}\Om_\Ke F\|_{L^2_uL^2_{\ub}L^2(S)}^2 \ls 2^{2\bar{N}}\ep^2.$$
Substituting this into \eqref{nab4logOm.3.1} and \eqref{nab4logOm.3.3}, we then obtain
$$\|\ub^{\f 12+\de} \varpi^{\bar{N}}\nab^3 (\Om^2  \nab_4\widetilde{\log\Om})\|_{L^{\i}_uL^2_{\ub}L^2(S)}^2 + \bar{N} \|\ub^{\f 12+\de} \varpi^{\bar{N}}\Om_\Ke\nab^3 (\Om^2  \nab_4\widetilde{\log\Om})\|_{L^{2}_uL^2_{\ub}L^2(S)}^2 \leq 2^{2\bar{N}}\bar{N}\ep^2.$$
At this point, we fix $\bar{N}$. Combining this with \eqref{nab4logOm.main.2}, we thus conclude the proof of \eqref{prop.nab4logOm.main}.

\textbf{Proof of \eqref{prop.nab4logOm} and \eqref{prop.nab4nablogOm}.} Applying Sobolev embedding (Proposition~\ref{Sobolev}) to \eqref{prop.nab4logOm.main}, we obtain
\begin{equation}\label{prop.nab4logOm.prelim}
\sum_{i\leq 1}\|\ub^{\f 12+\de}\nab^i(\Om^2\nab_4\widetilde{\log\Om})\|_{L^{\i}_uL^2_{\ub}L^{\i}(S)}\ls \ep.
\end{equation}
Using the fact $\Om\ls \Om_\Ke$, this implies
\begin{equation}\label{prop.nab4logOm.1}
\|\ub^{\f 12+\de}\Om_\Ke^2\nab_4\widetilde{\log\Om}\|_{L^{\i}_uL^2_{\ub}L^{\i}(S)}\ls \ep.
\end{equation}
Moreover, bounding $\nab\log\Om$ using Theorem~\ref{main.quantitative.thm}, controlling the commutator $[\nab_4,\nab]$ using Proposition~\ref{commute} and the estimates in Theorem~\ref{main.quantitative.thm}, and using $\Om\ls \Om_\Ke$, \eqref{prop.nab4logOm.prelim} also implies
\begin{equation}\label{prop.nab4nablogOm.1}
\|\ub^{\f 12+\de}\Om_\Ke^2\nab_4\nab\widetilde{\log\Om}\|_{L^{\i}_uL^2_{\ub}L^{\i}(S)}\ls \ep.
\end{equation}
Using Cauchy--Schwarz inequality, \eqref{prop.nab4logOm} and \eqref{prop.nab4nablogOm} then follow from \eqref{prop.nab4logOm.1} and \eqref{prop.nab4nablogOm.1} respectively. \qedhere
\end{proof}

\subsection{Preliminary continuity estimates}\label{sec.prelim.cont}

In this subsection, our goal (see Lemma \ref{cont.ext.prelim} below) will be to show that some of the geometric quantities, which are originally defined on $\mathcal U_\infty$ can be extended continuously to $\CH$ (with respect to the differential structure corresponding to $(u,\ub_{\CH},\th^1,\th^2)$ coordinate system). 

We begin with an improvement (see Corollary \ref{Morrey.lemma}) of the Sobolev inequality in Proposition \ref{Sobolev}, which also gives a bound in the H\"older space in addition to a bound in $L^\infty$. Such an improved inequality is known as Morrey's inequality. We first define the $C^\alpha$ norm:
\begin{definition}
Let $S$ be a closed smooth $2$-manifold, $\gamma$ be a continuous Riemannian metric on $S$ and $\alpha\in (0,1)$. Then given a function $\xi:S\to \mathbb R$, we define the $C^{\alpha}$-norm by
$$\|\xi\|_{C^{\alpha}(S,\gamma)}\doteq \sup_{x,y\in S}\f{\left|\xi(x)-\xi(y) \right|_{\gamma(x)}}{\left(dist_\gamma(x,y)\right)^{\alpha}}+\|\xi\|_{L^\infty(S,\gamma)},$$
where $dist_\gamma(x,y)$ is the distance between the points $x$ and $y$ induced by the Riemannian metric $\gamma$.
\end{definition}

We will use the following standard Morrey's inequality on smooth Riemannian manifolds. For the purposes of our application, we also note that the constant in Morrey's inequality can be uniformly controlled under sufficiently small $C^0$ perturbations of the metric:
\begin{proposition}\label{Sobolev.Holder}
For any closed smooth $2$-manifold $S$ with a smooth Riemannian metric $\gamma$ and $0<\alpha<\f 12$, the following estimate holds for all functions $\xi: S\to \mathbb R$ with a constant $C=C(\gamma,\alpha)>0$ depending on the metric $\gamma$:
\begin{equation}\label{Sobolev.Holder.ineq}
\| \xi\|_{C^{\alpha}(S,\gamma)}\leq C(\sum_{i\leq 1}\|\nab^i\xi\|_{L^4(S,\gamma)}),
\end{equation}
Moreover, given a smooth metric $\gamma_1$, there exists $\ep_{\gamma_1}>0$ depending on $\gamma_1$ such that for every $C^0$ metric $\gamma_2$ satisfying $(\gamma_1^{-1})^{AB}(\gamma_1^{-1})^{CD}((\gamma_1)_{AC}-(\gamma_2)_{AC})((\gamma_1)_{BD}-(\gamma_2)_{BD})<\ep_{\gamma_1}$, \eqref{Sobolev.Holder.ineq} holds for $\gamma$ replaced by $\gamma_2$ with a constant $C$ depending on $\gamma_1$ and $\alpha$. 
\end{proposition}
\begin{proof}
The fact that \eqref{Sobolev.Holder.ineq} holds for a smooth Riemannian metric $\gamma$ is a standard result and can be found for instance in Theorem 2.8 in \cite{Hebey}. To prove the statement on $C^0$ perturbations, notice that \eqref{Sobolev.Holder.ineq} only depends on the metric $\gamma$ through the distance $dist_\gamma(x,y)$ in the definition of of the $C^\alpha$ norm, the norm of $|\nab\xi|_\gamma$ and the volume form on the right hand side. In all instances, the dependence on $\gamma$ is continuous under $C^0$ perturbations.
\end{proof}

Proposition \ref{Sobolev.Holder} immediately implies that in our setting, we can apply the Morrey's inequality with a uniform constant. More precisely, we have
\begin{corollary}\label{Morrey.lemma}
Let $0<\alpha<\f 12$. Choosing $\ep_0$ smaller if necessary, the following estimate\footnote{We have used the convention (consistent with the rest of the paper but slightly different from Proposition \ref{Sobolev.Holder}) that the explicit dependence of the norms on the metric $\gamma$ is suppressed.} holds for all functions $\xi:S_{u,\ub}\to \mathbb R$ with an implicit constant independent of $u$ and $\ub$ (but depending on $\alpha$):
$$\| \xi\|_{C^{\alpha}(S_{u,\ub})}\ls \sum_{i\leq 2}\|\nab^i\xi\|_{L^2(S_{u,\ub})}.$$
\end{corollary}
\begin{proof}
Using the first estimate in Proposition \ref{Sobolev}, it suffices to establish the estimate
\begin{equation}\label{Sobolev.Holder.ineq.M}
\| \xi\|_{C^{\alpha}(S_{u,\ub})}\leq C(\sum_{i\leq 1}\|\nab^i\xi\|_{L^4(S_{u,\ub})}),
\end{equation}
with a constant $C$ independent of $u$ and $\ub$. To this end, let us first note that the estimate \eqref{Sobolev.Holder.ineq.M} holds on the background Kerr spacetime with a uniform constant $C$ by using Corollary \ref{Morrey.lemma} and the fact that the metrics $(\gamma_\Ke)(u,\ub)$ converge smoothly to a smooth metric as $r_*=u+\ub\to \pm\infty$. Applying the $C^0$ stability statement in Proposition \ref{Sobolev.Holder} and the estimates for $\gamma-\gamma_\Ke$ in the $\NS$ energy proven in Theorem~\ref{main.quantitative.thm} (together with Proposition~\ref{Sobolev}), we thus obtain \eqref{Sobolev.Holder.ineq.M} on $\mathcal U_\infty$.
\end{proof}

With the above preparations, we finally state and prove the main result of this subsection, which is a statement that various geometric quantities can be extended continuously to $\CH$:
\begin{proposition}\label{cont.ext.prelim}
$\gamma_{AB}$, $\f{\Om^2}{\Om_\Ke^2}$, $\eta_A$, $\etab_A$ and $\nab_A\log\Om$ extend continuously to $\CH$ with respect to the differential structure corresponding to $(u,\ub_{\CH},\th^1,\th^2)$ coordinate system.
\end{proposition}
\begin{proof}
Let $p=(u_\infty,(\ub_{\CH})_\infty,\th^1_\infty,\th^2_\infty)\in \CH.$ Consider a sequence of points $(u_n,(\ub_{\CH})_n,\th^1_n,\th^2_n)$ in $\mathcal U_\infty$ converging to $p$. For notational convenience, let us denote also $\vartheta_n=(\th^1_n,\th^2_n)$.

Take $X_1=\alp\f{\rd}{\rd\th^1}$ and $X_2=\alp\f{\rd}{\rd\th^2}$ for some smooth cutoff function $\alp$ which $= 1$ at $(\th^1_\infty,\th^2_\infty)$ and is compactly supported in a neighbourhood of $(\th^1_\infty,\th^2_\infty)$. We will prove the continuous extendibility of $\gamma(X_i,X_j)$, $\f{\Om^2}{\Om_\Ke^2}$, $\eta(X_i)$, $\etab(X_i)$ and $X_i(\log\Om)$ for all $i,j=1,2$. (Notice that we are doing this since Corollary \ref{Morrey.lemma} applies only to scalars.) Let us denote schematically by $\Xi$ the quantities above. We will 
\begin{itemize}
\item define the extension of $\Xi$ at $p\in \CH$ by $\Xi(p)=\lim_{n\to \infty}\Xi(u_n,(\ub_{\CH})_n,\vartheta_n)$,  
\item prove that this definition is independent of the choice of the sequence of points in $\mathcal U_\infty$, and 
\item prove that the extension is indeed continuous. 
\end{itemize}
To carry this out, we first need to prove some estimates.

{\bf Comparing the distance function on different $2$-spheres.} We claim that for $\ep_0$ (and hence $\ep$) sufficiently small, there exists $C>0$ (depending on $M$, $a$) such that if $N$ is sufficiently large, then for every $n, \,m,\,k,\, \ell\geq N$,
\begin{equation}\label{dist.est}
dist_{(S_{u_\ell,(\ub_{\CH})_{\ell}},\gamma)}(\vartheta_n, \vartheta_m)\leq C dist_{(S_{u_k,(\ub_{\CH})_k},\gamma)}(\vartheta_n, \vartheta_m),
\end{equation}
where 
\begin{equation*}
\begin{split}
&dist_{(S_{u_\ell,(\ub_{\CH})_\ell},\gamma)}(\vartheta_n, \vartheta_m)\\
=&\inf\{\mbox{Length}(\sigma): \sigma \mbox{ is a curve on } S_{u_\ell,(\ub_{\CH})_\ell} \mbox{ connecting }(u_\ell,(\ub_{\CH})_{\ell}, \th^1_n,\th^2_n)\mbox{ and }(u_\ell,(\ub_{\CH})_{\ell}, \th^1_m,\th^2_m)\}.
\end{split}
\end{equation*}
That this is the case follows from the continuity of the background metric $\gamma_\Ke$ and the fact that according to the estimates on $\NS$, we have $|\gamma-\gamma_{\Ke}|_{\gamma_\Ke}\ls \epsilon$.

{\bf Estimates of the $L^\infty_u L^\infty_{\ub} C^\alpha(S)$ norm of $\Xi$ for $\alpha\in (0,\f 12)$.} According to Corollary~\ref{Morrey.lemma}, in order to bound the $L^\infty_u L^\infty_{\ub} C^\alpha(S)$ norm of $\Xi$, it suffices to control the $L^\i_u L^\i_{\ub} L^2(S)$ norm of $\Xi$, $\nab\Xi$ and $\nab^2\Xi$.

To this end, we first note that by the bounds on $\NS$ proven in Theorem~\ref{main.quantitative.thm},
\begin{equation}\label{cont.nab.est}
\sum_{i\leq 2}\|\nab^i (\gamma,\f{\Om^2}{\Om_\Ke^2},\eta,\etab,\nab\log\Om)\|_{L^\infty_{\ub} L^{\i}_{u}L^2(S)}\ls 1.
\end{equation}
Then, using the fact that $\gamma_\Ke$ is smooth up to $\CH$ and also using the estimates for $\gamma-\gamma_\Ke$ and its derivatives in the $\NS$ energy proven in Theorem~\ref{main.quantitative.thm}, we have\footnote{Recall that $X_i$ are smooth vector fields so it suffices to control the Christoffel symbols.}
\begin{equation}\label{nab.X.est}
\|\nab^2 X_i\|_{L^{\i}_uL^{\i}_{\ub} L^2(S)}+\|\nab X_i\|_{L^{\i}_uL^{\i}_{\ub} L^4(S)}\ls 1.
\end{equation}
Combining \eqref{cont.nab.est} and \eqref{nab.X.est} and using the product rule, we have
\begin{equation*}
\begin{split}
\sum_{i\leq 2} \|\nab^i\Xi\|_{L^{\i}_uL^{\i}_{\ub}L^2(S)}\ls 1.
\end{split}
\end{equation*}
Therefore, by Corollary \ref{Morrey.lemma}, we have
\begin{equation}\label{cont.Calp.est}
\|\Xi\|_{L^\infty_{\ub} L^{\i}_{u}C^\alpha(S)}\ls 1
\end{equation}
for every $\alpha\in (0,\f 12)$ (with an implicit constant depending on $\alpha$).

{\bf Estimates of the $L^\infty_{\ub} L^2_{u} L^\i(S)$ norm of $\nab_3\Xi$.}
First, we claim that by Theorem~\ref{main.quantitative.thm},
\begin{equation}\label{cont.nab3.est}
\|\nab_3 (\gamma,\f{\Om^2}{\Om_\Ke^2},\eta,\etab,\nab\log\Om)\|_{L^\infty_{\ub} L^2_{u}L^\i(S)}\ls 1.
\end{equation}
To see this, we bound each of these terms: \underline{Estimates for $\nab_3\gamma$.} This is trivial by \eqref{nab.compatibility}. \underline{Estimates for $\nab_3\f{\Om^2}{\Om_\Ke^2}$.} Notice that $\nab_3\f{\Om^2}{\Om_\Ke^2}=2\f{\Om^2}{\Om_\Ke^2}(\nab_3\log\Om-\nab_3\log\Om_\Ke)=2\f{\Om^2}{\Om_\Ke^2}(\omb_\Ke-\omb)$. We can thus estimate this using the bounds for $\NS$. \underline{Estimates for $\nab_3\etab$.} We use the equation for $\nab_3\etab$ in \eqref{null.str2}, the terms $\chib\cdot\eta$ and $\chib\cdot\etab$ on the right hand side can be bounded in $L^\infty_{\ub} L^{2}_{u}L^\i(S)$ using the bounds on $\NH$ and $\NS$ proven in Theorem~\ref{main.quantitative.thm} (and Proposition~\ref{Sobolev}) and the estimates for the Kerr Ricci coefficients in Proposition~\ref{Kerr.Ricci.bound}. (Notice that we use here in particular the estimate for $\Om_\Ke^2$ in Proposition~\ref{Om.Kerr.bounds}, which guarantees that the background $\chib_\Ke$ is in $L^{\i}_{\ub}L^2_uL^{\i}(S)$.) The term $\betab$ can be controlled similarly after using the schematic Codazzi equation for $\betab$ in \eqref{sch.Codazzi}. \underline{Estimates for $\nab_3\nab\log\Om$.} We rewrite this using \eqref{gauge.con} as $\nab_3\nab\log\Om=-\nab \omb+[\nab_3,\nab]\log\Om$. Now since $|\nab \omb_\Ke|\ls \Om_\Ke^2$ and is therefore bounded in $L^\infty_{\ub} L^2_{u} L^\i(S)$, in order to control $\nab\omb$, it suffices to bound $\nab\widetilde{\omb}$. This can in turn be bounded in $L^\infty_{\ub} L^2_{u} L^\i(S)$ using the bounds on $\NH$ established in Theorem~\ref{main.quantitative.thm} and the Sobolev embedding theorem (Proposition \ref{Sobolev}). On the other hand, using the formula in Proposition \ref{commute} for $[\nab_3,\nab]$ and \eqref{gauge.con} (which guarantees $\eta-\zeta=0$), it is then easy to see that $[\nab_3,\nab]\log\Om \eqs \chib\psi$, which can therefore be estimated in $L^\infty_{\ub} L^2_{u} L^\i(S)$ similarly as above. \underline{Estimate for $\nab_3\eta$.} This follows from \eqref{gauge.con} and the bounds for $\nab_3\etab$ and $\nab_3\nab\log\Om$ that we have just derived.

On the other hand, by \eqref{nab3.def}, in order to control $\nab_3 X_i$, it suffices to bound $\chib$. According to the estimates on $\NH$ for $\widetilde\chib$ and the bounds in Proposition~\ref{Kerr.Ricci.bound} for $\chib_\Ke$, $\chib$ is bounded in $L^{\i}_{\ub}L^{2}_uL^{\i}(S)$. Consequently,
\begin{equation}\label{nab3.X.est}
\|\nab_3 X_i\|_{L^{\i}_{\ub}L^2_uL^{\i}(S)}\ls 1.
\end{equation}
Therefore, by \eqref{cont.nab.est}, Proposition~\ref{Sobolev}, \eqref{cont.nab3.est} and \eqref{nab3.X.est}, it holds that
\begin{equation}\label{nab3.Xi.est}
\|\nab_3\Xi\|_{L^\infty_{\ub} L^2_u L^\i(S)}\ls 1.
\end{equation}
{\bf Estimates of the $L^\infty_u L^1_{\ub} L^\i(S)$ norm of the $\Om^2_{\Ke}\nab_4\Xi$.}
We first claim that 
\begin{equation}\label{cont.nab4.est}
\|\Om_\Ke^2\nab_4 (\gamma,\f{\Om^2}{\Om_\Ke^2},\eta,\etab,\nab\log\Om)\|_{L^\infty_u L^1_{\ub} L^\i(S)}\ls 1.
\end{equation}
To show this, we consider every term: \underline{Estimates for $\nab_4\gamma$.} This is trivial by \eqref{nab.compatibility}. \underline{Estimates for $\nab_4\f{\Om^2}{\Om_\Ke^2}$.} Note that $\nab_4\f{\Om^2}{\Om_\Ke^2}=2\f{\Om^2}{\Om_\Ke^2}(\nab_4\log\Om-\nab_4\log\Om_\Ke)$. The estimate thus follows from $\f{\Om}{\Om_\Ke}\ls 1$ and \eqref{prop.nab4logOm}. \underline{Estimates for $\nab_4\eta$.} We use the equation for $\nab_4\eta$ in \eqref{null.str2}. The terms $\chi\cdot\eta$ and $\chi\cdot\etab$ can be controlled in $L^\infty_u L^1_{\ub} L^\i(S)$ using the bound on $\NS$ established in Theorem~\ref{main.quantitative.thm} and Proposition~\ref{Kerr.der.Ricci.bound}. The term $\beta$ can be similarly bounded using \eqref{sch.Codazzi} and estimating $\nab\chi$ and $\psi\chi$ using the bound on $\NS$ in Theorem~\ref{main.quantitative.thm} and Proposition~\ref{Kerr.der.Ricci.bound}. \underline{Estimates for $\nab_4\nab\log\Om$.} This follows directly from \eqref{prop.nab4nablogOm} and Proposition~\ref{Om.der.Kerr.bounds}. \underline{Estimates for $\nab_4\etab$.} This follows from \eqref{gauge.con} and the bounds for $\nab_4\eta$ and $\nab_4\nab\log\Om$ that we have just derived.

In addition to the estimate \eqref{cont.nab4.est}, we need to bound $\Om_\Ke^2\nab_4 X_i$ in $L^\infty_u L^1_{\ub} L^{\i}(S)$. By \eqref{nab4.def}, since $X_i$ is a smooth vector field for $i=1,2$, it suffices to estimate $\Om^2_\Ke \chi$ and $(\f{\rd b^A}{\rd \th^B})$ in $L^\infty_u L^1_{\ub} L^{\i}(S)$. Furthermore, by \eqref{D.b.improved} and Propositions~\ref{Om.Kerr.bounds} and \ref{Kerr.Ricci.bound}, the corresponding Kerr value are bounded. Hence,
\begin{equation}\label{nab4.X.est}
\|\Om^2_{\Ke}\nab_4 X_i\|_{L^{\i}_u L^1_{\ub} L^{\i}(S)}\ls 1+ \|\Om^2_\Ke \widetilde{\chi}\|_{L^{\i}_u L^1_{\ub} L^{\i}(S)}+\sum_{i\leq 1}\|\nab^i \tb\|_{L^{\i}_u L^1_{\ub} L^{\i}(S)}\ls 1,
\end{equation}
where in the last inequality, we used the bound in $\NH$ and $\NS$ (proven in Theorem~\ref{main.quantitative.thm}), together with the Cauchy--Schwarz inequality and Sobolev embedding (Proposition \ref{Sobolev}).

Therefore, by \eqref{cont.nab.est}, Proposition~\ref{Sobolev}, \eqref{cont.nab4.est} and \eqref{nab4.X.est}, the following holds:
\begin{equation}\label{nab4.Xi.est}
\|\Om_\Ke^2\nab_4\Xi\|_{L^\infty_u L^1_{\ub} L^\i(S)}\ls 1.
\end{equation}
{\bf Conclusion of the proof.} Let $(u_n,(\ub_{\CH})_n,\vartheta_n)\in \mathcal U_\infty$ for $n\in \mathbb N$ be as introduced in the beginning of the proof, it follows from the triangle inequality that
\begin{equation*}
\begin{split}
&\left|\Xi(u_n,(\ub_{\CH})_n,\vartheta_n)-\Xi(u_m,(\ub_{\CH})_m,\vartheta_m)\right|\\
\leq & \underbrace{\left|\Xi(u_n,(\ub_{\CH})_n,\vartheta_n)-\Xi(u_n,(\ub_{\CH})_n,\vartheta_m)\right|}_{\doteq I_{n,m}}+\underbrace{\left|\Xi(u_{m},(\ub_{\CH})_n,\vartheta_{m})-\Xi(u_n,(\ub_{\CH})_n,\vartheta_m)\right|}_{\doteq II_{n,m}}\\
&+\underbrace{\left|\Xi(u_{m},(\ub_{\CH})_n,\vartheta_{m})-\Xi(u_{m},(\ub_{\CH})_{m},\vartheta_m)\right|}_{\doteq III_{n,m}}.
\end{split}
\end{equation*}
%\begin{equation*}
%\begin{split}
%&\left|\Xi(u_n,(\ub_{\CH})_n,\vartheta_n)-\Xi(u_\i,(\ub_{\CH})_\i,\vartheta_\i)\right|\\
%\leq & \underbrace{\left|\Xi(u_n,(\ub_{\CH})_n,\vartheta_n)-\Xi(u_n,(\ub_{\CH})_n,\vartheta_\i)\right|}_{\doteq I_n}+\underbrace{\left|\Xi(u_{\i},(\ub_{\CH})_n,\vartheta_{\i})-\Xi(u_n,(\ub_{\CH})_n,\vartheta_\i)\right|}_{\doteq II_n}\\
%&+\underbrace{\left|\Xi(u_{\i},(\ub_{\CH})_n,\vartheta_{\i})-\Xi(u_{\i},(\ub_{\CH})_{\i},\vartheta_\i)\right|}_{\doteq III_n}.
%\end{split}
%\end{equation*}
$I_{n,m}\to 0$ as $n,m\to\infty$ thanks to \eqref{dist.est} and \eqref{cont.Calp.est}. $II_{n,m}\to 0$ as $n,m\to\infty$ thanks to \eqref{nab3.Xi.est}, the fundamental theorem of calculus and H\"older's inequality. Finally, to show that $III_{n,m}\to 0$ as $n,m\to \infty$, recall the definition of $\ub_{\CH}$ in the beginning of the subsection (see \eqref{CH.def.last}) from which we deduce $d\ub_{\CH}\sim \Om_\Ke^2\, d\ub$ on every finite $u$-interval (with an implicit constant depending on the interval). Therefore, \eqref{nab4.Xi.est} together with the fundamental theorem of calculus imply that $III_{n,m}\to 0$ as $n,m\to \infty$. Consequently, we have
\begin{equation}\label{Xi.Cauchy}
\left|\Xi(u_n,(\ub_{\CH})_n,\vartheta_n)-\Xi(u_m,(\ub_{\CH})_m,\vartheta_m)\right|\to 0 \mbox{ as }n,m\to \infty.
\end{equation}
Recall that $p=(u_\i,(\ub_{\CH})_\i,\vartheta_\i)\in \CH$. Now, we define 
\begin{equation}\label{Xi.limit.def}
\Xi(p)\doteq \lim_{n\to \infty}\Xi(u_n,(\ub_{\CH})_n,\vartheta_n).
\end{equation}
The computation \eqref{Xi.Cauchy} shows that the limit indeed exists. Moreover, since \eqref{Xi.Cauchy} holds for all sequences $(u_n,(\ub_{\CH})_n,\vartheta_n)$ in $\mathcal U_\infty$ converging to $p\in \CH$, the definition \eqref{Xi.limit.def} is independent of the choice of the sequence. Finally, it remains to show the continuity of (the extension of) $\Xi$. Consider $(u_n,(\ub_{\CH})_n,\vartheta_n)$ in $\mathcal U_\infty\cup \CH$ converging to $p \in \CH$. (We can assume that $p\in \CH$ since the metric is smooth in $\mathcal U_\infty$.) By \eqref{Xi.limit.def} and the discussion above, for any subsequence $(u_{n_k},(\ub_{\CH})_{n_k},\vartheta_{n_k})$ which is in $\mathcal U_\infty$, we clearly have $\Xi(p)=\lim_{k\to \infty}\Xi(u_{n_k},(\ub_{\CH})_{n_k},\vartheta_{n_k})$. We therefore assume that $(u_n,(\ub_{\CH})_n,\vartheta_n)\in \CH$ for all $n\in \mathbb N$. For each $n\in \mathbb N$, we can find a sequence $(u_{n,\ell},(\ub_{\CH})_{n,\ell},\vartheta_{n,\ell})\in \mathcal U_\infty$ such that $(u_{n,\ell},(\ub_{\CH})_{n,\ell},\vartheta_{n,\ell})\to (u_{n},(\ub_{\CH})_{n},\vartheta_{n})$ as $\ell\to \infty$. %By \eqref{Xi.limit.def}, $\Xi(u_n,(\ub_{\CH})_n,\vartheta_n)=\lim_{\ell\to \infty}\Xi(u_{n,\ell},(\ub_{\CH})_{n,\ell},\vartheta_{n,\ell})$. 
In particular, for every $n\in \mathbb N$, there exists $L_n$ such that $|\Xi(u_{n,\ell},(\ub_{\CH})_{n,\ell},\vartheta_{n,\ell})-\Xi(u_n,(\ub_{\CH})_n,\vartheta_n)|\leq 2^{-n}$ whenever $\ell \geq L_n$. We now extract a ``diagonal'' subsequence $(u_{n,\ell_n},(\ub_{\CH})_{n,\ell_n},\vartheta_{n,\ell_n})\to p$ with $\ell_n\geq L_n$. By \eqref{Xi.limit.def}, we have
$$\Xi(p)=\lim_{n\to \infty}\Xi(u_{n,\ell_n},(\ub_{\CH})_{n,\ell_n},\vartheta_{n,\ell_n})=\lim_{n\to \infty}\Xi(u_{n},(\ub_{\CH})_{n},\vartheta_{n})+O(\lim_{n\to\infty} 2^{-n})=\lim_{n\to \infty}\Xi(u_{n},(\ub_{\CH})_{n},\vartheta_{n}).$$
Hence, the extension of $\Xi$ is continuous, as desired.
\end{proof}

We end this subsection with another H\"older-type estimate, which holds for $b$ and its angular derivatives. The statement is weaker than that for the quantities in Proposition~\ref{cont.ext.prelim}, but will be sufficient for our purposes.
\begin{lemma}
For any $\alpha\in (0,\f 12)$, the following estimates hold with implicit constants which are independent of $u$ and $\ub$:
\begin{equation}\label{b.cont.est.1}
(\int_{-u+C_R}^\infty \ub^{1+2\de} \sup_{dist_{(S_{u,\ub},\gamma)}(\vartheta_1,\vartheta_2)<a}|b(u,\ub,\vartheta_1) - b(u,\ub,\vartheta_2)|_\gamma^2\, d\ub)^{\f 12}\ls a
\end{equation}
and
\begin{equation}\label{b.cont.est.2}
(\int_{-u+C_R}^\infty \ub^{1+2\de} \sup_{dist_{(S_{u,\ub},\gamma)}(\vartheta_1,\vartheta_2)<a}|(\nab b)(u,\ub,\vartheta_1)-(\nab b)(u,\ub,\vartheta_2)|_\gamma^2 \, d\ub)^{\f 12} \ls a^{\alpha}.
\end{equation}
\end{lemma}
\begin{proof}
We first note that the above estimate holds when $b$ is replaced by $b_\Ke$. (This is not trivial because $b_\Ke$ does not decay as $\ub\to \infty$. Nevertheless, since $(b_\Ke)^{\phi_*}=\frac{4Mar}{\Sigma R^2}$ and $(b_\Ke)^{\th_*}=0$ (by \eqref{Kerr.metric.comp}), by an explicit computation, it can be shown that $b_\Ke$ can be decomposed as $b_\Ke = b_{\Ke,1}+b_{\Ke,2}$, where $b_{\Ke,1}$ is independent of $\th_*$, $\phi_*$ and $|b_{\Ke,2}|_\gamma \ls |\Delta|$. Together with \eqref{D.b.improved}, this then implies the desired statement.) Given this, \eqref{b.cont.est.1} is an immediate consequence of the estimates for $\nab^i\tb$ (for $i\leq 3$) and $\nab^i\tg$ (for $i\leq 3$) in $\NH$ (as established in Theorem~\ref{main.quantitative.thm}) and the mean value theorem. Similarly, \eqref{b.cont.est.2} is then an immediate consequence of the estimates for $\nab^i\tb$ (for $i\leq 3$) and $\nab^i\tg$ (for $i\leq 3$) in $\NH$ (as established in Theorem~\ref{main.quantitative.thm}) and Corollary~\ref{Morrey.lemma}.
\end{proof}

\subsection{A new coordinate system}\label{sec.CH.coord}

Given a coordinate system $(u,\ub_{\CH},\th^1,\th^2)$ where $(u,\th^1,\th^2)$ is as before (with initial data as in \eqref{coord.init}) and $\ub_{\CH}$ as in \eqref{CH.def.last} (with $\ub$ defined with initial data as in \eqref{coord.init}), we now introduce a new coordinate system $(\th^1_{\CH},\th^2_{\CH})$ on the $2$-spheres so that (as we will show later) the metric extends continuously to $\CH$ in the $(u,\ub_{\CH},\th^1_{\CH},\th^2_{\CH})$ coordinate system. Note that at this point we do not specify the precise system of coordinates $(\th^1,\th^2)$ on the $2$-spheres $S_{u,\ub}$, as long as they are constructed with the initial data \eqref{coord.init} and satisfy $\Lb\th^A=0$ for $A=1,2$. We will later introduce \emph{specific} coordinate systems on the $2$-spheres starting in Lemma~\ref{change.of.var.coord}.

Define a new coordinate system $(u,\ub_{\CH},\th^1_{\CH},\th^2_{\CH})$ for $\mathcal U_{\infty}\cup\{u=u_f\}\cup \CH\cup S_{u_f,\CH}$ by taking $\ub_{\CH}$ as in Section~\ref{sec.CH} and defining $\th^A_{\CH}$ by the following procedure: First, on the hypersurface $\{(u,\ub,\th^1,\th^2):u=u_f\}$, solve the ode\footnote{For some of the later computations, it is more convenient to state this ode in the $(u,\ub,\th^1,\th^2)$ coordinate system. In the $(u,\ub_{\CH},\th^1,\th^2)$ coordinate system, this ode reads
$$\left(e^{-\f{r_+-r_-}{r_-^2+a^2}\ub}\f{\rd\th^A_{\CH}}{\rd \ub_{\CH}}+ b^B\f{\rd\th^A_{\CH}}{\rd\th^B}\right)(u,\ub_{\CH},\th^1,\th^2)=0.$$
}
\begin{equation}\label{final.th.1}
\left(\f{\rd\th^A_{\CH}}{\rd \ub}+ b^B\f{\rd\th^A_{\CH}}{\rd\th^B}\right)(u_f,\ub,\th^1,\th^2)= (\underline{b}_\Ke)^A(u_f,\ub,\th^1,\th^2),\quad \mbox{ for $A=1,2$},
\end{equation}
with initial data $\th^A_{\CH}=\th^A$ for $A=1,2$ on the $2$-sphere $\{(u,\ub,\th^1,\th^2):u=u_f,\,\ub=\ub_0\}$ for some fixed $\ub_0\in\mathbb R$ satisfying $\f{4Mar_-}{r_-^2+a^2}\ub_0\in 2\pi \mathbb Z$. Here, $\underline{b}_\Ke$ is the $S$-tangent vector field given in the $(u,\ub,\th_*,\phi_*)$ coordinate system by
\begin{equation}\label{b.Ke.def}
\underline{b}_\Ke = \left(\f{4Mar_\Ke}{\Sigma_\Ke R_\Ke^2} - \f{4Mar_\Ke}{(r_-^2+a^2)^2}\right)\f{\rd}{\rd\phi_*},
\end{equation}
where $r_\Ke$ is the Boyer--Lindquist $r$-coordinate, and $\Sigma_{\Ke}$, $R_{\Ke}^2$ are as in \eqref{BL}.
Then, in the entire region $\{(u,\ub,\th^1,\th^2):u\leq u_f,\, \ub_0\leq \ub,\, u+\ub\geq C_R\}$, we require $\th^A_{\CH}$ to satisfy\footnote{$\f{\rd}{\rd u}$ here is to be understood in the $(u,\ub,\th^A)$ coordinate system.}
\begin{equation}\label{final.th.2}
\f{\rd\th^A_{\CH}}{\rd u}=0,\quad \mbox{ for $A=1,2$}.
\end{equation}
Let us note that \eqref{final.th.1} ensures that in the new coordinate system, the new $b^A$ (which we will denote as $b_{\CH}^A$) coincides with that on the background Kerr spacetime on the (null) hypersurface $\{u=u_f\}$ (see discussions after Corollary~\ref{cor:coord.map}). \eqref{final.th.2} then ensures that the coordinate function $\th^A_{\CH}$ is transported along integral curves of $\Lb$ (as is required for the general class of metrics that we study; cf.~Section~\ref{thegeneralclass}).

In order to solve the ode \eqref{final.th.1} for $(\th^1_{\CH},\th^2_{\CH})$, we will introduce multiple coordinate charts. The following lemma shows that the definition is consistent on the intersection of the coordinate charts. In view of \eqref{final.th.2}, we only need to check the consistency on $\{u=u_f\}$.
\begin{lemma}\label{change.of.coord.consistent}
Suppose we have two systems of coordinates on $\mathbb S^2$: $(\th^1_{(1)}, \th^2_{(1)})$ in $\mathcal V_1$ and $(\th^1_{(2)}, \th^2_{(2)})$ in $\mathcal V_2$. On $\mathcal V_1\cap \mathcal V_2$, we have a diffeomorphism $\Psi:(\th^1_{(1)}, \th^2_{(1)})\mapsto (\th^1_{(2)}, \th^2_{(2)})$. Suppose for $i=1,2$, we define $(\th^1_{(i),\CH}, \th^2_{(i),\CH})$ by solving the following ode on $\{u=u_f\}$
$$\left(\f{\rd\th^A_{(i),\CH}}{\rd \ub}+ b^B\f{\rd\th^A_{(i),\CH}}{\rd\th_{(i)}^B}\right)(u,\ub,\th_{(i)}^1,\th_{(i)}^2)= (\underline{b}_\Ke)_{(i)}^A,\quad \mbox{ for $A=1,2$},$$
with initial data 
\begin{equation}\label{change.of.coord.ID}
(\th^1_{(2),\CH,init}, \th^2_{(2),\CH,init}) = \Psi((\th^1_{(1),\CH,init}, \th^2_{(1),\CH,init}))
\end{equation}
on the $2$-sphere $S_{u_f,\ub_0}$, where $(\underline{b}_\Ke)_{(i)}^A$ is the coefficient of the vector field $\underline{b}_\Ke$ (cf.~\eqref{b.Ke.def}) in the basis $\{\f{\rd}{\rd\th^1_{(i)}}, \f{\rd}{\rd\th^2_{(i)}}\}$.
Then, as long as $(\th^1_{(1),\CH}, \th^2_{(1),\CH}) \in \mathcal V_1\cap \mathcal V_2$, we have
$$(\th^1_{(2),\CH}, \th^2_{(2),\CH}) = \Psi ((\th^1_{(1),\CH}, \th^2_{(1),\CH})).$$
In particular, for $\ub \geq -u_f+C_R$, \eqref{final.th.1} gives rise to a well-defined map $\vartheta \in S_{u_f,\ub_0} \mapsto \vartheta_{\CH}(u_f,\ub)\in S_{u_f,\ub}$.
\end{lemma}
\begin{proof}
Define $(\th^1_{(1),\CH}, \th^2_{(1),\CH})$ as above.
Consider now 
$$(\tilde{\th}^1_{(2),\CH}, \tilde{\th}^2_{(2),\CH}) = \Psi ((\th^1_{(1),\CH}, \th^2_{(1),\CH})),$$
which in particular implies that
$$\f{\rd\tilde{\th}^A_{(2),\CH}}{\rd\th_{(1),\CH}^C} = \f{\rd\th^A_{(2)}}{\rd\th_{(1)}^C}.$$
Using this, one checks that
\begin{equation*}
\begin{split}
&\left(\f{\rd\tilde{\th}^A_{(2),\CH}}{\rd \ub}+ b^B_{(2)}\f{\rd\tilde{\th}^A_{(2),\CH}}{\rd\th_{(2)}^B}\right)(u,\ub,\th_{(2)}^1,\th_{(2)}^2)\\
= &\left(\f{\rd\tilde{\th}^A_{(2),\CH}}{\rd\th_{(1),\CH}^C}\f{\rd\th^C_{(1),\CH}}{\rd \ub}+ b^B\f{\rd\th^C_{(1),\CH}}{\rd\th_{(1)}^B}\f{\rd\tilde{\th}^A_{(2),\CH}}{\rd\th_{(1),\CH}^C}\right)(u,\ub,\th_{(2)}^1,\th_{(2)}^2)\\
= &\f{\rd\tilde{\th}^A_{(2),\CH}}{\rd\th_{(1),\CH}^C}(\underline{b}_\Ke)_{(1)}^C = \f{\rd\th^A_{(2)}}{\rd\th_{(1)}^C}(\underline{b}_\Ke)_{(1)}^C = (\underline{b}_\Ke)_{(2)}^A,\quad \mbox{ for $A=1,2$}.
\end{split}
\end{equation*}
Now since the initial data satisfy \eqref{change.of.coord.ID}, we conclude using the uniqueness of solutions to odes.
\end{proof}

Since \eqref{final.th.2} is clearly independent of the choice of coordinate systems, this immediately gives
\begin{corollary}\label{cor:coord.map}
For $(u,\ub)\in \mathcal W_{\infty} = \{(u,\ub): u< u_f,\, -u+C_R\leq \ub\}$, \eqref{final.th.1} and \eqref{final.th.2} give rise to a well-defined map $\vartheta \in S_{u_f,\ub_0} \mapsto \vartheta_{\CH}(u,\ub)\in S_{u,\ub}$.

Moreover, after identifying $\vartheta \in S_{u,\ub}$ with the point in $S_{u_f,\ub_0}$ with the same coordinate values $(\th^1,\th^2)$, this gives rise to a well-defined map $\vartheta(u,\ub) \in S_{u,\ub} \mapsto \vartheta_{\CH}(u,\ub)\in S_{u,\ub}$ for any $(u,\ub)\in \mathcal W_{\infty}$.
\end{corollary}

In order to prove estimates regarding the map $\vartheta \in S_{u,\ub} \mapsto \vartheta_{\CH}(u,\ub)\in S_{u,\ub}$ in Corollary~\ref{cor:coord.map}, we will compare it with the corresponding map
\begin{equation}\label{Kerr.coord.map}
\vartheta \in S_{u,\ub} \mapsto \vartheta_{\CH,\Ke}(u,\ub) \in S_{u,\ub}
\end{equation}
on $\mathcal M_{Kerr}$, which is defined in coordinates by $(\th_*,\phi_*)=\vartheta \mapsto \vartheta_{\CH,\Ke}=(\th_*,\phi_{*,\CH})$; see Section~\ref{sec.coord.near.CH}. Let us note that\footnote{Now this is to be understood possibly in a different coordinate system $(\th^1,\th^2)$ on the $2$-spheres. As in Lemma~\ref{change.of.coord.consistent}, the ode \eqref{Kerr.coord.transport} holds in more general coordinate systems.} $\vartheta_{\CH,\Ke}= (\th^1_{\CH,\Ke},\th^2_{\CH,\Ke})$ satisfies
\begin{equation}\label{Kerr.coord.transport}
\left(\f{\rd\th^A_{\CH,\Ke}}{\rd \ub}+ b_\Ke^B\f{\rd\th^A_{\CH,\Ke}}{\rd\th^B}\right)(u_f,\ub,\th^1,\th^2)= (\underline{b}_\Ke)^A(u_f,\ub,\th^1,\th^2), \quad\mbox{for $u=u_f$}
\end{equation}
and
$$\f{\rd \th_{\CH,\Ke}^A}{\rd u} =0,\quad \mbox{everywhere in $\mathcal U_\infty$},$$
for $A=1,2$, with $(\th^1_{\CH,\Ke},\th^2_{\CH,\Ke}) = (\th^1,\th^2)$ whenever $\f{4Mar_-}{r_-^2+a^2}\ub\in 2\pi \mathbb Z$.

Next, we will find appropriate coordinate charts in which we can prove estimates. For this purpose, let us recall the coordinate systems defined via the stereographic projection as introduced before Proposition~\ref{prop:metric.higher}. More precisely, we first pull back the $(\th_*,\phi_*)$ on $\mathcal M_{Kerr}$ (see Section~\ref{Kerr.dbn}) to $\mathcal U_{\infty}$ using the identification introduced in Section~\ref{sec.identification}. Then let $\mathcal V_1\doteq \mathbb S^2\setminus\{\th_*=0\}$ and $\mathcal V_2\doteq \mathbb S^2\setminus\{\th_*=\pi\}$ respectively. In $\mathcal V_1$, let $(\th_{(1)}^1,\th_{(1)}^2) = (\cot \f{\th_*}{2}\cos\phi_*, \cot \f{\th_*}{2}\sin\phi_*)$ and in $\mathcal V_2$, let $(\th_{(2)}^1,\th_{(2)}^2) = (\tan \f{\th_*}{2}\cos\phi_*, \tan \f{\th_*}{2}\sin\phi_*)$. Using Corollary~\ref{cor:coord.map} and \eqref{Kerr.coord.map}, we can then also introduce $(\th_{(i),\CH}^1, \th_{(i),\CH}^2)$ and $(\th_{(i),\CH,\Ke}^1, \th_{(i),\CH,\Ke}^2)$ in $\mathcal V_i$ for $i=1,2$.

From now on and in the rest of this subsection, unless otherwise stated, we will consider the \emph{specific} coordinates systems $(\th_{(1)}^1,\th_{(1)}^2)$ and $(\th_{(2)}^1,\th_{(2)}^2)$ (and $(\th_{(i),\CH}^1, \th_{(i),\CH}^2)$ and $(\th_{(i),\CH,\Ke}^1, \th_{(i),\CH,\Ke}^2)$) in $\mathcal V_1$ and $\mathcal V_2$ respectively that we introduced above\footnote{Let us also note when there is no risk of confusion, we will sometimes drop the subscripts $_{(i)}$. See already the proof of Lemma~\ref{change.of.var.coord}.}.

\begin{lemma}\label{change.of.var.coord}
Let $\mathcal V_1$ and $\mathcal V_2$ be subsets of $\mathbb S^2$ as above. There exists $\ub_0$ such that for $i=1,2$, there exist precompact open sets $\mathcal V_i'$ and $\mathcal V_i''$ satisfying $\mathcal V_i'\subset \mathcal V_i''\subset \mathcal V_i$ with the following properties
\begin{itemize}
\item $\mathcal V_1'\cup \mathcal V_2' = \mathbb S^2$.
\item Given a point $\vartheta = (\th^1_{(i)},\th^2_{(i)}) \in S_{u,\ub} \cap \mathcal V_i'$ and $(u,\ub)\in \mathcal W_\infty =\{(u,\ub): u\leq u_f,\, -u+C_R\leq \ub\}$, then 
\begin{itemize}
\item for $\vartheta_{\CH} = (\th^1_{(i),\CH},\th^2_{(i),\CH})$ the corresponding point in $S_{u,\ub}$ defined by Corollary~\ref{cor:coord.map}, 
\item for $\vartheta_{\CH,\Ke} = (\th^1_{(i),\CH,\Ke},\th^2_{(i),\CH,\Ke})$ the corresponding point in $S_{u,\ub}$ defined as in \eqref{Kerr.coord.map},
\end{itemize}
it holds that $\vartheta_{\CH} \in S_{u,\ub} \cap \mathcal V_i''$, and\footnote{As we will note in the proof, we also have $\vartheta_{\CH,\Ke} \in S_{u,\ub} \cap \mathcal V_i'$ and therefore this difference makes sense in the $\mathcal V_i$ coordinate patch.}
$$\max_{A=1,2} |\th^A_{(i),\CH,\Ke} - \th^A_{(i),\CH}|\ls \ep.$$
\end{itemize}
\end{lemma}
\begin{proof}
Let $\mathcal V_1'\subset \mathcal V_1''\subset \mathcal V_1$ be defined as 
$$\mathcal V_1'\doteq \{(\th_{(1)}^1, \th_{(1)}^2): (\th_{(1)}^1)^2 + (\th_{(1)}^2)^2 < 2 \},\quad \mathcal V_1''\doteq \{(\th_{(1)}^1, \th_{(1)}^2): (\th_{(1)}^1)^2 + (\th_{(1)}^2)^2 < 3 \},$$
and similarly let $\mathcal V_2'\subset \mathcal V_2''\subset \mathcal V_2$ be defined as 
$$\mathcal V_2'\doteq \{(\th_{(2)}^1, \th_{(2)}^2): (\th_{(2)}^1)^2 + (\th_{(2)}^2)^2 < 2 \},\quad \mathcal V_2''\doteq \{(\th_{(2)}^1, \th_{(2)}^2): (\th_{(2)}^1)^2 + (\th_{(2)}^2)^2 < 3 \}.$$
An easy explicit computation shows that $\mathcal V_1'\cup \mathcal V_2'$ indeed covers the whole sphere.

For the rest of the proof, in order to simplify the notation, let us fix $i$ and suppress the subscripts ${ }_{(i)}$ in the coordinates (or the $b$'s, etc.).

To proceed, note that using the estimates for $\gamma-\gamma_\Ke$ in $\NS$ established in Theorem~\ref{main.quantitative.thm} and Sobolev embedding (Proposition~\ref{Sobolev}), it is easy to see that in any of $\mathcal V_1'$, $\mathcal V_1''$, $\mathcal V_2'$ and $\mathcal V_2''$, $\gamma_{AB}$ is a matrix with uniformly upper bounded entries and uniformly lower bounded determinant. 

By the definition of the Kerr $\th_{\CH,\Ke}$ coordinates in Section~\ref{sec.coord.near.CH}, since the coordinate change only rotates the azimuthal direction, the map $\vartheta\mapsto \vartheta_{\CH,\Ke}(u,\ub,\vartheta)$ maps $\mathcal V_i'$ to $\mathcal V_i'$ (and similarly maps $\mathcal V_i''$ to $\mathcal V_i''$). It therefore suffices to control the difference $\th^A_{\CH}-\th^A_{\CH,\Ke}$. For this, we first note the following equation
 \begin{equation*}
\begin{split}
\f{\rd}{\rd \ub}(\th^A_{\CH}-\th^A_{\CH,\Ke})+ b^B\f{\rd}{\rd\th^B}(\th^A_{\CH}-\th^A_{\CH,\Ke}) +(b^B-b^B_{\Ke}) \f{\rd\th^A_{\CH,\Ke}}{\rd\th^B}=  0,
\end{split}
\end{equation*}
for $A=1,2$ on $\{u=u_f\}$ (which follows from \eqref{final.th.1} and \eqref{Kerr.coord.transport}). Recall that $\vartheta_{\CH,\Ke} = \vartheta$ on $S_{u_f,\ub}$ for $\f{4Mar_-}{r_-^2+a^2}\ub\in 2\pi \mathbb Z$. Hence, by definition, $\th^A_{\CH}-\th^A_{\CH,\Ke} = 0$ on $S_{u_f,\ub_0} \cap \mathcal V_i'$. Therefore, as long as $(\th^1,\th^2)\in \mathcal V_i'$, the estimates on $b-b_\Ke$ in $\NS$ (which is controlled in Theorem~\ref{main.quantitative.thm}), the estimates on $\f{\rd\th^A_{\CH,\Ke}}{\rd\th^B}$ (which follow from considering the explicit map \eqref{Kerr.coord.map} defined in Section~\ref{sec.coord.near.CH}), together with the bounds on the components of $\gamma$ in coordinates imply that
$$|\th^A_{\CH,\Ke} - \th^A_{\CH}|\ls \ep.$$
Since $(\th^1_{\CH,\Ke},\th^2_{\CH,\Ke})\in \mathcal V_i'$, it follows that for $\ep_0$ (and hence $\ep$) sufficiently small, we have $(\th^1_{\CH},\th^2_{\CH})\in \mathcal V_i''$.
\end{proof}

From now on until the end of the section, we will take\footnote{Note that this is different from the convention in Section~\ref{sec:HR}, where $\mathcal V_i'$ is taken to be an \emph{arbitrary} precompact open subset of $\mathcal V_i$.} $\mathcal V_i'$ and $\mathcal V_i''$ to be the sets defined in the proof of Lemma~\ref{change.of.var.coord}.

We claim that the coordinate transformation $(\th_{(i)}^1,\th_{(i)}^2)\mapsto (\th_{(i),\CH}^1,\th_{(i),\CH}^2)$ gives a regular coordinate transformation in\footnote{We emphasise that the coordinate transformation is \underline{not} regular up to the boundary $\{\ub_{\CH}=1\}$. It is precisely for this reason that we perform this coordinate transformation. Indeed, while we will show (see Theorem \ref{metric.cont.ext}) that the metric is continuous up to the boundary in the $(u,\ub_{\CH},\th^1_{(i),\CH},\th^2_{(i),\CH})$ coordinate system, in general it may \underline{not} be continuous up to the boundary in the $(u,\ub_{\CH},\th^1_{(i)},\th_{(i)}^2)$ coordinate system.} $\mathcal W\times \mathcal V_i$. More precisely, we have the following lemma:

\begin{proposition}\label{prop.dthdth.diff}
For $\vartheta\in \mathcal V_i'$ (where $i=1,2$), we have
\begin{equation}\label{est.dthdth.diff}
\left|\f{\rd\th_{(i),\CH}^A}{\rd\th_{(i)}^B}-\f{\rd\th_{(i),\CH,\Ke}^A}{\rd\th_{(i)}^B}\right|(u,\ub,\vartheta)\ls \ep.
\end{equation}
Moreover,
\begin{itemize}
\item the map $(u,\ub,\vartheta)\mapsto (u,\ub_{\CH},\vartheta_{\CH})$ is a diffeomorphism, and
\item $\f{\rd\th_{(i),\CH}^A}{\rd\th_{(i)}^B}$ extends continuously to the Cauchy horizon $\CH$ with respect to the differential structure given by the $(u,\ub_{\CH},\th^1_{(i)},\th^2_{(i)})$ coordinates.
\end{itemize}
\end{proposition}
\begin{proof}
\textbf{Proof of \eqref{est.dthdth.diff}.} Clearly we only need to prove this estimate on the hypersurface $\{u=u_f\}$ since all of $\th_{(i)}^A$, $\th_{(i),\CH}^A$ and $\th_{(i),\CH,\Ke}^A$ are all transported along the integral curves of $\Lb$.

By \eqref{final.th.1} and \eqref{Kerr.coord.transport}, we have the following transport equation on $\{u=u_f\}$ for $i=1,2$:
\begin{equation}\label{dth.dth.transport.eqn}
\begin{split}
(\f{\rd}{\rd \ub}+ b^C\f{\rd}{\rd \th_{(i)}^C})(\f{\rd\th_{(i),\CH}^A}{\rd\th_{(i)}^B}-\f{\rd\th_{(i),\CH,\Ke}^A}{\rd\th_{(i)}^B})
= &-\f{\rd b^C}{\rd\th_{(i)}^B}\f{\rd \th^A_{(i),\CH}}{\rd\th_{(i)}^C}+\f{\rd b_\Ke^C}{\rd\th_{(i)}^B}\f{\rd \th^A_{(i),\CH,\Ke}}{\rd\th_{(i)}^C}\\
%= &-(\f{\rd b^C}{\rd\th^B}-\f{\rd b_\Ke^C}{\rd\th^B})\f{\rd \th^A_{\CH}}{\rd\th^C}-\f{\rd b_\Ke^C}{\rd\th^B}(\f{\rd\th_{\CH}^A}{\rd\th^C}-\f{\rd\th_{\CH,\Ke}^A}{\rd\th^C})\\
= &-(\f{\rd b^C}{\rd\th_{(i)}^B}-\f{\rd b_\Ke^C}{\rd\th_{(i)}^B})\f{\rd \th^A_{(i),\CH,\Ke}}{\rd\th_{(i)}^C}-\f{\rd b^C}{\rd\th_{(i)}^B}(\f{\rd\th_{(i),\CH}^A}{\rd\th_{(i)}^C}-\f{\rd\th_{(i),\CH,\Ke}^A}{\rd\th_{(i)}^C}).
\end{split}
\end{equation}
Moreover, the Kerr $\th_{(i),\CH,\Ke}^A$ coordinates in Section~\ref{sec.coord.near.CH} are chosen such that $\f{\rd\th_{(i),\CH,\Ke}^A}{\rd\th_{(i)}^B} = \de^A_B$ whenever $\f{4Mar_-}{r_-^2+a^2}\ub\in 2\pi \mathbb Z$. Hence, on the sphere $S_{u_f,\ub_0}$, we have 
$$\f{\rd\th_{(i),\CH}^A}{\rd\th_{(i)}^B} = \f{\rd\th_{(i),\CH,\Ke}^A}{\rd\th_{(i)}^B} = \de^A_B.$$

As we noted in the proof of Lemma~\ref{change.of.var.coord}, in a coordinate chart $\mathcal V_i''$, the components of $\gamma$ and $\gamma^{-1}$ are bounded; and $\f{\rd \th^A_{(i),\CH,\Ke}}{\rd\th_{(i)}^C}$ is also bounded componentwise. Therefore, integrating \eqref{dth.dth.transport.eqn} along the integral curve of $\f{\rd}{\rd\ub}+b^C\f{\rd}{\rd\th_{(i)}^C}$, we obtain that for any $\ub \geq \ub_0$ and $\vartheta \in \mathcal V_i'$,
\begin{equation}\label{est.dth.dth.one.patch}
\begin{split}
&\left|\f{\rd\th_{(i),\CH}^A}{\rd\th_{(i)}^B}-\f{\rd\th_{(i),\CH,\Ke}^A}{\rd\th_{(i)}^B}\right|(u_f,\ub, \vartheta)\\
\ls &\int_{-u_f+C_R}^{\ub} (e^{-\f{r_+-r_-}{r_-^2+a^2}(u_f+\ub')} + \sum_{j\leq 1} \|\nab^j \tb\|_{L^\i(S_{u_f,\ub'})})\sup_{\vartheta'\in S_{u_f,\ub'}\cap \mathcal V_i''} \left|\f{\rd\th_{(i),\CH}^A}{\rd\th_{(i)}^B}-\f{\rd\th_{(i),\CH,\Ke}^A}{\rd\th_{(i)}^B}\right|(u_f,\ub',\vartheta')\, d\ub' \\
&+  \sum_{j\leq 1}\|\ub^{\f 12+\de}\nab^j\tb\|_{L^\i_u L^2_{\ub}L^\i(S)} (\int_{-u_f+C_R}^{\ub}(\ub')^{-1-2\de} \,d\ub')^{\f 12}\\
\ls &\ep +\int_{-u_f+C_R}^{\ub} (e^{-\f{r_+-r_-}{r_-^2+a^2}(u_f+\ub')} + \sum_{j\leq 1}\|\nab^j \tb\|_{L^\i(S_{u_f,\ub'})})\sup_{\vartheta'\in S_{u_f,\ub'}\cap \mathcal V_i''} \left|\f{\rd\th_{(i),\CH}^A}{\rd\th_{(i)}^B}-\f{\rd\th_{(i),\CH,\Ke}^A}{\rd\th_{(i)}^B}\right|(u_f,\ub',\vartheta')\, d\ub',
\end{split}
\end{equation}
where in the last line we have used the estimates for $\NH$ established in Theorem~\ref{main.quantitative.thm}.
Notice that
$$\mathcal V_1''\setminus \mathcal V_1' \subset \mathcal V_2', \quad \mathcal V_2''\setminus \mathcal V_2' \subset \mathcal V_1'.$$
Moreover, since
$$\f{\rd\th_{(1),\CH}^A}{\rd\th_{(1)}^B} = \f{\rd\th_{(2),\CH}^C}{\rd\th_{(2)}^D}\f{\rd\th_{(1)}^A}{\rd\th_{(2)}^C} \f{\rd\th_{(2)}^D}{\rd\th_{(1)}^B}, \quad \f{\rd\th_{(1),\CH,\Ke}^A}{\rd\th_{(1)}^B} = \f{\rd\th_{(2),\CH,\Ke}^C}{\rd\th_{(2)}^D}\f{\rd\th_{(1)}^A}{\rd\th_{(2)}^C} \f{\rd\th_{(2)}^D}{\rd\th_{(1)}^B},$$
the boundedness of the transition maps implies that on $\mathcal V_1''\cap \mathcal V_2''$,
\begin{equation}\label{comparison.in.the.transition}
\left|\f{\rd\th_{(1),\CH}^A}{\rd\th_{(1)}^B}-\f{\rd\th_{\Ke,(1),\CH}^A}{\rd\th^B_{(1)}}\right|\ls \left|\f{\rd\th_{(2),\CH}^A}{\rd\th_{(2)}^B}-\f{\rd\th_{\Ke,(2),\CH}^A}{\rd\th^B_{(2)}}\right|\ls \left|\f{\rd\th_{(1),\CH}^A}{\rd\th_{(1)}^B}-\f{\rd\th_{\Ke,(1),\CH}^A}{\rd\th^B_{(1)}}\right|.
\end{equation}
Now define
$$A(\ub) \doteq \sup_{\vartheta \in \mathcal V_1}\left|\f{\rd\th_{(1),\CH}^A}{\rd\th_{(1)}^B}-\f{\rd\th_{(1),\CH,\Ke}^A}{\rd\th^B_{(1)}}\right|(u_f,\ub, \vartheta) + \sup_{\vartheta \in \mathcal V_2}\left|\f{\rd\th_{(2),\CH}^A}{\rd\th_{(2)}^B}-\f{\rd\th_{(2),\CH,\Ke}^A}{\rd\th^B_{(2)}}\right|(u_f,\ub, \vartheta),$$
sum over the estimates \eqref{est.dth.dth.one.patch} for both $\mathcal V_1$ and $\mathcal V_2$ and applying \eqref{comparison.in.the.transition} and Gr\"onwall's inequality, we obtain
\begin{equation*}
\begin{split}
A(\ub) \ls &\: \ep + \int_{-u_f+C_R}^{\ub} (e^{-\f{r_+-r_-}{r_-^2+a^2}(u_f+\ub')} + \sum_{j\leq 1}\|\nab^j \tb\|_{L^\i(S_{u_f,\ub'})}) A(\ub')\,d\ub'\\
\ls & \: \ep \exp(\int_{-u_f+C_R}^{\ub} e^{-\f{r_+-r_-}{r_-^2+a^2}(u_f+\ub')}\, d\ub' + \sum_{j\leq 1} \|\nab^j \tb\|_{L^\i_uL^1_{\ub}L^\i(S)})\ls \ep,
\end{split}
\end{equation*}
where in the last inequality we have used the estimates for $\NH$ established in Theorem~\ref{main.quantitative.thm} and Proposition~\ref{Sobolev}.

\textbf{$(u,\ub,\vartheta)\mapsto (u,\ub_{\CH},\vartheta_{\CH})$ is a diffeomorphism.} Clearly the map is smooth by construction. It suffices therefore to check its invertibility. Take $\vartheta\in \mathcal V_i$ so that we compute using the $(\th^1_{(i)},\th^2_{(i)})$ coordinates on the $2$-spheres. A direct computation (with subscripts ${ }_{(i)}$ suppressed for notational simplicity) shows that
\[ \left( \begin{array}{cccc}
\f{\rd u}{\rd u} & \f{\rd u}{\rd \ub_{\CH}} & \f{\rd u}{\rd \th^1} &\f{\rd u}{\rd \th^2}\\
\f{\rd \ub_{\CH}}{\rd u} & \f{\rd \ub_{\CH}}{\rd \ub_{\CH}} & \f{\rd \ub_{\CH}}{\rd \th^1} &\f{\rd \ub_{\CH}}{\rd \th^2}\\
\f{\rd \th^1_{\CH}}{\rd u} & \f{\rd \th^1_{\CH}}{\rd \ub_{\CH}} & \f{\rd \th^1_{\CH}}{\rd \th^1} &\f{\rd \th^1_{\CH}}{\rd \th^2}\\
\f{\rd \th^2_{\CH}}{\rd u} & \f{\rd \th^2_{\CH}}{\rd \ub_{\CH}} & \f{\rd \th^2_{\CH}}{\rd \th^1} &\f{\rd \th^2_{\CH}}{\rd \th^2}
\end{array} \right)
=
\left( \begin{array}{cccc}
1 & 0 & 0 & 0 \\
0 & 1 & 0 & 0 \\
0 & e^{\f{r_+-r_-}{r_-^2+a^2}\ub}\f{\rd\th^1_{\CH}}{\rd\ub} & \f{\rd \th^1_{\CH}}{\rd \th^1} &\f{\rd \th^1_{\CH}}{\rd \th^2}\\
0 & e^{\f{r_+-r_-}{r_-^2+a^2}\ub}\f{\rd\th^2_{\CH}}{\rd\ub} & \f{\rd \th^2_{\CH}}{\rd \th^1} &\f{\rd \th^2_{\CH}}{\rd \th^2}
\end{array} \right).
\]
Therefore, by the inverse function theorem, the map is invertible if the matrix $(\f{\rd\th^A_{\CH}}{\rd\th^B})$ is invertible. On the other hand, the invertibility of $(\f{\rd\th^A_{\CH}}{\rd\th^B})$ follows directly from \eqref{est.dthdth.diff} and the explicit formula for the Kerr $\th_{\CH,\Ke}^A$ coordinates defined in Section~\ref{sec.coord.near.CH}.

\textbf{Continuous extendiblity of $\f{\rd\th_{(i),\CH}^A}{\rd\th_{(i)}^B}$.} Again, we first show this on $\{u=u_f\}$. As before, we fix $i$ and suppress the subscripts ${ }_{(i)}$ in our notations. To prove that $\f{\rd\th_{\CH}^A}{\rd\th^B}$ is continuously extendible to $\CH$ (with respect to the differentiable structure given by $(u,\ub_{\CH},\th^1,\th^2)$), consider a sequence $((\ub_{\CH})_n,\vartheta_n)\to ( (\ub_{\CH})_\i,\vartheta_\i)$, where $(u_f,(\ub_{\CH})_\i,\vartheta_\i)\in \CH$ and $\th_n,\,\th_\infty\in \mathcal V_i$. In the following, it will be more convenient to use the $\ub$ coordinate instead of $\ub_{\CH}$. Note that we have $\ub_n\to \infty$.

We will integrate \eqref{dth.dth.transport.eqn} along the integral curve of $\f{\rd}{\rd\ub}+b^A\f{\rd}{\rd\th^A}$. For this purpose, it is useful to denote by $\tilde{\vartheta}_n(\ub)$ the unique value of $\vartheta$ such that $(u_f,\ub,\tilde{\vartheta}_n(\ub))$ lies on the integral curve of $\f{\rd}{\rd\ub}+b^A\f{\rd}{\rd\th^A}$ which passes through $(u_f,\ub_n,\vartheta_n)$. Now, by \eqref{b.cont.est.1} and Gr\"onwall's inequality, we estimate
$$dist_{(S_{u,\ub},\gamma)}(\vartheta_n, \vartheta_m) \ls dist_{(S_{u,\ub},\gamma)}(\tilde{\vartheta}_n(\ub), \tilde{\vartheta}_m(\ub))\ls dist_{(S_{u,\ub},\gamma)}(\vartheta_n, \vartheta_m).$$

We now write down the solution to \eqref{dth.dth.transport.eqn} in terms of an integral along the integral curve of $\f{\rd}{\rd\ub}+b^A\f{\rd}{\rd\th^A}$ (which we parametrize by $\ub$) as follows\footnote{Here, the exponentials are to be understood as exponentials of matrices.}:
\begin{equation}\label{dth.dth.n.eqn}
\begin{split}
&\f{\rd\th_{\CH}^A}{\rd\th^B}(\ub_n,\vartheta_n) -\f{\rd\th_{\CH,\Ke}^A}{\rd\th^B}(\ub_n,\vartheta_n)\\
= &\de^A_B - e^{-\int^{\ub_n}_{\ub_0} \f{\rd b^C}{\rd\th^B}(\ub', \tilde{\vartheta}_n(\ub'))\, d\ub'} \int_{\ub_0}^{\ub_n} e^{\int_{\ub_0}^{\ub'} \f{\rd b^C}{\rd\th^B}(\ub'', \tilde{\vartheta}_n(\ub''))\, d\ub''}(\f{\rd b^C}{\rd\th^B}-\f{\rd b_\Ke^C}{\rd\th^B})\f{\rd \th^A_{\CH,\Ke}}{\rd\th^C}(\ub', \tilde{\vartheta}_n(\ub'))\, d\ub'.
\end{split}
\end{equation}
This allows us to estimate $\f{\rd\th_{\CH}^A}{\rd\th^B}(u_f,\ub_n,\vartheta_n) - \f{\rd\th_{\CH}^A}{\rd\th^B}(u_f,\ub_m,\vartheta_m)$. First, by the triangle inequality,
\begin{equation}\label{dth.dth.n.I.II}
\begin{split}
& |\f{\rd\th_{\CH}^A}{\rd\th^B}(u_f,\ub_n,\vartheta_n) - \f{\rd\th_{\CH}^A}{\rd\th^B}(u_f,\ub_m,\vartheta_m)|\\
\ls &\underbrace{|\f{\rd\th_{\CH}^A}{\rd\th^B}(u_f,\ub_n,\vartheta_n) - \f{\rd\th_{\CH}^A}{\rd\th^B}(u_f,\ub_n,\vartheta_m)|}_{\doteq I}+\underbrace{|\f{\rd\th_{\CH}^A}{\rd\th^B}(u_f,\ub_n,\vartheta_m) - \f{\rd\th_{\CH}^A}{\rd\th^B}(u_f,\ub_m,\vartheta_m)|}_{\doteq II}.
\end{split}
\end{equation}
To bound $I$, we use \eqref{dth.dth.n.eqn}. We first prove some preliminary estimates. According to \eqref{D.b.improved}, $|\f{\rd b^C}{\rd\th^B}(\ub', \tilde{\vartheta}_n(\ub'))|\ls \Om_\Ke^2$ (componentwise). Hence, using the estimates for $\NH$ established in Theorem~\ref{main.quantitative.thm}, 
\begin{equation}\label{exponential.int.b.bd}
\begin{split}
 | e^{\int^{\infty}_{\ub_0} |\f{\rd b^C}{\rd\th^B}(\ub', \tilde{\vartheta}_n(\ub'))|\, d\ub'}|\ls &|e^{\int^{\infty}_{\ub_0} \Om_\Ke^2\, d\ub'}| | e^{\sum_{i\leq 1}\| \nab^i\tb \|_{L^1_{\ub}L^{\i}(S)}}|\ls 1.
\end{split}
\end{equation}
On the other hand, by \eqref{b.cont.est.2}, it holds that for any $\alp\in (0,\f 12)$,
\begin{equation}\label{d.b.diff.th}
\begin{split}
 \int_{\ub_0}^{\infty} \sup_C|\f{\rd b^C}{\rd\th^B}(\ub', \tilde{\vartheta}_n(\ub'))-\f{\rd b^C}{\rd\th^B}(\ub', \tilde{\vartheta}_m(\ub'))|\, d\ub'
\ls & (dist_{(S_{u,\ub},\gamma)}(\vartheta_n, \vartheta_m))^{\alp},
\end{split}
\end{equation}
which then also implies
\begin{equation}\label{d.b.diff.th.exp}
\begin{split}
 | e^{-\int^{\ub_n}_{\ub_0} \f{\rd b^C}{\rd\th^B}(\ub', \tilde{\vartheta}_n(\ub'))\, d\ub'} - e^{-\int^{\ub_n}_{\ub_0} \f{\rd b^C}{\rd\th^B}(\ub', \tilde{\vartheta}_m(\ub'))\, d\ub'}|
\ls & (dist_{(S_{u,\ub},\gamma)}(\vartheta_n, \vartheta_m))^{\alp}.
\end{split}
\end{equation}
By the equation \eqref{dth.dth.n.eqn}, the estimates \eqref{exponential.int.b.bd}, \eqref{d.b.diff.th}, \eqref{d.b.diff.th.exp} and the fact that the change of variable map $\f{\rd\th_{\CH,\Ke}^A}{\rd\th^B}$ is uniformly $C^1$ in $\th$, we estimate $I$ in \eqref{dth.dth.n.I.II} as follows:
\begin{equation}\label{dth.dth.n.I}
\begin{split}
&|\f{\rd\th_{\CH}^A}{\rd\th^B}(u_f,\ub_n,\vartheta_n) - \f{\rd\th_{\CH}^A}{\rd\th^B}(u_f,\ub_n,\vartheta_m)|\\
\ls & \int_{\ub_0}^{\infty} \sup_C|\f{\rd b^C}{\rd\th^B}(\ub', \tilde{\vartheta}_n(\ub'))-\f{\rd b^C}{\rd\th^B}(\ub', \tilde{\vartheta}_m(\ub'))|\, d\ub'+ dist_{(S_{u,\ub},\gamma)}(\vartheta_n, \vartheta_m)\\
\ls & (dist_{(S_{u,\ub},\gamma)}(\vartheta_n, \vartheta_m))^{\alp},
\end{split}
\end{equation}
for any $\alp \in (0,\f 12)$ (with implicit constant depending on $\alp$). On the other hand, using again \eqref{dth.dth.n.eqn}, the term $II$ in \eqref{dth.dth.n.I.II} can be controlled using \eqref{exponential.int.b.bd}, the bound for $\NH$ in Theorem~\ref{main.quantitative.thm}, the boundedness of $\f{\rd \th^A_{\CH,\Ke}}{\rd\th^C}$, and trivial estimates for the Kerr background terms as follows:
\begin{equation}\label{dth.dth.n.II}
\begin{split}
&|\f{\rd\th_{\CH}^A}{\rd\th^B}(u_f,\ub_n,\vartheta_m) - \f{\rd\th_{\CH}^A}{\rd\th^B}(u_f,\ub_m,\vartheta_m)|\\
\ls &|\int_{\ub_m}^{\ub_n} (\f{\rd b^C}{\rd\th^B}-\f{\rd b_\Ke^C}{\rd\th^B})\f{\rd \th^A_{\CH,\Ke}}{\rd\th^C}(\ub', \tilde{\vartheta}_m(\ub'))\, d\ub'|+|\f{\rd\th_{\CH,\Ke}^A}{\rd\th^B}(\ub_n,\vartheta_m)-\f{\rd\th_{\CH,\Ke}^A}{\rd\th^B}(\ub_m,\vartheta_m)| \\
\ls & |\ub_m^{-2\de}-\ub_n^{-2\de}|^{\f 12} (1+\|\ub^{\f 12+\de} \nab \tb\|_{L^\i_u L^2_{\ub}L^\i(S)}) \ls |\ub_m^{-2\de}-\ub_n^{-2\de}|^{\f 12}.
\end{split}
\end{equation}
Combining \eqref{dth.dth.n.I.II}, \eqref{dth.dth.n.I} and \eqref{dth.dth.n.II}, we therefore conclude that
$$\f{\rd\th_{\CH}^A}{\rd\th^B}(u_f,\ub_\i=\infty,\vartheta_\infty) = \lim_{n\to\infty}\f{\rd\th_{\CH}^A}{\rd\th^B}(u_f,\ub_n,\vartheta_n)$$
is well-defined and independent of the choice of the sequence $(\ub_n,\vartheta_n)$. In view of \eqref{final.th.2}, for any $u\leq u_f$, we can also define
$$\f{\rd\th_{\CH}^A}{\rd\th^B}(u,\ub_\i=\infty,\vartheta_\infty) = \lim_{n\to\infty}\f{\rd\th_{\CH}^A}{\rd\th^B}(u,\ub_n,\vartheta_n),$$
where $(\ub_n,\vartheta_n)\to (\ub_\i,\vartheta_\i)$.

We have thus constructed an extension of $\f{\rd\th_{\CH}^A}{\rd\th^B}$ to $\CH$. Finally, to show that the extension is indeed continuous, we can argue exactly as in the last part of the proof of Proposition~\ref{cont.ext.prelim}; we omit the details. \qedhere

\end{proof}

\begin{lemma}\label{lemma.change.of.coord}
For $i=1,2$ and $(u,\ub,\vartheta)\in \mathcal W_{\infty}\times \mathcal V_i$, the following estimate holds:
\begin{equation}\label{LthACH}
\left|\f{\rd\th^A_{(i),\CH}}{\rd \ub}+ b^B\f{\rd\th^A_{(i),\CH}}{\rd\th_{(i)}^B}\right|(u,\ub,\vartheta) \ls e^{-\f{r_+-r_-}{r_-^2+a^2}(\ub+u)}.
\end{equation}
Moreover,
\begin{equation}\label{ext.cov}
e^{\f{r_+-r_-}{r_-^2+a^2}(\ub+u)}\left(\f{\rd\th^A_{(i),\CH}}{\rd \ub}+ b^B\f{\rd\th^A_{(i),\CH}}{\rd\th_{(i)}^B}\right)(u,\ub_{\CH},\vartheta)
\end{equation}
extends continuously to $\CH$ with respect to the differential structure given by the $(u,\ub_{\CH},\th^1_{(i)},\th_{(i)}^2)$ coordinates.
\end{lemma}
\begin{proof}
To simplify the notations, we will fix $i$ and suppress the subscripts ${ }_{(i)}$.

\textbf{Proof of \eqref{LthACH}.} Notice that since $\f{\rd\th^A_{\CH}}{\rd u}=0$ by \eqref{final.th.2}, we have, by Proposition \ref{metric.der.Ricci},
\begin{equation}\label{LbLthACH}
\f{\rd}{\rd u}\left(\f{\rd\th^A_{\CH}}{\rd \ub}+ b^B\f{\rd\th^A_{\CH}}{\rd\th^B}\right)=\f{\rd b^B}{\rd u}\f{\rd\th_{\CH}^A}{\rd\th^B}=2\Om^2(\eta^B-\etab^B)\f{\rd\th_{\CH}^A}{\rd\th^B}.
\end{equation}
To proceed, we note that the initial condition \eqref{final.th.1} and the fact that $\|b_\Ke\|_{L^\infty(S_{u,\ub})}\ls e^{-\f{r_+-r_-}{r_-^2+a^2}(\ub+u)}$ imply that for any $\vartheta\in S_{u_f,\ub}$,
$$\left|\f{\rd\th^A_{\CH}}{\rd \ub}+ b^B\f{\rd\th^A_{\CH}}{\rd\th^B}\right|(u_f,\ub,\vartheta)\ls e^{-\f{r_+-r_-}{r_-^2+a^2}(\ub+u_f)}.$$ 
We then use the following very rough bounds: (1) $|\f{\rd\th_{\CH}^A}{\rd\th^B}|\ls 1$ by Proposition~\ref{prop.dthdth.diff}, (2) $\|\eta\|_{L^{\i}_uL^{\i}_{\ub}L^{\i}(S)}+\|\etab\|_{L^{\i}_uL^{\i}_{\ub}L^{\i}(S)}\ls 1$ by the bounds on $\NS$ and on the background Kerr value and that (3) $\Om^2\ls \Om_\Ke^2 \ls e^{-\f{r_+-r_-}{r_-^2+a^2}(\ub+u)}$, which again follows from the bounds on $\NS$. Integrating these bounds and using the initial condition \eqref{final.th.1}, we thus obtain \eqref{LthACH}:
$$\left|\f{\rd\th^A_{\CH}}{\rd \ub}+ b^B\f{\rd\th^A_{\CH}}{\rd\th^B}\right|_{\gamma}\ls e^{-\f{r_+-r_-}{r_-^2+a^2}(\ub+u)}.$$

\textbf{Continuous extendibility of \eqref{ext.cov}.} Finally, consider \eqref{LbLthACH} again and note that $\mbox{(RHS of \eqref{LbLthACH})}=2\Om^2(\eta^B-\etab^B)\f{\rd\th_{\CH}^A}{\rd\th^B}$ satisfies the property that $\Om_\Ke^{-2}\mbox{(RHS of \eqref{LbLthACH})}$ is continuously extendible to $\CH$ due to Proposition~\ref{cont.ext.prelim} and the continuous extendibility of $\f{\rd\th_{\CH}^A}{\rd\th^B}$ that we established in Proposition~\ref{prop.dthdth.diff} above. Consequently, \eqref{LbLthACH} implies that 
$$e^{\f{r_+-r_-}{r_-^2+a^2}(\ub+u)}\left(\f{\rd\th^A_{\CH}}{\rd \ub}+ b^B\f{\rd\th^A_{\CH}}{\rd\th^B}\right)$$ 
is also continuously extendible to $\CH$. This concludes the proof of the lemma.
\end{proof}

The following statement, which has already been discussed in Section~\ref{sec:two.diff.str}, follows as an easy corollary of Proposition~\ref{prop.dthdth.diff}. This will allow us to infer that the continuous extendibility statements in Proposition~\ref{cont.ext.prelim} and Lemma~\ref{lemma.change.of.coord} also apply in the $(u,\ub_{\CH},\th^1_{(i),\CH},\th^2_{(i),\CH})$ coordinate system.
\begin{corollary}\label{prop:sphere.eq}
For $i=1,2$, if an $S$-tangent tensor field $\phi$ is continuously extendible to the boundary $\CH$ in the $(u,\ub_{\CH},\th_{(i)}^1,\th_{(i)}^2)$ coordinate system, then $\phi$ is also continuously extendible to $\CH$ in the $(u,\ub_{\CH},\th^1_{(i),\CH},\th^2_{(i),\CH})$ coordinate system.
\end{corollary}

\subsection{Proof of continuity of the metric}\label{sec.metric.cont.ext}

We are now ready to show that the metric is continuously extendible to $\CH$. In the following theorem, we take $(\th^1_{(i),\CH},\th^2_{(i),\CH})$ to be the specific coordinates defined using the stereographic projection in Section~\ref{sec.CH.coord}.
\begin{theorem}\label{metric.cont.ext}
For $i=1,2$, the metric $g$ in the $(u,\ub_{\CH},\th^1_{(i),\CH},\th^2_{(i),\CH})$ coordinate system in $\mathcal W_{\infty}\times \mathcal V_i$ extends continuously up to the Cauchy horizon $\CH$.
\end{theorem}

\begin{proof}
As before, we fix $i$ and suppress the subscripts ${ }_{(i)}$ in our notations. With $\gamma$, $b^A$ and $\Om^2$ denoting the metric component in the $(u,\ub,\th^1,\th^2)$ coordinate system, i.e.
$$g=-2\Omega^2(du\otimes d\ub+d\ub\otimes du)+\gamma_{AB}(d\th^A-b^A d\ub)\otimes (d\th^B-b^B d\ub),$$
the metric in the $(u,\ub_{\CH},\th_{\CH}^1,\th_{\CH}^2)$ coordinate system in the region\footnote{Recall \eqref{CH.def.last} in Section~\ref{sec.CH}.} $\{(u,\ub_{\CH}):u\leq u_f,\, \ub_-\leq \ub,\, u+\ub\geq C_R\}\times \mathbb S^2$, takes the form
$$g=-2\Omega^2_{\CH}(du\otimes d\ub_{\CH}+d\ub_{\CH}\otimes du)+(\gamma_{\CH})_{AB}(d\th_{\CH}^A-(b_{\CH})^A d\ub_{\CH})\otimes (d\th_{\CH}^B-(b_{\CH})^B d\ub_{\CH}),$$
where 
\begin{equation}\label{CH.metric.coeff.1}
(\gamma_{\CH})_{AB}=\gamma_{A'B'}((\f{\rd\th_{\CH}}{\rd\th})^{-1})_A{ }^{A'}((\f{\rd\th_{\CH}}{\rd\th})^{-1})_B{ }^{B'},
\end{equation}
 \begin{equation}\label{CH.metric.coeff.2}
\Omega^2_{\CH}=\Om^2 e^{\f{r_+-r_-}{r_-^2+a^2}\ub},\quad (b_{\CH})^A=e^{\f{r_+-r_-}{r_-^2+a^2}\ub}\left(\f{\rd\th^A_{\CH}}{\rd \ub}+ b^B\f{\rd\th_{\CH}^A}{\rd\th^B}\right).
\end{equation}
We now check that the metric is continuous up to $\CH$. First note that by Corollary~\ref{prop:sphere.eq}, for $S$-tangent tensors, it suffices to check continuous extendibility with respect to the $(u,\ub_{\CH},\th^1,\th^2)$ coordinates. Now for $\gamma_{\CH}$, the continuous extendibility follows from the continuity of $\gamma$ (Proposition~\ref{cont.ext.prelim}) and the continuity of $(\f{\rd\th_{\CH}}{\rd\th})^{-1}$ (Lemma \ref{lemma.change.of.coord}). For $\Om^2_{\CH}$, first note that by the estimate in Proposition~\ref{Om.Kerr.bounds} and the smoothness of the Kerr metric, $\Om_{\Ke}^2 e^{\f{r_+-r_-}{r_-^2+a^2}\ub}$ extends smoothly to $\CH$. Therefore, by Proposition~\ref{cont.ext.prelim}, $\Om^2 e^{\f{r_+-r_-}{r_-^2+a^2}\ub}=\f{\Om^2}{\Om_\Ke^2} (\Om_{\Ke}^2 e^{\f{r_+-r_-}{r_-^2+a^2}\ub})$ extends continuously to $\CH$. Finally, $(b_{\CH})^A$ extends continuously to $\CH$ by Lemma \ref{lemma.change.of.coord}.\qedhere

\end{proof}

\subsection{$\protect C^0$-closeness to the Kerr metric}\label{sec.C0.closeness}

In this subsection, we show that the metric we have constructed in Theorem~\ref{metric.cont.ext} is in fact $C^0$-close to the background Kerr metric, all the way up to and including the Cauchy horizon $\CH$. Let us note that we have already proven that (as in the statement of Theorem~\ref{aprioriestimates}) the spacetime is close to Kerr in the sense of \eqref{main.apriori.est}. On the other hand, one may be interested in a closeness statement in terms of the $C^0$ norm of the metric component, in a coordinate system for which the metric extends continuously up to the Cauchy horizon. For this purpose, we consider the metric in the coordinate system as in Theorem~\ref{metric.cont.ext}, with metric components given in \eqref{CH.metric.coeff.1} and \eqref{CH.metric.coeff.2}.

In order to compare this metric with Kerr, we will need to use a \emph{different} identification of ($\mathcal U_{\infty}\cup \CH,g)$ with $(\mathcal M_{Kerr}\cup \CH,g_{a,M})$. For this purpose, recall the various coordinate systems on the $2$-spheres $\th^A_{(i)}$, $\th^A_{(i),\CH}$ and $\th^A_{(i),\CH,\Ke}$ (see in particular Corollary~\ref{cor:coord.map} and \eqref{cor:coord.map}). Now, instead of identifying the two spacetimes via identifying the $(u,\ub_{\CH},\th^1_{(i)},\th^2_{(i)})$ coordinates (as we did up to this point), we identify the points $(u,\ub_{\CH},\th_{(i),\CH}^1,\th_{(i),\CH}^2) \in \mathcal U_{\infty}\cup \CH$ with $(u,\ub_{\CH},\th_{(i),\CH,\Ke}^1,\th_{(i),\CH,\Ke}^2) \in \mathcal M_{Kerr}\cup \CH$. 

In order to use the bounds we have obtained for $\NS$ to estimate the difference between the metric components in the $(u,\ub_{\CH},\th_{(i),\CH}^1,\th_{(i),\CH}^2)$ coordinate system under the new identification, we first need to bound the distance between this change of coordinates.

Abusing notation, we define $\vartheta(\vartheta_{\CH})$ to be the inverse of the map $\vartheta\mapsto \vartheta_{\CH}$. We also define $\vartheta_\Ke(\vartheta_{\CH})$ to be the inverse of $\vartheta\mapsto \vartheta_{\CH,\Ke}$. (Notice that we have made an identification of $\vartheta_{\CH}$ and $\vartheta_{\CH,\Ke}$.) We will also use the notation $\th^A(\vartheta_{\CH})$ and $\th^A_{\Ke}(\vartheta_{\CH})$ to denote the corresponding values of the coordinate functions in the $\mathcal V_i''$ coordinate patch of these maps when $\vartheta_{\CH}\in \mathcal V_i'$. With the estimates that we have obtained thus far, it is easy to check that (in a similar manner as in the proof of Lemma~\ref{change.of.var.coord})
\begin{equation}\label{th.diff.given.thCH}
|\th^A(\vartheta_{\CH})-\th_\Ke^A(\vartheta_{\CH})|\ls \ep.
\end{equation}
This then implies the following estimates (note, however, the weights $e^{-\f{r_+-r_-}{r_-^2+a^2}u}$ in some of the estimates, which $\to \infty$ as $u\to -\infty$):
\begin{proposition}
$\gamma_{\CH}$, $\Om^2_{\CH}$ and $b^A_{\CH}$ satisfy the following estimates:
$$\sup_{u,\ub} (\|\gamma_{\CH} - \gamma_{\CH,\Ke}\|_{L^\i(S_{u,\ub})} +\|\log \Om_{\CH} - \log\Om_{\CH,\Ke}\|_{L^\i(S_{u,\ub})}) \ls \ep,$$
$$\sup_{\ub} \|\Om^2_{\CH} - \Om^2_{\CH,\Ke}\|_{L^\i(S_{u,\ub})} \ls \ep e^{-\f{r_+-r_-}{r_-^2+a^2}u},\quad \sup_{\ub} \|b_{\CH} - b_{\CH,\Ke}\|_{L^\i(S_{u,\ub})} \ls \ep e^{-\f{r_+-r_-}{r_-^2+a^2}u}.$$
\end{proposition}
\begin{proof}
Unless otherwise specified, we will compute in the $\mathcal V_i'$ coordinate patch for a fixed $i$ and will from now on suppress all subscripts ${ }_{(i)}$. By \eqref{CH.metric.coeff.1}, 
$$(\gamma_{\CH})_{AB}=\gamma_{A'B'}((\f{\rd\th_{\CH}}{\rd\th})^{-1})_A{ }^{A'}((\f{\rd\th_{\CH}}{\rd\th})^{-1})_B{ }^{B'},$$
and similarly we have
$$(\gamma_{\CH,\Ke})_{AB}=(\gamma_\Ke)_{A'B'}((\f{\rd\th_{\CH,\Ke}}{\rd\th})^{-1})_A{ }^{A'}((\f{\rd\th_{\CH,\Ke}}{\rd\th})^{-1})_B{ }^{B'}.$$
Consider a fixed $(u,\ub,\vartheta_{\CH})$, where $\vartheta_{\CH}\in \mathcal V_i'$. Thus $\vartheta(\vartheta_{\CH})\in\mathcal V_i''$ and $\vartheta_{\Ke}(\vartheta_{\CH}) \in \mathcal V_i'\subset \mathcal V_i''$ and we can perform coordinate calculations as follows:
\begin{equation*}
\begin{split}
& \:|(\gamma_{\CH})_{AB}(u,\ub,\vartheta_{\CH}) - (\gamma_{\CH,\Ke})_{AB}(u,\ub,\vartheta_{\CH})|\\
= & \:|\gamma_{A'B'}((\f{\rd\th_{\CH}}{\rd\th})^{-1})_A{ }^{A'}((\f{\rd\th_{\CH}}{\rd\th})^{-1})_B{ }^{B'}(u,\ub,\vartheta(\vartheta_{\CH})) \\
&\: - (\gamma_\Ke)_{A'B'}((\f{\rd\th_{\CH,\Ke}}{\rd\th})^{-1})_A{ }^{A'}((\f{\rd\th_{\CH,\Ke}}{\rd\th})^{-1})_B{ }^{B'}(u,\ub,\vartheta_\Ke(\vartheta_{\CH}))|\\
\ls & \:\sup_{\vartheta\in S_{u,\ub}}(\sup_{A'B'}\underbrace{|\gamma_{A'B'} - (\gamma_\Ke)_{A'B'}|(u,\ub,\vartheta)}_{\doteq I} + \sup_{CC'}\underbrace{|(\f{\rd\th_{\CH}}{\rd\th})^{-1})_C{ }^{C'}-(\f{\rd\th_{\CH,\Ke}}{\rd\th})^{-1})_C{ }^{C'}|(u,\ub,\vartheta)}_{\doteq II})\\
& \:+ |\underbrace{(\gamma_\Ke)_{A'B'}((\f{\rd\th_{\CH,\Ke}}{\rd\th})^{-1})_A{ }^{A'}((\f{\rd\th_{\CH,\Ke}}{\rd\th})^{-1})_B{ }^{B'}(u,\ub,\vartheta(\vartheta_{\CH}))}_{\doteq III_A}\\
&\:\qquad -\underbrace{(\gamma_\Ke)_{A'B'}((\f{\rd\th_{\CH,\Ke}}{\rd\th})^{-1})_A{ }^{A'}((\f{\rd\th_{\CH,\Ke}}{\rd\th})^{-1})_B{ }^{B'}(u,\ub,\vartheta_\Ke(\vartheta_{\CH}))|}_{\doteq III_B}\\
\ls &\: \ep,
\end{split}
\end{equation*}
where $I$ is controlled using the bound for $\tg$ in the $\NS$ energy (established in Theorem~\ref{main.quantitative.thm}), $II$ is estimated using \eqref{est.dthdth.diff}, and $III_A-III_B$ is bounded using \eqref{th.diff.given.thCH} and the mean value theorem (and the uniform smoothness of all the Kerr quantities). Since the estimate is independent of $u$, $\ub$ and $\vartheta_{\CH}$, we obtain the desired conclusion.

For $\log\Om_{\CH}-\log\Om_{\CH,\Ke}$, note that by \eqref{CH.metric.coeff.2} and the computations in Section~\ref{sec.coord.near.CH},
$$\log \Omega_{\CH}=\log \Om + \f{r_+-r_-}{2(r_-^2+a^2)}\ub,\quad \log \Omega_{\CH,\Ke}=\log \Om_{\Ke} + \f{r_+-r_-}{2(r_-^2+a^2)}\ub.$$
The desired estimate for $\log\Om_{\CH}-\log\Om_{\CH,\Ke}$ hence follow similarly as before using \eqref{th.diff.given.thCH}, the estimates for $\NS$ and the mean value theorem; we omit the details. 

Once we have obtained the estimate for $\log\Om_{\CH}-\log\Om_{\CH,\Ke}$, the desired bound for $\Om_{\CH}^2-\Om_{\CH,\Ke}^2$ simply follows from the calculus inequality $|e^{\vartheta}-1|\leq |\vartheta|e^{|\vartheta|}$ and the bound $\Om_{\CH,\Ke}^2\ls e^{-\f{r_+-r_-}{r_-^2+a^2}u}$.

For $b_{\CH}$, since we have not yet obtained an estimate for $\f{\rd\th^A_{\CH}}{\rd \ub} - \f{\rd\th^A_{\CH,\Ke}}{\rd \ub}$, instead of using \eqref{CH.metric.coeff.2}, we apply Proposition~\ref{metric.der.Ricci} (to \eqref{CH.metric.coeff.1} and \eqref{CH.metric.coeff.2}), which reads
$$\f{\rd b_{\CH}^A}{\rd u} = 2\Omega^2(\eta^A-\etab^A)=4\Omega_{\CH}^2(\zeta_{\CH}^A-\nab^A\log\Om_{\CH}).$$
By \eqref{final.th.1} and the definition of the Kerr coordinates near the Cauchy horizon in Section~\ref{sec.coord.near.CH}, $b_{\CH}^A = b_{\CH,\Ke}$ on $\{u=u_f\}$. Therefore, to obtain the desired conclusion it suffices to bound
$$\int_{u}^{u_f} \left(\Omega_{\CH}^2(\zeta_{\CH}^A-\nab^A\log\Om_{\CH}) - \Omega_{\CH,\Ke}^2(\zeta_{\CH,\Ke}^A-\nab^A\log\Om_{\CH,\Ke})\right)(u',\ub_{\CH},\vartheta_{\CH}) \, du'$$
by $\ep e^{-\f{r_+-r_-}{r_-^2+a^2}u}$. 
Now note that using the same argument as we controlled $\gamma_{\CH}-\gamma_{\CH,\Ke}$ above, we have
$$\sup_{u,\ub}(\|\zeta_{\CH} - \zeta_{\CH,\Ke}\|_{L^{\i}(S_{u,\ub})} + \|\nab\log\Om_{\CH} - \nab\log\Om_{\CH,\Ke}\|_{L^{\i}(S_{u,\ub})} )\ls \ep.$$
Combining this with the estimates for $\Om_{\CH}^2-\Om_{\CH,\Ke}^2$ we obtained above, this yields
\begin{equation*}
\begin{split}
&\: \left|\int_{u}^{u_f} \left(\Omega_{\CH}^2(\zeta_{\CH}^A-\nab^A\log\Om_{\CH}) - \Omega_{\CH,\Ke}^2(\zeta_{\CH,\Ke}^A-\nab^A\log\Om_{\CH,\Ke})\right)(u',\ub_{\CH},\vartheta_{\CH}) \, du'\right|\\
\ls &\: \ep \int_{u}^{u_f}  e^{-\f{r_+-r_-}{r_-^2+a^2}u'}\, du' \ls \ep e^{-\f{r_+-r_-}{r_-^2+a^2}u},
\end{split}
\end{equation*}
which implies the desired bound for $b_{\CH}-b_{\CH,\Ke}$. \qedhere
\end{proof}

\subsection{Conclusion of the proof of Theorem~\ref{aprioriestimates}}\label{sec.completion.everything}

We are now ready to conclude the proof of Theorem~\ref{aprioriestimates}. Given Theorems~\ref{I.final} and \ref{metric.cont.ext}, it remains to check that $(\mathcal U_\infty,g)$ is indeed the maximal globally hyperbolic future development of $\Sigma_0\cap \{\tau> -u_f+C_R\}$:
\begin{proposition}\label{lem:GH}
The solution $(\mathcal U_\infty,g)$ is the maximal globally hyperbolic future development of $\Sigma_0\cap \{\tau> -u_f+C_R\}$.
\end{proposition}
\begin{proof}
$(\mathcal U_\infty,g)$ is by definition a future development. Hence, it suffices to prove maximality\footnote{Note that in \cite{geroch}, maximality is to be understood in the sense that every other future development embeds isometrically into it. Nevertheless, the result of \cite{geroch} implies that it suffices to check that it cannot be future extended as a globally hyperbolic future development.}. Suppose not, i.e.~there exists a proper globally hyperbolic future extension $(\underline{\mathcal M},\underline{g})$ of $(\mathcal U_\infty,g)$. Consider a (smooth) future-directed causal curve $\Gamma$ which passes from $(\mathcal U_\infty,g)$ to $(\underline{\mathcal M},\underline{g})$ such that $\Gamma\restriction_{\mathcal U_{\infty}}$ is a future-directed inextendible causal curve in $(\mathcal U_\infty,g)$. Since $\Gamma\restriction_{\mathcal U_{\infty}}$ is inextendible, it follows that either $\sup_\Gamma u=u_f$ or $\sup_\Gamma \ub = \infty$ (otherwise $\Gamma\restriction_{\mathcal U_{\infty}}$ is contained in a compact subset of $\mathcal U_{\ub_f}$, which contradicts inextendibility). In particular, there exists a sequence of points $\{q_n\}_{n=1}^\infty\subset \mathcal U_{\ub_f}$ and a point $q \in \rd\mathcal U_{\infty}$ with $q_{n+1} \in J^+(q_n)$ and $q_n\to q$ such that either $u(q_n)\to u_f$ or $\ub(q_n)\to\infty$. However, $J^-(q)\cap (\Sigma_0\cap \{\tau> -u_f+C_R\})$ (as a subset of $\underline{\mathcal M}$) is then non-compact, contradicting global hyperbolicity.
\end{proof}

\begin{proof}[Proof of Theorem~\ref{aprioriestimates}]
This follows from Theorems~\ref{I.final} and \ref{metric.cont.ext} and Proposition~\ref{lem:GH}.
\end{proof}

\section{Relation with weak null singularities}\label{sec:weaknull}

In this final section, we discuss the relation between the spacetimes that we have constructed in Theorem \ref{PRWTO} and the spacetimes with essential weak null singularities constructed in~\cite{LukWeakNull}. Recall from the discussion in the introduction (see Conjecture~\ref{ChrSCC}) that conjecturally a generic subclass of the class of initial data considered in Theorem~\ref{PRWTO} lead to an essential weak null singularity. While we have not proved that any of the spacetimes we constructed are actually singular, we show in this section that the estimates that we have proven in this paper for the spacetime metric are consistent with this expectation. In particular, we show that the upper bounds that we have obtained in the course of the proof of Theorem~\ref{PRWTO} are comparable to those assumed in~\cite{LukWeakNull} (see Theorem \ref{thm.wns}).

To proceed, we consider the spacetime $(\mathcal U_\infty\cup\CH, g)$ in the $(u,\ub_{\CH}, \th^1_{(i),\CH}, \th^2_{(i),\CH})$ coordinate system as in Theorem~\ref{metric.cont.ext}. Recall from the proof of Theorem~\ref{metric.cont.ext} that (after suppressing the subscripts ${ }_{(i)}$ in the notations)
$$g=-2\Omega^2_{\CH}(du\otimes d\ub_{\CH}+d\ub_{\CH}\otimes du)+(\gamma_{\CH})_{AB}(d\th_{\CH}^A-(b_{\CH})^A d\ub_{\CH})\otimes (d\th_{\CH}^B-(b_{\CH})^B d\ub_{\CH}),$$
where 
$$(\gamma_{\CH})_{AB}=\gamma_{A'B'}((\f{\rd\th_{\CH}}{\rd\th})^{-1})_A{ }^{A'}((\f{\rd\th_{\CH}}{\rd\th})^{-1})_B{ }^{B'},$$
$$ \Omega^2_{\CH}=\Om^2 e^{\f{r_+-r_-}{r_-^2+a^2}\ub},\quad (b_{\CH})^A=e^{\f{r_+-r_-}{r_-^2+a^2}\ub}\left(\f{\rd\th^A_{\CH}}{\rd \ub}+ b^B\f{\rd\th_{\CH}^A}{\rd\th^B}\right).$$

Our goal in this section is to translate the estimates obtained in Theorem~\ref{aprioriestimates} to estimates in the $(u,\ub_{\CH}, \th^1_{(i),\CH}, \th^2_{(i),\CH})$ coordinate system. Since we are only concerned with \emph{local} behaviour of the metric near $\CH$, in this section, \textbf{\emph{all implicit constants in $\ls$ are allowed to depend on $u<u_f$, in addition to $M$, $a$, $\de$ and $C_R$}}. For any given $u<u_f$, however, all the estimates are required to hold uniformly in $\ub$, $\th^1_{(i)}$ and $\th^2_{(i)}$.

First, we note that by Theorem~\ref{metric.cont.ext}, $(u,\ub_{\CH}, \th^1_{(i),\CH}, \th^2_{(i),\CH})$ is a system of coordinates such that the metric components are continuous up to $\CH$. Moreover, on any finite interval of $u$, $\Om_{\CH}$ satisfies $1\ls \Om^2_{\CH}\ls 1$ (with implicit constants depending on the $u$-interval) and $\det \gamma \gtrsim 1$.

We now consider the behaviour of the Ricci coefficients defined with respect to the following regular null frame (with $e_3$, $e_4$ given according to \eqref{e3.e4.def}):\footnote{Note that this is a different null frame from that defined in \cite{LukWeakNull}. Nevertheless, if one introduces the null frame
$$(e_3')_{\CH}=\f{1}{\Om_{\CH}}\f{\rd}{\rd u},\;(e_4')_{\CH}=\f{1}{\Om_{\CH}}\left(\f{\rd}{\rd\ub_{\CH}}+(b_{\CH})^A\f{\rd}{\rd\th^A_{\CH}}\right),\;(e_A)_{\CH}=\f{\rd}{\rd\th^A_{\CH}}$$
as in \cite{LukWeakNull}, we would obtain similar estimates since $1\ls \Om^2_{\CH}\ls 1$ and the regular derivatives of $\Om^2_{\CH}$ are also bounded.
} \footnote{Here, $\f{\rd}{\rd\ub_{\CH}}$ is understood as the coordinate vector field in the $(u,\ub_{\CH},\th^1_{\CH}, \th^2_{\CH})$ coordinate system.}
$$(e_3)_{\CH}=\f{\rd}{\rd u},\;(e_4)_{\CH}=\f{1}{\Om_{\CH}^2}\left(\f{\rd}{\rd\ub_{\CH}}+(b_{\CH})^A\f{\rd}{\rd\th^A_{\CH}}\right), \;(e_A)_{\CH}=\f{\rd}{\rd\th^A_{\CH}}$$
and then define the Ricci coefficients according to \eqref{Ricci.coeff.def} with this null frame, which we will label as $\chi_{\CH}$, $\eta_{\CH}$, $\etab_{\CH}$ and $\chib_{\CH}$. Nevertheless, one can check by direct computations and using the estimates that we have obtained that $\chi_{\CH}$, $\eta_{\CH}$, $\etab_{\CH}$ and $\chib_{\CH}$ are in fact comparable to $\chi$, $\eta$, $\etab$ and $\chib$ respectively. We will therefore just derive the upper bounds for the Ricci coefficients defined with respect to $e_3$, $e_4$ and $e_A$ such that the bounds are in terms of the variables $u$ and $\ub_{\CH}$. The following is the main result of this section, which shows that locally near $\CH$, the Ricci coefficients (and the Gauss curvature) satisfy similar upper bounds as in~\cite{LukWeakNull} (see, however, Remarks \ref{chi.caveat} and \ref{ang.der.caveat} below):
\begin{theorem}\label{thm.wns}
Given the spacetime $(\mathcal U_{\infty}\cup \CH,g)$ we constructed in Theorem~\ref{metric.cont.ext}, the Ricci coefficients satisfy the following upper bounds for every fixed $u<u_f$ with an implicit constant depending on $u$ (in addition to $M$, $a$, $\de$ and $C_R$):
\begin{equation}\label{thm.wns.1}
\begin{split}
\sum_{i\leq 2}&\|\ub_{\CH}\nab^i\chi\|_{L^\infty_{\ub_{\CH}}L^2(S_{u,\ub_{\CH}})}\\
&+\sum_{i\leq 3}\|\ub_{\CH}^{\f12}\log_+^{\f 12+\de}(\f{1}{\ub_{\CH}})\nab^i\chi\|_{L^2_{\ub_{\CH}}L^2(S_{u,\ub_{\CH}})}\ls  1,
\end{split}
\end{equation}
and
$$\sum_{i\leq 2}\|\nab^i(\eta,\etab,\chib)\|_{L^\infty_{\ub_{\CH}}L^2(S_{u,\ub})}+\sum_{i\leq 1}\|\nab^i K\|_{L^\infty_{\ub_{\CH}}L^2(S_{u,\ub})}\ls 1.$$
Here, $L^2_{\ub_{\CH}}$ and $L^{\i}_{\ub_{\CH}}$ is taken with respect to the measure $d\ub_{\CH}$, where $\ub(\ub_{\CH})\in [-u+C_R,\infty)$ and $\ub(\ub_{\CH})$ is defined as in \eqref{CH.def.last} (for large $\ub$). Also, we have used the notation $\log_+ x=\begin{cases}
\log x\quad\mbox{if }x\geq e\\
1\quad\mbox{otherwise}.
\end{cases}
$
\end{theorem}
\begin{proof}
First, it is straightforward to check that the Kerr metric is smooth up to $\CH$. Therefore, in order to prove the stated estimates, it suffices to consider $\tilde{\chi}$, $\tilde{\eta}$, $\tilde{\etab}$ and $\tilde{\chib}$. Recall from Theorem \ref{Om.Kerr.bounds} that near $\CH$, we have
$$\Om_\Ke^2\sim e^{-\f{r_+-r_-}{r_-^2+a^2}(\ub+u)}.$$
Therefore, in view of \eqref{CH.def.last}, for every $u$, we have (with the implicit constant depending on $u$)
$$\ub_{\CH}\sim \Om_\Ke^2.$$
Now the bounds 
$$\sum_{i\leq 2}\left(\|\ub_{\CH}\nab^i\chi\|_{L^\infty_{\ub_{\CH}}L^2(S_{u,\ub_{\CH}})}+\|\nab^i(\eta,\etab,\chib)\|_{L^\infty_{\ub_{\CH}}L^2(S_{u,\ub})}\right)+\sum_{i\leq 1}\|\nab^i K\|_{L^\infty_{\ub_{\CH}}L^2(S_{u,\ub})}\ls 1$$
follow straightforwardly from the boundedness of $\NS$ established in Theorem~\ref{aprioriestimates}. Finally, for the weighted-$L^2_{\ub_{\CH}}L^2(S_{u,\ub_{\CH}})$ norm of $\nab^i\chi$, we note that
$$\Om_\Ke^4\, d\ub \sim \ub_{\CH}\,d\ub_{\CH}$$
and the estimate therefore follows from the boundedness of $\NH$ established in Theorem~\ref{aprioriestimates}.
\end{proof}

\begin{remark}[Conjectural \underline{lower} bound on $\chi$]
\underline{Conjecturally}, for $(\mathcal U_\i\cup \CH,g)$ arising from a \underline{generic} subclass of initial data (of the class of initial data that we consider), the metric $g$ is not in $W^{1,2}_{loc}$. This will correspond to $\chi$ blowing up as $\ub_{\CH}\to 0$. While this is not demonstrated in the present paper, one may \underline{conjecture} based on an analogy with the linear wave equation (see Section~\ref{LWERNK}) that \underline{generically}, \eqref{thm.wns.1} in the $i=0$ case does \underline{not} hold when $\de$ is chosen to be too large, i.e.~there exists $q>0$ such that 
$$\|\ub_{\CH}^{\f12}\log_{+}^{q}(\f{1}{\ub_{\CH}})\chi\|_{L^2_{\ub_{\CH}}L^2(S_{u,\ub_{\CH}})}=\infty$$
for every $u<u_f$.
\end{remark}

\begin{remark}[Different norms of $\chi$]\label{chi.caveat}
We note that if we take the boundedness of 
$$\sum_{i\leq 2}\|\ub_{\CH}\nab^i\chi\|_{L^\infty_{\ub_{\CH}}L^2(S_{u,\ub_{\CH}})}$$ 
and then take the $L^2_{\ub_{\CH}}$ norm of $\nab^i\chi$ for $i\leq 2$, we do \underline{not} obtain the boundedness of 
\begin{equation}\label{L2.bd.chi.local}
\sum_{i\leq 2}\|\ub_{\CH}^{\f12}\log_+^{\f 12+\de}(\f{1}{\ub_{\CH}})\nab^i\chi\|_{L^2_{\ub_{\CH}}L^2(S_{u,\ub_{\CH}})}.
\end{equation} 
In other words, in the $L^\infty_{\ub_{\CH}}L^2(S_{u,\ub_{\CH}})$ norm, we have slightly weaker control\footnote{One should note that if we track the norms more carefully in the proof, we can also bound the slightly stronger norm $$\sum_{i\leq 2}\|\ub_{\CH}\log_+^{\de}(\f{1}{\ub_{\CH}})\nab^i\chi\|_{L^\infty_{\ub_{\CH}}L^2(S_{u,\ub_{\CH}})},$$ but it is still weaker than \eqref{L2.bd.chi.local}.} regarding the potential blow-up profile. This is slightly different from the setup in \cite{LukWeakNull}, where $\nab^i\chi$ is controlled in a weighted $L^\i_{\ub}L^2(S)$ space in which the inverse of the weight is integrable. However, it is not difficult to see that with the slightly weaker upper bounds as we have in this paper, one can also derive the main theorems in \cite{LukWeakNull} with only straightforward modifications of the proof.
\end{remark}

\begin{remark}[Regularity in the angular directions]\label{ang.der.caveat}
Although the strength of the singular weight function as $\ub_{\CH}\to 0$ in Theorem \ref{thm.wns} is similar to that in \cite{LukWeakNull}, the spacetimes constructed in \cite{LukWeakNull} are more regular in the angular $\nab$ directions. We note however that this does not change the nature of the analysis significantly. Moreover, if one were to require also the boundedness of higher angular derivatives (with the same weights in $\ub$) in the initial data energy $\mathcal D$ in \eqref{data.D.def}, then it is straightforward to see using the methods in the present paper that one can also control higher angular derivatives near $\CH$.
\end{remark}

\begin{remark}[Degeneracy of the top order norms]
Notice that while at lower order of derivatives, the quantities $(K,\eta,\etab)$ are bounded\footnote{More precisely, by the estimates of $\NS$, we have that $\|(\eta,\etab)\|_{L^\i(S_{u,\ub})}$ and $\|K\|_{L^p(S_{u,\ub})}$ ($p<\infty$) are uniformly bounded in $u$ and $\ub$.} as $\ub_{\CH}\to 0$ for every fixed $u< u_f$, at the top order of derivatives (i.e.~$2$ derivatives for $K$ and $3$ derivatives of $\eta$ and $\etab$), the $L^2_uL^2(S_{u,\ub_{\CH}})$ norms given by $\NH$ are in fact \underline{degenerate} as $\ub_{\CH}\to 0$. More precisely, in the double null foliation $(u,\ub_{\CH})$, the control for $(\nab^2K,\nab^3 \eta,\nab^3\etab)$ in the $L^2_uL^2(S_{u,\ub_{\CH}})$ norms given by $\NH$ translates to the boundedness of 
$$\|\ub_{\CH}^{\f 12}\log_+^{\f 12+\de}(\f{1}{\ub_{\CH}})(\nab^2 K, \nab^3\eta, \nab^3\etab)\|_{L^2_uL^2(S_{u,\ub_{\CH}})}.$$
This is consistent with the degeneracy of the top order norms in \cite{LukWeakNull}. On the other hand, as in \cite{LukWeakNull}, in order to close our estimates, it is important that we have \underline{non-degenerate} \underline{spacetime} norms. More precisely, in the double null foliation $(u,\ub_{\CH})$, the boundedness of $\NI$ implies that at the top order
$$\|\log_+^{\f 12+\de}(\f{1}{\ub_{\CH}})(\nab^2 K, \nab^3\eta, \nab^3\etab)\|_{L^2_u L^2_{\ub_{\CH}} L^2(S_{u,\ub_{\CH}})}$$
is bounded. Notice that in the regular coordinates, not only are there no degeneracies, we in fact have a favourable weight $\log_+^{\f 12+\de}(\f{1}{\ub_{\CH}})$.
\end{remark}

\appendix

\section{Geometry of Kerr spacetime}\label{sec.Kerr.geometry}

Let $\mathcal M_{Kerr}$ be a smooth four-dimensional manifold diffeomorphic to $\mathbb R^2\times \mathbb S^2$. From now on, $\mathcal M_{Kerr}$ will be the manifold for the \emph{black hole interior} (not to be confused with the maximal globally hyperbolic development of Kerr data discussed in Section~\ref{KCHintro}). Let $M, a\in \mathbb R$ be such that $0<|a|<M$. The Kerr metric $g_{a,M}$ with mass $M$ and specific angular momentum $a$ is a smooth metric on $\mathcal M_{Kerr}$, which takes the following form in the Boyer-Lindquist local coordinates\footnote{$(\th,\phi)$ are spherical coordinates on $\mathbb S^2$ and have a coordinate singularity at $\th=0,\pi$. It is a standard fact that regular coordinates can be introduced there. We omit the details.} $(t,r,\th,\phi)$:
\begin{equation}\label{BL}
\begin{split}
g_{a,M}=&-(1-\frac {2Mr}{\Sigma})\, dt\otimes dt+\frac{\Sigma}{\Delta}\, dr\otimes dr+\Sigma\, d\th\otimes d\th\\
&+ R^2\sin^2\th\, d\phi\otimes d\phi-\frac{2Mar\sin^2\th}{\Sigma}(d\phi\otimes dt+dt\otimes d\phi),
\end{split}
\end{equation}
where\footnote{Here, $\Sigma$ is not to be confused with the notation for the spacelike hypersurfaces $\Sigma_0$ and $\Sigma_\Ke$ (see for example Sections~\ref{precise.PRWTO} and \ref{sec.Kerr.spacelike}).}
$$\Sigma=r^2+a^2\cos^2\th,\quad R^2=r^2+a^2+\frac{2Ma^2r\sin^2\th}{\Sigma},\quad\Delta=r^2+a^2-2Mr$$
and $t\in \mathbb R$, $r\in (r_-,r_+)$, $\th\in (0,\pi)$ and $\phi\in [0,2\pi)$. Here, $r_-$ and $r_+$ are the smaller and larger roots of $\Delta=r^2+a^2-2Mr$ respectively. In the Boyer-Lindquist $(t,r,\th,\phi)$-coordinate system, $g_{a,M}^{-1}$ is given by
\begin{equation}\label{BL.inverse}
\begin{split}
g_{a,M}^{-1}=&-\f{R^2}{\Delta}\f{\rd}{\rd t}\otimes\f{\rd}{\rd t}-\frac{2Mar}{\Sigma\Delta}(\f{\rd}{\rd t}\otimes\f{\rd}{\rd \phi}+\f{\rd}{\rd \phi}\otimes\f{\rd}{\rd t})+\f{1}{\Delta\sin^2\th}(1-\frac {2Mr}{\Sigma})\f{\rd}{\rd \phi}\otimes\f{\rd}{\rd \phi}\\
&+\f{\Delta}{\Sigma}\f{\rd}{\rd r}\otimes\f{\rd}{\rd r}+\f{1}{\Sigma}\f{\rd}{\rd \th}\otimes\f{\rd}{\rd \th}.
\end{split}
\end{equation}
{\bf\emph{In this paper, we only consider the case $0< |a|< M$. From now on, we fix the values $M$ and $a$.}}

We will later define the \emph{event horizon} $\mathcal H^+$ and the \emph{Cauchy horizon} $\mathcal C\mathcal H^+$ so that $\overline{\mathcal M}_{Kerr}\doteq \mathcal M_{Kerr}\cup \mathcal H^+\cup\mathcal C\mathcal H^+$ is a manifold-with-corner (see Section \ref{sec.def.horizon}).

We briefly outline the organisation of this section. Our main goal is to introduce the double null foliation in $\mathcal M_{Kerr}$ so that we can use the geometric setup introduced in Section~\ref{secsetup}. In \textbf{Section~\ref{sec.PIcoord}}, we introduce the Pretorius--Israel coordinate system $(t,r_*,\th_*,\phi)$ from \cite{Pretorius}. In \textbf{Section~\ref{sec.basic.r.th}} and \textbf{Section~\ref{sec.r.th.higher}}, we then prove estimates regarding the change of variable $(t,r,\th,\phi)\mapsto (t,r_*,\th_*,\phi)$. Using this, we then introduce in \textbf{Section~\ref{Kerr.dbn}} a double null coordinate system $(u,\ub,\th_*,\phi_*)$ so that the metric $g_{a,M}$ takes the form as in Section \ref{coordinates}. In \textbf{Section~\ref{sec.def.horizon}}, we introduce the event horizon and the Cauchy horizon, which can be viewed as boundaries of $\mathcal M_{Kerr}$. In \textbf{Section~\ref{sec.Kerr.RC.CC.est}}, we derive estimates for the Ricci coefficients. In \textbf{Section~\ref{sec.Kerr.cuv}}, we then obtain bounds for the renormalised null curvature components. Finally, in \textbf{Section~\ref{sec.Kerr.spacelike}}, we consider spacelike hypersurfaces in $\mathcal M_{Kerr}$. 

\subsection{The Pretorius--Israel coordinate system}\label{sec.PIcoord}

Following\footnote{In \cite{Pretorius}, the coordinate system is defined in the exterior region by an appropriate normalisation as $r_*\to \infty$. In the present paper, since we are only interested in the interior of the Kerr black hole, the definition is slightly different, but all the crucial algebraic manipulations are already contained in \cite{Pretorius}.} \cite{Pretorius}, we now define two new functions $r_*=r_*(r,\th)$ and $\th_*=\th_*(r,\th)$. We will show later that $(t,r_*,\th_*,\phi)$ forms a system of coordinates in $(\mathcal M_{Kerr},g_{a,M})$.

We begin by defining $\th_*$. For\footnote{While $r\in (r_-,r_+)$ in $\mathcal M_{Kerr}$, we show that $\th_*$ can in fact be defined and is smooth for $r\in [r_-,r_+]$. In particular, after we define the event horizon and Cauchy horizon in Section \ref{sec.def.horizon}, it can be easily seen that $\sin^2\th_*$ extends to a smooth function on $\mathcal M_{Kerr}\cup \H\cup \CH$. } $r\in [r_-,r_+]$, $\th\in (0,\f{\pi}{2})$, define $\th_*(r,\th)\in [\th,\f{\pi}{2})$ implicitly by
\begin{equation}\label{imp.def}
F(\th_*,r,\th)\doteq \int_{\th_*}^{\th} \frac{d\th'}{a\sqrt{\sin^2\th_*-\sin^2\th'}}+\int_{r}^{r_+}\frac{dr'}{\sqrt{(r'^2+a^2)^2-a^2 \sin^2\th_*\Delta'}}=0.
\end{equation}
where $\Delta'=r'^2-2Mr'+a^2$. We will see that this is indeed well defined, i.e.~there exists a unique $\th_*$ satisfying \eqref{imp.def}. The proof of this fact will be postponed to the next subsection (see Proposition \ref{th.well.defined}).

For $\th\in (\f{\pi}2,\pi)$, we define $\th_*$ by
\begin{equation}\label{imp.def.2}
\th_*(r,\th)=\pi-\th_*(r,\pi-\th).
\end{equation}
Moreover, let
\begin{equation}\label{imp.def.endpoints}
\th_*(r,0)=0,\quad\th_*(r,\frac{\pi}{2})=\frac{\pi}{2},\quad \th_*(r,\pi)=\pi,
\end{equation}
for $r\in [r_-,r_+]$.

We then define\footnote{The notation $\rho$ defined here is only used in this subsection to be consistent with the conventions in \cite{Pretorius}. We note that this is not related to the curvature component $\rho$ defined in \eqref{curv.comp.def}, which will be the convention used in all other parts of the paper.} $\rho$ by
\begin{equation}\label{rho.def}
\begin{split}
\rho=&\int \frac{r^2+a^2}{\Delta}\,dr+\int^{r_+}_r \frac{\left((r')^2+a^2-\sqrt{((r')^2+a^2)^2-a^2\sin^2\th_*\Delta'}\right)}{\Delta'}dr'\\
&+\int_{\th_*}^{\th}a\sqrt{\sin^2\th_*-\sin^2\th'}\,d\th'.
\end{split}
\end{equation}
Here, $\int \frac{r^2+a^2}{\Delta}\,dr$ is some fixed anti-derivative\footnote{The choice of the ambiguous additive constant will not be important in the rest of the paper.} of $\frac{r^2+a^2}{\Delta}$. Notice that the second integral converges since
$$|(r')^2+a^2-\sqrt{((r')^2+a^2)^2-a^2\sin^2\th_*\Delta'}|\ls \f{a^2\sin^2\th_*|\Delta'|}{((r')^2+a^2)},$$
for some implicit constant depending only on $M$ and $a$ and $\f{a^2\sin^2\th_*}{((r')^2+a^2)}$ clearly has a convergent integral. Also, $\rho$ ranges from $-\infty$ to $+\infty$ as $r$ varies from $r_+$ to $r_-$. As defined above, $\rho$ is a function of $r$, $\th$ and $\th_*$. Recall that $\th_*=\th_*(r,\th)$ as defined in \eqref{imp.def}, \eqref{imp.def.2} and \eqref{imp.def.endpoints}, we can define $r_*$ as a function of $r$ and $\th$ as follows:
\begin{equation}\label{r*.def}
r_*(r,\th)=\rho(r,\th,\th_*(r,\th)).
\end{equation}

We have thus defined a new coordinate system\footnote{Strictly speaking, at this point, we have not shown that this forms a coordinate system. That this is indeed the case will be proved in Proposition \ref{det.est} in the next subsection. Moreover, we will show in Section \ref{sec.r.th.higher} that the change of coordinate is smooth and therefore the $(t,r_*,\th_*,\phi)$ coordinate system gives rise to the same differential structure as that of $(t,r,\th,\phi)$.} $(t,r_*,\th_*,\phi)$. Notice that by the definition of $\rho$ in \eqref{rho.def} and the definition of $\th_*$ in \eqref{imp.def}, $r_*$ satisfies\footnote{To justify this calculation, strictly speaking we need $\f{\rd\th_*}{\rd r}$ and $\f{\rd\th_*}{\rd\th}$ to be finite. That this is indeed the case can be inferred from the estimates in Propositions \ref{det.est} and \ref{Kerr.pd.est} that we will prove later.}:
\begin{equation}\label{rstar.der.1}
\f{\rd r_*}{\rd r}=\f{\sqrt{(r^2+a^2)^2-a^2\sin^2\th_*\Delta}}{\Delta}
\end{equation}
and
\begin{equation}\label{rstar.der.2}
\f{\rd r_*}{\rd \th}=a\sqrt{\sin^2\th_*-\sin^2\th}.
\end{equation}
Therefore,
\begin{equation*}
\Delta(\f{\rd r_*}{\rd r} )^2+(\f{\rd r_*}{\rd\th})^2=\f{(r^2+a^2)^2}{\Delta}-a^2\sin^2\th.
\end{equation*}
In particular, by \eqref{BL.inverse}, 
\begin{equation}\label{Eikonal.r}
\begin{split}
(g_{a,M}^{-1})^{\mu\nu}\rd_\mu r_* \rd_\nu r_*=&(g_{a,M}^{-1})^{rr}(\f{\rd r_*}{\rd r} )^2+(g_{a,M}^{-1})^{\th\th}(\f{\rd r_*}{\rd\th})^2 \\
=&\f{\Delta}{\Sigma}(\f{\rd r_*}{\rd r} )^2+\f{1}{\Sigma}(\f{\rd r_*}{\rd\th})^2
=\f{(r^2+a^2)^2}{\Sigma\Delta}-\f{a^2\sin^2\th}{\Sigma}<0
\end{split}
\end{equation}
and $\f{\rd}{\rd r_*}$ is a future-directed timelike vector field. Recalling \eqref{BL.inverse} and noting that $(g_{a,M}^{-1})^{tt}=-\f{R^2}{\Delta}$, \eqref{Eikonal.r} also implies that 
\begin{equation}\label{Kerr.u.ub.def}
\ub=\f12(r_*+t),\quad u=\f12(r^*-t)
\end{equation}
are null variables, i.e.~they verify the eikonal equation:
\begin{equation}\label{Eikonal.r.1}
(g_{a,M}^{-1})^{\mu\nu}\rd_\mu \ub\rd_{\nu} \ub=(g_{a,M}^{-1})^{\mu\nu}\rd_\mu u\rd_{\nu} u=0.
\end{equation}
This can be seen by a direct computation:
\begin{equation*}
\begin{split}
&\f{(r^2+a^2)^2}{\Sigma\Delta}-\f{a^2\sin^2\th}{\Sigma}-\f{R^2}{\Delta}\\
=&\f{(r^2+a^2)^2-a^2\sin^2\th(r^2-2Mr+a^2)-(r^2+a^2)(r^2+a^2\cos^2\th)-2Ma^2r\sin^2\th}{\Sigma\Delta}\\
=&\f{(r^2+a^2)a^2\sin^2\th-a^2\sin^2\th(r^2+a^2)}{\Sigma\Delta}=0.
\end{split}
\end{equation*}

\subsection{Basic estimates on $r_*$ and $\th_*$}\label{sec.basic.r.th}

We now move on to obtaining estimates on $r_*$, $\th_*$ and their derivatives. In particular, we will also prove that $\th_*$ is well-defined via the relation \eqref{imp.def}. Our goal is to show that while $r_*$ and $\th_*$ are only defined implicitly, we can nonetheless obtain strong enough estimates on $r_*$, $\th_*$ and their derivatives for the purpose of the proof of our main theorem. In view of \eqref{imp.def.2}, it suffices to consider $\th\in (0,\f{\pi}{2})$. We will therefore make this assumption in the remainder of this subsection.

Before we proceed, we introduce a convention: {\bf\emph{in what follows, we will use the notation $A \ls B$ to denote the statement that $A\leq CB$ for some constant $C>0$. Unless otherwise stated, this constant will only be allowed to depend on $M$ and $a$.}} (Note in particular that we also allow this convention to be used when $A$, $B$ are both negative.)

We now show that $\th_*$, defined for $\th\in (0,\f{\pi}{2})$ in \eqref{imp.def}, is indeed well-defined:
\begin{proposition}\label{th.well.defined}
$\th_*$ is well-defined: for every $r\in [r_-,r_+]$ and $\th\in (0,\f{\pi}{2})$, there exists a unique $\th_*\in[\th,\f{\pi}{2})$ satisfying \eqref{imp.def}. Moreover, $\th_*$ satisfies the estimate
\begin{equation}\label{th.well.defined.est}
\f{\sin \th_*}{\sin\th}+\f{\cos\th}{\cos\th_*}\ls 1.
\end{equation}
Here, the implicit constant depends only on $M$ and $a$. In particular, according to \eqref{imp.def.endpoints}, $\th_*$ is continuous at the endpoints $\th=0$ and $\th=\f{\pi}{2}$.
\end{proposition}
\begin{proof}
It suffices to show that $F(\th_*;r,\th)$ defined by
$$F(\th_*;r,\th)\doteq \int_{\th_*}^{\th} \frac{d\th'}{a\sqrt{\sin^2\th_*-\sin^2\th'}}+\int_{r}^{r_+}\frac{dr'}{\sqrt{(r'^2+a^2)^2-a^2 \sin^2\th_*\Delta'}}$$
in \eqref{imp.def} has a unique zero in $(\th,\f{\pi}{2})$. Since $r\leq r_+$, we have
$$F(\th;r,\th)\geq 0.$$
Moreover, notice that since
$$\lim_{\th_*\to\f{\pi}{2}}\int_{\th_*}^{\th}\f{d\th'}{a\sqrt{\sin^2\th_*-\sin^2\th'}}\to -\infty,$$
we have
$$\lim_{\th_*\to \f{\pi}{2}}F(\th_*;r,\th)<0.$$
Since for every fixed $r$ and $\th$, $F(\th_*;r,\th)$ is continuous in $\th_*$, there exists a $\th_*\in (\th,\f{\pi}{2}]$ satisfying \eqref{imp.def} by the intermediate value theorem. To show the uniqueness of $\th_*$, we claim that $F(\th_*;r,\th)$ is monotonically decreasing in $\th_*$. Recalling that $\Delta'<0$, the monotonicity is obvious for the second term in the definition of $F$. For the first term, since
$$\int_0^{\th} \f{d\th'}{a\sqrt{\sin^2\th_*-\sin^2\th'}}$$
is clearly decreasing in $\th_*$, it suffices to show that
$$\int_0^{\th_*} \f{d\th'}{a\sqrt{\sin^2\th_*-\sin^2\th'}}$$
is increasing in $\th_*$. Let $\tilde{\th_*}\geq \th_*$. Introducing the change of variable $\f{\sin^2\tilde\th'}{\sin^2\th'}=\f{\sin^2\tilde\th_*}{\sin^2\th_*}$, we have
\begin{equation}\label{F.2.monotonicity}
\begin{split}
&\int_0^{\tilde\th_*}\f{d\tilde\th'}{a\sqrt{\sin^2\tilde\th_*-\sin^2\tilde\th'}}-\int_0^{\th_*}\f{d\th'}{a\sqrt{\sin^2\th_*-\sin^2\th'}}\\
=&\int_0^{\th_*} \big(\f{\cos\th'}{\cos\tilde\th'}-1\big)\f{d\th'}{a\sqrt{\sin^2\th_*-\sin^2\th'}}\geq 0
\end{split}
\end{equation}
as desired. This shows that indeed $\th_*$ is well-defined.

We now turn to prove the assertion on the upper bound of $\sin\th_*$. We have the following upper bound for the second integral in the definition of $F$:
\begin{equation*}
\begin{split}
&\int_{r}^{r_+}\frac{dr'}{\sqrt{(r'^2+a^2)^2-a^2 \sin^2\th_*\Delta'}}\\
\leq &\int_{r_-}^{r_+}\frac{dr'}{r'^2+a^2}\\
=&\f{1}{a}\tan^{-1}\big(\f{r_+}{a}\big)-\f 1a \tan^{-1}\big(\f{r_-}{a}\big).
\end{split}
\end{equation*}
Therefore, for every fixed\footnote{Indeed, the estimate degenerates as $\f aM\to 0$.} $M$, $a$ with $0<|a|<M$, there exists a constant $c_{M,a}>0$ such that
\begin{equation}\label{well.defined.upper.bound}
\int_{r}^{r_+}\frac{dr'}{\sqrt{(r'^2+a^2)^2-a^2 \sin^2\th_*\Delta'}}\leq \f 1a\big(\f{\pi}{2}-c_{M,a}\big).
\end{equation}
On the other hand, introducing the change of variables $\sin\th'=\sin\th_*\sin\tilde{\th}$, we obtain
\begin{equation}\label{well.defined.lower.bound}
\begin{split}
&\int_{\th}^{\th_*} \frac{d\th'}{a\sqrt{\sin^2\th_*-\sin^2\th'}}\\
=&\int_{\sin^{-1}(\f{\sin\th}{\sin\th_*})}^{\f{\pi}{2}}\frac{d\tilde{\th}}{a\cos\th'}\\
\geq &\f{1}{a}\big(\f{\pi}{2}-\sin^{-1}(\f{\sin\th}{\sin\th_*})\big).
\end{split}
\end{equation}
By the definition \eqref{imp.def}, we have
$$\int_{r}^{r_+}\frac{dr'}{\sqrt{(r'^2+a^2)^2-a^2 \sin^2\th_*\Delta'}}=\int_{\th}^{\th_*} \frac{d\th'}{a\sqrt{\sin^2\th_*-\sin^2\th'}}.$$
Therefore, combining \eqref{well.defined.upper.bound} and \eqref{well.defined.lower.bound}, we obtain
$$\f{\sin\th}{\sin\th_*}\geq \sin c_{M,a},$$
as desired.

Finally, we turn to the proof of the lower bound of $\cos\th_*$. The proof is similar to that for the upper bound for $\sin\th_*$ except that there is more room for the argument. Introduce the change of variable $\cos\th'=\cos\th_*\sec\tilde{\th}$, we have
\begin{equation}\label{well.defined.3}
\begin{split}
&\int_{\th}^{\th_*} \frac{d\th'}{a\sqrt{\sin^2\th_*-\sin^2\th'}}\\
=&\int_{\th}^{\th_*} \frac{d\th'}{a\sqrt{\cos^2\th'-\cos^2\th_*}}\\
=&\int_0^{\sec^{-1}\big(\f{\cos\th}{\cos\th_*}\big)} \frac{\sec\tilde\th d\tilde\th}{a\sin\th'}\\
\geq &\f1a \big[\log |\sec x+\tan x| \big]^{x=\sec^{-1}\big(\f{\cos\th}{\cos\th_*}\big)}_{x=0}\\
=&\f 1a\log \left| \f{\cos\th}{\cos\th_*}+\sqrt{\f{\cos^2\th}{\cos^2\th_*}-1}\right|.
\end{split}
\end{equation}
Combining \eqref{well.defined.upper.bound} and \eqref{well.defined.3}, we thus have
$$\log \left| \f{\cos\th}{\cos\th_*}+\sqrt{\f{\cos^2\th}{\cos^2\th_*}-1}\right|\leq \big(\f{\pi}{2}-c_{M,a}\big).$$
Therefore, we obtain the bound for $\f{\cos\th}{\cos\th_*}$, as desired.
\end{proof}

While we have shown that $\th_*$ is indeed well-defined, we have not yet proven that $(t,r_*,\th_*,\phi)$ forms a coordinate system. To that end, we will need upper and lower bounds on $\f{\rd}{\rd \th_*}\restriction_{r, \th fixed} F(\th_*;r,\th)$. In fact, we will also need to estimate higher derivatives to obtain bounds for the Ricci coefficients on Kerr spacetime with respect to the double null coordinates $(u,\ub,\th_*,\phi)$. First, we compute $\f{\rd}{\rd \th_*}\restriction_{r, \th fixed} F(\th_*;r,\th)$:
\begin{proposition}\label{dF.explicit}
The $\th_*$ derivative of $F$ is given by the following formula:
\begin{equation*}
\begin{split}
&\f{\rd}{\rd\th_*}\restriction_{r, \th fixed} F(\th_*;r,\th)\\
=& -\f{1}{a\sin\th_*\cos\th_*}\int_\th^{\th_*} \sqrt{\sin^2\th_*-\sin^2\th'}d\th'-\f{\sin (2\th)}{a\sin (2\th_*)\sqrt{\sin^2\th_*-\sin^2\th}}\\
&+\int_{r}^{r_+}\frac{a^2\Delta'\sin 2\th_*\,dr'}{2\big((r'^2+a^2)^2-a^2 \sin^2\th_*\Delta'\big)^{\f32}}.
\end{split}
\end{equation*}
Here, the partial derivative $\f{\rd}{\rd\th_*}\restriction_{r, \th fixed}$ is understood such that $r$ and $\th$ are both fixed.
\end{proposition}
\begin{proof}
%We first make an important notational remark which should have already been clear from the statement of the proposition: we use $\f{\rd}{\rd \th_*}\restriction_{r, \th fixed}$ to denote the derivative in $\th_*$ while $r$ and $\th$ are held fixed. This is in contrast to the notation $\f{\rd}{\rd \th_*}$ in the rest of this section where $r_*$ is held fixed instead and $r$, $\th$ are viewed as functions of $r_*$ and $\th_*$. 
Using the following formula for derivatives of elliptic integrals:
$$\f{d}{dm} \int_0^z \f{dt}{\sqrt{1-m\sin^2 t}}=\f{1}{2(1-m)}\int_0^z \f{\cos^2 t \, dt}{\sqrt{1- m\sin^2t}}-\f{\sin(2z)}{4(1-m)\sqrt{1-m\sin^2z}}, $$
we have that\footnote{We have used the above formula for $m=\f{1}{\sin^2\th_*}$.} for every $\ep\in (0,\th_*-\th)$,
\begin{equation*}
\begin{split}
&\f{\rd}{\rd \th_*}\restriction_{r, \th fixed} \int_{\th}^{\th_*-\ep} \f{d\th'}{a\sqrt{\sin^2\th_*-\sin^2\th'}}\\
=& -\f{\cos\th_*}{a\sin\th_*} \int_{\th}^{\th_*-\ep} \f{d\th'}{\sqrt{\sin^2\th_*-\sin^2\th'}}+\f{1}{a\sin\th_*}\f{\rd}{\rd\th_*}\restriction_{r, \th fixed}\int_{\th}^{\th_*-\ep} \f{d\th'}{\sqrt{1-\f{\sin^2\th'}{\sin^2\th_*}}}\\
=&-\f{\cos\th_*}{a\sin\th_*} \int_{\th}^{\th_*-\ep} \f{d\th'}{\sqrt{\sin^2\th_*-\sin^2\th'}}+\f{1}{a\sin\th_*\cos\th_*}\int_{\th}^{\th_*-\ep} \f{\cos^2\th'\,d\th'}{\sqrt{\sin^2\th_*-\sin^2\th'}}\\
&-\f{\sin(2(\th_*-\ep))}{2a\sin\th_*\cos\th_*\sqrt{\sin^2\th_*-\sin^2(\th_*-\ep)}}+\f{\sin(2\th)}{2a\sin\th_*\cos\th_*\sqrt{\sin^2\th_*-\sin^2\th}}\\
&+\f{1}{a\sqrt{\sin^2\th_*-\sin^2(\th_*-\ep)}}\\
=& \f{1}{a\sin\th_*\cos\th_*}\int_\th^{\th_*-\ep} \sqrt{\sin^2\th_*-\sin^2\th'}d\th'+\f{\sin (2\th)}{a\sin (2\th_*)\sqrt{\sin^2\th_*-\sin^2\th}}\\
&+\f{1}{a\sqrt{\sin^2\th_*-\sin^2(\th_*-\ep)}}(1-\f{\sin(2(\th_*-\ep))}{\sin (2\th_*)}).
\end{split}
\end{equation*}
For the term in the last line, notice that
\begin{equation*}
\begin{split}
&\f{1}{a\sqrt{\sin^2\th_*-\sin^2(\th_*-\ep)}}(1-\f{\sin(2(\th_*-\ep))}{\sin (2\th_*)})\\
=&\f{1}{a\sin (2\th_*)\sqrt{\sin^2\th_*-\sin^2(\th_*-\ep)}}(\sin (2\th_*)-\sin(2(\th_*-\ep)))\\
=&\f{2\cos\th_*(\sin\th_*-\sin(\th_*-\ep))+2\sin(\th_*-\ep)(\cos\th_*-\cos(\th_*-\ep))}{a\sin (2\th_*)\sqrt{\sin^2\th_*-\sin^2(\th_*-\ep)}}\\
=&\f{2\cos\th_*}{a\sin(2\th_*)}\sqrt{\f{\sin\th_*-\sin(\th_*-\ep)}{\sin\th_*+\sin(\th_*-\ep)}}-\f{2\sin(\th_*-\ep)}{a\sin(2\th_*)}\sqrt{\f{\cos(\th_*-\ep)-\cos\th_*}{\cos(\th_*-\ep)+\cos\th_*}},\\
\end{split}
\end{equation*}
which converges uniformly to $0$ for $\th_*$ in any compact subset of $(0,\f{\pi}{2})$ as $\ep\to 0$.
Therefore,
\begin{equation*}
\begin{split}
&\lim_{\ep\to 0}\f{\rd}{\rd \th_*}\restriction_{r, \th fixed} \int_{\th}^{\th_*-\ep} \f{d\th'}{a\sqrt{\sin^2\th_*-\sin^2\th'}}\\
=& \f{1}{a\sin\th_*\cos\th_*}\int_\th^{\th_*} \sqrt{\sin^2\th_*-\sin^2\th'}d\th'+\f{\sin (2\th)}{a\sin (2\th_*)\sqrt{\sin^2\th_*-\sin^2\th}},
\end{split}
\end{equation*}
where the convergence is uniform for $\th_*$ in any compact subset of $(0,\f{\pi}{2})$. Finally, since we also have
$$\lim_{\ep\to 0} \int_{\th}^{\th_*-\ep} \f{d\th'}{a\sqrt{\sin^2\th_*-\sin^2\th'}}=\int_{\th}^{\th_*} \f{d\th'}{a\sqrt{\sin^2\th_*-\sin^2\th'}},$$
where the convergence is again uniform for $\th_*$ in any compact subset of $(0,\f{\pi}{2})$, we conclude that
\begin{equation}\label{dF.coordfixed}
\begin{split}
&\f{\rd}{\rd \th_*}\restriction_{r, \th fixed} \int_{\th}^{\th_*} \f{d\th'}{a\sqrt{\sin^2\th_*-\sin^2\th'}}\\
=& \f{1}{a\sin\th_*\cos\th_*}\int_\th^{\th_*} \sqrt{\sin^2\th_*-\sin^2\th'}d\th'+\f{\sin (2\th)}{a\sin (2\th_*)\sqrt{\sin^2\th_*-\sin^2\th}}.
\end{split}
\end{equation}
Substituting this into the definition of $F$, we thus obtain the desired conclusion.
%\begin{equation*}
%\begin{split}
%&\f{\rd}{\rd \th_*} \int_{\th}^{\th_*} \f{d\th'}{a\sqrt{\sin^2\th_*-\sin^2\th'}}\\
%=& -\f{\cot\th_*\sec\th_*}{a} [\mathcal F(\th_*,\f{1}{\sin^2\th_*})-\mathcal F(\th,\f{1}{\sin^2\th_*})]\\
%&+\f{1}{a\sin\th_*}(-\f{2\cos\th_*}{\sin^2\th_*})[\f{\mathcal E(\th_*,\f{1}{\sin^2\th_*})}{-\f{2\cos^2\th_*}{\sin^4\th_*}}-\f{\mathcal F(\th_*,\f{1}{\sin^2\th_*})}{\f{2}{\sin^2\th_*}}]\\
%&+\f{1}{a\sin\th_*}(-\f{2\cos\th_*}{\sin^2\th_*})[\f{\mathcal E(\th,\f{1}{\sin^2\th_*})}{-\f{2\cos^2\th_*}{\sin^4\th_*}}-\f{\mathcal F(\th,\f{1}{\sin^2\th_*})}{\f{2}{\sin^2\th_*}}-\f{\sin(2\th)}{-\f{4\cos^2\th_*}{\sin^2\th_*}\sqrt{1-\f{\sin^2\th}{\sin^2\th_*}}}]\\
%=& \f{1}{a\sin\th_*\cos\th_*}\int_\th^{\th_*} \sqrt{\sin^2\th_*-\sin^2\th'}d\th'+\f{\sin (2\th)}{a\sin (2\th_*)\sqrt{\sin^2\th_*-\sin^2\th}}.
%\end{split}
%\end{equation*}
\end{proof}
From this we obtain the following bounds:
\begin{proposition}\label{dF.lower.bd}
Suppose $F(\th_*; r,\th) =0$. Then
$$-\f{C_{M,a}}{\sqrt{\sin^2\th_*-\sin^2\th}}\leq \f{\rd}{\rd\th_*}\restriction_{r, \th fixed} F(\th_*;r,\th)\leq -\f{\sin(2\th)}{a\sin(2\th_*)\sqrt{\sin^2\th_*-\sin^2\th}}$$
for some constant $C_{M,a}>0$ depending on $M$ and $a$.
\end{proposition}
\begin{proof}
Using Proposition \ref{dF.explicit}, it suffices to notice that 
$$-\f{1}{a\sin\th_*\cos\th_*}\int_\th^{\th_*} \sqrt{\sin^2\th_*-\sin^2\th'}d\th'+\int_{r}^{r_+}\frac{a^2\Delta'\sin 2\th_*\,dr'}{2\big((r'^2+a^2)^2-a^2 \sin^2\th_*\Delta'\big)^{\f32}}$$
is negative and with uniformly bounded absolute values.
\end{proof}
In order to lighten the notation, we define from now on
\begin{equation}\label{G.def}
G(r_*,\th_*)\doteq \f{\rd}{\rd\th_*}\restriction_{r, \th fixed} F(\th_*;r,\th),
\end{equation}
where the formula for $G$ is given in Proposition \ref{dF.explicit}. With the bounds in Proposition \ref{dF.lower.bd}, we now show that the map $\Psi$ taking $(r_*,\th_*)\mapsto (r,\th)$ is invertible and prove bounds for the partial derivatives of the inverse. First notice that by differentiating \eqref{imp.def} with respect to $\th$, we obtain
\begin{equation}\label{dth*.dth}
\begin{split}
\f{\rd\th_*}{\rd \th}=-\f{1}{a G\sqrt{\sin^2\th_*-\sin^2\th}}.
\end{split}
\end{equation}
Differentiating \eqref{imp.def} with respect to $r$, we have
\begin{equation*}
\begin{split}
\f{\rd\th_*}{\rd r}=\f{1}{G\sqrt{(r^2+a^2)^2-a^2\sin^2\th_*\Delta}}.
\end{split}
\end{equation*}
Using this we can now compute the determinant of the Jacobian for the change of variables:
\begin{equation}\label{det.def}
\begin{split}
D=\left| \begin{array}{cc}
\f{\rd r_*}{\rd r} & \f{\rd r_*}{\rd \th}  \\
\f{\rd \th_*}{\rd r} & \f{\rd\th_*}{\rd \th} \end{array} \right|=&-\f{(r^2+a^2)^2-a^2\sin^2\th\Delta}{a\Delta G\sqrt{\sin^2\th_*-\sin^2\th}\sqrt{(r^2+a^2)^2-a^2\sin^2\th_*\Delta}}.
\end{split}
\end{equation}

The previous estimates are sufficient to give both upper and lower bounds for $D$:
\begin{proposition}\label{det.est}
For $r_-< r< r_+$, $D<0$ and obeys the following estimate:
$$|\Delta|\ls |D|^{-1}\ls  |\Delta|.$$
In particular, this implies that $(t,r_*,\th_*,\phi)$ is indeed a coordinate system for $(\mathcal M_{Kerr},g_{a,M})$.
\end{proposition}
\begin{proof}
This is a direct consequence of \eqref{det.def} and Propositions \ref{th.well.defined} and \ref{dF.lower.bd}.
\end{proof}
We compute the first partial derivatives of the inverse map $\Psi^{-1}$ which maps $(r_*,\th_*)\mapsto (r,\th)$. Using in particular Proposition~\ref{det.est}, we have the following identities and bounds for these partial derivatives.
\begin{proposition}\label{Kerr.pd.est}
The partial derivatives of $(r,\th)$ with respect to $(r_*,\th_*)$ are given by
$$\f{\rd r}{\rd r_*}=\f{\Delta \sqrt{(r^2+a^2)^2-a^2\sin^2\th_*\Delta}}{(r^2+a^2)^2-a^2\sin^2\th\Delta},$$
$$\f{\rd r}{\rd \th_*}=\f{a^2\Delta (\sin^2\th_*-\sin^2\th)\sqrt{(r^2+a^2)^2-a^2\sin^2\th_*\Delta}G}{(r^2+a^2)^2-a^2\sin^2\th\Delta},$$
$$\f{\rd \th}{\rd r_*}=\f{a\Delta \sqrt{\sin^2\th_*-\sin^2\th}}{(r^2+a^2)^2-a^2\sin^2\th\Delta},$$
$$\f{\rd \th}{\rd \th_*}=-\f{a \sqrt{\sin^2\th_*-\sin^2\th}((r^2+a^2)^2-a^2\sin^2\th_*\Delta)G}{(r^2+a^2)^2-a^2\sin^2\th\Delta}.$$
In particular, this implies that in the region $\{r_-\leq r\leq r_+\}$, we have
$$|\f{\rd r}{\rd r_*}|\ls |\Delta|,\quad |\f{\rd r}{\rd \th_*}|\ls |\Delta|\sin(2\th_*),$$
$$|\f{\rd \th}{\rd r_*}|\ls |\Delta|\sin(2\th_*),\quad |\f{\rd \th}{\rd \th_*}|\ls 1.$$
\end{proposition}
\begin{proof}
The partial derivatives of $r$ and $\th$ are direct calculations. The only potentially non-trivial bound is the following:
\begin{equation*}
\begin{split}
\sqrt{\sin^2\th_*-\sin^2\th}\ls \sin(2\th_*).
\end{split}
\end{equation*}
Clearly this holds away from $\th_*=0$ and $\th_*=\f{\pi}{2}$. Recalling that $\sin(2\th_*)=2\sin\th_*\cos\th_*$, the desired estimate near $\th_*=0$ and $\th_*=\f{\pi}{2}$ follows from the bound \eqref{th.well.defined.est} and the fact that
$$\sin^2\th_*-\sin^2\th=\cos^2\th-\cos^2\th_*.$$
\end{proof}
\subsection{Estimates for higher derivatives of $r$ and $\th$ as functions of $r_*$ and $\th_*$}\label{sec.r.th.higher}
After having derived the expressions for the derivatives of $r$ and $\th$ (recall Proposition \ref{Kerr.pd.est}), the main goal of this subsection, which is stated as Proposition \ref{r.th.higher.order} below, is to show that all higher derivatives of $r$ and $\th$ with respect to $r_*$ and $\th_*$ are bounded:
\begin{proposition}\label{r.th.higher.order}
For every fixed $k\geq 2$, $r$ and $\th$ are $C^k$ functions\footnote{In fact, more is true: The presence of the $\f{1}{\sin(2\th_*)}$ weights in the estimates shows that $\f{\rd\th}{\rd\th_*}$ and $r$ are in fact smooth functions with respect to the natural differential structure on the $2$-spheres which are defined as the intersections of the level sets of $t$ and $r_*$.} of $r_*$ and $\th_*$. Moreover, the following bounds hold with implicit constants depending on $k$ (in addition to $M$ and $a$):
\begin{equation}\label{r.th.higher.order.1}
\sum_{1\leq k_1\leq k-1}\left|\left(\f{1}{\sin(2\th_*)}\f{\rd}{\rd \th_*}\right)^{k_1}\left(\f{\rd\th}{\rd\th_*}\right)\right|\ls 1,
\end{equation}
\begin{equation}\label{r.th.higher.order.2}
\sum_{1\leq k_1\leq k} \left|\left(\f{\rd}{\rd r_*}\right)^{k_1} \th\right|\ls |\Delta|\sin(2\th_*),
\end{equation}
\begin{equation}\label{r.th.higher.order.3}
\sum_{\substack{1\leq k_1+k_2\leq k-1\\k_2\geq 1}}\left|\left(\f{1}{\sin(2\th_*)}\f{\rd}{\rd \th_*}\right)^{k_1}\left(\f{\rd}{\rd r_*}\right)^{k_2} \left(\f{\rd\th}{\rd\th_*}\right)\right|\ls |\Delta|,
\end{equation}
\begin{equation}\label{r.th.higher.order.4}
\sum_{1\leq k_1+k_2\leq k}\left|\left(\f{\rd}{\rd r_*}\right)^{k_1}\left(\f{1}{\sin(2\th_*)}\f{\rd}{\rd \th_*}\right)^{k_2} r\right|\ls |\Delta|.
\end{equation}
\end{proposition}

Let us first comment on the main difficulties in establishing Proposition \ref{r.th.higher.order}. It is clear that the functions $\th$ and $r$ are smooth away from $\th_*=0,\f{\pi}{2}$. Therefore, on the one hand, the challenge is to show that the derivatives are uniformly bounded as $\th_*\to 0$ and $\th_*\to \f{\pi}{2}$. On the other hand, we also need to prove bounds with appropriate powers of $|\Delta|$ to capture the behaviour as $r_*\to \pm\infty$.

Before we begin the proof of Proposition \ref{r.th.higher.order}, we need to compute the derivatives of various expressions appearing in Proposition \ref{Kerr.pd.est}. First, we note that by Proposition \ref{Kerr.pd.est} and the expression of $G$ given by \eqref{G.def} and \eqref{dF.explicit}, we have
\begin{equation}\label{dth.dth*}
\begin{split}
\f{\rd \th}{\rd \th_*}=& \f{(r^2+a^2)^2-a^2\sin^2\th_*\Delta}{(r^2+a^2)^2-a^2\sin^2\th\Delta}\left(\f{\sqrt{\sin^2\th_*-\sin^2\th}}{\sin\th_*\cos\th_*}\int_\th^{\th_*} \sqrt{\sin^2\th_*-\sin^2\th'}d\th'\right.\\
&\left.\quad+\f{\sin (2\th)}{\sin (2\th_*)}-\sqrt{\sin^2\th_*-\sin^2\th}\int_{r}^{r_+}\frac{a^3\Delta'\sin 2\th_*\,dr'}{2\big((r'^2+a^2)^2-a^2 \sin^2\th_*\Delta'\big)^{\f32}}\right).
\end{split}
\end{equation}
We will refer to \eqref{dth.dth*} frequently in the computations below.

In the next few lemmas (Lemmas \ref{prop.Kerr.th.int}, \ref{prop.Kerr.th.int.r}, \ref{lemma.sqrtsindiff.der} and \ref{lemma.sinfrac.der}), we will compute the derivatives of various expressions in Proposition \ref{Kerr.pd.est}. We first compute the derivatives of the terms involving the integrals of powers of $\sqrt{\sin^2\th_*-\sin^2\th'}$. The key thing to note is that by \eqref{th.well.defined.est}, the right hand side of the expression in Lemma \ref{prop.Kerr.th.int} is bounded. Moreover, the integral $\f{4(\f n2+1)}{\sin^{n+3}(2\th_*)}\int_{\th}^{\th_*}\left(\sin^2\th_*-\sin^2\th'\right)^{\f {n+2}2} \,d\th'$ that appears on the right hand side is exactly that of the form whose derivative we control. We can therefore inductively bound all the derivatives of the integrals of this form.
\begin{lemma}\label{prop.Kerr.th.int}
For any integer $n\geq 1$, the following identity holds:
\begin{equation*}
\begin{split}
&\left(\f{1}{\sin(2\th_*)}\f{\rd}{\rd\th_*}\right)\left(\f{1}{\sin^{n+1}(2\th_*)}\int_{\th}^{\th_*} \left(\sin^2\th_*-\sin^2\th'\right)^{\f n2}\,d\th'\right)\\
=&\f{4(\f n2+1)}{\sin^{n+3}(2\th_*)}\int_{\th}^{\th_*}\left(\sin^2\th_*-\sin^2\th'\right)^{\f {n+2}2} \,d\th'\\
&-\left(\f{\sqrt{\sin^2\th_*-\sin^2\th}}{\sin\th_*\cos\th_*}\int_\th^{\th_*} \f{\sqrt{\sin^2\th_*-\sin^2\th'}}{\sin^2(2\th_*)}d\th'-\f{\sqrt{\sin^2\th_*-\sin^2\th}}{\sin(2\th_*)}\int_{r}^{r_+}\frac{a^3\Delta'\,dr'}{2\big((r'^2+a^2)^2-a^2 \sin^2\th_*\Delta'\big)^{\f32}}\right)\\
&\qquad\times\left(\f{(r^2+a^2)^2-a^2\sin^2\th_*\Delta}{(r^2+a^2)^2-a^2\sin^2\th\Delta}\right)\times\f{\left(\sin^2\th_*-\sin^2\th\right)^{\f n2}}{\sin^{n}(2\th_*)}\\
&-\f{\left(\sin^2\th_*-\sin^2\th\right)^{\f n2}}{\sin^{n}(2\th_*)}\left(\f{a^2(\sin^2\th_*-\sin^2\th)\Delta}{\sin^2(2\th_*)\left((r^2+a^2)^2-a^2\sin^2\th\Delta\right)}\right)\left(\f{\sin(2\th)}{\sin(2\th_*)}\right).
\end{split}
\end{equation*}
\end{lemma}
\begin{proof}
Differentiation gives
\begin{equation}\label{Kerr.th.int.1}
\begin{split}
&\f{\rd}{\rd\th_*}\left(\f{1}{\sin^{n+1}(2\th_*)}\int_{\th}^{\th_*} \left(\sin^2\th_*-\sin^2\th'\right)^{\f n2}\,d\th'\right)\\
=&-\f{2(n+1)\cos(2\th_*)}{\sin^{n+2}(2\th_*)}\int_{\th}^{\th_*} \left(\sin^2\th_*-\sin^2\th'\right)^{\f n2}\,d\th'+ \f{n\sin\th_*\cos\th_*}{\sin^{n+1}(2\th_*)}\int_{\th}^{\th_*} \left(\sin^2\th_*-\sin^2\th'\right)^{\f n2-1}\,d\th'\\
&-(\f{\rd\th}{\rd\th_*})\f{1}{\sin^{n+1}(2\th_*)}\left(\sin^2\th_*-\sin^2\th\right)^{\f n2}.
\end{split}
\end{equation}
We first consider the contribution of the first two terms on the right hand side of \eqref{Kerr.th.int.1}:
\begin{equation}\label{Kerr.th.int.2}
\begin{split}
&-\f{2(n+1)\cos(2\th_*)}{\sin^{n+2}(2\th_*)}\int_{\th}^{\th_*} \left(\sin^2\th_*-\sin^2\th'\right)^{\f n2}\,d\th'+ \f{n\sin\th_*\cos\th_*}{\sin^{n+1}(2\th_*)}\int_{\th}^{\th_*} \left(\sin^2\th_*-\sin^2\th'\right)^{\f n2-1}\,d\th'\\
=&\f{1}{\sin^{n+2}(2\th_*)}\int_{\th}^{\th_*}\left(\sin^2\th_*-\sin^2\th'\right)^{\f n2-1} \left(-2(n+1)\cos(2\th_*)(\sin^2\th_*-\sin^2\th')+\f n2\sin^2(2\th_*)\right)\,d\th'
\end{split}
\end{equation}
We now consider the following expression, which appeared in the second bracket in the integral on the right hand side of \eqref{Kerr.th.int.2}:
\begin{equation}\label{Kerr.th.int.3}
\begin{split}
&-2(n+1)\cos(2\th_*)(\sin^2\th_*-\sin^2\th')+\f n2\sin^2(2\th_*)\\
=&-2(n+1)\cos(2\th_*)(\sin^2\th_*-\sin^2\th')+\f n2\sin^2(2\th')+\f n2 (\sin^2(2\th_*)-\sin^2(2\th'))\\
=&-(n+1)\cos(2\th_*)(\cos(2\th')-\cos(2\th_*))+\f n2\sin^2(2\th')+\f n2 (\cos^2(2\th')-\cos^2(2\th_*))\\
=&-(n+1)\cos(2\th_*)(\cos(2\th')-\cos(2\th_*))+\f n2\sin^2(2\th')+\f n2 (\cos(2\th')+\cos(2\th_*))(\cos(2\th')-\cos(2\th_*))\\
=&\left(-(n+1)\cos(2\th_*)+\f n2 (\cos(2\th')+\cos(2\th_*))\right)\left(\cos(2\th')-\cos(2\th_*)\right)+\f n2\sin^2(2\th')\\
=&\left(-(\f n2 +1)\cos(2\th_*)+\f n2 \cos(2\th')\right)\left(\cos(2\th')-\cos(2\th_*)\right)+\f n2\sin^2(2\th')\\
=&-\cos(2\th')(\cos(2\th')-\cos(2\th_*))+\f n2\sin^2(2\th')+(\f n2+1)(\cos(2\th')-\cos(2\th_*))^2\\
=&-\cos(2\th')(\cos(2\th')-\cos(2\th_*))+\f n2\sin^2(2\th')+4(\f n2+1)(\sin^2\th_*-\sin^2\th')^2.
\end{split}
\end{equation}
Substituting \eqref{Kerr.th.int.3} into \eqref{Kerr.th.int.2}, we obtain
\begin{equation}\label{Kerr.th.int.4}
\begin{split}
&-\f{2(n+1)\cos(2\th_*)}{\sin^{n+2}(2\th_*)}\int_{\th}^{\th_*} \left(\sin^2\th_*-\sin^2\th'\right)^{\f n2}\,d\th'+ \f{n\sin\th_*\cos\th_*}{\sin^{n+1}(2\th_*)}\int_{\th}^{\th_*} \left(\sin^2\th_*-\sin^2\th'\right)^{\f n2-1}\,d\th'\\
=&\f{1}{\sin^{n+2}(2\th_*)}\int_{\th}^{\th_*}\left(\sin^2\th_*-\sin^2\th'\right)^{\f n2-1} \left(-\cos(2\th')(\cos(2\th')-\cos(2\th_*))+\f n2\sin^2(2\th')\right)\,d\th'\\
&+\f{4(\f n2+1)}{\sin^{n+2}(2\th_*)}\int_{\th}^{\th_*}\left(\sin^2\th_*-\sin^2\th'\right)^{\f {n+2}2} \,d\th'.
\end{split}
\end{equation}
On the other hand, we have
\begin{equation}\label{Kerr.th.int.5}
\begin{split}
&\f{\rd}{\rd\th'} \left(\sin(2\th')\left(\sin^2\th_*-\sin^2\th'\right)^{\f n2}\right)\\
=&2\cos(2\th')\left(\sin^2\th_*-\sin^2\th'\right)^{\f n2}-\f{n}{2}\sin^2(2\th')\left(\sin^2\th_*-\sin^2\th'\right)^{\f n2-1}\\
=&\left(\sin^2\th_*-\sin^2\th'\right)^{\f n2-1}\left(\cos(2\th')\left(\cos(2\th')-\cos(2\th_*)\right)-\f{n}{2}\sin^2(2\th')\right),
\end{split}
\end{equation}
which implies (for $n\geq 1$) that
\begin{equation}\label{Kerr.th.int.6}
\begin{split}
&\int_{\th}^{\th_*}\left(\sin^2\th_*-\sin^2\th'\right)^{\f n2-1}\left(\cos(2\th')\left(\cos(2\th')-\cos(2\th_*)\right)-\f{n}{2}\sin^2(2\th')\right)\,d\th'\\
=&-\sin(2\th)\left(\sin^2\th_*-\sin^2\th\right)^{\f n2}.
\end{split}
\end{equation}
Combining \eqref{Kerr.th.int.1}, \eqref{Kerr.th.int.4} and \eqref{Kerr.th.int.6}, we thus obtain
\begin{equation}\label{Kerr.th.int.7}
\begin{split}
&\f{\rd}{\rd\th_*}\left(\f{1}{\sin^{n+1}(2\th_*)}\int_{\th}^{\th_*} \left(\sin^2\th_*-\sin^2\th'\right)^{\f n2}\,d\th'\right)\\
=&\f{4(\f n2+1)}{\sin^{n+2}(2\th_*)}\int_{\th}^{\th_*}\left(\sin^2\th_*-\sin^2\th'\right)^{\f {n+2}2} \,d\th'-\left((\f{\rd\th}{\rd\th_*})-\f{\sin(2\th)}{\sin(2\th_*)}\right)\f{\left(\sin^2\th_*-\sin^2\th\right)^{\f n2}}{\sin^{n+1}(2\th_*)}.
\end{split}
\end{equation}
Recalling \eqref{dth.dth*}, we have
\begin{equation}\label{dth.dth*.lo}
\begin{split}
\f{\rd \th}{\rd \th_*}-\f{\sin(2\th)}{\sin(2\th_*)}=& \f{(r^2+a^2)^2-a^2\sin^2\th_*\Delta}{(r^2+a^2)^2-a^2\sin^2\th\Delta}\left(\f{\sqrt{\sin^2\th_*-\sin^2\th}}{\sin\th_*\cos\th_*}\int_\th^{\th_*} \sqrt{\sin^2\th_*-\sin^2\th'}d\th'\right.\\
&\left.\quad-\sqrt{\sin^2\th_*-\sin^2\th}\int_{r}^{r_+}\frac{a^3\Delta'\sin 2\th_*\,dr'}{2\big((r'^2+a^2)^2-a^2 \sin^2\th_*\Delta'\big)^{\f32}}\right)\\
%&+(1-\f{(r^2+a^2)^2-a^2\sin^2\th_*\Delta}{(r^2+a^2)^2-a^2\sin^2\th\Delta})\f{\sin(2\th)}{\sin(2\th_*)}\\
%=& \f{(r^2+a^2)^2-a^2\sin^2\th_*\Delta}{(r^2+a^2)^2-a^2\sin^2\th\Delta}\left(\f{\sqrt{\sin^2\th_*-\sin^2\th}}{\sin\th_*\cos\th_*}\int_\th^{\th_*} \sqrt{\sin^2\th_*-\sin^2\th'}d\th'\right.\\
%&\left.\quad-\sqrt{\sin^2\th_*-\sin^2\th}\int_{r}^{r_+}\frac{a^3\Delta'\sin 2\th_*\,dr'}{2\big((r'^2+a^2)^2-a^2 \sin^2\th_*\Delta'\big)^{\f32}}\right)\\
&+\left(\f{a^2(\sin^2\th_*-\sin^2\th)\Delta}{(r^2+a^2)^2-a^2\sin^2\th\Delta}\right)\left(\f{\sin(2\th)}{\sin(2\th_*)}\right),
\end{split}
\end{equation}
which together with \eqref{Kerr.th.int.7} imply
\begin{equation*}
\begin{split}
&\f{\rd}{\rd\th_*}\left(\f{1}{\sin^{n+1}(2\th_*)}\int_{\th}^{\th_*} \left(\sin^2\th_*-\sin^2\th'\right)^{\f n2}\,d\th'\right)\\
=&\f{4(\f n2+1)}{\sin^{n+2}(2\th_*)}\int_{\th}^{\th_*}\left(\sin^2\th_*-\sin^2\th'\right)^{\f {n+2}2} \,d\th'\\
&-\left(\f{\sqrt{\sin^2\th_*-\sin^2\th}}{\sin\th_*\cos\th_*}\int_\th^{\th_*} \sqrt{\sin^2\th_*-\sin^2\th'}d\th'-\sqrt{\sin^2\th_*-\sin^2\th}\int_{r}^{r_+}\frac{a^3\Delta'\sin 2\th_*\,dr'}{2\big((r'^2+a^2)^2-a^2 \sin^2\th_*\Delta'\big)^{\f32}}\right)\\
&\qquad\times\left(\f{(r^2+a^2)^2-a^2\sin^2\th_*\Delta}{(r^2+a^2)^2-a^2\sin^2\th\Delta}\right)\times\f{\left(\sin^2\th_*-\sin^2\th\right)^{\f n2}}{\sin^{n+1}(2\th_*)}\\
&-\f{\left(\sin^2\th_*-\sin^2\th\right)^{\f n2}}{\sin^{n+1}(2\th_*)}\left(\f{a^2(\sin^2\th_*-\sin^2\th)\Delta}{(r^2+a^2)^2-a^2\sin^2\th\Delta}\right)\left(\f{\sin(2\th)}{\sin(2\th_*)}\right).
\end{split}
\end{equation*}
Dividing by $\sin(2\th_*)$ thus gives the desired result.
\end{proof}
We then compute the $\f{\rd}{\rd r_*}$ derivative of the integral appearing in Lemma \ref{prop.Kerr.th.int}, which is significantly easier:
\begin{lemma}\label{prop.Kerr.th.int.r}
For any integer $n\geq 1$, the following identity holds:
\begin{equation*}
\begin{split}
&\f{\rd}{\rd r_*}\left(\f{1}{\sin^{n+1}(2\th_*)}\int_{\th}^{\th_*} \left(\sin^2\th_*-\sin^2\th'\right)^{\f n2}\,d\th'\right)=-\f{a\Delta \left(\sin^2\th_*-\sin^2\th\right)^{\f {n+1}2}}{\sin^{n+1}(2\th_*)\left((r^2+a^2)^2-a^2\sin^2\th\Delta\right)}.
\end{split}
\end{equation*}
\end{lemma}
\begin{proof}
This is a direct application of Proposition \ref{Kerr.pd.est}.
\end{proof}
The next to be computed are the derivatives of $\f{\sqrt{\sin^2\th_*-\sin^2\th}}{\sin(2\th_*)}$:
\begin{lemma}\label{lemma.sqrtsindiff.der}
The following identities hold:
\begin{equation*}
\begin{split}
&\f{1}{\sin (2\th_*)}\f{\rd}{\rd\th_*} \f{\sqrt{\sin^2\th_*-\sin^2\th}}{\sin(2\th_*)}\\
=&\f{2(\sin^2\th_*-\sin^2\th)^{\f32}}{\sin^3(2\th_*)}-\left(\f{\sin^2(2\th)}{2\sin^3(2\th_*)\sqrt{\sin^2\th_*-\sin^2\th}}\right)\left(\f{a^2(\sin^2\th_*-\sin^2\th)\Delta}{(r^2+a^2)^2-a^2\sin^2\th\Delta}\right)\\
&-\f{\sin(2\th)}{2\sin^2(2\th_*)\sqrt{\sin^2\th_*-\sin^2\th}}\f{(r^2+a^2)^2-a^2\sin^2\th_*\Delta}{(r^2+a^2)^2-a^2\sin^2\th\Delta}\left(\f{\sqrt{\sin^2\th_*-\sin^2\th}}{\sin\th_*\cos\th_*}\int_\th^{\th_*} \sqrt{\sin^2\th_*-\sin^2\th'}d\th'\right.\\
&\left.\qquad-\sqrt{\sin^2\th_*-\sin^2\th}\int_{r}^{r_+}\frac{a^3\Delta'\sin 2\th_*\,dr'}{2\big((r'^2+a^2)^2-a^2 \sin^2\th_*\Delta'\big)^{\f32}}\right)
\end{split}
\end{equation*}
and
\begin{equation*}
\begin{split}
\f{\rd}{\rd r_*} \f{\sqrt{\sin^2\th_*-\sin^2\th}}{\sin(2\th_*)}
=&\f{a\Delta \sin\th\cos\th}{\sin(2\th_*)\left((r^2+a^2)^2-a^2\sin^2\th\Delta\right)}.
\end{split}
\end{equation*}
\end{lemma}
\begin{proof}
We first compute the $\f{\rd}{\rd\th_*}$ derivative as follows.
\begin{equation}\label{prop.sqrtsindiff.der.1}
\begin{split}
&\f{1}{\sin (2\th_*)}\f{\rd}{\rd\th_*} \f{\sqrt{\sin^2\th_*-\sin^2\th}}{\sin(2\th_*)}\\
=&-\f{2\cos(2\th_*)\sqrt{\sin^2\th_*-\sin^2\th}}{\sin^3(2\th_*)}+\f{\sin\th_*\cos\th_*-\sin\th\cos\th\f{\rd\th}{\rd\th_*}}{\sin^2(2\th_*)\sqrt{\sin^2\th_*-\sin^2\th}}.
\end{split}
\end{equation}
Recall now the expression \eqref{dth.dth*} and notice that except for the term\footnote{Notice that more precisely, the term is $\f{(r^2+a^2)^2-a^2\sin^2\th_*\Delta}{(r^2+a^2)^2-a^2\sin^2\th\Delta}\f{\sin(2\th)}{\sin(2\th_*)}$, but this differs from $\f{\sin(2\th)}{\sin(2\th_*)}$ by a term that can be dominated by $\sin^2(2\th_*)$.} $\f{\sin(2\th)}{\sin(2\th_*)}$, all other terms can be bounded from above by $\sin^2(2\th_*)$. We therefore rewrite \eqref{prop.sqrtsindiff.der.1} as follows:
\begin{equation}\label{prop.sqrtsindiff.der.2}
\begin{split}
&\f{1}{\sin (2\th_*)}\f{\rd}{\rd\th_*} \f{\sqrt{\sin^2\th_*-\sin^2\th}}{\sin(2\th_*)}\\
=&\f{-2\cos(2\th_*)\left(\sin^2\th_*-\sin^2\th\right)+2\sin^2\th_*\cos^2\th_*-\sin\th\cos\th\left(\sin(2\th)+\sin(2\th_*)\left(\f{\rd\th}{\rd\th_*}-\f{\sin(2\th)}{\sin(2\th_*)}\right)\right)}{\sin^3(2\th_*)\sqrt{\sin^2\th_*-\sin^2\th}}.
\end{split}
\end{equation}
We now consider the main contribution from the numerator:
\begin{equation}\label{prop.sqrtsindiff.der.3}
\begin{split}
&-2\cos(2\th_*)(\sin^2\th_*-\sin^2\th)+2\sin^2\th_*\cos^2\th_*-\sin\th\cos\th\sin(2\th)\\
=&-2(1-2\sin^2\th_*)(\sin^2\th_*-\sin^2\th)+2\sin^2\th_*(1-\sin^2\th_*)-2\sin^2\th(1-\sin^2\th)\\
=&4\sin^4\th_*-4\sin^2\th\sin^2\th_*-2\sin^4\th_*+2\sin^4\th\\
=&2(\sin^2\th_*-\sin^2\th)^2.
\end{split}
\end{equation}
Substituting \eqref{prop.sqrtsindiff.der.3} into \eqref{prop.sqrtsindiff.der.2}, we thus obtain 
\begin{equation*}
\begin{split}
\f{1}{\sin (2\th_*)}\f{\rd}{\rd\th_*} \f{\sqrt{\sin^2\th_*-\sin^2\th}}{\sin(2\th_*)}
=&\f{2(\sin^2\th_*-\sin^2\th)^{\f32}}{\sin^3(2\th_*)}-\f{\sin(2\th)\left(\f{\rd\th}{\rd\th_*}-\f{\sin(2\th)}{\sin(2\th_*)}\right)}{2\sin^2(2\th_*)\sqrt{\sin^2\th_*-\sin^2\th}}.
\end{split}
\end{equation*}
Recalling \eqref{dth.dth*.lo}, this thus implies the first formula.

We now compute the $\f{\rd}{\rd r_*}$ derivative:
\begin{equation*}
\begin{split}
\f{\rd}{\rd r_*} \f{\sqrt{\sin^2\th_*-\sin^2\th}}{\sin(2\th_*)}
=&\f{a\Delta \sin\th\cos\th}{\sin(2\th_*)\left((r^2+a^2)^2-a^2\sin^2\th\Delta\right)}
\end{split}
\end{equation*}
as desired.\qedhere

\end{proof}
As our last lemma, we compute the derivatives of $\f{\sin(2\th)}{\sin(2\th_*)}$.
\begin{lemma}\label{lemma.sinfrac.der}
The following identities hold:
\begin{equation*}
\begin{split}
&\left(\f{1}{\sin(2\th_*)}\f{\rd}{\rd\th_*}\right)\left(\f{\sin(2\th)}{\sin(2\th_*)}\right)\\
=&\f{4\sin(2\th)(\sin^2\th_*-\sin^2\th)}{\sin^3(2\th_*)}+\f{2\sin (2\th)\cos(2\th)}{\sin^3 (2\th_*)}\f{a^2\Delta(\sin^2\th-\sin^2\th_*)}{(r^2+a^2)^2-a^2\sin^2\th\Delta}\\
& +\f{2\cos(2\th)}{\sin^2(2\th_*)}\f{(r^2+a^2)^2-a^2\sin^2\th_*\Delta}{(r^2+a^2)^2-a^2\sin^2\th\Delta}\left(\f{\sqrt{\sin^2\th_*-\sin^2\th}}{\sin\th_*\cos\th_*}\int_\th^{\th_*} \sqrt{\sin^2\th_*-\sin^2\th'}d\th'\right.\\
&\left.\quad-\sqrt{\sin^2\th_*-\sin^2\th}\int_{r}^{r_+}\frac{a^3\Delta'\sin 2\th_*\,dr'}{2\big((r'^2+a^2)^2-a^2 \sin^2\th_*\Delta'\big)^{\f32}}\right).
\end{split}
\end{equation*}
and
\begin{equation*}
\begin{split}
\f{\rd}{\rd r_*}\f{\sin(2\th)}{\sin(2\th_*)}=\f{\sqrt{\sin^2\th_*-\sin^2\th}}{\sin(2\th_*)}\f{2a\cos(2\th)\Delta}{(r^2+a^2)^2-a^2\sin^2\th\Delta}.
\end{split}
\end{equation*}
\end{lemma}
\begin{proof}
We compute using \eqref{dth.dth*}:
\begin{equation*}
\begin{split}
&\f{\rd}{\rd\th_*}\f{\sin(2\th)}{\sin(2\th_*)}\\
=&\f{2\cos(2\th)\sin(2\th_*)\f{\rd\th}{\rd\th_*}-2\cos(2\th_*)\sin(2\th)}{\sin^2(2\th_*)}\\
=&\f{2\sin(2\th)(\cos(2\th)-\cos(2\th_*))}{\sin^2(2\th_*)}+\f{2\sin (2\th)\cos(2\th)}{\sin^2 (2\th_*)}\f{a^2\Delta(\sin^2\th-\sin^2\th_*)}{(r^2+a^2)^2-a^2\sin^2\th\Delta}\\
& +\f{2\cos(2\th)}{\sin(2\th_*)}\f{(r^2+a^2)^2-a^2\sin^2\th_*\Delta}{(r^2+a^2)^2-a^2\sin^2\th\Delta}\left(\f{\sqrt{\sin^2\th_*-\sin^2\th}}{\sin\th_*\cos\th_*}\int_\th^{\th_*} \sqrt{\sin^2\th_*-\sin^2\th'}d\th'\right.\\
&\left.\quad-\sqrt{\sin^2\th_*-\sin^2\th}\int_{r}^{r_+}\frac{a^3\Delta'\sin 2\th_*\,dr'}{2\big((r'^2+a^2)^2-a^2 \sin^2\th_*\Delta'\big)^{\f32}}\right).
\end{split}
\end{equation*}
Noting that
$$\cos(2\th)-\cos(2\th_*)=2(\sin^2\th_*-\sin^2\th),$$
we thus obtain the first formula after dividing by $\sin(2\th_*)$.

Finally, we compute the $\f{\rd}{\rd r_*}$ derivative of $\f{\sin(2\th)}{\sin(2\th_*)}$ using the expression for $\f{\rd \th}{\rd r_*}$ in Proposition \ref{Kerr.pd.est}:
\begin{equation*}
\begin{split}
\f{\rd}{\rd r_*}\f{\sin(2\th)}{\sin(2\th_*)}
=&\f{2\cos(2\th)}{\sin(2\th_*)}\f{a\Delta\sqrt{\sin^2\th_*-\sin^2\th}}{(r^2+a^2)^2-a^2\sin^2\th\Delta}=\f{\sqrt{\sin^2\th_*-\sin^2\th}}{\sin(2\th_*)}\f{2a\cos(2\th)\Delta}{(r^2+a^2)^2-a^2\sin^2\th\Delta}.
\end{split}
\end{equation*}
\end{proof}

After all the preliminary calculations (Lemmas \ref{prop.Kerr.th.int}, \ref{prop.Kerr.th.int.r}, \ref{lemma.sqrtsindiff.der} and \ref{lemma.sinfrac.der}), we now return to the proof of Proposition \ref{r.th.higher.order}:

\begin{proof}[Proof of Proposition \ref{r.th.higher.order}]
Recalling the expressions for the partial derivatives of $r$ and $\th$ in Proposition \ref{Kerr.pd.est}, the main step is to prove the following bounds
\begin{equation}\label{r.th.higher.induct.1}
\sum_{\substack{0\leq k_1+k_2\leq k\\0\leq n\leq 2k-1}}\left|\left(\f{1}{\sin(2\th_*)}\f{\rd}{\rd\th_*}\right)^{k_1}\left(\f{\rd}{\rd r_*}\right)^{k_2}\left(\f{1}{\sin^{n+1}(2\th_*)}\int_\th^{\th_*}\left(\sin^2\th_*-\sin^2\th'\right)^{\f n2} d\th'\right)\right|\ls |\Delta|^{\min\{k_2,1\}},
\end{equation}
\begin{equation}\label{r.th.higher.induct.2}
\sum_{0\leq k_1+k_2\leq k}\left|\left(\f{1}{\sin(2\th_*)}\f{\rd}{\rd\th_*}\right)^{k_1}\left(\f{\rd}{\rd r_*}\right)^{k_2}\left(\f{\sqrt{\sin^2\th_*-\sin^2\th}}{\sin(2\th_*)}\right)\right|\ls |\Delta|^{\min\{k_2,1\}},
\end{equation}
\begin{equation}\label{r.th.higher.induct.3}
\sum_{0\leq k_1+k_2\leq k}\left|\left(\f{1}{\sin(2\th_*)}\f{\rd}{\rd\th_*}\right)^{k_1}\left(\f{\rd}{\rd r_*}\right)^{k_2}\left(\f{\sin(2\th)}{\sin(2\th_*)}\right)\right|\ls |\Delta|^{\min\{k_2,1\}}.
\end{equation}
for every $k\geq 0$ (with implicit constants depending on $k$). Here, we note explicitly that 
$$|\Delta|^{\min\{k_2,1\}}=
\begin{cases}
1\quad &\mbox{if } k_2=0\\
|\Delta|\quad &\mbox{if } k_2>0.
\end{cases}
$$

\noindent{\bf Proof of \eqref{r.th.higher.induct.1}-\eqref{r.th.higher.induct.3}.} We will prove this by induction. For the base case, simply notice that the estimates hold for $k=0$ in view of \eqref{th.well.defined.est}.

For the induction step, we assume that there exists $k_0\geq 1$ such that \eqref{r.th.higher.induct.1}, \eqref{r.th.higher.induct.2} and \eqref{r.th.higher.induct.3}  hold\footnote{where the constants are of course dependent on $k_0$, in addition to $M$ and $a$.} for all $k\leq k_0-1$. Our goal is to show that \eqref{r.th.higher.induct.1}-\eqref{r.th.higher.induct.3} hold for $k=k_0$. On the other hand, this is a direct consequence of the formulas of the derivatives in Lemmas \ref{prop.Kerr.th.int}, \ref{prop.Kerr.th.int.r}, \ref{lemma.sqrtsindiff.der} and \ref{lemma.sinfrac.der}, the induction hypotheses, Proposition \ref{Kerr.pd.est} and equation \eqref{th.well.defined.est}.

Given \eqref{r.th.higher.induct.1}-\eqref{r.th.higher.induct.3}, the remainder of the proof is quite straightforward. We briefly sketch the relevant estimates:

\noindent{\bf Proof of \eqref{r.th.higher.order.1}.} Simply differentiate \eqref{dth.dth*} and use the bounds \eqref{r.th.higher.induct.1}-\eqref{r.th.higher.induct.3}.

\noindent{\bf Proof of \eqref{r.th.higher.order.2}.} This follows from an induction argument involving differentiating repeatedly in $(\f{\rd}{\rd r_*})$ the equation 
$$\f{\rd \th}{\rd r_*}=\f{a\Delta \sqrt{\sin^2\th_*-\sin^2\th}}{(r^2+a^2)^2-a^2\sin^2\th\Delta}=\f{\sqrt{\sin^2\th_*-\sin^2\th}}{\sin(2\th_*)}\f{a\Delta \sin(2\th_*)}{(r^2+a^2)^2-a^2\sin^2\th\Delta}$$
and using \eqref{r.th.higher.induct.2}.

\noindent{\bf Proof of \eqref{r.th.higher.order.3}.} This can be proven in a similar manner as \eqref{r.th.higher.order.1}.

\noindent{\bf Proof of \eqref{r.th.higher.order.4}.} This follows from differentiating the expressions in Proposition \ref{Kerr.pd.est} and using the bounds \eqref{r.th.higher.induct.1}-\eqref{r.th.higher.induct.3}.
\end{proof}

We have thus proven Proposition \ref{r.th.higher.order} and this concludes the present subsection.

\subsection{The Kerr metric in double null coordinates}\label{Kerr.dbn}
In this section, we use the calculations in the previous subsections to write the Kerr metric in double null coordinates as in Section \ref{secdnf}. First, notice that in the $(t,r_*,\th_*,\phi)$ coordinates, (see Section 4 in \cite{Pretorius}) the Kerr metric takes the form
$$g_{a,M}=\frac{\Delta}{R^2}(-dt\otimes dt+dr_*\otimes dr_*)+\frac{\ell^2}{R^2}d\th_*\otimes d\th_*+R^2\sin^2\th(d\phi-\frac{2Mar}{\Sigma R^2} dt)\otimes (d\phi-\frac{2Mar}{\Sigma R^2} dt).$$
Here, $\ell$ is defined as
$$\ell\doteq -G a\sqrt{\sin^2\th_*-\sin^2\th}\sqrt{(r^2+a^2)^2-a^2\sin^2\th_*\Delta}.$$
Recall from \eqref{Eikonal.r.1} that $\ub=\frac 12(r_*+t)$ and $u=\frac 12(r_*-t)$ satisfy the Eikonal equation.
We note the simple relations
\begin{equation*}
\begin{split}
dr_*=du+d\ub,\quad dt=d\ub-du.%dr_*=du+d\ub,\quad dt=du-d\ub.%,\quad\f{\rd}{\rd r_*}=\f12 (\f{\rd}{\rd u}+\f{\rd}{\rd\ub}),\quad\f{\rd}{\rd t}=\f12 (\f{\rd}{\rd\ub}-\f{\rd}{\rd u}).
\end{split}
\end{equation*}
The Kerr metric in the $(u,\ub,\th_*,\phi)$ coordinate system takes the form
$$g_{a,M}=2\frac{\Delta}{R^2}(du\otimes d\ub+d\ub\otimes du)+\frac{\ell^2}{R^2}d\th_*\otimes d\th_*+R^2\sin^2\th(d\phi-\frac{2Mar}{\Sigma R^2} (d\ub-du))\otimes (d\phi-\frac{2Mar}{\Sigma R^2} (d\ub-du)). $$
%$$g_{a,M}=2\frac{\Delta}{R^2}(du\otimes d\ub+d\ub\otimes du)+\frac{\ell^2}{R^2}d\th_*\otimes d\th_*+R^2\sin^2\th(d\phi-\frac{2Mar}{\Sigma R^2} dt)\otimes (d\phi-\frac{2Mar}{\Sigma R^2} dt). $$

Finally, to put the metric in the form of the canonical double null coordinates (see Section \ref{coordinates}), we follow \cite{vacuumscatter} and define a new coordinate function $\phi_*=\phi-h(r_*,\th_*)$ by solving the ordinary differential equation (for every fixed $\th_*\in (0,\pi)$)
\begin{equation}\label{h.ODE}
\f{d h}{d r_*}(r_*,\th_*)=-\frac{2Mar}{\Sigma R^2}
\end{equation}
%\begin{equation}\label{h.ODE}
%\f{\rd h}{\rd r_*}=\frac{\rd h}{\rd u}=\frac{\rd h}{\rd \ub}=\frac{2Mar}{\Sigma R^2}
%\end{equation}
with initial data $h(r_*=0;\th_*)=0$.

From now on we will denote the metric components, Ricci coefficients and null curvature components of the background Kerr solution using the subscript ${ }_{\Ke}$. With the definition of $h$ above, we can therefore put the metric into the form of the canonical double null coordinates: 
\begin{equation}\label{Kerr.doublenull}
g_{a,M}=-2\Omega_\Ke^2 (du\otimes d\ub+d\ub\otimes du)+(\gamma_\Ke)_{AB}(d\th^A-(b_\Ke)^A d\ub)\otimes(d\th^B-(b_{\Ke})^B d\ub).
\end{equation}
Here, $\th^1=\th_*$ and $\th^2=\phi_*$ and the metric components take the following values\footnote{Recall again that since $h$ is independent of $\phi_*$, $\frac{\rd h}{\rd\th_*}$ can be defined unambiguously independent of whether $\f{\rd}{\rd\th_*}$ is defined with respect to the $(u,\ub,\th_*,\phi)$ or the $(u,\ub,\th_*,\phi_*)$ coordinate system.}:
\begin{equation}\label{Kerr.metric.comp}
\begin{split}
\Om_\Ke^2=-\frac{\Delta}{R^2},\quad (b_\Ke)^{\phi_*}&=\frac{4Mar}{\Sigma R^2},\quad (b_\Ke)^{\th_*}=0,\\
(\gamma_\Ke)_{\phi_*\phi_*}=R^2\sin^2\th,\quad &(\gamma_\Ke)_{\th_*\th_*}=\frac{\ell^2}{R^2}+(\frac{\rd h}{\rd\th_*})^2 R^2\sin^2\th,\\
(\gamma_\Ke)_{\th_*\phi_*}=&(\gamma_\Ke)_{\phi_*\th_*}=(\frac{\rd h}{\rd\th_*}) R^2\sin^2\th.
\end{split}
\end{equation}
The inverse of $\gamma_\Ke$ is given by
\begin{equation}\label{gamma.Ke.inverse}
\begin{split}
(\gamma_\Ke^{-1})^{\phi_*\phi_*}=\frac{1}{R^2\sin^2\th}+(\frac{\rd h}{\rd\th_*})^2 &\frac{R^2}{\ell^2},\quad (\gamma_\Ke^{-1})^{\th_*\th_*}=\frac{R^2}{\ell^2},\\
(\gamma_\Ke^{-1})^{\th_*\phi_*}=(\gamma_\Ke^{-1})^{\phi_*\th_*}&=-(\frac{\rd h}{\rd\th_*})\frac{R^2}{\ell^2}.
\end{split}
\end{equation}
%We make a brief comment regarding the choice of the null coordinates: Notice that both $u$ and $\ub$ increases towards the future. Moreover, formally, the Cauchy horizon is given by $\{\ub=\infty\}$ while the event horizon is given by $\{u=-\infty\}$.
\begin{remark}
Notice that using
\begin{equation*}
\begin{split}
\f{\rd}{\rd \th_*}\big(\frac{4Mar}{\Sigma R^2}\big)=& \f{\rd r}{\rd \th_*}\f{\rd}{\rd r}\big(\frac{4Mar}{\Sigma R^2}\big)+\f{\rd \th}{\rd \th_*}\f{\rd}{\rd \th}\big(\frac{4Mar}{\Sigma R^2}\big)
\end{split}
\end{equation*}
together with
$$\left|\f{\rd}{\rd \th}\big(\frac{4Mar}{\Sigma R^2}\big)\right|=\left|\f{4Ma^3r\sin 2\th\Delta}{((r^2+a^2)^2-a^2\sin^2\th\Delta)^2}\right|\ls |\Delta||\sin (2\th)|$$
and the estimates derived in Proposition \ref{Kerr.pd.est}, we have
\begin{equation}\label{D.b}
\left|\f{\rd}{\rd \th_*}\big(\frac{4Mar}{\Sigma R^2}\big)\right|\ls |\Delta||\sin (2\th_*)|.
\end{equation}
As a consequence, the expression $\frac{\rd h}{\rd\th_*}$ that appears in the metric component, which can be obtained via solving the ode \eqref{h.ODE}, satisfies the following bound
$$\left|\frac{\rd h}{\rd\th_*}\right|\ls |\sin (2\th_*)|.$$
\end{remark}
In the remainder of this subsection, we derive some estimates for the metric components and their derivatives. We first consider $\gamma_{\Ke}$. The following proposition (whose proof is omitted) is a consequence of the expression of $\gamma$ in \eqref{Kerr.metric.comp}, the estimates in Propositions~\ref{Kerr.pd.est} and \ref{r.th.higher.order}:
\begin{proposition}\label{gamma.Kerr.bounds}
For every $u$ and $\ub$, $\gamma_{\Ke}$ is a smooth metric on $S_{u,\ub}$. Moreover, the following holds:
\begin{itemize}
\item The following components\footnote{For the last term, notice that \emph{up to terms that are $O(\sin^2\th_*)$ near the axis}, 
$$\ell^2 = G^2 a^2 (\sin^2\th_*-\sin\th) ((r^2+a^2)^2 - a^2\sin^2\th_*\Delta) \sim (r^2+a^2)^2\f{\sin^2(2\th)}{\sin^2(2\th_*)},\quad R^2\sim r^2+a^2.$$
Hence, we have
$$(\gamma_{\Ke})_{\th_*\th_*} \sim \f{\ell^2}{R^2}\sim \f{(r^2+a^2)\sin^2(2\th)}{\sin^2(2\th_*)},\quad \f{(\gamma_{\Ke})_{\phi_*\phi_*}}{\sin^2\th_*} =\f{R^2\sin^2\th}{\sin^2\th_*}\sim \f{(r^2+a^2)\sin^2 \th}{\sin^2\th_*},$$
which, together with Propositions~\ref{Kerr.pd.est} and \ref{r.th.higher.order}, give the desired result.
}
$$(\gamma_{\Ke})_{\th_*\th_*},\quad\f{(\gamma_{\Ke})_{\th_*\phi_*}}{\sin^3 \th_*},\quad\f{(\gamma_{\Ke})_{\phi_*\phi_*}}{\sin^2\th_*},\quad\f{1}{\sin^2\th_*}((\gamma_{\Ke})_{\th_*\th_*}-\f{(\gamma_{\Ke})_{\phi_*\phi_*}}{\sin^2\th_*})$$ 
are smooth functions of $\sin^2 \th_*$ such that all the derivatives with respect to $\sin^2\th_*$ are bounded uniformly in $\mathcal M_{Kerr}$.
\item The following $\f{\rd}{\rd r_*}$ derivatives of $\gamma$:
$$\f{\rd}{\rd r_*}(\gamma_{\Ke})_{\th_*\th_*},\quad\f{\rd}{\rd r_*}\left(\f{(\gamma_{\Ke})_{\th_*\phi_*}}{\sin^3 \th_*}\right),\quad\f{\rd}{\rd r_*}\left(\f{(\gamma_{\Ke})_{\phi_*\phi_*}}{\sin^2\th_*}\right),\quad\f{\rd}{\rd r_*}\left(\f{1}{\sin^2\th_*}((\gamma_{\Ke})_{\th_*\th_*}-\f{(\gamma_{\Ke})_{\phi_*\phi_*}}{\sin^2\th_*})\right)$$
are smooth functions of $\sin^2 \th_*$ such that all the derivatives with respect to $\sin^2\th_*$ are bounded uniformly by $|\Delta|$ in $\mathcal M_{Kerr}$.
\end{itemize}
\end{proposition}

Before we turn to the estimates for $b_\Ke$ and $\Om_\Ke$, it is convenient to introduce some conventions:
\begin{definition}\label{def.Kerr.norm}
\begin{enumerate}
\item Given a rank $r$ tensor field $\phi_{A_1...A_r}$ on $\mathcal M_{Kerr}$ which is tangential to $2$-spheres $S_{u,\ub}$ for every $u$, $\ub$, we define
$$|\phi|_{\gamma_\Ke}\doteq  \left((\gamma_\Ke^{-1})^{A_1 B_1}...(\gamma_\Ke^{-1})^{A_r B_r}\phi_{A_1...A_r}\phi_{B_1...B_r}\right)^{\f12} .$$
\item Given a smooth real-valued function $\phi$ on $\mathcal M_{Kerr}$ and $i\in \mathbb N$, we define $|(\nab_\Ke)^i\phi|_{\gamma_\Ke}$ by considering $(\nab_\Ke)^i\phi$ as a rank $i$ $S_{u,\ub}$-tangent tensor field, where $\nab_\Ke$ is the Levi--Civita connection induced by $\gamma_{\Ke}$ (see also Definition \ref{def.slashed}).
\end{enumerate}
\end{definition}

\begin{proposition}\label{b.Kerr.bounds}
$b_\Ke^{\phi_*}$ is a uniformly bounded smooth function of $\sin^2\th_*$ and $b_{\Ke}^{\th_*}=0$. In addition, $b_\Ke$ satisfies the following estimates 
\begin{equation}\label{b.Kerr.bounds.1}
\sum_{0\leq i\leq I} |(\nab_\Ke)^i b_\Ke |_{\gamma_\Ke} \ls_{I} 1.
\end{equation}
Moreover, the coordinate $\f{1}{\sin (2\th_*)}\f{\rd }{\rd\th_*}$ derivative of $b_\Ke$ satisfies better estimates\footnote{Note that this fact is important because in the Christoffel symbol for $\nab_4$, it is precisely the coordinate derivative of $b$ that appears, see \eqref{nab4.def}.} near the horizons so that for every $k\in \mathbb N\cup \{0\}$
\begin{equation}\label{D.b.improved}
\left|\left(\f{1}{\sin (2\th_*)}\f{\rd }{\rd\th_*}\right)^k b_\Ke^{\phi_*} \right|\ls_k |\Delta|^{\min\{k,1\}}.
\end{equation}
\end{proposition}
\begin{proof}
That $b_\Ke^{\phi_*}$ is a uniformly bounded smooth function of $\sin^2\th_*$, $b_{\Ke}^{\th_*}=0$ and the bound \eqref{b.Kerr.bounds.1} holds all follow from the expressions of $b_\Ke$ in \eqref{Kerr.metric.comp} and the estimates in Propositions~\ref{Kerr.pd.est} and \ref{r.th.higher.order}.

Finally, \eqref{D.b.improved} follows from additionally taking into account the argument leading to \eqref{D.b}.
\end{proof}

Next, we prove some upper and lower bounds on $\Om_\Ke^2$. We note explicitly that $\Om_\Ke^2\to 0$ both when $r\to r_-$ and $r\to r_+$ (which corresponds to the event horizon and the Cauchy horizon respectively, see Section \ref{sec.def.horizon}). Moreover, when $(r-r_-)$ is small, $\Om_\Ke^2$ is decreasing as $\ub \to \infty$ and increasing as $u\to -\infty$; while when $(r_+-r)$ is small, $\Om_\Ke^2$ is increasing as $\ub\to \infty$ and decreasing as $u\to-\infty$.
\begin{proposition}\label{Om.Kerr.bounds}
To the \underline{future} of the hypersurface $\{(u,\ub,\th_*,\phi_*):u+\ub=C_R\}$, where $C_R$ is any\footnote{We explicitly mention that $C_R$ is allowed to be negative, positive or zero.} real number, the following upper and lower bounds hold for $\Om_\Ke^2$:
$$e^{-\f{r_+-r_-}{r_-^2+a^2}(u+\ub)} \ls_{C_R} \Om_\Ke^2\ls_{C_R} e^{-\f{r_+-r_-}{r_-^2+a^2}(u+\ub)},$$
where the implicit constants depend on $C_R$, in addition to $M$ and $a$. Similarly, to the \underline{past} of the hypersurface $\{(u,\ub,\th_*,\phi_*):u+\ub=C_R\}$, where $C_R$ is any real number, the following upper and lower bounds hold for $\Om_\Ke^2$:
$$e^{\f{r_+-r_-}{r_+^2+a^2}(u+\ub)} \ls_{C_R} \Om_\Ke^2\ls_{C_R} e^{\f{r_+-r_-}{r_+^2+a^2}(u+\ub)}.$$
Here, again the implicit constants depend on $C_R$, in addition to $M$ and $a$. Moreover, $|\Delta|$ is comparable to $\Om_\Ke^2$ everywhere in $\mathcal M_{Kerr}$ and therefore also satisfies the estimates above with $\Om_\Ke^2$ replaced by $|\Delta|$.
\end{proposition}
\begin{proof}
We first consider the region to the future of $\{u+\ub=C_R\}$, i.e.~$\{u+\ub\geq C_R\}$. By Proposition \ref{Kerr.pd.est}, we have
$$\f{\rd r}{\rd r_*}=\f{\Delta \sqrt{(r^2+a^2)^2-a^2\sin^2\th_*\Delta}}{(r^2+a^2)^2-a^2\sin^2\th\Delta}.$$
Notice that as $r\to r_-$, the right hand side satisfies
$$\f{\Delta \sqrt{(r^2+a^2)^2-a^2\sin^2\th_*\Delta}}{(r^2+a^2)^2-a^2\sin^2\th\Delta} = -\f{(r_+-r_-)(r-r_-)}{r_-^2+a^2}+O((r-r_-)^2).$$
Therefore, for every fixed $\th_*$, as $r_*\to \infty$ (i.e.~as $r\to r_-$), we have\footnote{Notice that these quantities are all negative.}
$$-\f{r_+-r_-}{r_-^2+a^2}r_*\ls \log (r-r_-)\ls -\f{r_+-r_-}{r_-^2+a^2}r_*.$$
Moreover, the constants can be chosen to be independent of $\th_*$. 
Therefore, there exists $(r_*)_0$ sufficiently large such that the following holds for $r_*\geq (r_*)_0$:
$$e^{-\f{r_+-r_-}{r_-^2+a^2}r_*}\ls (r-r_-)\ls e^{-\f{r_+-r_-}{r_-^2+a^2}r_*}.$$
Recall that $r_*=u+\ub$. Moreover, in the region $C_R\leq r_*\leq (r_*)_0$, the quantity $r-r_-$ is bounded below away from zero. Hence, by changing the implicit constants (depending on $C_R$) if necessary, we have
$$e^{-\f{r_+-r_-}{r_-^2+a^2}(u+\ub)}\ls (r-r_-)\ls e^{-\f{r_+-r_-}{r_-^2+a^2}(u+\ub)}.$$
To obtain the first conclusion of the proposition, we recall that $\Om_\Ke^2=-\f{\Delta}{R^2}$ and observe that 
$$(r-r_-)\ls (-\Delta) \ls (r-r_-),\quad \mbox{and}\quad 1\ls R^2\ls 1$$
in the region $\{u+\ub\geq C_R\}$. Finally, for the second assertion of the proposition, i.e.~the bound in the region $\{u+\ub\leq C_R\}$, notice that as $r\to r_+$,
$$\f{\Delta \sqrt{(r^2+a^2)^2-a^2\sin^2\th_*\Delta}}{(r^2+a^2)^2-a^2\sin^2\th\Delta}=-\f{(r_+-r_-)(r_+-r)}{r_+^2+a^2}+O((r_+-r)^2)$$
and the rest of the proof proceeds analogously as before.
\end{proof}

In addition to the above asymptotics of $\Om_\Ke^2$ as $r\to r_{\pm}$, we also need estimates for the derivatives of $\log\Om_\Ke$. Using the conventions in Definition~\ref{def.Kerr.norm}, we now state the estimates for the derivatives of $\log\Om_{\Ke}$ on Kerr spacetime.
\begin{proposition}\label{Om.der.Kerr.bounds}
For every $I\in \mathbb N$, the angular derivatives of $\log\Om_{\Ke}$ satisfy the following estimate everywhere in $\mathcal M_{Kerr}$:
$$\sum_{1\leq i\leq I}|(\nab_\Ke)^i\log\Om_{\Ke}|_{\gamma_\Ke}\ls_{I} 1$$
Here, the implicit constants depend on $I$, in addition to $M$ and $a$.

Moreover, for every $C_R\in \mathbb R$ and every $I\in \mathbb N$, the following holds for the $\f{\rd}{\rd r_*}$ derivative of $\log\Om_\Ke$ to the \underline{past} of the hypersurface $\{(u,\ub,\th_*,\phi_*):u+\ub=C_R\}$:
\begin{equation}\label{logOm.der.1}
\sum_{0\leq i\leq I} \left|(\nab_\Ke)^i(\f{\rd \log\Om_\Ke}{\rd r_*}-\f 12 \f{r_+-r_-}{r_+^2+a^2})\right|\ls_{I, C_R} |\Delta|
\end{equation}
and the following holds for the $\f{\rd}{\rd r_*}$ derivative of $\log\Om_\Ke$ to the \underline{future} of the hypersurface $\{(u,\ub,\th_*,\phi_*):u+\ub=C_R\}$:
\begin{equation}\label{logOm.der.2}
\sum_{0\leq i\leq I} \left|(\nab_\Ke)^i(\f{\rd \log\Om_\Ke}{\rd r_*}+\f 12 \f{r_+-r_-}{r_-^2+a^2})\right|\ls_{I, C_R} |\Delta|.
\end{equation}
In \eqref{logOm.der.1} and \eqref{logOm.der.2}, the implicit constants depend on $I$ and $C_R$, in addition to $M$ and $a$.
\end{proposition}
\begin{proof}
We first consider angular derivatives of $\log\Om_\Ke$. By \eqref{Kerr.metric.comp}, Propositions \ref{Kerr.pd.est} and \ref{r.th.higher.order}, it follows that $\nab^i\log\Om_\Ke$ is uniformly bounded on any compact set of $r_*$. To show that they are in fact uniformly bounded for all $r_*$, we note that $R^2$ is bounded above and below and that $\Delta$ is a function of $r$ alone (in the $(t,r,\th,\phi)$ coordinate system). Therefore, the estimates follow after using Propositions \ref{Kerr.pd.est} and \ref{r.th.higher.order} again.

For the $\f{\rd}{\rd r_*}$ derivatives, we need to compute more carefully. We first compute the limit of $\f{\rd}{\rd r_*}\log\Om_{\Ke}$ as $r_*\to \pm\infty$. Using the fact that $\Delta\to 0$ and $r\to r_{\pm}$, we have\footnote{Here, $\f{\rd}{\rd r}$ is to be understood in the $(r,\th)$-coordinate system}
$$\f{\rd}{\rd r}\restriction_{r=r_-} \f{\Delta}{R^2}=\f{1}{R^2}\f{\rd\Delta}{\rd r}\restriction_{r=r_-} =-\f{r_+-r_-}{R^2},\quad \f{\rd}{\rd r}\restriction_{r=r_+} \f{\Delta}{R^2}=\f{1}{R^2}\f{\rd\Delta}{\rd r}\restriction_{r=r_+} =\f{r_+-r_-}{R^2},$$
$$\f{\rd}{\rd\th}\restriction_{r=r_-} \f{\Delta}{R^2}=\f{\rd}{\rd\th}\restriction_{r=r_+} \f{\Delta}{R^2}=0.$$
Since $\f{1}{\Delta}\f{\rd r}{\rd r_*}\to \f{1}{r_{\pm}^2+a^2}$ as $r\to r_{\pm}$ and $|\f{\rd\th}{\rd r_*}|\ls |\Delta|\sin(2\th_*)$, this implies that 
$$\f{\rd}{\rd r_*}\restriction_{r=r_-}\log\Om_{\Ke}=\f 12 \f{R^2}{r_-^2+a^2}(\f{\rd}{\rd r} \f{\Delta}{R^2})\restriction_{r=r_-}=-\f 12\f{r_+-r_-}{r_-^2+a^2},$$ 
$$\f{\rd}{\rd r_*}\restriction_{r=r_+}\log\Om_{\Ke}=\f 12 \f{R^2}{r_+^2+a^2}(\f{\rd}{\rd r} \f{\Delta}{R^2})\restriction_{r=r_+}=\f 12\f{r_+-r_-}{r_+^2+a^2}.$$
Once we have computed the limits as $r_*\to \pm\infty$, the desired estimates follow from noting that $\Om_{\Ke}$ is a smooth function of $r$ and $\th$ and using $(r-r_-)\ls|\Delta|$ together with the bounds in Propositions \ref{Kerr.pd.est} and \ref{r.th.higher.order}.
\end{proof}

%Before we end this subsection, we also have the following bounds for $b_\Ke$. 
%\begin{proposition}\label{b.background}
%Using the convention introduced in Definition \ref{def.Kerr.norm}, $b_\Ke$ satisfies the following estimate
%\begin{equation}\label{b.background.1}
%|b_\Ke |_{\gamma_\Ke}\ls 1.
%\end{equation}
%Moreover, for every $I\in \mathbb N$, the angular derivatives of $b$ satisfy the following estimate:
%\begin{equation}\label{b.background.2}
%\sum_{1\leq i\leq I}|(\nab_\Ke)^i b_\Ke|_{\gamma_\Ke}\ls_I |\Delta|.
%\end{equation}
%\end{proposition}
%\begin{proof}
%\eqref{b.background.1} follows immediately from \eqref{Kerr.metric.comp}. The estimate \eqref{b.background.2}, which states that the estimate is stronger when at least one angular derivative is taken, follows from \eqref{Kerr.metric.comp} and \eqref{D.b}, together with the bounds in Propositions \ref{Kerr.pd.est} and \ref{r.th.higher.order}.
%\end{proof}

\subsection{The event horizon and the Cauchy horizon}\label{sec.def.horizon}

Equipped with the definitions of the null coordinate functions $u$ and $\ub$, we now define in this section the event horizon $\H$ and the Cauchy horizon $\CH$. In particular, we will define $\overline{\mathcal M}_{Kerr}=\mathcal M_{Kerr}\cup \H\cup\CH$ to be a manifold-with-corner.

\subsubsection{Another system of double null coordinates for $r$ close to $r_+$}\label{sec.coord.near.EH}

To define the event horizon and the Cauchy horizon, we introduce two more systems of double null coordinates.
First, define $u_{\H}$, $\ub_{\H}$ and $\phi_{*,\H}$ as follows
\begin{itemize}
\item Define $u_{\H}:\mathbb R\to (0,\infty)$ to be a smooth and strictly increasing function satisfying 
\begin{itemize}
\item $u_{\H}(u)=u$ for $u\geq 1$,
\item $u_{\H}(u)\to 0$ as $u\to -\infty$, and 
\item there exists $u_+\leq 1$ such that $\f{du_{\H}(u)}{du}=e^{\f{r_+-r_-}{r_+^2+a^2}u}$ for $u\leq u_+$.
\end{itemize}
Slightly abusing notation, let $u_{\H}:\mathcal M_{Kerr}\to (0,\infty)$ be defined as $u_{\H}(u)$, where $u:\mathcal M\to \mathbb R$ is the null variable we defined in \eqref{Kerr.u.ub.def}.
\item Similarly, define $\ub_{\H}:\mathbb R\to (0,\infty)$ to be a smooth and strictly increasing function such that 
\begin{itemize}
\item $\ub_{\H}(\ub)$ agrees with $\ub$ for $\ub\geq 1$, 
\item $\ub_{\H}\to 0$ as $\ub\to -\infty$, and 
\item there exists $\ub_+\leq 1$ such that $\f{d\ub_{\H}(\ub)}{d\ub}=e^{\f{r_+-r_-}{r_+^2+a^2}\ub}$ for $\ub\leq \ub_+$.
\end{itemize}
Slightly abusing notation as before, let $\ub_{\H}:\mathcal M_{Kerr}\to (0,\infty)$ be defined as $\ub_{\H}(\ub)$, where $\ub:\mathcal M\to \mathbb R$ is the null variable we defined in \eqref{Kerr.u.ub.def}.
\item Let $\phi_{*,\H}:\mathcal M_{Kerr} \to [0,2\pi)$ be defined by
$$\phi_{*,\H} = \phi_* -\f{4Mar_+}{(r_+^2+a^2)^2}\ub,$$
where the equality is to be understood modulo $2\pi$.
\end{itemize}

\subsubsection{Another system of double null coordinates for $r$ close to $r_-$}\label{sec.coord.near.CH}
Near $r_-$, we define yet another different system of coordinates as follows:
\begin{itemize}
\item Define $u_{\CH}:\mathbb R\to [1,\infty)$ to be a smooth and strictly increasing function satisfying 
\begin{itemize}
\item $u_{\CH}(u) = u$ for $u\leq -1$,
\item $u_{\CH}\to -0$ as $u\to \infty$, and
\item there exists $u_-\geq -1$ such that $\f{d u_{\CH}(u)}{du}=e^{-\f{r_+-r_-}{r_-^2+a^2}u}$ for $u\geq u_-$.
\end{itemize}
Slightly abusing notation, let $u_{\CH}:\mathcal M_{Kerr}\to (-\infty,0)$ be defined as $u_{\CH}(u)$, where $u:\mathcal M\to \mathbb R$ is the null variable we defined in \eqref{Kerr.u.ub.def}.
\item Similarly, define $\ub_{\CH}:\mathbb R\to [1,\infty)$ to be a smooth and strictly increasing function such that 
\begin{itemize}
\item $\ub_{\CH}(\ub) = \ub$ for $\ub\leq -1$, 
\item $\ub_{\CH}\to 0$ as $\ub\to \infty$, and 
\item there exists $\ub_-\geq -1$ such that $\f{d \ub_{\CH}(\ub)}{d\ub}=e^{-\f{r_+-r_-}{r_-^2+a^2}\ub}$ for $\ub\geq \ub_-$.
\end{itemize}
Slightly abusing notation as before, let $\ub_{\CH}:\mathcal M_{Kerr}\to (-\infty,0)$ be defined as $\ub_{\CH}(\ub)$, where $\ub:\mathcal M\to \mathbb R$ is the null variable we defined in \eqref{Kerr.u.ub.def}.
\item Let $\phi_{*,\CH}:\mathcal M_{Kerr} \to [0,2\pi)$ be defined by
$$\phi_{*,\CH} = \phi_* -\f{4Mar_-}{(r_-^2+a^2)^2}\ub,$$
where the equality is to be understood modulo $2\pi$.
\end{itemize}

\subsubsection{Definition of the event horizon}\label{sec.def.EH}

After the above preliminaries, we now define the event horizon. Using the coordinate chart $(u_{\H},\ub_{\H},\th_*,\phi_{*,\H})$ defined in Section~\ref{sec.coord.near.EH}, attach the boundary $\H=\H_1\cup\H_2\cup \mathbb S^2_{\H}$ where $\H_1=\{u_{\H}=0\}\times (0,\infty)\times\mathbb S^2$, $\H_2=(0,\infty)\times \{\ub_{\H}=0\}\times\mathbb S^2$ and $\mathbb S^2_{\H}$ be the $2$-sphere with $(u_{\H},\ub_{\H})=(0,0)$, which is a common boundary to both $\H_1$ and $\H_2$. We will call $\H$ the \emph{event horizon}.

It can be checked that that $g_{a,M}$ extends to a smooth Lorentzian metric on $\mathcal M_{Kerr} \cup \H$. %Since this is a straightforward 

\subsubsection{Definition of the Cauchy horizon} \label{sec.def.CH}

The Cauchy horizon $\CH$ is defined in a similar manner. Using the coordinate chart $(u_{\CH}, \ub_{\CH}, \th_*,\phi_{*,\CH})$ defined in Section~\ref{sec.coord.near.CH}, attach the boundary $\CH=\CH_1\cup\CH_2\cup\mathbb S^2_{\CH}$ where $\CH_1=(-\infty,0)\times \{\ub_{\CH}=0\}\times\mathbb S^2$, $\CH_2=\{u_{\CH}=0\} \times (-\infty,0)\times \mathbb S^2$ and $\mathbb S^2_{\CH}$ be the $2$-sphere with $(u_{\CH},\ub_{\CH})=(0,0)$ which is a common boundary to both $\CH_1$ and $\CH_2$. We will call $\CH$ the \emph{Cauchy horizon}.

As in Section~\ref{sec.def.EH}, it can be checked that the Kerr metric on $\mathcal M_{Kerr}$ extends to a smooth Lorentzian metric on $\mathcal M_{Kerr}\cup\CH$.

\subsubsection{$\overline{\mathcal M}_{Kerr}$ as a manifold-with-corner}

Combining all the discussions above, we now define $\overline{\mathcal M}_{Kerr}=\mathcal M_{Kerr}\cup \H\cup\CH$ as a manifold-with-corner. Moreover, the Kerr metric on $\mathcal M_{Kerr}$ extends smoothly to $\overline{\mathcal M}_{Kerr}$. 

\subsubsection{Metric components in regular coordinates at $\protect\CH$}

This paper is mostly concerned with (subsets of) the region $\{(u,\ub,\th_*,\phi_*):u+\ub\geq C_R,\,u\leq -1\}$ for some fixed (but arbitrary) $C_R\in \mathbb R$. In this region\footnote{Note that $u_{\CH}(u) = u$ for $u\leq -1$.}, the $(u,\ub_{\CH},\th_*,\phi_*)$ is a regular coordinate system up to the Cauchy horizon. Restricting moreover to $\ub\geq \ub_-$, we compute
$$ d\ub = e^{\f{r_+-r_-}{r^2_-+a^2}\ub} d\ub_{\CH},$$ 
$$d \phi_{*} = d\phi_{*,\CH} + \f{4Mar_-}{(r_-^2+a^2)^2} d\ub = d\phi_{*,\CH} + \f{4Mar_-}{(r_-^2+a^2)^2}e^{\f{r_+-r_-}{r^2_-+a^2}\ub} d\ub_{\CH}. $$
Therefore, in the region $u\leq -1$ and $\ub\geq \ub_-$, the metric in the $(u, \ub_{\CH},\th_*,\phi_{*,\CH})$ coordinate system takes the form
\begin{equation*}
\begin{split}
g_{a,M} =& -2\Omega_{\CH,\Ke}^2 (du_{\CH}\otimes d\ub_{\CH}+d\ub_{\CH}\otimes du_{\CH})\\
&+(\gamma_{\CH,\Ke})_{AB}(d\th_{\CH}^A-(b_{\CH,\Ke})^A d\ub_{\CH})\otimes(d\th_{\CH}^B-(b_{\CH,\Ke})^B d\ub_{\CH}),
\end{split}
\end{equation*}
where we have used the notation $(\th_{\CH,\Ke}^1, \th_{\CH,\Ke}^2) = (\th_*, \phi_{*,\CH})$, and the metric components are given by
\begin{equation}\label{Kerr.metric.CH}
\begin{split}
\Om_{\CH,\Ke}^2=-\frac{\Delta}{R^2}e^{\f{r_+-r_-}{r^2_-+a^2}\ub},\quad (b_{\CH,\Ke})^{\phi_{*,\CH}}&= \left(\frac{4Mar}{\Sigma R^2}-\f{4Mar_-}{(r_-^2+a^2)^2}\right)e^{\f{r_+-r_-}{r^2_-+a^2}\ub},\quad (b_{\CH,\Ke})^{\th_*}=0,\\
(\gamma_{\CH,\Ke})_{\phi_{*,\CH}\phi_{*,\CH}}=R^2\sin^2\th,\quad &(\gamma_\Ke)_{\th_*\th_*}=\frac{\ell^2}{R^2}+(\frac{\rd h}{\rd\th_*})^2 R^2\sin^2\th,\\
(\gamma_{\CH,\Ke})_{\th_*\phi_{*,\CH}}=&(\gamma_{\CH,\Ke})_{\phi_{*,\CH}\th_*}=(\frac{\rd h}{\rd\th_*}) R^2\sin^2\th.
\end{split}
\end{equation}
Let us explicitly note that $\frac{4Mar}{\Sigma R^2}\to \f{4Mar_-}{(r_-^2+a^2)^2}$ sufficiently fast as $\ub\to \infty$ so that $(b_{\CH,\Ke})^{\phi_{*,\CH}}$ is bounded.

\subsection{Estimates for the Ricci coefficients on Kerr spacetime}\label{sec.Kerr.RC.CC.est}

\begin{figure}
\centering{
\def\svgwidth{9pc}
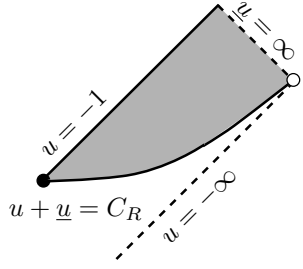}
\caption{The region of validity of our estimates}\label{appendnewlabel}
\end{figure}

We now discuss the Ricci coefficients and curvature components on Kerr spacetime adapted to the double null foliation introduced in Section \ref{Kerr.dbn}. Recall from Section~\ref{Kerr.dbn} that the Kerr metric takes the form \eqref{Kerr.doublenull} in the $(u,\ub,\th_*,\phi_*)$ coordinate system. In accordance with \eqref{L.Lb.def} and \eqref{e3.e4.def}, we then define
\begin{equation}\label{Kerr.equiv}
\Lb=\frac{\rd}{\rd u},\quad L=\frac{\rd}{\rd \ub}+\frac{4Mar}{\Sigma R^2}\frac{\rd}{\rd\phi_*}.
\end{equation}
and
$$e_3=\frac{\rd}{\rd u},\quad e_4=-\frac{R^2}{\Delta}\big(\frac{\rd}{\rd \ub}+\frac{4Mar}{\Sigma R^2}\frac{\rd}{\rd\phi_*}\big).$$
Notice that the vector fields $e_3$ and $e_4$ are regular near the part of the Cauchy horizon given formally by $\{(u,\ub,\th_*,\phi_*):\ub=\infty\}$, but are irregular either near $\{(u,\ub,\th_*,\phi_*):u=\infty\}$ or near the event horizon. For this reason that we confine our discussions in this section to the subset $\{(u,\ub,\th_*,\phi_*):u+\ub\geq C_R,\,u\leq -1\}$ of $\mathcal M_{Kerr}$; see Figure~\ref{appendnewlabel}.
. Note that our present paper indeed considers perturbations of the Kerr metric on (subsets of) such a set.

Given the vector fields $e_3$, $e_4$ and the double null foliation $(u,\ub)$, we define the Ricci coefficients and null curvature components according to \eqref{Ricci.coeff.def} and \eqref{curv.comp.def}. Using the above expressions of $e_3$ and $e_4$, we compute the Ricci coefficients using the following formulae\footnote{Here, we use the axisymmetry of the Kerr spacetime to suppress the $\frac{\rd}{\rd\phi_*}$ derivatives in the computation for $\chi_\Ke$.} (see Proposition \ref{metric.der.Ricci}):
\begin{equation}\label{chib.Ke}
(\chib_\Ke)_{AB}=\f 12 \f{\rd}{\rd u}(\gamma_\Ke)_{AB},
\end{equation}
\begin{equation}\label{chi.Ke}
(\chi_\Ke)_{AB}=-\f {R^2}{2\Delta} \left(\f{\rd}{\rd \ub}(\gamma_\Ke)_{AB}+\f{\rd}{\rd \th^A}\big(\frac{4Mar}{\Sigma R^2}\big)(\gamma_\Ke)_{\phi_* B}+\f{\rd}{\rd \th^B}\big(\frac{4Mar}{\Sigma R^2}\big)(\gamma_\Ke)_{\phi_* A}\right),
\end{equation}
\begin{equation}\label{zeta.Ke}
(\zeta_\Ke)^A=-\f{R^2}{4\Delta} \f{\rd}{\rd u}(b_\Ke)^A+(\gamma_{\Ke}^{-1})^{AB}\nab_B\log\Om_\Ke,
\end{equation}
\begin{equation}\label{omb.Ke}
\omb_\Ke=-\f{\rd}{\rd u}\log \Om_\Ke.
\end{equation}
The values of $\eta$ and $\etab$ can then be derived using the relations in \eqref{gauge.con} together with \eqref{Kerr.metric.comp} and \eqref{zeta.Ke}.

For the purpose of this paper, we will not use the precise formulae for the Ricci coefficients above, but instead it suffices to obtain estimates for them. We will only be concerned with bounds for the Ricci coefficients in the region $\{(u,\ub,\th_*,\phi_*):u+\ub\geq C_R,\,u\leq -1\}$ for some fixed (but arbitrary) $C_R\in \mathbb R$; see Figure~\ref{appendnewlabel} for a depiction of the region of interest. (Notice that the bounds we derive below degenerate as $C_R\to -\infty$.) \textbf{\emph{In the remaining propositions of this section (Propositions~\ref{Kerr.Ricci.bound}, \ref{Kerr.der.Ricci.bound}, \ref{prop.Kerr.K}, \ref{Kerr.curv.est} and \ref{Kerr.high.quantity}), the implicit constants in the inequalities $\ls$ depend on $M$, $a$ and $C_R$.}} We have estimates for the Ricci coefficients as well as for their higher order angular covariant derivatives. For clarity of the exposition, let us first give the estimates for the Ricci coefficients themselves:
\begin{proposition}\label{Kerr.Ricci.bound}
In the region $\{(u,\ub,\th_*,\phi_*):u+\ub\geq C_R,\,u\leq -1\}$, the components of $\chib_\Ke$ satisfy the following estimates:
$$|(\chib_\Ke)_{\phi_*\phi_*}|\ls_{C_R} |\Delta|\sin^2\th_*,\quad |(\chib_\Ke)_{\th_*\phi_*}|\ls_{C_R} |\Delta|\sin^3\th_*,\quad |(\chib_\Ke)_{\th_*\th_*}|\ls_{C_R} |\Delta|;$$
the components of $\chib_\Ke$ satisfy the following estimates:
$$|(\chi_\Ke)_{\phi_*\phi_*}|\ls_{C_R} \sin^2\th_*,\quad |(\chi_\Ke)_{\th_*\phi_*}|\ls_{C_R} \sin^3\th_*,\quad |(\chi_\Ke)_{\th_*\th_*}|\ls_{C_R} 1;$$
the components of $\zeta_\Ke$ verify 
$$|(\zeta_\Ke)_{\phi_*}|\ls_{C_R} \sin^2\th_*,\quad |(\zeta_\Ke)_{\th_*}|\ls_{C_R} \sin\th_*;$$
%$\nab_{\f{\rd}{\rd\phi_*}}\log \Om_\Ke=0$ and $\nab_{\f{\rd}{\rd\th_*}}\log \Om_\Ke$ obeys
%$$|\nab_{\f{\rd}{\rd\th_*}}\log \Om_\Ke|\ls_{C_R} 1;$$
and $\omb_\Ke$ verifies
$$\left|\omb_\Ke-\f 12\f{r_+-r_-}{r_-^2+a^2}\right|\ls_{C_R} |\Delta|.$$
In particular, using the convention in Definition \ref{def.Kerr.norm}, this implies
\begin{equation}\label{Ricci.bd.norm}
\begin{split}
|\chib_\Ke|_{\gamma_\Ke}\ls_{C_R} |\Delta|,&\quad |\chi_\Ke|_{\gamma_\Ke}\ls_{C_R} 1,\quad |\zeta_\Ke|_{\gamma_\Ke}\ls_{C_R} 1,\\
|\nab\log\Om_\Ke|_{\gamma_\Ke}\ls_{C_R} 1,&\quad \left|\omb_\Ke-\f 12\f{r_+-r_-}{r_-^2+a^2}\right|\ls_{C_R} |\Delta|.
\end{split}
\end{equation}
\end{proposition}
\begin{proof}
In the proof, we suppress the explicit dependence on $C_R$ in the notations.

{\bf Estimates for $\chib_\Ke$.} The estimates for $\chib_\Ke$ can be inferred from the formula \eqref{chib.Ke} and the estimates for the $\f{\rd}{\rd r_*}$ derivatives of components of $\gamma_\Ke$ in Proposition~\ref{gamma.Kerr.bounds}.

{\bf Estimates for $\chi_\Ke$.} We then turn to the estimates for $\chi_\Ke$. Recall from \eqref{chi.Ke} that $\chi_\Ke$ is given by
$$(\chi_\Ke)_{AB}=-\f {R^2}{2\Delta} \left(\f{\rd}{\rd \ub}(\gamma_\Ke)_{AB}+\f{\rd}{\rd \th^A}\big(\frac{4Mar}{\Sigma R^2}\big)(\gamma_\Ke)_{\phi_* B}+\f{\rd}{\rd \th^B}\big(\frac{4Mar}{\Sigma R^2}\big)(\gamma_\Ke)_{\phi_* A}\right).$$
The term $-\f {R^2}{2\Delta}\f{\rd}{\rd \ub}(\gamma_\Ke)_{AB}$ can be dealt with using the estimates for the $\f{\rd}{\rd r_*}$ derivatives of components of $\gamma_\Ke$ in Proposition~\ref{gamma.Kerr.bounds} (in a similar manner as the estimates for $\chib_\Ke$, except for an extra factor of $\f {R^2}{2\Delta}$).

For the term $\f{\rd}{\rd \th^A}\big(\frac{4Mar}{\Sigma R^2}\big)(\gamma_\Ke)_{\phi_* B}$, notice that this is only non-vanishing when $\th^A=\th_*$. Therefore, using \eqref{D.b}, we have
$$\left|\f{\rd}{\rd \ub}(\gamma_\Ke)_{\th_*\th_*}+2\f{\rd}{\rd \th_*}\big(\frac{4Mar}{\Sigma R^2}\big)(\gamma_\Ke)_{\phi_* \th_*}\right|\ls |\Delta|,$$
$$\left|\f{\rd}{\rd \ub}(\gamma_\Ke)_{\th_*\phi_*}+\f{\rd}{\rd \th_*}\big(\frac{4Mar}{\Sigma R^2}\big)(\gamma_\Ke)_{\phi_* \phi_*}\right|\ls |\Delta|\sin^3\th,$$
$$\left|\f{\rd}{\rd \ub}(\gamma_\Ke)_{\phi_*\phi_*}\right|\ls |\Delta|\sin^2\th,$$
from which the desired bounds for $\chi_\Ke$ follow.

{\bf Estimates for $\zeta_\Ke$.} We next estimate $\zeta_\Ke$. We first control $\zeta^{\phi_*}$ and $\zeta^{\th_*}$. By \eqref{zeta.Ke}, we need to control $-\f{R^2}{4\Delta} \f{\rd}{\rd u}(b_\Ke)^A=-\f{R^2}{4\Delta} \f{\rd}{\rd r_*}(b_\Ke)^A$ and $(\gamma_{\Ke}^{-1})^{AB}\nab_B\log\Om_\Ke$. For the first term, by \eqref{Kerr.metric.comp} we only have the $\phi_*$-component, which satisfies the bound
$$\left|\f{R^2}{4\Delta} \f{\rd}{\rd r_*}(b_\Ke)^{\phi_*}\right|=\left|\f{R^2}{4\Delta} \left(\f{\rd}{\rd r_*}\f{4Mar}{\Sigma R^2}\right)\right|\ls 1$$
thanks to the estimates in Proposition \ref{Kerr.pd.est}. For the second term, differentiating the expression of $\Om_\Ke$ in \eqref{Kerr.metric.comp}, using the expression of $\gamma_\Ke^{-1}$ in \eqref{gamma.Ke.inverse} and applying the estimates in Proposition \ref{Kerr.pd.est}, we obtain
$$|(\gamma_{\Ke}^{-1})^{\th_*\th_*}\nab_{\th_*}\log\Om_\Ke|\ls \sin\th_*,\quad |(\gamma_{\Ke}^{-1})^{\phi_*\th_*}\nab_{\th_*}\log\Om_\Ke|\ls \sin\th_*.$$
Combining the above observations yields $|\zeta^{\phi_*}|\ls 1$ and $|\zeta^{\th_*}|\ls \sin\th_*$, which in turn gives the desired estimates for indices down after using the properties of $\gamma_\Ke$ in \eqref{Kerr.metric.comp}.

{\bf Estimates for $\omb_\Ke$.} To derive the estimates for $\omb_{\Ke}$, recall from Proposition \ref{metric.der.Ricci} that
$$\omb_\Ke=-\f{\rd}{\rd u}\log\Om_\Ke=-\f{\rd}{\rd r_*}\log\Om_\Ke,$$
where the last equality holds due since $\log\Om_\Ke$ is independent of $t$. The desired estimate then follows thanks to \eqref{logOm.der.2}.

{\bf Estimates for the Ricci coefficients with respect to the $\gamma_\Ke$ norm.} Finally, to obtain \eqref{Ricci.bd.norm}, observe that by \eqref{gamma.Ke.inverse}, we have
$$|(\gamma_\Ke^{-1})^{\phi_*\phi_*}|\ls \f{1}{\sin^2\th_*},\quad |(\gamma_\Ke^{-1})^{\th_*\th_*}|+|(\gamma_\Ke^{-1})^{\th_*\phi_*}|\ls 1.$$
The conclusion then follows from the previous bounds together with the estimates in Proposition \ref{Om.der.Kerr.bounds}.
\end{proof}

\begin{remark}
Note that on the Cauchy horizon $\CH$, we have $\omb_\Ke=\f 12\f{r_+-r_-}{r_-^2+a^2}>0$. The positivity of $\omb_\Ke$ in a neighbourhood of the Cauchy horizon is the origin of the {\bf local blue shift effect}, which plays an important role in the present work.
\end{remark}

We also have the following bounds for the higher derivatives of the Ricci coefficients:
\begin{proposition}\label{Kerr.der.Ricci.bound}
Using the convention introduced in Definition \ref{def.Kerr.norm}, for every $I\in \mathbb N$, the following bounds hold for the angular covariant derivatives of the Ricci coefficients on Kerr spacetime in the region $\{(u,\ub,\th_*,\phi_*):u+\ub\geq C_R,\,u\leq -1\}$:
$$\sum_{0\leq i\leq I}|(\nab_\Ke)^i\chib_\Ke|_{\gamma_\Ke}\ls_{C_R,I} |\Delta|,\quad \sum_{0\leq i\leq I}(|(\nab_\Ke)^i\chi_\Ke|_{\gamma_\Ke}+|(\nab_\Ke)^i\zeta_\Ke|_{\gamma_\Ke}+|(\nab_\Ke)^{i+1}\log\Om_\Ke|_{\gamma_\Ke})\ls_{C_R,I} 1,$$
$$\sum_{1\leq i\leq I}|(\nab_\Ke)^i \omb_\Ke|_{\gamma_\Ke}\ls_{C_R,I} |\Delta|.$$
Here, the implicit constants depend on $C_R$ and $I$, in addition to $M$ and $a$.
\end{proposition}
\begin{remark}
We emphasise that the bounds for $\nab_\Ke^i\omb_\Ke$ hold only when $i\geq 1$. For $i=0$, we instead have the bound in Proposition \ref{Kerr.Ricci.bound}, so that $\omb_{\Ke}$ is a non-zero constant at the Cauchy horizon. 
\end{remark}
%\begin{remark}[Estimates for the null curvature components]
%Once we have the estimates for the Ricci coefficients in Propositions \ref{Kerr.Ricci.bound} and \ref{Kerr.der.Ricci.bound}, by virtue of the fact that the Kerr metric is a solution to the Einstein vacuum equations, we can also obtain bounds for the null curvature components. This will be carried out later in Proposition \ref{Kerr.curv.est}.
%\end{remark}

The Gauss curvature of the spheres $S_{u,\ub}$ can be bounded as follows:
\begin{proposition}\label{prop.Kerr.K}
For every $I\in \mathbb N$, the following bound holds for the angular covariant derivative of the Gauss curvature of the $2$-spheres $S_{u,\ub}$:
$$\sum_{0\leq i\leq I}\sup_{u,\ub}|(\nab_\Ke)^i K_{\Ke}|_{\gamma_\Ke}\ls_{I} 1,$$
where the implicit constant depends on $I$, in addition to $M$ and $a$.
\end{proposition}
\begin{proof}
According to \eqref{Kerr.metric.comp} and the estimates in Propositions \ref{Kerr.pd.est} and \ref{r.th.higher.order}, for every $u$ and $\ub$, the metric $\gamma_\Ke$ is smooth. Moreover, $\gamma_\Ke$ also has a smooth limit as $u+\ub\to\pm\infty$. Since for every $i\in \mathbb N\cup\{0\}$, $(\nab_\Ke)^i K_{\Ke}$ is intrinsic to the $2$-spheres $S_{u,\ub}$, i.e.~it only depends on $\gamma_\Ke$ and its tangential derivatives along the $2$-spheres, we obtain the desired estimate.
\end{proof}

In the next proposition, we will obtain a uniform upper bound on the isoperimetric constant on $S_{u,\ub}$. Before we proceed, we recall the definition of the isoperimetric constant. Given a Riemannian $2$-surface $(S,\gamma)$, we define the isoperimetric constant ${\bf I}(S,\gamma)$ by
$${\bf I}(S,\gamma)=\sup_{\substack{U\\\partial U \in C^1}} \f{\min\{\mbox{Area}(U),\mbox{Area}(U^c)\}}{(\mbox{Perimeter}(\partial U))^2}.$$
The importance of the isoperimetric constant in the context of this paper lies in the fact that it is intimately tied to the constants in the Sobolev embedding theorems, c.f.~Propositions~\ref{L4.prelim}, \ref{Linfty.prelim}, \ref{isoperimetric}, \ref{Sobolev}. We now state the bound we have for the isoperimetric constant on the Kerr spacetime:
\begin{proposition}\label{isoperimetric.Kerr}
There exists a constant ${\bf I}_{\Ke}>0$ (depending on $M$ and $a$) such that
$$\sup_{u,\ub}{\bf I}(S_{u,\ub},\gamma_{\Ke})\leq {\bf I}_{\Ke}.$$
\end{proposition}
\begin{proof}
First, notice that given two $2$-spheres $S_{u_1,\ub_1}$ and $S_{u_2,\ub_2}$ such that $u_1+\ub_1=u_2+\ub_2=r_*$, by identifying the value of the coordinate functions $\th_*$ and $\phi_*$ on $S_{u_1,\ub_1}$ and $S_{u_2,\ub_2}$, it is easy to see that $S_{u_1,\ub_1}$ and $S_{u_2,\ub_2}$ are in fact isometric. In particular, the isoperimetric constant is only a function of the $r_*$ value associated to the $2$-sphere. Second, since the isoperimetric constant is continuous in $r_*$, it is uniformly bounded for $r_*\in K$, where $K$ is any fixed compact set of $\mathbb R$. It therefore suffices to show that ${\bf I}(S_{u,\ub},\gamma_{\Ke})$ does not go to infinity as $r_*\to \pm \infty$, which is obvious since $\gamma_\Ke$ has a smooth limit as $r_*\to \infty$ or $r_*\to -\infty$.
\end{proof}

\subsection{Estimates for the null curvature components on Kerr spacetime}\label{sec.Kerr.cuv}

Since the Kerr metric is a solution to the Einstein vacuum equations \eqref{INTROvace}, after using \eqref{null.str3} in Section~\ref{seceqn}, Propositions \ref{Kerr.Ricci.bound} and \ref{Kerr.der.Ricci.bound} imply estimates for the renormalised null curvature components $\betab_\Ke$, $\sigmac_{\Ke}$ and $\beta_{\Ke}$. These can be obtained after writing the Einstein vacuum equations in the double null coordinate gauge. More precisely, in the region $\{(u,\ub,\th_*,\phi_*):u+\ub\geq C_R,\,u\leq -1\}$ for $C_R\in \mathbb R$, we have the following bounds:
\begin{proposition}\label{Kerr.curv.est}
Using the convention introduced in Definition \ref{def.Kerr.norm}, for every $C_R\in \mathbb R$ and $I\in \mathbb N$, the angular covariant derivatives of the null curvature components on Kerr spacetime satisfy the following estimates in the region $\{(u,\ub,\th_*,\phi_*):u+\ub\geq C_R,\,u\leq -1\}$:
$$\sum_{0\leq i\leq I}|(\nab_\Ke)^i\betab_\Ke|_{\gamma_\Ke}\ls_{C_R,I} |\Delta|,\quad \sum_{0\leq i\leq I}|(\nab_\Ke)^i\sigmac_\Ke|_{\gamma_\Ke}\ls_{C_R,I} 1,\quad \sum_{0\leq i\leq I}|(\nab_\Ke)^i\beta_\Ke|_{\gamma_\Ke}\ls_{C_R,I} 1.$$
Here, the implicit constants depend on $C_R$ and $I$, in addition to $M$ and $a$.
\end{proposition}
\begin{proof}
This is a direct consequence of the fact that the Kerr metric is a solution to the Einstein vacuum equations and the application of the constraint equations \eqref{null.str3}.
\end{proof}

Combining the estimates in Propositions~\ref{Kerr.der.Ricci.bound}, \ref{prop.Kerr.K} and \ref{Kerr.curv.est}, we also immediately infer the following bounds for $\mu_\Ke$, $\mub_\Ke$ and $\ombs_\Ke$ (cf.~\eqref{top.quantities.def}):
\begin{proposition}\label{Kerr.high.quantity}
Using the convention introduced in Definition \ref{def.Kerr.norm}, for every $C_R\in \mathbb R$ and $I\in \mathbb N$, the following estimates hold in the region $\{(u,\ub,\th_*,\phi_*):u+\ub\geq C_R,\,u\leq -1\}$:
$$\sum_{0\leq i\leq I}|(\nab_\Ke)^i\mu_\Ke|_{\gamma_\Ke}\ls_{C_R,I} 1,\quad \sum_{0\leq i\leq I}|(\nab_\Ke)^i\mub_\Ke|_{\gamma_\Ke}\ls_{C_R,I} 1,\quad \sum_{0\leq i\leq I}|(\nab_\Ke)^i\ombs_\Ke|_{\gamma_\Ke}\ls_{C_R,I} |\Delta|.$$
Here, the implicit constants depend on $C_R$ and $I$, in addition to $M$ and $a$.
\end{proposition}

\subsection{Spacelike hypersurfaces in $\mathcal M_{Kerr}$}\label{sec.Kerr.spacelike}

In Theorem~\ref{PRWTO}, the initial data are posed on a spacelike hypersurface. In order to compare the initial data with that of Kerr, we briefly discuss in this subsection the induced metric and second fundamental form on spacelike hypersurfaces in $\mathcal M_{Kerr}$. We will moreover do so in a coordinate system that is easy to compare with the double null coordinate system that we discussed earlier.

More precisely, we consider a smooth embedded spacelike hypersurface (with boundary)\footnote{In this subsection, $\Sigma_\Ke$ is used to denote the spacelike hypersurface defined below. This is not to be confused with the $\Sigma$ in \eqref{BL}.} $\Sigma_\Ke\subset \mathcal M_{Kerr}$ given by%\footnote{It is for convenience that we consider the specific spacelike hypersurfaces below. In fact, our methods would show that Theorem~\ref{PRWTO} also holds for perturbations of more general spacelike hypersurfaces
%$$\Sigma_\Ke=\{(u,\ub,\th_*,\phi_*): u-f(\ub,\th^1,\th^2)=0,\, \ub\geq 1\},$$ 
%as long as $f$ is smooth and satisfies
%$$(L_\Ke f+\Om_\Ke^2|\nab f|_{\gamma_{\Ke}}^2)(\ub,\th^1,\th^2)<-c< 0$$
%and 
%$$\ub+f(\ub,\th^1,\th^2)\geq C_R'.$$
%This would however be computationally more difficult, and we therefore only stick to the more specific spacelike hypersurfaces.
%}
$$\Sigma_\Ke=\{(u,\ub): u+\ub=C_R,\, \ub\geq 1\}\times\mathbb S^2,$$ 
where $C_R\in \mathbb R$ is a constant (but otherwise arbitrary), $u$ and $\ub$ are the null coordinates defined in Section~\ref{Kerr.dbn} and $\mathbb S^2$ corresponds to $2$-spheres of constant $u$, $\ub$ values.

Let $(\th^1,\th^2)$ be an arbitrary local coordinates system on $\mathbb S^2$ satisfying $\Lb \th^A=0$. (For instance $(\th^1=\th_*,\th^2=\phi_*)$ is one such system of local coordinates.) Introducing the functions $\tau\doteq \ub$, $\th^A_{\Sigma}\doteq \th^A$ for $A=1,2$, it is easy to see that $(\tau,\th^1_{\Sigma},\th^2_{\Sigma})$ forms a system of local coordinates on $\Sigma_\Ke$. In this coordinate system, the induced metric $\hat{g}_\Ke$ on $\Sigma_\Ke$ is given by
\begin{equation}\label{Kerr.SigmaKe}
\hat{g}_{\Ke}= \Phi_{\Ke} d\tau\otimes d\tau+ (w_{\Ke})_A (d\tau\otimes d\th_{\Sigma}^A+d\th_{\Sigma}^A\otimes d\tau)+(\gamma_{\Ke})_{AB} d\th_{\Sigma}^A\otimes d\th_{\Sigma}^B,
\end{equation}
where
$$\Phi_{\Ke}=(\gamma_{\Ke})_{AB} (b_{\Ke})^A (b_{\Ke})^B+4\Om_{\Ke}^2,\quad (w_{\Ke})_A=-(\gamma_{\Ke})_{AB} (b_\Ke)^B$$
and $(\gamma_{\Ke})_{AB}$ and $(b_\Ke)^A$ are as in \eqref{Kerr.metric.comp}.

Notice that\footnote{Here, we have used the convention that $\f{\rd}{\rd\tau}$, $\f{\rd}{\rd\th^1_\Sigma}$ and $\f{\rd}{\rd\th^2_\Sigma}$ are coordinate vector fields in the $(\tau,\th_\Sigma^1,\th_\Sigma^2)$-coordinate system on $\Sigma_\Ke$.}
$$\left\{\f{\rd}{\rd\tau}=-\f{\rd}{\rd u}+\f{\rd}{\rd\ub}, \;\f{\rd}{\rd\th^1_\Sigma}=\f{\rd}{\rd\th^1}, \; \f{\rd}{\rd\th^2_\Sigma}=\f{\rd}{\rd\th^2} \right\}$$
forms a local basis of the tangent space of $\Sigma_\Ke$.

Next, we claim that the vector field
\begin{equation*}
\begin{split}
N_\Ke=&\f{1}{2\Om_{\Ke}}\left( \f{\rd}{\rd u}+\f{\rd}{\rd\ub}+ b_{\Ke}^B\f{\rd}{\rd\th^B}\right)=\f{1}{2\Om_{\Ke}}\left(e_3+\Om_\Ke^2 e_4 \right)
\end{split}
\end{equation*}
is the future-directed unit normal to $\Sigma_\Ke$. To see this, we first check that it is orthogonal to the tangent vectors of $\Sigma_\Ke$, i.e.
\begin{equation*}
\begin{split}
g_\Ke\left(N_\Ke,\f{\rd}{\rd\th^A}\right) = 0,\quad g_\Ke\left(N_\Ke,-\f{\rd}{\rd u}+\f{\rd}{\rd\ub}\right) = \f{1}{2\Om_\Ke} g_\Ke(e_3+\Om^2_{\Ke} e_4,-e_3+\Om_\Ke^2 e_4)=0.
\end{split}
\end{equation*}
We then compute the norm of $N_\Ke$:
\begin{equation*}
\begin{split}
g_\Ke(N_\Ke,N_\Ke)
=&\f{1}{4\Om_{\Ke}^2}2\Om_\Ke^2 g(e_3,e_4) = -1.
\end{split}
\end{equation*}
Therefore, $N_\Ke$ is indeed the future-directed unit normal to $\Sigma_\Ke$. In particular, this also shows that $\Sigma_\Ke$ is a spacelike hypersurface. 

On the other hand, define $n_\Ke$ by
\begin{equation*}
\begin{split}
n_\Ke=&\f{1}{2\Om_\Ke}\left(\f{\rd}{\rd\tau}-(\gamma^{-1})^{AB} (w_{\Ke})_A \f{\rd}{\rd\th_\Sigma^B}\right)=\f{1}{2\Om_\Ke}\left(-\f{\rd}{\rd u}+\f{\rd}{\rd\ub}+b_{\Ke}^B\f{\rd}{\rd\th^B}\right),
\end{split}
\end{equation*}
which, by a similar computation as above, is tangent to $\Sigma_\Ke$ and is normal (with respect to the induced Riemannian metric on $\Sigma_\Ke$) to the $2$-spheres given by constant-$\tau$. By the formulae for $N_\Ke$ and $n_\Ke$, we obtain

\begin{equation}\label{Lb.Kerr.def}
\Lb_\Ke = \f{\rd}{\rd u}= \Om_\Ke(N_\Ke-n_\Ke)
\end{equation}
and
\begin{equation}\label{L.Kerr.def}
L_\Ke =\f{\rd}{\rd\ub}+(b_\Ke)^A\f{\rd}{\rd\th^A}=\Om_\Ke(N_\Ke+n_\Ke).
\end{equation}
Here, $L_\Ke$ and $\Lb_\Ke$ are the vector fields $L$ and $\Lb$ defined in \eqref{Kerr.equiv}, where we have used the subscript ${ }_\Ke$ to emphasise that its definition depends on the Kerr metric.

\bibliographystyle{DLplain}
\bibliography{KerrInterior}

\end{document}